%% file: main.tex
\documentclass[sigconf,screen,nonacm]{acmart}
\usepackage{amsmath,amsfonts}
\usepackage{algorithmic}
\usepackage{graphicx}
\usepackage{textcomp}
\usepackage{algorithm}
\usepackage{algorithmic}
\usepackage{multirow}
\usepackage{makecell}
\usepackage[switch]{lineno}
\usepackage[inline,shortlabels]{enumitem}
\usepackage{verbatim}
\usepackage{color, colortbl}
\usepackage{hhline}
\usepackage{xspace}

\usepackage{microtype}
\usepackage{xcolor}
\usepackage{hyperref}

\usepackage{shortcuts}

\newcommand*\ghidra{\textrm{Ghidra}\xspace}

\newcommand*\LLVM{\textrm{LLVM}\xspace}

\newcommand*\nop{\textrm{nop}\xspace}

\newcommand*\rose{\textrm{ROSE}\xspace}
\newcommand*\radare{\textrm{Radare2}\xspace}
\newcommand*\bap{\textrm{BAP}\xspace}

\copyrightyear{2020}
\acmYear{2020}
\setcopyright{rightsretained}
\acmConference[FEAST '20]{2020 Workshop on Forming an Ecosystem Around Software
  Transformation}{November 13, 2020}{Virtual Event, USA}
\acmBooktitle{2020 Workshop on Forming an Ecosystem Around Software
  Transformation (FEAST '20), November 13, 2020, Virtual Event, USA}
\acmPrice{}
\acmDOI{10.1145/3411502.3418429} %
\acmISBN{978-1-4503-8089-8/20/11}


\begin{document}

\author{Kaiyuan Li}
\email{squid@cmu.edu}
\affiliation{
  \institution{Carnegie Mellon University, CyLab}
}
\author{Maverick Woo}
\email{pooh@cmu.edu}
\affiliation{
  \institution{Carnegie Mellon University, CyLab}
}
\author{Limin Jia}
\email{liminjia@cmu.edu}
\affiliation{
  \institution{Carnegie Mellon University, CyLab}
}

\title{On the Generation of Disassembly Ground Truth and the Evaluation of
  Disassemblers}

\titlenote{%
  This manuscript is a revised and extended version of our publication that
  first appeared in the 2020 Workshop on Forming an Ecosystem Around Software
  Transformation (FEAST '20), November 13, 2020.
  The revisions are all wording changes at the sentence level, and the
  extensions appear in~\S\ref{sec:gt:implementation} and
  Appendix~\ref{sec:in-depth-Linux}.
}

\setlength{\tabcolsep}{2pt}

\input{evaluation.data.tex}
\input{appendix.data.tex}

\input{abstract.tex}

\maketitle %

\input{introduction.tex}

\input{related.tex}

\input{groundtruth.tex}

\input{evaluation.tex}

\input{futurework.tex}
\input{conclusion.tex}

\bibliographystyle{ACM-Reference-Format}
\bibliography{Pangine,misc}

\appendix
\input{appendix.tex}

\end{document}

%% file: evaluation.data.tex
\newcommand{\WinXsfWinLsRecall}{0\xspace}
\newcommand{\WinXsfWinBapRecall}{0\xspace}
\newcommand{\WinXsfWinGhiRecall}{3\xspace}
\newcommand{\WinXsfWinRdaRecall}{0\xspace}
\newcommand{\WinXsfWinRseRecall}{0\xspace}
\newcommand{\WinXsfWinLsPrecision}{0\xspace}
\newcommand{\WinXsfWinBapPrecision}{0\xspace}
\newcommand{\WinXsfWinGhiPrecision}{3\xspace}
\newcommand{\WinXsfWinRdaPrecision}{0\xspace}
\newcommand{\WinXsfWinRsePrecision}{0\xspace}
\newcommand{\WinXsfWinLsFone}{0\xspace}
\newcommand{\WinXsfWinBapFone}{0\xspace}
\newcommand{\WinXsfWinGhiFone}{3\xspace}
\newcommand{\WinXsfWinRdaFone}{0\xspace}
\newcommand{\WinXsfWinRseFone}{0\xspace}
\newcommand{\WinXsfAvgLsRecall}{0.99999\xspace}
\newcommand{\WinXsfAvgBapRecall}{0.71179\xspace}
\newcommand{\WinXsfAvgGhiRecall}{0.92938\xspace}
\newcommand{\WinXsfAvgRdaRecall}{0.77705\xspace}
\newcommand{\WinXsfAvgRseRecall}{0.75949\xspace}
\newcommand{\WinXsfAvgLsPrecision}{0.98316\xspace}
\newcommand{\WinXsfAvgBapPrecision}{0.99998\xspace}
\newcommand{\WinXsfAvgGhiPrecision}{1.00000\xspace}
\newcommand{\WinXsfAvgRdaPrecision}{0.99999\xspace}
\newcommand{\WinXsfAvgRsePrecision}{0.99964\xspace}
\newcommand{\WinXsfAvgLsFone}{0.99150\xspace}
\newcommand{\WinXsfAvgBapFone}{0.83163\xspace}
\newcommand{\WinXsfAvgGhiFone}{0.96340\xspace}
\newcommand{\WinXsfAvgRdaFone}{0.87454\xspace}
\newcommand{\WinXsfAvgRseFone}{0.86317\xspace}
\newcommand{\WinXsfFailLsFone}{0\xspace}
\newcommand{\WinXsfFailBapFone}{0\xspace}
\newcommand{\WinXsfFailGhiFone}{0\xspace}
\newcommand{\WinXsfFailRdaFone}{0\xspace}
\newcommand{\WinXsfFailRseFone}{0\xspace}
\newcommand{\WinXttWinLsRecall}{0\xspace}
\newcommand{\WinXttWinBapRecall}{0\xspace}
\newcommand{\WinXttWinGhiRecall}{2\xspace}
\newcommand{\WinXttWinRdaRecall}{0\xspace}
\newcommand{\WinXttWinRseRecall}{1\xspace}
\newcommand{\WinXttWinLsPrecision}{0\xspace}
\newcommand{\WinXttWinBapPrecision}{1\xspace}
\newcommand{\WinXttWinGhiPrecision}{1\xspace}
\newcommand{\WinXttWinRdaPrecision}{1\xspace}
\newcommand{\WinXttWinRsePrecision}{0\xspace}
\newcommand{\WinXttWinLsFone}{0\xspace}
\newcommand{\WinXttWinBapFone}{0\xspace}
\newcommand{\WinXttWinGhiFone}{3\xspace}
\newcommand{\WinXttWinRdaFone}{0\xspace}
\newcommand{\WinXttWinRseFone}{0\xspace}
\newcommand{\WinXttAvgLsRecall}{0.99989\xspace}
\newcommand{\WinXttAvgBapRecall}{0.66079\xspace}
\newcommand{\WinXttAvgGhiRecall}{0.95017\xspace}
\newcommand{\WinXttAvgRdaRecall}{0.83991\xspace}
\newcommand{\WinXttAvgRseRecall}{0.91219\xspace}
\newcommand{\WinXttAvgLsPrecision}{0.97509\xspace}
\newcommand{\WinXttAvgBapPrecision}{0.99965\xspace}
\newcommand{\WinXttAvgGhiPrecision}{0.99981\xspace}
\newcommand{\WinXttAvgRdaPrecision}{0.99971\xspace}
\newcommand{\WinXttAvgRsePrecision}{0.99905\xspace}
\newcommand{\WinXttAvgLsFone}{0.98733\xspace}
\newcommand{\WinXttAvgBapFone}{0.79565\xspace}
\newcommand{\WinXttAvgGhiFone}{0.97436\xspace}
\newcommand{\WinXttAvgRdaFone}{0.91287\xspace}
\newcommand{\WinXttAvgRseFone}{0.95365\xspace}
\newcommand{\WinXttFailLsFone}{0\xspace}
\newcommand{\WinXttFailBapFone}{0\xspace}
\newcommand{\WinXttFailGhiFone}{0\xspace}
\newcommand{\WinXttFailRdaFone}{0\xspace}
\newcommand{\WinXttFailRseFone}{0\xspace}
\newcommand{\LnxXsfWinLsRecall}{0\xspace}
\newcommand{\LnxXsfWinBapRecall}{0\xspace}
\newcommand{\LnxXsfWinGhiRecall}{2\xspace}
\newcommand{\LnxXsfWinRdaRecall}{1\xspace}
\newcommand{\LnxXsfWinRseRecall}{27\xspace}
\newcommand{\LnxXsfWinLsPrecision}{0\xspace}
\newcommand{\LnxXsfWinBapPrecision}{4\xspace}
\newcommand{\LnxXsfWinGhiPrecision}{18\xspace}
\newcommand{\LnxXsfWinRdaPrecision}{0\xspace}
\newcommand{\LnxXsfWinRsePrecision}{8\xspace}
\newcommand{\LnxXsfWinLsFone}{0\xspace}
\newcommand{\LnxXsfWinBapFone}{0\xspace}
\newcommand{\LnxXsfWinGhiFone}{2\xspace}
\newcommand{\LnxXsfWinRdaFone}{1\xspace}
\newcommand{\LnxXsfWinRseFone}{27\xspace}
\newcommand{\LnxXsfAvgLsRecall}{1.00000\xspace}
\newcommand{\LnxXsfAvgBapRecall}{0.80288\xspace}
\newcommand{\LnxXsfAvgGhiRecall}{0.92657\xspace}
\newcommand{\LnxXsfAvgRdaRecall}{0.83889\xspace}
\newcommand{\LnxXsfAvgRseRecall}{0.97905\xspace}
\newcommand{\LnxXsfAvgLsPrecision}{0.99959\xspace}
\newcommand{\LnxXsfAvgBapPrecision}{0.99998\xspace}
\newcommand{\LnxXsfAvgGhiPrecision}{0.99965\xspace}
\newcommand{\LnxXsfAvgRdaPrecision}{0.99988\xspace}
\newcommand{\LnxXsfAvgRsePrecision}{0.99999\xspace}
\newcommand{\LnxXsfAvgLsFone}{0.99980\xspace}
\newcommand{\LnxXsfAvgBapFone}{0.89066\xspace}
\newcommand{\LnxXsfAvgGhiFone}{0.96173\xspace}
\newcommand{\LnxXsfAvgRdaFone}{0.91234\xspace}
\newcommand{\LnxXsfAvgRseFone}{0.98941\xspace}
\newcommand{\LnxXsfFailLsFone}{0\xspace}
\newcommand{\LnxXsfFailBapFone}{0\xspace}
\newcommand{\LnxXsfFailGhiFone}{0\xspace}
\newcommand{\LnxXsfFailRdaFone}{11\xspace}
\newcommand{\LnxXsfFailRseFone}{0\xspace}
\newcommand{\LnxXttWinLsRecall}{0\xspace}
\newcommand{\LnxXttWinBapRecall}{6\xspace}
\newcommand{\LnxXttWinGhiRecall}{6\xspace}
\newcommand{\LnxXttWinRdaRecall}{0\xspace}
\newcommand{\LnxXttWinRseRecall}{18\xspace}
\newcommand{\LnxXttWinLsPrecision}{0\xspace}
\newcommand{\LnxXttWinBapPrecision}{20\xspace}
\newcommand{\LnxXttWinGhiPrecision}{5\xspace}
\newcommand{\LnxXttWinRdaPrecision}{1\xspace}
\newcommand{\LnxXttWinRsePrecision}{4\xspace}
\newcommand{\LnxXttWinLsFone}{0\xspace}
\newcommand{\LnxXttWinBapFone}{6\xspace}
\newcommand{\LnxXttWinGhiFone}{6\xspace}
\newcommand{\LnxXttWinRdaFone}{0\xspace}
\newcommand{\LnxXttWinRseFone}{18\xspace}
\newcommand{\LnxXttAvgLsRecall}{1.00000\xspace}
\newcommand{\LnxXttAvgBapRecall}{0.82989\xspace}
\newcommand{\LnxXttAvgGhiRecall}{0.75014\xspace}
\newcommand{\LnxXttAvgRdaRecall}{0.85280\xspace}
\newcommand{\LnxXttAvgRseRecall}{0.97241\xspace}
\newcommand{\LnxXttAvgLsPrecision}{0.99938\xspace}
\newcommand{\LnxXttAvgBapPrecision}{0.99997\xspace}
\newcommand{\LnxXttAvgGhiPrecision}{0.99813\xspace}
\newcommand{\LnxXttAvgRdaPrecision}{0.99994\xspace}
\newcommand{\LnxXttAvgRsePrecision}{0.99998\xspace}
\newcommand{\LnxXttAvgLsFone}{0.99969\xspace}
\newcommand{\LnxXttAvgBapFone}{0.90703\xspace}
\newcommand{\LnxXttAvgGhiFone}{0.85655\xspace}
\newcommand{\LnxXttAvgRdaFone}{0.92053\xspace}
\newcommand{\LnxXttAvgRseFone}{0.98601\xspace}
\newcommand{\LnxXttFailLsFone}{0\xspace}
\newcommand{\LnxXttFailBapFone}{0\xspace}
\newcommand{\LnxXttFailGhiFone}{0\xspace}
\newcommand{\LnxXttFailRdaFone}{0\xspace}
\newcommand{\LnxXttFailRseFone}{0\xspace}

\newcommand{\SqlUsrTimeBap}{166.65\xspace}
\newcommand{\SqlUsrTimeGhi}{214.43\xspace}
\newcommand{\SqlUsrTimeRda}{47.59\xspace}
\newcommand{\SqlUsrTimeRse}{274.96\xspace}
\newcommand{\SqlSysTimeBap}{1.41\xspace}
\newcommand{\SqlSysTimeGhi}{4.45\xspace}
\newcommand{\SqlSysTimeRda}{0.19\xspace}
\newcommand{\SqlSysTimeRse}{4.97\xspace}
\newcommand{\SqlClkTimeBap}{168.21\xspace}
\newcommand{\SqlClkTimeGhi}{101.03\xspace}
\newcommand{\SqlClkTimeRda}{47.79\xspace}
\newcommand{\SqlClkTimeRse}{65.03\xspace}
\newcommand{\SqlSpaceBap}{9,221,200\xspace}
\newcommand{\SqlSpaceGhi}{2,035,936\xspace}
\newcommand{\SqlSpaceRda}{477,228\xspace}
\newcommand{\SqlSpaceRse}{4,592,024\xspace}

\newcommand{\XsfELFBinSize}{980\xspace}
\newcommand{\XsfELFGTSize}{2,290\xspace}
\newcommand{\XsfELFMaxBin}{40,920\xspace}
\newcommand{\XsfELFMinBin}{58\xspace}
\newcommand{\XsfELFInsn}{85,839,363\xspace}
\newcommand{\XsfELFFunc}{595,917\xspace}
\newcommand{\XsfELFIndJ}{319,657\xspace}
\newcommand{\XsfELFMnem}{404\xspace}

\newcommand{\XttELFBinSize}{955\xspace}
\newcommand{\XttELFGTSize}{2,280\xspace}
\newcommand{\XttELFMaxBin}{19,276\xspace}
\newcommand{\XttELFMinBin}{62\xspace}
\newcommand{\XttELFInsn}{83,623,983\xspace}
\newcommand{\XttELFFunc}{574,454\xspace}
\newcommand{\XttELFIndJ}{225,273\xspace}
\newcommand{\XttELFMnem}{389\xspace}

\newcommand{\XsfCOFFBinSize}{46\xspace}
\newcommand{\XsfCOFFGTSize}{109\xspace}
\newcommand{\XsfCOFFMaxBin}{7,463\xspace}
\newcommand{\XsfCOFFMinBin}{152\xspace}
\newcommand{\XsfCOFFInsn}{3,542,001\xspace}
\newcommand{\XsfCOFFFunc}{37,850\xspace}
\newcommand{\XsfCOFFIndJ}{26,671\xspace}
\newcommand{\XsfCOFFMnem}{255\xspace}

\newcommand{\XttCOFFBinSize}{46\xspace}
\newcommand{\XttCOFFGTSize}{105\xspace}
\newcommand{\XttCOFFMaxBin}{7,394\xspace}
\newcommand{\XttCOFFMinBin}{152\xspace}
\newcommand{\XttCOFFInsn}{3,947,966\xspace}
\newcommand{\XttCOFFFunc}{38,449\xspace}
\newcommand{\XttCOFFIndJ}{25,544\xspace}
\newcommand{\XttCOFFMnem}{210\xspace}

%% file: appendix.data.tex
\newcommand{\XsfGccFOoSzpGT}{20410\xspace}
\newcommand{\XsfGccFOoSzpLsRecall}{1\xspace}
\newcommand{\XsfGccFOoSzpLsPrecision}{1\xspace}
\newcommand{\XsfGccFOoSzpLsFone}{1\xspace}
\newcommand{\XsfGccFOoSzpBapRecall}{0.976\xspace}
\newcommand{\XsfGccFOoSzpBapPrecision}{1\xspace}
\newcommand{\XsfGccFOoSzpBapFone}{0.988\xspace}
\newcommand{\XsfGccFOoSzpGhiRecall}{1\xspace}
\newcommand{\XsfGccFOoSzpGhiPrecision}{1\xspace}
\newcommand{\XsfGccFOoSzpGhiFone}{1\xspace}
\newcommand{\XsfGccFOoSzpRdaRecall}{1\xspace}
\newcommand{\XsfGccFOoSzpRdaPrecision}{1\xspace}
\newcommand{\XsfGccFOoSzpRdaFone}{1\xspace}
\newcommand{\XsfGccFOoSzpRseRecall}{1\xspace}
\newcommand{\XsfGccFOoSzpRsePrecision}{1\xspace}
\newcommand{\XsfGccFOoSzpRseFone}{1\xspace}
\newcommand{\XsfGccFOoCapGT}{315956\xspace}
\newcommand{\XsfGccFOoCapLsRecall}{1\xspace}
\newcommand{\XsfGccFOoCapLsPrecision}{1\xspace}
\newcommand{\XsfGccFOoCapLsFone}{1\xspace}
\newcommand{\XsfGccFOoCapBapRecall}{0.460\xspace}
\newcommand{\XsfGccFOoCapBapPrecision}{1\xspace}
\newcommand{\XsfGccFOoCapBapFone}{0.631\xspace}
\newcommand{\XsfGccFOoCapGhiRecall}{1\xspace}
\newcommand{\XsfGccFOoCapGhiPrecision}{1\xspace}
\newcommand{\XsfGccFOoCapGhiFone}{1\xspace}
\newcommand{\XsfGccFOoCapRdaRecall}{0.899\xspace}
\newcommand{\XsfGccFOoCapRdaPrecision}{1\xspace}
\newcommand{\XsfGccFOoCapRdaFone}{0.947\xspace}
\newcommand{\XsfGccFOoCapRseRecall}{1\xspace}
\newcommand{\XsfGccFOoCapRsePrecision}{1\xspace}
\newcommand{\XsfGccFOoCapRseFone}{1\xspace}
\newcommand{\XsfGccFOoExmGT}{181906\xspace}
\newcommand{\XsfGccFOoExmLsRecall}{1\xspace}
\newcommand{\XsfGccFOoExmLsPrecision}{1\xspace}
\newcommand{\XsfGccFOoExmLsFone}{1\xspace}
\newcommand{\XsfGccFOoExmBapRecall}{0.860\xspace}
\newcommand{\XsfGccFOoExmBapPrecision}{1\xspace}
\newcommand{\XsfGccFOoExmBapFone}{0.925\xspace}
\newcommand{\XsfGccFOoExmGhiRecall}{1\xspace}
\newcommand{\XsfGccFOoExmGhiPrecision}{1\xspace}
\newcommand{\XsfGccFOoExmGhiFone}{1\xspace}
\newcommand{\XsfGccFOoExmRdaRecall}{0.962\xspace}
\newcommand{\XsfGccFOoExmRdaPrecision}{1.000\xspace}
\newcommand{\XsfGccFOoExmRdaFone}{0.981\xspace}
\newcommand{\XsfGccFOoExmRseRecall}{1.000\xspace}
\newcommand{\XsfGccFOoExmRsePrecision}{1\xspace}
\newcommand{\XsfGccFOoExmRseFone}{1.000\xspace}
\newcommand{\XsfGccFOoLgtGT}{38930\xspace}
\newcommand{\XsfGccFOoLgtLsRecall}{1\xspace}
\newcommand{\XsfGccFOoLgtLsPrecision}{1\xspace}
\newcommand{\XsfGccFOoLgtLsFone}{1\xspace}
\newcommand{\XsfGccFOoLgtBapRecall}{0.875\xspace}
\newcommand{\XsfGccFOoLgtBapPrecision}{1\xspace}
\newcommand{\XsfGccFOoLgtBapFone}{0.933\xspace}
\newcommand{\XsfGccFOoLgtGhiRecall}{1\xspace}
\newcommand{\XsfGccFOoLgtGhiPrecision}{1\xspace}
\newcommand{\XsfGccFOoLgtGhiFone}{1\xspace}
\newcommand{\XsfGccFOoLgtRdaRecall}{0.999\xspace}
\newcommand{\XsfGccFOoLgtRdaPrecision}{1.000\xspace}
\newcommand{\XsfGccFOoLgtRdaFone}{0.999\xspace}
\newcommand{\XsfGccFOoLgtRseRecall}{1\xspace}
\newcommand{\XsfGccFOoLgtRsePrecision}{1\xspace}
\newcommand{\XsfGccFOoLgtRseFone}{1\xspace}
\newcommand{\XsfGccFOoBzpGT}{23842\xspace}
\newcommand{\XsfGccFOoBzpLsRecall}{1\xspace}
\newcommand{\XsfGccFOoBzpLsPrecision}{1\xspace}
\newcommand{\XsfGccFOoBzpLsFone}{1\xspace}
\newcommand{\XsfGccFOoBzpBapRecall}{0.786\xspace}
\newcommand{\XsfGccFOoBzpBapPrecision}{1\xspace}
\newcommand{\XsfGccFOoBzpBapFone}{0.880\xspace}
\newcommand{\XsfGccFOoBzpGhiRecall}{1\xspace}
\newcommand{\XsfGccFOoBzpGhiPrecision}{1\xspace}
\newcommand{\XsfGccFOoBzpGhiFone}{1\xspace}
\newcommand{\XsfGccFOoBzpRdaRecall}{0.888\xspace}
\newcommand{\XsfGccFOoBzpRdaPrecision}{1\xspace}
\newcommand{\XsfGccFOoBzpRdaFone}{0.941\xspace}
\newcommand{\XsfGccFOoBzpRseRecall}{1\xspace}
\newcommand{\XsfGccFOoBzpRsePrecision}{1\xspace}
\newcommand{\XsfGccFOoBzpRseFone}{1\xspace}
\newcommand{\XsfGccFOoGccGT}{1292369\xspace}
\newcommand{\XsfGccFOoGccLsRecall}{1\xspace}
\newcommand{\XsfGccFOoGccLsPrecision}{1\xspace}
\newcommand{\XsfGccFOoGccLsFone}{1\xspace}
\newcommand{\XsfGccFOoGccBapRecall}{0.713\xspace}
\newcommand{\XsfGccFOoGccBapPrecision}{1\xspace}
\newcommand{\XsfGccFOoGccBapFone}{0.832\xspace}
\newcommand{\XsfGccFOoGccGhiRecall}{0.997\xspace}
\newcommand{\XsfGccFOoGccGhiPrecision}{1.000\xspace}
\newcommand{\XsfGccFOoGccGhiFone}{0.998\xspace}
\newcommand{\XsfGccFOoGccRdaRecall}{0.884\xspace}
\newcommand{\XsfGccFOoGccRdaPrecision}{1\xspace}
\newcommand{\XsfGccFOoGccRdaFone}{0.938\xspace}
\newcommand{\XsfGccFOoGccRseRecall}{0.999\xspace}
\newcommand{\XsfGccFOoGccRsePrecision}{1\xspace}
\newcommand{\XsfGccFOoGccRseFone}{0.999\xspace}
\newcommand{\XsfGccFOoGzpGT}{12539\xspace}
\newcommand{\XsfGccFOoGzpLsRecall}{1\xspace}
\newcommand{\XsfGccFOoGzpLsPrecision}{1\xspace}
\newcommand{\XsfGccFOoGzpLsFone}{1\xspace}
\newcommand{\XsfGccFOoGzpBapRecall}{0.991\xspace}
\newcommand{\XsfGccFOoGzpBapPrecision}{1\xspace}
\newcommand{\XsfGccFOoGzpBapFone}{0.996\xspace}
\newcommand{\XsfGccFOoGzpGhiRecall}{1\xspace}
\newcommand{\XsfGccFOoGzpGhiPrecision}{1\xspace}
\newcommand{\XsfGccFOoGzpGhiFone}{1\xspace}
\newcommand{\XsfGccFOoGzpRdaRecall}{1\xspace}
\newcommand{\XsfGccFOoGzpRdaPrecision}{1\xspace}
\newcommand{\XsfGccFOoGzpRdaFone}{1\xspace}
\newcommand{\XsfGccFOoGzpRseRecall}{0.999\xspace}
\newcommand{\XsfGccFOoGzpRsePrecision}{1\xspace}
\newcommand{\XsfGccFOoGzpRseFone}{0.999\xspace}
\newcommand{\XsfGccFOoOggGT}{58285\xspace}
\newcommand{\XsfGccFOoOggLsRecall}{1\xspace}
\newcommand{\XsfGccFOoOggLsPrecision}{1\xspace}
\newcommand{\XsfGccFOoOggLsFone}{1\xspace}
\newcommand{\XsfGccFOoOggBapRecall}{0.978\xspace}
\newcommand{\XsfGccFOoOggBapPrecision}{1\xspace}
\newcommand{\XsfGccFOoOggBapFone}{0.989\xspace}
\newcommand{\XsfGccFOoOggGhiRecall}{1\xspace}
\newcommand{\XsfGccFOoOggGhiPrecision}{1\xspace}
\newcommand{\XsfGccFOoOggGhiFone}{1\xspace}
\newcommand{\XsfGccFOoOggRdaRecall}{0.998\xspace}
\newcommand{\XsfGccFOoOggRdaPrecision}{1\xspace}
\newcommand{\XsfGccFOoOggRdaFone}{0.999\xspace}
\newcommand{\XsfGccFOoOggRseRecall}{1\xspace}
\newcommand{\XsfGccFOoOggRsePrecision}{1\xspace}
\newcommand{\XsfGccFOoOggRseFone}{1\xspace}
\newcommand{\XsfGccFOoNgxGT}{164509\xspace}
\newcommand{\XsfGccFOoNgxLsRecall}{1\xspace}
\newcommand{\XsfGccFOoNgxLsPrecision}{1\xspace}
\newcommand{\XsfGccFOoNgxLsFone}{1\xspace}
\newcommand{\XsfGccFOoNgxBapRecall}{0.968\xspace}
\newcommand{\XsfGccFOoNgxBapPrecision}{1\xspace}
\newcommand{\XsfGccFOoNgxBapFone}{0.984\xspace}
\newcommand{\XsfGccFOoNgxGhiRecall}{1.000\xspace}
\newcommand{\XsfGccFOoNgxGhiPrecision}{1\xspace}
\newcommand{\XsfGccFOoNgxGhiFone}{1.000\xspace}
\newcommand{\XsfGccFOoNgxRdaRecall}{0.992\xspace}
\newcommand{\XsfGccFOoNgxRdaPrecision}{1.000\xspace}
\newcommand{\XsfGccFOoNgxRdaFone}{0.996\xspace}
\newcommand{\XsfGccFOoNgxRseRecall}{1\xspace}
\newcommand{\XsfGccFOoNgxRsePrecision}{1\xspace}
\newcommand{\XsfGccFOoNgxRseFone}{1\xspace}
\newcommand{\XsfGccFOoSshGT}{139163\xspace}
\newcommand{\XsfGccFOoSshLsRecall}{1\xspace}
\newcommand{\XsfGccFOoSshLsPrecision}{1\xspace}
\newcommand{\XsfGccFOoSshLsFone}{1\xspace}
\newcommand{\XsfGccFOoSshBapRecall}{0.951\xspace}
\newcommand{\XsfGccFOoSshBapPrecision}{1\xspace}
\newcommand{\XsfGccFOoSshBapFone}{0.975\xspace}
\newcommand{\XsfGccFOoSshGhiRecall}{1\xspace}
\newcommand{\XsfGccFOoSshGhiPrecision}{1\xspace}
\newcommand{\XsfGccFOoSshGhiFone}{1\xspace}
\newcommand{\XsfGccFOoSshRdaRecall}{0.998\xspace}
\newcommand{\XsfGccFOoSshRdaPrecision}{1.000\xspace}
\newcommand{\XsfGccFOoSshRdaFone}{0.999\xspace}
\newcommand{\XsfGccFOoSshRseRecall}{0.980\xspace}
\newcommand{\XsfGccFOoSshRsePrecision}{1\xspace}
\newcommand{\XsfGccFOoSshRseFone}{0.990\xspace}
\newcommand{\XsfGccFOoPcrGT}{5962\xspace}
\newcommand{\XsfGccFOoPcrLsRecall}{1\xspace}
\newcommand{\XsfGccFOoPcrLsPrecision}{1\xspace}
\newcommand{\XsfGccFOoPcrLsFone}{1\xspace}
\newcommand{\XsfGccFOoPcrBapRecall}{0.910\xspace}
\newcommand{\XsfGccFOoPcrBapPrecision}{1\xspace}
\newcommand{\XsfGccFOoPcrBapFone}{0.953\xspace}
\newcommand{\XsfGccFOoPcrGhiRecall}{1\xspace}
\newcommand{\XsfGccFOoPcrGhiPrecision}{1\xspace}
\newcommand{\XsfGccFOoPcrGhiFone}{1\xspace}
\newcommand{\XsfGccFOoPcrRdaRecall}{0.991\xspace}
\newcommand{\XsfGccFOoPcrRdaPrecision}{1\xspace}
\newcommand{\XsfGccFOoPcrRdaFone}{0.996\xspace}
\newcommand{\XsfGccFOoPcrRseRecall}{1\xspace}
\newcommand{\XsfGccFOoPcrRsePrecision}{1\xspace}
\newcommand{\XsfGccFOoPcrRseFone}{1\xspace}
\newcommand{\XsfGccFOoSqlGT}{225474\xspace}
\newcommand{\XsfGccFOoSqlLsRecall}{1\xspace}
\newcommand{\XsfGccFOoSqlLsPrecision}{1\xspace}
\newcommand{\XsfGccFOoSqlLsFone}{1\xspace}
\newcommand{\XsfGccFOoSqlBapRecall}{0.880\xspace}
\newcommand{\XsfGccFOoSqlBapPrecision}{1\xspace}
\newcommand{\XsfGccFOoSqlBapFone}{0.936\xspace}
\newcommand{\XsfGccFOoSqlGhiRecall}{1\xspace}
\newcommand{\XsfGccFOoSqlGhiPrecision}{1\xspace}
\newcommand{\XsfGccFOoSqlGhiFone}{1\xspace}
\newcommand{\XsfGccFOoSqlRdaRecall}{0.991\xspace}
\newcommand{\XsfGccFOoSqlRdaPrecision}{1\xspace}
\newcommand{\XsfGccFOoSqlRdaFone}{0.995\xspace}
\newcommand{\XsfGccFOoSqlRseRecall}{1\xspace}
\newcommand{\XsfGccFOoSqlRsePrecision}{1\xspace}
\newcommand{\XsfGccFOoSqlRseFone}{1\xspace}
\newcommand{\XsfGccFOoVimGT}{638606\xspace}
\newcommand{\XsfGccFOoVimLsRecall}{1\xspace}
\newcommand{\XsfGccFOoVimLsPrecision}{1\xspace}
\newcommand{\XsfGccFOoVimLsFone}{1\xspace}
\newcommand{\XsfGccFOoVimBapRecall}{0.947\xspace}
\newcommand{\XsfGccFOoVimBapPrecision}{1\xspace}
\newcommand{\XsfGccFOoVimBapFone}{0.973\xspace}
\newcommand{\XsfGccFOoVimGhiRecall}{1\xspace}
\newcommand{\XsfGccFOoVimGhiPrecision}{1\xspace}
\newcommand{\XsfGccFOoVimGhiFone}{1\xspace}
\newcommand{\XsfGccFOoVimRdaRecall}{0.989\xspace}
\newcommand{\XsfGccFOoVimRdaPrecision}{1.000\xspace}
\newcommand{\XsfGccFOoVimRdaFone}{0.995\xspace}
\newcommand{\XsfGccFOoVimRseRecall}{1\xspace}
\newcommand{\XsfGccFOoVimRsePrecision}{1\xspace}
\newcommand{\XsfGccFOoVimRseFone}{1\xspace}
\newcommand{\XsfGccFOoVsfGT}{24188\xspace}
\newcommand{\XsfGccFOoVsfLsRecall}{1\xspace}
\newcommand{\XsfGccFOoVsfLsPrecision}{1\xspace}
\newcommand{\XsfGccFOoVsfLsFone}{1\xspace}
\newcommand{\XsfGccFOoVsfBapRecall}{0.997\xspace}
\newcommand{\XsfGccFOoVsfBapPrecision}{1\xspace}
\newcommand{\XsfGccFOoVsfBapFone}{0.999\xspace}
\newcommand{\XsfGccFOoVsfGhiRecall}{1\xspace}
\newcommand{\XsfGccFOoVsfGhiPrecision}{1\xspace}
\newcommand{\XsfGccFOoVsfGhiFone}{1\xspace}
\newcommand{\XsfGccFOoVsfRdaRecall}{0.987\xspace}
\newcommand{\XsfGccFOoVsfRdaPrecision}{1.000\xspace}
\newcommand{\XsfGccFOoVsfRdaFone}{0.993\xspace}
\newcommand{\XsfGccFOoVsfRseRecall}{1.000\xspace}
\newcommand{\XsfGccFOoVsfRsePrecision}{1\xspace}
\newcommand{\XsfGccFOoVsfRseFone}{1.000\xspace}
\newcommand{\XsfGccFOaSzpGT}{12215\xspace}
\newcommand{\XsfGccFOaSzpLsRecall}{1\xspace}
\newcommand{\XsfGccFOaSzpLsPrecision}{1\xspace}
\newcommand{\XsfGccFOaSzpLsFone}{1\xspace}
\newcommand{\XsfGccFOaSzpBapRecall}{0.972\xspace}
\newcommand{\XsfGccFOaSzpBapPrecision}{1\xspace}
\newcommand{\XsfGccFOaSzpBapFone}{0.986\xspace}
\newcommand{\XsfGccFOaSzpGhiRecall}{1\xspace}
\newcommand{\XsfGccFOaSzpGhiPrecision}{1\xspace}
\newcommand{\XsfGccFOaSzpGhiFone}{1\xspace}
\newcommand{\XsfGccFOaSzpRdaRecall}{0.999\xspace}
\newcommand{\XsfGccFOaSzpRdaPrecision}{1\xspace}
\newcommand{\XsfGccFOaSzpRdaFone}{1.000\xspace}
\newcommand{\XsfGccFOaSzpRseRecall}{1\xspace}
\newcommand{\XsfGccFOaSzpRsePrecision}{1\xspace}
\newcommand{\XsfGccFOaSzpRseFone}{1\xspace}
\newcommand{\XsfGccFOaCapGT}{222585\xspace}
\newcommand{\XsfGccFOaCapLsRecall}{1\xspace}
\newcommand{\XsfGccFOaCapLsPrecision}{1\xspace}
\newcommand{\XsfGccFOaCapLsFone}{1\xspace}
\newcommand{\XsfGccFOaCapBapRecall}{0.419\xspace}
\newcommand{\XsfGccFOaCapBapPrecision}{1\xspace}
\newcommand{\XsfGccFOaCapBapFone}{0.590\xspace}
\newcommand{\XsfGccFOaCapGhiRecall}{0.875\xspace}
\newcommand{\XsfGccFOaCapGhiPrecision}{1\xspace}
\newcommand{\XsfGccFOaCapGhiFone}{0.933\xspace}
\newcommand{\XsfGccFOaCapRdaRecall}{0.860\xspace}
\newcommand{\XsfGccFOaCapRdaPrecision}{1\xspace}
\newcommand{\XsfGccFOaCapRdaFone}{0.925\xspace}
\newcommand{\XsfGccFOaCapRseRecall}{1.000\xspace}
\newcommand{\XsfGccFOaCapRsePrecision}{1\xspace}
\newcommand{\XsfGccFOaCapRseFone}{1.000\xspace}
\newcommand{\XsfGccFOaExmGT}{137718\xspace}
\newcommand{\XsfGccFOaExmLsRecall}{1\xspace}
\newcommand{\XsfGccFOaExmLsPrecision}{1\xspace}
\newcommand{\XsfGccFOaExmLsFone}{1\xspace}
\newcommand{\XsfGccFOaExmBapRecall}{0.842\xspace}
\newcommand{\XsfGccFOaExmBapPrecision}{1.000\xspace}
\newcommand{\XsfGccFOaExmBapFone}{0.914\xspace}
\newcommand{\XsfGccFOaExmGhiRecall}{0.998\xspace}
\newcommand{\XsfGccFOaExmGhiPrecision}{1\xspace}
\newcommand{\XsfGccFOaExmGhiFone}{0.999\xspace}
\newcommand{\XsfGccFOaExmRdaRecall}{0.971\xspace}
\newcommand{\XsfGccFOaExmRdaPrecision}{1.000\xspace}
\newcommand{\XsfGccFOaExmRdaFone}{0.985\xspace}
\newcommand{\XsfGccFOaExmRseRecall}{1.000\xspace}
\newcommand{\XsfGccFOaExmRsePrecision}{1\xspace}
\newcommand{\XsfGccFOaExmRseFone}{1.000\xspace}
\newcommand{\XsfGccFOaLgtGT}{25408\xspace}
\newcommand{\XsfGccFOaLgtLsRecall}{1\xspace}
\newcommand{\XsfGccFOaLgtLsPrecision}{1\xspace}
\newcommand{\XsfGccFOaLgtLsFone}{1\xspace}
\newcommand{\XsfGccFOaLgtBapRecall}{0.872\xspace}
\newcommand{\XsfGccFOaLgtBapPrecision}{1\xspace}
\newcommand{\XsfGccFOaLgtBapFone}{0.932\xspace}
\newcommand{\XsfGccFOaLgtGhiRecall}{1\xspace}
\newcommand{\XsfGccFOaLgtGhiPrecision}{1\xspace}
\newcommand{\XsfGccFOaLgtGhiFone}{1\xspace}
\newcommand{\XsfGccFOaLgtRdaRecall}{0.985\xspace}
\newcommand{\XsfGccFOaLgtRdaPrecision}{1.000\xspace}
\newcommand{\XsfGccFOaLgtRdaFone}{0.992\xspace}
\newcommand{\XsfGccFOaLgtRseRecall}{1\xspace}
\newcommand{\XsfGccFOaLgtRsePrecision}{1\xspace}
\newcommand{\XsfGccFOaLgtRseFone}{1\xspace}
\newcommand{\XsfGccFOaBzpGT}{14177\xspace}
\newcommand{\XsfGccFOaBzpLsRecall}{1\xspace}
\newcommand{\XsfGccFOaBzpLsPrecision}{1\xspace}
\newcommand{\XsfGccFOaBzpLsFone}{1\xspace}
\newcommand{\XsfGccFOaBzpBapRecall}{0.828\xspace}
\newcommand{\XsfGccFOaBzpBapPrecision}{1\xspace}
\newcommand{\XsfGccFOaBzpBapFone}{0.906\xspace}
\newcommand{\XsfGccFOaBzpGhiRecall}{1\xspace}
\newcommand{\XsfGccFOaBzpGhiPrecision}{1\xspace}
\newcommand{\XsfGccFOaBzpGhiFone}{1\xspace}
\newcommand{\XsfGccFOaBzpRdaRecall}{0.860\xspace}
\newcommand{\XsfGccFOaBzpRdaPrecision}{1\xspace}
\newcommand{\XsfGccFOaBzpRdaFone}{0.925\xspace}
\newcommand{\XsfGccFOaBzpRseRecall}{1\xspace}
\newcommand{\XsfGccFOaBzpRsePrecision}{1\xspace}
\newcommand{\XsfGccFOaBzpRseFone}{1\xspace}
\newcommand{\XsfGccFOaGccGT}{806133\xspace}
\newcommand{\XsfGccFOaGccLsRecall}{1\xspace}
\newcommand{\XsfGccFOaGccLsPrecision}{1\xspace}
\newcommand{\XsfGccFOaGccLsFone}{1\xspace}
\newcommand{\XsfGccFOaGccBapRecall}{0.694\xspace}
\newcommand{\XsfGccFOaGccBapPrecision}{1.000\xspace}
\newcommand{\XsfGccFOaGccBapFone}{0.819\xspace}
\newcommand{\XsfGccFOaGccGhiRecall}{0.987\xspace}
\newcommand{\XsfGccFOaGccGhiPrecision}{1\xspace}
\newcommand{\XsfGccFOaGccGhiFone}{0.994\xspace}
\newcommand{\XsfGccFOaGccRdaRecall}{0.879\xspace}
\newcommand{\XsfGccFOaGccRdaPrecision}{1\xspace}
\newcommand{\XsfGccFOaGccRdaFone}{0.936\xspace}
\newcommand{\XsfGccFOaGccRseRecall}{0.998\xspace}
\newcommand{\XsfGccFOaGccRsePrecision}{1\xspace}
\newcommand{\XsfGccFOaGccRseFone}{0.999\xspace}
\newcommand{\XsfGccFOaGzpGT}{9420\xspace}
\newcommand{\XsfGccFOaGzpLsRecall}{1\xspace}
\newcommand{\XsfGccFOaGzpLsPrecision}{1\xspace}
\newcommand{\XsfGccFOaGzpLsFone}{1\xspace}
\newcommand{\XsfGccFOaGzpBapRecall}{0.985\xspace}
\newcommand{\XsfGccFOaGzpBapPrecision}{1\xspace}
\newcommand{\XsfGccFOaGzpBapFone}{0.992\xspace}
\newcommand{\XsfGccFOaGzpGhiRecall}{1\xspace}
\newcommand{\XsfGccFOaGzpGhiPrecision}{1\xspace}
\newcommand{\XsfGccFOaGzpGhiFone}{1\xspace}
\newcommand{\XsfGccFOaGzpRdaRecall}{1\xspace}
\newcommand{\XsfGccFOaGzpRdaPrecision}{1\xspace}
\newcommand{\XsfGccFOaGzpRdaFone}{1\xspace}
\newcommand{\XsfGccFOaGzpRseRecall}{1\xspace}
\newcommand{\XsfGccFOaGzpRsePrecision}{1\xspace}
\newcommand{\XsfGccFOaGzpRseFone}{1\xspace}
\newcommand{\XsfGccFOaOggGT}{35339\xspace}
\newcommand{\XsfGccFOaOggLsRecall}{1\xspace}
\newcommand{\XsfGccFOaOggLsPrecision}{1\xspace}
\newcommand{\XsfGccFOaOggLsFone}{1\xspace}
\newcommand{\XsfGccFOaOggBapRecall}{0.978\xspace}
\newcommand{\XsfGccFOaOggBapPrecision}{1.000\xspace}
\newcommand{\XsfGccFOaOggBapFone}{0.989\xspace}
\newcommand{\XsfGccFOaOggGhiRecall}{1\xspace}
\newcommand{\XsfGccFOaOggGhiPrecision}{1\xspace}
\newcommand{\XsfGccFOaOggGhiFone}{1\xspace}
\newcommand{\XsfGccFOaOggRdaRecall}{1\xspace}
\newcommand{\XsfGccFOaOggRdaPrecision}{1\xspace}
\newcommand{\XsfGccFOaOggRdaFone}{1\xspace}
\newcommand{\XsfGccFOaOggRseRecall}{1\xspace}
\newcommand{\XsfGccFOaOggRsePrecision}{1\xspace}
\newcommand{\XsfGccFOaOggRseFone}{1\xspace}
\newcommand{\XsfGccFOaNgxGT}{98849\xspace}
\newcommand{\XsfGccFOaNgxLsRecall}{1\xspace}
\newcommand{\XsfGccFOaNgxLsPrecision}{1\xspace}
\newcommand{\XsfGccFOaNgxLsFone}{1\xspace}
\newcommand{\XsfGccFOaNgxBapRecall}{0.962\xspace}
\newcommand{\XsfGccFOaNgxBapPrecision}{1.000\xspace}
\newcommand{\XsfGccFOaNgxBapFone}{0.981\xspace}
\newcommand{\XsfGccFOaNgxGhiRecall}{1\xspace}
\newcommand{\XsfGccFOaNgxGhiPrecision}{1\xspace}
\newcommand{\XsfGccFOaNgxGhiFone}{1\xspace}
\newcommand{\XsfGccFOaNgxRdaRecall}{1.000\xspace}
\newcommand{\XsfGccFOaNgxRdaPrecision}{1.000\xspace}
\newcommand{\XsfGccFOaNgxRdaFone}{1.000\xspace}
\newcommand{\XsfGccFOaNgxRseRecall}{1\xspace}
\newcommand{\XsfGccFOaNgxRsePrecision}{1\xspace}
\newcommand{\XsfGccFOaNgxRseFone}{1\xspace}
\newcommand{\XsfGccFOaSshGT}{97507\xspace}
\newcommand{\XsfGccFOaSshLsRecall}{1\xspace}
\newcommand{\XsfGccFOaSshLsPrecision}{1\xspace}
\newcommand{\XsfGccFOaSshLsFone}{1\xspace}
\newcommand{\XsfGccFOaSshBapRecall}{0.941\xspace}
\newcommand{\XsfGccFOaSshBapPrecision}{1\xspace}
\newcommand{\XsfGccFOaSshBapFone}{0.970\xspace}
\newcommand{\XsfGccFOaSshGhiRecall}{1\xspace}
\newcommand{\XsfGccFOaSshGhiPrecision}{1\xspace}
\newcommand{\XsfGccFOaSshGhiFone}{1\xspace}
\newcommand{\XsfGccFOaSshRdaRecall}{0.969\xspace}
\newcommand{\XsfGccFOaSshRdaPrecision}{1.000\xspace}
\newcommand{\XsfGccFOaSshRdaFone}{0.984\xspace}
\newcommand{\XsfGccFOaSshRseRecall}{0.984\xspace}
\newcommand{\XsfGccFOaSshRsePrecision}{1\xspace}
\newcommand{\XsfGccFOaSshRseFone}{0.992\xspace}
\newcommand{\XsfGccFOaPcrGT}{4495\xspace}
\newcommand{\XsfGccFOaPcrLsRecall}{1\xspace}
\newcommand{\XsfGccFOaPcrLsPrecision}{1\xspace}
\newcommand{\XsfGccFOaPcrLsFone}{1\xspace}
\newcommand{\XsfGccFOaPcrBapRecall}{0.889\xspace}
\newcommand{\XsfGccFOaPcrBapPrecision}{1\xspace}
\newcommand{\XsfGccFOaPcrBapFone}{0.941\xspace}
\newcommand{\XsfGccFOaPcrGhiRecall}{1\xspace}
\newcommand{\XsfGccFOaPcrGhiPrecision}{1\xspace}
\newcommand{\XsfGccFOaPcrGhiFone}{1\xspace}
\newcommand{\XsfGccFOaPcrRdaRecall}{1\xspace}
\newcommand{\XsfGccFOaPcrRdaPrecision}{1\xspace}
\newcommand{\XsfGccFOaPcrRdaFone}{1\xspace}
\newcommand{\XsfGccFOaPcrRseRecall}{1\xspace}
\newcommand{\XsfGccFOaPcrRsePrecision}{1\xspace}
\newcommand{\XsfGccFOaPcrRseFone}{1\xspace}
\newcommand{\XsfGccFOaSqlGT}{150702\xspace}
\newcommand{\XsfGccFOaSqlLsRecall}{1\xspace}
\newcommand{\XsfGccFOaSqlLsPrecision}{1\xspace}
\newcommand{\XsfGccFOaSqlLsFone}{1\xspace}
\newcommand{\XsfGccFOaSqlBapRecall}{0.849\xspace}
\newcommand{\XsfGccFOaSqlBapPrecision}{1\xspace}
\newcommand{\XsfGccFOaSqlBapFone}{0.918\xspace}
\newcommand{\XsfGccFOaSqlGhiRecall}{1\xspace}
\newcommand{\XsfGccFOaSqlGhiPrecision}{1\xspace}
\newcommand{\XsfGccFOaSqlGhiFone}{1\xspace}
\newcommand{\XsfGccFOaSqlRdaRecall}{0.999\xspace}
\newcommand{\XsfGccFOaSqlRdaPrecision}{1\xspace}
\newcommand{\XsfGccFOaSqlRdaFone}{0.999\xspace}
\newcommand{\XsfGccFOaSqlRseRecall}{1\xspace}
\newcommand{\XsfGccFOaSqlRsePrecision}{1\xspace}
\newcommand{\XsfGccFOaSqlRseFone}{1\xspace}
\newcommand{\XsfGccFOaVimGT}{456084\xspace}
\newcommand{\XsfGccFOaVimLsRecall}{1\xspace}
\newcommand{\XsfGccFOaVimLsPrecision}{1\xspace}
\newcommand{\XsfGccFOaVimLsFone}{1\xspace}
\newcommand{\XsfGccFOaVimBapRecall}{0.919\xspace}
\newcommand{\XsfGccFOaVimBapPrecision}{1.000\xspace}
\newcommand{\XsfGccFOaVimBapFone}{0.958\xspace}
\newcommand{\XsfGccFOaVimGhiRecall}{1.000\xspace}
\newcommand{\XsfGccFOaVimGhiPrecision}{1\xspace}
\newcommand{\XsfGccFOaVimGhiFone}{1.000\xspace}
\newcommand{\XsfGccFOaVimRdaRecall}{0.990\xspace}
\newcommand{\XsfGccFOaVimRdaPrecision}{1.000\xspace}
\newcommand{\XsfGccFOaVimRdaFone}{0.995\xspace}
\newcommand{\XsfGccFOaVimRseRecall}{1.000\xspace}
\newcommand{\XsfGccFOaVimRsePrecision}{1\xspace}
\newcommand{\XsfGccFOaVimRseFone}{1.000\xspace}
\newcommand{\XsfGccFOaVsfGT}{16881\xspace}
\newcommand{\XsfGccFOaVsfLsRecall}{1\xspace}
\newcommand{\XsfGccFOaVsfLsPrecision}{1\xspace}
\newcommand{\XsfGccFOaVsfLsFone}{1\xspace}
\newcommand{\XsfGccFOaVsfBapRecall}{0.995\xspace}
\newcommand{\XsfGccFOaVsfBapPrecision}{1\xspace}
\newcommand{\XsfGccFOaVsfBapFone}{0.998\xspace}
\newcommand{\XsfGccFOaVsfGhiRecall}{1\xspace}
\newcommand{\XsfGccFOaVsfGhiPrecision}{1\xspace}
\newcommand{\XsfGccFOaVsfGhiFone}{1\xspace}
\newcommand{\XsfGccFOaVsfRdaRecall}{0.993\xspace}
\newcommand{\XsfGccFOaVsfRdaPrecision}{1.000\xspace}
\newcommand{\XsfGccFOaVsfRdaFone}{0.996\xspace}
\newcommand{\XsfGccFOaVsfRseRecall}{1.000\xspace}
\newcommand{\XsfGccFOaVsfRsePrecision}{1\xspace}
\newcommand{\XsfGccFOaVsfRseFone}{1.000\xspace}
\newcommand{\XsfGccFObSzpGT}{11909\xspace}
\newcommand{\XsfGccFObSzpLsRecall}{1\xspace}
\newcommand{\XsfGccFObSzpLsPrecision}{1\xspace}
\newcommand{\XsfGccFObSzpLsFone}{1\xspace}
\newcommand{\XsfGccFObSzpBapRecall}{0.975\xspace}
\newcommand{\XsfGccFObSzpBapPrecision}{1\xspace}
\newcommand{\XsfGccFObSzpBapFone}{0.988\xspace}
\newcommand{\XsfGccFObSzpGhiRecall}{1\xspace}
\newcommand{\XsfGccFObSzpGhiPrecision}{1\xspace}
\newcommand{\XsfGccFObSzpGhiFone}{1\xspace}
\newcommand{\XsfGccFObSzpRdaRecall}{0.994\xspace}
\newcommand{\XsfGccFObSzpRdaPrecision}{1\xspace}
\newcommand{\XsfGccFObSzpRdaFone}{0.997\xspace}
\newcommand{\XsfGccFObSzpRseRecall}{1\xspace}
\newcommand{\XsfGccFObSzpRsePrecision}{1\xspace}
\newcommand{\XsfGccFObSzpRseFone}{1\xspace}
\newcommand{\XsfGccFObCapGT}{182477\xspace}
\newcommand{\XsfGccFObCapLsRecall}{1\xspace}
\newcommand{\XsfGccFObCapLsPrecision}{1\xspace}
\newcommand{\XsfGccFObCapLsFone}{1\xspace}
\newcommand{\XsfGccFObCapBapRecall}{0.486\xspace}
\newcommand{\XsfGccFObCapBapPrecision}{1\xspace}
\newcommand{\XsfGccFObCapBapFone}{0.654\xspace}
\newcommand{\XsfGccFObCapGhiRecall}{0.869\xspace}
\newcommand{\XsfGccFObCapGhiPrecision}{1\xspace}
\newcommand{\XsfGccFObCapGhiFone}{0.930\xspace}
\newcommand{\XsfGccFObCapRdaRecall}{0.826\xspace}
\newcommand{\XsfGccFObCapRdaPrecision}{1\xspace}
\newcommand{\XsfGccFObCapRdaFone}{0.905\xspace}
\newcommand{\XsfGccFObCapRseRecall}{0.999\xspace}
\newcommand{\XsfGccFObCapRsePrecision}{1\xspace}
\newcommand{\XsfGccFObCapRseFone}{1.000\xspace}
\newcommand{\XsfGccFObExmGT}{138556\xspace}
\newcommand{\XsfGccFObExmLsRecall}{1\xspace}
\newcommand{\XsfGccFObExmLsPrecision}{1\xspace}
\newcommand{\XsfGccFObExmLsFone}{1\xspace}
\newcommand{\XsfGccFObExmBapRecall}{0.863\xspace}
\newcommand{\XsfGccFObExmBapPrecision}{1.000\xspace}
\newcommand{\XsfGccFObExmBapFone}{0.926\xspace}
\newcommand{\XsfGccFObExmGhiRecall}{0.995\xspace}
\newcommand{\XsfGccFObExmGhiPrecision}{1\xspace}
\newcommand{\XsfGccFObExmGhiFone}{0.998\xspace}
\newcommand{\XsfGccFObExmRdaRecall}{0.919\xspace}
\newcommand{\XsfGccFObExmRdaPrecision}{1.000\xspace}
\newcommand{\XsfGccFObExmRdaFone}{0.958\xspace}
\newcommand{\XsfGccFObExmRseRecall}{1.000\xspace}
\newcommand{\XsfGccFObExmRsePrecision}{1\xspace}
\newcommand{\XsfGccFObExmRseFone}{1.000\xspace}
\newcommand{\XsfGccFObLgtGT}{25106\xspace}
\newcommand{\XsfGccFObLgtLsRecall}{1\xspace}
\newcommand{\XsfGccFObLgtLsPrecision}{1\xspace}
\newcommand{\XsfGccFObLgtLsFone}{1\xspace}
\newcommand{\XsfGccFObLgtBapRecall}{0.890\xspace}
\newcommand{\XsfGccFObLgtBapPrecision}{1\xspace}
\newcommand{\XsfGccFObLgtBapFone}{0.942\xspace}
\newcommand{\XsfGccFObLgtGhiRecall}{0.972\xspace}
\newcommand{\XsfGccFObLgtGhiPrecision}{1\xspace}
\newcommand{\XsfGccFObLgtGhiFone}{0.986\xspace}
\newcommand{\XsfGccFObLgtRdaRecall}{0.986\xspace}
\newcommand{\XsfGccFObLgtRdaPrecision}{1.000\xspace}
\newcommand{\XsfGccFObLgtRdaFone}{0.993\xspace}
\newcommand{\XsfGccFObLgtRseRecall}{1\xspace}
\newcommand{\XsfGccFObLgtRsePrecision}{1\xspace}
\newcommand{\XsfGccFObLgtRseFone}{1\xspace}
\newcommand{\XsfGccFObBzpGT}{14276\xspace}
\newcommand{\XsfGccFObBzpLsRecall}{1\xspace}
\newcommand{\XsfGccFObBzpLsPrecision}{1\xspace}
\newcommand{\XsfGccFObBzpLsFone}{1\xspace}
\newcommand{\XsfGccFObBzpBapRecall}{0.929\xspace}
\newcommand{\XsfGccFObBzpBapPrecision}{1\xspace}
\newcommand{\XsfGccFObBzpBapFone}{0.963\xspace}
\newcommand{\XsfGccFObBzpGhiRecall}{1\xspace}
\newcommand{\XsfGccFObBzpGhiPrecision}{1\xspace}
\newcommand{\XsfGccFObBzpGhiFone}{1\xspace}
\newcommand{\XsfGccFObBzpRdaRecall}{0.847\xspace}
\newcommand{\XsfGccFObBzpRdaPrecision}{1\xspace}
\newcommand{\XsfGccFObBzpRdaFone}{0.917\xspace}
\newcommand{\XsfGccFObBzpRseRecall}{1\xspace}
\newcommand{\XsfGccFObBzpRsePrecision}{1\xspace}
\newcommand{\XsfGccFObBzpRseFone}{1\xspace}
\newcommand{\XsfGccFObGccGT}{832758\xspace}
\newcommand{\XsfGccFObGccLsRecall}{1\xspace}
\newcommand{\XsfGccFObGccLsPrecision}{1\xspace}
\newcommand{\XsfGccFObGccLsFone}{1\xspace}
\newcommand{\XsfGccFObGccBapRecall}{0.723\xspace}
\newcommand{\XsfGccFObGccBapPrecision}{1.000\xspace}
\newcommand{\XsfGccFObGccBapFone}{0.840\xspace}
\newcommand{\XsfGccFObGccGhiRecall}{0.990\xspace}
\newcommand{\XsfGccFObGccGhiPrecision}{1\xspace}
\newcommand{\XsfGccFObGccGhiFone}{0.995\xspace}
\newcommand{\XsfGccFObGccRdaRecall}{0.834\xspace}
\newcommand{\XsfGccFObGccRdaPrecision}{1.000\xspace}
\newcommand{\XsfGccFObGccRdaFone}{0.909\xspace}
\newcommand{\XsfGccFObGccRseRecall}{0.999\xspace}
\newcommand{\XsfGccFObGccRsePrecision}{1\xspace}
\newcommand{\XsfGccFObGccRseFone}{1.000\xspace}
\newcommand{\XsfGccFObGzpGT}{9650\xspace}
\newcommand{\XsfGccFObGzpLsRecall}{1\xspace}
\newcommand{\XsfGccFObGzpLsPrecision}{1\xspace}
\newcommand{\XsfGccFObGzpLsFone}{1\xspace}
\newcommand{\XsfGccFObGzpBapRecall}{0.985\xspace}
\newcommand{\XsfGccFObGzpBapPrecision}{1\xspace}
\newcommand{\XsfGccFObGzpBapFone}{0.993\xspace}
\newcommand{\XsfGccFObGzpGhiRecall}{1\xspace}
\newcommand{\XsfGccFObGzpGhiPrecision}{1\xspace}
\newcommand{\XsfGccFObGzpGhiFone}{1\xspace}
\newcommand{\XsfGccFObGzpRdaRecall}{1\xspace}
\newcommand{\XsfGccFObGzpRdaPrecision}{1\xspace}
\newcommand{\XsfGccFObGzpRdaFone}{1\xspace}
\newcommand{\XsfGccFObGzpRseRecall}{1\xspace}
\newcommand{\XsfGccFObGzpRsePrecision}{1\xspace}
\newcommand{\XsfGccFObGzpRseFone}{1\xspace}
\newcommand{\XsfGccFObOggGT}{37123\xspace}
\newcommand{\XsfGccFObOggLsRecall}{1\xspace}
\newcommand{\XsfGccFObOggLsPrecision}{1\xspace}
\newcommand{\XsfGccFObOggLsFone}{1\xspace}
\newcommand{\XsfGccFObOggBapRecall}{0.980\xspace}
\newcommand{\XsfGccFObOggBapPrecision}{1.000\xspace}
\newcommand{\XsfGccFObOggBapFone}{0.990\xspace}
\newcommand{\XsfGccFObOggGhiRecall}{1\xspace}
\newcommand{\XsfGccFObOggGhiPrecision}{1\xspace}
\newcommand{\XsfGccFObOggGhiFone}{1\xspace}
\newcommand{\XsfGccFObOggRdaRecall}{1\xspace}
\newcommand{\XsfGccFObOggRdaPrecision}{1\xspace}
\newcommand{\XsfGccFObOggRdaFone}{1\xspace}
\newcommand{\XsfGccFObOggRseRecall}{1\xspace}
\newcommand{\XsfGccFObOggRsePrecision}{1\xspace}
\newcommand{\XsfGccFObOggRseFone}{1\xspace}
\newcommand{\XsfGccFObNgxGT}{97678\xspace}
\newcommand{\XsfGccFObNgxLsRecall}{1\xspace}
\newcommand{\XsfGccFObNgxLsPrecision}{1\xspace}
\newcommand{\XsfGccFObNgxLsFone}{1\xspace}
\newcommand{\XsfGccFObNgxBapRecall}{0.966\xspace}
\newcommand{\XsfGccFObNgxBapPrecision}{1.000\xspace}
\newcommand{\XsfGccFObNgxBapFone}{0.983\xspace}
\newcommand{\XsfGccFObNgxGhiRecall}{1\xspace}
\newcommand{\XsfGccFObNgxGhiPrecision}{1\xspace}
\newcommand{\XsfGccFObNgxGhiFone}{1\xspace}
\newcommand{\XsfGccFObNgxRdaRecall}{0.973\xspace}
\newcommand{\XsfGccFObNgxRdaPrecision}{1.000\xspace}
\newcommand{\XsfGccFObNgxRdaFone}{0.986\xspace}
\newcommand{\XsfGccFObNgxRseRecall}{1\xspace}
\newcommand{\XsfGccFObNgxRsePrecision}{1\xspace}
\newcommand{\XsfGccFObNgxRseFone}{1\xspace}
\newcommand{\XsfGccFObSshGT}{97615\xspace}
\newcommand{\XsfGccFObSshLsRecall}{1\xspace}
\newcommand{\XsfGccFObSshLsPrecision}{1\xspace}
\newcommand{\XsfGccFObSshLsFone}{1\xspace}
\newcommand{\XsfGccFObSshBapRecall}{0.955\xspace}
\newcommand{\XsfGccFObSshBapPrecision}{1.000\xspace}
\newcommand{\XsfGccFObSshBapFone}{0.977\xspace}
\newcommand{\XsfGccFObSshGhiRecall}{1\xspace}
\newcommand{\XsfGccFObSshGhiPrecision}{1\xspace}
\newcommand{\XsfGccFObSshGhiFone}{1\xspace}
\newcommand{\XsfGccFObSshRdaRecall}{0.975\xspace}
\newcommand{\XsfGccFObSshRdaPrecision}{1.000\xspace}
\newcommand{\XsfGccFObSshRdaFone}{0.987\xspace}
\newcommand{\XsfGccFObSshRseRecall}{0.991\xspace}
\newcommand{\XsfGccFObSshRsePrecision}{1\xspace}
\newcommand{\XsfGccFObSshRseFone}{0.996\xspace}
\newcommand{\XsfGccFObPcrGT}{4351\xspace}
\newcommand{\XsfGccFObPcrLsRecall}{1\xspace}
\newcommand{\XsfGccFObPcrLsPrecision}{1\xspace}
\newcommand{\XsfGccFObPcrLsFone}{1\xspace}
\newcommand{\XsfGccFObPcrBapRecall}{0.913\xspace}
\newcommand{\XsfGccFObPcrBapPrecision}{1\xspace}
\newcommand{\XsfGccFObPcrBapFone}{0.955\xspace}
\newcommand{\XsfGccFObPcrGhiRecall}{1\xspace}
\newcommand{\XsfGccFObPcrGhiPrecision}{1\xspace}
\newcommand{\XsfGccFObPcrGhiFone}{1\xspace}
\newcommand{\XsfGccFObPcrRdaRecall}{1\xspace}
\newcommand{\XsfGccFObPcrRdaPrecision}{1\xspace}
\newcommand{\XsfGccFObPcrRdaFone}{1\xspace}
\newcommand{\XsfGccFObPcrRseRecall}{1\xspace}
\newcommand{\XsfGccFObPcrRsePrecision}{1\xspace}
\newcommand{\XsfGccFObPcrRseFone}{1\xspace}
\newcommand{\XsfGccFObSqlGT}{166599\xspace}
\newcommand{\XsfGccFObSqlLsRecall}{1\xspace}
\newcommand{\XsfGccFObSqlLsPrecision}{1\xspace}
\newcommand{\XsfGccFObSqlLsFone}{1\xspace}
\newcommand{\XsfGccFObSqlBapRecall}{0.855\xspace}
\newcommand{\XsfGccFObSqlBapPrecision}{1.000\xspace}
\newcommand{\XsfGccFObSqlBapFone}{0.922\xspace}
\newcommand{\XsfGccFObSqlGhiRecall}{0.949\xspace}
\newcommand{\XsfGccFObSqlGhiPrecision}{1\xspace}
\newcommand{\XsfGccFObSqlGhiFone}{0.974\xspace}
\newcommand{\XsfGccFObSqlRdaRecall}{0.989\xspace}
\newcommand{\XsfGccFObSqlRdaPrecision}{1\xspace}
\newcommand{\XsfGccFObSqlRdaFone}{0.994\xspace}
\newcommand{\XsfGccFObSqlRseRecall}{1\xspace}
\newcommand{\XsfGccFObSqlRsePrecision}{1\xspace}
\newcommand{\XsfGccFObSqlRseFone}{1\xspace}
\newcommand{\XsfGccFObVimGT}{468904\xspace}
\newcommand{\XsfGccFObVimLsRecall}{1\xspace}
\newcommand{\XsfGccFObVimLsPrecision}{1\xspace}
\newcommand{\XsfGccFObVimLsFone}{1\xspace}
\newcommand{\XsfGccFObVimBapRecall}{0.925\xspace}
\newcommand{\XsfGccFObVimBapPrecision}{1.000\xspace}
\newcommand{\XsfGccFObVimBapFone}{0.961\xspace}
\newcommand{\XsfGccFObVimGhiRecall}{0.982\xspace}
\newcommand{\XsfGccFObVimGhiPrecision}{1\xspace}
\newcommand{\XsfGccFObVimGhiFone}{0.991\xspace}
\newcommand{\XsfGccFObVimRdaRecall}{0.000\xspace}
\newcommand{\XsfGccFObVimRdaPrecision}{NaN\xspace}
\newcommand{\XsfGccFObVimRdaFone}{0.000\xspace}
\newcommand{\XsfGccFObVimRseRecall}{1.000\xspace}
\newcommand{\XsfGccFObVimRsePrecision}{1\xspace}
\newcommand{\XsfGccFObVimRseFone}{1.000\xspace}
\newcommand{\XsfGccFObVsfGT}{17759\xspace}
\newcommand{\XsfGccFObVsfLsRecall}{1\xspace}
\newcommand{\XsfGccFObVsfLsPrecision}{1\xspace}
\newcommand{\XsfGccFObVsfLsFone}{1\xspace}
\newcommand{\XsfGccFObVsfBapRecall}{0.996\xspace}
\newcommand{\XsfGccFObVsfBapPrecision}{1\xspace}
\newcommand{\XsfGccFObVsfBapFone}{0.998\xspace}
\newcommand{\XsfGccFObVsfGhiRecall}{1\xspace}
\newcommand{\XsfGccFObVsfGhiPrecision}{1\xspace}
\newcommand{\XsfGccFObVsfGhiFone}{1\xspace}
\newcommand{\XsfGccFObVsfRdaRecall}{0.987\xspace}
\newcommand{\XsfGccFObVsfRdaPrecision}{0.999\xspace}
\newcommand{\XsfGccFObVsfRdaFone}{0.993\xspace}
\newcommand{\XsfGccFObVsfRseRecall}{1\xspace}
\newcommand{\XsfGccFObVsfRsePrecision}{1\xspace}
\newcommand{\XsfGccFObVsfRseFone}{1\xspace}
\newcommand{\XsfGccFOcSzpGT}{14601\xspace}
\newcommand{\XsfGccFOcSzpLsRecall}{1\xspace}
\newcommand{\XsfGccFOcSzpLsPrecision}{1\xspace}
\newcommand{\XsfGccFOcSzpLsFone}{1\xspace}
\newcommand{\XsfGccFOcSzpBapRecall}{0.961\xspace}
\newcommand{\XsfGccFOcSzpBapPrecision}{1\xspace}
\newcommand{\XsfGccFOcSzpBapFone}{0.980\xspace}
\newcommand{\XsfGccFOcSzpGhiRecall}{1\xspace}
\newcommand{\XsfGccFOcSzpGhiPrecision}{1\xspace}
\newcommand{\XsfGccFOcSzpGhiFone}{1\xspace}
\newcommand{\XsfGccFOcSzpRdaRecall}{0.985\xspace}
\newcommand{\XsfGccFOcSzpRdaPrecision}{1\xspace}
\newcommand{\XsfGccFOcSzpRdaFone}{0.992\xspace}
\newcommand{\XsfGccFOcSzpRseRecall}{1\xspace}
\newcommand{\XsfGccFOcSzpRsePrecision}{1\xspace}
\newcommand{\XsfGccFOcSzpRseFone}{1\xspace}
\newcommand{\XsfGccFOcCapGT}{199413\xspace}
\newcommand{\XsfGccFOcCapLsRecall}{1\xspace}
\newcommand{\XsfGccFOcCapLsPrecision}{1\xspace}
\newcommand{\XsfGccFOcCapLsFone}{1\xspace}
\newcommand{\XsfGccFOcCapBapRecall}{0.491\xspace}
\newcommand{\XsfGccFOcCapBapPrecision}{1\xspace}
\newcommand{\XsfGccFOcCapBapFone}{0.659\xspace}
\newcommand{\XsfGccFOcCapGhiRecall}{0.880\xspace}
\newcommand{\XsfGccFOcCapGhiPrecision}{1\xspace}
\newcommand{\XsfGccFOcCapGhiFone}{0.936\xspace}
\newcommand{\XsfGccFOcCapRdaRecall}{0.839\xspace}
\newcommand{\XsfGccFOcCapRdaPrecision}{1\xspace}
\newcommand{\XsfGccFOcCapRdaFone}{0.913\xspace}
\newcommand{\XsfGccFOcCapRseRecall}{0.999\xspace}
\newcommand{\XsfGccFOcCapRsePrecision}{1\xspace}
\newcommand{\XsfGccFOcCapRseFone}{0.999\xspace}
\newcommand{\XsfGccFOcExmGT}{167429\xspace}
\newcommand{\XsfGccFOcExmLsRecall}{1\xspace}
\newcommand{\XsfGccFOcExmLsPrecision}{1\xspace}
\newcommand{\XsfGccFOcExmLsFone}{1\xspace}
\newcommand{\XsfGccFOcExmBapRecall}{0.870\xspace}
\newcommand{\XsfGccFOcExmBapPrecision}{1.000\xspace}
\newcommand{\XsfGccFOcExmBapFone}{0.930\xspace}
\newcommand{\XsfGccFOcExmGhiRecall}{0.993\xspace}
\newcommand{\XsfGccFOcExmGhiPrecision}{1\xspace}
\newcommand{\XsfGccFOcExmGhiFone}{0.996\xspace}
\newcommand{\XsfGccFOcExmRdaRecall}{0.956\xspace}
\newcommand{\XsfGccFOcExmRdaPrecision}{1.000\xspace}
\newcommand{\XsfGccFOcExmRdaFone}{0.977\xspace}
\newcommand{\XsfGccFOcExmRseRecall}{1.000\xspace}
\newcommand{\XsfGccFOcExmRsePrecision}{1\xspace}
\newcommand{\XsfGccFOcExmRseFone}{1.000\xspace}
\newcommand{\XsfGccFOcLgtGT}{31246\xspace}
\newcommand{\XsfGccFOcLgtLsRecall}{1\xspace}
\newcommand{\XsfGccFOcLgtLsPrecision}{1\xspace}
\newcommand{\XsfGccFOcLgtLsFone}{1\xspace}
\newcommand{\XsfGccFOcLgtBapRecall}{0.903\xspace}
\newcommand{\XsfGccFOcLgtBapPrecision}{1\xspace}
\newcommand{\XsfGccFOcLgtBapFone}{0.949\xspace}
\newcommand{\XsfGccFOcLgtGhiRecall}{0.977\xspace}
\newcommand{\XsfGccFOcLgtGhiPrecision}{1\xspace}
\newcommand{\XsfGccFOcLgtGhiFone}{0.988\xspace}
\newcommand{\XsfGccFOcLgtRdaRecall}{0.968\xspace}
\newcommand{\XsfGccFOcLgtRdaPrecision}{1.000\xspace}
\newcommand{\XsfGccFOcLgtRdaFone}{0.984\xspace}
\newcommand{\XsfGccFOcLgtRseRecall}{1\xspace}
\newcommand{\XsfGccFOcLgtRsePrecision}{1\xspace}
\newcommand{\XsfGccFOcLgtRseFone}{1\xspace}
\newcommand{\XsfGccFOcBzpGT}{18859\xspace}
\newcommand{\XsfGccFOcBzpLsRecall}{1\xspace}
\newcommand{\XsfGccFOcBzpLsPrecision}{1\xspace}
\newcommand{\XsfGccFOcBzpLsFone}{1\xspace}
\newcommand{\XsfGccFOcBzpBapRecall}{0.944\xspace}
\newcommand{\XsfGccFOcBzpBapPrecision}{1\xspace}
\newcommand{\XsfGccFOcBzpBapFone}{0.971\xspace}
\newcommand{\XsfGccFOcBzpGhiRecall}{1\xspace}
\newcommand{\XsfGccFOcBzpGhiPrecision}{1\xspace}
\newcommand{\XsfGccFOcBzpGhiFone}{1\xspace}
\newcommand{\XsfGccFOcBzpRdaRecall}{0.862\xspace}
\newcommand{\XsfGccFOcBzpRdaPrecision}{1\xspace}
\newcommand{\XsfGccFOcBzpRdaFone}{0.926\xspace}
\newcommand{\XsfGccFOcBzpRseRecall}{1\xspace}
\newcommand{\XsfGccFOcBzpRsePrecision}{1\xspace}
\newcommand{\XsfGccFOcBzpRseFone}{1\xspace}
\newcommand{\XsfGccFOcGccGT}{1069090\xspace}
\newcommand{\XsfGccFOcGccLsRecall}{1\xspace}
\newcommand{\XsfGccFOcGccLsPrecision}{1\xspace}
\newcommand{\XsfGccFOcGccLsFone}{1\xspace}
\newcommand{\XsfGccFOcGccBapRecall}{0.771\xspace}
\newcommand{\XsfGccFOcGccBapPrecision}{1.000\xspace}
\newcommand{\XsfGccFOcGccBapFone}{0.871\xspace}
\newcommand{\XsfGccFOcGccGhiRecall}{0.990\xspace}
\newcommand{\XsfGccFOcGccGhiPrecision}{1\xspace}
\newcommand{\XsfGccFOcGccGhiFone}{0.995\xspace}
\newcommand{\XsfGccFOcGccRdaRecall}{0.854\xspace}
\newcommand{\XsfGccFOcGccRdaPrecision}{1\xspace}
\newcommand{\XsfGccFOcGccRdaFone}{0.922\xspace}
\newcommand{\XsfGccFOcGccRseRecall}{1.000\xspace}
\newcommand{\XsfGccFOcGccRsePrecision}{1\xspace}
\newcommand{\XsfGccFOcGccRseFone}{1.000\xspace}
\newcommand{\XsfGccFOcGzpGT}{17741\xspace}
\newcommand{\XsfGccFOcGzpLsRecall}{1\xspace}
\newcommand{\XsfGccFOcGzpLsPrecision}{1\xspace}
\newcommand{\XsfGccFOcGzpLsFone}{1\xspace}
\newcommand{\XsfGccFOcGzpBapRecall}{0.992\xspace}
\newcommand{\XsfGccFOcGzpBapPrecision}{1\xspace}
\newcommand{\XsfGccFOcGzpBapFone}{0.996\xspace}
\newcommand{\XsfGccFOcGzpGhiRecall}{1\xspace}
\newcommand{\XsfGccFOcGzpGhiPrecision}{1\xspace}
\newcommand{\XsfGccFOcGzpGhiFone}{1\xspace}
\newcommand{\XsfGccFOcGzpRdaRecall}{0.974\xspace}
\newcommand{\XsfGccFOcGzpRdaPrecision}{1\xspace}
\newcommand{\XsfGccFOcGzpRdaFone}{0.987\xspace}
\newcommand{\XsfGccFOcGzpRseRecall}{1\xspace}
\newcommand{\XsfGccFOcGzpRsePrecision}{1\xspace}
\newcommand{\XsfGccFOcGzpRseFone}{1\xspace}
\newcommand{\XsfGccFOcOggGT}{60544\xspace}
\newcommand{\XsfGccFOcOggLsRecall}{1\xspace}
\newcommand{\XsfGccFOcOggLsPrecision}{1\xspace}
\newcommand{\XsfGccFOcOggLsFone}{1\xspace}
\newcommand{\XsfGccFOcOggBapRecall}{0.988\xspace}
\newcommand{\XsfGccFOcOggBapPrecision}{1.000\xspace}
\newcommand{\XsfGccFOcOggBapFone}{0.994\xspace}
\newcommand{\XsfGccFOcOggGhiRecall}{1\xspace}
\newcommand{\XsfGccFOcOggGhiPrecision}{1\xspace}
\newcommand{\XsfGccFOcOggGhiFone}{1\xspace}
\newcommand{\XsfGccFOcOggRdaRecall}{1\xspace}
\newcommand{\XsfGccFOcOggRdaPrecision}{1\xspace}
\newcommand{\XsfGccFOcOggRdaFone}{1\xspace}
\newcommand{\XsfGccFOcOggRseRecall}{1\xspace}
\newcommand{\XsfGccFOcOggRsePrecision}{1\xspace}
\newcommand{\XsfGccFOcOggRseFone}{1\xspace}
\newcommand{\XsfGccFOcNgxGT}{110058\xspace}
\newcommand{\XsfGccFOcNgxLsRecall}{1\xspace}
\newcommand{\XsfGccFOcNgxLsPrecision}{1\xspace}
\newcommand{\XsfGccFOcNgxLsFone}{1\xspace}
\newcommand{\XsfGccFOcNgxBapRecall}{0.968\xspace}
\newcommand{\XsfGccFOcNgxBapPrecision}{1\xspace}
\newcommand{\XsfGccFOcNgxBapFone}{0.984\xspace}
\newcommand{\XsfGccFOcNgxGhiRecall}{1\xspace}
\newcommand{\XsfGccFOcNgxGhiPrecision}{1\xspace}
\newcommand{\XsfGccFOcNgxGhiFone}{1\xspace}
\newcommand{\XsfGccFOcNgxRdaRecall}{0.972\xspace}
\newcommand{\XsfGccFOcNgxRdaPrecision}{1.000\xspace}
\newcommand{\XsfGccFOcNgxRdaFone}{0.986\xspace}
\newcommand{\XsfGccFOcNgxRseRecall}{1\xspace}
\newcommand{\XsfGccFOcNgxRsePrecision}{1\xspace}
\newcommand{\XsfGccFOcNgxRseFone}{1\xspace}
\newcommand{\XsfGccFOcSshGT}{119056\xspace}
\newcommand{\XsfGccFOcSshLsRecall}{1\xspace}
\newcommand{\XsfGccFOcSshLsPrecision}{1\xspace}
\newcommand{\XsfGccFOcSshLsFone}{1\xspace}
\newcommand{\XsfGccFOcSshBapRecall}{0.956\xspace}
\newcommand{\XsfGccFOcSshBapPrecision}{1.000\xspace}
\newcommand{\XsfGccFOcSshBapFone}{0.977\xspace}
\newcommand{\XsfGccFOcSshGhiRecall}{0.999\xspace}
\newcommand{\XsfGccFOcSshGhiPrecision}{1\xspace}
\newcommand{\XsfGccFOcSshGhiFone}{0.999\xspace}
\newcommand{\XsfGccFOcSshRdaRecall}{0.962\xspace}
\newcommand{\XsfGccFOcSshRdaPrecision}{1.000\xspace}
\newcommand{\XsfGccFOcSshRdaFone}{0.981\xspace}
\newcommand{\XsfGccFOcSshRseRecall}{0.993\xspace}
\newcommand{\XsfGccFOcSshRsePrecision}{1\xspace}
\newcommand{\XsfGccFOcSshRseFone}{0.996\xspace}
\newcommand{\XsfGccFOcPcrGT}{4684\xspace}
\newcommand{\XsfGccFOcPcrLsRecall}{1\xspace}
\newcommand{\XsfGccFOcPcrLsPrecision}{1\xspace}
\newcommand{\XsfGccFOcPcrLsFone}{1\xspace}
\newcommand{\XsfGccFOcPcrBapRecall}{0.917\xspace}
\newcommand{\XsfGccFOcPcrBapPrecision}{1\xspace}
\newcommand{\XsfGccFOcPcrBapFone}{0.957\xspace}
\newcommand{\XsfGccFOcPcrGhiRecall}{1\xspace}
\newcommand{\XsfGccFOcPcrGhiPrecision}{1\xspace}
\newcommand{\XsfGccFOcPcrGhiFone}{1\xspace}
\newcommand{\XsfGccFOcPcrRdaRecall}{1\xspace}
\newcommand{\XsfGccFOcPcrRdaPrecision}{1\xspace}
\newcommand{\XsfGccFOcPcrRdaFone}{1\xspace}
\newcommand{\XsfGccFOcPcrRseRecall}{1\xspace}
\newcommand{\XsfGccFOcPcrRsePrecision}{1\xspace}
\newcommand{\XsfGccFOcPcrRseFone}{1\xspace}
\newcommand{\XsfGccFOcSqlGT}{207181\xspace}
\newcommand{\XsfGccFOcSqlLsRecall}{1\xspace}
\newcommand{\XsfGccFOcSqlLsPrecision}{1\xspace}
\newcommand{\XsfGccFOcSqlLsFone}{1\xspace}
\newcommand{\XsfGccFOcSqlBapRecall}{0.856\xspace}
\newcommand{\XsfGccFOcSqlBapPrecision}{1\xspace}
\newcommand{\XsfGccFOcSqlBapFone}{0.922\xspace}
\newcommand{\XsfGccFOcSqlGhiRecall}{0.946\xspace}
\newcommand{\XsfGccFOcSqlGhiPrecision}{1\xspace}
\newcommand{\XsfGccFOcSqlGhiFone}{0.972\xspace}
\newcommand{\XsfGccFOcSqlRdaRecall}{0.996\xspace}
\newcommand{\XsfGccFOcSqlRdaPrecision}{1.000\xspace}
\newcommand{\XsfGccFOcSqlRdaFone}{0.998\xspace}
\newcommand{\XsfGccFOcSqlRseRecall}{1\xspace}
\newcommand{\XsfGccFOcSqlRsePrecision}{1\xspace}
\newcommand{\XsfGccFOcSqlRseFone}{1\xspace}
\newcommand{\XsfGccFOcVimGT}{644404\xspace}
\newcommand{\XsfGccFOcVimLsRecall}{1\xspace}
\newcommand{\XsfGccFOcVimLsPrecision}{1\xspace}
\newcommand{\XsfGccFOcVimLsFone}{1\xspace}
\newcommand{\XsfGccFOcVimBapRecall}{0.937\xspace}
\newcommand{\XsfGccFOcVimBapPrecision}{1.000\xspace}
\newcommand{\XsfGccFOcVimBapFone}{0.968\xspace}
\newcommand{\XsfGccFOcVimGhiRecall}{0.987\xspace}
\newcommand{\XsfGccFOcVimGhiPrecision}{1\xspace}
\newcommand{\XsfGccFOcVimGhiFone}{0.993\xspace}
\newcommand{\XsfGccFOcVimRdaRecall}{0.955\xspace}
\newcommand{\XsfGccFOcVimRdaPrecision}{1.000\xspace}
\newcommand{\XsfGccFOcVimRdaFone}{0.977\xspace}
\newcommand{\XsfGccFOcVimRseRecall}{1.000\xspace}
\newcommand{\XsfGccFOcVimRsePrecision}{1\xspace}
\newcommand{\XsfGccFOcVimRseFone}{1.000\xspace}
\newcommand{\XsfGccFOcVsfGT}{23502\xspace}
\newcommand{\XsfGccFOcVsfLsRecall}{1\xspace}
\newcommand{\XsfGccFOcVsfLsPrecision}{1\xspace}
\newcommand{\XsfGccFOcVsfLsFone}{1\xspace}
\newcommand{\XsfGccFOcVsfBapRecall}{0.993\xspace}
\newcommand{\XsfGccFOcVsfBapPrecision}{1\xspace}
\newcommand{\XsfGccFOcVsfBapFone}{0.996\xspace}
\newcommand{\XsfGccFOcVsfGhiRecall}{1\xspace}
\newcommand{\XsfGccFOcVsfGhiPrecision}{1\xspace}
\newcommand{\XsfGccFOcVsfGhiFone}{1\xspace}
\newcommand{\XsfGccFOcVsfRdaRecall}{0.987\xspace}
\newcommand{\XsfGccFOcVsfRdaPrecision}{1.000\xspace}
\newcommand{\XsfGccFOcVsfRdaFone}{0.994\xspace}
\newcommand{\XsfGccFOcVsfRseRecall}{1.000\xspace}
\newcommand{\XsfGccFOcVsfRsePrecision}{1\xspace}
\newcommand{\XsfGccFOcVsfRseFone}{1.000\xspace}
\newcommand{\XsfGccFOdSzpGT}{14601\xspace}
\newcommand{\XsfGccFOdSzpLsRecall}{1\xspace}
\newcommand{\XsfGccFOdSzpLsPrecision}{1\xspace}
\newcommand{\XsfGccFOdSzpLsFone}{1\xspace}
\newcommand{\XsfGccFOdSzpBapRecall}{0.961\xspace}
\newcommand{\XsfGccFOdSzpBapPrecision}{1\xspace}
\newcommand{\XsfGccFOdSzpBapFone}{0.980\xspace}
\newcommand{\XsfGccFOdSzpGhiRecall}{1\xspace}
\newcommand{\XsfGccFOdSzpGhiPrecision}{1\xspace}
\newcommand{\XsfGccFOdSzpGhiFone}{1\xspace}
\newcommand{\XsfGccFOdSzpRdaRecall}{0.985\xspace}
\newcommand{\XsfGccFOdSzpRdaPrecision}{1\xspace}
\newcommand{\XsfGccFOdSzpRdaFone}{0.992\xspace}
\newcommand{\XsfGccFOdSzpRseRecall}{1\xspace}
\newcommand{\XsfGccFOdSzpRsePrecision}{1\xspace}
\newcommand{\XsfGccFOdSzpRseFone}{1\xspace}
\newcommand{\XsfGccFOdCapGT}{199413\xspace}
\newcommand{\XsfGccFOdCapLsRecall}{1\xspace}
\newcommand{\XsfGccFOdCapLsPrecision}{1\xspace}
\newcommand{\XsfGccFOdCapLsFone}{1\xspace}
\newcommand{\XsfGccFOdCapBapRecall}{0.491\xspace}
\newcommand{\XsfGccFOdCapBapPrecision}{1\xspace}
\newcommand{\XsfGccFOdCapBapFone}{0.659\xspace}
\newcommand{\XsfGccFOdCapGhiRecall}{0.879\xspace}
\newcommand{\XsfGccFOdCapGhiPrecision}{1\xspace}
\newcommand{\XsfGccFOdCapGhiFone}{0.936\xspace}
\newcommand{\XsfGccFOdCapRdaRecall}{0.839\xspace}
\newcommand{\XsfGccFOdCapRdaPrecision}{1\xspace}
\newcommand{\XsfGccFOdCapRdaFone}{0.913\xspace}
\newcommand{\XsfGccFOdCapRseRecall}{0.999\xspace}
\newcommand{\XsfGccFOdCapRsePrecision}{1\xspace}
\newcommand{\XsfGccFOdCapRseFone}{0.999\xspace}
\newcommand{\XsfGccFOdExmGT}{167309\xspace}
\newcommand{\XsfGccFOdExmLsRecall}{1\xspace}
\newcommand{\XsfGccFOdExmLsPrecision}{1\xspace}
\newcommand{\XsfGccFOdExmLsFone}{1\xspace}
\newcommand{\XsfGccFOdExmBapRecall}{0.870\xspace}
\newcommand{\XsfGccFOdExmBapPrecision}{1.000\xspace}
\newcommand{\XsfGccFOdExmBapFone}{0.931\xspace}
\newcommand{\XsfGccFOdExmGhiRecall}{0.992\xspace}
\newcommand{\XsfGccFOdExmGhiPrecision}{1\xspace}
\newcommand{\XsfGccFOdExmGhiFone}{0.996\xspace}
\newcommand{\XsfGccFOdExmRdaRecall}{0.959\xspace}
\newcommand{\XsfGccFOdExmRdaPrecision}{1.000\xspace}
\newcommand{\XsfGccFOdExmRdaFone}{0.979\xspace}
\newcommand{\XsfGccFOdExmRseRecall}{1.000\xspace}
\newcommand{\XsfGccFOdExmRsePrecision}{1\xspace}
\newcommand{\XsfGccFOdExmRseFone}{1.000\xspace}
\newcommand{\XsfGccFOdLgtGT}{31239\xspace}
\newcommand{\XsfGccFOdLgtLsRecall}{1\xspace}
\newcommand{\XsfGccFOdLgtLsPrecision}{1\xspace}
\newcommand{\XsfGccFOdLgtLsFone}{1\xspace}
\newcommand{\XsfGccFOdLgtBapRecall}{0.903\xspace}
\newcommand{\XsfGccFOdLgtBapPrecision}{1\xspace}
\newcommand{\XsfGccFOdLgtBapFone}{0.949\xspace}
\newcommand{\XsfGccFOdLgtGhiRecall}{0.977\xspace}
\newcommand{\XsfGccFOdLgtGhiPrecision}{1\xspace}
\newcommand{\XsfGccFOdLgtGhiFone}{0.988\xspace}
\newcommand{\XsfGccFOdLgtRdaRecall}{0.976\xspace}
\newcommand{\XsfGccFOdLgtRdaPrecision}{1.000\xspace}
\newcommand{\XsfGccFOdLgtRdaFone}{0.988\xspace}
\newcommand{\XsfGccFOdLgtRseRecall}{1\xspace}
\newcommand{\XsfGccFOdLgtRsePrecision}{1\xspace}
\newcommand{\XsfGccFOdLgtRseFone}{1\xspace}
\newcommand{\XsfGccFOdBzpGT}{18843\xspace}
\newcommand{\XsfGccFOdBzpLsRecall}{1\xspace}
\newcommand{\XsfGccFOdBzpLsPrecision}{1\xspace}
\newcommand{\XsfGccFOdBzpLsFone}{1\xspace}
\newcommand{\XsfGccFOdBzpBapRecall}{0.944\xspace}
\newcommand{\XsfGccFOdBzpBapPrecision}{1\xspace}
\newcommand{\XsfGccFOdBzpBapFone}{0.971\xspace}
\newcommand{\XsfGccFOdBzpGhiRecall}{1\xspace}
\newcommand{\XsfGccFOdBzpGhiPrecision}{1\xspace}
\newcommand{\XsfGccFOdBzpGhiFone}{1\xspace}
\newcommand{\XsfGccFOdBzpRdaRecall}{0.862\xspace}
\newcommand{\XsfGccFOdBzpRdaPrecision}{1\xspace}
\newcommand{\XsfGccFOdBzpRdaFone}{0.926\xspace}
\newcommand{\XsfGccFOdBzpRseRecall}{1\xspace}
\newcommand{\XsfGccFOdBzpRsePrecision}{1\xspace}
\newcommand{\XsfGccFOdBzpRseFone}{1\xspace}
\newcommand{\XsfGccFOdGccGT}{1068183\xspace}
\newcommand{\XsfGccFOdGccLsRecall}{1\xspace}
\newcommand{\XsfGccFOdGccLsPrecision}{1\xspace}
\newcommand{\XsfGccFOdGccLsFone}{1\xspace}
\newcommand{\XsfGccFOdGccBapRecall}{0.771\xspace}
\newcommand{\XsfGccFOdGccBapPrecision}{1.000\xspace}
\newcommand{\XsfGccFOdGccBapFone}{0.871\xspace}
\newcommand{\XsfGccFOdGccGhiRecall}{0.989\xspace}
\newcommand{\XsfGccFOdGccGhiPrecision}{1\xspace}
\newcommand{\XsfGccFOdGccGhiFone}{0.995\xspace}
\newcommand{\XsfGccFOdGccRdaRecall}{0.853\xspace}
\newcommand{\XsfGccFOdGccRdaPrecision}{1.000\xspace}
\newcommand{\XsfGccFOdGccRdaFone}{0.921\xspace}
\newcommand{\XsfGccFOdGccRseRecall}{1.000\xspace}
\newcommand{\XsfGccFOdGccRsePrecision}{1\xspace}
\newcommand{\XsfGccFOdGccRseFone}{1.000\xspace}
\newcommand{\XsfGccFOdGzpGT}{17661\xspace}
\newcommand{\XsfGccFOdGzpLsRecall}{1\xspace}
\newcommand{\XsfGccFOdGzpLsPrecision}{1\xspace}
\newcommand{\XsfGccFOdGzpLsFone}{1\xspace}
\newcommand{\XsfGccFOdGzpBapRecall}{0.992\xspace}
\newcommand{\XsfGccFOdGzpBapPrecision}{1\xspace}
\newcommand{\XsfGccFOdGzpBapFone}{0.996\xspace}
\newcommand{\XsfGccFOdGzpGhiRecall}{1\xspace}
\newcommand{\XsfGccFOdGzpGhiPrecision}{1\xspace}
\newcommand{\XsfGccFOdGzpGhiFone}{1\xspace}
\newcommand{\XsfGccFOdGzpRdaRecall}{0.964\xspace}
\newcommand{\XsfGccFOdGzpRdaPrecision}{1\xspace}
\newcommand{\XsfGccFOdGzpRdaFone}{0.982\xspace}
\newcommand{\XsfGccFOdGzpRseRecall}{1\xspace}
\newcommand{\XsfGccFOdGzpRsePrecision}{1\xspace}
\newcommand{\XsfGccFOdGzpRseFone}{1\xspace}
\newcommand{\XsfGccFOdOggGT}{61254\xspace}
\newcommand{\XsfGccFOdOggLsRecall}{1\xspace}
\newcommand{\XsfGccFOdOggLsPrecision}{1\xspace}
\newcommand{\XsfGccFOdOggLsFone}{1\xspace}
\newcommand{\XsfGccFOdOggBapRecall}{0.988\xspace}
\newcommand{\XsfGccFOdOggBapPrecision}{1.000\xspace}
\newcommand{\XsfGccFOdOggBapFone}{0.994\xspace}
\newcommand{\XsfGccFOdOggGhiRecall}{1\xspace}
\newcommand{\XsfGccFOdOggGhiPrecision}{1\xspace}
\newcommand{\XsfGccFOdOggGhiFone}{1\xspace}
\newcommand{\XsfGccFOdOggRdaRecall}{1\xspace}
\newcommand{\XsfGccFOdOggRdaPrecision}{1\xspace}
\newcommand{\XsfGccFOdOggRdaFone}{1\xspace}
\newcommand{\XsfGccFOdOggRseRecall}{1\xspace}
\newcommand{\XsfGccFOdOggRsePrecision}{1\xspace}
\newcommand{\XsfGccFOdOggRseFone}{1\xspace}
\newcommand{\XsfGccFOdNgxGT}{109988\xspace}
\newcommand{\XsfGccFOdNgxLsRecall}{1\xspace}
\newcommand{\XsfGccFOdNgxLsPrecision}{1\xspace}
\newcommand{\XsfGccFOdNgxLsFone}{1\xspace}
\newcommand{\XsfGccFOdNgxBapRecall}{0.968\xspace}
\newcommand{\XsfGccFOdNgxBapPrecision}{1\xspace}
\newcommand{\XsfGccFOdNgxBapFone}{0.984\xspace}
\newcommand{\XsfGccFOdNgxGhiRecall}{1\xspace}
\newcommand{\XsfGccFOdNgxGhiPrecision}{1\xspace}
\newcommand{\XsfGccFOdNgxGhiFone}{1\xspace}
\newcommand{\XsfGccFOdNgxRdaRecall}{0.972\xspace}
\newcommand{\XsfGccFOdNgxRdaPrecision}{1.000\xspace}
\newcommand{\XsfGccFOdNgxRdaFone}{0.986\xspace}
\newcommand{\XsfGccFOdNgxRseRecall}{1\xspace}
\newcommand{\XsfGccFOdNgxRsePrecision}{1\xspace}
\newcommand{\XsfGccFOdNgxRseFone}{1\xspace}
\newcommand{\XsfGccFOdSshGT}{119036\xspace}
\newcommand{\XsfGccFOdSshLsRecall}{1\xspace}
\newcommand{\XsfGccFOdSshLsPrecision}{1\xspace}
\newcommand{\XsfGccFOdSshLsFone}{1\xspace}
\newcommand{\XsfGccFOdSshBapRecall}{0.956\xspace}
\newcommand{\XsfGccFOdSshBapPrecision}{1.000\xspace}
\newcommand{\XsfGccFOdSshBapFone}{0.977\xspace}
\newcommand{\XsfGccFOdSshGhiRecall}{0.999\xspace}
\newcommand{\XsfGccFOdSshGhiPrecision}{1\xspace}
\newcommand{\XsfGccFOdSshGhiFone}{0.999\xspace}
\newcommand{\XsfGccFOdSshRdaRecall}{0.962\xspace}
\newcommand{\XsfGccFOdSshRdaPrecision}{1.000\xspace}
\newcommand{\XsfGccFOdSshRdaFone}{0.981\xspace}
\newcommand{\XsfGccFOdSshRseRecall}{0.993\xspace}
\newcommand{\XsfGccFOdSshRsePrecision}{1\xspace}
\newcommand{\XsfGccFOdSshRseFone}{0.996\xspace}
\newcommand{\XsfGccFOdPcrGT}{4684\xspace}
\newcommand{\XsfGccFOdPcrLsRecall}{1\xspace}
\newcommand{\XsfGccFOdPcrLsPrecision}{1\xspace}
\newcommand{\XsfGccFOdPcrLsFone}{1\xspace}
\newcommand{\XsfGccFOdPcrBapRecall}{0.917\xspace}
\newcommand{\XsfGccFOdPcrBapPrecision}{1\xspace}
\newcommand{\XsfGccFOdPcrBapFone}{0.957\xspace}
\newcommand{\XsfGccFOdPcrGhiRecall}{1\xspace}
\newcommand{\XsfGccFOdPcrGhiPrecision}{1\xspace}
\newcommand{\XsfGccFOdPcrGhiFone}{1\xspace}
\newcommand{\XsfGccFOdPcrRdaRecall}{1\xspace}
\newcommand{\XsfGccFOdPcrRdaPrecision}{0.999\xspace}
\newcommand{\XsfGccFOdPcrRdaFone}{0.999\xspace}
\newcommand{\XsfGccFOdPcrRseRecall}{1\xspace}
\newcommand{\XsfGccFOdPcrRsePrecision}{1\xspace}
\newcommand{\XsfGccFOdPcrRseFone}{1\xspace}
\newcommand{\XsfGccFOdSqlGT}{207191\xspace}
\newcommand{\XsfGccFOdSqlLsRecall}{1\xspace}
\newcommand{\XsfGccFOdSqlLsPrecision}{1\xspace}
\newcommand{\XsfGccFOdSqlLsFone}{1\xspace}
\newcommand{\XsfGccFOdSqlBapRecall}{0.852\xspace}
\newcommand{\XsfGccFOdSqlBapPrecision}{1\xspace}
\newcommand{\XsfGccFOdSqlBapFone}{0.920\xspace}
\newcommand{\XsfGccFOdSqlGhiRecall}{0.942\xspace}
\newcommand{\XsfGccFOdSqlGhiPrecision}{1\xspace}
\newcommand{\XsfGccFOdSqlGhiFone}{0.970\xspace}
\newcommand{\XsfGccFOdSqlRdaRecall}{0.996\xspace}
\newcommand{\XsfGccFOdSqlRdaPrecision}{1.000\xspace}
\newcommand{\XsfGccFOdSqlRdaFone}{0.998\xspace}
\newcommand{\XsfGccFOdSqlRseRecall}{1\xspace}
\newcommand{\XsfGccFOdSqlRsePrecision}{1\xspace}
\newcommand{\XsfGccFOdSqlRseFone}{1\xspace}
\newcommand{\XsfGccFOdVimGT}{644207\xspace}
\newcommand{\XsfGccFOdVimLsRecall}{1\xspace}
\newcommand{\XsfGccFOdVimLsPrecision}{1\xspace}
\newcommand{\XsfGccFOdVimLsFone}{1\xspace}
\newcommand{\XsfGccFOdVimBapRecall}{0.937\xspace}
\newcommand{\XsfGccFOdVimBapPrecision}{1.000\xspace}
\newcommand{\XsfGccFOdVimBapFone}{0.968\xspace}
\newcommand{\XsfGccFOdVimGhiRecall}{0.987\xspace}
\newcommand{\XsfGccFOdVimGhiPrecision}{1\xspace}
\newcommand{\XsfGccFOdVimGhiFone}{0.993\xspace}
\newcommand{\XsfGccFOdVimRdaRecall}{0.955\xspace}
\newcommand{\XsfGccFOdVimRdaPrecision}{1.000\xspace}
\newcommand{\XsfGccFOdVimRdaFone}{0.977\xspace}
\newcommand{\XsfGccFOdVimRseRecall}{1.000\xspace}
\newcommand{\XsfGccFOdVimRsePrecision}{1\xspace}
\newcommand{\XsfGccFOdVimRseFone}{1.000\xspace}
\newcommand{\XsfGccFOdVsfGT}{23502\xspace}
\newcommand{\XsfGccFOdVsfLsRecall}{1\xspace}
\newcommand{\XsfGccFOdVsfLsPrecision}{1\xspace}
\newcommand{\XsfGccFOdVsfLsFone}{1\xspace}
\newcommand{\XsfGccFOdVsfBapRecall}{0.993\xspace}
\newcommand{\XsfGccFOdVsfBapPrecision}{1\xspace}
\newcommand{\XsfGccFOdVsfBapFone}{0.996\xspace}
\newcommand{\XsfGccFOdVsfGhiRecall}{1\xspace}
\newcommand{\XsfGccFOdVsfGhiPrecision}{1\xspace}
\newcommand{\XsfGccFOdVsfGhiFone}{1\xspace}
\newcommand{\XsfGccFOdVsfRdaRecall}{0.987\xspace}
\newcommand{\XsfGccFOdVsfRdaPrecision}{1.000\xspace}
\newcommand{\XsfGccFOdVsfRdaFone}{0.994\xspace}
\newcommand{\XsfGccFOdVsfRseRecall}{1.000\xspace}
\newcommand{\XsfGccFOdVsfRsePrecision}{1\xspace}
\newcommand{\XsfGccFOdVsfRseFone}{1.000\xspace}
\newcommand{\XsfGccFOsSzpGT}{10705\xspace}
\newcommand{\XsfGccFOsSzpLsRecall}{1\xspace}
\newcommand{\XsfGccFOsSzpLsPrecision}{1\xspace}
\newcommand{\XsfGccFOsSzpLsFone}{1\xspace}
\newcommand{\XsfGccFOsSzpBapRecall}{0.977\xspace}
\newcommand{\XsfGccFOsSzpBapPrecision}{1\xspace}
\newcommand{\XsfGccFOsSzpBapFone}{0.988\xspace}
\newcommand{\XsfGccFOsSzpGhiRecall}{1\xspace}
\newcommand{\XsfGccFOsSzpGhiPrecision}{1\xspace}
\newcommand{\XsfGccFOsSzpGhiFone}{1\xspace}
\newcommand{\XsfGccFOsSzpRdaRecall}{0.995\xspace}
\newcommand{\XsfGccFOsSzpRdaPrecision}{1\xspace}
\newcommand{\XsfGccFOsSzpRdaFone}{0.997\xspace}
\newcommand{\XsfGccFOsSzpRseRecall}{1\xspace}
\newcommand{\XsfGccFOsSzpRsePrecision}{1\xspace}
\newcommand{\XsfGccFOsSzpRseFone}{1\xspace}
\newcommand{\XsfGccFOsCapGT}{178648\xspace}
\newcommand{\XsfGccFOsCapLsRecall}{1\xspace}
\newcommand{\XsfGccFOsCapLsPrecision}{1\xspace}
\newcommand{\XsfGccFOsCapLsFone}{1\xspace}
\newcommand{\XsfGccFOsCapBapRecall}{0.624\xspace}
\newcommand{\XsfGccFOsCapBapPrecision}{1\xspace}
\newcommand{\XsfGccFOsCapBapFone}{0.768\xspace}
\newcommand{\XsfGccFOsCapGhiRecall}{0.997\xspace}
\newcommand{\XsfGccFOsCapGhiPrecision}{1\xspace}
\newcommand{\XsfGccFOsCapGhiFone}{0.998\xspace}
\newcommand{\XsfGccFOsCapRdaRecall}{0.998\xspace}
\newcommand{\XsfGccFOsCapRdaPrecision}{1\xspace}
\newcommand{\XsfGccFOsCapRdaFone}{0.999\xspace}
\newcommand{\XsfGccFOsCapRseRecall}{1.000\xspace}
\newcommand{\XsfGccFOsCapRsePrecision}{1\xspace}
\newcommand{\XsfGccFOsCapRseFone}{1.000\xspace}
\newcommand{\XsfGccFOsExmGT}{122113\xspace}
\newcommand{\XsfGccFOsExmLsRecall}{1\xspace}
\newcommand{\XsfGccFOsExmLsPrecision}{1\xspace}
\newcommand{\XsfGccFOsExmLsFone}{1\xspace}
\newcommand{\XsfGccFOsExmBapRecall}{0.867\xspace}
\newcommand{\XsfGccFOsExmBapPrecision}{1\xspace}
\newcommand{\XsfGccFOsExmBapFone}{0.929\xspace}
\newcommand{\XsfGccFOsExmGhiRecall}{0.998\xspace}
\newcommand{\XsfGccFOsExmGhiPrecision}{1\xspace}
\newcommand{\XsfGccFOsExmGhiFone}{0.999\xspace}
\newcommand{\XsfGccFOsExmRdaRecall}{0.934\xspace}
\newcommand{\XsfGccFOsExmRdaPrecision}{1.000\xspace}
\newcommand{\XsfGccFOsExmRdaFone}{0.966\xspace}
\newcommand{\XsfGccFOsExmRseRecall}{1.000\xspace}
\newcommand{\XsfGccFOsExmRsePrecision}{1\xspace}
\newcommand{\XsfGccFOsExmRseFone}{1.000\xspace}
\newcommand{\XsfGccFOsLgtGT}{22549\xspace}
\newcommand{\XsfGccFOsLgtLsRecall}{1\xspace}
\newcommand{\XsfGccFOsLgtLsPrecision}{1\xspace}
\newcommand{\XsfGccFOsLgtLsFone}{1\xspace}
\newcommand{\XsfGccFOsLgtBapRecall}{0.914\xspace}
\newcommand{\XsfGccFOsLgtBapPrecision}{1\xspace}
\newcommand{\XsfGccFOsLgtBapFone}{0.955\xspace}
\newcommand{\XsfGccFOsLgtGhiRecall}{1\xspace}
\newcommand{\XsfGccFOsLgtGhiPrecision}{1\xspace}
\newcommand{\XsfGccFOsLgtGhiFone}{1\xspace}
\newcommand{\XsfGccFOsLgtRdaRecall}{1.000\xspace}
\newcommand{\XsfGccFOsLgtRdaPrecision}{1.000\xspace}
\newcommand{\XsfGccFOsLgtRdaFone}{1.000\xspace}
\newcommand{\XsfGccFOsLgtRseRecall}{1\xspace}
\newcommand{\XsfGccFOsLgtRsePrecision}{1\xspace}
\newcommand{\XsfGccFOsLgtRseFone}{1\xspace}
\newcommand{\XsfGccFOsBzpGT}{10941\xspace}
\newcommand{\XsfGccFOsBzpLsRecall}{1\xspace}
\newcommand{\XsfGccFOsBzpLsPrecision}{1\xspace}
\newcommand{\XsfGccFOsBzpLsFone}{1\xspace}
\newcommand{\XsfGccFOsBzpBapRecall}{0.798\xspace}
\newcommand{\XsfGccFOsBzpBapPrecision}{1\xspace}
\newcommand{\XsfGccFOsBzpBapFone}{0.887\xspace}
\newcommand{\XsfGccFOsBzpGhiRecall}{1\xspace}
\newcommand{\XsfGccFOsBzpGhiPrecision}{1\xspace}
\newcommand{\XsfGccFOsBzpGhiFone}{1\xspace}
\newcommand{\XsfGccFOsBzpRdaRecall}{0.994\xspace}
\newcommand{\XsfGccFOsBzpRdaPrecision}{1\xspace}
\newcommand{\XsfGccFOsBzpRdaFone}{0.997\xspace}
\newcommand{\XsfGccFOsBzpRseRecall}{1\xspace}
\newcommand{\XsfGccFOsBzpRsePrecision}{1\xspace}
\newcommand{\XsfGccFOsBzpRseFone}{1\xspace}
\newcommand{\XsfGccFOsGccGT}{715124\xspace}
\newcommand{\XsfGccFOsGccLsRecall}{1\xspace}
\newcommand{\XsfGccFOsGccLsPrecision}{1\xspace}
\newcommand{\XsfGccFOsGccLsFone}{1\xspace}
\newcommand{\XsfGccFOsGccBapRecall}{0.777\xspace}
\newcommand{\XsfGccFOsGccBapPrecision}{1.000\xspace}
\newcommand{\XsfGccFOsGccBapFone}{0.875\xspace}
\newcommand{\XsfGccFOsGccGhiRecall}{0.986\xspace}
\newcommand{\XsfGccFOsGccGhiPrecision}{1\xspace}
\newcommand{\XsfGccFOsGccGhiFone}{0.993\xspace}
\newcommand{\XsfGccFOsGccRdaRecall}{0.836\xspace}
\newcommand{\XsfGccFOsGccRdaPrecision}{1.000\xspace}
\newcommand{\XsfGccFOsGccRdaFone}{0.911\xspace}
\newcommand{\XsfGccFOsGccRseRecall}{0.997\xspace}
\newcommand{\XsfGccFOsGccRsePrecision}{1\xspace}
\newcommand{\XsfGccFOsGccRseFone}{0.998\xspace}
\newcommand{\XsfGccFOsGzpGT}{8184\xspace}
\newcommand{\XsfGccFOsGzpLsRecall}{1\xspace}
\newcommand{\XsfGccFOsGzpLsPrecision}{1\xspace}
\newcommand{\XsfGccFOsGzpLsFone}{1\xspace}
\newcommand{\XsfGccFOsGzpBapRecall}{0.985\xspace}
\newcommand{\XsfGccFOsGzpBapPrecision}{1\xspace}
\newcommand{\XsfGccFOsGzpBapFone}{0.992\xspace}
\newcommand{\XsfGccFOsGzpGhiRecall}{1\xspace}
\newcommand{\XsfGccFOsGzpGhiPrecision}{1\xspace}
\newcommand{\XsfGccFOsGzpGhiFone}{1\xspace}
\newcommand{\XsfGccFOsGzpRdaRecall}{1\xspace}
\newcommand{\XsfGccFOsGzpRdaPrecision}{1\xspace}
\newcommand{\XsfGccFOsGzpRdaFone}{1\xspace}
\newcommand{\XsfGccFOsGzpRseRecall}{1\xspace}
\newcommand{\XsfGccFOsGzpRsePrecision}{1\xspace}
\newcommand{\XsfGccFOsGzpRseFone}{1\xspace}
\newcommand{\XsfGccFOsOggGT}{30078\xspace}
\newcommand{\XsfGccFOsOggLsRecall}{1\xspace}
\newcommand{\XsfGccFOsOggLsPrecision}{1\xspace}
\newcommand{\XsfGccFOsOggLsFone}{1\xspace}
\newcommand{\XsfGccFOsOggBapRecall}{0.997\xspace}
\newcommand{\XsfGccFOsOggBapPrecision}{1.000\xspace}
\newcommand{\XsfGccFOsOggBapFone}{0.999\xspace}
\newcommand{\XsfGccFOsOggGhiRecall}{1\xspace}
\newcommand{\XsfGccFOsOggGhiPrecision}{1\xspace}
\newcommand{\XsfGccFOsOggGhiFone}{1\xspace}
\newcommand{\XsfGccFOsOggRdaRecall}{1\xspace}
\newcommand{\XsfGccFOsOggRdaPrecision}{1\xspace}
\newcommand{\XsfGccFOsOggRdaFone}{1\xspace}
\newcommand{\XsfGccFOsOggRseRecall}{1\xspace}
\newcommand{\XsfGccFOsOggRsePrecision}{1\xspace}
\newcommand{\XsfGccFOsOggRseFone}{1\xspace}
\newcommand{\XsfGccFOsNgxGT}{85741\xspace}
\newcommand{\XsfGccFOsNgxLsRecall}{1\xspace}
\newcommand{\XsfGccFOsNgxLsPrecision}{1\xspace}
\newcommand{\XsfGccFOsNgxLsFone}{1\xspace}
\newcommand{\XsfGccFOsNgxBapRecall}{0.976\xspace}
\newcommand{\XsfGccFOsNgxBapPrecision}{1.000\xspace}
\newcommand{\XsfGccFOsNgxBapFone}{0.988\xspace}
\newcommand{\XsfGccFOsNgxGhiRecall}{1\xspace}
\newcommand{\XsfGccFOsNgxGhiPrecision}{1\xspace}
\newcommand{\XsfGccFOsNgxGhiFone}{1\xspace}
\newcommand{\XsfGccFOsNgxRdaRecall}{0.981\xspace}
\newcommand{\XsfGccFOsNgxRdaPrecision}{1.000\xspace}
\newcommand{\XsfGccFOsNgxRdaFone}{0.990\xspace}
\newcommand{\XsfGccFOsNgxRseRecall}{1\xspace}
\newcommand{\XsfGccFOsNgxRsePrecision}{1\xspace}
\newcommand{\XsfGccFOsNgxRseFone}{1\xspace}
\newcommand{\XsfGccFOsSshGT}{88291\xspace}
\newcommand{\XsfGccFOsSshLsRecall}{1\xspace}
\newcommand{\XsfGccFOsSshLsPrecision}{1\xspace}
\newcommand{\XsfGccFOsSshLsFone}{1\xspace}
\newcommand{\XsfGccFOsSshBapRecall}{0.957\xspace}
\newcommand{\XsfGccFOsSshBapPrecision}{1\xspace}
\newcommand{\XsfGccFOsSshBapFone}{0.978\xspace}
\newcommand{\XsfGccFOsSshGhiRecall}{0.999\xspace}
\newcommand{\XsfGccFOsSshGhiPrecision}{1\xspace}
\newcommand{\XsfGccFOsSshGhiFone}{0.999\xspace}
\newcommand{\XsfGccFOsSshRdaRecall}{0.970\xspace}
\newcommand{\XsfGccFOsSshRdaPrecision}{1.000\xspace}
\newcommand{\XsfGccFOsSshRdaFone}{0.985\xspace}
\newcommand{\XsfGccFOsSshRseRecall}{0.987\xspace}
\newcommand{\XsfGccFOsSshRsePrecision}{1\xspace}
\newcommand{\XsfGccFOsSshRseFone}{0.993\xspace}
\newcommand{\XsfGccFOsPcrGT}{3889\xspace}
\newcommand{\XsfGccFOsPcrLsRecall}{1\xspace}
\newcommand{\XsfGccFOsPcrLsPrecision}{1\xspace}
\newcommand{\XsfGccFOsPcrLsFone}{1\xspace}
\newcommand{\XsfGccFOsPcrBapRecall}{0.947\xspace}
\newcommand{\XsfGccFOsPcrBapPrecision}{1\xspace}
\newcommand{\XsfGccFOsPcrBapFone}{0.973\xspace}
\newcommand{\XsfGccFOsPcrGhiRecall}{1\xspace}
\newcommand{\XsfGccFOsPcrGhiPrecision}{1\xspace}
\newcommand{\XsfGccFOsPcrGhiFone}{1\xspace}
\newcommand{\XsfGccFOsPcrRdaRecall}{1\xspace}
\newcommand{\XsfGccFOsPcrRdaPrecision}{1\xspace}
\newcommand{\XsfGccFOsPcrRdaFone}{1\xspace}
\newcommand{\XsfGccFOsPcrRseRecall}{1\xspace}
\newcommand{\XsfGccFOsPcrRsePrecision}{1\xspace}
\newcommand{\XsfGccFOsPcrRseFone}{1\xspace}
\newcommand{\XsfGccFOsSqlGT}{133739\xspace}
\newcommand{\XsfGccFOsSqlLsRecall}{1\xspace}
\newcommand{\XsfGccFOsSqlLsPrecision}{1\xspace}
\newcommand{\XsfGccFOsSqlLsFone}{1\xspace}
\newcommand{\XsfGccFOsSqlBapRecall}{0.871\xspace}
\newcommand{\XsfGccFOsSqlBapPrecision}{1\xspace}
\newcommand{\XsfGccFOsSqlBapFone}{0.931\xspace}
\newcommand{\XsfGccFOsSqlGhiRecall}{1\xspace}
\newcommand{\XsfGccFOsSqlGhiPrecision}{1\xspace}
\newcommand{\XsfGccFOsSqlGhiFone}{1\xspace}
\newcommand{\XsfGccFOsSqlRdaRecall}{0.994\xspace}
\newcommand{\XsfGccFOsSqlRdaPrecision}{1.000\xspace}
\newcommand{\XsfGccFOsSqlRdaFone}{0.997\xspace}
\newcommand{\XsfGccFOsSqlRseRecall}{1\xspace}
\newcommand{\XsfGccFOsSqlRsePrecision}{1\xspace}
\newcommand{\XsfGccFOsSqlRseFone}{1\xspace}
\newcommand{\XsfGccFOsVimGT}{407404\xspace}
\newcommand{\XsfGccFOsVimLsRecall}{1\xspace}
\newcommand{\XsfGccFOsVimLsPrecision}{1\xspace}
\newcommand{\XsfGccFOsVimLsFone}{1\xspace}
\newcommand{\XsfGccFOsVimBapRecall}{0.965\xspace}
\newcommand{\XsfGccFOsVimBapPrecision}{1.000\xspace}
\newcommand{\XsfGccFOsVimBapFone}{0.982\xspace}
\newcommand{\XsfGccFOsVimGhiRecall}{1.000\xspace}
\newcommand{\XsfGccFOsVimGhiPrecision}{1\xspace}
\newcommand{\XsfGccFOsVimGhiFone}{1.000\xspace}
\newcommand{\XsfGccFOsVimRdaRecall}{0.978\xspace}
\newcommand{\XsfGccFOsVimRdaPrecision}{1.000\xspace}
\newcommand{\XsfGccFOsVimRdaFone}{0.989\xspace}
\newcommand{\XsfGccFOsVimRseRecall}{1.000\xspace}
\newcommand{\XsfGccFOsVimRsePrecision}{1\xspace}
\newcommand{\XsfGccFOsVimRseFone}{1.000\xspace}
\newcommand{\XsfGccFOsVsfGT}{15728\xspace}
\newcommand{\XsfGccFOsVsfLsRecall}{1\xspace}
\newcommand{\XsfGccFOsVsfLsPrecision}{1\xspace}
\newcommand{\XsfGccFOsVsfLsFone}{1\xspace}
\newcommand{\XsfGccFOsVsfBapRecall}{0.998\xspace}
\newcommand{\XsfGccFOsVsfBapPrecision}{1.000\xspace}
\newcommand{\XsfGccFOsVsfBapFone}{0.999\xspace}
\newcommand{\XsfGccFOsVsfGhiRecall}{1\xspace}
\newcommand{\XsfGccFOsVsfGhiPrecision}{1\xspace}
\newcommand{\XsfGccFOsVsfGhiFone}{1\xspace}
\newcommand{\XsfGccFOsVsfRdaRecall}{0.991\xspace}
\newcommand{\XsfGccFOsVsfRdaPrecision}{1.000\xspace}
\newcommand{\XsfGccFOsVsfRdaFone}{0.996\xspace}
\newcommand{\XsfGccFOsVsfRseRecall}{1.000\xspace}
\newcommand{\XsfGccFOsVsfRsePrecision}{1\xspace}
\newcommand{\XsfGccFOsVsfRseFone}{1.000\xspace}
\newcommand{\XsfGccSOoSzpGT}{20539\xspace}
\newcommand{\XsfGccSOoSzpLsRecall}{1\xspace}
\newcommand{\XsfGccSOoSzpLsPrecision}{1\xspace}
\newcommand{\XsfGccSOoSzpLsFone}{1\xspace}
\newcommand{\XsfGccSOoSzpBapRecall}{0.976\xspace}
\newcommand{\XsfGccSOoSzpBapPrecision}{1\xspace}
\newcommand{\XsfGccSOoSzpBapFone}{0.988\xspace}
\newcommand{\XsfGccSOoSzpGhiRecall}{1\xspace}
\newcommand{\XsfGccSOoSzpGhiPrecision}{1\xspace}
\newcommand{\XsfGccSOoSzpGhiFone}{1\xspace}
\newcommand{\XsfGccSOoSzpRdaRecall}{1\xspace}
\newcommand{\XsfGccSOoSzpRdaPrecision}{1\xspace}
\newcommand{\XsfGccSOoSzpRdaFone}{1\xspace}
\newcommand{\XsfGccSOoSzpRseRecall}{1\xspace}
\newcommand{\XsfGccSOoSzpRsePrecision}{1\xspace}
\newcommand{\XsfGccSOoSzpRseFone}{1\xspace}
\newcommand{\XsfGccSOoCapGT}{318862\xspace}
\newcommand{\XsfGccSOoCapLsRecall}{1\xspace}
\newcommand{\XsfGccSOoCapLsPrecision}{1\xspace}
\newcommand{\XsfGccSOoCapLsFone}{1\xspace}
\newcommand{\XsfGccSOoCapBapRecall}{0.463\xspace}
\newcommand{\XsfGccSOoCapBapPrecision}{1\xspace}
\newcommand{\XsfGccSOoCapBapFone}{0.633\xspace}
\newcommand{\XsfGccSOoCapGhiRecall}{0.623\xspace}
\newcommand{\XsfGccSOoCapGhiPrecision}{0.999\xspace}
\newcommand{\XsfGccSOoCapGhiFone}{0.767\xspace}
\newcommand{\XsfGccSOoCapRdaRecall}{0.861\xspace}
\newcommand{\XsfGccSOoCapRdaPrecision}{1.000\xspace}
\newcommand{\XsfGccSOoCapRdaFone}{0.925\xspace}
\newcommand{\XsfGccSOoCapRseRecall}{0.530\xspace}
\newcommand{\XsfGccSOoCapRsePrecision}{1\xspace}
\newcommand{\XsfGccSOoCapRseFone}{0.693\xspace}
\newcommand{\XsfGccSOoExmGT}{183561\xspace}
\newcommand{\XsfGccSOoExmLsRecall}{1\xspace}
\newcommand{\XsfGccSOoExmLsPrecision}{1\xspace}
\newcommand{\XsfGccSOoExmLsFone}{1\xspace}
\newcommand{\XsfGccSOoExmBapRecall}{0.860\xspace}
\newcommand{\XsfGccSOoExmBapPrecision}{1\xspace}
\newcommand{\XsfGccSOoExmBapFone}{0.925\xspace}
\newcommand{\XsfGccSOoExmGhiRecall}{1\xspace}
\newcommand{\XsfGccSOoExmGhiPrecision}{1\xspace}
\newcommand{\XsfGccSOoExmGhiFone}{1\xspace}
\newcommand{\XsfGccSOoExmRdaRecall}{0.947\xspace}
\newcommand{\XsfGccSOoExmRdaPrecision}{1.000\xspace}
\newcommand{\XsfGccSOoExmRdaFone}{0.973\xspace}
\newcommand{\XsfGccSOoExmRseRecall}{0.956\xspace}
\newcommand{\XsfGccSOoExmRsePrecision}{1\xspace}
\newcommand{\XsfGccSOoExmRseFone}{0.977\xspace}
\newcommand{\XsfGccSOoLgtGT}{39246\xspace}
\newcommand{\XsfGccSOoLgtLsRecall}{1\xspace}
\newcommand{\XsfGccSOoLgtLsPrecision}{1\xspace}
\newcommand{\XsfGccSOoLgtLsFone}{1\xspace}
\newcommand{\XsfGccSOoLgtBapRecall}{0.874\xspace}
\newcommand{\XsfGccSOoLgtBapPrecision}{1\xspace}
\newcommand{\XsfGccSOoLgtBapFone}{0.933\xspace}
\newcommand{\XsfGccSOoLgtGhiRecall}{1\xspace}
\newcommand{\XsfGccSOoLgtGhiPrecision}{1\xspace}
\newcommand{\XsfGccSOoLgtGhiFone}{1\xspace}
\newcommand{\XsfGccSOoLgtRdaRecall}{0.952\xspace}
\newcommand{\XsfGccSOoLgtRdaPrecision}{0.999\xspace}
\newcommand{\XsfGccSOoLgtRdaFone}{0.975\xspace}
\newcommand{\XsfGccSOoLgtRseRecall}{0.978\xspace}
\newcommand{\XsfGccSOoLgtRsePrecision}{1\xspace}
\newcommand{\XsfGccSOoLgtRseFone}{0.989\xspace}
\newcommand{\XsfGccSOoBzpGT}{23963\xspace}
\newcommand{\XsfGccSOoBzpLsRecall}{1\xspace}
\newcommand{\XsfGccSOoBzpLsPrecision}{1\xspace}
\newcommand{\XsfGccSOoBzpLsFone}{1\xspace}
\newcommand{\XsfGccSOoBzpBapRecall}{0.787\xspace}
\newcommand{\XsfGccSOoBzpBapPrecision}{1\xspace}
\newcommand{\XsfGccSOoBzpBapFone}{0.881\xspace}
\newcommand{\XsfGccSOoBzpGhiRecall}{1\xspace}
\newcommand{\XsfGccSOoBzpGhiPrecision}{1\xspace}
\newcommand{\XsfGccSOoBzpGhiFone}{1\xspace}
\newcommand{\XsfGccSOoBzpRdaRecall}{0.888\xspace}
\newcommand{\XsfGccSOoBzpRdaPrecision}{1.000\xspace}
\newcommand{\XsfGccSOoBzpRdaFone}{0.941\xspace}
\newcommand{\XsfGccSOoBzpRseRecall}{0.999\xspace}
\newcommand{\XsfGccSOoBzpRsePrecision}{1\xspace}
\newcommand{\XsfGccSOoBzpRseFone}{0.999\xspace}
\newcommand{\XsfGccSOoGccGT}{1308204\xspace}
\newcommand{\XsfGccSOoGccLsRecall}{1\xspace}
\newcommand{\XsfGccSOoGccLsPrecision}{1\xspace}
\newcommand{\XsfGccSOoGccLsFone}{1\xspace}
\newcommand{\XsfGccSOoGccBapRecall}{0.715\xspace}
\newcommand{\XsfGccSOoGccBapPrecision}{1\xspace}
\newcommand{\XsfGccSOoGccBapFone}{0.834\xspace}
\newcommand{\XsfGccSOoGccGhiRecall}{0.652\xspace}
\newcommand{\XsfGccSOoGccGhiPrecision}{0.997\xspace}
\newcommand{\XsfGccSOoGccGhiFone}{0.788\xspace}
\newcommand{\XsfGccSOoGccRdaRecall}{0.839\xspace}
\newcommand{\XsfGccSOoGccRdaPrecision}{1.000\xspace}
\newcommand{\XsfGccSOoGccRdaFone}{0.912\xspace}
\newcommand{\XsfGccSOoGccRseRecall}{0.857\xspace}
\newcommand{\XsfGccSOoGccRsePrecision}{1\xspace}
\newcommand{\XsfGccSOoGccRseFone}{0.923\xspace}
\newcommand{\XsfGccSOoGzpGT}{13111\xspace}
\newcommand{\XsfGccSOoGzpLsRecall}{1\xspace}
\newcommand{\XsfGccSOoGzpLsPrecision}{1\xspace}
\newcommand{\XsfGccSOoGzpLsFone}{1\xspace}
\newcommand{\XsfGccSOoGzpBapRecall}{0.992\xspace}
\newcommand{\XsfGccSOoGzpBapPrecision}{1\xspace}
\newcommand{\XsfGccSOoGzpBapFone}{0.996\xspace}
\newcommand{\XsfGccSOoGzpGhiRecall}{1\xspace}
\newcommand{\XsfGccSOoGzpGhiPrecision}{1\xspace}
\newcommand{\XsfGccSOoGzpGhiFone}{1\xspace}
\newcommand{\XsfGccSOoGzpRdaRecall}{1\xspace}
\newcommand{\XsfGccSOoGzpRdaPrecision}{1.000\xspace}
\newcommand{\XsfGccSOoGzpRdaFone}{1.000\xspace}
\newcommand{\XsfGccSOoGzpRseRecall}{0.996\xspace}
\newcommand{\XsfGccSOoGzpRsePrecision}{1\xspace}
\newcommand{\XsfGccSOoGzpRseFone}{0.998\xspace}
\newcommand{\XsfGccSOoOggGT}{58272\xspace}
\newcommand{\XsfGccSOoOggLsRecall}{1\xspace}
\newcommand{\XsfGccSOoOggLsPrecision}{1\xspace}
\newcommand{\XsfGccSOoOggLsFone}{1\xspace}
\newcommand{\XsfGccSOoOggBapRecall}{0.978\xspace}
\newcommand{\XsfGccSOoOggBapPrecision}{1\xspace}
\newcommand{\XsfGccSOoOggBapFone}{0.989\xspace}
\newcommand{\XsfGccSOoOggGhiRecall}{1\xspace}
\newcommand{\XsfGccSOoOggGhiPrecision}{1\xspace}
\newcommand{\XsfGccSOoOggGhiFone}{1\xspace}
\newcommand{\XsfGccSOoOggRdaRecall}{0.998\xspace}
\newcommand{\XsfGccSOoOggRdaPrecision}{1\xspace}
\newcommand{\XsfGccSOoOggRdaFone}{0.999\xspace}
\newcommand{\XsfGccSOoOggRseRecall}{0.995\xspace}
\newcommand{\XsfGccSOoOggRsePrecision}{1\xspace}
\newcommand{\XsfGccSOoOggRseFone}{0.998\xspace}
\newcommand{\XsfGccSOoNgxGT}{165941\xspace}
\newcommand{\XsfGccSOoNgxLsRecall}{1\xspace}
\newcommand{\XsfGccSOoNgxLsPrecision}{1\xspace}
\newcommand{\XsfGccSOoNgxLsFone}{1\xspace}
\newcommand{\XsfGccSOoNgxBapRecall}{0.967\xspace}
\newcommand{\XsfGccSOoNgxBapPrecision}{1\xspace}
\newcommand{\XsfGccSOoNgxBapFone}{0.983\xspace}
\newcommand{\XsfGccSOoNgxGhiRecall}{1.000\xspace}
\newcommand{\XsfGccSOoNgxGhiPrecision}{1\xspace}
\newcommand{\XsfGccSOoNgxGhiFone}{1.000\xspace}
\newcommand{\XsfGccSOoNgxRdaRecall}{0.993\xspace}
\newcommand{\XsfGccSOoNgxRdaPrecision}{1.000\xspace}
\newcommand{\XsfGccSOoNgxRdaFone}{0.996\xspace}
\newcommand{\XsfGccSOoNgxRseRecall}{0.992\xspace}
\newcommand{\XsfGccSOoNgxRsePrecision}{1\xspace}
\newcommand{\XsfGccSOoNgxRseFone}{0.996\xspace}
\newcommand{\XsfGccSOoSshGT}{139322\xspace}
\newcommand{\XsfGccSOoSshLsRecall}{1\xspace}
\newcommand{\XsfGccSOoSshLsPrecision}{1\xspace}
\newcommand{\XsfGccSOoSshLsFone}{1\xspace}
\newcommand{\XsfGccSOoSshBapRecall}{0.951\xspace}
\newcommand{\XsfGccSOoSshBapPrecision}{1\xspace}
\newcommand{\XsfGccSOoSshBapFone}{0.975\xspace}
\newcommand{\XsfGccSOoSshGhiRecall}{1\xspace}
\newcommand{\XsfGccSOoSshGhiPrecision}{1\xspace}
\newcommand{\XsfGccSOoSshGhiFone}{1\xspace}
\newcommand{\XsfGccSOoSshRdaRecall}{0.998\xspace}
\newcommand{\XsfGccSOoSshRdaPrecision}{1\xspace}
\newcommand{\XsfGccSOoSshRdaFone}{0.999\xspace}
\newcommand{\XsfGccSOoSshRseRecall}{0.980\xspace}
\newcommand{\XsfGccSOoSshRsePrecision}{1\xspace}
\newcommand{\XsfGccSOoSshRseFone}{0.990\xspace}
\newcommand{\XsfGccSOoPcrGT}{6013\xspace}
\newcommand{\XsfGccSOoPcrLsRecall}{1\xspace}
\newcommand{\XsfGccSOoPcrLsPrecision}{1\xspace}
\newcommand{\XsfGccSOoPcrLsFone}{1\xspace}
\newcommand{\XsfGccSOoPcrBapRecall}{0.909\xspace}
\newcommand{\XsfGccSOoPcrBapPrecision}{1\xspace}
\newcommand{\XsfGccSOoPcrBapFone}{0.952\xspace}
\newcommand{\XsfGccSOoPcrGhiRecall}{1\xspace}
\newcommand{\XsfGccSOoPcrGhiPrecision}{1\xspace}
\newcommand{\XsfGccSOoPcrGhiFone}{1\xspace}
\newcommand{\XsfGccSOoPcrRdaRecall}{0.991\xspace}
\newcommand{\XsfGccSOoPcrRdaPrecision}{0.999\xspace}
\newcommand{\XsfGccSOoPcrRdaFone}{0.995\xspace}
\newcommand{\XsfGccSOoPcrRseRecall}{0.988\xspace}
\newcommand{\XsfGccSOoPcrRsePrecision}{1\xspace}
\newcommand{\XsfGccSOoPcrRseFone}{0.994\xspace}
\newcommand{\XsfGccSOoSqlGT}{226640\xspace}
\newcommand{\XsfGccSOoSqlLsRecall}{1\xspace}
\newcommand{\XsfGccSOoSqlLsPrecision}{1\xspace}
\newcommand{\XsfGccSOoSqlLsFone}{1\xspace}
\newcommand{\XsfGccSOoSqlBapRecall}{0.879\xspace}
\newcommand{\XsfGccSOoSqlBapPrecision}{1\xspace}
\newcommand{\XsfGccSOoSqlBapFone}{0.936\xspace}
\newcommand{\XsfGccSOoSqlGhiRecall}{1\xspace}
\newcommand{\XsfGccSOoSqlGhiPrecision}{1\xspace}
\newcommand{\XsfGccSOoSqlGhiFone}{1\xspace}
\newcommand{\XsfGccSOoSqlRdaRecall}{0.932\xspace}
\newcommand{\XsfGccSOoSqlRdaPrecision}{1.000\xspace}
\newcommand{\XsfGccSOoSqlRdaFone}{0.965\xspace}
\newcommand{\XsfGccSOoSqlRseRecall}{0.921\xspace}
\newcommand{\XsfGccSOoSqlRsePrecision}{1\xspace}
\newcommand{\XsfGccSOoSqlRseFone}{0.959\xspace}
\newcommand{\XsfGccSOoVimGT}{643305\xspace}
\newcommand{\XsfGccSOoVimLsRecall}{1\xspace}
\newcommand{\XsfGccSOoVimLsPrecision}{1\xspace}
\newcommand{\XsfGccSOoVimLsFone}{1\xspace}
\newcommand{\XsfGccSOoVimBapRecall}{0.947\xspace}
\newcommand{\XsfGccSOoVimBapPrecision}{1\xspace}
\newcommand{\XsfGccSOoVimBapFone}{0.973\xspace}
\newcommand{\XsfGccSOoVimGhiRecall}{0.748\xspace}
\newcommand{\XsfGccSOoVimGhiPrecision}{0.999\xspace}
\newcommand{\XsfGccSOoVimGhiFone}{0.855\xspace}
\newcommand{\XsfGccSOoVimRdaRecall}{0.981\xspace}
\newcommand{\XsfGccSOoVimRdaPrecision}{1.000\xspace}
\newcommand{\XsfGccSOoVimRdaFone}{0.990\xspace}
\newcommand{\XsfGccSOoVimRseRecall}{0.982\xspace}
\newcommand{\XsfGccSOoVimRsePrecision}{1\xspace}
\newcommand{\XsfGccSOoVimRseFone}{0.991\xspace}
\newcommand{\XsfGccSOoVsfGT}{24167\xspace}
\newcommand{\XsfGccSOoVsfLsRecall}{1\xspace}
\newcommand{\XsfGccSOoVsfLsPrecision}{1\xspace}
\newcommand{\XsfGccSOoVsfLsFone}{1\xspace}
\newcommand{\XsfGccSOoVsfBapRecall}{0.997\xspace}
\newcommand{\XsfGccSOoVsfBapPrecision}{1\xspace}
\newcommand{\XsfGccSOoVsfBapFone}{0.999\xspace}
\newcommand{\XsfGccSOoVsfGhiRecall}{1\xspace}
\newcommand{\XsfGccSOoVsfGhiPrecision}{1\xspace}
\newcommand{\XsfGccSOoVsfGhiFone}{1\xspace}
\newcommand{\XsfGccSOoVsfRdaRecall}{0.991\xspace}
\newcommand{\XsfGccSOoVsfRdaPrecision}{1.000\xspace}
\newcommand{\XsfGccSOoVsfRdaFone}{0.996\xspace}
\newcommand{\XsfGccSOoVsfRseRecall}{1.000\xspace}
\newcommand{\XsfGccSOoVsfRsePrecision}{1\xspace}
\newcommand{\XsfGccSOoVsfRseFone}{1.000\xspace}
\newcommand{\XsfGccSOaSzpGT}{12309\xspace}
\newcommand{\XsfGccSOaSzpLsRecall}{1\xspace}
\newcommand{\XsfGccSOaSzpLsPrecision}{1\xspace}
\newcommand{\XsfGccSOaSzpLsFone}{1\xspace}
\newcommand{\XsfGccSOaSzpBapRecall}{0.973\xspace}
\newcommand{\XsfGccSOaSzpBapPrecision}{1\xspace}
\newcommand{\XsfGccSOaSzpBapFone}{0.986\xspace}
\newcommand{\XsfGccSOaSzpGhiRecall}{1\xspace}
\newcommand{\XsfGccSOaSzpGhiPrecision}{1\xspace}
\newcommand{\XsfGccSOaSzpGhiFone}{1\xspace}
\newcommand{\XsfGccSOaSzpRdaRecall}{1\xspace}
\newcommand{\XsfGccSOaSzpRdaPrecision}{1\xspace}
\newcommand{\XsfGccSOaSzpRdaFone}{1\xspace}
\newcommand{\XsfGccSOaSzpRseRecall}{1\xspace}
\newcommand{\XsfGccSOaSzpRsePrecision}{1\xspace}
\newcommand{\XsfGccSOaSzpRseFone}{1\xspace}
\newcommand{\XsfGccSOaCapGT}{226265\xspace}
\newcommand{\XsfGccSOaCapLsRecall}{1\xspace}
\newcommand{\XsfGccSOaCapLsPrecision}{1\xspace}
\newcommand{\XsfGccSOaCapLsFone}{1\xspace}
\newcommand{\XsfGccSOaCapBapRecall}{0.416\xspace}
\newcommand{\XsfGccSOaCapBapPrecision}{1\xspace}
\newcommand{\XsfGccSOaCapBapFone}{0.588\xspace}
\newcommand{\XsfGccSOaCapGhiRecall}{0.593\xspace}
\newcommand{\XsfGccSOaCapGhiPrecision}{1\xspace}
\newcommand{\XsfGccSOaCapGhiFone}{0.745\xspace}
\newcommand{\XsfGccSOaCapRdaRecall}{0.473\xspace}
\newcommand{\XsfGccSOaCapRdaPrecision}{0.998\xspace}
\newcommand{\XsfGccSOaCapRdaFone}{0.642\xspace}
\newcommand{\XsfGccSOaCapRseRecall}{0.484\xspace}
\newcommand{\XsfGccSOaCapRsePrecision}{1\xspace}
\newcommand{\XsfGccSOaCapRseFone}{0.652\xspace}
\newcommand{\XsfGccSOaExmGT}{141294\xspace}
\newcommand{\XsfGccSOaExmLsRecall}{1\xspace}
\newcommand{\XsfGccSOaExmLsPrecision}{1\xspace}
\newcommand{\XsfGccSOaExmLsFone}{1\xspace}
\newcommand{\XsfGccSOaExmBapRecall}{0.845\xspace}
\newcommand{\XsfGccSOaExmBapPrecision}{1.000\xspace}
\newcommand{\XsfGccSOaExmBapFone}{0.916\xspace}
\newcommand{\XsfGccSOaExmGhiRecall}{0.998\xspace}
\newcommand{\XsfGccSOaExmGhiPrecision}{1\xspace}
\newcommand{\XsfGccSOaExmGhiFone}{0.999\xspace}
\newcommand{\XsfGccSOaExmRdaRecall}{0.897\xspace}
\newcommand{\XsfGccSOaExmRdaPrecision}{1.000\xspace}
\newcommand{\XsfGccSOaExmRdaFone}{0.946\xspace}
\newcommand{\XsfGccSOaExmRseRecall}{0.986\xspace}
\newcommand{\XsfGccSOaExmRsePrecision}{1\xspace}
\newcommand{\XsfGccSOaExmRseFone}{0.993\xspace}
\newcommand{\XsfGccSOaLgtGT}{25706\xspace}
\newcommand{\XsfGccSOaLgtLsRecall}{1\xspace}
\newcommand{\XsfGccSOaLgtLsPrecision}{1\xspace}
\newcommand{\XsfGccSOaLgtLsFone}{1\xspace}
\newcommand{\XsfGccSOaLgtBapRecall}{0.881\xspace}
\newcommand{\XsfGccSOaLgtBapPrecision}{1\xspace}
\newcommand{\XsfGccSOaLgtBapFone}{0.937\xspace}
\newcommand{\XsfGccSOaLgtGhiRecall}{1\xspace}
\newcommand{\XsfGccSOaLgtGhiPrecision}{1\xspace}
\newcommand{\XsfGccSOaLgtGhiFone}{1\xspace}
\newcommand{\XsfGccSOaLgtRdaRecall}{0.974\xspace}
\newcommand{\XsfGccSOaLgtRdaPrecision}{1.000\xspace}
\newcommand{\XsfGccSOaLgtRdaFone}{0.987\xspace}
\newcommand{\XsfGccSOaLgtRseRecall}{0.978\xspace}
\newcommand{\XsfGccSOaLgtRsePrecision}{1\xspace}
\newcommand{\XsfGccSOaLgtRseFone}{0.989\xspace}
\newcommand{\XsfGccSOaBzpGT}{14438\xspace}
\newcommand{\XsfGccSOaBzpLsRecall}{1\xspace}
\newcommand{\XsfGccSOaBzpLsPrecision}{1\xspace}
\newcommand{\XsfGccSOaBzpLsFone}{1\xspace}
\newcommand{\XsfGccSOaBzpBapRecall}{0.828\xspace}
\newcommand{\XsfGccSOaBzpBapPrecision}{1\xspace}
\newcommand{\XsfGccSOaBzpBapFone}{0.906\xspace}
\newcommand{\XsfGccSOaBzpGhiRecall}{1\xspace}
\newcommand{\XsfGccSOaBzpGhiPrecision}{1\xspace}
\newcommand{\XsfGccSOaBzpGhiFone}{1\xspace}
\newcommand{\XsfGccSOaBzpRdaRecall}{0.881\xspace}
\newcommand{\XsfGccSOaBzpRdaPrecision}{1\xspace}
\newcommand{\XsfGccSOaBzpRdaFone}{0.937\xspace}
\newcommand{\XsfGccSOaBzpRseRecall}{0.999\xspace}
\newcommand{\XsfGccSOaBzpRsePrecision}{1\xspace}
\newcommand{\XsfGccSOaBzpRseFone}{0.999\xspace}
\newcommand{\XsfGccSOaGccGT}{826871\xspace}
\newcommand{\XsfGccSOaGccLsRecall}{1\xspace}
\newcommand{\XsfGccSOaGccLsPrecision}{1\xspace}
\newcommand{\XsfGccSOaGccLsFone}{1\xspace}
\newcommand{\XsfGccSOaGccBapRecall}{0.691\xspace}
\newcommand{\XsfGccSOaGccBapPrecision}{1.000\xspace}
\newcommand{\XsfGccSOaGccBapFone}{0.817\xspace}
\newcommand{\XsfGccSOaGccGhiRecall}{0.594\xspace}
\newcommand{\XsfGccSOaGccGhiPrecision}{0.994\xspace}
\newcommand{\XsfGccSOaGccGhiFone}{0.744\xspace}
\newcommand{\XsfGccSOaGccRdaRecall}{0.793\xspace}
\newcommand{\XsfGccSOaGccRdaPrecision}{1.000\xspace}
\newcommand{\XsfGccSOaGccRdaFone}{0.884\xspace}
\newcommand{\XsfGccSOaGccRseRecall}{0.867\xspace}
\newcommand{\XsfGccSOaGccRsePrecision}{1.000\xspace}
\newcommand{\XsfGccSOaGccRseFone}{0.928\xspace}
\newcommand{\XsfGccSOaGzpGT}{9855\xspace}
\newcommand{\XsfGccSOaGzpLsRecall}{1\xspace}
\newcommand{\XsfGccSOaGzpLsPrecision}{1\xspace}
\newcommand{\XsfGccSOaGzpLsFone}{1\xspace}
\newcommand{\XsfGccSOaGzpBapRecall}{0.985\xspace}
\newcommand{\XsfGccSOaGzpBapPrecision}{1\xspace}
\newcommand{\XsfGccSOaGzpBapFone}{0.993\xspace}
\newcommand{\XsfGccSOaGzpGhiRecall}{1\xspace}
\newcommand{\XsfGccSOaGzpGhiPrecision}{1\xspace}
\newcommand{\XsfGccSOaGzpGhiFone}{1\xspace}
\newcommand{\XsfGccSOaGzpRdaRecall}{1\xspace}
\newcommand{\XsfGccSOaGzpRdaPrecision}{1\xspace}
\newcommand{\XsfGccSOaGzpRdaFone}{1\xspace}
\newcommand{\XsfGccSOaGzpRseRecall}{0.993\xspace}
\newcommand{\XsfGccSOaGzpRsePrecision}{1\xspace}
\newcommand{\XsfGccSOaGzpRseFone}{0.997\xspace}
\newcommand{\XsfGccSOaOggGT}{35922\xspace}
\newcommand{\XsfGccSOaOggLsRecall}{1\xspace}
\newcommand{\XsfGccSOaOggLsPrecision}{1\xspace}
\newcommand{\XsfGccSOaOggLsFone}{1\xspace}
\newcommand{\XsfGccSOaOggBapRecall}{0.977\xspace}
\newcommand{\XsfGccSOaOggBapPrecision}{1.000\xspace}
\newcommand{\XsfGccSOaOggBapFone}{0.988\xspace}
\newcommand{\XsfGccSOaOggGhiRecall}{1\xspace}
\newcommand{\XsfGccSOaOggGhiPrecision}{1\xspace}
\newcommand{\XsfGccSOaOggGhiFone}{1\xspace}
\newcommand{\XsfGccSOaOggRdaRecall}{0.999\xspace}
\newcommand{\XsfGccSOaOggRdaPrecision}{1.000\xspace}
\newcommand{\XsfGccSOaOggRdaFone}{0.999\xspace}
\newcommand{\XsfGccSOaOggRseRecall}{0.995\xspace}
\newcommand{\XsfGccSOaOggRsePrecision}{1\xspace}
\newcommand{\XsfGccSOaOggRseFone}{0.997\xspace}
\newcommand{\XsfGccSOaNgxGT}{99967\xspace}
\newcommand{\XsfGccSOaNgxLsRecall}{1\xspace}
\newcommand{\XsfGccSOaNgxLsPrecision}{1\xspace}
\newcommand{\XsfGccSOaNgxLsFone}{1\xspace}
\newcommand{\XsfGccSOaNgxBapRecall}{0.967\xspace}
\newcommand{\XsfGccSOaNgxBapPrecision}{1.000\xspace}
\newcommand{\XsfGccSOaNgxBapFone}{0.983\xspace}
\newcommand{\XsfGccSOaNgxGhiRecall}{1\xspace}
\newcommand{\XsfGccSOaNgxGhiPrecision}{1\xspace}
\newcommand{\XsfGccSOaNgxGhiFone}{1\xspace}
\newcommand{\XsfGccSOaNgxRdaRecall}{0.960\xspace}
\newcommand{\XsfGccSOaNgxRdaPrecision}{1.000\xspace}
\newcommand{\XsfGccSOaNgxRdaFone}{0.979\xspace}
\newcommand{\XsfGccSOaNgxRseRecall}{0.991\xspace}
\newcommand{\XsfGccSOaNgxRsePrecision}{1.000\xspace}
\newcommand{\XsfGccSOaNgxRseFone}{0.995\xspace}
\newcommand{\XsfGccSOaSshGT}{97382\xspace}
\newcommand{\XsfGccSOaSshLsRecall}{1\xspace}
\newcommand{\XsfGccSOaSshLsPrecision}{1\xspace}
\newcommand{\XsfGccSOaSshLsFone}{1\xspace}
\newcommand{\XsfGccSOaSshBapRecall}{0.941\xspace}
\newcommand{\XsfGccSOaSshBapPrecision}{1\xspace}
\newcommand{\XsfGccSOaSshBapFone}{0.970\xspace}
\newcommand{\XsfGccSOaSshGhiRecall}{1\xspace}
\newcommand{\XsfGccSOaSshGhiPrecision}{1\xspace}
\newcommand{\XsfGccSOaSshGhiFone}{1\xspace}
\newcommand{\XsfGccSOaSshRdaRecall}{0.952\xspace}
\newcommand{\XsfGccSOaSshRdaPrecision}{1\xspace}
\newcommand{\XsfGccSOaSshRdaFone}{0.975\xspace}
\newcommand{\XsfGccSOaSshRseRecall}{0.992\xspace}
\newcommand{\XsfGccSOaSshRsePrecision}{1\xspace}
\newcommand{\XsfGccSOaSshRseFone}{0.996\xspace}
\newcommand{\XsfGccSOaPcrGT}{4633\xspace}
\newcommand{\XsfGccSOaPcrLsRecall}{1\xspace}
\newcommand{\XsfGccSOaPcrLsPrecision}{1\xspace}
\newcommand{\XsfGccSOaPcrLsFone}{1\xspace}
\newcommand{\XsfGccSOaPcrBapRecall}{0.878\xspace}
\newcommand{\XsfGccSOaPcrBapPrecision}{1\xspace}
\newcommand{\XsfGccSOaPcrBapFone}{0.935\xspace}
\newcommand{\XsfGccSOaPcrGhiRecall}{1\xspace}
\newcommand{\XsfGccSOaPcrGhiPrecision}{1\xspace}
\newcommand{\XsfGccSOaPcrGhiFone}{1\xspace}
\newcommand{\XsfGccSOaPcrRdaRecall}{0.968\xspace}
\newcommand{\XsfGccSOaPcrRdaPrecision}{0.998\xspace}
\newcommand{\XsfGccSOaPcrRdaFone}{0.983\xspace}
\newcommand{\XsfGccSOaPcrRseRecall}{0.985\xspace}
\newcommand{\XsfGccSOaPcrRsePrecision}{1\xspace}
\newcommand{\XsfGccSOaPcrRseFone}{0.993\xspace}
\newcommand{\XsfGccSOaSqlGT}{152147\xspace}
\newcommand{\XsfGccSOaSqlLsRecall}{1\xspace}
\newcommand{\XsfGccSOaSqlLsPrecision}{1\xspace}
\newcommand{\XsfGccSOaSqlLsFone}{1\xspace}
\newcommand{\XsfGccSOaSqlBapRecall}{0.849\xspace}
\newcommand{\XsfGccSOaSqlBapPrecision}{1\xspace}
\newcommand{\XsfGccSOaSqlBapFone}{0.918\xspace}
\newcommand{\XsfGccSOaSqlGhiRecall}{1\xspace}
\newcommand{\XsfGccSOaSqlGhiPrecision}{1\xspace}
\newcommand{\XsfGccSOaSqlGhiFone}{1\xspace}
\newcommand{\XsfGccSOaSqlRdaRecall}{0.938\xspace}
\newcommand{\XsfGccSOaSqlRdaPrecision}{1.000\xspace}
\newcommand{\XsfGccSOaSqlRdaFone}{0.968\xspace}
\newcommand{\XsfGccSOaSqlRseRecall}{0.926\xspace}
\newcommand{\XsfGccSOaSqlRsePrecision}{1.000\xspace}
\newcommand{\XsfGccSOaSqlRseFone}{0.961\xspace}
\newcommand{\XsfGccSOaVimGT}{463329\xspace}
\newcommand{\XsfGccSOaVimLsRecall}{1\xspace}
\newcommand{\XsfGccSOaVimLsPrecision}{1\xspace}
\newcommand{\XsfGccSOaVimLsFone}{1\xspace}
\newcommand{\XsfGccSOaVimBapRecall}{0.920\xspace}
\newcommand{\XsfGccSOaVimBapPrecision}{1.000\xspace}
\newcommand{\XsfGccSOaVimBapFone}{0.958\xspace}
\newcommand{\XsfGccSOaVimGhiRecall}{0.754\xspace}
\newcommand{\XsfGccSOaVimGhiPrecision}{0.999\xspace}
\newcommand{\XsfGccSOaVimGhiFone}{0.859\xspace}
\newcommand{\XsfGccSOaVimRdaRecall}{0.947\xspace}
\newcommand{\XsfGccSOaVimRdaPrecision}{1.000\xspace}
\newcommand{\XsfGccSOaVimRdaFone}{0.973\xspace}
\newcommand{\XsfGccSOaVimRseRecall}{0.987\xspace}
\newcommand{\XsfGccSOaVimRsePrecision}{1\xspace}
\newcommand{\XsfGccSOaVimRseFone}{0.993\xspace}
\newcommand{\XsfGccSOaVsfGT}{17153\xspace}
\newcommand{\XsfGccSOaVsfLsRecall}{1\xspace}
\newcommand{\XsfGccSOaVsfLsPrecision}{1\xspace}
\newcommand{\XsfGccSOaVsfLsFone}{1\xspace}
\newcommand{\XsfGccSOaVsfBapRecall}{0.996\xspace}
\newcommand{\XsfGccSOaVsfBapPrecision}{1.000\xspace}
\newcommand{\XsfGccSOaVsfBapFone}{0.998\xspace}
\newcommand{\XsfGccSOaVsfGhiRecall}{1\xspace}
\newcommand{\XsfGccSOaVsfGhiPrecision}{1\xspace}
\newcommand{\XsfGccSOaVsfGhiFone}{1\xspace}
\newcommand{\XsfGccSOaVsfRdaRecall}{0.982\xspace}
\newcommand{\XsfGccSOaVsfRdaPrecision}{1.000\xspace}
\newcommand{\XsfGccSOaVsfRdaFone}{0.991\xspace}
\newcommand{\XsfGccSOaVsfRseRecall}{1.000\xspace}
\newcommand{\XsfGccSOaVsfRsePrecision}{1\xspace}
\newcommand{\XsfGccSOaVsfRseFone}{1.000\xspace}
\newcommand{\XsfGccSObSzpGT}{12156\xspace}
\newcommand{\XsfGccSObSzpLsRecall}{1\xspace}
\newcommand{\XsfGccSObSzpLsPrecision}{1\xspace}
\newcommand{\XsfGccSObSzpLsFone}{1\xspace}
\newcommand{\XsfGccSObSzpBapRecall}{0.976\xspace}
\newcommand{\XsfGccSObSzpBapPrecision}{1\xspace}
\newcommand{\XsfGccSObSzpBapFone}{0.988\xspace}
\newcommand{\XsfGccSObSzpGhiRecall}{1\xspace}
\newcommand{\XsfGccSObSzpGhiPrecision}{1\xspace}
\newcommand{\XsfGccSObSzpGhiFone}{1\xspace}
\newcommand{\XsfGccSObSzpRdaRecall}{0.993\xspace}
\newcommand{\XsfGccSObSzpRdaPrecision}{1\xspace}
\newcommand{\XsfGccSObSzpRdaFone}{0.996\xspace}
\newcommand{\XsfGccSObSzpRseRecall}{1\xspace}
\newcommand{\XsfGccSObSzpRsePrecision}{1\xspace}
\newcommand{\XsfGccSObSzpRseFone}{1\xspace}
\newcommand{\XsfGccSObCapGT}{174572\xspace}
\newcommand{\XsfGccSObCapLsRecall}{1\xspace}
\newcommand{\XsfGccSObCapLsPrecision}{1\xspace}
\newcommand{\XsfGccSObCapLsFone}{1\xspace}
\newcommand{\XsfGccSObCapBapRecall}{0.503\xspace}
\newcommand{\XsfGccSObCapBapPrecision}{1\xspace}
\newcommand{\XsfGccSObCapBapFone}{0.669\xspace}
\newcommand{\XsfGccSObCapGhiRecall}{0.801\xspace}
\newcommand{\XsfGccSObCapGhiPrecision}{1\xspace}
\newcommand{\XsfGccSObCapGhiFone}{0.890\xspace}
\newcommand{\XsfGccSObCapRdaRecall}{0.535\xspace}
\newcommand{\XsfGccSObCapRdaPrecision}{0.996\xspace}
\newcommand{\XsfGccSObCapRdaFone}{0.696\xspace}
\newcommand{\XsfGccSObCapRseRecall}{0.679\xspace}
\newcommand{\XsfGccSObCapRsePrecision}{1\xspace}
\newcommand{\XsfGccSObCapRseFone}{0.809\xspace}
\newcommand{\XsfGccSObExmGT}{142689\xspace}
\newcommand{\XsfGccSObExmLsRecall}{1\xspace}
\newcommand{\XsfGccSObExmLsPrecision}{1\xspace}
\newcommand{\XsfGccSObExmLsFone}{1\xspace}
\newcommand{\XsfGccSObExmBapRecall}{0.862\xspace}
\newcommand{\XsfGccSObExmBapPrecision}{1.000\xspace}
\newcommand{\XsfGccSObExmBapFone}{0.926\xspace}
\newcommand{\XsfGccSObExmGhiRecall}{0.967\xspace}
\newcommand{\XsfGccSObExmGhiPrecision}{1\xspace}
\newcommand{\XsfGccSObExmGhiFone}{0.983\xspace}
\newcommand{\XsfGccSObExmRdaRecall}{0.880\xspace}
\newcommand{\XsfGccSObExmRdaPrecision}{1.000\xspace}
\newcommand{\XsfGccSObExmRdaFone}{0.936\xspace}
\newcommand{\XsfGccSObExmRseRecall}{0.994\xspace}
\newcommand{\XsfGccSObExmRsePrecision}{1\xspace}
\newcommand{\XsfGccSObExmRseFone}{0.997\xspace}
\newcommand{\XsfGccSObLgtGT}{25616\xspace}
\newcommand{\XsfGccSObLgtLsRecall}{1\xspace}
\newcommand{\XsfGccSObLgtLsPrecision}{1\xspace}
\newcommand{\XsfGccSObLgtLsFone}{1\xspace}
\newcommand{\XsfGccSObLgtBapRecall}{0.881\xspace}
\newcommand{\XsfGccSObLgtBapPrecision}{1\xspace}
\newcommand{\XsfGccSObLgtBapFone}{0.937\xspace}
\newcommand{\XsfGccSObLgtGhiRecall}{0.970\xspace}
\newcommand{\XsfGccSObLgtGhiPrecision}{1\xspace}
\newcommand{\XsfGccSObLgtGhiFone}{0.985\xspace}
\newcommand{\XsfGccSObLgtRdaRecall}{0.981\xspace}
\newcommand{\XsfGccSObLgtRdaPrecision}{1\xspace}
\newcommand{\XsfGccSObLgtRdaFone}{0.990\xspace}
\newcommand{\XsfGccSObLgtRseRecall}{0.994\xspace}
\newcommand{\XsfGccSObLgtRsePrecision}{1\xspace}
\newcommand{\XsfGccSObLgtRseFone}{0.997\xspace}
\newcommand{\XsfGccSObBzpGT}{14383\xspace}
\newcommand{\XsfGccSObBzpLsRecall}{1\xspace}
\newcommand{\XsfGccSObBzpLsPrecision}{1\xspace}
\newcommand{\XsfGccSObBzpLsFone}{1\xspace}
\newcommand{\XsfGccSObBzpBapRecall}{0.836\xspace}
\newcommand{\XsfGccSObBzpBapPrecision}{1\xspace}
\newcommand{\XsfGccSObBzpBapFone}{0.911\xspace}
\newcommand{\XsfGccSObBzpGhiRecall}{1\xspace}
\newcommand{\XsfGccSObBzpGhiPrecision}{1\xspace}
\newcommand{\XsfGccSObBzpGhiFone}{1\xspace}
\newcommand{\XsfGccSObBzpRdaRecall}{0.852\xspace}
\newcommand{\XsfGccSObBzpRdaPrecision}{1\xspace}
\newcommand{\XsfGccSObBzpRdaFone}{0.920\xspace}
\newcommand{\XsfGccSObBzpRseRecall}{0.999\xspace}
\newcommand{\XsfGccSObBzpRsePrecision}{1\xspace}
\newcommand{\XsfGccSObBzpRseFone}{0.999\xspace}
\newcommand{\XsfGccSObGccGT}{842580\xspace}
\newcommand{\XsfGccSObGccLsRecall}{1\xspace}
\newcommand{\XsfGccSObGccLsPrecision}{1\xspace}
\newcommand{\XsfGccSObGccLsFone}{1\xspace}
\newcommand{\XsfGccSObGccBapRecall}{0.724\xspace}
\newcommand{\XsfGccSObGccBapPrecision}{1.000\xspace}
\newcommand{\XsfGccSObGccBapFone}{0.840\xspace}
\newcommand{\XsfGccSObGccGhiRecall}{0.622\xspace}
\newcommand{\XsfGccSObGccGhiPrecision}{0.994\xspace}
\newcommand{\XsfGccSObGccGhiFone}{0.766\xspace}
\newcommand{\XsfGccSObGccRdaRecall}{0.776\xspace}
\newcommand{\XsfGccSObGccRdaPrecision}{0.999\xspace}
\newcommand{\XsfGccSObGccRdaFone}{0.874\xspace}
\newcommand{\XsfGccSObGccRseRecall}{0.976\xspace}
\newcommand{\XsfGccSObGccRsePrecision}{1.000\xspace}
\newcommand{\XsfGccSObGccRseFone}{0.988\xspace}
\newcommand{\XsfGccSObGzpGT}{10058\xspace}
\newcommand{\XsfGccSObGzpLsRecall}{1\xspace}
\newcommand{\XsfGccSObGzpLsPrecision}{1\xspace}
\newcommand{\XsfGccSObGzpLsFone}{1\xspace}
\newcommand{\XsfGccSObGzpBapRecall}{0.986\xspace}
\newcommand{\XsfGccSObGzpBapPrecision}{1\xspace}
\newcommand{\XsfGccSObGzpBapFone}{0.993\xspace}
\newcommand{\XsfGccSObGzpGhiRecall}{1\xspace}
\newcommand{\XsfGccSObGzpGhiPrecision}{1\xspace}
\newcommand{\XsfGccSObGzpGhiFone}{1\xspace}
\newcommand{\XsfGccSObGzpRdaRecall}{1\xspace}
\newcommand{\XsfGccSObGzpRdaPrecision}{1\xspace}
\newcommand{\XsfGccSObGzpRdaFone}{1\xspace}
\newcommand{\XsfGccSObGzpRseRecall}{0.998\xspace}
\newcommand{\XsfGccSObGzpRsePrecision}{1\xspace}
\newcommand{\XsfGccSObGzpRseFone}{0.999\xspace}
\newcommand{\XsfGccSObOggGT}{37140\xspace}
\newcommand{\XsfGccSObOggLsRecall}{1\xspace}
\newcommand{\XsfGccSObOggLsPrecision}{1\xspace}
\newcommand{\XsfGccSObOggLsFone}{1\xspace}
\newcommand{\XsfGccSObOggBapRecall}{0.980\xspace}
\newcommand{\XsfGccSObOggBapPrecision}{1.000\xspace}
\newcommand{\XsfGccSObOggBapFone}{0.990\xspace}
\newcommand{\XsfGccSObOggGhiRecall}{0.997\xspace}
\newcommand{\XsfGccSObOggGhiPrecision}{1\xspace}
\newcommand{\XsfGccSObOggGhiFone}{0.999\xspace}
\newcommand{\XsfGccSObOggRdaRecall}{0.987\xspace}
\newcommand{\XsfGccSObOggRdaPrecision}{1.000\xspace}
\newcommand{\XsfGccSObOggRdaFone}{0.993\xspace}
\newcommand{\XsfGccSObOggRseRecall}{1\xspace}
\newcommand{\XsfGccSObOggRsePrecision}{1\xspace}
\newcommand{\XsfGccSObOggRseFone}{1\xspace}
\newcommand{\XsfGccSObNgxGT}{98225\xspace}
\newcommand{\XsfGccSObNgxLsRecall}{1\xspace}
\newcommand{\XsfGccSObNgxLsPrecision}{1\xspace}
\newcommand{\XsfGccSObNgxLsFone}{1\xspace}
\newcommand{\XsfGccSObNgxBapRecall}{0.972\xspace}
\newcommand{\XsfGccSObNgxBapPrecision}{1.000\xspace}
\newcommand{\XsfGccSObNgxBapFone}{0.986\xspace}
\newcommand{\XsfGccSObNgxGhiRecall}{1\xspace}
\newcommand{\XsfGccSObNgxGhiPrecision}{1\xspace}
\newcommand{\XsfGccSObNgxGhiFone}{1\xspace}
\newcommand{\XsfGccSObNgxRdaRecall}{0.966\xspace}
\newcommand{\XsfGccSObNgxRdaPrecision}{0.996\xspace}
\newcommand{\XsfGccSObNgxRdaFone}{0.981\xspace}
\newcommand{\XsfGccSObNgxRseRecall}{0.997\xspace}
\newcommand{\XsfGccSObNgxRsePrecision}{1\xspace}
\newcommand{\XsfGccSObNgxRseFone}{0.999\xspace}
\newcommand{\XsfGccSObSshGT}{97439\xspace}
\newcommand{\XsfGccSObSshLsRecall}{1\xspace}
\newcommand{\XsfGccSObSshLsPrecision}{1\xspace}
\newcommand{\XsfGccSObSshLsFone}{1\xspace}
\newcommand{\XsfGccSObSshBapRecall}{0.954\xspace}
\newcommand{\XsfGccSObSshBapPrecision}{1\xspace}
\newcommand{\XsfGccSObSshBapFone}{0.976\xspace}
\newcommand{\XsfGccSObSshGhiRecall}{0.986\xspace}
\newcommand{\XsfGccSObSshGhiPrecision}{1\xspace}
\newcommand{\XsfGccSObSshGhiFone}{0.993\xspace}
\newcommand{\XsfGccSObSshRdaRecall}{0.915\xspace}
\newcommand{\XsfGccSObSshRdaPrecision}{1\xspace}
\newcommand{\XsfGccSObSshRdaFone}{0.955\xspace}
\newcommand{\XsfGccSObSshRseRecall}{0.991\xspace}
\newcommand{\XsfGccSObSshRsePrecision}{1\xspace}
\newcommand{\XsfGccSObSshRseFone}{0.996\xspace}
\newcommand{\XsfGccSObPcrGT}{4566\xspace}
\newcommand{\XsfGccSObPcrLsRecall}{1\xspace}
\newcommand{\XsfGccSObPcrLsPrecision}{1\xspace}
\newcommand{\XsfGccSObPcrLsFone}{1\xspace}
\newcommand{\XsfGccSObPcrBapRecall}{0.903\xspace}
\newcommand{\XsfGccSObPcrBapPrecision}{1\xspace}
\newcommand{\XsfGccSObPcrBapFone}{0.949\xspace}
\newcommand{\XsfGccSObPcrGhiRecall}{1\xspace}
\newcommand{\XsfGccSObPcrGhiPrecision}{1\xspace}
\newcommand{\XsfGccSObPcrGhiFone}{1\xspace}
\newcommand{\XsfGccSObPcrRdaRecall}{0.982\xspace}
\newcommand{\XsfGccSObPcrRdaPrecision}{0.998\xspace}
\newcommand{\XsfGccSObPcrRdaFone}{0.990\xspace}
\newcommand{\XsfGccSObPcrRseRecall}{0.982\xspace}
\newcommand{\XsfGccSObPcrRsePrecision}{1\xspace}
\newcommand{\XsfGccSObPcrRseFone}{0.991\xspace}
\newcommand{\XsfGccSObSqlGT}{167305\xspace}
\newcommand{\XsfGccSObSqlLsRecall}{1\xspace}
\newcommand{\XsfGccSObSqlLsPrecision}{1\xspace}
\newcommand{\XsfGccSObSqlLsFone}{1\xspace}
\newcommand{\XsfGccSObSqlBapRecall}{0.858\xspace}
\newcommand{\XsfGccSObSqlBapPrecision}{1\xspace}
\newcommand{\XsfGccSObSqlBapFone}{0.924\xspace}
\newcommand{\XsfGccSObSqlGhiRecall}{0.999\xspace}
\newcommand{\XsfGccSObSqlGhiPrecision}{1\xspace}
\newcommand{\XsfGccSObSqlGhiFone}{0.999\xspace}
\newcommand{\XsfGccSObSqlRdaRecall}{0.934\xspace}
\newcommand{\XsfGccSObSqlRdaPrecision}{1.000\xspace}
\newcommand{\XsfGccSObSqlRdaFone}{0.966\xspace}
\newcommand{\XsfGccSObSqlRseRecall}{0.986\xspace}
\newcommand{\XsfGccSObSqlRsePrecision}{1\xspace}
\newcommand{\XsfGccSObSqlRseFone}{0.993\xspace}
\newcommand{\XsfGccSObVimGT}{476511\xspace}
\newcommand{\XsfGccSObVimLsRecall}{1\xspace}
\newcommand{\XsfGccSObVimLsPrecision}{1\xspace}
\newcommand{\XsfGccSObVimLsFone}{1\xspace}
\newcommand{\XsfGccSObVimBapRecall}{0.925\xspace}
\newcommand{\XsfGccSObVimBapPrecision}{1.000\xspace}
\newcommand{\XsfGccSObVimBapFone}{0.961\xspace}
\newcommand{\XsfGccSObVimGhiRecall}{0.719\xspace}
\newcommand{\XsfGccSObVimGhiPrecision}{0.999\xspace}
\newcommand{\XsfGccSObVimGhiFone}{0.836\xspace}
\newcommand{\XsfGccSObVimRdaRecall}{0.943\xspace}
\newcommand{\XsfGccSObVimRdaPrecision}{1.000\xspace}
\newcommand{\XsfGccSObVimRdaFone}{0.971\xspace}
\newcommand{\XsfGccSObVimRseRecall}{0.996\xspace}
\newcommand{\XsfGccSObVimRsePrecision}{1.000\xspace}
\newcommand{\XsfGccSObVimRseFone}{0.998\xspace}
\newcommand{\XsfGccSObVsfGT}{17721\xspace}
\newcommand{\XsfGccSObVsfLsRecall}{1\xspace}
\newcommand{\XsfGccSObVsfLsPrecision}{1\xspace}
\newcommand{\XsfGccSObVsfLsFone}{1\xspace}
\newcommand{\XsfGccSObVsfBapRecall}{0.993\xspace}
\newcommand{\XsfGccSObVsfBapPrecision}{1.000\xspace}
\newcommand{\XsfGccSObVsfBapFone}{0.996\xspace}
\newcommand{\XsfGccSObVsfGhiRecall}{1\xspace}
\newcommand{\XsfGccSObVsfGhiPrecision}{1\xspace}
\newcommand{\XsfGccSObVsfGhiFone}{1\xspace}
\newcommand{\XsfGccSObVsfRdaRecall}{0.978\xspace}
\newcommand{\XsfGccSObVsfRdaPrecision}{1.000\xspace}
\newcommand{\XsfGccSObVsfRdaFone}{0.989\xspace}
\newcommand{\XsfGccSObVsfRseRecall}{1\xspace}
\newcommand{\XsfGccSObVsfRsePrecision}{1\xspace}
\newcommand{\XsfGccSObVsfRseFone}{1\xspace}
\newcommand{\XsfGccSOcSzpGT}{15982\xspace}
\newcommand{\XsfGccSOcSzpLsRecall}{1\xspace}
\newcommand{\XsfGccSOcSzpLsPrecision}{1\xspace}
\newcommand{\XsfGccSOcSzpLsFone}{1\xspace}
\newcommand{\XsfGccSOcSzpBapRecall}{0.966\xspace}
\newcommand{\XsfGccSOcSzpBapPrecision}{1\xspace}
\newcommand{\XsfGccSOcSzpBapFone}{0.983\xspace}
\newcommand{\XsfGccSOcSzpGhiRecall}{0.993\xspace}
\newcommand{\XsfGccSOcSzpGhiPrecision}{1\xspace}
\newcommand{\XsfGccSOcSzpGhiFone}{0.997\xspace}
\newcommand{\XsfGccSOcSzpRdaRecall}{0.989\xspace}
\newcommand{\XsfGccSOcSzpRdaPrecision}{1.000\xspace}
\newcommand{\XsfGccSOcSzpRdaFone}{0.994\xspace}
\newcommand{\XsfGccSOcSzpRseRecall}{1\xspace}
\newcommand{\XsfGccSOcSzpRsePrecision}{1\xspace}
\newcommand{\XsfGccSOcSzpRseFone}{1\xspace}
\newcommand{\XsfGccSOcCapGT}{197315\xspace}
\newcommand{\XsfGccSOcCapLsRecall}{1\xspace}
\newcommand{\XsfGccSOcCapLsPrecision}{1\xspace}
\newcommand{\XsfGccSOcCapLsFone}{1\xspace}
\newcommand{\XsfGccSOcCapBapRecall}{0.498\xspace}
\newcommand{\XsfGccSOcCapBapPrecision}{1\xspace}
\newcommand{\XsfGccSOcCapBapFone}{0.665\xspace}
\newcommand{\XsfGccSOcCapGhiRecall}{0.785\xspace}
\newcommand{\XsfGccSOcCapGhiPrecision}{1\xspace}
\newcommand{\XsfGccSOcCapGhiFone}{0.880\xspace}
\newcommand{\XsfGccSOcCapRdaRecall}{0.551\xspace}
\newcommand{\XsfGccSOcCapRdaPrecision}{0.997\xspace}
\newcommand{\XsfGccSOcCapRdaFone}{0.710\xspace}
\newcommand{\XsfGccSOcCapRseRecall}{0.706\xspace}
\newcommand{\XsfGccSOcCapRsePrecision}{1\xspace}
\newcommand{\XsfGccSOcCapRseFone}{0.827\xspace}
\newcommand{\XsfGccSOcExmGT}{176237\xspace}
\newcommand{\XsfGccSOcExmLsRecall}{1\xspace}
\newcommand{\XsfGccSOcExmLsPrecision}{1\xspace}
\newcommand{\XsfGccSOcExmLsFone}{1\xspace}
\newcommand{\XsfGccSOcExmBapRecall}{0.868\xspace}
\newcommand{\XsfGccSOcExmBapPrecision}{1.000\xspace}
\newcommand{\XsfGccSOcExmBapFone}{0.930\xspace}
\newcommand{\XsfGccSOcExmGhiRecall}{0.975\xspace}
\newcommand{\XsfGccSOcExmGhiPrecision}{1\xspace}
\newcommand{\XsfGccSOcExmGhiFone}{0.987\xspace}
\newcommand{\XsfGccSOcExmRdaRecall}{0.882\xspace}
\newcommand{\XsfGccSOcExmRdaPrecision}{1.000\xspace}
\newcommand{\XsfGccSOcExmRdaFone}{0.938\xspace}
\newcommand{\XsfGccSOcExmRseRecall}{0.991\xspace}
\newcommand{\XsfGccSOcExmRsePrecision}{1\xspace}
\newcommand{\XsfGccSOcExmRseFone}{0.996\xspace}
\newcommand{\XsfGccSOcLgtGT}{31119\xspace}
\newcommand{\XsfGccSOcLgtLsRecall}{1\xspace}
\newcommand{\XsfGccSOcLgtLsPrecision}{1\xspace}
\newcommand{\XsfGccSOcLgtLsFone}{1\xspace}
\newcommand{\XsfGccSOcLgtBapRecall}{0.894\xspace}
\newcommand{\XsfGccSOcLgtBapPrecision}{1.000\xspace}
\newcommand{\XsfGccSOcLgtBapFone}{0.944\xspace}
\newcommand{\XsfGccSOcLgtGhiRecall}{0.977\xspace}
\newcommand{\XsfGccSOcLgtGhiPrecision}{1\xspace}
\newcommand{\XsfGccSOcLgtGhiFone}{0.988\xspace}
\newcommand{\XsfGccSOcLgtRdaRecall}{0.975\xspace}
\newcommand{\XsfGccSOcLgtRdaPrecision}{0.999\xspace}
\newcommand{\XsfGccSOcLgtRdaFone}{0.987\xspace}
\newcommand{\XsfGccSOcLgtRseRecall}{0.993\xspace}
\newcommand{\XsfGccSOcLgtRsePrecision}{1\xspace}
\newcommand{\XsfGccSOcLgtRseFone}{0.997\xspace}
\newcommand{\XsfGccSOcBzpGT}{20982\xspace}
\newcommand{\XsfGccSOcBzpLsRecall}{1\xspace}
\newcommand{\XsfGccSOcBzpLsPrecision}{1\xspace}
\newcommand{\XsfGccSOcBzpLsFone}{1\xspace}
\newcommand{\XsfGccSOcBzpBapRecall}{0.815\xspace}
\newcommand{\XsfGccSOcBzpBapPrecision}{1\xspace}
\newcommand{\XsfGccSOcBzpBapFone}{0.898\xspace}
\newcommand{\XsfGccSOcBzpGhiRecall}{1\xspace}
\newcommand{\XsfGccSOcBzpGhiPrecision}{1\xspace}
\newcommand{\XsfGccSOcBzpGhiFone}{1\xspace}
\newcommand{\XsfGccSOcBzpRdaRecall}{0.832\xspace}
\newcommand{\XsfGccSOcBzpRdaPrecision}{1\xspace}
\newcommand{\XsfGccSOcBzpRdaFone}{0.908\xspace}
\newcommand{\XsfGccSOcBzpRseRecall}{0.999\xspace}
\newcommand{\XsfGccSOcBzpRsePrecision}{1\xspace}
\newcommand{\XsfGccSOcBzpRseFone}{1.000\xspace}
\newcommand{\XsfGccSOcGccGT}{1084991\xspace}
\newcommand{\XsfGccSOcGccLsRecall}{1\xspace}
\newcommand{\XsfGccSOcGccLsPrecision}{1\xspace}
\newcommand{\XsfGccSOcGccLsFone}{1\xspace}
\newcommand{\XsfGccSOcGccBapRecall}{0.776\xspace}
\newcommand{\XsfGccSOcGccBapPrecision}{1.000\xspace}
\newcommand{\XsfGccSOcGccBapFone}{0.874\xspace}
\newcommand{\XsfGccSOcGccGhiRecall}{0.612\xspace}
\newcommand{\XsfGccSOcGccGhiPrecision}{0.996\xspace}
\newcommand{\XsfGccSOcGccGhiFone}{0.758\xspace}
\newcommand{\XsfGccSOcGccRdaRecall}{0.802\xspace}
\newcommand{\XsfGccSOcGccRdaPrecision}{0.999\xspace}
\newcommand{\XsfGccSOcGccRdaFone}{0.890\xspace}
\newcommand{\XsfGccSOcGccRseRecall}{0.981\xspace}
\newcommand{\XsfGccSOcGccRsePrecision}{1.000\xspace}
\newcommand{\XsfGccSOcGccRseFone}{0.990\xspace}
\newcommand{\XsfGccSOcGzpGT}{18564\xspace}
\newcommand{\XsfGccSOcGzpLsRecall}{1\xspace}
\newcommand{\XsfGccSOcGzpLsPrecision}{1\xspace}
\newcommand{\XsfGccSOcGzpLsFone}{1\xspace}
\newcommand{\XsfGccSOcGzpBapRecall}{0.993\xspace}
\newcommand{\XsfGccSOcGzpBapPrecision}{1\xspace}
\newcommand{\XsfGccSOcGzpBapFone}{0.996\xspace}
\newcommand{\XsfGccSOcGzpGhiRecall}{1\xspace}
\newcommand{\XsfGccSOcGzpGhiPrecision}{1\xspace}
\newcommand{\XsfGccSOcGzpGhiFone}{1\xspace}
\newcommand{\XsfGccSOcGzpRdaRecall}{0.836\xspace}
\newcommand{\XsfGccSOcGzpRdaPrecision}{1\xspace}
\newcommand{\XsfGccSOcGzpRdaFone}{0.911\xspace}
\newcommand{\XsfGccSOcGzpRseRecall}{0.999\xspace}
\newcommand{\XsfGccSOcGzpRsePrecision}{1\xspace}
\newcommand{\XsfGccSOcGzpRseFone}{0.999\xspace}
\newcommand{\XsfGccSOcOggGT}{64469\xspace}
\newcommand{\XsfGccSOcOggLsRecall}{1\xspace}
\newcommand{\XsfGccSOcOggLsPrecision}{1\xspace}
\newcommand{\XsfGccSOcOggLsFone}{1\xspace}
\newcommand{\XsfGccSOcOggBapRecall}{0.988\xspace}
\newcommand{\XsfGccSOcOggBapPrecision}{1.000\xspace}
\newcommand{\XsfGccSOcOggBapFone}{0.994\xspace}
\newcommand{\XsfGccSOcOggGhiRecall}{1\xspace}
\newcommand{\XsfGccSOcOggGhiPrecision}{1\xspace}
\newcommand{\XsfGccSOcOggGhiFone}{1\xspace}
\newcommand{\XsfGccSOcOggRdaRecall}{0.993\xspace}
\newcommand{\XsfGccSOcOggRdaPrecision}{0.999\xspace}
\newcommand{\XsfGccSOcOggRdaFone}{0.996\xspace}
\newcommand{\XsfGccSOcOggRseRecall}{1\xspace}
\newcommand{\XsfGccSOcOggRsePrecision}{1\xspace}
\newcommand{\XsfGccSOcOggRseFone}{1\xspace}
\newcommand{\XsfGccSOcNgxGT}{115251\xspace}
\newcommand{\XsfGccSOcNgxLsRecall}{1\xspace}
\newcommand{\XsfGccSOcNgxLsPrecision}{1\xspace}
\newcommand{\XsfGccSOcNgxLsFone}{1\xspace}
\newcommand{\XsfGccSOcNgxBapRecall}{0.974\xspace}
\newcommand{\XsfGccSOcNgxBapPrecision}{1.000\xspace}
\newcommand{\XsfGccSOcNgxBapFone}{0.987\xspace}
\newcommand{\XsfGccSOcNgxGhiRecall}{1\xspace}
\newcommand{\XsfGccSOcNgxGhiPrecision}{1\xspace}
\newcommand{\XsfGccSOcNgxGhiFone}{1\xspace}
\newcommand{\XsfGccSOcNgxRdaRecall}{0.972\xspace}
\newcommand{\XsfGccSOcNgxRdaPrecision}{0.997\xspace}
\newcommand{\XsfGccSOcNgxRdaFone}{0.984\xspace}
\newcommand{\XsfGccSOcNgxRseRecall}{0.998\xspace}
\newcommand{\XsfGccSOcNgxRsePrecision}{1\xspace}
\newcommand{\XsfGccSOcNgxRseFone}{0.999\xspace}
\newcommand{\XsfGccSOcSshGT}{123117\xspace}
\newcommand{\XsfGccSOcSshLsRecall}{1\xspace}
\newcommand{\XsfGccSOcSshLsPrecision}{1\xspace}
\newcommand{\XsfGccSOcSshLsFone}{1\xspace}
\newcommand{\XsfGccSOcSshBapRecall}{0.952\xspace}
\newcommand{\XsfGccSOcSshBapPrecision}{1\xspace}
\newcommand{\XsfGccSOcSshBapFone}{0.976\xspace}
\newcommand{\XsfGccSOcSshGhiRecall}{0.986\xspace}
\newcommand{\XsfGccSOcSshGhiPrecision}{1\xspace}
\newcommand{\XsfGccSOcSshGhiFone}{0.993\xspace}
\newcommand{\XsfGccSOcSshRdaRecall}{0.893\xspace}
\newcommand{\XsfGccSOcSshRdaPrecision}{1.000\xspace}
\newcommand{\XsfGccSOcSshRdaFone}{0.944\xspace}
\newcommand{\XsfGccSOcSshRseRecall}{0.993\xspace}
\newcommand{\XsfGccSOcSshRsePrecision}{1\xspace}
\newcommand{\XsfGccSOcSshRseFone}{0.997\xspace}
\newcommand{\XsfGccSOcPcrGT}{5329\xspace}
\newcommand{\XsfGccSOcPcrLsRecall}{1\xspace}
\newcommand{\XsfGccSOcPcrLsPrecision}{1\xspace}
\newcommand{\XsfGccSOcPcrLsFone}{1\xspace}
\newcommand{\XsfGccSOcPcrBapRecall}{0.874\xspace}
\newcommand{\XsfGccSOcPcrBapPrecision}{1\xspace}
\newcommand{\XsfGccSOcPcrBapFone}{0.933\xspace}
\newcommand{\XsfGccSOcPcrGhiRecall}{1\xspace}
\newcommand{\XsfGccSOcPcrGhiPrecision}{1\xspace}
\newcommand{\XsfGccSOcPcrGhiFone}{1\xspace}
\newcommand{\XsfGccSOcPcrRdaRecall}{0.971\xspace}
\newcommand{\XsfGccSOcPcrRdaPrecision}{0.990\xspace}
\newcommand{\XsfGccSOcPcrRdaFone}{0.980\xspace}
\newcommand{\XsfGccSOcPcrRseRecall}{0.991\xspace}
\newcommand{\XsfGccSOcPcrRsePrecision}{1\xspace}
\newcommand{\XsfGccSOcPcrRseFone}{0.996\xspace}
\newcommand{\XsfGccSOcSqlGT}{204743\xspace}
\newcommand{\XsfGccSOcSqlLsRecall}{1\xspace}
\newcommand{\XsfGccSOcSqlLsPrecision}{1\xspace}
\newcommand{\XsfGccSOcSqlLsFone}{1\xspace}
\newcommand{\XsfGccSOcSqlBapRecall}{0.854\xspace}
\newcommand{\XsfGccSOcSqlBapPrecision}{1.000\xspace}
\newcommand{\XsfGccSOcSqlBapFone}{0.921\xspace}
\newcommand{\XsfGccSOcSqlGhiRecall}{0.942\xspace}
\newcommand{\XsfGccSOcSqlGhiPrecision}{1\xspace}
\newcommand{\XsfGccSOcSqlGhiFone}{0.970\xspace}
\newcommand{\XsfGccSOcSqlRdaRecall}{0.915\xspace}
\newcommand{\XsfGccSOcSqlRdaPrecision}{1.000\xspace}
\newcommand{\XsfGccSOcSqlRdaFone}{0.956\xspace}
\newcommand{\XsfGccSOcSqlRseRecall}{0.989\xspace}
\newcommand{\XsfGccSOcSqlRsePrecision}{1\xspace}
\newcommand{\XsfGccSOcSqlRseFone}{0.994\xspace}
\newcommand{\XsfGccSOcVimGT}{677690\xspace}
\newcommand{\XsfGccSOcVimLsRecall}{1\xspace}
\newcommand{\XsfGccSOcVimLsPrecision}{1\xspace}
\newcommand{\XsfGccSOcVimLsFone}{1\xspace}
\newcommand{\XsfGccSOcVimBapRecall}{0.935\xspace}
\newcommand{\XsfGccSOcVimBapPrecision}{1.000\xspace}
\newcommand{\XsfGccSOcVimBapFone}{0.966\xspace}
\newcommand{\XsfGccSOcVimGhiRecall}{0.693\xspace}
\newcommand{\XsfGccSOcVimGhiPrecision}{0.999\xspace}
\newcommand{\XsfGccSOcVimGhiFone}{0.818\xspace}
\newcommand{\XsfGccSOcVimRdaRecall}{0.946\xspace}
\newcommand{\XsfGccSOcVimRdaPrecision}{1.000\xspace}
\newcommand{\XsfGccSOcVimRdaFone}{0.972\xspace}
\newcommand{\XsfGccSOcVimRseRecall}{0.995\xspace}
\newcommand{\XsfGccSOcVimRsePrecision}{1.000\xspace}
\newcommand{\XsfGccSOcVimRseFone}{0.997\xspace}
\newcommand{\XsfGccSOcVsfGT}{23935\xspace}
\newcommand{\XsfGccSOcVsfLsRecall}{1\xspace}
\newcommand{\XsfGccSOcVsfLsPrecision}{1\xspace}
\newcommand{\XsfGccSOcVsfLsFone}{1\xspace}
\newcommand{\XsfGccSOcVsfBapRecall}{0.992\xspace}
\newcommand{\XsfGccSOcVsfBapPrecision}{1.000\xspace}
\newcommand{\XsfGccSOcVsfBapFone}{0.996\xspace}
\newcommand{\XsfGccSOcVsfGhiRecall}{1\xspace}
\newcommand{\XsfGccSOcVsfGhiPrecision}{1\xspace}
\newcommand{\XsfGccSOcVsfGhiFone}{1\xspace}
\newcommand{\XsfGccSOcVsfRdaRecall}{0.979\xspace}
\newcommand{\XsfGccSOcVsfRdaPrecision}{1.000\xspace}
\newcommand{\XsfGccSOcVsfRdaFone}{0.989\xspace}
\newcommand{\XsfGccSOcVsfRseRecall}{1\xspace}
\newcommand{\XsfGccSOcVsfRsePrecision}{1\xspace}
\newcommand{\XsfGccSOcVsfRseFone}{1\xspace}
\newcommand{\XsfGccSOdSzpGT}{15982\xspace}
\newcommand{\XsfGccSOdSzpLsRecall}{1\xspace}
\newcommand{\XsfGccSOdSzpLsPrecision}{1\xspace}
\newcommand{\XsfGccSOdSzpLsFone}{1\xspace}
\newcommand{\XsfGccSOdSzpBapRecall}{0.966\xspace}
\newcommand{\XsfGccSOdSzpBapPrecision}{1\xspace}
\newcommand{\XsfGccSOdSzpBapFone}{0.983\xspace}
\newcommand{\XsfGccSOdSzpGhiRecall}{0.993\xspace}
\newcommand{\XsfGccSOdSzpGhiPrecision}{1\xspace}
\newcommand{\XsfGccSOdSzpGhiFone}{0.997\xspace}
\newcommand{\XsfGccSOdSzpRdaRecall}{0.989\xspace}
\newcommand{\XsfGccSOdSzpRdaPrecision}{1.000\xspace}
\newcommand{\XsfGccSOdSzpRdaFone}{0.994\xspace}
\newcommand{\XsfGccSOdSzpRseRecall}{1\xspace}
\newcommand{\XsfGccSOdSzpRsePrecision}{1\xspace}
\newcommand{\XsfGccSOdSzpRseFone}{1\xspace}
\newcommand{\XsfGccSOdCapGT}{197315\xspace}
\newcommand{\XsfGccSOdCapLsRecall}{1\xspace}
\newcommand{\XsfGccSOdCapLsPrecision}{1\xspace}
\newcommand{\XsfGccSOdCapLsFone}{1\xspace}
\newcommand{\XsfGccSOdCapBapRecall}{0.498\xspace}
\newcommand{\XsfGccSOdCapBapPrecision}{1\xspace}
\newcommand{\XsfGccSOdCapBapFone}{0.665\xspace}
\newcommand{\XsfGccSOdCapGhiRecall}{0.783\xspace}
\newcommand{\XsfGccSOdCapGhiPrecision}{1\xspace}
\newcommand{\XsfGccSOdCapGhiFone}{0.878\xspace}
\newcommand{\XsfGccSOdCapRdaRecall}{0.551\xspace}
\newcommand{\XsfGccSOdCapRdaPrecision}{0.997\xspace}
\newcommand{\XsfGccSOdCapRdaFone}{0.710\xspace}
\newcommand{\XsfGccSOdCapRseRecall}{0.706\xspace}
\newcommand{\XsfGccSOdCapRsePrecision}{1\xspace}
\newcommand{\XsfGccSOdCapRseFone}{0.827\xspace}
\newcommand{\XsfGccSOdExmGT}{176119\xspace}
\newcommand{\XsfGccSOdExmLsRecall}{1\xspace}
\newcommand{\XsfGccSOdExmLsPrecision}{1\xspace}
\newcommand{\XsfGccSOdExmLsFone}{1\xspace}
\newcommand{\XsfGccSOdExmBapRecall}{0.870\xspace}
\newcommand{\XsfGccSOdExmBapPrecision}{1.000\xspace}
\newcommand{\XsfGccSOdExmBapFone}{0.931\xspace}
\newcommand{\XsfGccSOdExmGhiRecall}{0.975\xspace}
\newcommand{\XsfGccSOdExmGhiPrecision}{1\xspace}
\newcommand{\XsfGccSOdExmGhiFone}{0.987\xspace}
\newcommand{\XsfGccSOdExmRdaRecall}{0.883\xspace}
\newcommand{\XsfGccSOdExmRdaPrecision}{1.000\xspace}
\newcommand{\XsfGccSOdExmRdaFone}{0.938\xspace}
\newcommand{\XsfGccSOdExmRseRecall}{0.991\xspace}
\newcommand{\XsfGccSOdExmRsePrecision}{1\xspace}
\newcommand{\XsfGccSOdExmRseFone}{0.996\xspace}
\newcommand{\XsfGccSOdLgtGT}{31112\xspace}
\newcommand{\XsfGccSOdLgtLsRecall}{1\xspace}
\newcommand{\XsfGccSOdLgtLsPrecision}{1\xspace}
\newcommand{\XsfGccSOdLgtLsFone}{1\xspace}
\newcommand{\XsfGccSOdLgtBapRecall}{0.894\xspace}
\newcommand{\XsfGccSOdLgtBapPrecision}{1.000\xspace}
\newcommand{\XsfGccSOdLgtBapFone}{0.944\xspace}
\newcommand{\XsfGccSOdLgtGhiRecall}{0.977\xspace}
\newcommand{\XsfGccSOdLgtGhiPrecision}{1\xspace}
\newcommand{\XsfGccSOdLgtGhiFone}{0.988\xspace}
\newcommand{\XsfGccSOdLgtRdaRecall}{0.970\xspace}
\newcommand{\XsfGccSOdLgtRdaPrecision}{1.000\xspace}
\newcommand{\XsfGccSOdLgtRdaFone}{0.985\xspace}
\newcommand{\XsfGccSOdLgtRseRecall}{0.993\xspace}
\newcommand{\XsfGccSOdLgtRsePrecision}{1\xspace}
\newcommand{\XsfGccSOdLgtRseFone}{0.997\xspace}
\newcommand{\XsfGccSOdBzpGT}{21388\xspace}
\newcommand{\XsfGccSOdBzpLsRecall}{1\xspace}
\newcommand{\XsfGccSOdBzpLsPrecision}{1\xspace}
\newcommand{\XsfGccSOdBzpLsFone}{1\xspace}
\newcommand{\XsfGccSOdBzpBapRecall}{0.819\xspace}
\newcommand{\XsfGccSOdBzpBapPrecision}{1\xspace}
\newcommand{\XsfGccSOdBzpBapFone}{0.900\xspace}
\newcommand{\XsfGccSOdBzpGhiRecall}{1\xspace}
\newcommand{\XsfGccSOdBzpGhiPrecision}{1\xspace}
\newcommand{\XsfGccSOdBzpGhiFone}{1\xspace}
\newcommand{\XsfGccSOdBzpRdaRecall}{0.844\xspace}
\newcommand{\XsfGccSOdBzpRdaPrecision}{1.000\xspace}
\newcommand{\XsfGccSOdBzpRdaFone}{0.915\xspace}
\newcommand{\XsfGccSOdBzpRseRecall}{0.999\xspace}
\newcommand{\XsfGccSOdBzpRsePrecision}{1\xspace}
\newcommand{\XsfGccSOdBzpRseFone}{1.000\xspace}
\newcommand{\XsfGccSOdGccGT}{1085103\xspace}
\newcommand{\XsfGccSOdGccLsRecall}{1\xspace}
\newcommand{\XsfGccSOdGccLsPrecision}{1\xspace}
\newcommand{\XsfGccSOdGccLsFone}{1\xspace}
\newcommand{\XsfGccSOdGccBapRecall}{0.776\xspace}
\newcommand{\XsfGccSOdGccBapPrecision}{1.000\xspace}
\newcommand{\XsfGccSOdGccBapFone}{0.874\xspace}
\newcommand{\XsfGccSOdGccGhiRecall}{0.627\xspace}
\newcommand{\XsfGccSOdGccGhiPrecision}{0.996\xspace}
\newcommand{\XsfGccSOdGccGhiFone}{0.770\xspace}
\newcommand{\XsfGccSOdGccRdaRecall}{0.801\xspace}
\newcommand{\XsfGccSOdGccRdaPrecision}{0.999\xspace}
\newcommand{\XsfGccSOdGccRdaFone}{0.889\xspace}
\newcommand{\XsfGccSOdGccRseRecall}{0.981\xspace}
\newcommand{\XsfGccSOdGccRsePrecision}{1.000\xspace}
\newcommand{\XsfGccSOdGccRseFone}{0.990\xspace}
\newcommand{\XsfGccSOdGzpGT}{18499\xspace}
\newcommand{\XsfGccSOdGzpLsRecall}{1\xspace}
\newcommand{\XsfGccSOdGzpLsPrecision}{1\xspace}
\newcommand{\XsfGccSOdGzpLsFone}{1\xspace}
\newcommand{\XsfGccSOdGzpBapRecall}{0.992\xspace}
\newcommand{\XsfGccSOdGzpBapPrecision}{1\xspace}
\newcommand{\XsfGccSOdGzpBapFone}{0.996\xspace}
\newcommand{\XsfGccSOdGzpGhiRecall}{1\xspace}
\newcommand{\XsfGccSOdGzpGhiPrecision}{1\xspace}
\newcommand{\XsfGccSOdGzpGhiFone}{1\xspace}
\newcommand{\XsfGccSOdGzpRdaRecall}{0.868\xspace}
\newcommand{\XsfGccSOdGzpRdaPrecision}{1\xspace}
\newcommand{\XsfGccSOdGzpRdaFone}{0.930\xspace}
\newcommand{\XsfGccSOdGzpRseRecall}{0.999\xspace}
\newcommand{\XsfGccSOdGzpRsePrecision}{1\xspace}
\newcommand{\XsfGccSOdGzpRseFone}{0.999\xspace}
\newcommand{\XsfGccSOdOggGT}{65157\xspace}
\newcommand{\XsfGccSOdOggLsRecall}{1\xspace}
\newcommand{\XsfGccSOdOggLsPrecision}{1\xspace}
\newcommand{\XsfGccSOdOggLsFone}{1\xspace}
\newcommand{\XsfGccSOdOggBapRecall}{0.989\xspace}
\newcommand{\XsfGccSOdOggBapPrecision}{1.000\xspace}
\newcommand{\XsfGccSOdOggBapFone}{0.994\xspace}
\newcommand{\XsfGccSOdOggGhiRecall}{1\xspace}
\newcommand{\XsfGccSOdOggGhiPrecision}{1\xspace}
\newcommand{\XsfGccSOdOggGhiFone}{1\xspace}
\newcommand{\XsfGccSOdOggRdaRecall}{0.993\xspace}
\newcommand{\XsfGccSOdOggRdaPrecision}{0.999\xspace}
\newcommand{\XsfGccSOdOggRdaFone}{0.996\xspace}
\newcommand{\XsfGccSOdOggRseRecall}{1\xspace}
\newcommand{\XsfGccSOdOggRsePrecision}{1\xspace}
\newcommand{\XsfGccSOdOggRseFone}{1\xspace}
\newcommand{\XsfGccSOdNgxGT}{115199\xspace}
\newcommand{\XsfGccSOdNgxLsRecall}{1\xspace}
\newcommand{\XsfGccSOdNgxLsPrecision}{1\xspace}
\newcommand{\XsfGccSOdNgxLsFone}{1\xspace}
\newcommand{\XsfGccSOdNgxBapRecall}{0.974\xspace}
\newcommand{\XsfGccSOdNgxBapPrecision}{1.000\xspace}
\newcommand{\XsfGccSOdNgxBapFone}{0.987\xspace}
\newcommand{\XsfGccSOdNgxGhiRecall}{1\xspace}
\newcommand{\XsfGccSOdNgxGhiPrecision}{1\xspace}
\newcommand{\XsfGccSOdNgxGhiFone}{1\xspace}
\newcommand{\XsfGccSOdNgxRdaRecall}{0.973\xspace}
\newcommand{\XsfGccSOdNgxRdaPrecision}{0.997\xspace}
\newcommand{\XsfGccSOdNgxRdaFone}{0.985\xspace}
\newcommand{\XsfGccSOdNgxRseRecall}{0.998\xspace}
\newcommand{\XsfGccSOdNgxRsePrecision}{1\xspace}
\newcommand{\XsfGccSOdNgxRseFone}{0.999\xspace}
\newcommand{\XsfGccSOdSshGT}{123096\xspace}
\newcommand{\XsfGccSOdSshLsRecall}{1\xspace}
\newcommand{\XsfGccSOdSshLsPrecision}{1\xspace}
\newcommand{\XsfGccSOdSshLsFone}{1\xspace}
\newcommand{\XsfGccSOdSshBapRecall}{0.952\xspace}
\newcommand{\XsfGccSOdSshBapPrecision}{1\xspace}
\newcommand{\XsfGccSOdSshBapFone}{0.976\xspace}
\newcommand{\XsfGccSOdSshGhiRecall}{0.986\xspace}
\newcommand{\XsfGccSOdSshGhiPrecision}{1\xspace}
\newcommand{\XsfGccSOdSshGhiFone}{0.993\xspace}
\newcommand{\XsfGccSOdSshRdaRecall}{0.893\xspace}
\newcommand{\XsfGccSOdSshRdaPrecision}{1.000\xspace}
\newcommand{\XsfGccSOdSshRdaFone}{0.944\xspace}
\newcommand{\XsfGccSOdSshRseRecall}{0.993\xspace}
\newcommand{\XsfGccSOdSshRsePrecision}{1\xspace}
\newcommand{\XsfGccSOdSshRseFone}{0.997\xspace}
\newcommand{\XsfGccSOdPcrGT}{5329\xspace}
\newcommand{\XsfGccSOdPcrLsRecall}{1\xspace}
\newcommand{\XsfGccSOdPcrLsPrecision}{1\xspace}
\newcommand{\XsfGccSOdPcrLsFone}{1\xspace}
\newcommand{\XsfGccSOdPcrBapRecall}{0.874\xspace}
\newcommand{\XsfGccSOdPcrBapPrecision}{1\xspace}
\newcommand{\XsfGccSOdPcrBapFone}{0.933\xspace}
\newcommand{\XsfGccSOdPcrGhiRecall}{1\xspace}
\newcommand{\XsfGccSOdPcrGhiPrecision}{1\xspace}
\newcommand{\XsfGccSOdPcrGhiFone}{1\xspace}
\newcommand{\XsfGccSOdPcrRdaRecall}{0.971\xspace}
\newcommand{\XsfGccSOdPcrRdaPrecision}{0.990\xspace}
\newcommand{\XsfGccSOdPcrRdaFone}{0.980\xspace}
\newcommand{\XsfGccSOdPcrRseRecall}{0.991\xspace}
\newcommand{\XsfGccSOdPcrRsePrecision}{1\xspace}
\newcommand{\XsfGccSOdPcrRseFone}{0.996\xspace}
\newcommand{\XsfGccSOdSqlGT}{204697\xspace}
\newcommand{\XsfGccSOdSqlLsRecall}{1\xspace}
\newcommand{\XsfGccSOdSqlLsPrecision}{1\xspace}
\newcommand{\XsfGccSOdSqlLsFone}{1\xspace}
\newcommand{\XsfGccSOdSqlBapRecall}{0.854\xspace}
\newcommand{\XsfGccSOdSqlBapPrecision}{1.000\xspace}
\newcommand{\XsfGccSOdSqlBapFone}{0.921\xspace}
\newcommand{\XsfGccSOdSqlGhiRecall}{0.942\xspace}
\newcommand{\XsfGccSOdSqlGhiPrecision}{1\xspace}
\newcommand{\XsfGccSOdSqlGhiFone}{0.970\xspace}
\newcommand{\XsfGccSOdSqlRdaRecall}{0.915\xspace}
\newcommand{\XsfGccSOdSqlRdaPrecision}{1.000\xspace}
\newcommand{\XsfGccSOdSqlRdaFone}{0.956\xspace}
\newcommand{\XsfGccSOdSqlRseRecall}{0.989\xspace}
\newcommand{\XsfGccSOdSqlRsePrecision}{1\xspace}
\newcommand{\XsfGccSOdSqlRseFone}{0.994\xspace}
\newcommand{\XsfGccSOdVimGT}{677872\xspace}
\newcommand{\XsfGccSOdVimLsRecall}{1\xspace}
\newcommand{\XsfGccSOdVimLsPrecision}{1\xspace}
\newcommand{\XsfGccSOdVimLsFone}{1\xspace}
\newcommand{\XsfGccSOdVimBapRecall}{0.938\xspace}
\newcommand{\XsfGccSOdVimBapPrecision}{1.000\xspace}
\newcommand{\XsfGccSOdVimBapFone}{0.968\xspace}
\newcommand{\XsfGccSOdVimGhiRecall}{0.691\xspace}
\newcommand{\XsfGccSOdVimGhiPrecision}{0.999\xspace}
\newcommand{\XsfGccSOdVimGhiFone}{0.817\xspace}
\newcommand{\XsfGccSOdVimRdaRecall}{0.945\xspace}
\newcommand{\XsfGccSOdVimRdaPrecision}{1.000\xspace}
\newcommand{\XsfGccSOdVimRdaFone}{0.972\xspace}
\newcommand{\XsfGccSOdVimRseRecall}{0.994\xspace}
\newcommand{\XsfGccSOdVimRsePrecision}{1.000\xspace}
\newcommand{\XsfGccSOdVimRseFone}{0.997\xspace}
\newcommand{\XsfGccSOdVsfGT}{23942\xspace}
\newcommand{\XsfGccSOdVsfLsRecall}{1\xspace}
\newcommand{\XsfGccSOdVsfLsPrecision}{1\xspace}
\newcommand{\XsfGccSOdVsfLsFone}{1\xspace}
\newcommand{\XsfGccSOdVsfBapRecall}{0.992\xspace}
\newcommand{\XsfGccSOdVsfBapPrecision}{1.000\xspace}
\newcommand{\XsfGccSOdVsfBapFone}{0.996\xspace}
\newcommand{\XsfGccSOdVsfGhiRecall}{1\xspace}
\newcommand{\XsfGccSOdVsfGhiPrecision}{1\xspace}
\newcommand{\XsfGccSOdVsfGhiFone}{1\xspace}
\newcommand{\XsfGccSOdVsfRdaRecall}{0.979\xspace}
\newcommand{\XsfGccSOdVsfRdaPrecision}{1.000\xspace}
\newcommand{\XsfGccSOdVsfRdaFone}{0.989\xspace}
\newcommand{\XsfGccSOdVsfRseRecall}{1\xspace}
\newcommand{\XsfGccSOdVsfRsePrecision}{1\xspace}
\newcommand{\XsfGccSOdVsfRseFone}{1\xspace}
\newcommand{\XsfGccSOsSzpGT}{10787\xspace}
\newcommand{\XsfGccSOsSzpLsRecall}{1\xspace}
\newcommand{\XsfGccSOsSzpLsPrecision}{1\xspace}
\newcommand{\XsfGccSOsSzpLsFone}{1\xspace}
\newcommand{\XsfGccSOsSzpBapRecall}{0.977\xspace}
\newcommand{\XsfGccSOsSzpBapPrecision}{1\xspace}
\newcommand{\XsfGccSOsSzpBapFone}{0.988\xspace}
\newcommand{\XsfGccSOsSzpGhiRecall}{1\xspace}
\newcommand{\XsfGccSOsSzpGhiPrecision}{1\xspace}
\newcommand{\XsfGccSOsSzpGhiFone}{1\xspace}
\newcommand{\XsfGccSOsSzpRdaRecall}{0.987\xspace}
\newcommand{\XsfGccSOsSzpRdaPrecision}{1.000\xspace}
\newcommand{\XsfGccSOsSzpRdaFone}{0.993\xspace}
\newcommand{\XsfGccSOsSzpRseRecall}{1\xspace}
\newcommand{\XsfGccSOsSzpRsePrecision}{1\xspace}
\newcommand{\XsfGccSOsSzpRseFone}{1\xspace}
\newcommand{\XsfGccSOsCapGT}{176898\xspace}
\newcommand{\XsfGccSOsCapLsRecall}{1\xspace}
\newcommand{\XsfGccSOsCapLsPrecision}{1\xspace}
\newcommand{\XsfGccSOsCapLsFone}{1\xspace}
\newcommand{\XsfGccSOsCapBapRecall}{0.627\xspace}
\newcommand{\XsfGccSOsCapBapPrecision}{1\xspace}
\newcommand{\XsfGccSOsCapBapFone}{0.771\xspace}
\newcommand{\XsfGccSOsCapGhiRecall}{0.922\xspace}
\newcommand{\XsfGccSOsCapGhiPrecision}{1\xspace}
\newcommand{\XsfGccSOsCapGhiFone}{0.959\xspace}
\newcommand{\XsfGccSOsCapRdaRecall}{0.851\xspace}
\newcommand{\XsfGccSOsCapRdaPrecision}{0.999\xspace}
\newcommand{\XsfGccSOsCapRdaFone}{0.919\xspace}
\newcommand{\XsfGccSOsCapRseRecall}{0.712\xspace}
\newcommand{\XsfGccSOsCapRsePrecision}{1\xspace}
\newcommand{\XsfGccSOsCapRseFone}{0.832\xspace}
\newcommand{\XsfGccSOsExmGT}{124619\xspace}
\newcommand{\XsfGccSOsExmLsRecall}{1\xspace}
\newcommand{\XsfGccSOsExmLsPrecision}{1\xspace}
\newcommand{\XsfGccSOsExmLsFone}{1\xspace}
\newcommand{\XsfGccSOsExmBapRecall}{0.866\xspace}
\newcommand{\XsfGccSOsExmBapPrecision}{1\xspace}
\newcommand{\XsfGccSOsExmBapFone}{0.928\xspace}
\newcommand{\XsfGccSOsExmGhiRecall}{0.998\xspace}
\newcommand{\XsfGccSOsExmGhiPrecision}{1\xspace}
\newcommand{\XsfGccSOsExmGhiFone}{0.999\xspace}
\newcommand{\XsfGccSOsExmRdaRecall}{0.948\xspace}
\newcommand{\XsfGccSOsExmRdaPrecision}{1.000\xspace}
\newcommand{\XsfGccSOsExmRdaFone}{0.973\xspace}
\newcommand{\XsfGccSOsExmRseRecall}{0.996\xspace}
\newcommand{\XsfGccSOsExmRsePrecision}{1\xspace}
\newcommand{\XsfGccSOsExmRseFone}{0.998\xspace}
\newcommand{\XsfGccSOsLgtGT}{22703\xspace}
\newcommand{\XsfGccSOsLgtLsRecall}{1\xspace}
\newcommand{\XsfGccSOsLgtLsPrecision}{1\xspace}
\newcommand{\XsfGccSOsLgtLsFone}{1\xspace}
\newcommand{\XsfGccSOsLgtBapRecall}{0.910\xspace}
\newcommand{\XsfGccSOsLgtBapPrecision}{1.000\xspace}
\newcommand{\XsfGccSOsLgtBapFone}{0.953\xspace}
\newcommand{\XsfGccSOsLgtGhiRecall}{0.998\xspace}
\newcommand{\XsfGccSOsLgtGhiPrecision}{1\xspace}
\newcommand{\XsfGccSOsLgtGhiFone}{0.999\xspace}
\newcommand{\XsfGccSOsLgtRdaRecall}{0.960\xspace}
\newcommand{\XsfGccSOsLgtRdaPrecision}{0.999\xspace}
\newcommand{\XsfGccSOsLgtRdaFone}{0.979\xspace}
\newcommand{\XsfGccSOsLgtRseRecall}{0.988\xspace}
\newcommand{\XsfGccSOsLgtRsePrecision}{1\xspace}
\newcommand{\XsfGccSOsLgtRseFone}{0.994\xspace}
\newcommand{\XsfGccSOsBzpGT}{10632\xspace}
\newcommand{\XsfGccSOsBzpLsRecall}{1\xspace}
\newcommand{\XsfGccSOsBzpLsPrecision}{1\xspace}
\newcommand{\XsfGccSOsBzpLsFone}{1\xspace}
\newcommand{\XsfGccSOsBzpBapRecall}{0.802\xspace}
\newcommand{\XsfGccSOsBzpBapPrecision}{1\xspace}
\newcommand{\XsfGccSOsBzpBapFone}{0.890\xspace}
\newcommand{\XsfGccSOsBzpGhiRecall}{1\xspace}
\newcommand{\XsfGccSOsBzpGhiPrecision}{1\xspace}
\newcommand{\XsfGccSOsBzpGhiFone}{1\xspace}
\newcommand{\XsfGccSOsBzpRdaRecall}{0.993\xspace}
\newcommand{\XsfGccSOsBzpRdaPrecision}{1\xspace}
\newcommand{\XsfGccSOsBzpRdaFone}{0.996\xspace}
\newcommand{\XsfGccSOsBzpRseRecall}{1\xspace}
\newcommand{\XsfGccSOsBzpRsePrecision}{1\xspace}
\newcommand{\XsfGccSOsBzpRseFone}{1\xspace}
\newcommand{\XsfGccSOsGccGT}{720043\xspace}
\newcommand{\XsfGccSOsGccLsRecall}{1\xspace}
\newcommand{\XsfGccSOsGccLsPrecision}{1\xspace}
\newcommand{\XsfGccSOsGccLsFone}{1\xspace}
\newcommand{\XsfGccSOsGccBapRecall}{0.779\xspace}
\newcommand{\XsfGccSOsGccBapPrecision}{1.000\xspace}
\newcommand{\XsfGccSOsGccBapFone}{0.876\xspace}
\newcommand{\XsfGccSOsGccGhiRecall}{0.684\xspace}
\newcommand{\XsfGccSOsGccGhiPrecision}{0.997\xspace}
\newcommand{\XsfGccSOsGccGhiFone}{0.811\xspace}
\newcommand{\XsfGccSOsGccRdaRecall}{0.816\xspace}
\newcommand{\XsfGccSOsGccRdaPrecision}{1.000\xspace}
\newcommand{\XsfGccSOsGccRdaFone}{0.899\xspace}
\newcommand{\XsfGccSOsGccRseRecall}{0.892\xspace}
\newcommand{\XsfGccSOsGccRsePrecision}{1.000\xspace}
\newcommand{\XsfGccSOsGccRseFone}{0.943\xspace}
\newcommand{\XsfGccSOsGzpGT}{8525\xspace}
\newcommand{\XsfGccSOsGzpLsRecall}{1\xspace}
\newcommand{\XsfGccSOsGzpLsPrecision}{1\xspace}
\newcommand{\XsfGccSOsGzpLsFone}{1\xspace}
\newcommand{\XsfGccSOsGzpBapRecall}{0.986\xspace}
\newcommand{\XsfGccSOsGzpBapPrecision}{1\xspace}
\newcommand{\XsfGccSOsGzpBapFone}{0.993\xspace}
\newcommand{\XsfGccSOsGzpGhiRecall}{1\xspace}
\newcommand{\XsfGccSOsGzpGhiPrecision}{1\xspace}
\newcommand{\XsfGccSOsGzpGhiFone}{1\xspace}
\newcommand{\XsfGccSOsGzpRdaRecall}{1\xspace}
\newcommand{\XsfGccSOsGzpRdaPrecision}{1\xspace}
\newcommand{\XsfGccSOsGzpRdaFone}{1\xspace}
\newcommand{\XsfGccSOsGzpRseRecall}{0.993\xspace}
\newcommand{\XsfGccSOsGzpRsePrecision}{1\xspace}
\newcommand{\XsfGccSOsGzpRseFone}{0.996\xspace}
\newcommand{\XsfGccSOsOggGT}{29620\xspace}
\newcommand{\XsfGccSOsOggLsRecall}{1\xspace}
\newcommand{\XsfGccSOsOggLsPrecision}{1\xspace}
\newcommand{\XsfGccSOsOggLsFone}{1\xspace}
\newcommand{\XsfGccSOsOggBapRecall}{0.997\xspace}
\newcommand{\XsfGccSOsOggBapPrecision}{1.000\xspace}
\newcommand{\XsfGccSOsOggBapFone}{0.999\xspace}
\newcommand{\XsfGccSOsOggGhiRecall}{1\xspace}
\newcommand{\XsfGccSOsOggGhiPrecision}{1\xspace}
\newcommand{\XsfGccSOsOggGhiFone}{1\xspace}
\newcommand{\XsfGccSOsOggRdaRecall}{1.000\xspace}
\newcommand{\XsfGccSOsOggRdaPrecision}{1.000\xspace}
\newcommand{\XsfGccSOsOggRdaFone}{1.000\xspace}
\newcommand{\XsfGccSOsOggRseRecall}{1\xspace}
\newcommand{\XsfGccSOsOggRsePrecision}{1\xspace}
\newcommand{\XsfGccSOsOggRseFone}{1\xspace}
\newcommand{\XsfGccSOsNgxGT}{86643\xspace}
\newcommand{\XsfGccSOsNgxLsRecall}{1\xspace}
\newcommand{\XsfGccSOsNgxLsPrecision}{1\xspace}
\newcommand{\XsfGccSOsNgxLsFone}{1\xspace}
\newcommand{\XsfGccSOsNgxBapRecall}{0.976\xspace}
\newcommand{\XsfGccSOsNgxBapPrecision}{1.000\xspace}
\newcommand{\XsfGccSOsNgxBapFone}{0.988\xspace}
\newcommand{\XsfGccSOsNgxGhiRecall}{1\xspace}
\newcommand{\XsfGccSOsNgxGhiPrecision}{1\xspace}
\newcommand{\XsfGccSOsNgxGhiFone}{1\xspace}
\newcommand{\XsfGccSOsNgxRdaRecall}{0.976\xspace}
\newcommand{\XsfGccSOsNgxRdaPrecision}{0.998\xspace}
\newcommand{\XsfGccSOsNgxRdaFone}{0.987\xspace}
\newcommand{\XsfGccSOsNgxRseRecall}{0.995\xspace}
\newcommand{\XsfGccSOsNgxRsePrecision}{1\xspace}
\newcommand{\XsfGccSOsNgxRseFone}{0.997\xspace}
\newcommand{\XsfGccSOsSshGT}{88244\xspace}
\newcommand{\XsfGccSOsSshLsRecall}{1\xspace}
\newcommand{\XsfGccSOsSshLsPrecision}{1\xspace}
\newcommand{\XsfGccSOsSshLsFone}{1\xspace}
\newcommand{\XsfGccSOsSshBapRecall}{0.959\xspace}
\newcommand{\XsfGccSOsSshBapPrecision}{1\xspace}
\newcommand{\XsfGccSOsSshBapFone}{0.979\xspace}
\newcommand{\XsfGccSOsSshGhiRecall}{0.999\xspace}
\newcommand{\XsfGccSOsSshGhiPrecision}{1\xspace}
\newcommand{\XsfGccSOsSshGhiFone}{0.999\xspace}
\newcommand{\XsfGccSOsSshRdaRecall}{0.959\xspace}
\newcommand{\XsfGccSOsSshRdaPrecision}{1\xspace}
\newcommand{\XsfGccSOsSshRdaFone}{0.979\xspace}
\newcommand{\XsfGccSOsSshRseRecall}{0.988\xspace}
\newcommand{\XsfGccSOsSshRsePrecision}{1\xspace}
\newcommand{\XsfGccSOsSshRseFone}{0.994\xspace}
\newcommand{\XsfGccSOsPcrGT}{3996\xspace}
\newcommand{\XsfGccSOsPcrLsRecall}{1\xspace}
\newcommand{\XsfGccSOsPcrLsPrecision}{1\xspace}
\newcommand{\XsfGccSOsPcrLsFone}{1\xspace}
\newcommand{\XsfGccSOsPcrBapRecall}{0.943\xspace}
\newcommand{\XsfGccSOsPcrBapPrecision}{1\xspace}
\newcommand{\XsfGccSOsPcrBapFone}{0.971\xspace}
\newcommand{\XsfGccSOsPcrGhiRecall}{1\xspace}
\newcommand{\XsfGccSOsPcrGhiPrecision}{1\xspace}
\newcommand{\XsfGccSOsPcrGhiFone}{1\xspace}
\newcommand{\XsfGccSOsPcrRdaRecall}{1.000\xspace}
\newcommand{\XsfGccSOsPcrRdaPrecision}{0.998\xspace}
\newcommand{\XsfGccSOsPcrRdaFone}{0.999\xspace}
\newcommand{\XsfGccSOsPcrRseRecall}{1\xspace}
\newcommand{\XsfGccSOsPcrRsePrecision}{1\xspace}
\newcommand{\XsfGccSOsPcrRseFone}{1\xspace}
\newcommand{\XsfGccSOsSqlGT}{133840\xspace}
\newcommand{\XsfGccSOsSqlLsRecall}{1\xspace}
\newcommand{\XsfGccSOsSqlLsPrecision}{1\xspace}
\newcommand{\XsfGccSOsSqlLsFone}{1\xspace}
\newcommand{\XsfGccSOsSqlBapRecall}{0.873\xspace}
\newcommand{\XsfGccSOsSqlBapPrecision}{1\xspace}
\newcommand{\XsfGccSOsSqlBapFone}{0.932\xspace}
\newcommand{\XsfGccSOsSqlGhiRecall}{1\xspace}
\newcommand{\XsfGccSOsSqlGhiPrecision}{1\xspace}
\newcommand{\XsfGccSOsSqlGhiFone}{1\xspace}
\newcommand{\XsfGccSOsSqlRdaRecall}{0.944\xspace}
\newcommand{\XsfGccSOsSqlRdaPrecision}{1.000\xspace}
\newcommand{\XsfGccSOsSqlRdaFone}{0.971\xspace}
\newcommand{\XsfGccSOsSqlRseRecall}{0.944\xspace}
\newcommand{\XsfGccSOsSqlRsePrecision}{1.000\xspace}
\newcommand{\XsfGccSOsSqlRseFone}{0.971\xspace}
\newcommand{\XsfGccSOsVimGT}{411040\xspace}
\newcommand{\XsfGccSOsVimLsRecall}{1\xspace}
\newcommand{\XsfGccSOsVimLsPrecision}{1\xspace}
\newcommand{\XsfGccSOsVimLsFone}{1\xspace}
\newcommand{\XsfGccSOsVimBapRecall}{0.963\xspace}
\newcommand{\XsfGccSOsVimBapPrecision}{1.000\xspace}
\newcommand{\XsfGccSOsVimBapFone}{0.981\xspace}
\newcommand{\XsfGccSOsVimGhiRecall}{0.767\xspace}
\newcommand{\XsfGccSOsVimGhiPrecision}{0.999\xspace}
\newcommand{\XsfGccSOsVimGhiFone}{0.868\xspace}
\newcommand{\XsfGccSOsVimRdaRecall}{0.971\xspace}
\newcommand{\XsfGccSOsVimRdaPrecision}{1.000\xspace}
\newcommand{\XsfGccSOsVimRdaFone}{0.985\xspace}
\newcommand{\XsfGccSOsVimRseRecall}{0.994\xspace}
\newcommand{\XsfGccSOsVimRsePrecision}{1\xspace}
\newcommand{\XsfGccSOsVimRseFone}{0.997\xspace}
\newcommand{\XsfGccSOsVsfGT}{15596\xspace}
\newcommand{\XsfGccSOsVsfLsRecall}{1\xspace}
\newcommand{\XsfGccSOsVsfLsPrecision}{1\xspace}
\newcommand{\XsfGccSOsVsfLsFone}{1\xspace}
\newcommand{\XsfGccSOsVsfBapRecall}{0.998\xspace}
\newcommand{\XsfGccSOsVsfBapPrecision}{1.000\xspace}
\newcommand{\XsfGccSOsVsfBapFone}{0.999\xspace}
\newcommand{\XsfGccSOsVsfGhiRecall}{1\xspace}
\newcommand{\XsfGccSOsVsfGhiPrecision}{1\xspace}
\newcommand{\XsfGccSOsVsfGhiFone}{1\xspace}
\newcommand{\XsfGccSOsVsfRdaRecall}{0.991\xspace}
\newcommand{\XsfGccSOsVsfRdaPrecision}{1\xspace}
\newcommand{\XsfGccSOsVsfRdaFone}{0.996\xspace}
\newcommand{\XsfGccSOsVsfRseRecall}{1.000\xspace}
\newcommand{\XsfGccSOsVsfRsePrecision}{1\xspace}
\newcommand{\XsfGccSOsVsfRseFone}{1.000\xspace}
\newcommand{\XsfClangTOoSzpGT}{20170\xspace}
\newcommand{\XsfClangTOoSzpLsRecall}{1\xspace}
\newcommand{\XsfClangTOoSzpLsPrecision}{1\xspace}
\newcommand{\XsfClangTOoSzpLsFone}{1\xspace}
\newcommand{\XsfClangTOoSzpBapRecall}{0.980\xspace}
\newcommand{\XsfClangTOoSzpBapPrecision}{1\xspace}
\newcommand{\XsfClangTOoSzpBapFone}{0.990\xspace}
\newcommand{\XsfClangTOoSzpGhiRecall}{1\xspace}
\newcommand{\XsfClangTOoSzpGhiPrecision}{1\xspace}
\newcommand{\XsfClangTOoSzpGhiFone}{1\xspace}
\newcommand{\XsfClangTOoSzpRdaRecall}{0.980\xspace}
\newcommand{\XsfClangTOoSzpRdaPrecision}{1\xspace}
\newcommand{\XsfClangTOoSzpRdaFone}{0.990\xspace}
\newcommand{\XsfClangTOoSzpRseRecall}{1\xspace}
\newcommand{\XsfClangTOoSzpRsePrecision}{1\xspace}
\newcommand{\XsfClangTOoSzpRseFone}{1\xspace}
\newcommand{\XsfClangTOoCapGT}{298190\xspace}
\newcommand{\XsfClangTOoCapLsRecall}{1\xspace}
\newcommand{\XsfClangTOoCapLsPrecision}{1\xspace}
\newcommand{\XsfClangTOoCapLsFone}{1\xspace}
\newcommand{\XsfClangTOoCapBapRecall}{0.592\xspace}
\newcommand{\XsfClangTOoCapBapPrecision}{1\xspace}
\newcommand{\XsfClangTOoCapBapFone}{0.744\xspace}
\newcommand{\XsfClangTOoCapGhiRecall}{1\xspace}
\newcommand{\XsfClangTOoCapGhiPrecision}{1\xspace}
\newcommand{\XsfClangTOoCapGhiFone}{1\xspace}
\newcommand{\XsfClangTOoCapRdaRecall}{0.422\xspace}
\newcommand{\XsfClangTOoCapRdaPrecision}{1\xspace}
\newcommand{\XsfClangTOoCapRdaFone}{0.594\xspace}
\newcommand{\XsfClangTOoCapRseRecall}{1\xspace}
\newcommand{\XsfClangTOoCapRsePrecision}{1\xspace}
\newcommand{\XsfClangTOoCapRseFone}{1\xspace}
\newcommand{\XsfClangTOoExmGT}{179768\xspace}
\newcommand{\XsfClangTOoExmLsRecall}{1\xspace}
\newcommand{\XsfClangTOoExmLsPrecision}{1\xspace}
\newcommand{\XsfClangTOoExmLsFone}{1\xspace}
\newcommand{\XsfClangTOoExmBapRecall}{0.866\xspace}
\newcommand{\XsfClangTOoExmBapPrecision}{1\xspace}
\newcommand{\XsfClangTOoExmBapFone}{0.928\xspace}
\newcommand{\XsfClangTOoExmGhiRecall}{0.985\xspace}
\newcommand{\XsfClangTOoExmGhiPrecision}{1\xspace}
\newcommand{\XsfClangTOoExmGhiFone}{0.993\xspace}
\newcommand{\XsfClangTOoExmRdaRecall}{0.830\xspace}
\newcommand{\XsfClangTOoExmRdaPrecision}{1.000\xspace}
\newcommand{\XsfClangTOoExmRdaFone}{0.907\xspace}
\newcommand{\XsfClangTOoExmRseRecall}{1.000\xspace}
\newcommand{\XsfClangTOoExmRsePrecision}{1\xspace}
\newcommand{\XsfClangTOoExmRseFone}{1.000\xspace}
\newcommand{\XsfClangTOoLgtGT}{39130\xspace}
\newcommand{\XsfClangTOoLgtLsRecall}{1\xspace}
\newcommand{\XsfClangTOoLgtLsPrecision}{1\xspace}
\newcommand{\XsfClangTOoLgtLsFone}{1\xspace}
\newcommand{\XsfClangTOoLgtBapRecall}{0.915\xspace}
\newcommand{\XsfClangTOoLgtBapPrecision}{1\xspace}
\newcommand{\XsfClangTOoLgtBapFone}{0.956\xspace}
\newcommand{\XsfClangTOoLgtGhiRecall}{1\xspace}
\newcommand{\XsfClangTOoLgtGhiPrecision}{1\xspace}
\newcommand{\XsfClangTOoLgtGhiFone}{1\xspace}
\newcommand{\XsfClangTOoLgtRdaRecall}{0.915\xspace}
\newcommand{\XsfClangTOoLgtRdaPrecision}{1.000\xspace}
\newcommand{\XsfClangTOoLgtRdaFone}{0.956\xspace}
\newcommand{\XsfClangTOoLgtRseRecall}{1\xspace}
\newcommand{\XsfClangTOoLgtRsePrecision}{1\xspace}
\newcommand{\XsfClangTOoLgtRseFone}{1\xspace}
\newcommand{\XsfClangTOoBzpGT}{21415\xspace}
\newcommand{\XsfClangTOoBzpLsRecall}{1\xspace}
\newcommand{\XsfClangTOoBzpLsPrecision}{1\xspace}
\newcommand{\XsfClangTOoBzpLsFone}{1\xspace}
\newcommand{\XsfClangTOoBzpBapRecall}{0.782\xspace}
\newcommand{\XsfClangTOoBzpBapPrecision}{1\xspace}
\newcommand{\XsfClangTOoBzpBapFone}{0.877\xspace}
\newcommand{\XsfClangTOoBzpGhiRecall}{0.979\xspace}
\newcommand{\XsfClangTOoBzpGhiPrecision}{1\xspace}
\newcommand{\XsfClangTOoBzpGhiFone}{0.989\xspace}
\newcommand{\XsfClangTOoBzpRdaRecall}{0.775\xspace}
\newcommand{\XsfClangTOoBzpRdaPrecision}{1.000\xspace}
\newcommand{\XsfClangTOoBzpRdaFone}{0.873\xspace}
\newcommand{\XsfClangTOoBzpRseRecall}{1\xspace}
\newcommand{\XsfClangTOoBzpRsePrecision}{1\xspace}
\newcommand{\XsfClangTOoBzpRseFone}{1\xspace}
\newcommand{\XsfClangTOoGccGT}{1288674\xspace}
\newcommand{\XsfClangTOoGccLsRecall}{1\xspace}
\newcommand{\XsfClangTOoGccLsPrecision}{1\xspace}
\newcommand{\XsfClangTOoGccLsFone}{1\xspace}
\newcommand{\XsfClangTOoGccBapRecall}{0.796\xspace}
\newcommand{\XsfClangTOoGccBapPrecision}{1\xspace}
\newcommand{\XsfClangTOoGccBapFone}{0.886\xspace}
\newcommand{\XsfClangTOoGccGhiRecall}{0.996\xspace}
\newcommand{\XsfClangTOoGccGhiPrecision}{1\xspace}
\newcommand{\XsfClangTOoGccGhiFone}{0.998\xspace}
\newcommand{\XsfClangTOoGccRdaRecall}{0.742\xspace}
\newcommand{\XsfClangTOoGccRdaPrecision}{1.000\xspace}
\newcommand{\XsfClangTOoGccRdaFone}{0.852\xspace}
\newcommand{\XsfClangTOoGccRseRecall}{1\xspace}
\newcommand{\XsfClangTOoGccRsePrecision}{1\xspace}
\newcommand{\XsfClangTOoGccRseFone}{1\xspace}
\newcommand{\XsfClangTOoGzpGT}{13150\xspace}
\newcommand{\XsfClangTOoGzpLsRecall}{1\xspace}
\newcommand{\XsfClangTOoGzpLsPrecision}{1\xspace}
\newcommand{\XsfClangTOoGzpLsFone}{1\xspace}
\newcommand{\XsfClangTOoGzpBapRecall}{0.992\xspace}
\newcommand{\XsfClangTOoGzpBapPrecision}{1\xspace}
\newcommand{\XsfClangTOoGzpBapFone}{0.996\xspace}
\newcommand{\XsfClangTOoGzpGhiRecall}{1\xspace}
\newcommand{\XsfClangTOoGzpGhiPrecision}{1\xspace}
\newcommand{\XsfClangTOoGzpGhiFone}{1\xspace}
\newcommand{\XsfClangTOoGzpRdaRecall}{0.992\xspace}
\newcommand{\XsfClangTOoGzpRdaPrecision}{1\xspace}
\newcommand{\XsfClangTOoGzpRdaFone}{0.996\xspace}
\newcommand{\XsfClangTOoGzpRseRecall}{0.999\xspace}
\newcommand{\XsfClangTOoGzpRsePrecision}{1\xspace}
\newcommand{\XsfClangTOoGzpRseFone}{0.999\xspace}
\newcommand{\XsfClangTOoOggGT}{51120\xspace}
\newcommand{\XsfClangTOoOggLsRecall}{1\xspace}
\newcommand{\XsfClangTOoOggLsPrecision}{1\xspace}
\newcommand{\XsfClangTOoOggLsFone}{1\xspace}
\newcommand{\XsfClangTOoOggBapRecall}{0.994\xspace}
\newcommand{\XsfClangTOoOggBapPrecision}{1\xspace}
\newcommand{\XsfClangTOoOggBapFone}{0.997\xspace}
\newcommand{\XsfClangTOoOggGhiRecall}{1\xspace}
\newcommand{\XsfClangTOoOggGhiPrecision}{1\xspace}
\newcommand{\XsfClangTOoOggGhiFone}{1\xspace}
\newcommand{\XsfClangTOoOggRdaRecall}{0.994\xspace}
\newcommand{\XsfClangTOoOggRdaPrecision}{1\xspace}
\newcommand{\XsfClangTOoOggRdaFone}{0.997\xspace}
\newcommand{\XsfClangTOoOggRseRecall}{1\xspace}
\newcommand{\XsfClangTOoOggRsePrecision}{1\xspace}
\newcommand{\XsfClangTOoOggRseFone}{1\xspace}
\newcommand{\XsfClangTOoNgxGT}{159940\xspace}
\newcommand{\XsfClangTOoNgxLsRecall}{1\xspace}
\newcommand{\XsfClangTOoNgxLsPrecision}{1\xspace}
\newcommand{\XsfClangTOoNgxLsFone}{1\xspace}
\newcommand{\XsfClangTOoNgxBapRecall}{0.967\xspace}
\newcommand{\XsfClangTOoNgxBapPrecision}{1\xspace}
\newcommand{\XsfClangTOoNgxBapFone}{0.983\xspace}
\newcommand{\XsfClangTOoNgxGhiRecall}{1.000\xspace}
\newcommand{\XsfClangTOoNgxGhiPrecision}{1\xspace}
\newcommand{\XsfClangTOoNgxGhiFone}{1.000\xspace}
\newcommand{\XsfClangTOoNgxRdaRecall}{0.966\xspace}
\newcommand{\XsfClangTOoNgxRdaPrecision}{1\xspace}
\newcommand{\XsfClangTOoNgxRdaFone}{0.982\xspace}
\newcommand{\XsfClangTOoNgxRseRecall}{1\xspace}
\newcommand{\XsfClangTOoNgxRsePrecision}{1\xspace}
\newcommand{\XsfClangTOoNgxRseFone}{1\xspace}
\newcommand{\XsfClangTOoSshGT}{174011\xspace}
\newcommand{\XsfClangTOoSshLsRecall}{1\xspace}
\newcommand{\XsfClangTOoSshLsPrecision}{1\xspace}
\newcommand{\XsfClangTOoSshLsFone}{1\xspace}
\newcommand{\XsfClangTOoSshBapRecall}{0.964\xspace}
\newcommand{\XsfClangTOoSshBapPrecision}{1\xspace}
\newcommand{\XsfClangTOoSshBapFone}{0.981\xspace}
\newcommand{\XsfClangTOoSshGhiRecall}{0.986\xspace}
\newcommand{\XsfClangTOoSshGhiPrecision}{1\xspace}
\newcommand{\XsfClangTOoSshGhiFone}{0.993\xspace}
\newcommand{\XsfClangTOoSshRdaRecall}{0.847\xspace}
\newcommand{\XsfClangTOoSshRdaPrecision}{0.999\xspace}
\newcommand{\XsfClangTOoSshRdaFone}{0.917\xspace}
\newcommand{\XsfClangTOoSshRseRecall}{0.988\xspace}
\newcommand{\XsfClangTOoSshRsePrecision}{1\xspace}
\newcommand{\XsfClangTOoSshRseFone}{0.994\xspace}
\newcommand{\XsfClangTOoPcrGT}{6280\xspace}
\newcommand{\XsfClangTOoPcrLsRecall}{1\xspace}
\newcommand{\XsfClangTOoPcrLsPrecision}{1\xspace}
\newcommand{\XsfClangTOoPcrLsFone}{1\xspace}
\newcommand{\XsfClangTOoPcrBapRecall}{0.911\xspace}
\newcommand{\XsfClangTOoPcrBapPrecision}{1\xspace}
\newcommand{\XsfClangTOoPcrBapFone}{0.953\xspace}
\newcommand{\XsfClangTOoPcrGhiRecall}{1\xspace}
\newcommand{\XsfClangTOoPcrGhiPrecision}{1\xspace}
\newcommand{\XsfClangTOoPcrGhiFone}{1\xspace}
\newcommand{\XsfClangTOoPcrRdaRecall}{0.910\xspace}
\newcommand{\XsfClangTOoPcrRdaPrecision}{1\xspace}
\newcommand{\XsfClangTOoPcrRdaFone}{0.953\xspace}
\newcommand{\XsfClangTOoPcrRseRecall}{1\xspace}
\newcommand{\XsfClangTOoPcrRsePrecision}{1\xspace}
\newcommand{\XsfClangTOoPcrRseFone}{1\xspace}
\newcommand{\XsfClangTOoSqlGT}{213759\xspace}
\newcommand{\XsfClangTOoSqlLsRecall}{1\xspace}
\newcommand{\XsfClangTOoSqlLsPrecision}{1\xspace}
\newcommand{\XsfClangTOoSqlLsFone}{1\xspace}
\newcommand{\XsfClangTOoSqlBapRecall}{0.892\xspace}
\newcommand{\XsfClangTOoSqlBapPrecision}{1\xspace}
\newcommand{\XsfClangTOoSqlBapFone}{0.943\xspace}
\newcommand{\XsfClangTOoSqlGhiRecall}{1\xspace}
\newcommand{\XsfClangTOoSqlGhiPrecision}{1\xspace}
\newcommand{\XsfClangTOoSqlGhiFone}{1\xspace}
\newcommand{\XsfClangTOoSqlRdaRecall}{0.884\xspace}
\newcommand{\XsfClangTOoSqlRdaPrecision}{1\xspace}
\newcommand{\XsfClangTOoSqlRdaFone}{0.939\xspace}
\newcommand{\XsfClangTOoSqlRseRecall}{1\xspace}
\newcommand{\XsfClangTOoSqlRsePrecision}{1\xspace}
\newcommand{\XsfClangTOoSqlRseFone}{1\xspace}
\newcommand{\XsfClangTOoVimGT}{621689\xspace}
\newcommand{\XsfClangTOoVimLsRecall}{1\xspace}
\newcommand{\XsfClangTOoVimLsPrecision}{1\xspace}
\newcommand{\XsfClangTOoVimLsFone}{1\xspace}
\newcommand{\XsfClangTOoVimBapRecall}{0.964\xspace}
\newcommand{\XsfClangTOoVimBapPrecision}{1\xspace}
\newcommand{\XsfClangTOoVimBapFone}{0.982\xspace}
\newcommand{\XsfClangTOoVimGhiRecall}{1\xspace}
\newcommand{\XsfClangTOoVimGhiPrecision}{1\xspace}
\newcommand{\XsfClangTOoVimGhiFone}{1\xspace}
\newcommand{\XsfClangTOoVimRdaRecall}{0.924\xspace}
\newcommand{\XsfClangTOoVimRdaPrecision}{1.000\xspace}
\newcommand{\XsfClangTOoVimRdaFone}{0.961\xspace}
\newcommand{\XsfClangTOoVimRseRecall}{1.000\xspace}
\newcommand{\XsfClangTOoVimRsePrecision}{1\xspace}
\newcommand{\XsfClangTOoVimRseFone}{1.000\xspace}
\newcommand{\XsfClangTOoVsfGT}{23665\xspace}
\newcommand{\XsfClangTOoVsfLsRecall}{1\xspace}
\newcommand{\XsfClangTOoVsfLsPrecision}{1\xspace}
\newcommand{\XsfClangTOoVsfLsFone}{1\xspace}
\newcommand{\XsfClangTOoVsfBapRecall}{0.994\xspace}
\newcommand{\XsfClangTOoVsfBapPrecision}{1\xspace}
\newcommand{\XsfClangTOoVsfBapFone}{0.997\xspace}
\newcommand{\XsfClangTOoVsfGhiRecall}{0.997\xspace}
\newcommand{\XsfClangTOoVsfGhiPrecision}{1\xspace}
\newcommand{\XsfClangTOoVsfGhiFone}{0.999\xspace}
\newcommand{\XsfClangTOoVsfRdaRecall}{0.991\xspace}
\newcommand{\XsfClangTOoVsfRdaPrecision}{1.000\xspace}
\newcommand{\XsfClangTOoVsfRdaFone}{0.995\xspace}
\newcommand{\XsfClangTOoVsfRseRecall}{1.000\xspace}
\newcommand{\XsfClangTOoVsfRsePrecision}{1\xspace}
\newcommand{\XsfClangTOoVsfRseFone}{1.000\xspace}
\newcommand{\XsfClangTOaSzpGT}{12019\xspace}
\newcommand{\XsfClangTOaSzpLsRecall}{1\xspace}
\newcommand{\XsfClangTOaSzpLsPrecision}{1\xspace}
\newcommand{\XsfClangTOaSzpLsFone}{1\xspace}
\newcommand{\XsfClangTOaSzpBapRecall}{0.979\xspace}
\newcommand{\XsfClangTOaSzpBapPrecision}{1\xspace}
\newcommand{\XsfClangTOaSzpBapFone}{0.989\xspace}
\newcommand{\XsfClangTOaSzpGhiRecall}{1\xspace}
\newcommand{\XsfClangTOaSzpGhiPrecision}{1\xspace}
\newcommand{\XsfClangTOaSzpGhiFone}{1\xspace}
\newcommand{\XsfClangTOaSzpRdaRecall}{1\xspace}
\newcommand{\XsfClangTOaSzpRdaPrecision}{1\xspace}
\newcommand{\XsfClangTOaSzpRdaFone}{1\xspace}
\newcommand{\XsfClangTOaSzpRseRecall}{1\xspace}
\newcommand{\XsfClangTOaSzpRsePrecision}{1\xspace}
\newcommand{\XsfClangTOaSzpRseFone}{1\xspace}
\newcommand{\XsfClangTOaCapGT}{182063\xspace}
\newcommand{\XsfClangTOaCapLsRecall}{1\xspace}
\newcommand{\XsfClangTOaCapLsPrecision}{1\xspace}
\newcommand{\XsfClangTOaCapLsFone}{1\xspace}
\newcommand{\XsfClangTOaCapBapRecall}{0.375\xspace}
\newcommand{\XsfClangTOaCapBapPrecision}{1\xspace}
\newcommand{\XsfClangTOaCapBapFone}{0.546\xspace}
\newcommand{\XsfClangTOaCapGhiRecall}{1\xspace}
\newcommand{\XsfClangTOaCapGhiPrecision}{1\xspace}
\newcommand{\XsfClangTOaCapGhiFone}{1\xspace}
\newcommand{\XsfClangTOaCapRdaRecall}{1\xspace}
\newcommand{\XsfClangTOaCapRdaPrecision}{1\xspace}
\newcommand{\XsfClangTOaCapRdaFone}{1\xspace}
\newcommand{\XsfClangTOaCapRseRecall}{1.000\xspace}
\newcommand{\XsfClangTOaCapRsePrecision}{1\xspace}
\newcommand{\XsfClangTOaCapRseFone}{1.000\xspace}
\newcommand{\XsfClangTOaExmGT}{130464\xspace}
\newcommand{\XsfClangTOaExmLsRecall}{1\xspace}
\newcommand{\XsfClangTOaExmLsPrecision}{1\xspace}
\newcommand{\XsfClangTOaExmLsFone}{1\xspace}
\newcommand{\XsfClangTOaExmBapRecall}{0.858\xspace}
\newcommand{\XsfClangTOaExmBapPrecision}{1.000\xspace}
\newcommand{\XsfClangTOaExmBapFone}{0.923\xspace}
\newcommand{\XsfClangTOaExmGhiRecall}{0.986\xspace}
\newcommand{\XsfClangTOaExmGhiPrecision}{1\xspace}
\newcommand{\XsfClangTOaExmGhiFone}{0.993\xspace}
\newcommand{\XsfClangTOaExmRdaRecall}{0.976\xspace}
\newcommand{\XsfClangTOaExmRdaPrecision}{1.000\xspace}
\newcommand{\XsfClangTOaExmRdaFone}{0.988\xspace}
\newcommand{\XsfClangTOaExmRseRecall}{1.000\xspace}
\newcommand{\XsfClangTOaExmRsePrecision}{1\xspace}
\newcommand{\XsfClangTOaExmRseFone}{1.000\xspace}
\newcommand{\XsfClangTOaLgtGT}{23571\xspace}
\newcommand{\XsfClangTOaLgtLsRecall}{1\xspace}
\newcommand{\XsfClangTOaLgtLsPrecision}{1\xspace}
\newcommand{\XsfClangTOaLgtLsFone}{1\xspace}
\newcommand{\XsfClangTOaLgtBapRecall}{0.915\xspace}
\newcommand{\XsfClangTOaLgtBapPrecision}{1.000\xspace}
\newcommand{\XsfClangTOaLgtBapFone}{0.956\xspace}
\newcommand{\XsfClangTOaLgtGhiRecall}{1\xspace}
\newcommand{\XsfClangTOaLgtGhiPrecision}{1\xspace}
\newcommand{\XsfClangTOaLgtGhiFone}{1\xspace}
\newcommand{\XsfClangTOaLgtRdaRecall}{0.991\xspace}
\newcommand{\XsfClangTOaLgtRdaPrecision}{1.000\xspace}
\newcommand{\XsfClangTOaLgtRdaFone}{0.996\xspace}
\newcommand{\XsfClangTOaLgtRseRecall}{1\xspace}
\newcommand{\XsfClangTOaLgtRsePrecision}{1\xspace}
\newcommand{\XsfClangTOaLgtRseFone}{1\xspace}
\newcommand{\XsfClangTOaBzpGT}{13036\xspace}
\newcommand{\XsfClangTOaBzpLsRecall}{1\xspace}
\newcommand{\XsfClangTOaBzpLsPrecision}{1\xspace}
\newcommand{\XsfClangTOaBzpLsFone}{1\xspace}
\newcommand{\XsfClangTOaBzpBapRecall}{0.892\xspace}
\newcommand{\XsfClangTOaBzpBapPrecision}{1\xspace}
\newcommand{\XsfClangTOaBzpBapFone}{0.943\xspace}
\newcommand{\XsfClangTOaBzpGhiRecall}{1\xspace}
\newcommand{\XsfClangTOaBzpGhiPrecision}{1\xspace}
\newcommand{\XsfClangTOaBzpGhiFone}{1\xspace}
\newcommand{\XsfClangTOaBzpRdaRecall}{0.972\xspace}
\newcommand{\XsfClangTOaBzpRdaPrecision}{1\xspace}
\newcommand{\XsfClangTOaBzpRdaFone}{0.986\xspace}
\newcommand{\XsfClangTOaBzpRseRecall}{1.000\xspace}
\newcommand{\XsfClangTOaBzpRsePrecision}{1\xspace}
\newcommand{\XsfClangTOaBzpRseFone}{1.000\xspace}
\newcommand{\XsfClangTOaGccGT}{771565\xspace}
\newcommand{\XsfClangTOaGccLsRecall}{1\xspace}
\newcommand{\XsfClangTOaGccLsPrecision}{1\xspace}
\newcommand{\XsfClangTOaGccLsFone}{1\xspace}
\newcommand{\XsfClangTOaGccBapRecall}{0.754\xspace}
\newcommand{\XsfClangTOaGccBapPrecision}{1.000\xspace}
\newcommand{\XsfClangTOaGccBapFone}{0.860\xspace}
\newcommand{\XsfClangTOaGccGhiRecall}{0.992\xspace}
\newcommand{\XsfClangTOaGccGhiPrecision}{1\xspace}
\newcommand{\XsfClangTOaGccGhiFone}{0.996\xspace}
\newcommand{\XsfClangTOaGccRdaRecall}{0.837\xspace}
\newcommand{\XsfClangTOaGccRdaPrecision}{1.000\xspace}
\newcommand{\XsfClangTOaGccRdaFone}{0.911\xspace}
\newcommand{\XsfClangTOaGccRseRecall}{0.998\xspace}
\newcommand{\XsfClangTOaGccRsePrecision}{1\xspace}
\newcommand{\XsfClangTOaGccRseFone}{0.999\xspace}
\newcommand{\XsfClangTOaGzpGT}{8856\xspace}
\newcommand{\XsfClangTOaGzpLsRecall}{1\xspace}
\newcommand{\XsfClangTOaGzpLsPrecision}{1\xspace}
\newcommand{\XsfClangTOaGzpLsFone}{1\xspace}
\newcommand{\XsfClangTOaGzpBapRecall}{0.991\xspace}
\newcommand{\XsfClangTOaGzpBapPrecision}{1\xspace}
\newcommand{\XsfClangTOaGzpBapFone}{0.995\xspace}
\newcommand{\XsfClangTOaGzpGhiRecall}{1\xspace}
\newcommand{\XsfClangTOaGzpGhiPrecision}{1\xspace}
\newcommand{\XsfClangTOaGzpGhiFone}{1\xspace}
\newcommand{\XsfClangTOaGzpRdaRecall}{1\xspace}
\newcommand{\XsfClangTOaGzpRdaPrecision}{1\xspace}
\newcommand{\XsfClangTOaGzpRdaFone}{1\xspace}
\newcommand{\XsfClangTOaGzpRseRecall}{0.998\xspace}
\newcommand{\XsfClangTOaGzpRsePrecision}{1\xspace}
\newcommand{\XsfClangTOaGzpRseFone}{0.999\xspace}
\newcommand{\XsfClangTOaOggGT}{33410\xspace}
\newcommand{\XsfClangTOaOggLsRecall}{1\xspace}
\newcommand{\XsfClangTOaOggLsPrecision}{1\xspace}
\newcommand{\XsfClangTOaOggLsFone}{1\xspace}
\newcommand{\XsfClangTOaOggBapRecall}{0.981\xspace}
\newcommand{\XsfClangTOaOggBapPrecision}{1.000\xspace}
\newcommand{\XsfClangTOaOggBapFone}{0.990\xspace}
\newcommand{\XsfClangTOaOggGhiRecall}{1\xspace}
\newcommand{\XsfClangTOaOggGhiPrecision}{1\xspace}
\newcommand{\XsfClangTOaOggGhiFone}{1\xspace}
\newcommand{\XsfClangTOaOggRdaRecall}{1\xspace}
\newcommand{\XsfClangTOaOggRdaPrecision}{1.000\xspace}
\newcommand{\XsfClangTOaOggRdaFone}{1.000\xspace}
\newcommand{\XsfClangTOaOggRseRecall}{1\xspace}
\newcommand{\XsfClangTOaOggRsePrecision}{1\xspace}
\newcommand{\XsfClangTOaOggRseFone}{1\xspace}
\newcommand{\XsfClangTOaNgxGT}{93929\xspace}
\newcommand{\XsfClangTOaNgxLsRecall}{1\xspace}
\newcommand{\XsfClangTOaNgxLsPrecision}{1\xspace}
\newcommand{\XsfClangTOaNgxLsFone}{1\xspace}
\newcommand{\XsfClangTOaNgxBapRecall}{0.969\xspace}
\newcommand{\XsfClangTOaNgxBapPrecision}{1.000\xspace}
\newcommand{\XsfClangTOaNgxBapFone}{0.984\xspace}
\newcommand{\XsfClangTOaNgxGhiRecall}{1\xspace}
\newcommand{\XsfClangTOaNgxGhiPrecision}{1\xspace}
\newcommand{\XsfClangTOaNgxGhiFone}{1\xspace}
\newcommand{\XsfClangTOaNgxRdaRecall}{0.998\xspace}
\newcommand{\XsfClangTOaNgxRdaPrecision}{1.000\xspace}
\newcommand{\XsfClangTOaNgxRdaFone}{0.999\xspace}
\newcommand{\XsfClangTOaNgxRseRecall}{1\xspace}
\newcommand{\XsfClangTOaNgxRsePrecision}{1\xspace}
\newcommand{\XsfClangTOaNgxRseFone}{1\xspace}
\newcommand{\XsfClangTOaSshGT}{96521\xspace}
\newcommand{\XsfClangTOaSshLsRecall}{1\xspace}
\newcommand{\XsfClangTOaSshLsPrecision}{1\xspace}
\newcommand{\XsfClangTOaSshLsFone}{1\xspace}
\newcommand{\XsfClangTOaSshBapRecall}{0.954\xspace}
\newcommand{\XsfClangTOaSshBapPrecision}{1\xspace}
\newcommand{\XsfClangTOaSshBapFone}{0.976\xspace}
\newcommand{\XsfClangTOaSshGhiRecall}{1\xspace}
\newcommand{\XsfClangTOaSshGhiPrecision}{1\xspace}
\newcommand{\XsfClangTOaSshGhiFone}{1\xspace}
\newcommand{\XsfClangTOaSshRdaRecall}{0.961\xspace}
\newcommand{\XsfClangTOaSshRdaPrecision}{1.000\xspace}
\newcommand{\XsfClangTOaSshRdaFone}{0.980\xspace}
\newcommand{\XsfClangTOaSshRseRecall}{0.988\xspace}
\newcommand{\XsfClangTOaSshRsePrecision}{1\xspace}
\newcommand{\XsfClangTOaSshRseFone}{0.994\xspace}
\newcommand{\XsfClangTOaPcrGT}{4217\xspace}
\newcommand{\XsfClangTOaPcrLsRecall}{1\xspace}
\newcommand{\XsfClangTOaPcrLsPrecision}{1\xspace}
\newcommand{\XsfClangTOaPcrLsFone}{1\xspace}
\newcommand{\XsfClangTOaPcrBapRecall}{0.888\xspace}
\newcommand{\XsfClangTOaPcrBapPrecision}{1\xspace}
\newcommand{\XsfClangTOaPcrBapFone}{0.941\xspace}
\newcommand{\XsfClangTOaPcrGhiRecall}{1\xspace}
\newcommand{\XsfClangTOaPcrGhiPrecision}{1\xspace}
\newcommand{\XsfClangTOaPcrGhiFone}{1\xspace}
\newcommand{\XsfClangTOaPcrRdaRecall}{1\xspace}
\newcommand{\XsfClangTOaPcrRdaPrecision}{1\xspace}
\newcommand{\XsfClangTOaPcrRdaFone}{1\xspace}
\newcommand{\XsfClangTOaPcrRseRecall}{1\xspace}
\newcommand{\XsfClangTOaPcrRsePrecision}{1\xspace}
\newcommand{\XsfClangTOaPcrRseFone}{1\xspace}
\newcommand{\XsfClangTOaSqlGT}{149669\xspace}
\newcommand{\XsfClangTOaSqlLsRecall}{1\xspace}
\newcommand{\XsfClangTOaSqlLsPrecision}{1\xspace}
\newcommand{\XsfClangTOaSqlLsFone}{1\xspace}
\newcommand{\XsfClangTOaSqlBapRecall}{0.881\xspace}
\newcommand{\XsfClangTOaSqlBapPrecision}{1.000\xspace}
\newcommand{\XsfClangTOaSqlBapFone}{0.937\xspace}
\newcommand{\XsfClangTOaSqlGhiRecall}{1\xspace}
\newcommand{\XsfClangTOaSqlGhiPrecision}{1\xspace}
\newcommand{\XsfClangTOaSqlGhiFone}{1\xspace}
\newcommand{\XsfClangTOaSqlRdaRecall}{1\xspace}
\newcommand{\XsfClangTOaSqlRdaPrecision}{1.000\xspace}
\newcommand{\XsfClangTOaSqlRdaFone}{1.000\xspace}
\newcommand{\XsfClangTOaSqlRseRecall}{1.000\xspace}
\newcommand{\XsfClangTOaSqlRsePrecision}{1\xspace}
\newcommand{\XsfClangTOaSqlRseFone}{1.000\xspace}
\newcommand{\XsfClangTOaVimGT}{439673\xspace}
\newcommand{\XsfClangTOaVimLsRecall}{1\xspace}
\newcommand{\XsfClangTOaVimLsPrecision}{1\xspace}
\newcommand{\XsfClangTOaVimLsFone}{1\xspace}
\newcommand{\XsfClangTOaVimBapRecall}{0.929\xspace}
\newcommand{\XsfClangTOaVimBapPrecision}{1.000\xspace}
\newcommand{\XsfClangTOaVimBapFone}{0.963\xspace}
\newcommand{\XsfClangTOaVimGhiRecall}{1.000\xspace}
\newcommand{\XsfClangTOaVimGhiPrecision}{1\xspace}
\newcommand{\XsfClangTOaVimGhiFone}{1.000\xspace}
\newcommand{\XsfClangTOaVimRdaRecall}{0.989\xspace}
\newcommand{\XsfClangTOaVimRdaPrecision}{1.000\xspace}
\newcommand{\XsfClangTOaVimRdaFone}{0.994\xspace}
\newcommand{\XsfClangTOaVimRseRecall}{1.000\xspace}
\newcommand{\XsfClangTOaVimRsePrecision}{1\xspace}
\newcommand{\XsfClangTOaVimRseFone}{1.000\xspace}
\newcommand{\XsfClangTOaVsfGT}{17469\xspace}
\newcommand{\XsfClangTOaVsfLsRecall}{1\xspace}
\newcommand{\XsfClangTOaVsfLsPrecision}{1\xspace}
\newcommand{\XsfClangTOaVsfLsFone}{1\xspace}
\newcommand{\XsfClangTOaVsfBapRecall}{0.990\xspace}
\newcommand{\XsfClangTOaVsfBapPrecision}{1\xspace}
\newcommand{\XsfClangTOaVsfBapFone}{0.995\xspace}
\newcommand{\XsfClangTOaVsfGhiRecall}{1\xspace}
\newcommand{\XsfClangTOaVsfGhiPrecision}{1\xspace}
\newcommand{\XsfClangTOaVsfGhiFone}{1\xspace}
\newcommand{\XsfClangTOaVsfRdaRecall}{0.961\xspace}
\newcommand{\XsfClangTOaVsfRdaPrecision}{1.000\xspace}
\newcommand{\XsfClangTOaVsfRdaFone}{0.980\xspace}
\newcommand{\XsfClangTOaVsfRseRecall}{0.993\xspace}
\newcommand{\XsfClangTOaVsfRsePrecision}{1\xspace}
\newcommand{\XsfClangTOaVsfRseFone}{0.996\xspace}
\newcommand{\XsfClangTObSzpGT}{14443\xspace}
\newcommand{\XsfClangTObSzpLsRecall}{1\xspace}
\newcommand{\XsfClangTObSzpLsPrecision}{1\xspace}
\newcommand{\XsfClangTObSzpLsFone}{1\xspace}
\newcommand{\XsfClangTObSzpBapRecall}{0.966\xspace}
\newcommand{\XsfClangTObSzpBapPrecision}{1\xspace}
\newcommand{\XsfClangTObSzpBapFone}{0.983\xspace}
\newcommand{\XsfClangTObSzpGhiRecall}{0.975\xspace}
\newcommand{\XsfClangTObSzpGhiPrecision}{1\xspace}
\newcommand{\XsfClangTObSzpGhiFone}{0.988\xspace}
\newcommand{\XsfClangTObSzpRdaRecall}{1\xspace}
\newcommand{\XsfClangTObSzpRdaPrecision}{1\xspace}
\newcommand{\XsfClangTObSzpRdaFone}{1\xspace}
\newcommand{\XsfClangTObSzpRseRecall}{1\xspace}
\newcommand{\XsfClangTObSzpRsePrecision}{1\xspace}
\newcommand{\XsfClangTObSzpRseFone}{1\xspace}
\newcommand{\XsfClangTObCapGT}{168405\xspace}
\newcommand{\XsfClangTObCapLsRecall}{1\xspace}
\newcommand{\XsfClangTObCapLsPrecision}{1\xspace}
\newcommand{\XsfClangTObCapLsFone}{1\xspace}
\newcommand{\XsfClangTObCapBapRecall}{0.352\xspace}
\newcommand{\XsfClangTObCapBapPrecision}{1\xspace}
\newcommand{\XsfClangTObCapBapFone}{0.521\xspace}
\newcommand{\XsfClangTObCapGhiRecall}{0.841\xspace}
\newcommand{\XsfClangTObCapGhiPrecision}{1\xspace}
\newcommand{\XsfClangTObCapGhiFone}{0.914\xspace}
\newcommand{\XsfClangTObCapRdaRecall}{1\xspace}
\newcommand{\XsfClangTObCapRdaPrecision}{1\xspace}
\newcommand{\XsfClangTObCapRdaFone}{1\xspace}
\newcommand{\XsfClangTObCapRseRecall}{1.000\xspace}
\newcommand{\XsfClangTObCapRsePrecision}{1\xspace}
\newcommand{\XsfClangTObCapRseFone}{1.000\xspace}
\newcommand{\XsfClangTObExmGT}{149770\xspace}
\newcommand{\XsfClangTObExmLsRecall}{1\xspace}
\newcommand{\XsfClangTObExmLsPrecision}{1\xspace}
\newcommand{\XsfClangTObExmLsFone}{1\xspace}
\newcommand{\XsfClangTObExmBapRecall}{0.839\xspace}
\newcommand{\XsfClangTObExmBapPrecision}{1\xspace}
\newcommand{\XsfClangTObExmBapFone}{0.912\xspace}
\newcommand{\XsfClangTObExmGhiRecall}{0.987\xspace}
\newcommand{\XsfClangTObExmGhiPrecision}{1\xspace}
\newcommand{\XsfClangTObExmGhiFone}{0.993\xspace}
\newcommand{\XsfClangTObExmRdaRecall}{0.967\xspace}
\newcommand{\XsfClangTObExmRdaPrecision}{1.000\xspace}
\newcommand{\XsfClangTObExmRdaFone}{0.983\xspace}
\newcommand{\XsfClangTObExmRseRecall}{1.000\xspace}
\newcommand{\XsfClangTObExmRsePrecision}{1\xspace}
\newcommand{\XsfClangTObExmRseFone}{1.000\xspace}
\newcommand{\XsfClangTObLgtGT}{25740\xspace}
\newcommand{\XsfClangTObLgtLsRecall}{1\xspace}
\newcommand{\XsfClangTObLgtLsPrecision}{1\xspace}
\newcommand{\XsfClangTObLgtLsFone}{1\xspace}
\newcommand{\XsfClangTObLgtBapRecall}{0.908\xspace}
\newcommand{\XsfClangTObLgtBapPrecision}{1.000\xspace}
\newcommand{\XsfClangTObLgtBapFone}{0.952\xspace}
\newcommand{\XsfClangTObLgtGhiRecall}{1\xspace}
\newcommand{\XsfClangTObLgtGhiPrecision}{1\xspace}
\newcommand{\XsfClangTObLgtGhiFone}{1\xspace}
\newcommand{\XsfClangTObLgtRdaRecall}{1\xspace}
\newcommand{\XsfClangTObLgtRdaPrecision}{1\xspace}
\newcommand{\XsfClangTObLgtRdaFone}{1\xspace}
\newcommand{\XsfClangTObLgtRseRecall}{1\xspace}
\newcommand{\XsfClangTObLgtRsePrecision}{1\xspace}
\newcommand{\XsfClangTObLgtRseFone}{1\xspace}
\newcommand{\XsfClangTObBzpGT}{17453\xspace}
\newcommand{\XsfClangTObBzpLsRecall}{1\xspace}
\newcommand{\XsfClangTObBzpLsPrecision}{1\xspace}
\newcommand{\XsfClangTObBzpLsFone}{1\xspace}
\newcommand{\XsfClangTObBzpBapRecall}{0.971\xspace}
\newcommand{\XsfClangTObBzpBapPrecision}{1\xspace}
\newcommand{\XsfClangTObBzpBapFone}{0.985\xspace}
\newcommand{\XsfClangTObBzpGhiRecall}{1\xspace}
\newcommand{\XsfClangTObBzpGhiPrecision}{1\xspace}
\newcommand{\XsfClangTObBzpGhiFone}{1\xspace}
\newcommand{\XsfClangTObBzpRdaRecall}{0.942\xspace}
\newcommand{\XsfClangTObBzpRdaPrecision}{1\xspace}
\newcommand{\XsfClangTObBzpRdaFone}{0.970\xspace}
\newcommand{\XsfClangTObBzpRseRecall}{1\xspace}
\newcommand{\XsfClangTObBzpRsePrecision}{1\xspace}
\newcommand{\XsfClangTObBzpRseFone}{1\xspace}
\newcommand{\XsfClangTObGccGT}{1111356\xspace}
\newcommand{\XsfClangTObGccLsRecall}{1\xspace}
\newcommand{\XsfClangTObGccLsPrecision}{1\xspace}
\newcommand{\XsfClangTObGccLsFone}{1\xspace}
\newcommand{\XsfClangTObGccBapRecall}{0.731\xspace}
\newcommand{\XsfClangTObGccBapPrecision}{1.000\xspace}
\newcommand{\XsfClangTObGccBapFone}{0.845\xspace}
\newcommand{\XsfClangTObGccGhiRecall}{0.963\xspace}
\newcommand{\XsfClangTObGccGhiPrecision}{1\xspace}
\newcommand{\XsfClangTObGccGhiFone}{0.981\xspace}
\newcommand{\XsfClangTObGccRdaRecall}{0.849\xspace}
\newcommand{\XsfClangTObGccRdaPrecision}{1.000\xspace}
\newcommand{\XsfClangTObGccRdaFone}{0.918\xspace}
\newcommand{\XsfClangTObGccRseRecall}{0.999\xspace}
\newcommand{\XsfClangTObGccRsePrecision}{1\xspace}
\newcommand{\XsfClangTObGccRseFone}{1.000\xspace}
\newcommand{\XsfClangTObGzpGT}{12568\xspace}
\newcommand{\XsfClangTObGzpLsRecall}{1\xspace}
\newcommand{\XsfClangTObGzpLsPrecision}{1\xspace}
\newcommand{\XsfClangTObGzpLsFone}{1\xspace}
\newcommand{\XsfClangTObGzpBapRecall}{0.985\xspace}
\newcommand{\XsfClangTObGzpBapPrecision}{1\xspace}
\newcommand{\XsfClangTObGzpBapFone}{0.993\xspace}
\newcommand{\XsfClangTObGzpGhiRecall}{1\xspace}
\newcommand{\XsfClangTObGzpGhiPrecision}{1\xspace}
\newcommand{\XsfClangTObGzpGhiFone}{1\xspace}
\newcommand{\XsfClangTObGzpRdaRecall}{1\xspace}
\newcommand{\XsfClangTObGzpRdaPrecision}{1\xspace}
\newcommand{\XsfClangTObGzpRdaFone}{1\xspace}
\newcommand{\XsfClangTObGzpRseRecall}{0.999\xspace}
\newcommand{\XsfClangTObGzpRsePrecision}{1\xspace}
\newcommand{\XsfClangTObGzpRseFone}{0.999\xspace}
\newcommand{\XsfClangTObOggGT}{47434\xspace}
\newcommand{\XsfClangTObOggLsRecall}{1\xspace}
\newcommand{\XsfClangTObOggLsPrecision}{1\xspace}
\newcommand{\XsfClangTObOggLsFone}{1\xspace}
\newcommand{\XsfClangTObOggBapRecall}{0.984\xspace}
\newcommand{\XsfClangTObOggBapPrecision}{1.000\xspace}
\newcommand{\XsfClangTObOggBapFone}{0.992\xspace}
\newcommand{\XsfClangTObOggGhiRecall}{1\xspace}
\newcommand{\XsfClangTObOggGhiPrecision}{1\xspace}
\newcommand{\XsfClangTObOggGhiFone}{1\xspace}
\newcommand{\XsfClangTObOggRdaRecall}{1\xspace}
\newcommand{\XsfClangTObOggRdaPrecision}{1\xspace}
\newcommand{\XsfClangTObOggRdaFone}{1\xspace}
\newcommand{\XsfClangTObOggRseRecall}{1\xspace}
\newcommand{\XsfClangTObOggRsePrecision}{1\xspace}
\newcommand{\XsfClangTObOggRseFone}{1\xspace}
\newcommand{\XsfClangTObNgxGT}{98238\xspace}
\newcommand{\XsfClangTObNgxLsRecall}{1\xspace}
\newcommand{\XsfClangTObNgxLsPrecision}{1\xspace}
\newcommand{\XsfClangTObNgxLsFone}{1\xspace}
\newcommand{\XsfClangTObNgxBapRecall}{0.968\xspace}
\newcommand{\XsfClangTObNgxBapPrecision}{1.000\xspace}
\newcommand{\XsfClangTObNgxBapFone}{0.984\xspace}
\newcommand{\XsfClangTObNgxGhiRecall}{1\xspace}
\newcommand{\XsfClangTObNgxGhiPrecision}{1\xspace}
\newcommand{\XsfClangTObNgxGhiFone}{1\xspace}
\newcommand{\XsfClangTObNgxRdaRecall}{0.998\xspace}
\newcommand{\XsfClangTObNgxRdaPrecision}{1\xspace}
\newcommand{\XsfClangTObNgxRdaFone}{0.999\xspace}
\newcommand{\XsfClangTObNgxRseRecall}{1\xspace}
\newcommand{\XsfClangTObNgxRsePrecision}{1\xspace}
\newcommand{\XsfClangTObNgxRseFone}{1\xspace}
\newcommand{\XsfClangTObSshGT}{107113\xspace}
\newcommand{\XsfClangTObSshLsRecall}{1\xspace}
\newcommand{\XsfClangTObSshLsPrecision}{1\xspace}
\newcommand{\XsfClangTObSshLsFone}{1\xspace}
\newcommand{\XsfClangTObSshBapRecall}{0.943\xspace}
\newcommand{\XsfClangTObSshBapPrecision}{1\xspace}
\newcommand{\XsfClangTObSshBapFone}{0.971\xspace}
\newcommand{\XsfClangTObSshGhiRecall}{1\xspace}
\newcommand{\XsfClangTObSshGhiPrecision}{1\xspace}
\newcommand{\XsfClangTObSshGhiFone}{1\xspace}
\newcommand{\XsfClangTObSshRdaRecall}{0.963\xspace}
\newcommand{\XsfClangTObSshRdaPrecision}{1.000\xspace}
\newcommand{\XsfClangTObSshRdaFone}{0.981\xspace}
\newcommand{\XsfClangTObSshRseRecall}{0.989\xspace}
\newcommand{\XsfClangTObSshRsePrecision}{1\xspace}
\newcommand{\XsfClangTObSshRseFone}{0.994\xspace}
\newcommand{\XsfClangTObPcrGT}{4839\xspace}
\newcommand{\XsfClangTObPcrLsRecall}{1\xspace}
\newcommand{\XsfClangTObPcrLsPrecision}{1\xspace}
\newcommand{\XsfClangTObPcrLsFone}{1\xspace}
\newcommand{\XsfClangTObPcrBapRecall}{0.799\xspace}
\newcommand{\XsfClangTObPcrBapPrecision}{1\xspace}
\newcommand{\XsfClangTObPcrBapFone}{0.888\xspace}
\newcommand{\XsfClangTObPcrGhiRecall}{1\xspace}
\newcommand{\XsfClangTObPcrGhiPrecision}{1\xspace}
\newcommand{\XsfClangTObPcrGhiFone}{1\xspace}
\newcommand{\XsfClangTObPcrRdaRecall}{1\xspace}
\newcommand{\XsfClangTObPcrRdaPrecision}{1\xspace}
\newcommand{\XsfClangTObPcrRdaFone}{1\xspace}
\newcommand{\XsfClangTObPcrRseRecall}{1\xspace}
\newcommand{\XsfClangTObPcrRsePrecision}{1\xspace}
\newcommand{\XsfClangTObPcrRseFone}{1\xspace}
\newcommand{\XsfClangTObSqlGT}{255016\xspace}
\newcommand{\XsfClangTObSqlLsRecall}{1\xspace}
\newcommand{\XsfClangTObSqlLsPrecision}{1\xspace}
\newcommand{\XsfClangTObSqlLsFone}{1\xspace}
\newcommand{\XsfClangTObSqlBapRecall}{0.810\xspace}
\newcommand{\XsfClangTObSqlBapPrecision}{1.000\xspace}
\newcommand{\XsfClangTObSqlBapFone}{0.895\xspace}
\newcommand{\XsfClangTObSqlGhiRecall}{0.941\xspace}
\newcommand{\XsfClangTObSqlGhiPrecision}{1\xspace}
\newcommand{\XsfClangTObSqlGhiFone}{0.970\xspace}
\newcommand{\XsfClangTObSqlRdaRecall}{0.974\xspace}
\newcommand{\XsfClangTObSqlRdaPrecision}{1\xspace}
\newcommand{\XsfClangTObSqlRdaFone}{0.987\xspace}
\newcommand{\XsfClangTObSqlRseRecall}{0.996\xspace}
\newcommand{\XsfClangTObSqlRsePrecision}{1\xspace}
\newcommand{\XsfClangTObSqlRseFone}{0.998\xspace}
\newcommand{\XsfClangTObVimGT}{540475\xspace}
\newcommand{\XsfClangTObVimLsRecall}{1\xspace}
\newcommand{\XsfClangTObVimLsPrecision}{1\xspace}
\newcommand{\XsfClangTObVimLsFone}{1\xspace}
\newcommand{\XsfClangTObVimBapRecall}{0.868\xspace}
\newcommand{\XsfClangTObVimBapPrecision}{1.000\xspace}
\newcommand{\XsfClangTObVimBapFone}{0.929\xspace}
\newcommand{\XsfClangTObVimGhiRecall}{0.996\xspace}
\newcommand{\XsfClangTObVimGhiPrecision}{1\xspace}
\newcommand{\XsfClangTObVimGhiFone}{0.998\xspace}
\newcommand{\XsfClangTObVimRdaRecall}{0.000\xspace}
\newcommand{\XsfClangTObVimRdaPrecision}{NaN\xspace}
\newcommand{\XsfClangTObVimRdaFone}{0.000\xspace}
\newcommand{\XsfClangTObVimRseRecall}{1.000\xspace}
\newcommand{\XsfClangTObVimRsePrecision}{1\xspace}
\newcommand{\XsfClangTObVimRseFone}{1.000\xspace}
\newcommand{\XsfClangTObVsfGT}{19238\xspace}
\newcommand{\XsfClangTObVsfLsRecall}{1\xspace}
\newcommand{\XsfClangTObVsfLsPrecision}{1\xspace}
\newcommand{\XsfClangTObVsfLsFone}{1\xspace}
\newcommand{\XsfClangTObVsfBapRecall}{0.989\xspace}
\newcommand{\XsfClangTObVsfBapPrecision}{1\xspace}
\newcommand{\XsfClangTObVsfBapFone}{0.994\xspace}
\newcommand{\XsfClangTObVsfGhiRecall}{1\xspace}
\newcommand{\XsfClangTObVsfGhiPrecision}{1\xspace}
\newcommand{\XsfClangTObVsfGhiFone}{1\xspace}
\newcommand{\XsfClangTObVsfRdaRecall}{0.994\xspace}
\newcommand{\XsfClangTObVsfRdaPrecision}{1.000\xspace}
\newcommand{\XsfClangTObVsfRdaFone}{0.997\xspace}
\newcommand{\XsfClangTObVsfRseRecall}{0.994\xspace}
\newcommand{\XsfClangTObVsfRsePrecision}{1\xspace}
\newcommand{\XsfClangTObVsfRseFone}{0.997\xspace}
\newcommand{\XsfClangTOcSzpGT}{14727\xspace}
\newcommand{\XsfClangTOcSzpLsRecall}{1\xspace}
\newcommand{\XsfClangTOcSzpLsPrecision}{1\xspace}
\newcommand{\XsfClangTOcSzpLsFone}{1\xspace}
\newcommand{\XsfClangTOcSzpBapRecall}{0.967\xspace}
\newcommand{\XsfClangTOcSzpBapPrecision}{1\xspace}
\newcommand{\XsfClangTOcSzpBapFone}{0.983\xspace}
\newcommand{\XsfClangTOcSzpGhiRecall}{0.975\xspace}
\newcommand{\XsfClangTOcSzpGhiPrecision}{1\xspace}
\newcommand{\XsfClangTOcSzpGhiFone}{0.987\xspace}
\newcommand{\XsfClangTOcSzpRdaRecall}{1\xspace}
\newcommand{\XsfClangTOcSzpRdaPrecision}{1\xspace}
\newcommand{\XsfClangTOcSzpRdaFone}{1\xspace}
\newcommand{\XsfClangTOcSzpRseRecall}{1\xspace}
\newcommand{\XsfClangTOcSzpRsePrecision}{1\xspace}
\newcommand{\XsfClangTOcSzpRseFone}{1\xspace}
\newcommand{\XsfClangTOcCapGT}{171472\xspace}
\newcommand{\XsfClangTOcCapLsRecall}{1\xspace}
\newcommand{\XsfClangTOcCapLsPrecision}{1\xspace}
\newcommand{\XsfClangTOcCapLsFone}{1\xspace}
\newcommand{\XsfClangTOcCapBapRecall}{0.350\xspace}
\newcommand{\XsfClangTOcCapBapPrecision}{1\xspace}
\newcommand{\XsfClangTOcCapBapFone}{0.519\xspace}
\newcommand{\XsfClangTOcCapGhiRecall}{0.843\xspace}
\newcommand{\XsfClangTOcCapGhiPrecision}{1\xspace}
\newcommand{\XsfClangTOcCapGhiFone}{0.915\xspace}
\newcommand{\XsfClangTOcCapRdaRecall}{1\xspace}
\newcommand{\XsfClangTOcCapRdaPrecision}{1\xspace}
\newcommand{\XsfClangTOcCapRdaFone}{1\xspace}
\newcommand{\XsfClangTOcCapRseRecall}{1.000\xspace}
\newcommand{\XsfClangTOcCapRsePrecision}{1\xspace}
\newcommand{\XsfClangTOcCapRseFone}{1.000\xspace}
\newcommand{\XsfClangTOcExmGT}{156841\xspace}
\newcommand{\XsfClangTOcExmLsRecall}{1\xspace}
\newcommand{\XsfClangTOcExmLsPrecision}{1\xspace}
\newcommand{\XsfClangTOcExmLsFone}{1\xspace}
\newcommand{\XsfClangTOcExmBapRecall}{0.832\xspace}
\newcommand{\XsfClangTOcExmBapPrecision}{1\xspace}
\newcommand{\XsfClangTOcExmBapFone}{0.909\xspace}
\newcommand{\XsfClangTOcExmGhiRecall}{0.945\xspace}
\newcommand{\XsfClangTOcExmGhiPrecision}{1\xspace}
\newcommand{\XsfClangTOcExmGhiFone}{0.972\xspace}
\newcommand{\XsfClangTOcExmRdaRecall}{0.965\xspace}
\newcommand{\XsfClangTOcExmRdaPrecision}{1.000\xspace}
\newcommand{\XsfClangTOcExmRdaFone}{0.982\xspace}
\newcommand{\XsfClangTOcExmRseRecall}{1.000\xspace}
\newcommand{\XsfClangTOcExmRsePrecision}{1\xspace}
\newcommand{\XsfClangTOcExmRseFone}{1.000\xspace}
\newcommand{\XsfClangTOcLgtGT}{26530\xspace}
\newcommand{\XsfClangTOcLgtLsRecall}{1\xspace}
\newcommand{\XsfClangTOcLgtLsPrecision}{1\xspace}
\newcommand{\XsfClangTOcLgtLsFone}{1\xspace}
\newcommand{\XsfClangTOcLgtBapRecall}{0.904\xspace}
\newcommand{\XsfClangTOcLgtBapPrecision}{1.000\xspace}
\newcommand{\XsfClangTOcLgtBapFone}{0.949\xspace}
\newcommand{\XsfClangTOcLgtGhiRecall}{1\xspace}
\newcommand{\XsfClangTOcLgtGhiPrecision}{1\xspace}
\newcommand{\XsfClangTOcLgtGhiFone}{1\xspace}
\newcommand{\XsfClangTOcLgtRdaRecall}{1\xspace}
\newcommand{\XsfClangTOcLgtRdaPrecision}{1.000\xspace}
\newcommand{\XsfClangTOcLgtRdaFone}{1.000\xspace}
\newcommand{\XsfClangTOcLgtRseRecall}{1\xspace}
\newcommand{\XsfClangTOcLgtRsePrecision}{1\xspace}
\newcommand{\XsfClangTOcLgtRseFone}{1\xspace}
\newcommand{\XsfClangTOcBzpGT}{18052\xspace}
\newcommand{\XsfClangTOcBzpLsRecall}{1\xspace}
\newcommand{\XsfClangTOcBzpLsPrecision}{1\xspace}
\newcommand{\XsfClangTOcBzpLsFone}{1\xspace}
\newcommand{\XsfClangTOcBzpBapRecall}{0.972\xspace}
\newcommand{\XsfClangTOcBzpBapPrecision}{1\xspace}
\newcommand{\XsfClangTOcBzpBapFone}{0.986\xspace}
\newcommand{\XsfClangTOcBzpGhiRecall}{1\xspace}
\newcommand{\XsfClangTOcBzpGhiPrecision}{1\xspace}
\newcommand{\XsfClangTOcBzpGhiFone}{1\xspace}
\newcommand{\XsfClangTOcBzpRdaRecall}{0.927\xspace}
\newcommand{\XsfClangTOcBzpRdaPrecision}{1\xspace}
\newcommand{\XsfClangTOcBzpRdaFone}{0.962\xspace}
\newcommand{\XsfClangTOcBzpRseRecall}{1\xspace}
\newcommand{\XsfClangTOcBzpRsePrecision}{1\xspace}
\newcommand{\XsfClangTOcBzpRseFone}{1\xspace}
\newcommand{\XsfClangTOcGccGT}{1204605\xspace}
\newcommand{\XsfClangTOcGccLsRecall}{1\xspace}
\newcommand{\XsfClangTOcGccLsPrecision}{1\xspace}
\newcommand{\XsfClangTOcGccLsFone}{1\xspace}
\newcommand{\XsfClangTOcGccBapRecall}{0.729\xspace}
\newcommand{\XsfClangTOcGccBapPrecision}{1.000\xspace}
\newcommand{\XsfClangTOcGccBapFone}{0.843\xspace}
\newcommand{\XsfClangTOcGccGhiRecall}{0.962\xspace}
\newcommand{\XsfClangTOcGccGhiPrecision}{1\xspace}
\newcommand{\XsfClangTOcGccGhiFone}{0.981\xspace}
\newcommand{\XsfClangTOcGccRdaRecall}{0.842\xspace}
\newcommand{\XsfClangTOcGccRdaPrecision}{1.000\xspace}
\newcommand{\XsfClangTOcGccRdaFone}{0.914\xspace}
\newcommand{\XsfClangTOcGccRseRecall}{0.999\xspace}
\newcommand{\XsfClangTOcGccRsePrecision}{1\xspace}
\newcommand{\XsfClangTOcGccRseFone}{1.000\xspace}
\newcommand{\XsfClangTOcGzpGT}{14402\xspace}
\newcommand{\XsfClangTOcGzpLsRecall}{1\xspace}
\newcommand{\XsfClangTOcGzpLsPrecision}{1\xspace}
\newcommand{\XsfClangTOcGzpLsFone}{1\xspace}
\newcommand{\XsfClangTOcGzpBapRecall}{0.987\xspace}
\newcommand{\XsfClangTOcGzpBapPrecision}{1\xspace}
\newcommand{\XsfClangTOcGzpBapFone}{0.994\xspace}
\newcommand{\XsfClangTOcGzpGhiRecall}{1\xspace}
\newcommand{\XsfClangTOcGzpGhiPrecision}{1\xspace}
\newcommand{\XsfClangTOcGzpGhiFone}{1\xspace}
\newcommand{\XsfClangTOcGzpRdaRecall}{1\xspace}
\newcommand{\XsfClangTOcGzpRdaPrecision}{1\xspace}
\newcommand{\XsfClangTOcGzpRdaFone}{1\xspace}
\newcommand{\XsfClangTOcGzpRseRecall}{0.999\xspace}
\newcommand{\XsfClangTOcGzpRsePrecision}{1\xspace}
\newcommand{\XsfClangTOcGzpRseFone}{0.999\xspace}
\newcommand{\XsfClangTOcOggGT}{53788\xspace}
\newcommand{\XsfClangTOcOggLsRecall}{1\xspace}
\newcommand{\XsfClangTOcOggLsPrecision}{1\xspace}
\newcommand{\XsfClangTOcOggLsFone}{1\xspace}
\newcommand{\XsfClangTOcOggBapRecall}{0.986\xspace}
\newcommand{\XsfClangTOcOggBapPrecision}{1.000\xspace}
\newcommand{\XsfClangTOcOggBapFone}{0.993\xspace}
\newcommand{\XsfClangTOcOggGhiRecall}{1\xspace}
\newcommand{\XsfClangTOcOggGhiPrecision}{1\xspace}
\newcommand{\XsfClangTOcOggGhiFone}{1\xspace}
\newcommand{\XsfClangTOcOggRdaRecall}{1\xspace}
\newcommand{\XsfClangTOcOggRdaPrecision}{1\xspace}
\newcommand{\XsfClangTOcOggRdaFone}{1\xspace}
\newcommand{\XsfClangTOcOggRseRecall}{1\xspace}
\newcommand{\XsfClangTOcOggRsePrecision}{1\xspace}
\newcommand{\XsfClangTOcOggRseFone}{1\xspace}
\newcommand{\XsfClangTOcNgxGT}{101846\xspace}
\newcommand{\XsfClangTOcNgxLsRecall}{1\xspace}
\newcommand{\XsfClangTOcNgxLsPrecision}{1\xspace}
\newcommand{\XsfClangTOcNgxLsFone}{1\xspace}
\newcommand{\XsfClangTOcNgxBapRecall}{0.966\xspace}
\newcommand{\XsfClangTOcNgxBapPrecision}{1.000\xspace}
\newcommand{\XsfClangTOcNgxBapFone}{0.983\xspace}
\newcommand{\XsfClangTOcNgxGhiRecall}{1\xspace}
\newcommand{\XsfClangTOcNgxGhiPrecision}{1\xspace}
\newcommand{\XsfClangTOcNgxGhiFone}{1\xspace}
\newcommand{\XsfClangTOcNgxRdaRecall}{0.998\xspace}
\newcommand{\XsfClangTOcNgxRdaPrecision}{1\xspace}
\newcommand{\XsfClangTOcNgxRdaFone}{0.999\xspace}
\newcommand{\XsfClangTOcNgxRseRecall}{1\xspace}
\newcommand{\XsfClangTOcNgxRsePrecision}{1\xspace}
\newcommand{\XsfClangTOcNgxRseFone}{1\xspace}
\newcommand{\XsfClangTOcSshGT}{113250\xspace}
\newcommand{\XsfClangTOcSshLsRecall}{1\xspace}
\newcommand{\XsfClangTOcSshLsPrecision}{1\xspace}
\newcommand{\XsfClangTOcSshLsFone}{1\xspace}
\newcommand{\XsfClangTOcSshBapRecall}{0.945\xspace}
\newcommand{\XsfClangTOcSshBapPrecision}{1\xspace}
\newcommand{\XsfClangTOcSshBapFone}{0.972\xspace}
\newcommand{\XsfClangTOcSshGhiRecall}{1\xspace}
\newcommand{\XsfClangTOcSshGhiPrecision}{1\xspace}
\newcommand{\XsfClangTOcSshGhiFone}{1\xspace}
\newcommand{\XsfClangTOcSshRdaRecall}{0.945\xspace}
\newcommand{\XsfClangTOcSshRdaPrecision}{1.000\xspace}
\newcommand{\XsfClangTOcSshRdaFone}{0.972\xspace}
\newcommand{\XsfClangTOcSshRseRecall}{0.989\xspace}
\newcommand{\XsfClangTOcSshRsePrecision}{1\xspace}
\newcommand{\XsfClangTOcSshRseFone}{0.995\xspace}
\newcommand{\XsfClangTOcPcrGT}{5614\xspace}
\newcommand{\XsfClangTOcPcrLsRecall}{1\xspace}
\newcommand{\XsfClangTOcPcrLsPrecision}{1\xspace}
\newcommand{\XsfClangTOcPcrLsFone}{1\xspace}
\newcommand{\XsfClangTOcPcrBapRecall}{0.784\xspace}
\newcommand{\XsfClangTOcPcrBapPrecision}{1\xspace}
\newcommand{\XsfClangTOcPcrBapFone}{0.879\xspace}
\newcommand{\XsfClangTOcPcrGhiRecall}{1\xspace}
\newcommand{\XsfClangTOcPcrGhiPrecision}{1\xspace}
\newcommand{\XsfClangTOcPcrGhiFone}{1\xspace}
\newcommand{\XsfClangTOcPcrRdaRecall}{1\xspace}
\newcommand{\XsfClangTOcPcrRdaPrecision}{1\xspace}
\newcommand{\XsfClangTOcPcrRdaFone}{1\xspace}
\newcommand{\XsfClangTOcPcrRseRecall}{1\xspace}
\newcommand{\XsfClangTOcPcrRsePrecision}{1\xspace}
\newcommand{\XsfClangTOcPcrRseFone}{1\xspace}
\newcommand{\XsfClangTOcSqlGT}{290371\xspace}
\newcommand{\XsfClangTOcSqlLsRecall}{1\xspace}
\newcommand{\XsfClangTOcSqlLsPrecision}{1\xspace}
\newcommand{\XsfClangTOcSqlLsFone}{1\xspace}
\newcommand{\XsfClangTOcSqlBapRecall}{0.810\xspace}
\newcommand{\XsfClangTOcSqlBapPrecision}{1.000\xspace}
\newcommand{\XsfClangTOcSqlBapFone}{0.895\xspace}
\newcommand{\XsfClangTOcSqlGhiRecall}{0.944\xspace}
\newcommand{\XsfClangTOcSqlGhiPrecision}{1\xspace}
\newcommand{\XsfClangTOcSqlGhiFone}{0.971\xspace}
\newcommand{\XsfClangTOcSqlRdaRecall}{0.971\xspace}
\newcommand{\XsfClangTOcSqlRdaPrecision}{1\xspace}
\newcommand{\XsfClangTOcSqlRdaFone}{0.985\xspace}
\newcommand{\XsfClangTOcSqlRseRecall}{0.997\xspace}
\newcommand{\XsfClangTOcSqlRsePrecision}{1\xspace}
\newcommand{\XsfClangTOcSqlRseFone}{0.998\xspace}
\newcommand{\XsfClangTOcVimGT}{580239\xspace}
\newcommand{\XsfClangTOcVimLsRecall}{1\xspace}
\newcommand{\XsfClangTOcVimLsPrecision}{1\xspace}
\newcommand{\XsfClangTOcVimLsFone}{1\xspace}
\newcommand{\XsfClangTOcVimBapRecall}{0.872\xspace}
\newcommand{\XsfClangTOcVimBapPrecision}{1.000\xspace}
\newcommand{\XsfClangTOcVimBapFone}{0.932\xspace}
\newcommand{\XsfClangTOcVimGhiRecall}{1.000\xspace}
\newcommand{\XsfClangTOcVimGhiPrecision}{1\xspace}
\newcommand{\XsfClangTOcVimGhiFone}{1.000\xspace}
\newcommand{\XsfClangTOcVimRdaRecall}{0.968\xspace}
\newcommand{\XsfClangTOcVimRdaPrecision}{1.000\xspace}
\newcommand{\XsfClangTOcVimRdaFone}{0.984\xspace}
\newcommand{\XsfClangTOcVimRseRecall}{1.000\xspace}
\newcommand{\XsfClangTOcVimRsePrecision}{1\xspace}
\newcommand{\XsfClangTOcVimRseFone}{1.000\xspace}
\newcommand{\XsfClangTOcVsfGT}{19844\xspace}
\newcommand{\XsfClangTOcVsfLsRecall}{1\xspace}
\newcommand{\XsfClangTOcVsfLsPrecision}{1\xspace}
\newcommand{\XsfClangTOcVsfLsFone}{1\xspace}
\newcommand{\XsfClangTOcVsfBapRecall}{0.989\xspace}
\newcommand{\XsfClangTOcVsfBapPrecision}{1\xspace}
\newcommand{\XsfClangTOcVsfBapFone}{0.995\xspace}
\newcommand{\XsfClangTOcVsfGhiRecall}{1\xspace}
\newcommand{\XsfClangTOcVsfGhiPrecision}{1\xspace}
\newcommand{\XsfClangTOcVsfGhiFone}{1\xspace}
\newcommand{\XsfClangTOcVsfRdaRecall}{0.994\xspace}
\newcommand{\XsfClangTOcVsfRdaPrecision}{0.999\xspace}
\newcommand{\XsfClangTOcVsfRdaFone}{0.997\xspace}
\newcommand{\XsfClangTOcVsfRseRecall}{0.994\xspace}
\newcommand{\XsfClangTOcVsfRsePrecision}{1\xspace}
\newcommand{\XsfClangTOcVsfRseFone}{0.997\xspace}
\newcommand{\XsfClangTOdSzpGT}{14727\xspace}
\newcommand{\XsfClangTOdSzpLsRecall}{1\xspace}
\newcommand{\XsfClangTOdSzpLsPrecision}{1\xspace}
\newcommand{\XsfClangTOdSzpLsFone}{1\xspace}
\newcommand{\XsfClangTOdSzpBapRecall}{0.967\xspace}
\newcommand{\XsfClangTOdSzpBapPrecision}{1\xspace}
\newcommand{\XsfClangTOdSzpBapFone}{0.983\xspace}
\newcommand{\XsfClangTOdSzpGhiRecall}{0.975\xspace}
\newcommand{\XsfClangTOdSzpGhiPrecision}{1\xspace}
\newcommand{\XsfClangTOdSzpGhiFone}{0.987\xspace}
\newcommand{\XsfClangTOdSzpRdaRecall}{1\xspace}
\newcommand{\XsfClangTOdSzpRdaPrecision}{1\xspace}
\newcommand{\XsfClangTOdSzpRdaFone}{1\xspace}
\newcommand{\XsfClangTOdSzpRseRecall}{1\xspace}
\newcommand{\XsfClangTOdSzpRsePrecision}{1\xspace}
\newcommand{\XsfClangTOdSzpRseFone}{1\xspace}
\newcommand{\XsfClangTOdCapGT}{171472\xspace}
\newcommand{\XsfClangTOdCapLsRecall}{1\xspace}
\newcommand{\XsfClangTOdCapLsPrecision}{1\xspace}
\newcommand{\XsfClangTOdCapLsFone}{1\xspace}
\newcommand{\XsfClangTOdCapBapRecall}{0.350\xspace}
\newcommand{\XsfClangTOdCapBapPrecision}{1\xspace}
\newcommand{\XsfClangTOdCapBapFone}{0.519\xspace}
\newcommand{\XsfClangTOdCapGhiRecall}{0.856\xspace}
\newcommand{\XsfClangTOdCapGhiPrecision}{1\xspace}
\newcommand{\XsfClangTOdCapGhiFone}{0.922\xspace}
\newcommand{\XsfClangTOdCapRdaRecall}{1\xspace}
\newcommand{\XsfClangTOdCapRdaPrecision}{1\xspace}
\newcommand{\XsfClangTOdCapRdaFone}{1\xspace}
\newcommand{\XsfClangTOdCapRseRecall}{1.000\xspace}
\newcommand{\XsfClangTOdCapRsePrecision}{1\xspace}
\newcommand{\XsfClangTOdCapRseFone}{1.000\xspace}
\newcommand{\XsfClangTOdExmGT}{156835\xspace}
\newcommand{\XsfClangTOdExmLsRecall}{1\xspace}
\newcommand{\XsfClangTOdExmLsPrecision}{1\xspace}
\newcommand{\XsfClangTOdExmLsFone}{1\xspace}
\newcommand{\XsfClangTOdExmBapRecall}{0.832\xspace}
\newcommand{\XsfClangTOdExmBapPrecision}{1\xspace}
\newcommand{\XsfClangTOdExmBapFone}{0.909\xspace}
\newcommand{\XsfClangTOdExmGhiRecall}{0.965\xspace}
\newcommand{\XsfClangTOdExmGhiPrecision}{1\xspace}
\newcommand{\XsfClangTOdExmGhiFone}{0.982\xspace}
\newcommand{\XsfClangTOdExmRdaRecall}{0.966\xspace}
\newcommand{\XsfClangTOdExmRdaPrecision}{1.000\xspace}
\newcommand{\XsfClangTOdExmRdaFone}{0.983\xspace}
\newcommand{\XsfClangTOdExmRseRecall}{1.000\xspace}
\newcommand{\XsfClangTOdExmRsePrecision}{1\xspace}
\newcommand{\XsfClangTOdExmRseFone}{1.000\xspace}
\newcommand{\XsfClangTOdLgtGT}{26530\xspace}
\newcommand{\XsfClangTOdLgtLsRecall}{1\xspace}
\newcommand{\XsfClangTOdLgtLsPrecision}{1\xspace}
\newcommand{\XsfClangTOdLgtLsFone}{1\xspace}
\newcommand{\XsfClangTOdLgtBapRecall}{0.904\xspace}
\newcommand{\XsfClangTOdLgtBapPrecision}{1.000\xspace}
\newcommand{\XsfClangTOdLgtBapFone}{0.949\xspace}
\newcommand{\XsfClangTOdLgtGhiRecall}{1\xspace}
\newcommand{\XsfClangTOdLgtGhiPrecision}{1\xspace}
\newcommand{\XsfClangTOdLgtGhiFone}{1\xspace}
\newcommand{\XsfClangTOdLgtRdaRecall}{1.000\xspace}
\newcommand{\XsfClangTOdLgtRdaPrecision}{1.000\xspace}
\newcommand{\XsfClangTOdLgtRdaFone}{1.000\xspace}
\newcommand{\XsfClangTOdLgtRseRecall}{1\xspace}
\newcommand{\XsfClangTOdLgtRsePrecision}{1\xspace}
\newcommand{\XsfClangTOdLgtRseFone}{1\xspace}
\newcommand{\XsfClangTOdBzpGT}{18054\xspace}
\newcommand{\XsfClangTOdBzpLsRecall}{1\xspace}
\newcommand{\XsfClangTOdBzpLsPrecision}{1\xspace}
\newcommand{\XsfClangTOdBzpLsFone}{1\xspace}
\newcommand{\XsfClangTOdBzpBapRecall}{0.972\xspace}
\newcommand{\XsfClangTOdBzpBapPrecision}{1\xspace}
\newcommand{\XsfClangTOdBzpBapFone}{0.986\xspace}
\newcommand{\XsfClangTOdBzpGhiRecall}{1\xspace}
\newcommand{\XsfClangTOdBzpGhiPrecision}{1\xspace}
\newcommand{\XsfClangTOdBzpGhiFone}{1\xspace}
\newcommand{\XsfClangTOdBzpRdaRecall}{0.904\xspace}
\newcommand{\XsfClangTOdBzpRdaPrecision}{1.000\xspace}
\newcommand{\XsfClangTOdBzpRdaFone}{0.949\xspace}
\newcommand{\XsfClangTOdBzpRseRecall}{1\xspace}
\newcommand{\XsfClangTOdBzpRsePrecision}{1\xspace}
\newcommand{\XsfClangTOdBzpRseFone}{1\xspace}
\newcommand{\XsfClangTOdGccGT}{1204636\xspace}
\newcommand{\XsfClangTOdGccLsRecall}{1\xspace}
\newcommand{\XsfClangTOdGccLsPrecision}{1\xspace}
\newcommand{\XsfClangTOdGccLsFone}{1\xspace}
\newcommand{\XsfClangTOdGccBapRecall}{0.729\xspace}
\newcommand{\XsfClangTOdGccBapPrecision}{1.000\xspace}
\newcommand{\XsfClangTOdGccBapFone}{0.843\xspace}
\newcommand{\XsfClangTOdGccGhiRecall}{0.964\xspace}
\newcommand{\XsfClangTOdGccGhiPrecision}{1\xspace}
\newcommand{\XsfClangTOdGccGhiFone}{0.982\xspace}
\newcommand{\XsfClangTOdGccRdaRecall}{0.842\xspace}
\newcommand{\XsfClangTOdGccRdaPrecision}{1\xspace}
\newcommand{\XsfClangTOdGccRdaFone}{0.914\xspace}
\newcommand{\XsfClangTOdGccRseRecall}{0.999\xspace}
\newcommand{\XsfClangTOdGccRsePrecision}{1\xspace}
\newcommand{\XsfClangTOdGccRseFone}{1.000\xspace}
\newcommand{\XsfClangTOdGzpGT}{14402\xspace}
\newcommand{\XsfClangTOdGzpLsRecall}{1\xspace}
\newcommand{\XsfClangTOdGzpLsPrecision}{1\xspace}
\newcommand{\XsfClangTOdGzpLsFone}{1\xspace}
\newcommand{\XsfClangTOdGzpBapRecall}{0.987\xspace}
\newcommand{\XsfClangTOdGzpBapPrecision}{1\xspace}
\newcommand{\XsfClangTOdGzpBapFone}{0.994\xspace}
\newcommand{\XsfClangTOdGzpGhiRecall}{1\xspace}
\newcommand{\XsfClangTOdGzpGhiPrecision}{1\xspace}
\newcommand{\XsfClangTOdGzpGhiFone}{1\xspace}
\newcommand{\XsfClangTOdGzpRdaRecall}{1\xspace}
\newcommand{\XsfClangTOdGzpRdaPrecision}{1\xspace}
\newcommand{\XsfClangTOdGzpRdaFone}{1\xspace}
\newcommand{\XsfClangTOdGzpRseRecall}{0.999\xspace}
\newcommand{\XsfClangTOdGzpRsePrecision}{1\xspace}
\newcommand{\XsfClangTOdGzpRseFone}{0.999\xspace}
\newcommand{\XsfClangTOdOggGT}{54445\xspace}
\newcommand{\XsfClangTOdOggLsRecall}{1\xspace}
\newcommand{\XsfClangTOdOggLsPrecision}{1\xspace}
\newcommand{\XsfClangTOdOggLsFone}{1\xspace}
\newcommand{\XsfClangTOdOggBapRecall}{0.986\xspace}
\newcommand{\XsfClangTOdOggBapPrecision}{1.000\xspace}
\newcommand{\XsfClangTOdOggBapFone}{0.993\xspace}
\newcommand{\XsfClangTOdOggGhiRecall}{1\xspace}
\newcommand{\XsfClangTOdOggGhiPrecision}{1\xspace}
\newcommand{\XsfClangTOdOggGhiFone}{1\xspace}
\newcommand{\XsfClangTOdOggRdaRecall}{1\xspace}
\newcommand{\XsfClangTOdOggRdaPrecision}{1\xspace}
\newcommand{\XsfClangTOdOggRdaFone}{1\xspace}
\newcommand{\XsfClangTOdOggRseRecall}{1\xspace}
\newcommand{\XsfClangTOdOggRsePrecision}{1\xspace}
\newcommand{\XsfClangTOdOggRseFone}{1\xspace}
\newcommand{\XsfClangTOdNgxGT}{101846\xspace}
\newcommand{\XsfClangTOdNgxLsRecall}{1\xspace}
\newcommand{\XsfClangTOdNgxLsPrecision}{1\xspace}
\newcommand{\XsfClangTOdNgxLsFone}{1\xspace}
\newcommand{\XsfClangTOdNgxBapRecall}{0.966\xspace}
\newcommand{\XsfClangTOdNgxBapPrecision}{1.000\xspace}
\newcommand{\XsfClangTOdNgxBapFone}{0.983\xspace}
\newcommand{\XsfClangTOdNgxGhiRecall}{1\xspace}
\newcommand{\XsfClangTOdNgxGhiPrecision}{1\xspace}
\newcommand{\XsfClangTOdNgxGhiFone}{1\xspace}
\newcommand{\XsfClangTOdNgxRdaRecall}{0.998\xspace}
\newcommand{\XsfClangTOdNgxRdaPrecision}{1\xspace}
\newcommand{\XsfClangTOdNgxRdaFone}{0.999\xspace}
\newcommand{\XsfClangTOdNgxRseRecall}{1\xspace}
\newcommand{\XsfClangTOdNgxRsePrecision}{1\xspace}
\newcommand{\XsfClangTOdNgxRseFone}{1\xspace}
\newcommand{\XsfClangTOdSshGT}{113250\xspace}
\newcommand{\XsfClangTOdSshLsRecall}{1\xspace}
\newcommand{\XsfClangTOdSshLsPrecision}{1\xspace}
\newcommand{\XsfClangTOdSshLsFone}{1\xspace}
\newcommand{\XsfClangTOdSshBapRecall}{0.945\xspace}
\newcommand{\XsfClangTOdSshBapPrecision}{1\xspace}
\newcommand{\XsfClangTOdSshBapFone}{0.972\xspace}
\newcommand{\XsfClangTOdSshGhiRecall}{1\xspace}
\newcommand{\XsfClangTOdSshGhiPrecision}{1\xspace}
\newcommand{\XsfClangTOdSshGhiFone}{1\xspace}
\newcommand{\XsfClangTOdSshRdaRecall}{0.945\xspace}
\newcommand{\XsfClangTOdSshRdaPrecision}{1.000\xspace}
\newcommand{\XsfClangTOdSshRdaFone}{0.972\xspace}
\newcommand{\XsfClangTOdSshRseRecall}{0.989\xspace}
\newcommand{\XsfClangTOdSshRsePrecision}{1\xspace}
\newcommand{\XsfClangTOdSshRseFone}{0.995\xspace}
\newcommand{\XsfClangTOdPcrGT}{5614\xspace}
\newcommand{\XsfClangTOdPcrLsRecall}{1\xspace}
\newcommand{\XsfClangTOdPcrLsPrecision}{1\xspace}
\newcommand{\XsfClangTOdPcrLsFone}{1\xspace}
\newcommand{\XsfClangTOdPcrBapRecall}{0.784\xspace}
\newcommand{\XsfClangTOdPcrBapPrecision}{1\xspace}
\newcommand{\XsfClangTOdPcrBapFone}{0.879\xspace}
\newcommand{\XsfClangTOdPcrGhiRecall}{1\xspace}
\newcommand{\XsfClangTOdPcrGhiPrecision}{1\xspace}
\newcommand{\XsfClangTOdPcrGhiFone}{1\xspace}
\newcommand{\XsfClangTOdPcrRdaRecall}{1\xspace}
\newcommand{\XsfClangTOdPcrRdaPrecision}{1.000\xspace}
\newcommand{\XsfClangTOdPcrRdaFone}{1.000\xspace}
\newcommand{\XsfClangTOdPcrRseRecall}{1\xspace}
\newcommand{\XsfClangTOdPcrRsePrecision}{1\xspace}
\newcommand{\XsfClangTOdPcrRseFone}{1\xspace}
\newcommand{\XsfClangTOdSqlGT}{290355\xspace}
\newcommand{\XsfClangTOdSqlLsRecall}{1\xspace}
\newcommand{\XsfClangTOdSqlLsPrecision}{1\xspace}
\newcommand{\XsfClangTOdSqlLsFone}{1\xspace}
\newcommand{\XsfClangTOdSqlBapRecall}{0.810\xspace}
\newcommand{\XsfClangTOdSqlBapPrecision}{1.000\xspace}
\newcommand{\XsfClangTOdSqlBapFone}{0.895\xspace}
\newcommand{\XsfClangTOdSqlGhiRecall}{0.944\xspace}
\newcommand{\XsfClangTOdSqlGhiPrecision}{1\xspace}
\newcommand{\XsfClangTOdSqlGhiFone}{0.971\xspace}
\newcommand{\XsfClangTOdSqlRdaRecall}{0.970\xspace}
\newcommand{\XsfClangTOdSqlRdaPrecision}{1.000\xspace}
\newcommand{\XsfClangTOdSqlRdaFone}{0.985\xspace}
\newcommand{\XsfClangTOdSqlRseRecall}{0.997\xspace}
\newcommand{\XsfClangTOdSqlRsePrecision}{1\xspace}
\newcommand{\XsfClangTOdSqlRseFone}{0.998\xspace}
\newcommand{\XsfClangTOdVimGT}{580227\xspace}
\newcommand{\XsfClangTOdVimLsRecall}{1\xspace}
\newcommand{\XsfClangTOdVimLsPrecision}{1\xspace}
\newcommand{\XsfClangTOdVimLsFone}{1\xspace}
\newcommand{\XsfClangTOdVimBapRecall}{0.872\xspace}
\newcommand{\XsfClangTOdVimBapPrecision}{1.000\xspace}
\newcommand{\XsfClangTOdVimBapFone}{0.932\xspace}
\newcommand{\XsfClangTOdVimGhiRecall}{0.997\xspace}
\newcommand{\XsfClangTOdVimGhiPrecision}{1\xspace}
\newcommand{\XsfClangTOdVimGhiFone}{0.999\xspace}
\newcommand{\XsfClangTOdVimRdaRecall}{0.969\xspace}
\newcommand{\XsfClangTOdVimRdaPrecision}{1.000\xspace}
\newcommand{\XsfClangTOdVimRdaFone}{0.984\xspace}
\newcommand{\XsfClangTOdVimRseRecall}{1.000\xspace}
\newcommand{\XsfClangTOdVimRsePrecision}{1\xspace}
\newcommand{\XsfClangTOdVimRseFone}{1.000\xspace}
\newcommand{\XsfClangTOdVsfGT}{19844\xspace}
\newcommand{\XsfClangTOdVsfLsRecall}{1\xspace}
\newcommand{\XsfClangTOdVsfLsPrecision}{1\xspace}
\newcommand{\XsfClangTOdVsfLsFone}{1\xspace}
\newcommand{\XsfClangTOdVsfBapRecall}{0.989\xspace}
\newcommand{\XsfClangTOdVsfBapPrecision}{1\xspace}
\newcommand{\XsfClangTOdVsfBapFone}{0.995\xspace}
\newcommand{\XsfClangTOdVsfGhiRecall}{1\xspace}
\newcommand{\XsfClangTOdVsfGhiPrecision}{1\xspace}
\newcommand{\XsfClangTOdVsfGhiFone}{1\xspace}
\newcommand{\XsfClangTOdVsfRdaRecall}{0.994\xspace}
\newcommand{\XsfClangTOdVsfRdaPrecision}{0.999\xspace}
\newcommand{\XsfClangTOdVsfRdaFone}{0.997\xspace}
\newcommand{\XsfClangTOdVsfRseRecall}{0.994\xspace}
\newcommand{\XsfClangTOdVsfRsePrecision}{1\xspace}
\newcommand{\XsfClangTOdVsfRseFone}{0.997\xspace}
\newcommand{\XsfClangTOsSzpGT}{11965\xspace}
\newcommand{\XsfClangTOsSzpLsRecall}{1\xspace}
\newcommand{\XsfClangTOsSzpLsPrecision}{1\xspace}
\newcommand{\XsfClangTOsSzpLsFone}{1\xspace}
\newcommand{\XsfClangTOsSzpBapRecall}{0.974\xspace}
\newcommand{\XsfClangTOsSzpBapPrecision}{1\xspace}
\newcommand{\XsfClangTOsSzpBapFone}{0.987\xspace}
\newcommand{\XsfClangTOsSzpGhiRecall}{1\xspace}
\newcommand{\XsfClangTOsSzpGhiPrecision}{1\xspace}
\newcommand{\XsfClangTOsSzpGhiFone}{1\xspace}
\newcommand{\XsfClangTOsSzpRdaRecall}{1\xspace}
\newcommand{\XsfClangTOsSzpRdaPrecision}{1\xspace}
\newcommand{\XsfClangTOsSzpRdaFone}{1\xspace}
\newcommand{\XsfClangTOsSzpRseRecall}{1\xspace}
\newcommand{\XsfClangTOsSzpRsePrecision}{1\xspace}
\newcommand{\XsfClangTOsSzpRseFone}{1\xspace}
\newcommand{\XsfClangTOsCapGT}{157569\xspace}
\newcommand{\XsfClangTOsCapLsRecall}{1\xspace}
\newcommand{\XsfClangTOsCapLsPrecision}{1\xspace}
\newcommand{\XsfClangTOsCapLsFone}{1\xspace}
\newcommand{\XsfClangTOsCapBapRecall}{0.360\xspace}
\newcommand{\XsfClangTOsCapBapPrecision}{1.000\xspace}
\newcommand{\XsfClangTOsCapBapFone}{0.530\xspace}
\newcommand{\XsfClangTOsCapGhiRecall}{0.844\xspace}
\newcommand{\XsfClangTOsCapGhiPrecision}{1\xspace}
\newcommand{\XsfClangTOsCapGhiFone}{0.915\xspace}
\newcommand{\XsfClangTOsCapRdaRecall}{1\xspace}
\newcommand{\XsfClangTOsCapRdaPrecision}{1\xspace}
\newcommand{\XsfClangTOsCapRdaFone}{1\xspace}
\newcommand{\XsfClangTOsCapRseRecall}{1.000\xspace}
\newcommand{\XsfClangTOsCapRsePrecision}{1\xspace}
\newcommand{\XsfClangTOsCapRseFone}{1.000\xspace}
\newcommand{\XsfClangTOsExmGT}{131306\xspace}
\newcommand{\XsfClangTOsExmLsRecall}{1\xspace}
\newcommand{\XsfClangTOsExmLsPrecision}{1\xspace}
\newcommand{\XsfClangTOsExmLsFone}{1\xspace}
\newcommand{\XsfClangTOsExmBapRecall}{0.838\xspace}
\newcommand{\XsfClangTOsExmBapPrecision}{1.000\xspace}
\newcommand{\XsfClangTOsExmBapFone}{0.912\xspace}
\newcommand{\XsfClangTOsExmGhiRecall}{0.965\xspace}
\newcommand{\XsfClangTOsExmGhiPrecision}{1\xspace}
\newcommand{\XsfClangTOsExmGhiFone}{0.982\xspace}
\newcommand{\XsfClangTOsExmRdaRecall}{0.969\xspace}
\newcommand{\XsfClangTOsExmRdaPrecision}{1.000\xspace}
\newcommand{\XsfClangTOsExmRdaFone}{0.984\xspace}
\newcommand{\XsfClangTOsExmRseRecall}{1.000\xspace}
\newcommand{\XsfClangTOsExmRsePrecision}{1\xspace}
\newcommand{\XsfClangTOsExmRseFone}{1.000\xspace}
\newcommand{\XsfClangTOsLgtGT}{23411\xspace}
\newcommand{\XsfClangTOsLgtLsRecall}{1\xspace}
\newcommand{\XsfClangTOsLgtLsPrecision}{1\xspace}
\newcommand{\XsfClangTOsLgtLsFone}{1\xspace}
\newcommand{\XsfClangTOsLgtBapRecall}{0.900\xspace}
\newcommand{\XsfClangTOsLgtBapPrecision}{1.000\xspace}
\newcommand{\XsfClangTOsLgtBapFone}{0.947\xspace}
\newcommand{\XsfClangTOsLgtGhiRecall}{1\xspace}
\newcommand{\XsfClangTOsLgtGhiPrecision}{1\xspace}
\newcommand{\XsfClangTOsLgtGhiFone}{1\xspace}
\newcommand{\XsfClangTOsLgtRdaRecall}{1\xspace}
\newcommand{\XsfClangTOsLgtRdaPrecision}{1.000\xspace}
\newcommand{\XsfClangTOsLgtRdaFone}{1.000\xspace}
\newcommand{\XsfClangTOsLgtRseRecall}{1\xspace}
\newcommand{\XsfClangTOsLgtRsePrecision}{1\xspace}
\newcommand{\XsfClangTOsLgtRseFone}{1\xspace}
\newcommand{\XsfClangTOsBzpGT}{15200\xspace}
\newcommand{\XsfClangTOsBzpLsRecall}{1\xspace}
\newcommand{\XsfClangTOsBzpLsPrecision}{1\xspace}
\newcommand{\XsfClangTOsBzpLsFone}{1\xspace}
\newcommand{\XsfClangTOsBzpBapRecall}{0.971\xspace}
\newcommand{\XsfClangTOsBzpBapPrecision}{1\xspace}
\newcommand{\XsfClangTOsBzpBapFone}{0.985\xspace}
\newcommand{\XsfClangTOsBzpGhiRecall}{1\xspace}
\newcommand{\XsfClangTOsBzpGhiPrecision}{1\xspace}
\newcommand{\XsfClangTOsBzpGhiFone}{1\xspace}
\newcommand{\XsfClangTOsBzpRdaRecall}{0.824\xspace}
\newcommand{\XsfClangTOsBzpRdaPrecision}{1\xspace}
\newcommand{\XsfClangTOsBzpRdaFone}{0.903\xspace}
\newcommand{\XsfClangTOsBzpRseRecall}{1\xspace}
\newcommand{\XsfClangTOsBzpRsePrecision}{1\xspace}
\newcommand{\XsfClangTOsBzpRseFone}{1\xspace}
\newcommand{\XsfClangTOsGccGT}{827103\xspace}
\newcommand{\XsfClangTOsGccLsRecall}{1\xspace}
\newcommand{\XsfClangTOsGccLsPrecision}{1\xspace}
\newcommand{\XsfClangTOsGccLsFone}{1\xspace}
\newcommand{\XsfClangTOsGccBapRecall}{0.720\xspace}
\newcommand{\XsfClangTOsGccBapPrecision}{1.000\xspace}
\newcommand{\XsfClangTOsGccBapFone}{0.837\xspace}
\newcommand{\XsfClangTOsGccGhiRecall}{0.968\xspace}
\newcommand{\XsfClangTOsGccGhiPrecision}{1\xspace}
\newcommand{\XsfClangTOsGccGhiFone}{0.984\xspace}
\newcommand{\XsfClangTOsGccRdaRecall}{0.842\xspace}
\newcommand{\XsfClangTOsGccRdaPrecision}{1.000\xspace}
\newcommand{\XsfClangTOsGccRdaFone}{0.914\xspace}
\newcommand{\XsfClangTOsGccRseRecall}{1.000\xspace}
\newcommand{\XsfClangTOsGccRsePrecision}{1\xspace}
\newcommand{\XsfClangTOsGccRseFone}{1.000\xspace}
\newcommand{\XsfClangTOsGzpGT}{8956\xspace}
\newcommand{\XsfClangTOsGzpLsRecall}{1\xspace}
\newcommand{\XsfClangTOsGzpLsPrecision}{1\xspace}
\newcommand{\XsfClangTOsGzpLsFone}{1\xspace}
\newcommand{\XsfClangTOsGzpBapRecall}{0.987\xspace}
\newcommand{\XsfClangTOsGzpBapPrecision}{1\xspace}
\newcommand{\XsfClangTOsGzpBapFone}{0.993\xspace}
\newcommand{\XsfClangTOsGzpGhiRecall}{1\xspace}
\newcommand{\XsfClangTOsGzpGhiPrecision}{1\xspace}
\newcommand{\XsfClangTOsGzpGhiFone}{1\xspace}
\newcommand{\XsfClangTOsGzpRdaRecall}{1\xspace}
\newcommand{\XsfClangTOsGzpRdaPrecision}{1\xspace}
\newcommand{\XsfClangTOsGzpRdaFone}{1\xspace}
\newcommand{\XsfClangTOsGzpRseRecall}{0.996\xspace}
\newcommand{\XsfClangTOsGzpRsePrecision}{1\xspace}
\newcommand{\XsfClangTOsGzpRseFone}{0.998\xspace}
\newcommand{\XsfClangTOsOggGT}{34684\xspace}
\newcommand{\XsfClangTOsOggLsRecall}{1\xspace}
\newcommand{\XsfClangTOsOggLsPrecision}{1\xspace}
\newcommand{\XsfClangTOsOggLsFone}{1\xspace}
\newcommand{\XsfClangTOsOggBapRecall}{0.982\xspace}
\newcommand{\XsfClangTOsOggBapPrecision}{1.000\xspace}
\newcommand{\XsfClangTOsOggBapFone}{0.991\xspace}
\newcommand{\XsfClangTOsOggGhiRecall}{1\xspace}
\newcommand{\XsfClangTOsOggGhiPrecision}{1\xspace}
\newcommand{\XsfClangTOsOggGhiFone}{1\xspace}
\newcommand{\XsfClangTOsOggRdaRecall}{1\xspace}
\newcommand{\XsfClangTOsOggRdaPrecision}{1\xspace}
\newcommand{\XsfClangTOsOggRdaFone}{1\xspace}
\newcommand{\XsfClangTOsOggRseRecall}{1\xspace}
\newcommand{\XsfClangTOsOggRsePrecision}{1\xspace}
\newcommand{\XsfClangTOsOggRseFone}{1\xspace}
\newcommand{\XsfClangTOsNgxGT}{92091\xspace}
\newcommand{\XsfClangTOsNgxLsRecall}{1\xspace}
\newcommand{\XsfClangTOsNgxLsPrecision}{1\xspace}
\newcommand{\XsfClangTOsNgxLsFone}{1\xspace}
\newcommand{\XsfClangTOsNgxBapRecall}{0.970\xspace}
\newcommand{\XsfClangTOsNgxBapPrecision}{1.000\xspace}
\newcommand{\XsfClangTOsNgxBapFone}{0.985\xspace}
\newcommand{\XsfClangTOsNgxGhiRecall}{1\xspace}
\newcommand{\XsfClangTOsNgxGhiPrecision}{1\xspace}
\newcommand{\XsfClangTOsNgxGhiFone}{1\xspace}
\newcommand{\XsfClangTOsNgxRdaRecall}{0.998\xspace}
\newcommand{\XsfClangTOsNgxRdaPrecision}{1.000\xspace}
\newcommand{\XsfClangTOsNgxRdaFone}{0.999\xspace}
\newcommand{\XsfClangTOsNgxRseRecall}{1\xspace}
\newcommand{\XsfClangTOsNgxRsePrecision}{1\xspace}
\newcommand{\XsfClangTOsNgxRseFone}{1\xspace}
\newcommand{\XsfClangTOsSshGT}{96149\xspace}
\newcommand{\XsfClangTOsSshLsRecall}{1\xspace}
\newcommand{\XsfClangTOsSshLsPrecision}{1\xspace}
\newcommand{\XsfClangTOsSshLsFone}{1\xspace}
\newcommand{\XsfClangTOsSshBapRecall}{0.947\xspace}
\newcommand{\XsfClangTOsSshBapPrecision}{1\xspace}
\newcommand{\XsfClangTOsSshBapFone}{0.973\xspace}
\newcommand{\XsfClangTOsSshGhiRecall}{1\xspace}
\newcommand{\XsfClangTOsSshGhiPrecision}{1\xspace}
\newcommand{\XsfClangTOsSshGhiFone}{1\xspace}
\newcommand{\XsfClangTOsSshRdaRecall}{0.961\xspace}
\newcommand{\XsfClangTOsSshRdaPrecision}{1.000\xspace}
\newcommand{\XsfClangTOsSshRdaFone}{0.980\xspace}
\newcommand{\XsfClangTOsSshRseRecall}{0.988\xspace}
\newcommand{\XsfClangTOsSshRsePrecision}{1\xspace}
\newcommand{\XsfClangTOsSshRseFone}{0.994\xspace}
\newcommand{\XsfClangTOsPcrGT}{4123\xspace}
\newcommand{\XsfClangTOsPcrLsRecall}{1\xspace}
\newcommand{\XsfClangTOsPcrLsPrecision}{1\xspace}
\newcommand{\XsfClangTOsPcrLsFone}{1\xspace}
\newcommand{\XsfClangTOsPcrBapRecall}{0.870\xspace}
\newcommand{\XsfClangTOsPcrBapPrecision}{1\xspace}
\newcommand{\XsfClangTOsPcrBapFone}{0.930\xspace}
\newcommand{\XsfClangTOsPcrGhiRecall}{1\xspace}
\newcommand{\XsfClangTOsPcrGhiPrecision}{1\xspace}
\newcommand{\XsfClangTOsPcrGhiFone}{1\xspace}
\newcommand{\XsfClangTOsPcrRdaRecall}{1\xspace}
\newcommand{\XsfClangTOsPcrRdaPrecision}{1\xspace}
\newcommand{\XsfClangTOsPcrRdaFone}{1\xspace}
\newcommand{\XsfClangTOsPcrRseRecall}{1\xspace}
\newcommand{\XsfClangTOsPcrRsePrecision}{1\xspace}
\newcommand{\XsfClangTOsPcrRseFone}{1\xspace}
\newcommand{\XsfClangTOsSqlGT}{161695\xspace}
\newcommand{\XsfClangTOsSqlLsRecall}{1\xspace}
\newcommand{\XsfClangTOsSqlLsPrecision}{1\xspace}
\newcommand{\XsfClangTOsSqlLsFone}{1\xspace}
\newcommand{\XsfClangTOsSqlBapRecall}{0.841\xspace}
\newcommand{\XsfClangTOsSqlBapPrecision}{1\xspace}
\newcommand{\XsfClangTOsSqlBapFone}{0.913\xspace}
\newcommand{\XsfClangTOsSqlGhiRecall}{1\xspace}
\newcommand{\XsfClangTOsSqlGhiPrecision}{1\xspace}
\newcommand{\XsfClangTOsSqlGhiFone}{1\xspace}
\newcommand{\XsfClangTOsSqlRdaRecall}{0.967\xspace}
\newcommand{\XsfClangTOsSqlRdaPrecision}{1.000\xspace}
\newcommand{\XsfClangTOsSqlRdaFone}{0.983\xspace}
\newcommand{\XsfClangTOsSqlRseRecall}{0.994\xspace}
\newcommand{\XsfClangTOsSqlRsePrecision}{1\xspace}
\newcommand{\XsfClangTOsSqlRseFone}{0.997\xspace}
\newcommand{\XsfClangTOsVimGT}{444272\xspace}
\newcommand{\XsfClangTOsVimLsRecall}{1\xspace}
\newcommand{\XsfClangTOsVimLsPrecision}{1\xspace}
\newcommand{\XsfClangTOsVimLsFone}{1\xspace}
\newcommand{\XsfClangTOsVimBapRecall}{0.908\xspace}
\newcommand{\XsfClangTOsVimBapPrecision}{1.000\xspace}
\newcommand{\XsfClangTOsVimBapFone}{0.952\xspace}
\newcommand{\XsfClangTOsVimGhiRecall}{1.000\xspace}
\newcommand{\XsfClangTOsVimGhiPrecision}{1\xspace}
\newcommand{\XsfClangTOsVimGhiFone}{1.000\xspace}
\newcommand{\XsfClangTOsVimRdaRecall}{0.982\xspace}
\newcommand{\XsfClangTOsVimRdaPrecision}{1.000\xspace}
\newcommand{\XsfClangTOsVimRdaFone}{0.991\xspace}
\newcommand{\XsfClangTOsVimRseRecall}{1.000\xspace}
\newcommand{\XsfClangTOsVimRsePrecision}{1\xspace}
\newcommand{\XsfClangTOsVimRseFone}{1.000\xspace}
\newcommand{\XsfClangTOsVsfGT}{17347\xspace}
\newcommand{\XsfClangTOsVsfLsRecall}{1\xspace}
\newcommand{\XsfClangTOsVsfLsPrecision}{1\xspace}
\newcommand{\XsfClangTOsVsfLsFone}{1\xspace}
\newcommand{\XsfClangTOsVsfBapRecall}{0.988\xspace}
\newcommand{\XsfClangTOsVsfBapPrecision}{1\xspace}
\newcommand{\XsfClangTOsVsfBapFone}{0.994\xspace}
\newcommand{\XsfClangTOsVsfGhiRecall}{1\xspace}
\newcommand{\XsfClangTOsVsfGhiPrecision}{1\xspace}
\newcommand{\XsfClangTOsVsfGhiFone}{1\xspace}
\newcommand{\XsfClangTOsVsfRdaRecall}{0.908\xspace}
\newcommand{\XsfClangTOsVsfRdaPrecision}{0.999\xspace}
\newcommand{\XsfClangTOsVsfRdaFone}{0.951\xspace}
\newcommand{\XsfClangTOsVsfRseRecall}{0.993\xspace}
\newcommand{\XsfClangTOsVsfRsePrecision}{1\xspace}
\newcommand{\XsfClangTOsVsfRseFone}{0.997\xspace}
\newcommand{\XsfClangSOoSzpGT}{20179\xspace}
\newcommand{\XsfClangSOoSzpLsRecall}{1\xspace}
\newcommand{\XsfClangSOoSzpLsPrecision}{1\xspace}
\newcommand{\XsfClangSOoSzpLsFone}{1\xspace}
\newcommand{\XsfClangSOoSzpBapRecall}{0.980\xspace}
\newcommand{\XsfClangSOoSzpBapPrecision}{1\xspace}
\newcommand{\XsfClangSOoSzpBapFone}{0.990\xspace}
\newcommand{\XsfClangSOoSzpGhiRecall}{1\xspace}
\newcommand{\XsfClangSOoSzpGhiPrecision}{1\xspace}
\newcommand{\XsfClangSOoSzpGhiFone}{1\xspace}
\newcommand{\XsfClangSOoSzpRdaRecall}{0.980\xspace}
\newcommand{\XsfClangSOoSzpRdaPrecision}{1\xspace}
\newcommand{\XsfClangSOoSzpRdaFone}{0.990\xspace}
\newcommand{\XsfClangSOoSzpRseRecall}{1\xspace}
\newcommand{\XsfClangSOoSzpRsePrecision}{1\xspace}
\newcommand{\XsfClangSOoSzpRseFone}{1\xspace}
\newcommand{\XsfClangSOoCapGT}{292610\xspace}
\newcommand{\XsfClangSOoCapLsRecall}{1\xspace}
\newcommand{\XsfClangSOoCapLsPrecision}{1\xspace}
\newcommand{\XsfClangSOoCapLsFone}{1\xspace}
\newcommand{\XsfClangSOoCapBapRecall}{0.476\xspace}
\newcommand{\XsfClangSOoCapBapPrecision}{1\xspace}
\newcommand{\XsfClangSOoCapBapFone}{0.645\xspace}
\newcommand{\XsfClangSOoCapGhiRecall}{1\xspace}
\newcommand{\XsfClangSOoCapGhiPrecision}{1\xspace}
\newcommand{\XsfClangSOoCapGhiFone}{1\xspace}
\newcommand{\XsfClangSOoCapRdaRecall}{0.406\xspace}
\newcommand{\XsfClangSOoCapRdaPrecision}{1\xspace}
\newcommand{\XsfClangSOoCapRdaFone}{0.578\xspace}
\newcommand{\XsfClangSOoCapRseRecall}{1\xspace}
\newcommand{\XsfClangSOoCapRsePrecision}{1\xspace}
\newcommand{\XsfClangSOoCapRseFone}{1\xspace}
\newcommand{\XsfClangSOoExmGT}{179115\xspace}
\newcommand{\XsfClangSOoExmLsRecall}{1\xspace}
\newcommand{\XsfClangSOoExmLsPrecision}{1\xspace}
\newcommand{\XsfClangSOoExmLsFone}{1\xspace}
\newcommand{\XsfClangSOoExmBapRecall}{0.853\xspace}
\newcommand{\XsfClangSOoExmBapPrecision}{1\xspace}
\newcommand{\XsfClangSOoExmBapFone}{0.920\xspace}
\newcommand{\XsfClangSOoExmGhiRecall}{0.984\xspace}
\newcommand{\XsfClangSOoExmGhiPrecision}{1\xspace}
\newcommand{\XsfClangSOoExmGhiFone}{0.992\xspace}
\newcommand{\XsfClangSOoExmRdaRecall}{0.816\xspace}
\newcommand{\XsfClangSOoExmRdaPrecision}{1.000\xspace}
\newcommand{\XsfClangSOoExmRdaFone}{0.899\xspace}
\newcommand{\XsfClangSOoExmRseRecall}{1.000\xspace}
\newcommand{\XsfClangSOoExmRsePrecision}{1\xspace}
\newcommand{\XsfClangSOoExmRseFone}{1.000\xspace}
\newcommand{\XsfClangSOoLgtGT}{38836\xspace}
\newcommand{\XsfClangSOoLgtLsRecall}{1\xspace}
\newcommand{\XsfClangSOoLgtLsPrecision}{1\xspace}
\newcommand{\XsfClangSOoLgtLsFone}{1\xspace}
\newcommand{\XsfClangSOoLgtBapRecall}{0.873\xspace}
\newcommand{\XsfClangSOoLgtBapPrecision}{1\xspace}
\newcommand{\XsfClangSOoLgtBapFone}{0.932\xspace}
\newcommand{\XsfClangSOoLgtGhiRecall}{1\xspace}
\newcommand{\XsfClangSOoLgtGhiPrecision}{1\xspace}
\newcommand{\XsfClangSOoLgtGhiFone}{1\xspace}
\newcommand{\XsfClangSOoLgtRdaRecall}{0.874\xspace}
\newcommand{\XsfClangSOoLgtRdaPrecision}{1.000\xspace}
\newcommand{\XsfClangSOoLgtRdaFone}{0.932\xspace}
\newcommand{\XsfClangSOoLgtRseRecall}{1\xspace}
\newcommand{\XsfClangSOoLgtRsePrecision}{1\xspace}
\newcommand{\XsfClangSOoLgtRseFone}{1\xspace}
\newcommand{\XsfClangSOoBzpGT}{21325\xspace}
\newcommand{\XsfClangSOoBzpLsRecall}{1\xspace}
\newcommand{\XsfClangSOoBzpLsPrecision}{1\xspace}
\newcommand{\XsfClangSOoBzpLsFone}{1\xspace}
\newcommand{\XsfClangSOoBzpBapRecall}{0.779\xspace}
\newcommand{\XsfClangSOoBzpBapPrecision}{1\xspace}
\newcommand{\XsfClangSOoBzpBapFone}{0.876\xspace}
\newcommand{\XsfClangSOoBzpGhiRecall}{0.979\xspace}
\newcommand{\XsfClangSOoBzpGhiPrecision}{1\xspace}
\newcommand{\XsfClangSOoBzpGhiFone}{0.989\xspace}
\newcommand{\XsfClangSOoBzpRdaRecall}{0.771\xspace}
\newcommand{\XsfClangSOoBzpRdaPrecision}{1\xspace}
\newcommand{\XsfClangSOoBzpRdaFone}{0.871\xspace}
\newcommand{\XsfClangSOoBzpRseRecall}{1\xspace}
\newcommand{\XsfClangSOoBzpRsePrecision}{1\xspace}
\newcommand{\XsfClangSOoBzpRseFone}{1\xspace}
\newcommand{\XsfClangSOoGccGT}{1269944\xspace}
\newcommand{\XsfClangSOoGccLsRecall}{1\xspace}
\newcommand{\XsfClangSOoGccLsPrecision}{1\xspace}
\newcommand{\XsfClangSOoGccLsFone}{1\xspace}
\newcommand{\XsfClangSOoGccBapRecall}{0.714\xspace}
\newcommand{\XsfClangSOoGccBapPrecision}{1\xspace}
\newcommand{\XsfClangSOoGccBapFone}{0.833\xspace}
\newcommand{\XsfClangSOoGccGhiRecall}{0.995\xspace}
\newcommand{\XsfClangSOoGccGhiPrecision}{1\xspace}
\newcommand{\XsfClangSOoGccGhiFone}{0.998\xspace}
\newcommand{\XsfClangSOoGccRdaRecall}{0.646\xspace}
\newcommand{\XsfClangSOoGccRdaPrecision}{1.000\xspace}
\newcommand{\XsfClangSOoGccRdaFone}{0.785\xspace}
\newcommand{\XsfClangSOoGccRseRecall}{1\xspace}
\newcommand{\XsfClangSOoGccRsePrecision}{1\xspace}
\newcommand{\XsfClangSOoGccRseFone}{1\xspace}
\newcommand{\XsfClangSOoGzpGT}{13272\xspace}
\newcommand{\XsfClangSOoGzpLsRecall}{1\xspace}
\newcommand{\XsfClangSOoGzpLsPrecision}{1\xspace}
\newcommand{\XsfClangSOoGzpLsFone}{1\xspace}
\newcommand{\XsfClangSOoGzpBapRecall}{0.992\xspace}
\newcommand{\XsfClangSOoGzpBapPrecision}{1\xspace}
\newcommand{\XsfClangSOoGzpBapFone}{0.996\xspace}
\newcommand{\XsfClangSOoGzpGhiRecall}{1\xspace}
\newcommand{\XsfClangSOoGzpGhiPrecision}{1\xspace}
\newcommand{\XsfClangSOoGzpGhiFone}{1\xspace}
\newcommand{\XsfClangSOoGzpRdaRecall}{0.992\xspace}
\newcommand{\XsfClangSOoGzpRdaPrecision}{1\xspace}
\newcommand{\XsfClangSOoGzpRdaFone}{0.996\xspace}
\newcommand{\XsfClangSOoGzpRseRecall}{0.999\xspace}
\newcommand{\XsfClangSOoGzpRsePrecision}{1\xspace}
\newcommand{\XsfClangSOoGzpRseFone}{0.999\xspace}
\newcommand{\XsfClangSOoOggGT}{51554\xspace}
\newcommand{\XsfClangSOoOggLsRecall}{1\xspace}
\newcommand{\XsfClangSOoOggLsPrecision}{1\xspace}
\newcommand{\XsfClangSOoOggLsFone}{1\xspace}
\newcommand{\XsfClangSOoOggBapRecall}{0.974\xspace}
\newcommand{\XsfClangSOoOggBapPrecision}{1\xspace}
\newcommand{\XsfClangSOoOggBapFone}{0.987\xspace}
\newcommand{\XsfClangSOoOggGhiRecall}{1\xspace}
\newcommand{\XsfClangSOoOggGhiPrecision}{1\xspace}
\newcommand{\XsfClangSOoOggGhiFone}{1\xspace}
\newcommand{\XsfClangSOoOggRdaRecall}{0.974\xspace}
\newcommand{\XsfClangSOoOggRdaPrecision}{1\xspace}
\newcommand{\XsfClangSOoOggRdaFone}{0.987\xspace}
\newcommand{\XsfClangSOoOggRseRecall}{1\xspace}
\newcommand{\XsfClangSOoOggRsePrecision}{1\xspace}
\newcommand{\XsfClangSOoOggRseFone}{1\xspace}
\newcommand{\XsfClangSOoNgxGT}{159762\xspace}
\newcommand{\XsfClangSOoNgxLsRecall}{1\xspace}
\newcommand{\XsfClangSOoNgxLsPrecision}{1\xspace}
\newcommand{\XsfClangSOoNgxLsFone}{1\xspace}
\newcommand{\XsfClangSOoNgxBapRecall}{0.961\xspace}
\newcommand{\XsfClangSOoNgxBapPrecision}{1\xspace}
\newcommand{\XsfClangSOoNgxBapFone}{0.980\xspace}
\newcommand{\XsfClangSOoNgxGhiRecall}{1.000\xspace}
\newcommand{\XsfClangSOoNgxGhiPrecision}{1\xspace}
\newcommand{\XsfClangSOoNgxGhiFone}{1.000\xspace}
\newcommand{\XsfClangSOoNgxRdaRecall}{0.960\xspace}
\newcommand{\XsfClangSOoNgxRdaPrecision}{1\xspace}
\newcommand{\XsfClangSOoNgxRdaFone}{0.980\xspace}
\newcommand{\XsfClangSOoNgxRseRecall}{1\xspace}
\newcommand{\XsfClangSOoNgxRsePrecision}{1\xspace}
\newcommand{\XsfClangSOoNgxRseFone}{1\xspace}
\newcommand{\XsfClangSOoSshGT}{144626\xspace}
\newcommand{\XsfClangSOoSshLsRecall}{1\xspace}
\newcommand{\XsfClangSOoSshLsPrecision}{1\xspace}
\newcommand{\XsfClangSOoSshLsFone}{1\xspace}
\newcommand{\XsfClangSOoSshBapRecall}{0.949\xspace}
\newcommand{\XsfClangSOoSshBapPrecision}{1\xspace}
\newcommand{\XsfClangSOoSshBapFone}{0.974\xspace}
\newcommand{\XsfClangSOoSshGhiRecall}{0.981\xspace}
\newcommand{\XsfClangSOoSshGhiPrecision}{1\xspace}
\newcommand{\XsfClangSOoSshGhiFone}{0.990\xspace}
\newcommand{\XsfClangSOoSshRdaRecall}{0.938\xspace}
\newcommand{\XsfClangSOoSshRdaPrecision}{1\xspace}
\newcommand{\XsfClangSOoSshRdaFone}{0.968\xspace}
\newcommand{\XsfClangSOoSshRseRecall}{0.980\xspace}
\newcommand{\XsfClangSOoSshRsePrecision}{1\xspace}
\newcommand{\XsfClangSOoSshRseFone}{0.990\xspace}
\newcommand{\XsfClangSOoPcrGT}{6143\xspace}
\newcommand{\XsfClangSOoPcrLsRecall}{1\xspace}
\newcommand{\XsfClangSOoPcrLsPrecision}{1\xspace}
\newcommand{\XsfClangSOoPcrLsFone}{1\xspace}
\newcommand{\XsfClangSOoPcrBapRecall}{0.893\xspace}
\newcommand{\XsfClangSOoPcrBapPrecision}{1\xspace}
\newcommand{\XsfClangSOoPcrBapFone}{0.944\xspace}
\newcommand{\XsfClangSOoPcrGhiRecall}{1\xspace}
\newcommand{\XsfClangSOoPcrGhiPrecision}{1\xspace}
\newcommand{\XsfClangSOoPcrGhiFone}{1\xspace}
\newcommand{\XsfClangSOoPcrRdaRecall}{0.893\xspace}
\newcommand{\XsfClangSOoPcrRdaPrecision}{1\xspace}
\newcommand{\XsfClangSOoPcrRdaFone}{0.944\xspace}
\newcommand{\XsfClangSOoPcrRseRecall}{1\xspace}
\newcommand{\XsfClangSOoPcrRsePrecision}{1\xspace}
\newcommand{\XsfClangSOoPcrRseFone}{1\xspace}
\newcommand{\XsfClangSOoSqlGT}{214103\xspace}
\newcommand{\XsfClangSOoSqlLsRecall}{1\xspace}
\newcommand{\XsfClangSOoSqlLsPrecision}{1\xspace}
\newcommand{\XsfClangSOoSqlLsFone}{1\xspace}
\newcommand{\XsfClangSOoSqlBapRecall}{0.875\xspace}
\newcommand{\XsfClangSOoSqlBapPrecision}{1\xspace}
\newcommand{\XsfClangSOoSqlBapFone}{0.933\xspace}
\newcommand{\XsfClangSOoSqlGhiRecall}{1\xspace}
\newcommand{\XsfClangSOoSqlGhiPrecision}{1\xspace}
\newcommand{\XsfClangSOoSqlGhiFone}{1\xspace}
\newcommand{\XsfClangSOoSqlRdaRecall}{0.869\xspace}
\newcommand{\XsfClangSOoSqlRdaPrecision}{1.000\xspace}
\newcommand{\XsfClangSOoSqlRdaFone}{0.930\xspace}
\newcommand{\XsfClangSOoSqlRseRecall}{1\xspace}
\newcommand{\XsfClangSOoSqlRsePrecision}{1\xspace}
\newcommand{\XsfClangSOoSqlRseFone}{1\xspace}
\newcommand{\XsfClangSOoVimGT}{619160\xspace}
\newcommand{\XsfClangSOoVimLsRecall}{1\xspace}
\newcommand{\XsfClangSOoVimLsPrecision}{1\xspace}
\newcommand{\XsfClangSOoVimLsFone}{1\xspace}
\newcommand{\XsfClangSOoVimBapRecall}{0.945\xspace}
\newcommand{\XsfClangSOoVimBapPrecision}{1\xspace}
\newcommand{\XsfClangSOoVimBapFone}{0.972\xspace}
\newcommand{\XsfClangSOoVimGhiRecall}{1.000\xspace}
\newcommand{\XsfClangSOoVimGhiPrecision}{1\xspace}
\newcommand{\XsfClangSOoVimGhiFone}{1.000\xspace}
\newcommand{\XsfClangSOoVimRdaRecall}{0.911\xspace}
\newcommand{\XsfClangSOoVimRdaPrecision}{1.000\xspace}
\newcommand{\XsfClangSOoVimRdaFone}{0.953\xspace}
\newcommand{\XsfClangSOoVimRseRecall}{1.000\xspace}
\newcommand{\XsfClangSOoVimRsePrecision}{1\xspace}
\newcommand{\XsfClangSOoVimRseFone}{1.000\xspace}
\newcommand{\XsfClangSOoVsfGT}{23377\xspace}
\newcommand{\XsfClangSOoVsfLsRecall}{1\xspace}
\newcommand{\XsfClangSOoVsfLsPrecision}{1\xspace}
\newcommand{\XsfClangSOoVsfLsFone}{1\xspace}
\newcommand{\XsfClangSOoVsfBapRecall}{0.993\xspace}
\newcommand{\XsfClangSOoVsfBapPrecision}{1\xspace}
\newcommand{\XsfClangSOoVsfBapFone}{0.997\xspace}
\newcommand{\XsfClangSOoVsfGhiRecall}{0.997\xspace}
\newcommand{\XsfClangSOoVsfGhiPrecision}{1\xspace}
\newcommand{\XsfClangSOoVsfGhiFone}{0.999\xspace}
\newcommand{\XsfClangSOoVsfRdaRecall}{0.985\xspace}
\newcommand{\XsfClangSOoVsfRdaPrecision}{1.000\xspace}
\newcommand{\XsfClangSOoVsfRdaFone}{0.992\xspace}
\newcommand{\XsfClangSOoVsfRseRecall}{1.000\xspace}
\newcommand{\XsfClangSOoVsfRsePrecision}{1\xspace}
\newcommand{\XsfClangSOoVsfRseFone}{1.000\xspace}
\newcommand{\XsfClangSOaSzpGT}{12075\xspace}
\newcommand{\XsfClangSOaSzpLsRecall}{1\xspace}
\newcommand{\XsfClangSOaSzpLsPrecision}{1\xspace}
\newcommand{\XsfClangSOaSzpLsFone}{1\xspace}
\newcommand{\XsfClangSOaSzpBapRecall}{0.979\xspace}
\newcommand{\XsfClangSOaSzpBapPrecision}{1\xspace}
\newcommand{\XsfClangSOaSzpBapFone}{0.989\xspace}
\newcommand{\XsfClangSOaSzpGhiRecall}{1\xspace}
\newcommand{\XsfClangSOaSzpGhiPrecision}{1\xspace}
\newcommand{\XsfClangSOaSzpGhiFone}{1\xspace}
\newcommand{\XsfClangSOaSzpRdaRecall}{1\xspace}
\newcommand{\XsfClangSOaSzpRdaPrecision}{1\xspace}
\newcommand{\XsfClangSOaSzpRdaFone}{1\xspace}
\newcommand{\XsfClangSOaSzpRseRecall}{1\xspace}
\newcommand{\XsfClangSOaSzpRsePrecision}{1\xspace}
\newcommand{\XsfClangSOaSzpRseFone}{1\xspace}
\newcommand{\XsfClangSOaCapGT}{176180\xspace}
\newcommand{\XsfClangSOaCapLsRecall}{1\xspace}
\newcommand{\XsfClangSOaCapLsPrecision}{1\xspace}
\newcommand{\XsfClangSOaCapLsFone}{1\xspace}
\newcommand{\XsfClangSOaCapBapRecall}{0.320\xspace}
\newcommand{\XsfClangSOaCapBapPrecision}{1\xspace}
\newcommand{\XsfClangSOaCapBapFone}{0.485\xspace}
\newcommand{\XsfClangSOaCapGhiRecall}{1\xspace}
\newcommand{\XsfClangSOaCapGhiPrecision}{1\xspace}
\newcommand{\XsfClangSOaCapGhiFone}{1\xspace}
\newcommand{\XsfClangSOaCapRdaRecall}{0.858\xspace}
\newcommand{\XsfClangSOaCapRdaPrecision}{1\xspace}
\newcommand{\XsfClangSOaCapRdaFone}{0.923\xspace}
\newcommand{\XsfClangSOaCapRseRecall}{1.000\xspace}
\newcommand{\XsfClangSOaCapRsePrecision}{1\xspace}
\newcommand{\XsfClangSOaCapRseFone}{1.000\xspace}
\newcommand{\XsfClangSOaExmGT}{130621\xspace}
\newcommand{\XsfClangSOaExmLsRecall}{1\xspace}
\newcommand{\XsfClangSOaExmLsPrecision}{1\xspace}
\newcommand{\XsfClangSOaExmLsFone}{1\xspace}
\newcommand{\XsfClangSOaExmBapRecall}{0.854\xspace}
\newcommand{\XsfClangSOaExmBapPrecision}{1\xspace}
\newcommand{\XsfClangSOaExmBapFone}{0.921\xspace}
\newcommand{\XsfClangSOaExmGhiRecall}{0.986\xspace}
\newcommand{\XsfClangSOaExmGhiPrecision}{1\xspace}
\newcommand{\XsfClangSOaExmGhiFone}{0.993\xspace}
\newcommand{\XsfClangSOaExmRdaRecall}{0.974\xspace}
\newcommand{\XsfClangSOaExmRdaPrecision}{1\xspace}
\newcommand{\XsfClangSOaExmRdaFone}{0.987\xspace}
\newcommand{\XsfClangSOaExmRseRecall}{1.000\xspace}
\newcommand{\XsfClangSOaExmRsePrecision}{1\xspace}
\newcommand{\XsfClangSOaExmRseFone}{1.000\xspace}
\newcommand{\XsfClangSOaLgtGT}{23620\xspace}
\newcommand{\XsfClangSOaLgtLsRecall}{1\xspace}
\newcommand{\XsfClangSOaLgtLsPrecision}{1\xspace}
\newcommand{\XsfClangSOaLgtLsFone}{1\xspace}
\newcommand{\XsfClangSOaLgtBapRecall}{0.884\xspace}
\newcommand{\XsfClangSOaLgtBapPrecision}{1\xspace}
\newcommand{\XsfClangSOaLgtBapFone}{0.938\xspace}
\newcommand{\XsfClangSOaLgtGhiRecall}{1\xspace}
\newcommand{\XsfClangSOaLgtGhiPrecision}{1\xspace}
\newcommand{\XsfClangSOaLgtGhiFone}{1\xspace}
\newcommand{\XsfClangSOaLgtRdaRecall}{0.998\xspace}
\newcommand{\XsfClangSOaLgtRdaPrecision}{1.000\xspace}
\newcommand{\XsfClangSOaLgtRdaFone}{0.999\xspace}
\newcommand{\XsfClangSOaLgtRseRecall}{1\xspace}
\newcommand{\XsfClangSOaLgtRsePrecision}{1\xspace}
\newcommand{\XsfClangSOaLgtRseFone}{1\xspace}
\newcommand{\XsfClangSOaBzpGT}{11541\xspace}
\newcommand{\XsfClangSOaBzpLsRecall}{1\xspace}
\newcommand{\XsfClangSOaBzpLsPrecision}{1\xspace}
\newcommand{\XsfClangSOaBzpLsFone}{1\xspace}
\newcommand{\XsfClangSOaBzpBapRecall}{0.909\xspace}
\newcommand{\XsfClangSOaBzpBapPrecision}{1\xspace}
\newcommand{\XsfClangSOaBzpBapFone}{0.952\xspace}
\newcommand{\XsfClangSOaBzpGhiRecall}{1\xspace}
\newcommand{\XsfClangSOaBzpGhiPrecision}{1\xspace}
\newcommand{\XsfClangSOaBzpGhiFone}{1\xspace}
\newcommand{\XsfClangSOaBzpRdaRecall}{0.971\xspace}
\newcommand{\XsfClangSOaBzpRdaPrecision}{1.000\xspace}
\newcommand{\XsfClangSOaBzpRdaFone}{0.985\xspace}
\newcommand{\XsfClangSOaBzpRseRecall}{1\xspace}
\newcommand{\XsfClangSOaBzpRsePrecision}{1\xspace}
\newcommand{\XsfClangSOaBzpRseFone}{1\xspace}
\newcommand{\XsfClangSOaGccGT}{757769\xspace}
\newcommand{\XsfClangSOaGccLsRecall}{1\xspace}
\newcommand{\XsfClangSOaGccLsPrecision}{1\xspace}
\newcommand{\XsfClangSOaGccLsFone}{1\xspace}
\newcommand{\XsfClangSOaGccBapRecall}{0.721\xspace}
\newcommand{\XsfClangSOaGccBapPrecision}{1.000\xspace}
\newcommand{\XsfClangSOaGccBapFone}{0.838\xspace}
\newcommand{\XsfClangSOaGccGhiRecall}{0.990\xspace}
\newcommand{\XsfClangSOaGccGhiPrecision}{1\xspace}
\newcommand{\XsfClangSOaGccGhiFone}{0.995\xspace}
\newcommand{\XsfClangSOaGccRdaRecall}{0.000\xspace}
\newcommand{\XsfClangSOaGccRdaPrecision}{NaN\xspace}
\newcommand{\XsfClangSOaGccRdaFone}{0.000\xspace}
\newcommand{\XsfClangSOaGccRseRecall}{0.999\xspace}
\newcommand{\XsfClangSOaGccRsePrecision}{1.000\xspace}
\newcommand{\XsfClangSOaGccRseFone}{0.999\xspace}
\newcommand{\XsfClangSOaGzpGT}{8809\xspace}
\newcommand{\XsfClangSOaGzpLsRecall}{1\xspace}
\newcommand{\XsfClangSOaGzpLsPrecision}{1\xspace}
\newcommand{\XsfClangSOaGzpLsFone}{1\xspace}
\newcommand{\XsfClangSOaGzpBapRecall}{0.982\xspace}
\newcommand{\XsfClangSOaGzpBapPrecision}{1\xspace}
\newcommand{\XsfClangSOaGzpBapFone}{0.991\xspace}
\newcommand{\XsfClangSOaGzpGhiRecall}{1\xspace}
\newcommand{\XsfClangSOaGzpGhiPrecision}{1\xspace}
\newcommand{\XsfClangSOaGzpGhiFone}{1\xspace}
\newcommand{\XsfClangSOaGzpRdaRecall}{1\xspace}
\newcommand{\XsfClangSOaGzpRdaPrecision}{1\xspace}
\newcommand{\XsfClangSOaGzpRdaFone}{1\xspace}
\newcommand{\XsfClangSOaGzpRseRecall}{0.998\xspace}
\newcommand{\XsfClangSOaGzpRsePrecision}{1\xspace}
\newcommand{\XsfClangSOaGzpRseFone}{0.999\xspace}
\newcommand{\XsfClangSOaOggGT}{32393\xspace}
\newcommand{\XsfClangSOaOggLsRecall}{1\xspace}
\newcommand{\XsfClangSOaOggLsPrecision}{1\xspace}
\newcommand{\XsfClangSOaOggLsFone}{1\xspace}
\newcommand{\XsfClangSOaOggBapRecall}{0.975\xspace}
\newcommand{\XsfClangSOaOggBapPrecision}{1.000\xspace}
\newcommand{\XsfClangSOaOggBapFone}{0.987\xspace}
\newcommand{\XsfClangSOaOggGhiRecall}{1\xspace}
\newcommand{\XsfClangSOaOggGhiPrecision}{1\xspace}
\newcommand{\XsfClangSOaOggGhiFone}{1\xspace}
\newcommand{\XsfClangSOaOggRdaRecall}{1\xspace}
\newcommand{\XsfClangSOaOggRdaPrecision}{1\xspace}
\newcommand{\XsfClangSOaOggRdaFone}{1\xspace}
\newcommand{\XsfClangSOaOggRseRecall}{1\xspace}
\newcommand{\XsfClangSOaOggRsePrecision}{1\xspace}
\newcommand{\XsfClangSOaOggRseFone}{1\xspace}
\newcommand{\XsfClangSOaNgxGT}{93140\xspace}
\newcommand{\XsfClangSOaNgxLsRecall}{1\xspace}
\newcommand{\XsfClangSOaNgxLsPrecision}{1\xspace}
\newcommand{\XsfClangSOaNgxLsFone}{1\xspace}
\newcommand{\XsfClangSOaNgxBapRecall}{0.968\xspace}
\newcommand{\XsfClangSOaNgxBapPrecision}{1.000\xspace}
\newcommand{\XsfClangSOaNgxBapFone}{0.984\xspace}
\newcommand{\XsfClangSOaNgxGhiRecall}{1\xspace}
\newcommand{\XsfClangSOaNgxGhiPrecision}{1\xspace}
\newcommand{\XsfClangSOaNgxGhiFone}{1\xspace}
\newcommand{\XsfClangSOaNgxRdaRecall}{0.996\xspace}
\newcommand{\XsfClangSOaNgxRdaPrecision}{1\xspace}
\newcommand{\XsfClangSOaNgxRdaFone}{0.998\xspace}
\newcommand{\XsfClangSOaNgxRseRecall}{1\xspace}
\newcommand{\XsfClangSOaNgxRsePrecision}{1\xspace}
\newcommand{\XsfClangSOaNgxRseFone}{1\xspace}
\newcommand{\XsfClangSOaSshGT}{95192\xspace}
\newcommand{\XsfClangSOaSshLsRecall}{1\xspace}
\newcommand{\XsfClangSOaSshLsPrecision}{1\xspace}
\newcommand{\XsfClangSOaSshLsFone}{1\xspace}
\newcommand{\XsfClangSOaSshBapRecall}{0.949\xspace}
\newcommand{\XsfClangSOaSshBapPrecision}{1\xspace}
\newcommand{\XsfClangSOaSshBapFone}{0.974\xspace}
\newcommand{\XsfClangSOaSshGhiRecall}{1\xspace}
\newcommand{\XsfClangSOaSshGhiPrecision}{1\xspace}
\newcommand{\XsfClangSOaSshGhiFone}{1\xspace}
\newcommand{\XsfClangSOaSshRdaRecall}{0.953\xspace}
\newcommand{\XsfClangSOaSshRdaPrecision}{1.000\xspace}
\newcommand{\XsfClangSOaSshRdaFone}{0.976\xspace}
\newcommand{\XsfClangSOaSshRseRecall}{0.982\xspace}
\newcommand{\XsfClangSOaSshRsePrecision}{1\xspace}
\newcommand{\XsfClangSOaSshRseFone}{0.991\xspace}
\newcommand{\XsfClangSOaPcrGT}{4129\xspace}
\newcommand{\XsfClangSOaPcrLsRecall}{1\xspace}
\newcommand{\XsfClangSOaPcrLsPrecision}{1\xspace}
\newcommand{\XsfClangSOaPcrLsFone}{1\xspace}
\newcommand{\XsfClangSOaPcrBapRecall}{0.888\xspace}
\newcommand{\XsfClangSOaPcrBapPrecision}{1\xspace}
\newcommand{\XsfClangSOaPcrBapFone}{0.941\xspace}
\newcommand{\XsfClangSOaPcrGhiRecall}{1\xspace}
\newcommand{\XsfClangSOaPcrGhiPrecision}{1\xspace}
\newcommand{\XsfClangSOaPcrGhiFone}{1\xspace}
\newcommand{\XsfClangSOaPcrRdaRecall}{1\xspace}
\newcommand{\XsfClangSOaPcrRdaPrecision}{1\xspace}
\newcommand{\XsfClangSOaPcrRdaFone}{1\xspace}
\newcommand{\XsfClangSOaPcrRseRecall}{1\xspace}
\newcommand{\XsfClangSOaPcrRsePrecision}{1\xspace}
\newcommand{\XsfClangSOaPcrRseFone}{1\xspace}
\newcommand{\XsfClangSOaSqlGT}{148257\xspace}
\newcommand{\XsfClangSOaSqlLsRecall}{1\xspace}
\newcommand{\XsfClangSOaSqlLsPrecision}{1\xspace}
\newcommand{\XsfClangSOaSqlLsFone}{1\xspace}
\newcommand{\XsfClangSOaSqlBapRecall}{0.873\xspace}
\newcommand{\XsfClangSOaSqlBapPrecision}{1\xspace}
\newcommand{\XsfClangSOaSqlBapFone}{0.932\xspace}
\newcommand{\XsfClangSOaSqlGhiRecall}{0.963\xspace}
\newcommand{\XsfClangSOaSqlGhiPrecision}{1\xspace}
\newcommand{\XsfClangSOaSqlGhiFone}{0.981\xspace}
\newcommand{\XsfClangSOaSqlRdaRecall}{0.962\xspace}
\newcommand{\XsfClangSOaSqlRdaPrecision}{1\xspace}
\newcommand{\XsfClangSOaSqlRdaFone}{0.981\xspace}
\newcommand{\XsfClangSOaSqlRseRecall}{0.984\xspace}
\newcommand{\XsfClangSOaSqlRsePrecision}{1.000\xspace}
\newcommand{\XsfClangSOaSqlRseFone}{0.992\xspace}
\newcommand{\XsfClangSOaVimGT}{433076\xspace}
\newcommand{\XsfClangSOaVimLsRecall}{1\xspace}
\newcommand{\XsfClangSOaVimLsPrecision}{1\xspace}
\newcommand{\XsfClangSOaVimLsFone}{1\xspace}
\newcommand{\XsfClangSOaVimBapRecall}{0.919\xspace}
\newcommand{\XsfClangSOaVimBapPrecision}{1.000\xspace}
\newcommand{\XsfClangSOaVimBapFone}{0.958\xspace}
\newcommand{\XsfClangSOaVimGhiRecall}{0.996\xspace}
\newcommand{\XsfClangSOaVimGhiPrecision}{1\xspace}
\newcommand{\XsfClangSOaVimGhiFone}{0.998\xspace}
\newcommand{\XsfClangSOaVimRdaRecall}{0.986\xspace}
\newcommand{\XsfClangSOaVimRdaPrecision}{1.000\xspace}
\newcommand{\XsfClangSOaVimRdaFone}{0.993\xspace}
\newcommand{\XsfClangSOaVimRseRecall}{0.999\xspace}
\newcommand{\XsfClangSOaVimRsePrecision}{1\xspace}
\newcommand{\XsfClangSOaVimRseFone}{1.000\xspace}
\newcommand{\XsfClangSOaVsfGT}{17386\xspace}
\newcommand{\XsfClangSOaVsfLsRecall}{1\xspace}
\newcommand{\XsfClangSOaVsfLsPrecision}{1\xspace}
\newcommand{\XsfClangSOaVsfLsFone}{1\xspace}
\newcommand{\XsfClangSOaVsfBapRecall}{0.989\xspace}
\newcommand{\XsfClangSOaVsfBapPrecision}{1\xspace}
\newcommand{\XsfClangSOaVsfBapFone}{0.995\xspace}
\newcommand{\XsfClangSOaVsfGhiRecall}{1\xspace}
\newcommand{\XsfClangSOaVsfGhiPrecision}{1\xspace}
\newcommand{\XsfClangSOaVsfGhiFone}{1\xspace}
\newcommand{\XsfClangSOaVsfRdaRecall}{0.993\xspace}
\newcommand{\XsfClangSOaVsfRdaPrecision}{1.000\xspace}
\newcommand{\XsfClangSOaVsfRdaFone}{0.997\xspace}
\newcommand{\XsfClangSOaVsfRseRecall}{0.993\xspace}
\newcommand{\XsfClangSOaVsfRsePrecision}{1\xspace}
\newcommand{\XsfClangSOaVsfRseFone}{0.996\xspace}
\newcommand{\XsfClangSObSzpGT}{14895\xspace}
\newcommand{\XsfClangSObSzpLsRecall}{1\xspace}
\newcommand{\XsfClangSObSzpLsPrecision}{1\xspace}
\newcommand{\XsfClangSObSzpLsFone}{1\xspace}
\newcommand{\XsfClangSObSzpBapRecall}{0.967\xspace}
\newcommand{\XsfClangSObSzpBapPrecision}{1\xspace}
\newcommand{\XsfClangSObSzpBapFone}{0.983\xspace}
\newcommand{\XsfClangSObSzpGhiRecall}{1\xspace}
\newcommand{\XsfClangSObSzpGhiPrecision}{1\xspace}
\newcommand{\XsfClangSObSzpGhiFone}{1\xspace}
\newcommand{\XsfClangSObSzpRdaRecall}{1.000\xspace}
\newcommand{\XsfClangSObSzpRdaPrecision}{1\xspace}
\newcommand{\XsfClangSObSzpRdaFone}{1.000\xspace}
\newcommand{\XsfClangSObSzpRseRecall}{1\xspace}
\newcommand{\XsfClangSObSzpRsePrecision}{1\xspace}
\newcommand{\XsfClangSObSzpRseFone}{1\xspace}
\newcommand{\XsfClangSObCapGT}{166591\xspace}
\newcommand{\XsfClangSObCapLsRecall}{1\xspace}
\newcommand{\XsfClangSObCapLsPrecision}{1\xspace}
\newcommand{\XsfClangSObCapLsFone}{1\xspace}
\newcommand{\XsfClangSObCapBapRecall}{0.297\xspace}
\newcommand{\XsfClangSObCapBapPrecision}{1.000\xspace}
\newcommand{\XsfClangSObCapBapFone}{0.459\xspace}
\newcommand{\XsfClangSObCapGhiRecall}{0.837\xspace}
\newcommand{\XsfClangSObCapGhiPrecision}{1\xspace}
\newcommand{\XsfClangSObCapGhiFone}{0.911\xspace}
\newcommand{\XsfClangSObCapRdaRecall}{0.835\xspace}
\newcommand{\XsfClangSObCapRdaPrecision}{1\xspace}
\newcommand{\XsfClangSObCapRdaFone}{0.910\xspace}
\newcommand{\XsfClangSObCapRseRecall}{1\xspace}
\newcommand{\XsfClangSObCapRsePrecision}{1\xspace}
\newcommand{\XsfClangSObCapRseFone}{1\xspace}
\newcommand{\XsfClangSObExmGT}{149918\xspace}
\newcommand{\XsfClangSObExmLsRecall}{1\xspace}
\newcommand{\XsfClangSObExmLsPrecision}{1\xspace}
\newcommand{\XsfClangSObExmLsFone}{1\xspace}
\newcommand{\XsfClangSObExmBapRecall}{0.830\xspace}
\newcommand{\XsfClangSObExmBapPrecision}{1.000\xspace}
\newcommand{\XsfClangSObExmBapFone}{0.907\xspace}
\newcommand{\XsfClangSObExmGhiRecall}{0.945\xspace}
\newcommand{\XsfClangSObExmGhiPrecision}{1\xspace}
\newcommand{\XsfClangSObExmGhiFone}{0.972\xspace}
\newcommand{\XsfClangSObExmRdaRecall}{0.961\xspace}
\newcommand{\XsfClangSObExmRdaPrecision}{1.000\xspace}
\newcommand{\XsfClangSObExmRdaFone}{0.980\xspace}
\newcommand{\XsfClangSObExmRseRecall}{1.000\xspace}
\newcommand{\XsfClangSObExmRsePrecision}{1.000\xspace}
\newcommand{\XsfClangSObExmRseFone}{1.000\xspace}
\newcommand{\XsfClangSObLgtGT}{25842\xspace}
\newcommand{\XsfClangSObLgtLsRecall}{1\xspace}
\newcommand{\XsfClangSObLgtLsPrecision}{1\xspace}
\newcommand{\XsfClangSObLgtLsFone}{1\xspace}
\newcommand{\XsfClangSObLgtBapRecall}{0.877\xspace}
\newcommand{\XsfClangSObLgtBapPrecision}{1\xspace}
\newcommand{\XsfClangSObLgtBapFone}{0.934\xspace}
\newcommand{\XsfClangSObLgtGhiRecall}{1\xspace}
\newcommand{\XsfClangSObLgtGhiPrecision}{1\xspace}
\newcommand{\XsfClangSObLgtGhiFone}{1\xspace}
\newcommand{\XsfClangSObLgtRdaRecall}{0.990\xspace}
\newcommand{\XsfClangSObLgtRdaPrecision}{1\xspace}
\newcommand{\XsfClangSObLgtRdaFone}{0.995\xspace}
\newcommand{\XsfClangSObLgtRseRecall}{1\xspace}
\newcommand{\XsfClangSObLgtRsePrecision}{1\xspace}
\newcommand{\XsfClangSObLgtRseFone}{1\xspace}
\newcommand{\XsfClangSObBzpGT}{17404\xspace}
\newcommand{\XsfClangSObBzpLsRecall}{1\xspace}
\newcommand{\XsfClangSObBzpLsPrecision}{1\xspace}
\newcommand{\XsfClangSObBzpLsFone}{1\xspace}
\newcommand{\XsfClangSObBzpBapRecall}{0.979\xspace}
\newcommand{\XsfClangSObBzpBapPrecision}{1\xspace}
\newcommand{\XsfClangSObBzpBapFone}{0.989\xspace}
\newcommand{\XsfClangSObBzpGhiRecall}{1\xspace}
\newcommand{\XsfClangSObBzpGhiPrecision}{1\xspace}
\newcommand{\XsfClangSObBzpGhiFone}{1\xspace}
\newcommand{\XsfClangSObBzpRdaRecall}{0.878\xspace}
\newcommand{\XsfClangSObBzpRdaPrecision}{1.000\xspace}
\newcommand{\XsfClangSObBzpRdaFone}{0.935\xspace}
\newcommand{\XsfClangSObBzpRseRecall}{1\xspace}
\newcommand{\XsfClangSObBzpRsePrecision}{1\xspace}
\newcommand{\XsfClangSObBzpRseFone}{1\xspace}
\newcommand{\XsfClangSObGccGT}{1118521\xspace}
\newcommand{\XsfClangSObGccLsRecall}{1\xspace}
\newcommand{\XsfClangSObGccLsPrecision}{1\xspace}
\newcommand{\XsfClangSObGccLsFone}{1\xspace}
\newcommand{\XsfClangSObGccBapRecall}{0.706\xspace}
\newcommand{\XsfClangSObGccBapPrecision}{1.000\xspace}
\newcommand{\XsfClangSObGccBapFone}{0.828\xspace}
\newcommand{\XsfClangSObGccGhiRecall}{0.962\xspace}
\newcommand{\XsfClangSObGccGhiPrecision}{1\xspace}
\newcommand{\XsfClangSObGccGhiFone}{0.981\xspace}
\newcommand{\XsfClangSObGccRdaRecall}{0.801\xspace}
\newcommand{\XsfClangSObGccRdaPrecision}{1.000\xspace}
\newcommand{\XsfClangSObGccRdaFone}{0.889\xspace}
\newcommand{\XsfClangSObGccRseRecall}{0.999\xspace}
\newcommand{\XsfClangSObGccRsePrecision}{1\xspace}
\newcommand{\XsfClangSObGccRseFone}{1.000\xspace}
\newcommand{\XsfClangSObGzpGT}{12686\xspace}
\newcommand{\XsfClangSObGzpLsRecall}{1\xspace}
\newcommand{\XsfClangSObGzpLsPrecision}{1\xspace}
\newcommand{\XsfClangSObGzpLsFone}{1\xspace}
\newcommand{\XsfClangSObGzpBapRecall}{0.988\xspace}
\newcommand{\XsfClangSObGzpBapPrecision}{1\xspace}
\newcommand{\XsfClangSObGzpBapFone}{0.994\xspace}
\newcommand{\XsfClangSObGzpGhiRecall}{1\xspace}
\newcommand{\XsfClangSObGzpGhiPrecision}{1\xspace}
\newcommand{\XsfClangSObGzpGhiFone}{1\xspace}
\newcommand{\XsfClangSObGzpRdaRecall}{1\xspace}
\newcommand{\XsfClangSObGzpRdaPrecision}{1\xspace}
\newcommand{\XsfClangSObGzpRdaFone}{1\xspace}
\newcommand{\XsfClangSObGzpRseRecall}{0.999\xspace}
\newcommand{\XsfClangSObGzpRsePrecision}{1\xspace}
\newcommand{\XsfClangSObGzpRseFone}{0.999\xspace}
\newcommand{\XsfClangSObOggGT}{53867\xspace}
\newcommand{\XsfClangSObOggLsRecall}{1\xspace}
\newcommand{\XsfClangSObOggLsPrecision}{1\xspace}
\newcommand{\XsfClangSObOggLsFone}{1\xspace}
\newcommand{\XsfClangSObOggBapRecall}{0.969\xspace}
\newcommand{\XsfClangSObOggBapPrecision}{1.000\xspace}
\newcommand{\XsfClangSObOggBapFone}{0.984\xspace}
\newcommand{\XsfClangSObOggGhiRecall}{1\xspace}
\newcommand{\XsfClangSObOggGhiPrecision}{1\xspace}
\newcommand{\XsfClangSObOggGhiFone}{1\xspace}
\newcommand{\XsfClangSObOggRdaRecall}{1\xspace}
\newcommand{\XsfClangSObOggRdaPrecision}{1.000\xspace}
\newcommand{\XsfClangSObOggRdaFone}{1.000\xspace}
\newcommand{\XsfClangSObOggRseRecall}{1\xspace}
\newcommand{\XsfClangSObOggRsePrecision}{1\xspace}
\newcommand{\XsfClangSObOggRseFone}{1\xspace}
\newcommand{\XsfClangSObNgxGT}{99597\xspace}
\newcommand{\XsfClangSObNgxLsRecall}{1\xspace}
\newcommand{\XsfClangSObNgxLsPrecision}{1\xspace}
\newcommand{\XsfClangSObNgxLsFone}{1\xspace}
\newcommand{\XsfClangSObNgxBapRecall}{0.966\xspace}
\newcommand{\XsfClangSObNgxBapPrecision}{1.000\xspace}
\newcommand{\XsfClangSObNgxBapFone}{0.983\xspace}
\newcommand{\XsfClangSObNgxGhiRecall}{0.999\xspace}
\newcommand{\XsfClangSObNgxGhiPrecision}{1\xspace}
\newcommand{\XsfClangSObNgxGhiFone}{1.000\xspace}
\newcommand{\XsfClangSObNgxRdaRecall}{0.996\xspace}
\newcommand{\XsfClangSObNgxRdaPrecision}{1\xspace}
\newcommand{\XsfClangSObNgxRdaFone}{0.998\xspace}
\newcommand{\XsfClangSObNgxRseRecall}{1\xspace}
\newcommand{\XsfClangSObNgxRsePrecision}{1\xspace}
\newcommand{\XsfClangSObNgxRseFone}{1\xspace}
\newcommand{\XsfClangSObSshGT}{115759\xspace}
\newcommand{\XsfClangSObSshLsRecall}{1\xspace}
\newcommand{\XsfClangSObSshLsPrecision}{1\xspace}
\newcommand{\XsfClangSObSshLsFone}{1\xspace}
\newcommand{\XsfClangSObSshBapRecall}{0.919\xspace}
\newcommand{\XsfClangSObSshBapPrecision}{1\xspace}
\newcommand{\XsfClangSObSshBapFone}{0.958\xspace}
\newcommand{\XsfClangSObSshGhiRecall}{1.000\xspace}
\newcommand{\XsfClangSObSshGhiPrecision}{1\xspace}
\newcommand{\XsfClangSObSshGhiFone}{1.000\xspace}
\newcommand{\XsfClangSObSshRdaRecall}{0.885\xspace}
\newcommand{\XsfClangSObSshRdaPrecision}{1.000\xspace}
\newcommand{\XsfClangSObSshRdaFone}{0.939\xspace}
\newcommand{\XsfClangSObSshRseRecall}{0.966\xspace}
\newcommand{\XsfClangSObSshRsePrecision}{1\xspace}
\newcommand{\XsfClangSObSshRseFone}{0.983\xspace}
\newcommand{\XsfClangSObPcrGT}{4921\xspace}
\newcommand{\XsfClangSObPcrLsRecall}{1\xspace}
\newcommand{\XsfClangSObPcrLsPrecision}{1\xspace}
\newcommand{\XsfClangSObPcrLsFone}{1\xspace}
\newcommand{\XsfClangSObPcrBapRecall}{0.844\xspace}
\newcommand{\XsfClangSObPcrBapPrecision}{1\xspace}
\newcommand{\XsfClangSObPcrBapFone}{0.916\xspace}
\newcommand{\XsfClangSObPcrGhiRecall}{1\xspace}
\newcommand{\XsfClangSObPcrGhiPrecision}{1\xspace}
\newcommand{\XsfClangSObPcrGhiFone}{1\xspace}
\newcommand{\XsfClangSObPcrRdaRecall}{1\xspace}
\newcommand{\XsfClangSObPcrRdaPrecision}{1\xspace}
\newcommand{\XsfClangSObPcrRdaFone}{1\xspace}
\newcommand{\XsfClangSObPcrRseRecall}{1\xspace}
\newcommand{\XsfClangSObPcrRsePrecision}{1\xspace}
\newcommand{\XsfClangSObPcrRseFone}{1\xspace}
\newcommand{\XsfClangSObSqlGT}{260837\xspace}
\newcommand{\XsfClangSObSqlLsRecall}{1\xspace}
\newcommand{\XsfClangSObSqlLsPrecision}{1\xspace}
\newcommand{\XsfClangSObSqlLsFone}{1\xspace}
\newcommand{\XsfClangSObSqlBapRecall}{0.803\xspace}
\newcommand{\XsfClangSObSqlBapPrecision}{1.000\xspace}
\newcommand{\XsfClangSObSqlBapFone}{0.891\xspace}
\newcommand{\XsfClangSObSqlGhiRecall}{0.948\xspace}
\newcommand{\XsfClangSObSqlGhiPrecision}{1\xspace}
\newcommand{\XsfClangSObSqlGhiFone}{0.973\xspace}
\newcommand{\XsfClangSObSqlRdaRecall}{0.907\xspace}
\newcommand{\XsfClangSObSqlRdaPrecision}{1\xspace}
\newcommand{\XsfClangSObSqlRdaFone}{0.951\xspace}
\newcommand{\XsfClangSObSqlRseRecall}{0.999\xspace}
\newcommand{\XsfClangSObSqlRsePrecision}{1\xspace}
\newcommand{\XsfClangSObSqlRseFone}{1.000\xspace}
\newcommand{\XsfClangSObVimGT}{539442\xspace}
\newcommand{\XsfClangSObVimLsRecall}{1\xspace}
\newcommand{\XsfClangSObVimLsPrecision}{1\xspace}
\newcommand{\XsfClangSObVimLsFone}{1\xspace}
\newcommand{\XsfClangSObVimBapRecall}{0.850\xspace}
\newcommand{\XsfClangSObVimBapPrecision}{1.000\xspace}
\newcommand{\XsfClangSObVimBapFone}{0.919\xspace}
\newcommand{\XsfClangSObVimGhiRecall}{0.992\xspace}
\newcommand{\XsfClangSObVimGhiPrecision}{1\xspace}
\newcommand{\XsfClangSObVimGhiFone}{0.996\xspace}
\newcommand{\XsfClangSObVimRdaRecall}{0.969\xspace}
\newcommand{\XsfClangSObVimRdaPrecision}{1.000\xspace}
\newcommand{\XsfClangSObVimRdaFone}{0.984\xspace}
\newcommand{\XsfClangSObVimRseRecall}{1.000\xspace}
\newcommand{\XsfClangSObVimRsePrecision}{1\xspace}
\newcommand{\XsfClangSObVimRseFone}{1.000\xspace}
\newcommand{\XsfClangSObVsfGT}{19022\xspace}
\newcommand{\XsfClangSObVsfLsRecall}{1\xspace}
\newcommand{\XsfClangSObVsfLsPrecision}{1\xspace}
\newcommand{\XsfClangSObVsfLsFone}{1\xspace}
\newcommand{\XsfClangSObVsfBapRecall}{0.988\xspace}
\newcommand{\XsfClangSObVsfBapPrecision}{1\xspace}
\newcommand{\XsfClangSObVsfBapFone}{0.994\xspace}
\newcommand{\XsfClangSObVsfGhiRecall}{1\xspace}
\newcommand{\XsfClangSObVsfGhiPrecision}{1\xspace}
\newcommand{\XsfClangSObVsfGhiFone}{1\xspace}
\newcommand{\XsfClangSObVsfRdaRecall}{0.983\xspace}
\newcommand{\XsfClangSObVsfRdaPrecision}{1.000\xspace}
\newcommand{\XsfClangSObVsfRdaFone}{0.991\xspace}
\newcommand{\XsfClangSObVsfRseRecall}{0.994\xspace}
\newcommand{\XsfClangSObVsfRsePrecision}{1\xspace}
\newcommand{\XsfClangSObVsfRseFone}{0.997\xspace}
\newcommand{\XsfClangSOcSzpGT}{15776\xspace}
\newcommand{\XsfClangSOcSzpLsRecall}{1\xspace}
\newcommand{\XsfClangSOcSzpLsPrecision}{1\xspace}
\newcommand{\XsfClangSOcSzpLsFone}{1\xspace}
\newcommand{\XsfClangSOcSzpBapRecall}{0.969\xspace}
\newcommand{\XsfClangSOcSzpBapPrecision}{1\xspace}
\newcommand{\XsfClangSOcSzpBapFone}{0.984\xspace}
\newcommand{\XsfClangSOcSzpGhiRecall}{1\xspace}
\newcommand{\XsfClangSOcSzpGhiPrecision}{1\xspace}
\newcommand{\XsfClangSOcSzpGhiFone}{1\xspace}
\newcommand{\XsfClangSOcSzpRdaRecall}{0.995\xspace}
\newcommand{\XsfClangSOcSzpRdaPrecision}{1\xspace}
\newcommand{\XsfClangSOcSzpRdaFone}{0.998\xspace}
\newcommand{\XsfClangSOcSzpRseRecall}{1\xspace}
\newcommand{\XsfClangSOcSzpRsePrecision}{1\xspace}
\newcommand{\XsfClangSOcSzpRseFone}{1\xspace}
\newcommand{\XsfClangSOcCapGT}{170817\xspace}
\newcommand{\XsfClangSOcCapLsRecall}{1\xspace}
\newcommand{\XsfClangSOcCapLsPrecision}{1\xspace}
\newcommand{\XsfClangSOcCapLsFone}{1\xspace}
\newcommand{\XsfClangSOcCapBapRecall}{0.292\xspace}
\newcommand{\XsfClangSOcCapBapPrecision}{1.000\xspace}
\newcommand{\XsfClangSOcCapBapFone}{0.452\xspace}
\newcommand{\XsfClangSOcCapGhiRecall}{0.826\xspace}
\newcommand{\XsfClangSOcCapGhiPrecision}{1\xspace}
\newcommand{\XsfClangSOcCapGhiFone}{0.905\xspace}
\newcommand{\XsfClangSOcCapRdaRecall}{0.839\xspace}
\newcommand{\XsfClangSOcCapRdaPrecision}{1\xspace}
\newcommand{\XsfClangSOcCapRdaFone}{0.912\xspace}
\newcommand{\XsfClangSOcCapRseRecall}{1.000\xspace}
\newcommand{\XsfClangSOcCapRsePrecision}{1\xspace}
\newcommand{\XsfClangSOcCapRseFone}{1.000\xspace}
\newcommand{\XsfClangSOcExmGT}{162179\xspace}
\newcommand{\XsfClangSOcExmLsRecall}{1\xspace}
\newcommand{\XsfClangSOcExmLsPrecision}{1\xspace}
\newcommand{\XsfClangSOcExmLsFone}{1\xspace}
\newcommand{\XsfClangSOcExmBapRecall}{0.825\xspace}
\newcommand{\XsfClangSOcExmBapPrecision}{1.000\xspace}
\newcommand{\XsfClangSOcExmBapFone}{0.904\xspace}
\newcommand{\XsfClangSOcExmGhiRecall}{0.941\xspace}
\newcommand{\XsfClangSOcExmGhiPrecision}{1\xspace}
\newcommand{\XsfClangSOcExmGhiFone}{0.970\xspace}
\newcommand{\XsfClangSOcExmRdaRecall}{0.961\xspace}
\newcommand{\XsfClangSOcExmRdaPrecision}{1.000\xspace}
\newcommand{\XsfClangSOcExmRdaFone}{0.980\xspace}
\newcommand{\XsfClangSOcExmRseRecall}{1.000\xspace}
\newcommand{\XsfClangSOcExmRsePrecision}{1.000\xspace}
\newcommand{\XsfClangSOcExmRseFone}{1.000\xspace}
\newcommand{\XsfClangSOcLgtGT}{27773\xspace}
\newcommand{\XsfClangSOcLgtLsRecall}{1\xspace}
\newcommand{\XsfClangSOcLgtLsPrecision}{1\xspace}
\newcommand{\XsfClangSOcLgtLsFone}{1\xspace}
\newcommand{\XsfClangSOcLgtBapRecall}{0.859\xspace}
\newcommand{\XsfClangSOcLgtBapPrecision}{1\xspace}
\newcommand{\XsfClangSOcLgtBapFone}{0.924\xspace}
\newcommand{\XsfClangSOcLgtGhiRecall}{0.999\xspace}
\newcommand{\XsfClangSOcLgtGhiPrecision}{1\xspace}
\newcommand{\XsfClangSOcLgtGhiFone}{0.999\xspace}
\newcommand{\XsfClangSOcLgtRdaRecall}{0.991\xspace}
\newcommand{\XsfClangSOcLgtRdaPrecision}{1.000\xspace}
\newcommand{\XsfClangSOcLgtRdaFone}{0.995\xspace}
\newcommand{\XsfClangSOcLgtRseRecall}{1\xspace}
\newcommand{\XsfClangSOcLgtRsePrecision}{1\xspace}
\newcommand{\XsfClangSOcLgtRseFone}{1\xspace}
\newcommand{\XsfClangSOcBzpGT}{18754\xspace}
\newcommand{\XsfClangSOcBzpLsRecall}{1\xspace}
\newcommand{\XsfClangSOcBzpLsPrecision}{1\xspace}
\newcommand{\XsfClangSOcBzpLsFone}{1\xspace}
\newcommand{\XsfClangSOcBzpBapRecall}{0.978\xspace}
\newcommand{\XsfClangSOcBzpBapPrecision}{1\xspace}
\newcommand{\XsfClangSOcBzpBapFone}{0.989\xspace}
\newcommand{\XsfClangSOcBzpGhiRecall}{1\xspace}
\newcommand{\XsfClangSOcBzpGhiPrecision}{1\xspace}
\newcommand{\XsfClangSOcBzpGhiFone}{1\xspace}
\newcommand{\XsfClangSOcBzpRdaRecall}{0.882\xspace}
\newcommand{\XsfClangSOcBzpRdaPrecision}{1\xspace}
\newcommand{\XsfClangSOcBzpRdaFone}{0.937\xspace}
\newcommand{\XsfClangSOcBzpRseRecall}{1\xspace}
\newcommand{\XsfClangSOcBzpRsePrecision}{1\xspace}
\newcommand{\XsfClangSOcBzpRseFone}{1\xspace}
\newcommand{\XsfClangSOcGccGT}{1209700\xspace}
\newcommand{\XsfClangSOcGccLsRecall}{1\xspace}
\newcommand{\XsfClangSOcGccLsPrecision}{1\xspace}
\newcommand{\XsfClangSOcGccLsFone}{1\xspace}
\newcommand{\XsfClangSOcGccBapRecall}{0.713\xspace}
\newcommand{\XsfClangSOcGccBapPrecision}{1.000\xspace}
\newcommand{\XsfClangSOcGccBapFone}{0.833\xspace}
\newcommand{\XsfClangSOcGccGhiRecall}{0.956\xspace}
\newcommand{\XsfClangSOcGccGhiPrecision}{1\xspace}
\newcommand{\XsfClangSOcGccGhiFone}{0.978\xspace}
\newcommand{\XsfClangSOcGccRdaRecall}{0.803\xspace}
\newcommand{\XsfClangSOcGccRdaPrecision}{1\xspace}
\newcommand{\XsfClangSOcGccRdaFone}{0.891\xspace}
\newcommand{\XsfClangSOcGccRseRecall}{0.999\xspace}
\newcommand{\XsfClangSOcGccRsePrecision}{1\xspace}
\newcommand{\XsfClangSOcGccRseFone}{1.000\xspace}
\newcommand{\XsfClangSOcGzpGT}{15080\xspace}
\newcommand{\XsfClangSOcGzpLsRecall}{1\xspace}
\newcommand{\XsfClangSOcGzpLsPrecision}{1\xspace}
\newcommand{\XsfClangSOcGzpLsFone}{1\xspace}
\newcommand{\XsfClangSOcGzpBapRecall}{0.989\xspace}
\newcommand{\XsfClangSOcGzpBapPrecision}{1\xspace}
\newcommand{\XsfClangSOcGzpBapFone}{0.995\xspace}
\newcommand{\XsfClangSOcGzpGhiRecall}{1\xspace}
\newcommand{\XsfClangSOcGzpGhiPrecision}{1\xspace}
\newcommand{\XsfClangSOcGzpGhiFone}{1\xspace}
\newcommand{\XsfClangSOcGzpRdaRecall}{1\xspace}
\newcommand{\XsfClangSOcGzpRdaPrecision}{1\xspace}
\newcommand{\XsfClangSOcGzpRdaFone}{1\xspace}
\newcommand{\XsfClangSOcGzpRseRecall}{0.999\xspace}
\newcommand{\XsfClangSOcGzpRsePrecision}{1\xspace}
\newcommand{\XsfClangSOcGzpRseFone}{0.999\xspace}
\newcommand{\XsfClangSOcOggGT}{59801\xspace}
\newcommand{\XsfClangSOcOggLsRecall}{1\xspace}
\newcommand{\XsfClangSOcOggLsPrecision}{1\xspace}
\newcommand{\XsfClangSOcOggLsFone}{1\xspace}
\newcommand{\XsfClangSOcOggBapRecall}{0.971\xspace}
\newcommand{\XsfClangSOcOggBapPrecision}{1.000\xspace}
\newcommand{\XsfClangSOcOggBapFone}{0.986\xspace}
\newcommand{\XsfClangSOcOggGhiRecall}{1\xspace}
\newcommand{\XsfClangSOcOggGhiPrecision}{1\xspace}
\newcommand{\XsfClangSOcOggGhiFone}{1\xspace}
\newcommand{\XsfClangSOcOggRdaRecall}{0.986\xspace}
\newcommand{\XsfClangSOcOggRdaPrecision}{1\xspace}
\newcommand{\XsfClangSOcOggRdaFone}{0.993\xspace}
\newcommand{\XsfClangSOcOggRseRecall}{1\xspace}
\newcommand{\XsfClangSOcOggRsePrecision}{1\xspace}
\newcommand{\XsfClangSOcOggRseFone}{1\xspace}
\newcommand{\XsfClangSOcNgxGT}{104953\xspace}
\newcommand{\XsfClangSOcNgxLsRecall}{1\xspace}
\newcommand{\XsfClangSOcNgxLsPrecision}{1\xspace}
\newcommand{\XsfClangSOcNgxLsFone}{1\xspace}
\newcommand{\XsfClangSOcNgxBapRecall}{0.961\xspace}
\newcommand{\XsfClangSOcNgxBapPrecision}{1.000\xspace}
\newcommand{\XsfClangSOcNgxBapFone}{0.980\xspace}
\newcommand{\XsfClangSOcNgxGhiRecall}{0.993\xspace}
\newcommand{\XsfClangSOcNgxGhiPrecision}{1\xspace}
\newcommand{\XsfClangSOcNgxGhiFone}{0.996\xspace}
\newcommand{\XsfClangSOcNgxRdaRecall}{0.997\xspace}
\newcommand{\XsfClangSOcNgxRdaPrecision}{1\xspace}
\newcommand{\XsfClangSOcNgxRdaFone}{0.999\xspace}
\newcommand{\XsfClangSOcNgxRseRecall}{1\xspace}
\newcommand{\XsfClangSOcNgxRsePrecision}{1\xspace}
\newcommand{\XsfClangSOcNgxRseFone}{1\xspace}
\newcommand{\XsfClangSOcSshGT}{122339\xspace}
\newcommand{\XsfClangSOcSshLsRecall}{1\xspace}
\newcommand{\XsfClangSOcSshLsPrecision}{1\xspace}
\newcommand{\XsfClangSOcSshLsFone}{1\xspace}
\newcommand{\XsfClangSOcSshBapRecall}{0.917\xspace}
\newcommand{\XsfClangSOcSshBapPrecision}{1\xspace}
\newcommand{\XsfClangSOcSshBapFone}{0.957\xspace}
\newcommand{\XsfClangSOcSshGhiRecall}{1.000\xspace}
\newcommand{\XsfClangSOcSshGhiPrecision}{1\xspace}
\newcommand{\XsfClangSOcSshGhiFone}{1.000\xspace}
\newcommand{\XsfClangSOcSshRdaRecall}{0.882\xspace}
\newcommand{\XsfClangSOcSshRdaPrecision}{1.000\xspace}
\newcommand{\XsfClangSOcSshRdaFone}{0.937\xspace}
\newcommand{\XsfClangSOcSshRseRecall}{0.967\xspace}
\newcommand{\XsfClangSOcSshRsePrecision}{1\xspace}
\newcommand{\XsfClangSOcSshRseFone}{0.983\xspace}
\newcommand{\XsfClangSOcPcrGT}{5540\xspace}
\newcommand{\XsfClangSOcPcrLsRecall}{1\xspace}
\newcommand{\XsfClangSOcPcrLsPrecision}{1\xspace}
\newcommand{\XsfClangSOcPcrLsFone}{1\xspace}
\newcommand{\XsfClangSOcPcrBapRecall}{0.825\xspace}
\newcommand{\XsfClangSOcPcrBapPrecision}{1\xspace}
\newcommand{\XsfClangSOcPcrBapFone}{0.904\xspace}
\newcommand{\XsfClangSOcPcrGhiRecall}{0.971\xspace}
\newcommand{\XsfClangSOcPcrGhiPrecision}{1\xspace}
\newcommand{\XsfClangSOcPcrGhiFone}{0.985\xspace}
\newcommand{\XsfClangSOcPcrRdaRecall}{1\xspace}
\newcommand{\XsfClangSOcPcrRdaPrecision}{1.000\xspace}
\newcommand{\XsfClangSOcPcrRdaFone}{1.000\xspace}
\newcommand{\XsfClangSOcPcrRseRecall}{1\xspace}
\newcommand{\XsfClangSOcPcrRsePrecision}{1\xspace}
\newcommand{\XsfClangSOcPcrRseFone}{1\xspace}
\newcommand{\XsfClangSOcSqlGT}{307814\xspace}
\newcommand{\XsfClangSOcSqlLsRecall}{1\xspace}
\newcommand{\XsfClangSOcSqlLsPrecision}{1\xspace}
\newcommand{\XsfClangSOcSqlLsFone}{1\xspace}
\newcommand{\XsfClangSOcSqlBapRecall}{0.788\xspace}
\newcommand{\XsfClangSOcSqlBapPrecision}{1.000\xspace}
\newcommand{\XsfClangSOcSqlBapFone}{0.881\xspace}
\newcommand{\XsfClangSOcSqlGhiRecall}{0.917\xspace}
\newcommand{\XsfClangSOcSqlGhiPrecision}{1\xspace}
\newcommand{\XsfClangSOcSqlGhiFone}{0.957\xspace}
\newcommand{\XsfClangSOcSqlRdaRecall}{0.907\xspace}
\newcommand{\XsfClangSOcSqlRdaPrecision}{1\xspace}
\newcommand{\XsfClangSOcSqlRdaFone}{0.951\xspace}
\newcommand{\XsfClangSOcSqlRseRecall}{0.965\xspace}
\newcommand{\XsfClangSOcSqlRsePrecision}{1\xspace}
\newcommand{\XsfClangSOcSqlRseFone}{0.982\xspace}
\newcommand{\XsfClangSOcVimGT}{594181\xspace}
\newcommand{\XsfClangSOcVimLsRecall}{1\xspace}
\newcommand{\XsfClangSOcVimLsPrecision}{1\xspace}
\newcommand{\XsfClangSOcVimLsFone}{1\xspace}
\newcommand{\XsfClangSOcVimBapRecall}{0.854\xspace}
\newcommand{\XsfClangSOcVimBapPrecision}{1.000\xspace}
\newcommand{\XsfClangSOcVimBapFone}{0.921\xspace}
\newcommand{\XsfClangSOcVimGhiRecall}{0.990\xspace}
\newcommand{\XsfClangSOcVimGhiPrecision}{1\xspace}
\newcommand{\XsfClangSOcVimGhiFone}{0.995\xspace}
\newcommand{\XsfClangSOcVimRdaRecall}{0.961\xspace}
\newcommand{\XsfClangSOcVimRdaPrecision}{1.000\xspace}
\newcommand{\XsfClangSOcVimRdaFone}{0.980\xspace}
\newcommand{\XsfClangSOcVimRseRecall}{1.000\xspace}
\newcommand{\XsfClangSOcVimRsePrecision}{1\xspace}
\newcommand{\XsfClangSOcVimRseFone}{1.000\xspace}
\newcommand{\XsfClangSOcVsfGT}{20887\xspace}
\newcommand{\XsfClangSOcVsfLsRecall}{1\xspace}
\newcommand{\XsfClangSOcVsfLsPrecision}{1\xspace}
\newcommand{\XsfClangSOcVsfLsFone}{1\xspace}
\newcommand{\XsfClangSOcVsfBapRecall}{0.989\xspace}
\newcommand{\XsfClangSOcVsfBapPrecision}{1\xspace}
\newcommand{\XsfClangSOcVsfBapFone}{0.995\xspace}
\newcommand{\XsfClangSOcVsfGhiRecall}{0.999\xspace}
\newcommand{\XsfClangSOcVsfGhiPrecision}{1\xspace}
\newcommand{\XsfClangSOcVsfGhiFone}{0.999\xspace}
\newcommand{\XsfClangSOcVsfRdaRecall}{0.984\xspace}
\newcommand{\XsfClangSOcVsfRdaPrecision}{1.000\xspace}
\newcommand{\XsfClangSOcVsfRdaFone}{0.992\xspace}
\newcommand{\XsfClangSOcVsfRseRecall}{0.994\xspace}
\newcommand{\XsfClangSOcVsfRsePrecision}{1\xspace}
\newcommand{\XsfClangSOcVsfRseFone}{0.997\xspace}
\newcommand{\XsfClangSOdSzpGT}{15776\xspace}
\newcommand{\XsfClangSOdSzpLsRecall}{1\xspace}
\newcommand{\XsfClangSOdSzpLsPrecision}{1\xspace}
\newcommand{\XsfClangSOdSzpLsFone}{1\xspace}
\newcommand{\XsfClangSOdSzpBapRecall}{0.969\xspace}
\newcommand{\XsfClangSOdSzpBapPrecision}{1\xspace}
\newcommand{\XsfClangSOdSzpBapFone}{0.984\xspace}
\newcommand{\XsfClangSOdSzpGhiRecall}{1\xspace}
\newcommand{\XsfClangSOdSzpGhiPrecision}{1\xspace}
\newcommand{\XsfClangSOdSzpGhiFone}{1\xspace}
\newcommand{\XsfClangSOdSzpRdaRecall}{0.995\xspace}
\newcommand{\XsfClangSOdSzpRdaPrecision}{1\xspace}
\newcommand{\XsfClangSOdSzpRdaFone}{0.998\xspace}
\newcommand{\XsfClangSOdSzpRseRecall}{1\xspace}
\newcommand{\XsfClangSOdSzpRsePrecision}{1\xspace}
\newcommand{\XsfClangSOdSzpRseFone}{1\xspace}
\newcommand{\XsfClangSOdCapGT}{170817\xspace}
\newcommand{\XsfClangSOdCapLsRecall}{1\xspace}
\newcommand{\XsfClangSOdCapLsPrecision}{1\xspace}
\newcommand{\XsfClangSOdCapLsFone}{1\xspace}
\newcommand{\XsfClangSOdCapBapRecall}{0.292\xspace}
\newcommand{\XsfClangSOdCapBapPrecision}{1.000\xspace}
\newcommand{\XsfClangSOdCapBapFone}{0.452\xspace}
\newcommand{\XsfClangSOdCapGhiRecall}{0.826\xspace}
\newcommand{\XsfClangSOdCapGhiPrecision}{1\xspace}
\newcommand{\XsfClangSOdCapGhiFone}{0.905\xspace}
\newcommand{\XsfClangSOdCapRdaRecall}{0.839\xspace}
\newcommand{\XsfClangSOdCapRdaPrecision}{1\xspace}
\newcommand{\XsfClangSOdCapRdaFone}{0.912\xspace}
\newcommand{\XsfClangSOdCapRseRecall}{1.000\xspace}
\newcommand{\XsfClangSOdCapRsePrecision}{1\xspace}
\newcommand{\XsfClangSOdCapRseFone}{1.000\xspace}
\newcommand{\XsfClangSOdExmGT}{162174\xspace}
\newcommand{\XsfClangSOdExmLsRecall}{1\xspace}
\newcommand{\XsfClangSOdExmLsPrecision}{1\xspace}
\newcommand{\XsfClangSOdExmLsFone}{1\xspace}
\newcommand{\XsfClangSOdExmBapRecall}{0.825\xspace}
\newcommand{\XsfClangSOdExmBapPrecision}{1.000\xspace}
\newcommand{\XsfClangSOdExmBapFone}{0.904\xspace}
\newcommand{\XsfClangSOdExmGhiRecall}{0.939\xspace}
\newcommand{\XsfClangSOdExmGhiPrecision}{1\xspace}
\newcommand{\XsfClangSOdExmGhiFone}{0.969\xspace}
\newcommand{\XsfClangSOdExmRdaRecall}{0.961\xspace}
\newcommand{\XsfClangSOdExmRdaPrecision}{1.000\xspace}
\newcommand{\XsfClangSOdExmRdaFone}{0.980\xspace}
\newcommand{\XsfClangSOdExmRseRecall}{1.000\xspace}
\newcommand{\XsfClangSOdExmRsePrecision}{1.000\xspace}
\newcommand{\XsfClangSOdExmRseFone}{1.000\xspace}
\newcommand{\XsfClangSOdLgtGT}{27773\xspace}
\newcommand{\XsfClangSOdLgtLsRecall}{1\xspace}
\newcommand{\XsfClangSOdLgtLsPrecision}{1\xspace}
\newcommand{\XsfClangSOdLgtLsFone}{1\xspace}
\newcommand{\XsfClangSOdLgtBapRecall}{0.859\xspace}
\newcommand{\XsfClangSOdLgtBapPrecision}{1\xspace}
\newcommand{\XsfClangSOdLgtBapFone}{0.924\xspace}
\newcommand{\XsfClangSOdLgtGhiRecall}{0.999\xspace}
\newcommand{\XsfClangSOdLgtGhiPrecision}{1\xspace}
\newcommand{\XsfClangSOdLgtGhiFone}{0.999\xspace}
\newcommand{\XsfClangSOdLgtRdaRecall}{0.933\xspace}
\newcommand{\XsfClangSOdLgtRdaPrecision}{1.000\xspace}
\newcommand{\XsfClangSOdLgtRdaFone}{0.965\xspace}
\newcommand{\XsfClangSOdLgtRseRecall}{1\xspace}
\newcommand{\XsfClangSOdLgtRsePrecision}{1\xspace}
\newcommand{\XsfClangSOdLgtRseFone}{1\xspace}
\newcommand{\XsfClangSOdBzpGT}{18756\xspace}
\newcommand{\XsfClangSOdBzpLsRecall}{1\xspace}
\newcommand{\XsfClangSOdBzpLsPrecision}{1\xspace}
\newcommand{\XsfClangSOdBzpLsFone}{1\xspace}
\newcommand{\XsfClangSOdBzpBapRecall}{0.978\xspace}
\newcommand{\XsfClangSOdBzpBapPrecision}{1\xspace}
\newcommand{\XsfClangSOdBzpBapFone}{0.989\xspace}
\newcommand{\XsfClangSOdBzpGhiRecall}{1\xspace}
\newcommand{\XsfClangSOdBzpGhiPrecision}{1\xspace}
\newcommand{\XsfClangSOdBzpGhiFone}{1\xspace}
\newcommand{\XsfClangSOdBzpRdaRecall}{0.882\xspace}
\newcommand{\XsfClangSOdBzpRdaPrecision}{1.000\xspace}
\newcommand{\XsfClangSOdBzpRdaFone}{0.937\xspace}
\newcommand{\XsfClangSOdBzpRseRecall}{1\xspace}
\newcommand{\XsfClangSOdBzpRsePrecision}{1\xspace}
\newcommand{\XsfClangSOdBzpRseFone}{1\xspace}
\newcommand{\XsfClangSOdGccGT}{1209652\xspace}
\newcommand{\XsfClangSOdGccLsRecall}{1\xspace}
\newcommand{\XsfClangSOdGccLsPrecision}{1\xspace}
\newcommand{\XsfClangSOdGccLsFone}{1\xspace}
\newcommand{\XsfClangSOdGccBapRecall}{0.713\xspace}
\newcommand{\XsfClangSOdGccBapPrecision}{1.000\xspace}
\newcommand{\XsfClangSOdGccBapFone}{0.833\xspace}
\newcommand{\XsfClangSOdGccGhiRecall}{0.960\xspace}
\newcommand{\XsfClangSOdGccGhiPrecision}{1\xspace}
\newcommand{\XsfClangSOdGccGhiFone}{0.980\xspace}
\newcommand{\XsfClangSOdGccRdaRecall}{0.803\xspace}
\newcommand{\XsfClangSOdGccRdaPrecision}{1.000\xspace}
\newcommand{\XsfClangSOdGccRdaFone}{0.891\xspace}
\newcommand{\XsfClangSOdGccRseRecall}{0.999\xspace}
\newcommand{\XsfClangSOdGccRsePrecision}{1\xspace}
\newcommand{\XsfClangSOdGccRseFone}{1.000\xspace}
\newcommand{\XsfClangSOdGzpGT}{15080\xspace}
\newcommand{\XsfClangSOdGzpLsRecall}{1\xspace}
\newcommand{\XsfClangSOdGzpLsPrecision}{1\xspace}
\newcommand{\XsfClangSOdGzpLsFone}{1\xspace}
\newcommand{\XsfClangSOdGzpBapRecall}{0.989\xspace}
\newcommand{\XsfClangSOdGzpBapPrecision}{1\xspace}
\newcommand{\XsfClangSOdGzpBapFone}{0.995\xspace}
\newcommand{\XsfClangSOdGzpGhiRecall}{1\xspace}
\newcommand{\XsfClangSOdGzpGhiPrecision}{1\xspace}
\newcommand{\XsfClangSOdGzpGhiFone}{1\xspace}
\newcommand{\XsfClangSOdGzpRdaRecall}{1\xspace}
\newcommand{\XsfClangSOdGzpRdaPrecision}{1\xspace}
\newcommand{\XsfClangSOdGzpRdaFone}{1\xspace}
\newcommand{\XsfClangSOdGzpRseRecall}{0.999\xspace}
\newcommand{\XsfClangSOdGzpRsePrecision}{1\xspace}
\newcommand{\XsfClangSOdGzpRseFone}{0.999\xspace}
\newcommand{\XsfClangSOdOggGT}{60271\xspace}
\newcommand{\XsfClangSOdOggLsRecall}{1\xspace}
\newcommand{\XsfClangSOdOggLsPrecision}{1\xspace}
\newcommand{\XsfClangSOdOggLsFone}{1\xspace}
\newcommand{\XsfClangSOdOggBapRecall}{0.972\xspace}
\newcommand{\XsfClangSOdOggBapPrecision}{1.000\xspace}
\newcommand{\XsfClangSOdOggBapFone}{0.986\xspace}
\newcommand{\XsfClangSOdOggGhiRecall}{1\xspace}
\newcommand{\XsfClangSOdOggGhiPrecision}{1\xspace}
\newcommand{\XsfClangSOdOggGhiFone}{1\xspace}
\newcommand{\XsfClangSOdOggRdaRecall}{0.986\xspace}
\newcommand{\XsfClangSOdOggRdaPrecision}{1\xspace}
\newcommand{\XsfClangSOdOggRdaFone}{0.993\xspace}
\newcommand{\XsfClangSOdOggRseRecall}{1\xspace}
\newcommand{\XsfClangSOdOggRsePrecision}{1\xspace}
\newcommand{\XsfClangSOdOggRseFone}{1\xspace}
\newcommand{\XsfClangSOdNgxGT}{104953\xspace}
\newcommand{\XsfClangSOdNgxLsRecall}{1\xspace}
\newcommand{\XsfClangSOdNgxLsPrecision}{1\xspace}
\newcommand{\XsfClangSOdNgxLsFone}{1\xspace}
\newcommand{\XsfClangSOdNgxBapRecall}{0.961\xspace}
\newcommand{\XsfClangSOdNgxBapPrecision}{1.000\xspace}
\newcommand{\XsfClangSOdNgxBapFone}{0.980\xspace}
\newcommand{\XsfClangSOdNgxGhiRecall}{0.993\xspace}
\newcommand{\XsfClangSOdNgxGhiPrecision}{1\xspace}
\newcommand{\XsfClangSOdNgxGhiFone}{0.996\xspace}
\newcommand{\XsfClangSOdNgxRdaRecall}{0.997\xspace}
\newcommand{\XsfClangSOdNgxRdaPrecision}{1\xspace}
\newcommand{\XsfClangSOdNgxRdaFone}{0.999\xspace}
\newcommand{\XsfClangSOdNgxRseRecall}{1\xspace}
\newcommand{\XsfClangSOdNgxRsePrecision}{1\xspace}
\newcommand{\XsfClangSOdNgxRseFone}{1\xspace}
\newcommand{\XsfClangSOdSshGT}{122339\xspace}
\newcommand{\XsfClangSOdSshLsRecall}{1\xspace}
\newcommand{\XsfClangSOdSshLsPrecision}{1\xspace}
\newcommand{\XsfClangSOdSshLsFone}{1\xspace}
\newcommand{\XsfClangSOdSshBapRecall}{0.917\xspace}
\newcommand{\XsfClangSOdSshBapPrecision}{1\xspace}
\newcommand{\XsfClangSOdSshBapFone}{0.957\xspace}
\newcommand{\XsfClangSOdSshGhiRecall}{1.000\xspace}
\newcommand{\XsfClangSOdSshGhiPrecision}{1\xspace}
\newcommand{\XsfClangSOdSshGhiFone}{1.000\xspace}
\newcommand{\XsfClangSOdSshRdaRecall}{0.882\xspace}
\newcommand{\XsfClangSOdSshRdaPrecision}{1.000\xspace}
\newcommand{\XsfClangSOdSshRdaFone}{0.937\xspace}
\newcommand{\XsfClangSOdSshRseRecall}{0.967\xspace}
\newcommand{\XsfClangSOdSshRsePrecision}{1\xspace}
\newcommand{\XsfClangSOdSshRseFone}{0.983\xspace}
\newcommand{\XsfClangSOdPcrGT}{5540\xspace}
\newcommand{\XsfClangSOdPcrLsRecall}{1\xspace}
\newcommand{\XsfClangSOdPcrLsPrecision}{1\xspace}
\newcommand{\XsfClangSOdPcrLsFone}{1\xspace}
\newcommand{\XsfClangSOdPcrBapRecall}{0.825\xspace}
\newcommand{\XsfClangSOdPcrBapPrecision}{1\xspace}
\newcommand{\XsfClangSOdPcrBapFone}{0.904\xspace}
\newcommand{\XsfClangSOdPcrGhiRecall}{0.971\xspace}
\newcommand{\XsfClangSOdPcrGhiPrecision}{1\xspace}
\newcommand{\XsfClangSOdPcrGhiFone}{0.985\xspace}
\newcommand{\XsfClangSOdPcrRdaRecall}{0.991\xspace}
\newcommand{\XsfClangSOdPcrRdaPrecision}{0.999\xspace}
\newcommand{\XsfClangSOdPcrRdaFone}{0.995\xspace}
\newcommand{\XsfClangSOdPcrRseRecall}{1\xspace}
\newcommand{\XsfClangSOdPcrRsePrecision}{1\xspace}
\newcommand{\XsfClangSOdPcrRseFone}{1\xspace}
\newcommand{\XsfClangSOdSqlGT}{307827\xspace}
\newcommand{\XsfClangSOdSqlLsRecall}{1\xspace}
\newcommand{\XsfClangSOdSqlLsPrecision}{1\xspace}
\newcommand{\XsfClangSOdSqlLsFone}{1\xspace}
\newcommand{\XsfClangSOdSqlBapRecall}{0.788\xspace}
\newcommand{\XsfClangSOdSqlBapPrecision}{1.000\xspace}
\newcommand{\XsfClangSOdSqlBapFone}{0.881\xspace}
\newcommand{\XsfClangSOdSqlGhiRecall}{0.917\xspace}
\newcommand{\XsfClangSOdSqlGhiPrecision}{1\xspace}
\newcommand{\XsfClangSOdSqlGhiFone}{0.957\xspace}
\newcommand{\XsfClangSOdSqlRdaRecall}{0.908\xspace}
\newcommand{\XsfClangSOdSqlRdaPrecision}{1.000\xspace}
\newcommand{\XsfClangSOdSqlRdaFone}{0.952\xspace}
\newcommand{\XsfClangSOdSqlRseRecall}{0.965\xspace}
\newcommand{\XsfClangSOdSqlRsePrecision}{1\xspace}
\newcommand{\XsfClangSOdSqlRseFone}{0.982\xspace}
\newcommand{\XsfClangSOdVimGT}{594215\xspace}
\newcommand{\XsfClangSOdVimLsRecall}{1\xspace}
\newcommand{\XsfClangSOdVimLsPrecision}{1\xspace}
\newcommand{\XsfClangSOdVimLsFone}{1\xspace}
\newcommand{\XsfClangSOdVimBapRecall}{0.854\xspace}
\newcommand{\XsfClangSOdVimBapPrecision}{1.000\xspace}
\newcommand{\XsfClangSOdVimBapFone}{0.921\xspace}
\newcommand{\XsfClangSOdVimGhiRecall}{0.990\xspace}
\newcommand{\XsfClangSOdVimGhiPrecision}{1\xspace}
\newcommand{\XsfClangSOdVimGhiFone}{0.995\xspace}
\newcommand{\XsfClangSOdVimRdaRecall}{0.961\xspace}
\newcommand{\XsfClangSOdVimRdaPrecision}{1.000\xspace}
\newcommand{\XsfClangSOdVimRdaFone}{0.980\xspace}
\newcommand{\XsfClangSOdVimRseRecall}{1.000\xspace}
\newcommand{\XsfClangSOdVimRsePrecision}{1\xspace}
\newcommand{\XsfClangSOdVimRseFone}{1.000\xspace}
\newcommand{\XsfClangSOdVsfGT}{20885\xspace}
\newcommand{\XsfClangSOdVsfLsRecall}{1\xspace}
\newcommand{\XsfClangSOdVsfLsPrecision}{1\xspace}
\newcommand{\XsfClangSOdVsfLsFone}{1\xspace}
\newcommand{\XsfClangSOdVsfBapRecall}{0.989\xspace}
\newcommand{\XsfClangSOdVsfBapPrecision}{1\xspace}
\newcommand{\XsfClangSOdVsfBapFone}{0.995\xspace}
\newcommand{\XsfClangSOdVsfGhiRecall}{0.999\xspace}
\newcommand{\XsfClangSOdVsfGhiPrecision}{1\xspace}
\newcommand{\XsfClangSOdVsfGhiFone}{0.999\xspace}
\newcommand{\XsfClangSOdVsfRdaRecall}{0.984\xspace}
\newcommand{\XsfClangSOdVsfRdaPrecision}{1.000\xspace}
\newcommand{\XsfClangSOdVsfRdaFone}{0.992\xspace}
\newcommand{\XsfClangSOdVsfRseRecall}{0.994\xspace}
\newcommand{\XsfClangSOdVsfRsePrecision}{1\xspace}
\newcommand{\XsfClangSOdVsfRseFone}{0.997\xspace}
\newcommand{\XsfClangSOsSzpGT}{11725\xspace}
\newcommand{\XsfClangSOsSzpLsRecall}{1\xspace}
\newcommand{\XsfClangSOsSzpLsPrecision}{1\xspace}
\newcommand{\XsfClangSOsSzpLsFone}{1\xspace}
\newcommand{\XsfClangSOsSzpBapRecall}{0.974\xspace}
\newcommand{\XsfClangSOsSzpBapPrecision}{1\xspace}
\newcommand{\XsfClangSOsSzpBapFone}{0.987\xspace}
\newcommand{\XsfClangSOsSzpGhiRecall}{1\xspace}
\newcommand{\XsfClangSOsSzpGhiPrecision}{1\xspace}
\newcommand{\XsfClangSOsSzpGhiFone}{1\xspace}
\newcommand{\XsfClangSOsSzpRdaRecall}{1\xspace}
\newcommand{\XsfClangSOsSzpRdaPrecision}{1\xspace}
\newcommand{\XsfClangSOsSzpRdaFone}{1\xspace}
\newcommand{\XsfClangSOsSzpRseRecall}{1\xspace}
\newcommand{\XsfClangSOsSzpRsePrecision}{1\xspace}
\newcommand{\XsfClangSOsSzpRseFone}{1\xspace}
\newcommand{\XsfClangSOsCapGT}{149690\xspace}
\newcommand{\XsfClangSOsCapLsRecall}{1\xspace}
\newcommand{\XsfClangSOsCapLsPrecision}{1\xspace}
\newcommand{\XsfClangSOsCapLsFone}{1\xspace}
\newcommand{\XsfClangSOsCapBapRecall}{0.383\xspace}
\newcommand{\XsfClangSOsCapBapPrecision}{1\xspace}
\newcommand{\XsfClangSOsCapBapFone}{0.553\xspace}
\newcommand{\XsfClangSOsCapGhiRecall}{0.824\xspace}
\newcommand{\XsfClangSOsCapGhiPrecision}{1\xspace}
\newcommand{\XsfClangSOsCapGhiFone}{0.904\xspace}
\newcommand{\XsfClangSOsCapRdaRecall}{0.998\xspace}
\newcommand{\XsfClangSOsCapRdaPrecision}{1.000\xspace}
\newcommand{\XsfClangSOsCapRdaFone}{0.999\xspace}
\newcommand{\XsfClangSOsCapRseRecall}{1\xspace}
\newcommand{\XsfClangSOsCapRsePrecision}{1\xspace}
\newcommand{\XsfClangSOsCapRseFone}{1\xspace}
\newcommand{\XsfClangSOsExmGT}{130733\xspace}
\newcommand{\XsfClangSOsExmLsRecall}{1\xspace}
\newcommand{\XsfClangSOsExmLsPrecision}{1\xspace}
\newcommand{\XsfClangSOsExmLsFone}{1\xspace}
\newcommand{\XsfClangSOsExmBapRecall}{0.844\xspace}
\newcommand{\XsfClangSOsExmBapPrecision}{1\xspace}
\newcommand{\XsfClangSOsExmBapFone}{0.915\xspace}
\newcommand{\XsfClangSOsExmGhiRecall}{0.987\xspace}
\newcommand{\XsfClangSOsExmGhiPrecision}{1\xspace}
\newcommand{\XsfClangSOsExmGhiFone}{0.993\xspace}
\newcommand{\XsfClangSOsExmRdaRecall}{0.964\xspace}
\newcommand{\XsfClangSOsExmRdaPrecision}{1.000\xspace}
\newcommand{\XsfClangSOsExmRdaFone}{0.981\xspace}
\newcommand{\XsfClangSOsExmRseRecall}{1.000\xspace}
\newcommand{\XsfClangSOsExmRsePrecision}{1\xspace}
\newcommand{\XsfClangSOsExmRseFone}{1.000\xspace}
\newcommand{\XsfClangSOsLgtGT}{23319\xspace}
\newcommand{\XsfClangSOsLgtLsRecall}{1\xspace}
\newcommand{\XsfClangSOsLgtLsPrecision}{1\xspace}
\newcommand{\XsfClangSOsLgtLsFone}{1\xspace}
\newcommand{\XsfClangSOsLgtBapRecall}{0.901\xspace}
\newcommand{\XsfClangSOsLgtBapPrecision}{1\xspace}
\newcommand{\XsfClangSOsLgtBapFone}{0.948\xspace}
\newcommand{\XsfClangSOsLgtGhiRecall}{1\xspace}
\newcommand{\XsfClangSOsLgtGhiPrecision}{1\xspace}
\newcommand{\XsfClangSOsLgtGhiFone}{1\xspace}
\newcommand{\XsfClangSOsLgtRdaRecall}{0.985\xspace}
\newcommand{\XsfClangSOsLgtRdaPrecision}{1.000\xspace}
\newcommand{\XsfClangSOsLgtRdaFone}{0.992\xspace}
\newcommand{\XsfClangSOsLgtRseRecall}{1\xspace}
\newcommand{\XsfClangSOsLgtRsePrecision}{1\xspace}
\newcommand{\XsfClangSOsLgtRseFone}{1\xspace}
\newcommand{\XsfClangSOsBzpGT}{14512\xspace}
\newcommand{\XsfClangSOsBzpLsRecall}{1\xspace}
\newcommand{\XsfClangSOsBzpLsPrecision}{1\xspace}
\newcommand{\XsfClangSOsBzpLsFone}{1\xspace}
\newcommand{\XsfClangSOsBzpBapRecall}{0.974\xspace}
\newcommand{\XsfClangSOsBzpBapPrecision}{1\xspace}
\newcommand{\XsfClangSOsBzpBapFone}{0.987\xspace}
\newcommand{\XsfClangSOsBzpGhiRecall}{1\xspace}
\newcommand{\XsfClangSOsBzpGhiPrecision}{1\xspace}
\newcommand{\XsfClangSOsBzpGhiFone}{1\xspace}
\newcommand{\XsfClangSOsBzpRdaRecall}{0.839\xspace}
\newcommand{\XsfClangSOsBzpRdaPrecision}{1\xspace}
\newcommand{\XsfClangSOsBzpRdaFone}{0.913\xspace}
\newcommand{\XsfClangSOsBzpRseRecall}{1\xspace}
\newcommand{\XsfClangSOsBzpRsePrecision}{1\xspace}
\newcommand{\XsfClangSOsBzpRseFone}{1\xspace}
\newcommand{\XsfClangSOsGccGT}{807788\xspace}
\newcommand{\XsfClangSOsGccLsRecall}{1\xspace}
\newcommand{\XsfClangSOsGccLsPrecision}{1\xspace}
\newcommand{\XsfClangSOsGccLsFone}{1\xspace}
\newcommand{\XsfClangSOsGccBapRecall}{0.721\xspace}
\newcommand{\XsfClangSOsGccBapPrecision}{1.000\xspace}
\newcommand{\XsfClangSOsGccBapFone}{0.838\xspace}
\newcommand{\XsfClangSOsGccGhiRecall}{0.967\xspace}
\newcommand{\XsfClangSOsGccGhiPrecision}{1\xspace}
\newcommand{\XsfClangSOsGccGhiFone}{0.983\xspace}
\newcommand{\XsfClangSOsGccRdaRecall}{0.832\xspace}
\newcommand{\XsfClangSOsGccRdaPrecision}{1.000\xspace}
\newcommand{\XsfClangSOsGccRdaFone}{0.908\xspace}
\newcommand{\XsfClangSOsGccRseRecall}{0.999\xspace}
\newcommand{\XsfClangSOsGccRsePrecision}{1\xspace}
\newcommand{\XsfClangSOsGccRseFone}{1.000\xspace}
\newcommand{\XsfClangSOsGzpGT}{8868\xspace}
\newcommand{\XsfClangSOsGzpLsRecall}{1\xspace}
\newcommand{\XsfClangSOsGzpLsPrecision}{1\xspace}
\newcommand{\XsfClangSOsGzpLsFone}{1\xspace}
\newcommand{\XsfClangSOsGzpBapRecall}{0.991\xspace}
\newcommand{\XsfClangSOsGzpBapPrecision}{1\xspace}
\newcommand{\XsfClangSOsGzpBapFone}{0.996\xspace}
\newcommand{\XsfClangSOsGzpGhiRecall}{1\xspace}
\newcommand{\XsfClangSOsGzpGhiPrecision}{1\xspace}
\newcommand{\XsfClangSOsGzpGhiFone}{1\xspace}
\newcommand{\XsfClangSOsGzpRdaRecall}{1\xspace}
\newcommand{\XsfClangSOsGzpRdaPrecision}{1\xspace}
\newcommand{\XsfClangSOsGzpRdaFone}{1\xspace}
\newcommand{\XsfClangSOsGzpRseRecall}{1\xspace}
\newcommand{\XsfClangSOsGzpRsePrecision}{1\xspace}
\newcommand{\XsfClangSOsGzpRseFone}{1\xspace}
\newcommand{\XsfClangSOsOggGT}{33423\xspace}
\newcommand{\XsfClangSOsOggLsRecall}{1\xspace}
\newcommand{\XsfClangSOsOggLsPrecision}{1\xspace}
\newcommand{\XsfClangSOsOggLsFone}{1\xspace}
\newcommand{\XsfClangSOsOggBapRecall}{0.982\xspace}
\newcommand{\XsfClangSOsOggBapPrecision}{1.000\xspace}
\newcommand{\XsfClangSOsOggBapFone}{0.991\xspace}
\newcommand{\XsfClangSOsOggGhiRecall}{1\xspace}
\newcommand{\XsfClangSOsOggGhiPrecision}{1\xspace}
\newcommand{\XsfClangSOsOggGhiFone}{1\xspace}
\newcommand{\XsfClangSOsOggRdaRecall}{1\xspace}
\newcommand{\XsfClangSOsOggRdaPrecision}{1\xspace}
\newcommand{\XsfClangSOsOggRdaFone}{1\xspace}
\newcommand{\XsfClangSOsOggRseRecall}{1\xspace}
\newcommand{\XsfClangSOsOggRsePrecision}{1\xspace}
\newcommand{\XsfClangSOsOggRseFone}{1\xspace}
\newcommand{\XsfClangSOsNgxGT}{91425\xspace}
\newcommand{\XsfClangSOsNgxLsRecall}{1\xspace}
\newcommand{\XsfClangSOsNgxLsPrecision}{1\xspace}
\newcommand{\XsfClangSOsNgxLsFone}{1\xspace}
\newcommand{\XsfClangSOsNgxBapRecall}{0.970\xspace}
\newcommand{\XsfClangSOsNgxBapPrecision}{1.000\xspace}
\newcommand{\XsfClangSOsNgxBapFone}{0.985\xspace}
\newcommand{\XsfClangSOsNgxGhiRecall}{1\xspace}
\newcommand{\XsfClangSOsNgxGhiPrecision}{1\xspace}
\newcommand{\XsfClangSOsNgxGhiFone}{1\xspace}
\newcommand{\XsfClangSOsNgxRdaRecall}{0.997\xspace}
\newcommand{\XsfClangSOsNgxRdaPrecision}{1\xspace}
\newcommand{\XsfClangSOsNgxRdaFone}{0.999\xspace}
\newcommand{\XsfClangSOsNgxRseRecall}{1\xspace}
\newcommand{\XsfClangSOsNgxRsePrecision}{1\xspace}
\newcommand{\XsfClangSOsNgxRseFone}{1\xspace}
\newcommand{\XsfClangSOsSshGT}{93674\xspace}
\newcommand{\XsfClangSOsSshLsRecall}{1\xspace}
\newcommand{\XsfClangSOsSshLsPrecision}{1\xspace}
\newcommand{\XsfClangSOsSshLsFone}{1\xspace}
\newcommand{\XsfClangSOsSshBapRecall}{0.945\xspace}
\newcommand{\XsfClangSOsSshBapPrecision}{1\xspace}
\newcommand{\XsfClangSOsSshBapFone}{0.972\xspace}
\newcommand{\XsfClangSOsSshGhiRecall}{1\xspace}
\newcommand{\XsfClangSOsSshGhiPrecision}{1\xspace}
\newcommand{\XsfClangSOsSshGhiFone}{1\xspace}
\newcommand{\XsfClangSOsSshRdaRecall}{0.967\xspace}
\newcommand{\XsfClangSOsSshRdaPrecision}{1.000\xspace}
\newcommand{\XsfClangSOsSshRdaFone}{0.983\xspace}
\newcommand{\XsfClangSOsSshRseRecall}{0.987\xspace}
\newcommand{\XsfClangSOsSshRsePrecision}{1\xspace}
\newcommand{\XsfClangSOsSshRseFone}{0.993\xspace}
\newcommand{\XsfClangSOsPcrGT}{4164\xspace}
\newcommand{\XsfClangSOsPcrLsRecall}{1\xspace}
\newcommand{\XsfClangSOsPcrLsPrecision}{1\xspace}
\newcommand{\XsfClangSOsPcrLsFone}{1\xspace}
\newcommand{\XsfClangSOsPcrBapRecall}{0.886\xspace}
\newcommand{\XsfClangSOsPcrBapPrecision}{1\xspace}
\newcommand{\XsfClangSOsPcrBapFone}{0.940\xspace}
\newcommand{\XsfClangSOsPcrGhiRecall}{1\xspace}
\newcommand{\XsfClangSOsPcrGhiPrecision}{1\xspace}
\newcommand{\XsfClangSOsPcrGhiFone}{1\xspace}
\newcommand{\XsfClangSOsPcrRdaRecall}{1\xspace}
\newcommand{\XsfClangSOsPcrRdaPrecision}{1\xspace}
\newcommand{\XsfClangSOsPcrRdaFone}{1\xspace}
\newcommand{\XsfClangSOsPcrRseRecall}{1\xspace}
\newcommand{\XsfClangSOsPcrRsePrecision}{1\xspace}
\newcommand{\XsfClangSOsPcrRseFone}{1\xspace}
\newcommand{\XsfClangSOsSqlGT}{159968\xspace}
\newcommand{\XsfClangSOsSqlLsRecall}{1\xspace}
\newcommand{\XsfClangSOsSqlLsPrecision}{1\xspace}
\newcommand{\XsfClangSOsSqlLsFone}{1\xspace}
\newcommand{\XsfClangSOsSqlBapRecall}{0.840\xspace}
\newcommand{\XsfClangSOsSqlBapPrecision}{1\xspace}
\newcommand{\XsfClangSOsSqlBapFone}{0.913\xspace}
\newcommand{\XsfClangSOsSqlGhiRecall}{0.949\xspace}
\newcommand{\XsfClangSOsSqlGhiPrecision}{1\xspace}
\newcommand{\XsfClangSOsSqlGhiFone}{0.974\xspace}
\newcommand{\XsfClangSOsSqlRdaRecall}{0.943\xspace}
\newcommand{\XsfClangSOsSqlRdaPrecision}{1\xspace}
\newcommand{\XsfClangSOsSqlRdaFone}{0.971\xspace}
\newcommand{\XsfClangSOsSqlRseRecall}{1\xspace}
\newcommand{\XsfClangSOsSqlRsePrecision}{1\xspace}
\newcommand{\XsfClangSOsSqlRseFone}{1\xspace}
\newcommand{\XsfClangSOsVimGT}{437329\xspace}
\newcommand{\XsfClangSOsVimLsRecall}{1\xspace}
\newcommand{\XsfClangSOsVimLsPrecision}{1\xspace}
\newcommand{\XsfClangSOsVimLsFone}{1\xspace}
\newcommand{\XsfClangSOsVimBapRecall}{0.915\xspace}
\newcommand{\XsfClangSOsVimBapPrecision}{1.000\xspace}
\newcommand{\XsfClangSOsVimBapFone}{0.956\xspace}
\newcommand{\XsfClangSOsVimGhiRecall}{0.999\xspace}
\newcommand{\XsfClangSOsVimGhiPrecision}{1\xspace}
\newcommand{\XsfClangSOsVimGhiFone}{0.999\xspace}
\newcommand{\XsfClangSOsVimRdaRecall}{0.976\xspace}
\newcommand{\XsfClangSOsVimRdaPrecision}{1.000\xspace}
\newcommand{\XsfClangSOsVimRdaFone}{0.988\xspace}
\newcommand{\XsfClangSOsVimRseRecall}{1.000\xspace}
\newcommand{\XsfClangSOsVimRsePrecision}{1\xspace}
\newcommand{\XsfClangSOsVimRseFone}{1.000\xspace}
\newcommand{\XsfClangSOsVsfGT}{16989\xspace}
\newcommand{\XsfClangSOsVsfLsRecall}{1\xspace}
\newcommand{\XsfClangSOsVsfLsPrecision}{1\xspace}
\newcommand{\XsfClangSOsVsfLsFone}{1\xspace}
\newcommand{\XsfClangSOsVsfBapRecall}{0.988\xspace}
\newcommand{\XsfClangSOsVsfBapPrecision}{1\xspace}
\newcommand{\XsfClangSOsVsfBapFone}{0.994\xspace}
\newcommand{\XsfClangSOsVsfGhiRecall}{1\xspace}
\newcommand{\XsfClangSOsVsfGhiPrecision}{1\xspace}
\newcommand{\XsfClangSOsVsfGhiFone}{1\xspace}
\newcommand{\XsfClangSOsVsfRdaRecall}{0.984\xspace}
\newcommand{\XsfClangSOsVsfRdaPrecision}{0.999\xspace}
\newcommand{\XsfClangSOsVsfRdaFone}{0.992\xspace}
\newcommand{\XsfClangSOsVsfRseRecall}{0.993\xspace}
\newcommand{\XsfClangSOsVsfRsePrecision}{1\xspace}
\newcommand{\XsfClangSOsVsfRseFone}{0.996\xspace}
\newcommand{\XsfIccOoSzpGT}{23564\xspace}
\newcommand{\XsfIccOoSzpLsRecall}{1\xspace}
\newcommand{\XsfIccOoSzpLsPrecision}{1\xspace}
\newcommand{\XsfIccOoSzpLsFone}{1\xspace}
\newcommand{\XsfIccOoSzpBapRecall}{0.976\xspace}
\newcommand{\XsfIccOoSzpBapPrecision}{1\xspace}
\newcommand{\XsfIccOoSzpBapFone}{0.988\xspace}
\newcommand{\XsfIccOoSzpGhiRecall}{1\xspace}
\newcommand{\XsfIccOoSzpGhiPrecision}{1\xspace}
\newcommand{\XsfIccOoSzpGhiFone}{1\xspace}
\newcommand{\XsfIccOoSzpRdaRecall}{1\xspace}
\newcommand{\XsfIccOoSzpRdaPrecision}{1\xspace}
\newcommand{\XsfIccOoSzpRdaFone}{1\xspace}
\newcommand{\XsfIccOoSzpRseRecall}{1\xspace}
\newcommand{\XsfIccOoSzpRsePrecision}{1\xspace}
\newcommand{\XsfIccOoSzpRseFone}{1\xspace}
\newcommand{\XsfIccOoCapGT}{458350\xspace}
\newcommand{\XsfIccOoCapLsRecall}{1\xspace}
\newcommand{\XsfIccOoCapLsPrecision}{1\xspace}
\newcommand{\XsfIccOoCapLsFone}{1\xspace}
\newcommand{\XsfIccOoCapBapRecall}{0.250\xspace}
\newcommand{\XsfIccOoCapBapPrecision}{1\xspace}
\newcommand{\XsfIccOoCapBapFone}{0.400\xspace}
\newcommand{\XsfIccOoCapGhiRecall}{1.000\xspace}
\newcommand{\XsfIccOoCapGhiPrecision}{1\xspace}
\newcommand{\XsfIccOoCapGhiFone}{1.000\xspace}
\newcommand{\XsfIccOoCapRdaRecall}{0.902\xspace}
\newcommand{\XsfIccOoCapRdaPrecision}{1\xspace}
\newcommand{\XsfIccOoCapRdaFone}{0.949\xspace}
\newcommand{\XsfIccOoCapRseRecall}{1.000\xspace}
\newcommand{\XsfIccOoCapRsePrecision}{1\xspace}
\newcommand{\XsfIccOoCapRseFone}{1.000\xspace}
\newcommand{\XsfIccOoExmGT}{235211\xspace}
\newcommand{\XsfIccOoExmLsRecall}{1\xspace}
\newcommand{\XsfIccOoExmLsPrecision}{1\xspace}
\newcommand{\XsfIccOoExmLsFone}{1\xspace}
\newcommand{\XsfIccOoExmBapRecall}{0.837\xspace}
\newcommand{\XsfIccOoExmBapPrecision}{1\xspace}
\newcommand{\XsfIccOoExmBapFone}{0.911\xspace}
\newcommand{\XsfIccOoExmGhiRecall}{0.948\xspace}
\newcommand{\XsfIccOoExmGhiPrecision}{1\xspace}
\newcommand{\XsfIccOoExmGhiFone}{0.973\xspace}
\newcommand{\XsfIccOoExmRdaRecall}{0.940\xspace}
\newcommand{\XsfIccOoExmRdaPrecision}{1.000\xspace}
\newcommand{\XsfIccOoExmRdaFone}{0.969\xspace}
\newcommand{\XsfIccOoExmRseRecall}{0.985\xspace}
\newcommand{\XsfIccOoExmRsePrecision}{1\xspace}
\newcommand{\XsfIccOoExmRseFone}{0.993\xspace}
\newcommand{\XsfIccOoLgtGT}{49328\xspace}
\newcommand{\XsfIccOoLgtLsRecall}{1\xspace}
\newcommand{\XsfIccOoLgtLsPrecision}{1\xspace}
\newcommand{\XsfIccOoLgtLsFone}{1\xspace}
\newcommand{\XsfIccOoLgtBapRecall}{0.880\xspace}
\newcommand{\XsfIccOoLgtBapPrecision}{1\xspace}
\newcommand{\XsfIccOoLgtBapFone}{0.936\xspace}
\newcommand{\XsfIccOoLgtGhiRecall}{1.000\xspace}
\newcommand{\XsfIccOoLgtGhiPrecision}{1\xspace}
\newcommand{\XsfIccOoLgtGhiFone}{1.000\xspace}
\newcommand{\XsfIccOoLgtRdaRecall}{0.962\xspace}
\newcommand{\XsfIccOoLgtRdaPrecision}{1.000\xspace}
\newcommand{\XsfIccOoLgtRdaFone}{0.981\xspace}
\newcommand{\XsfIccOoLgtRseRecall}{0.996\xspace}
\newcommand{\XsfIccOoLgtRsePrecision}{1\xspace}
\newcommand{\XsfIccOoLgtRseFone}{0.998\xspace}
\newcommand{\XsfIccOoBzpGT}{27427\xspace}
\newcommand{\XsfIccOoBzpLsRecall}{1\xspace}
\newcommand{\XsfIccOoBzpLsPrecision}{1\xspace}
\newcommand{\XsfIccOoBzpLsFone}{1\xspace}
\newcommand{\XsfIccOoBzpBapRecall}{0.798\xspace}
\newcommand{\XsfIccOoBzpBapPrecision}{1\xspace}
\newcommand{\XsfIccOoBzpBapFone}{0.888\xspace}
\newcommand{\XsfIccOoBzpGhiRecall}{1\xspace}
\newcommand{\XsfIccOoBzpGhiPrecision}{1\xspace}
\newcommand{\XsfIccOoBzpGhiFone}{1\xspace}
\newcommand{\XsfIccOoBzpRdaRecall}{0.957\xspace}
\newcommand{\XsfIccOoBzpRdaPrecision}{1\xspace}
\newcommand{\XsfIccOoBzpRdaFone}{0.978\xspace}
\newcommand{\XsfIccOoBzpRseRecall}{1\xspace}
\newcommand{\XsfIccOoBzpRsePrecision}{1\xspace}
\newcommand{\XsfIccOoBzpRseFone}{1\xspace}
\newcommand{\XsfIccOoGccGT}{1691222\xspace}
\newcommand{\XsfIccOoGccLsRecall}{1\xspace}
\newcommand{\XsfIccOoGccLsPrecision}{1\xspace}
\newcommand{\XsfIccOoGccLsFone}{1\xspace}
\newcommand{\XsfIccOoGccBapRecall}{0.719\xspace}
\newcommand{\XsfIccOoGccBapPrecision}{1\xspace}
\newcommand{\XsfIccOoGccBapFone}{0.836\xspace}
\newcommand{\XsfIccOoGccGhiRecall}{0.986\xspace}
\newcommand{\XsfIccOoGccGhiPrecision}{1\xspace}
\newcommand{\XsfIccOoGccGhiFone}{0.993\xspace}
\newcommand{\XsfIccOoGccRdaRecall}{0.871\xspace}
\newcommand{\XsfIccOoGccRdaPrecision}{1.000\xspace}
\newcommand{\XsfIccOoGccRdaFone}{0.931\xspace}
\newcommand{\XsfIccOoGccRseRecall}{0.996\xspace}
\newcommand{\XsfIccOoGccRsePrecision}{1\xspace}
\newcommand{\XsfIccOoGccRseFone}{0.998\xspace}
\newcommand{\XsfIccOoGzpGT}{17045\xspace}
\newcommand{\XsfIccOoGzpLsRecall}{1\xspace}
\newcommand{\XsfIccOoGzpLsPrecision}{1\xspace}
\newcommand{\XsfIccOoGzpLsFone}{1\xspace}
\newcommand{\XsfIccOoGzpBapRecall}{0.991\xspace}
\newcommand{\XsfIccOoGzpBapPrecision}{1\xspace}
\newcommand{\XsfIccOoGzpBapFone}{0.996\xspace}
\newcommand{\XsfIccOoGzpGhiRecall}{1\xspace}
\newcommand{\XsfIccOoGzpGhiPrecision}{1\xspace}
\newcommand{\XsfIccOoGzpGhiFone}{1\xspace}
\newcommand{\XsfIccOoGzpRdaRecall}{1\xspace}
\newcommand{\XsfIccOoGzpRdaPrecision}{1.000\xspace}
\newcommand{\XsfIccOoGzpRdaFone}{1.000\xspace}
\newcommand{\XsfIccOoGzpRseRecall}{0.999\xspace}
\newcommand{\XsfIccOoGzpRsePrecision}{1\xspace}
\newcommand{\XsfIccOoGzpRseFone}{1.000\xspace}
\newcommand{\XsfIccOoOggGT}{64988\xspace}
\newcommand{\XsfIccOoOggLsRecall}{1\xspace}
\newcommand{\XsfIccOoOggLsPrecision}{1\xspace}
\newcommand{\XsfIccOoOggLsFone}{1\xspace}
\newcommand{\XsfIccOoOggBapRecall}{0.975\xspace}
\newcommand{\XsfIccOoOggBapPrecision}{1\xspace}
\newcommand{\XsfIccOoOggBapFone}{0.987\xspace}
\newcommand{\XsfIccOoOggGhiRecall}{1.000\xspace}
\newcommand{\XsfIccOoOggGhiPrecision}{1\xspace}
\newcommand{\XsfIccOoOggGhiFone}{1.000\xspace}
\newcommand{\XsfIccOoOggRdaRecall}{0.992\xspace}
\newcommand{\XsfIccOoOggRdaPrecision}{1\xspace}
\newcommand{\XsfIccOoOggRdaFone}{0.996\xspace}
\newcommand{\XsfIccOoOggRseRecall}{0.992\xspace}
\newcommand{\XsfIccOoOggRsePrecision}{1\xspace}
\newcommand{\XsfIccOoOggRseFone}{0.996\xspace}
\newcommand{\XsfIccOoNgxGT}{196775\xspace}
\newcommand{\XsfIccOoNgxLsRecall}{1\xspace}
\newcommand{\XsfIccOoNgxLsPrecision}{1\xspace}
\newcommand{\XsfIccOoNgxLsFone}{1\xspace}
\newcommand{\XsfIccOoNgxBapRecall}{0.963\xspace}
\newcommand{\XsfIccOoNgxBapPrecision}{1\xspace}
\newcommand{\XsfIccOoNgxBapFone}{0.981\xspace}
\newcommand{\XsfIccOoNgxGhiRecall}{0.999\xspace}
\newcommand{\XsfIccOoNgxGhiPrecision}{1\xspace}
\newcommand{\XsfIccOoNgxGhiFone}{0.999\xspace}
\newcommand{\XsfIccOoNgxRdaRecall}{0.993\xspace}
\newcommand{\XsfIccOoNgxRdaPrecision}{1.000\xspace}
\newcommand{\XsfIccOoNgxRdaFone}{0.997\xspace}
\newcommand{\XsfIccOoNgxRseRecall}{0.996\xspace}
\newcommand{\XsfIccOoNgxRsePrecision}{1\xspace}
\newcommand{\XsfIccOoNgxRseFone}{0.998\xspace}
\newcommand{\XsfIccOoSshGT}{182967\xspace}
\newcommand{\XsfIccOoSshLsRecall}{1\xspace}
\newcommand{\XsfIccOoSshLsPrecision}{1\xspace}
\newcommand{\XsfIccOoSshLsFone}{1\xspace}
\newcommand{\XsfIccOoSshBapRecall}{0.941\xspace}
\newcommand{\XsfIccOoSshBapPrecision}{1\xspace}
\newcommand{\XsfIccOoSshBapFone}{0.970\xspace}
\newcommand{\XsfIccOoSshGhiRecall}{0.997\xspace}
\newcommand{\XsfIccOoSshGhiPrecision}{1\xspace}
\newcommand{\XsfIccOoSshGhiFone}{0.999\xspace}
\newcommand{\XsfIccOoSshRdaRecall}{0.936\xspace}
\newcommand{\XsfIccOoSshRdaPrecision}{0.999\xspace}
\newcommand{\XsfIccOoSshRdaFone}{0.966\xspace}
\newcommand{\XsfIccOoSshRseRecall}{0.972\xspace}
\newcommand{\XsfIccOoSshRsePrecision}{1\xspace}
\newcommand{\XsfIccOoSshRseFone}{0.986\xspace}
\newcommand{\XsfIccOoPcrGT}{7461\xspace}
\newcommand{\XsfIccOoPcrLsRecall}{1\xspace}
\newcommand{\XsfIccOoPcrLsPrecision}{1\xspace}
\newcommand{\XsfIccOoPcrLsFone}{1\xspace}
\newcommand{\XsfIccOoPcrBapRecall}{0.943\xspace}
\newcommand{\XsfIccOoPcrBapPrecision}{1\xspace}
\newcommand{\XsfIccOoPcrBapFone}{0.971\xspace}
\newcommand{\XsfIccOoPcrGhiRecall}{1\xspace}
\newcommand{\XsfIccOoPcrGhiPrecision}{1\xspace}
\newcommand{\XsfIccOoPcrGhiFone}{1\xspace}
\newcommand{\XsfIccOoPcrRdaRecall}{0.999\xspace}
\newcommand{\XsfIccOoPcrRdaPrecision}{1\xspace}
\newcommand{\XsfIccOoPcrRdaFone}{1.000\xspace}
\newcommand{\XsfIccOoPcrRseRecall}{1\xspace}
\newcommand{\XsfIccOoPcrRsePrecision}{1\xspace}
\newcommand{\XsfIccOoPcrRseFone}{1\xspace}
\newcommand{\XsfIccOoSqlGT}{274121\xspace}
\newcommand{\XsfIccOoSqlLsRecall}{1\xspace}
\newcommand{\XsfIccOoSqlLsPrecision}{1\xspace}
\newcommand{\XsfIccOoSqlLsFone}{1\xspace}
\newcommand{\XsfIccOoSqlBapRecall}{0.871\xspace}
\newcommand{\XsfIccOoSqlBapPrecision}{1.000\xspace}
\newcommand{\XsfIccOoSqlBapFone}{0.931\xspace}
\newcommand{\XsfIccOoSqlGhiRecall}{0.964\xspace}
\newcommand{\XsfIccOoSqlGhiPrecision}{1\xspace}
\newcommand{\XsfIccOoSqlGhiFone}{0.981\xspace}
\newcommand{\XsfIccOoSqlRdaRecall}{0.983\xspace}
\newcommand{\XsfIccOoSqlRdaPrecision}{1\xspace}
\newcommand{\XsfIccOoSqlRdaFone}{0.991\xspace}
\newcommand{\XsfIccOoSqlRseRecall}{0.986\xspace}
\newcommand{\XsfIccOoSqlRsePrecision}{1\xspace}
\newcommand{\XsfIccOoSqlRseFone}{0.993\xspace}
\newcommand{\XsfIccOoVimGT}{791133\xspace}
\newcommand{\XsfIccOoVimLsRecall}{1\xspace}
\newcommand{\XsfIccOoVimLsPrecision}{1\xspace}
\newcommand{\XsfIccOoVimLsFone}{1\xspace}
\newcommand{\XsfIccOoVimBapRecall}{0.915\xspace}
\newcommand{\XsfIccOoVimBapPrecision}{1\xspace}
\newcommand{\XsfIccOoVimBapFone}{0.955\xspace}
\newcommand{\XsfIccOoVimGhiRecall}{0.995\xspace}
\newcommand{\XsfIccOoVimGhiPrecision}{1\xspace}
\newcommand{\XsfIccOoVimGhiFone}{0.998\xspace}
\newcommand{\XsfIccOoVimRdaRecall}{0.000\xspace}
\newcommand{\XsfIccOoVimRdaPrecision}{NaN\xspace}
\newcommand{\XsfIccOoVimRdaFone}{0.000\xspace}
\newcommand{\XsfIccOoVimRseRecall}{0.999\xspace}
\newcommand{\XsfIccOoVimRsePrecision}{1\xspace}
\newcommand{\XsfIccOoVimRseFone}{1.000\xspace}
\newcommand{\XsfIccOoVsfGT}{30872\xspace}
\newcommand{\XsfIccOoVsfLsRecall}{1\xspace}
\newcommand{\XsfIccOoVsfLsPrecision}{1\xspace}
\newcommand{\XsfIccOoVsfLsFone}{1\xspace}
\newcommand{\XsfIccOoVsfBapRecall}{0.998\xspace}
\newcommand{\XsfIccOoVsfBapPrecision}{1\xspace}
\newcommand{\XsfIccOoVsfBapFone}{0.999\xspace}
\newcommand{\XsfIccOoVsfGhiRecall}{1\xspace}
\newcommand{\XsfIccOoVsfGhiPrecision}{1\xspace}
\newcommand{\XsfIccOoVsfGhiFone}{1\xspace}
\newcommand{\XsfIccOoVsfRdaRecall}{0.988\xspace}
\newcommand{\XsfIccOoVsfRdaPrecision}{0.999\xspace}
\newcommand{\XsfIccOoVsfRdaFone}{0.993\xspace}
\newcommand{\XsfIccOoVsfRseRecall}{1.000\xspace}
\newcommand{\XsfIccOoVsfRsePrecision}{1\xspace}
\newcommand{\XsfIccOoVsfRseFone}{1.000\xspace}
\newcommand{\XsfIccOaSzpGT}{11163\xspace}
\newcommand{\XsfIccOaSzpLsRecall}{1\xspace}
\newcommand{\XsfIccOaSzpLsPrecision}{1\xspace}
\newcommand{\XsfIccOaSzpLsFone}{1\xspace}
\newcommand{\XsfIccOaSzpBapRecall}{0.976\xspace}
\newcommand{\XsfIccOaSzpBapPrecision}{1\xspace}
\newcommand{\XsfIccOaSzpBapFone}{0.988\xspace}
\newcommand{\XsfIccOaSzpGhiRecall}{1\xspace}
\newcommand{\XsfIccOaSzpGhiPrecision}{1\xspace}
\newcommand{\XsfIccOaSzpGhiFone}{1\xspace}
\newcommand{\XsfIccOaSzpRdaRecall}{0.976\xspace}
\newcommand{\XsfIccOaSzpRdaPrecision}{1\xspace}
\newcommand{\XsfIccOaSzpRdaFone}{0.988\xspace}
\newcommand{\XsfIccOaSzpRseRecall}{1\xspace}
\newcommand{\XsfIccOaSzpRsePrecision}{1\xspace}
\newcommand{\XsfIccOaSzpRseFone}{1\xspace}
\newcommand{\XsfIccOaCapGT}{183704\xspace}
\newcommand{\XsfIccOaCapLsRecall}{1\xspace}
\newcommand{\XsfIccOaCapLsPrecision}{1\xspace}
\newcommand{\XsfIccOaCapLsFone}{1\xspace}
\newcommand{\XsfIccOaCapBapRecall}{0.275\xspace}
\newcommand{\XsfIccOaCapBapPrecision}{1\xspace}
\newcommand{\XsfIccOaCapBapFone}{0.431\xspace}
\newcommand{\XsfIccOaCapGhiRecall}{0.875\xspace}
\newcommand{\XsfIccOaCapGhiPrecision}{1\xspace}
\newcommand{\XsfIccOaCapGhiFone}{0.934\xspace}
\newcommand{\XsfIccOaCapRdaRecall}{0.000\xspace}
\newcommand{\XsfIccOaCapRdaPrecision}{NaN\xspace}
\newcommand{\XsfIccOaCapRdaFone}{0.000\xspace}
\newcommand{\XsfIccOaCapRseRecall}{0.666\xspace}
\newcommand{\XsfIccOaCapRsePrecision}{1\xspace}
\newcommand{\XsfIccOaCapRseFone}{0.799\xspace}
\newcommand{\XsfIccOaExmGT}{130139\xspace}
\newcommand{\XsfIccOaExmLsRecall}{1\xspace}
\newcommand{\XsfIccOaExmLsPrecision}{1\xspace}
\newcommand{\XsfIccOaExmLsFone}{1\xspace}
\newcommand{\XsfIccOaExmBapRecall}{0.830\xspace}
\newcommand{\XsfIccOaExmBapPrecision}{1.000\xspace}
\newcommand{\XsfIccOaExmBapFone}{0.907\xspace}
\newcommand{\XsfIccOaExmGhiRecall}{0.961\xspace}
\newcommand{\XsfIccOaExmGhiPrecision}{1\xspace}
\newcommand{\XsfIccOaExmGhiFone}{0.980\xspace}
\newcommand{\XsfIccOaExmRdaRecall}{0.804\xspace}
\newcommand{\XsfIccOaExmRdaPrecision}{1.000\xspace}
\newcommand{\XsfIccOaExmRdaFone}{0.891\xspace}
\newcommand{\XsfIccOaExmRseRecall}{0.975\xspace}
\newcommand{\XsfIccOaExmRsePrecision}{1\xspace}
\newcommand{\XsfIccOaExmRseFone}{0.987\xspace}
\newcommand{\XsfIccOaLgtGT}{24578\xspace}
\newcommand{\XsfIccOaLgtLsRecall}{1\xspace}
\newcommand{\XsfIccOaLgtLsPrecision}{1\xspace}
\newcommand{\XsfIccOaLgtLsFone}{1\xspace}
\newcommand{\XsfIccOaLgtBapRecall}{0.874\xspace}
\newcommand{\XsfIccOaLgtBapPrecision}{1\xspace}
\newcommand{\XsfIccOaLgtBapFone}{0.933\xspace}
\newcommand{\XsfIccOaLgtGhiRecall}{0.999\xspace}
\newcommand{\XsfIccOaLgtGhiPrecision}{1\xspace}
\newcommand{\XsfIccOaLgtGhiFone}{1.000\xspace}
\newcommand{\XsfIccOaLgtRdaRecall}{0.871\xspace}
\newcommand{\XsfIccOaLgtRdaPrecision}{1.000\xspace}
\newcommand{\XsfIccOaLgtRdaFone}{0.931\xspace}
\newcommand{\XsfIccOaLgtRseRecall}{0.993\xspace}
\newcommand{\XsfIccOaLgtRsePrecision}{1\xspace}
\newcommand{\XsfIccOaLgtRseFone}{0.997\xspace}
\newcommand{\XsfIccOaBzpGT}{11071\xspace}
\newcommand{\XsfIccOaBzpLsRecall}{1\xspace}
\newcommand{\XsfIccOaBzpLsPrecision}{1\xspace}
\newcommand{\XsfIccOaBzpLsFone}{1\xspace}
\newcommand{\XsfIccOaBzpBapRecall}{0.795\xspace}
\newcommand{\XsfIccOaBzpBapPrecision}{1\xspace}
\newcommand{\XsfIccOaBzpBapFone}{0.886\xspace}
\newcommand{\XsfIccOaBzpGhiRecall}{1\xspace}
\newcommand{\XsfIccOaBzpGhiPrecision}{1\xspace}
\newcommand{\XsfIccOaBzpGhiFone}{1\xspace}
\newcommand{\XsfIccOaBzpRdaRecall}{0.787\xspace}
\newcommand{\XsfIccOaBzpRdaPrecision}{1\xspace}
\newcommand{\XsfIccOaBzpRdaFone}{0.881\xspace}
\newcommand{\XsfIccOaBzpRseRecall}{0.920\xspace}
\newcommand{\XsfIccOaBzpRsePrecision}{1\xspace}
\newcommand{\XsfIccOaBzpRseFone}{0.958\xspace}
\newcommand{\XsfIccOaGccGT}{756313\xspace}
\newcommand{\XsfIccOaGccLsRecall}{1\xspace}
\newcommand{\XsfIccOaGccLsPrecision}{1\xspace}
\newcommand{\XsfIccOaGccLsFone}{1\xspace}
\newcommand{\XsfIccOaGccBapRecall}{0.757\xspace}
\newcommand{\XsfIccOaGccBapPrecision}{1.000\xspace}
\newcommand{\XsfIccOaGccBapFone}{0.861\xspace}
\newcommand{\XsfIccOaGccGhiRecall}{0.927\xspace}
\newcommand{\XsfIccOaGccGhiPrecision}{1\xspace}
\newcommand{\XsfIccOaGccGhiFone}{0.962\xspace}
\newcommand{\XsfIccOaGccRdaRecall}{0.728\xspace}
\newcommand{\XsfIccOaGccRdaPrecision}{1.000\xspace}
\newcommand{\XsfIccOaGccRdaFone}{0.843\xspace}
\newcommand{\XsfIccOaGccRseRecall}{0.961\xspace}
\newcommand{\XsfIccOaGccRsePrecision}{1\xspace}
\newcommand{\XsfIccOaGccRseFone}{0.980\xspace}
\newcommand{\XsfIccOaGzpGT}{8906\xspace}
\newcommand{\XsfIccOaGzpLsRecall}{1\xspace}
\newcommand{\XsfIccOaGzpLsPrecision}{1\xspace}
\newcommand{\XsfIccOaGzpLsFone}{1\xspace}
\newcommand{\XsfIccOaGzpBapRecall}{0.986\xspace}
\newcommand{\XsfIccOaGzpBapPrecision}{1\xspace}
\newcommand{\XsfIccOaGzpBapFone}{0.993\xspace}
\newcommand{\XsfIccOaGzpGhiRecall}{1\xspace}
\newcommand{\XsfIccOaGzpGhiPrecision}{1\xspace}
\newcommand{\XsfIccOaGzpGhiFone}{1\xspace}
\newcommand{\XsfIccOaGzpRdaRecall}{0.986\xspace}
\newcommand{\XsfIccOaGzpRdaPrecision}{1\xspace}
\newcommand{\XsfIccOaGzpRdaFone}{0.993\xspace}
\newcommand{\XsfIccOaGzpRseRecall}{0.997\xspace}
\newcommand{\XsfIccOaGzpRsePrecision}{1\xspace}
\newcommand{\XsfIccOaGzpRseFone}{0.999\xspace}
\newcommand{\XsfIccOaOggGT}{32341\xspace}
\newcommand{\XsfIccOaOggLsRecall}{1\xspace}
\newcommand{\XsfIccOaOggLsPrecision}{1\xspace}
\newcommand{\XsfIccOaOggLsFone}{1\xspace}
\newcommand{\XsfIccOaOggBapRecall}{0.970\xspace}
\newcommand{\XsfIccOaOggBapPrecision}{1\xspace}
\newcommand{\XsfIccOaOggBapFone}{0.985\xspace}
\newcommand{\XsfIccOaOggGhiRecall}{1.000\xspace}
\newcommand{\XsfIccOaOggGhiPrecision}{1\xspace}
\newcommand{\XsfIccOaOggGhiFone}{1.000\xspace}
\newcommand{\XsfIccOaOggRdaRecall}{0.970\xspace}
\newcommand{\XsfIccOaOggRdaPrecision}{1\xspace}
\newcommand{\XsfIccOaOggRdaFone}{0.985\xspace}
\newcommand{\XsfIccOaOggRseRecall}{0.981\xspace}
\newcommand{\XsfIccOaOggRsePrecision}{1\xspace}
\newcommand{\XsfIccOaOggRseFone}{0.990\xspace}
\newcommand{\XsfIccOaNgxGT}{95949\xspace}
\newcommand{\XsfIccOaNgxLsRecall}{1\xspace}
\newcommand{\XsfIccOaNgxLsPrecision}{1\xspace}
\newcommand{\XsfIccOaNgxLsFone}{1\xspace}
\newcommand{\XsfIccOaNgxBapRecall}{0.967\xspace}
\newcommand{\XsfIccOaNgxBapPrecision}{1\xspace}
\newcommand{\XsfIccOaNgxBapFone}{0.983\xspace}
\newcommand{\XsfIccOaNgxGhiRecall}{0.995\xspace}
\newcommand{\XsfIccOaNgxGhiPrecision}{1\xspace}
\newcommand{\XsfIccOaNgxGhiFone}{0.998\xspace}
\newcommand{\XsfIccOaNgxRdaRecall}{0.965\xspace}
\newcommand{\XsfIccOaNgxRdaPrecision}{1.000\xspace}
\newcommand{\XsfIccOaNgxRdaFone}{0.982\xspace}
\newcommand{\XsfIccOaNgxRseRecall}{0.993\xspace}
\newcommand{\XsfIccOaNgxRsePrecision}{1.000\xspace}
\newcommand{\XsfIccOaNgxRseFone}{0.997\xspace}
\newcommand{\XsfIccOaSshGT}{95598\xspace}
\newcommand{\XsfIccOaSshLsRecall}{1\xspace}
\newcommand{\XsfIccOaSshLsPrecision}{1\xspace}
\newcommand{\XsfIccOaSshLsFone}{1\xspace}
\newcommand{\XsfIccOaSshBapRecall}{0.939\xspace}
\newcommand{\XsfIccOaSshBapPrecision}{1\xspace}
\newcommand{\XsfIccOaSshBapFone}{0.968\xspace}
\newcommand{\XsfIccOaSshGhiRecall}{0.980\xspace}
\newcommand{\XsfIccOaSshGhiPrecision}{1\xspace}
\newcommand{\XsfIccOaSshGhiFone}{0.990\xspace}
\newcommand{\XsfIccOaSshRdaRecall}{0.926\xspace}
\newcommand{\XsfIccOaSshRdaPrecision}{0.999\xspace}
\newcommand{\XsfIccOaSshRdaFone}{0.961\xspace}
\newcommand{\XsfIccOaSshRseRecall}{0.973\xspace}
\newcommand{\XsfIccOaSshRsePrecision}{1\xspace}
\newcommand{\XsfIccOaSshRseFone}{0.986\xspace}
\newcommand{\XsfIccOaPcrGT}{4070\xspace}
\newcommand{\XsfIccOaPcrLsRecall}{1\xspace}
\newcommand{\XsfIccOaPcrLsPrecision}{1\xspace}
\newcommand{\XsfIccOaPcrLsFone}{1\xspace}
\newcommand{\XsfIccOaPcrBapRecall}{0.912\xspace}
\newcommand{\XsfIccOaPcrBapPrecision}{1\xspace}
\newcommand{\XsfIccOaPcrBapFone}{0.954\xspace}
\newcommand{\XsfIccOaPcrGhiRecall}{1\xspace}
\newcommand{\XsfIccOaPcrGhiPrecision}{1\xspace}
\newcommand{\XsfIccOaPcrGhiFone}{1\xspace}
\newcommand{\XsfIccOaPcrRdaRecall}{0.851\xspace}
\newcommand{\XsfIccOaPcrRdaPrecision}{1\xspace}
\newcommand{\XsfIccOaPcrRdaFone}{0.919\xspace}
\newcommand{\XsfIccOaPcrRseRecall}{0.995\xspace}
\newcommand{\XsfIccOaPcrRsePrecision}{1\xspace}
\newcommand{\XsfIccOaPcrRseFone}{0.998\xspace}
\newcommand{\XsfIccOaSqlGT}{146124\xspace}
\newcommand{\XsfIccOaSqlLsRecall}{1\xspace}
\newcommand{\XsfIccOaSqlLsPrecision}{1\xspace}
\newcommand{\XsfIccOaSqlLsFone}{1\xspace}
\newcommand{\XsfIccOaSqlBapRecall}{0.852\xspace}
\newcommand{\XsfIccOaSqlBapPrecision}{1\xspace}
\newcommand{\XsfIccOaSqlBapFone}{0.920\xspace}
\newcommand{\XsfIccOaSqlGhiRecall}{0.951\xspace}
\newcommand{\XsfIccOaSqlGhiPrecision}{1\xspace}
\newcommand{\XsfIccOaSqlGhiFone}{0.975\xspace}
\newcommand{\XsfIccOaSqlRdaRecall}{0.852\xspace}
\newcommand{\XsfIccOaSqlRdaPrecision}{1.000\xspace}
\newcommand{\XsfIccOaSqlRdaFone}{0.920\xspace}
\newcommand{\XsfIccOaSqlRseRecall}{0.929\xspace}
\newcommand{\XsfIccOaSqlRsePrecision}{1\xspace}
\newcommand{\XsfIccOaSqlRseFone}{0.963\xspace}
\newcommand{\XsfIccOaVimGT}{441739\xspace}
\newcommand{\XsfIccOaVimLsRecall}{1\xspace}
\newcommand{\XsfIccOaVimLsPrecision}{1\xspace}
\newcommand{\XsfIccOaVimLsFone}{1\xspace}
\newcommand{\XsfIccOaVimBapRecall}{0.911\xspace}
\newcommand{\XsfIccOaVimBapPrecision}{1.000\xspace}
\newcommand{\XsfIccOaVimBapFone}{0.954\xspace}
\newcommand{\XsfIccOaVimGhiRecall}{0.998\xspace}
\newcommand{\XsfIccOaVimGhiPrecision}{1\xspace}
\newcommand{\XsfIccOaVimGhiFone}{0.999\xspace}
\newcommand{\XsfIccOaVimRdaRecall}{0.901\xspace}
\newcommand{\XsfIccOaVimRdaPrecision}{1.000\xspace}
\newcommand{\XsfIccOaVimRdaFone}{0.948\xspace}
\newcommand{\XsfIccOaVimRseRecall}{0.982\xspace}
\newcommand{\XsfIccOaVimRsePrecision}{1.000\xspace}
\newcommand{\XsfIccOaVimRseFone}{0.991\xspace}
\newcommand{\XsfIccOaVsfGT}{17391\xspace}
\newcommand{\XsfIccOaVsfLsRecall}{1\xspace}
\newcommand{\XsfIccOaVsfLsPrecision}{1\xspace}
\newcommand{\XsfIccOaVsfLsFone}{1\xspace}
\newcommand{\XsfIccOaVsfBapRecall}{0.988\xspace}
\newcommand{\XsfIccOaVsfBapPrecision}{1\xspace}
\newcommand{\XsfIccOaVsfBapFone}{0.994\xspace}
\newcommand{\XsfIccOaVsfGhiRecall}{1\xspace}
\newcommand{\XsfIccOaVsfGhiPrecision}{1\xspace}
\newcommand{\XsfIccOaVsfGhiFone}{1\xspace}
\newcommand{\XsfIccOaVsfRdaRecall}{0.980\xspace}
\newcommand{\XsfIccOaVsfRdaPrecision}{0.998\xspace}
\newcommand{\XsfIccOaVsfRdaFone}{0.989\xspace}
\newcommand{\XsfIccOaVsfRseRecall}{1.000\xspace}
\newcommand{\XsfIccOaVsfRsePrecision}{1\xspace}
\newcommand{\XsfIccOaVsfRseFone}{1.000\xspace}
\newcommand{\XsfIccObSzpGT}{25595\xspace}
\newcommand{\XsfIccObSzpLsRecall}{1\xspace}
\newcommand{\XsfIccObSzpLsPrecision}{0.994\xspace}
\newcommand{\XsfIccObSzpLsFone}{0.997\xspace}
\newcommand{\XsfIccObSzpBapRecall}{0.974\xspace}
\newcommand{\XsfIccObSzpBapPrecision}{1\xspace}
\newcommand{\XsfIccObSzpBapFone}{0.987\xspace}
\newcommand{\XsfIccObSzpGhiRecall}{1\xspace}
\newcommand{\XsfIccObSzpGhiPrecision}{1\xspace}
\newcommand{\XsfIccObSzpGhiFone}{1\xspace}
\newcommand{\XsfIccObSzpRdaRecall}{1\xspace}
\newcommand{\XsfIccObSzpRdaPrecision}{1\xspace}
\newcommand{\XsfIccObSzpRdaFone}{1\xspace}
\newcommand{\XsfIccObSzpRseRecall}{1\xspace}
\newcommand{\XsfIccObSzpRsePrecision}{1\xspace}
\newcommand{\XsfIccObSzpRseFone}{1\xspace}
\newcommand{\XsfIccObCapGT}{266475\xspace}
\newcommand{\XsfIccObCapLsRecall}{1\xspace}
\newcommand{\XsfIccObCapLsPrecision}{0.995\xspace}
\newcommand{\XsfIccObCapLsFone}{0.997\xspace}
\newcommand{\XsfIccObCapBapRecall}{0.299\xspace}
\newcommand{\XsfIccObCapBapPrecision}{1\xspace}
\newcommand{\XsfIccObCapBapFone}{0.460\xspace}
\newcommand{\XsfIccObCapGhiRecall}{1.000\xspace}
\newcommand{\XsfIccObCapGhiPrecision}{1\xspace}
\newcommand{\XsfIccObCapGhiFone}{1.000\xspace}
\newcommand{\XsfIccObCapRdaRecall}{0.909\xspace}
\newcommand{\XsfIccObCapRdaPrecision}{1.000\xspace}
\newcommand{\XsfIccObCapRdaFone}{0.952\xspace}
\newcommand{\XsfIccObCapRseRecall}{0.998\xspace}
\newcommand{\XsfIccObCapRsePrecision}{1\xspace}
\newcommand{\XsfIccObCapRseFone}{0.999\xspace}
\newcommand{\XsfIccObExmGT}{222763\xspace}
\newcommand{\XsfIccObExmLsRecall}{1\xspace}
\newcommand{\XsfIccObExmLsPrecision}{0.996\xspace}
\newcommand{\XsfIccObExmLsFone}{0.998\xspace}
\newcommand{\XsfIccObExmBapRecall}{0.844\xspace}
\newcommand{\XsfIccObExmBapPrecision}{1.000\xspace}
\newcommand{\XsfIccObExmBapFone}{0.916\xspace}
\newcommand{\XsfIccObExmGhiRecall}{0.947\xspace}
\newcommand{\XsfIccObExmGhiPrecision}{1.000\xspace}
\newcommand{\XsfIccObExmGhiFone}{0.973\xspace}
\newcommand{\XsfIccObExmRdaRecall}{0.949\xspace}
\newcommand{\XsfIccObExmRdaPrecision}{1.000\xspace}
\newcommand{\XsfIccObExmRdaFone}{0.974\xspace}
\newcommand{\XsfIccObExmRseRecall}{0.979\xspace}
\newcommand{\XsfIccObExmRsePrecision}{1\xspace}
\newcommand{\XsfIccObExmRseFone}{0.989\xspace}
\newcommand{\XsfIccObLgtGT}{36775\xspace}
\newcommand{\XsfIccObLgtLsRecall}{1\xspace}
\newcommand{\XsfIccObLgtLsPrecision}{0.989\xspace}
\newcommand{\XsfIccObLgtLsFone}{0.994\xspace}
\newcommand{\XsfIccObLgtBapRecall}{0.870\xspace}
\newcommand{\XsfIccObLgtBapPrecision}{0.998\xspace}
\newcommand{\XsfIccObLgtBapFone}{0.929\xspace}
\newcommand{\XsfIccObLgtGhiRecall}{0.998\xspace}
\newcommand{\XsfIccObLgtGhiPrecision}{0.998\xspace}
\newcommand{\XsfIccObLgtGhiFone}{0.998\xspace}
\newcommand{\XsfIccObLgtRdaRecall}{0.934\xspace}
\newcommand{\XsfIccObLgtRdaPrecision}{1.000\xspace}
\newcommand{\XsfIccObLgtRdaFone}{0.966\xspace}
\newcommand{\XsfIccObLgtRseRecall}{0.991\xspace}
\newcommand{\XsfIccObLgtRsePrecision}{1\xspace}
\newcommand{\XsfIccObLgtRseFone}{0.995\xspace}
\newcommand{\XsfIccObBzpGT}{24076\xspace}
\newcommand{\XsfIccObBzpLsRecall}{1\xspace}
\newcommand{\XsfIccObBzpLsPrecision}{0.997\xspace}
\newcommand{\XsfIccObBzpLsFone}{0.998\xspace}
\newcommand{\XsfIccObBzpBapRecall}{0.829\xspace}
\newcommand{\XsfIccObBzpBapPrecision}{0.999\xspace}
\newcommand{\XsfIccObBzpBapFone}{0.906\xspace}
\newcommand{\XsfIccObBzpGhiRecall}{1\xspace}
\newcommand{\XsfIccObBzpGhiPrecision}{1.000\xspace}
\newcommand{\XsfIccObBzpGhiFone}{1.000\xspace}
\newcommand{\XsfIccObBzpRdaRecall}{0.999\xspace}
\newcommand{\XsfIccObBzpRdaPrecision}{1\xspace}
\newcommand{\XsfIccObBzpRdaFone}{1.000\xspace}
\newcommand{\XsfIccObBzpRseRecall}{1\xspace}
\newcommand{\XsfIccObBzpRsePrecision}{1\xspace}
\newcommand{\XsfIccObBzpRseFone}{1\xspace}
\newcommand{\XsfIccObGccGT}{3304045\xspace}
\newcommand{\XsfIccObGccLsRecall}{1\xspace}
\newcommand{\XsfIccObGccLsPrecision}{0.997\xspace}
\newcommand{\XsfIccObGccLsFone}{0.999\xspace}
\newcommand{\XsfIccObGccBapRecall}{0.882\xspace}
\newcommand{\XsfIccObGccBapPrecision}{1.000\xspace}
\newcommand{\XsfIccObGccBapFone}{0.937\xspace}
\newcommand{\XsfIccObGccGhiRecall}{0.996\xspace}
\newcommand{\XsfIccObGccGhiPrecision}{1.000\xspace}
\newcommand{\XsfIccObGccGhiFone}{0.998\xspace}
\newcommand{\XsfIccObGccRdaRecall}{0.937\xspace}
\newcommand{\XsfIccObGccRdaPrecision}{1.000\xspace}
\newcommand{\XsfIccObGccRdaFone}{0.968\xspace}
\newcommand{\XsfIccObGccRseRecall}{0.995\xspace}
\newcommand{\XsfIccObGccRsePrecision}{1.000\xspace}
\newcommand{\XsfIccObGccRseFone}{0.998\xspace}
\newcommand{\XsfIccObGzpGT}{25474\xspace}
\newcommand{\XsfIccObGzpLsRecall}{1\xspace}
\newcommand{\XsfIccObGzpLsPrecision}{0.997\xspace}
\newcommand{\XsfIccObGzpLsFone}{0.998\xspace}
\newcommand{\XsfIccObGzpBapRecall}{0.993\xspace}
\newcommand{\XsfIccObGzpBapPrecision}{1.000\xspace}
\newcommand{\XsfIccObGzpBapFone}{0.996\xspace}
\newcommand{\XsfIccObGzpGhiRecall}{1\xspace}
\newcommand{\XsfIccObGzpGhiPrecision}{1.000\xspace}
\newcommand{\XsfIccObGzpGhiFone}{1.000\xspace}
\newcommand{\XsfIccObGzpRdaRecall}{0.991\xspace}
\newcommand{\XsfIccObGzpRdaPrecision}{1\xspace}
\newcommand{\XsfIccObGzpRdaFone}{0.995\xspace}
\newcommand{\XsfIccObGzpRseRecall}{1\xspace}
\newcommand{\XsfIccObGzpRsePrecision}{1\xspace}
\newcommand{\XsfIccObGzpRseFone}{1\xspace}
\newcommand{\XsfIccObOggGT}{101093\xspace}
\newcommand{\XsfIccObOggLsRecall}{1\xspace}
\newcommand{\XsfIccObOggLsPrecision}{0.996\xspace}
\newcommand{\XsfIccObOggLsFone}{0.998\xspace}
\newcommand{\XsfIccObOggBapRecall}{0.978\xspace}
\newcommand{\XsfIccObOggBapPrecision}{1.000\xspace}
\newcommand{\XsfIccObOggBapFone}{0.989\xspace}
\newcommand{\XsfIccObOggGhiRecall}{0.999\xspace}
\newcommand{\XsfIccObOggGhiPrecision}{1.000\xspace}
\newcommand{\XsfIccObOggGhiFone}{0.999\xspace}
\newcommand{\XsfIccObOggRdaRecall}{0.992\xspace}
\newcommand{\XsfIccObOggRdaPrecision}{1.000\xspace}
\newcommand{\XsfIccObOggRdaFone}{0.996\xspace}
\newcommand{\XsfIccObOggRseRecall}{0.992\xspace}
\newcommand{\XsfIccObOggRsePrecision}{1\xspace}
\newcommand{\XsfIccObOggRseFone}{0.996\xspace}
\newcommand{\XsfIccObNgxGT}{142377\xspace}
\newcommand{\XsfIccObNgxLsRecall}{1\xspace}
\newcommand{\XsfIccObNgxLsPrecision}{0.991\xspace}
\newcommand{\XsfIccObNgxLsFone}{0.995\xspace}
\newcommand{\XsfIccObNgxBapRecall}{0.953\xspace}
\newcommand{\XsfIccObNgxBapPrecision}{1.000\xspace}
\newcommand{\XsfIccObNgxBapFone}{0.976\xspace}
\newcommand{\XsfIccObNgxGhiRecall}{0.998\xspace}
\newcommand{\XsfIccObNgxGhiPrecision}{1.000\xspace}
\newcommand{\XsfIccObNgxGhiFone}{0.999\xspace}
\newcommand{\XsfIccObNgxRdaRecall}{0.000\xspace}
\newcommand{\XsfIccObNgxRdaPrecision}{NaN\xspace}
\newcommand{\XsfIccObNgxRdaFone}{0.000\xspace}
\newcommand{\XsfIccObNgxRseRecall}{0.978\xspace}
\newcommand{\XsfIccObNgxRsePrecision}{1\xspace}
\newcommand{\XsfIccObNgxRseFone}{0.989\xspace}
\newcommand{\XsfIccObSshGT}{193686\xspace}
\newcommand{\XsfIccObSshLsRecall}{1\xspace}
\newcommand{\XsfIccObSshLsPrecision}{0.992\xspace}
\newcommand{\XsfIccObSshLsFone}{0.996\xspace}
\newcommand{\XsfIccObSshBapRecall}{0.930\xspace}
\newcommand{\XsfIccObSshBapPrecision}{0.998\xspace}
\newcommand{\XsfIccObSshBapFone}{0.963\xspace}
\newcommand{\XsfIccObSshGhiRecall}{0.998\xspace}
\newcommand{\XsfIccObSshGhiPrecision}{0.998\xspace}
\newcommand{\XsfIccObSshGhiFone}{0.998\xspace}
\newcommand{\XsfIccObSshRdaRecall}{0.899\xspace}
\newcommand{\XsfIccObSshRdaPrecision}{0.998\xspace}
\newcommand{\XsfIccObSshRdaFone}{0.946\xspace}
\newcommand{\XsfIccObSshRseRecall}{0.980\xspace}
\newcommand{\XsfIccObSshRsePrecision}{0.998\xspace}
\newcommand{\XsfIccObSshRseFone}{0.989\xspace}
\newcommand{\XsfIccObPcrGT}{5677\xspace}
\newcommand{\XsfIccObPcrLsRecall}{1\xspace}
\newcommand{\XsfIccObPcrLsPrecision}{0.995\xspace}
\newcommand{\XsfIccObPcrLsFone}{0.998\xspace}
\newcommand{\XsfIccObPcrBapRecall}{0.944\xspace}
\newcommand{\XsfIccObPcrBapPrecision}{0.999\xspace}
\newcommand{\XsfIccObPcrBapFone}{0.971\xspace}
\newcommand{\XsfIccObPcrGhiRecall}{1\xspace}
\newcommand{\XsfIccObPcrGhiPrecision}{0.999\xspace}
\newcommand{\XsfIccObPcrGhiFone}{1.000\xspace}
\newcommand{\XsfIccObPcrRdaRecall}{1\xspace}
\newcommand{\XsfIccObPcrRdaPrecision}{1\xspace}
\newcommand{\XsfIccObPcrRdaFone}{1\xspace}
\newcommand{\XsfIccObPcrRseRecall}{1\xspace}
\newcommand{\XsfIccObPcrRsePrecision}{1\xspace}
\newcommand{\XsfIccObPcrRseFone}{1\xspace}
\newcommand{\XsfIccObSqlGT}{382649\xspace}
\newcommand{\XsfIccObSqlLsRecall}{1\xspace}
\newcommand{\XsfIccObSqlLsPrecision}{0.995\xspace}
\newcommand{\XsfIccObSqlLsFone}{0.998\xspace}
\newcommand{\XsfIccObSqlBapRecall}{0.898\xspace}
\newcommand{\XsfIccObSqlBapPrecision}{1.000\xspace}
\newcommand{\XsfIccObSqlBapFone}{0.946\xspace}
\newcommand{\XsfIccObSqlGhiRecall}{0.971\xspace}
\newcommand{\XsfIccObSqlGhiPrecision}{1.000\xspace}
\newcommand{\XsfIccObSqlGhiFone}{0.985\xspace}
\newcommand{\XsfIccObSqlRdaRecall}{0.990\xspace}
\newcommand{\XsfIccObSqlRdaPrecision}{1.000\xspace}
\newcommand{\XsfIccObSqlRdaFone}{0.995\xspace}
\newcommand{\XsfIccObSqlRseRecall}{0.991\xspace}
\newcommand{\XsfIccObSqlRsePrecision}{1\xspace}
\newcommand{\XsfIccObSqlRseFone}{0.996\xspace}
\newcommand{\XsfIccObVimGT}{885468\xspace}
\newcommand{\XsfIccObVimLsRecall}{1\xspace}
\newcommand{\XsfIccObVimLsPrecision}{0.993\xspace}
\newcommand{\XsfIccObVimLsFone}{0.997\xspace}
\newcommand{\XsfIccObVimBapRecall}{0.909\xspace}
\newcommand{\XsfIccObVimBapPrecision}{1.000\xspace}
\newcommand{\XsfIccObVimBapFone}{0.953\xspace}
\newcommand{\XsfIccObVimGhiRecall}{0.997\xspace}
\newcommand{\XsfIccObVimGhiPrecision}{1.000\xspace}
\newcommand{\XsfIccObVimGhiFone}{0.999\xspace}
\newcommand{\XsfIccObVimRdaRecall}{0.976\xspace}
\newcommand{\XsfIccObVimRdaPrecision}{1.000\xspace}
\newcommand{\XsfIccObVimRdaFone}{0.988\xspace}
\newcommand{\XsfIccObVimRseRecall}{0.998\xspace}
\newcommand{\XsfIccObVimRsePrecision}{1\xspace}
\newcommand{\XsfIccObVimRseFone}{0.999\xspace}
\newcommand{\XsfIccObVsfGT}{25766\xspace}
\newcommand{\XsfIccObVsfLsRecall}{1\xspace}
\newcommand{\XsfIccObVsfLsPrecision}{0.972\xspace}
\newcommand{\XsfIccObVsfLsFone}{0.986\xspace}
\newcommand{\XsfIccObVsfBapRecall}{0.988\xspace}
\newcommand{\XsfIccObVsfBapPrecision}{1.000\xspace}
\newcommand{\XsfIccObVsfBapFone}{0.994\xspace}
\newcommand{\XsfIccObVsfGhiRecall}{1\xspace}
\newcommand{\XsfIccObVsfGhiPrecision}{1.000\xspace}
\newcommand{\XsfIccObVsfGhiFone}{1.000\xspace}
\newcommand{\XsfIccObVsfRdaRecall}{0.980\xspace}
\newcommand{\XsfIccObVsfRdaPrecision}{0.998\xspace}
\newcommand{\XsfIccObVsfRdaFone}{0.989\xspace}
\newcommand{\XsfIccObVsfRseRecall}{1.000\xspace}
\newcommand{\XsfIccObVsfRsePrecision}{1.000\xspace}
\newcommand{\XsfIccObVsfRseFone}{1.000\xspace}
\newcommand{\XsfIccOcSzpGT}{26700\xspace}
\newcommand{\XsfIccOcSzpLsRecall}{1\xspace}
\newcommand{\XsfIccOcSzpLsPrecision}{0.994\xspace}
\newcommand{\XsfIccOcSzpLsFone}{0.997\xspace}
\newcommand{\XsfIccOcSzpBapRecall}{0.972\xspace}
\newcommand{\XsfIccOcSzpBapPrecision}{1\xspace}
\newcommand{\XsfIccOcSzpBapFone}{0.986\xspace}
\newcommand{\XsfIccOcSzpGhiRecall}{1\xspace}
\newcommand{\XsfIccOcSzpGhiPrecision}{1\xspace}
\newcommand{\XsfIccOcSzpGhiFone}{1\xspace}
\newcommand{\XsfIccOcSzpRdaRecall}{1\xspace}
\newcommand{\XsfIccOcSzpRdaPrecision}{1\xspace}
\newcommand{\XsfIccOcSzpRdaFone}{1\xspace}
\newcommand{\XsfIccOcSzpRseRecall}{1\xspace}
\newcommand{\XsfIccOcSzpRsePrecision}{1\xspace}
\newcommand{\XsfIccOcSzpRseFone}{1\xspace}
\newcommand{\XsfIccOcCapGT}{274363\xspace}
\newcommand{\XsfIccOcCapLsRecall}{1\xspace}
\newcommand{\XsfIccOcCapLsPrecision}{0.995\xspace}
\newcommand{\XsfIccOcCapLsFone}{0.997\xspace}
\newcommand{\XsfIccOcCapBapRecall}{0.306\xspace}
\newcommand{\XsfIccOcCapBapPrecision}{1\xspace}
\newcommand{\XsfIccOcCapBapFone}{0.468\xspace}
\newcommand{\XsfIccOcCapGhiRecall}{0.979\xspace}
\newcommand{\XsfIccOcCapGhiPrecision}{1\xspace}
\newcommand{\XsfIccOcCapGhiFone}{0.989\xspace}
\newcommand{\XsfIccOcCapRdaRecall}{0.000\xspace}
\newcommand{\XsfIccOcCapRdaPrecision}{NaN\xspace}
\newcommand{\XsfIccOcCapRdaFone}{0.000\xspace}
\newcommand{\XsfIccOcCapRseRecall}{0.998\xspace}
\newcommand{\XsfIccOcCapRsePrecision}{1\xspace}
\newcommand{\XsfIccOcCapRseFone}{0.999\xspace}
\newcommand{\XsfIccOcExmGT}{236759\xspace}
\newcommand{\XsfIccOcExmLsRecall}{1\xspace}
\newcommand{\XsfIccOcExmLsPrecision}{0.996\xspace}
\newcommand{\XsfIccOcExmLsFone}{0.998\xspace}
\newcommand{\XsfIccOcExmBapRecall}{0.846\xspace}
\newcommand{\XsfIccOcExmBapPrecision}{1.000\xspace}
\newcommand{\XsfIccOcExmBapFone}{0.916\xspace}
\newcommand{\XsfIccOcExmGhiRecall}{0.950\xspace}
\newcommand{\XsfIccOcExmGhiPrecision}{1.000\xspace}
\newcommand{\XsfIccOcExmGhiFone}{0.974\xspace}
\newcommand{\XsfIccOcExmRdaRecall}{0.948\xspace}
\newcommand{\XsfIccOcExmRdaPrecision}{1.000\xspace}
\newcommand{\XsfIccOcExmRdaFone}{0.973\xspace}
\newcommand{\XsfIccOcExmRseRecall}{0.977\xspace}
\newcommand{\XsfIccOcExmRsePrecision}{1\xspace}
\newcommand{\XsfIccOcExmRseFone}{0.989\xspace}
\newcommand{\XsfIccOcLgtGT}{38359\xspace}
\newcommand{\XsfIccOcLgtLsRecall}{1\xspace}
\newcommand{\XsfIccOcLgtLsPrecision}{0.990\xspace}
\newcommand{\XsfIccOcLgtLsFone}{0.995\xspace}
\newcommand{\XsfIccOcLgtBapRecall}{0.875\xspace}
\newcommand{\XsfIccOcLgtBapPrecision}{0.998\xspace}
\newcommand{\XsfIccOcLgtBapFone}{0.932\xspace}
\newcommand{\XsfIccOcLgtGhiRecall}{0.998\xspace}
\newcommand{\XsfIccOcLgtGhiPrecision}{0.998\xspace}
\newcommand{\XsfIccOcLgtGhiFone}{0.998\xspace}
\newcommand{\XsfIccOcLgtRdaRecall}{0.937\xspace}
\newcommand{\XsfIccOcLgtRdaPrecision}{1.000\xspace}
\newcommand{\XsfIccOcLgtRdaFone}{0.968\xspace}
\newcommand{\XsfIccOcLgtRseRecall}{0.991\xspace}
\newcommand{\XsfIccOcLgtRsePrecision}{1\xspace}
\newcommand{\XsfIccOcLgtRseFone}{0.996\xspace}
\newcommand{\XsfIccOcBzpGT}{27087\xspace}
\newcommand{\XsfIccOcBzpLsRecall}{1\xspace}
\newcommand{\XsfIccOcBzpLsPrecision}{0.997\xspace}
\newcommand{\XsfIccOcBzpLsFone}{0.999\xspace}
\newcommand{\XsfIccOcBzpBapRecall}{0.830\xspace}
\newcommand{\XsfIccOcBzpBapPrecision}{1.000\xspace}
\newcommand{\XsfIccOcBzpBapFone}{0.907\xspace}
\newcommand{\XsfIccOcBzpGhiRecall}{1\xspace}
\newcommand{\XsfIccOcBzpGhiPrecision}{1.000\xspace}
\newcommand{\XsfIccOcBzpGhiFone}{1.000\xspace}
\newcommand{\XsfIccOcBzpRdaRecall}{0.978\xspace}
\newcommand{\XsfIccOcBzpRdaPrecision}{1\xspace}
\newcommand{\XsfIccOcBzpRdaFone}{0.989\xspace}
\newcommand{\XsfIccOcBzpRseRecall}{1\xspace}
\newcommand{\XsfIccOcBzpRsePrecision}{1\xspace}
\newcommand{\XsfIccOcBzpRseFone}{1\xspace}
\newcommand{\XsfIccOcGccGT}{3682553\xspace}
\newcommand{\XsfIccOcGccLsRecall}{1\xspace}
\newcommand{\XsfIccOcGccLsPrecision}{0.997\xspace}
\newcommand{\XsfIccOcGccLsFone}{0.999\xspace}
\newcommand{\XsfIccOcGccBapRecall}{0.885\xspace}
\newcommand{\XsfIccOcGccBapPrecision}{1.000\xspace}
\newcommand{\XsfIccOcGccBapFone}{0.939\xspace}
\newcommand{\XsfIccOcGccGhiRecall}{0.994\xspace}
\newcommand{\XsfIccOcGccGhiPrecision}{1.000\xspace}
\newcommand{\XsfIccOcGccGhiFone}{0.997\xspace}
\newcommand{\XsfIccOcGccRdaRecall}{0.938\xspace}
\newcommand{\XsfIccOcGccRdaPrecision}{1.000\xspace}
\newcommand{\XsfIccOcGccRdaFone}{0.968\xspace}
\newcommand{\XsfIccOcGccRseRecall}{0.995\xspace}
\newcommand{\XsfIccOcGccRsePrecision}{1.000\xspace}
\newcommand{\XsfIccOcGccRseFone}{0.998\xspace}
\newcommand{\XsfIccOcGzpGT}{28842\xspace}
\newcommand{\XsfIccOcGzpLsRecall}{1\xspace}
\newcommand{\XsfIccOcGzpLsPrecision}{0.997\xspace}
\newcommand{\XsfIccOcGzpLsFone}{0.998\xspace}
\newcommand{\XsfIccOcGzpBapRecall}{0.994\xspace}
\newcommand{\XsfIccOcGzpBapPrecision}{1.000\xspace}
\newcommand{\XsfIccOcGzpBapFone}{0.997\xspace}
\newcommand{\XsfIccOcGzpGhiRecall}{1\xspace}
\newcommand{\XsfIccOcGzpGhiPrecision}{1.000\xspace}
\newcommand{\XsfIccOcGzpGhiFone}{1.000\xspace}
\newcommand{\XsfIccOcGzpRdaRecall}{0.992\xspace}
\newcommand{\XsfIccOcGzpRdaPrecision}{1\xspace}
\newcommand{\XsfIccOcGzpRdaFone}{0.996\xspace}
\newcommand{\XsfIccOcGzpRseRecall}{1\xspace}
\newcommand{\XsfIccOcGzpRsePrecision}{1\xspace}
\newcommand{\XsfIccOcGzpRseFone}{1\xspace}
\newcommand{\XsfIccOcOggGT}{111360\xspace}
\newcommand{\XsfIccOcOggLsRecall}{1\xspace}
\newcommand{\XsfIccOcOggLsPrecision}{0.996\xspace}
\newcommand{\XsfIccOcOggLsFone}{0.998\xspace}
\newcommand{\XsfIccOcOggBapRecall}{0.980\xspace}
\newcommand{\XsfIccOcOggBapPrecision}{1.000\xspace}
\newcommand{\XsfIccOcOggBapFone}{0.990\xspace}
\newcommand{\XsfIccOcOggGhiRecall}{0.999\xspace}
\newcommand{\XsfIccOcOggGhiPrecision}{1.000\xspace}
\newcommand{\XsfIccOcOggGhiFone}{1.000\xspace}
\newcommand{\XsfIccOcOggRdaRecall}{0.992\xspace}
\newcommand{\XsfIccOcOggRdaPrecision}{1.000\xspace}
\newcommand{\XsfIccOcOggRdaFone}{0.996\xspace}
\newcommand{\XsfIccOcOggRseRecall}{0.992\xspace}
\newcommand{\XsfIccOcOggRsePrecision}{1\xspace}
\newcommand{\XsfIccOcOggRseFone}{0.996\xspace}
\newcommand{\XsfIccOcNgxGT}{154673\xspace}
\newcommand{\XsfIccOcNgxLsRecall}{1\xspace}
\newcommand{\XsfIccOcNgxLsPrecision}{0.992\xspace}
\newcommand{\XsfIccOcNgxLsFone}{0.996\xspace}
\newcommand{\XsfIccOcNgxBapRecall}{0.938\xspace}
\newcommand{\XsfIccOcNgxBapPrecision}{1.000\xspace}
\newcommand{\XsfIccOcNgxBapFone}{0.968\xspace}
\newcommand{\XsfIccOcNgxGhiRecall}{0.998\xspace}
\newcommand{\XsfIccOcNgxGhiPrecision}{1.000\xspace}
\newcommand{\XsfIccOcNgxGhiFone}{0.999\xspace}
\newcommand{\XsfIccOcNgxRdaRecall}{0.000\xspace}
\newcommand{\XsfIccOcNgxRdaPrecision}{NaN\xspace}
\newcommand{\XsfIccOcNgxRdaFone}{0.000\xspace}
\newcommand{\XsfIccOcNgxRseRecall}{0.966\xspace}
\newcommand{\XsfIccOcNgxRsePrecision}{1\xspace}
\newcommand{\XsfIccOcNgxRseFone}{0.983\xspace}
\newcommand{\XsfIccOcSshGT}{202902\xspace}
\newcommand{\XsfIccOcSshLsRecall}{1\xspace}
\newcommand{\XsfIccOcSshLsPrecision}{0.993\xspace}
\newcommand{\XsfIccOcSshLsFone}{0.996\xspace}
\newcommand{\XsfIccOcSshBapRecall}{0.927\xspace}
\newcommand{\XsfIccOcSshBapPrecision}{0.998\xspace}
\newcommand{\XsfIccOcSshBapFone}{0.961\xspace}
\newcommand{\XsfIccOcSshGhiRecall}{0.996\xspace}
\newcommand{\XsfIccOcSshGhiPrecision}{0.998\xspace}
\newcommand{\XsfIccOcSshGhiFone}{0.997\xspace}
\newcommand{\XsfIccOcSshRdaRecall}{0.885\xspace}
\newcommand{\XsfIccOcSshRdaPrecision}{0.998\xspace}
\newcommand{\XsfIccOcSshRdaFone}{0.938\xspace}
\newcommand{\XsfIccOcSshRseRecall}{0.978\xspace}
\newcommand{\XsfIccOcSshRsePrecision}{0.998\xspace}
\newcommand{\XsfIccOcSshRseFone}{0.988\xspace}
\newcommand{\XsfIccOcPcrGT}{5930\xspace}
\newcommand{\XsfIccOcPcrLsRecall}{1\xspace}
\newcommand{\XsfIccOcPcrLsPrecision}{0.996\xspace}
\newcommand{\XsfIccOcPcrLsFone}{0.998\xspace}
\newcommand{\XsfIccOcPcrBapRecall}{0.910\xspace}
\newcommand{\XsfIccOcPcrBapPrecision}{0.999\xspace}
\newcommand{\XsfIccOcPcrBapFone}{0.953\xspace}
\newcommand{\XsfIccOcPcrGhiRecall}{1\xspace}
\newcommand{\XsfIccOcPcrGhiPrecision}{0.999\xspace}
\newcommand{\XsfIccOcPcrGhiFone}{1.000\xspace}
\newcommand{\XsfIccOcPcrRdaRecall}{1\xspace}
\newcommand{\XsfIccOcPcrRdaPrecision}{1\xspace}
\newcommand{\XsfIccOcPcrRdaFone}{1\xspace}
\newcommand{\XsfIccOcPcrRseRecall}{1\xspace}
\newcommand{\XsfIccOcPcrRsePrecision}{1\xspace}
\newcommand{\XsfIccOcPcrRseFone}{1\xspace}
\newcommand{\XsfIccOcSqlGT}{410176\xspace}
\newcommand{\XsfIccOcSqlLsRecall}{1\xspace}
\newcommand{\XsfIccOcSqlLsPrecision}{0.995\xspace}
\newcommand{\XsfIccOcSqlLsFone}{0.998\xspace}
\newcommand{\XsfIccOcSqlBapRecall}{0.898\xspace}
\newcommand{\XsfIccOcSqlBapPrecision}{1.000\xspace}
\newcommand{\XsfIccOcSqlBapFone}{0.946\xspace}
\newcommand{\XsfIccOcSqlGhiRecall}{0.970\xspace}
\newcommand{\XsfIccOcSqlGhiPrecision}{1.000\xspace}
\newcommand{\XsfIccOcSqlGhiFone}{0.985\xspace}
\newcommand{\XsfIccOcSqlRdaRecall}{0.990\xspace}
\newcommand{\XsfIccOcSqlRdaPrecision}{1.000\xspace}
\newcommand{\XsfIccOcSqlRdaFone}{0.995\xspace}
\newcommand{\XsfIccOcSqlRseRecall}{0.992\xspace}
\newcommand{\XsfIccOcSqlRsePrecision}{1\xspace}
\newcommand{\XsfIccOcSqlRseFone}{0.996\xspace}
\newcommand{\XsfIccOcVimGT}{984302\xspace}
\newcommand{\XsfIccOcVimLsRecall}{1\xspace}
\newcommand{\XsfIccOcVimLsPrecision}{0.994\xspace}
\newcommand{\XsfIccOcVimLsFone}{0.997\xspace}
\newcommand{\XsfIccOcVimBapRecall}{0.915\xspace}
\newcommand{\XsfIccOcVimBapPrecision}{1.000\xspace}
\newcommand{\XsfIccOcVimBapFone}{0.956\xspace}
\newcommand{\XsfIccOcVimGhiRecall}{0.997\xspace}
\newcommand{\XsfIccOcVimGhiPrecision}{1.000\xspace}
\newcommand{\XsfIccOcVimGhiFone}{0.998\xspace}
\newcommand{\XsfIccOcVimRdaRecall}{0.975\xspace}
\newcommand{\XsfIccOcVimRdaPrecision}{1.000\xspace}
\newcommand{\XsfIccOcVimRdaFone}{0.988\xspace}
\newcommand{\XsfIccOcVimRseRecall}{0.998\xspace}
\newcommand{\XsfIccOcVimRsePrecision}{1\xspace}
\newcommand{\XsfIccOcVimRseFone}{0.999\xspace}
\newcommand{\XsfIccOcVsfGT}{25821\xspace}
\newcommand{\XsfIccOcVsfLsRecall}{1\xspace}
\newcommand{\XsfIccOcVsfLsPrecision}{0.972\xspace}
\newcommand{\XsfIccOcVsfLsFone}{0.986\xspace}
\newcommand{\XsfIccOcVsfBapRecall}{0.988\xspace}
\newcommand{\XsfIccOcVsfBapPrecision}{1.000\xspace}
\newcommand{\XsfIccOcVsfBapFone}{0.994\xspace}
\newcommand{\XsfIccOcVsfGhiRecall}{1\xspace}
\newcommand{\XsfIccOcVsfGhiPrecision}{1.000\xspace}
\newcommand{\XsfIccOcVsfGhiFone}{1.000\xspace}
\newcommand{\XsfIccOcVsfRdaRecall}{0.971\xspace}
\newcommand{\XsfIccOcVsfRdaPrecision}{0.998\xspace}
\newcommand{\XsfIccOcVsfRdaFone}{0.985\xspace}
\newcommand{\XsfIccOcVsfRseRecall}{1.000\xspace}
\newcommand{\XsfIccOcVsfRsePrecision}{1.000\xspace}
\newcommand{\XsfIccOcVsfRseFone}{1.000\xspace}
\newcommand{\XsfIccOdSzpGT}{26700\xspace}
\newcommand{\XsfIccOdSzpLsRecall}{1\xspace}
\newcommand{\XsfIccOdSzpLsPrecision}{0.994\xspace}
\newcommand{\XsfIccOdSzpLsFone}{0.997\xspace}
\newcommand{\XsfIccOdSzpBapRecall}{0.972\xspace}
\newcommand{\XsfIccOdSzpBapPrecision}{1\xspace}
\newcommand{\XsfIccOdSzpBapFone}{0.986\xspace}
\newcommand{\XsfIccOdSzpGhiRecall}{1\xspace}
\newcommand{\XsfIccOdSzpGhiPrecision}{1\xspace}
\newcommand{\XsfIccOdSzpGhiFone}{1\xspace}
\newcommand{\XsfIccOdSzpRdaRecall}{1\xspace}
\newcommand{\XsfIccOdSzpRdaPrecision}{1\xspace}
\newcommand{\XsfIccOdSzpRdaFone}{1\xspace}
\newcommand{\XsfIccOdSzpRseRecall}{1\xspace}
\newcommand{\XsfIccOdSzpRsePrecision}{1\xspace}
\newcommand{\XsfIccOdSzpRseFone}{1\xspace}
\newcommand{\XsfIccOdCapGT}{274363\xspace}
\newcommand{\XsfIccOdCapLsRecall}{1\xspace}
\newcommand{\XsfIccOdCapLsPrecision}{0.995\xspace}
\newcommand{\XsfIccOdCapLsFone}{0.997\xspace}
\newcommand{\XsfIccOdCapBapRecall}{0.306\xspace}
\newcommand{\XsfIccOdCapBapPrecision}{1\xspace}
\newcommand{\XsfIccOdCapBapFone}{0.468\xspace}
\newcommand{\XsfIccOdCapGhiRecall}{1.000\xspace}
\newcommand{\XsfIccOdCapGhiPrecision}{1\xspace}
\newcommand{\XsfIccOdCapGhiFone}{1.000\xspace}
\newcommand{\XsfIccOdCapRdaRecall}{0.000\xspace}
\newcommand{\XsfIccOdCapRdaPrecision}{NaN\xspace}
\newcommand{\XsfIccOdCapRdaFone}{0.000\xspace}
\newcommand{\XsfIccOdCapRseRecall}{0.998\xspace}
\newcommand{\XsfIccOdCapRsePrecision}{1\xspace}
\newcommand{\XsfIccOdCapRseFone}{0.999\xspace}
\newcommand{\XsfIccOdExmGT}{236761\xspace}
\newcommand{\XsfIccOdExmLsRecall}{1\xspace}
\newcommand{\XsfIccOdExmLsPrecision}{0.996\xspace}
\newcommand{\XsfIccOdExmLsFone}{0.998\xspace}
\newcommand{\XsfIccOdExmBapRecall}{0.846\xspace}
\newcommand{\XsfIccOdExmBapPrecision}{1.000\xspace}
\newcommand{\XsfIccOdExmBapFone}{0.916\xspace}
\newcommand{\XsfIccOdExmGhiRecall}{0.948\xspace}
\newcommand{\XsfIccOdExmGhiPrecision}{1.000\xspace}
\newcommand{\XsfIccOdExmGhiFone}{0.973\xspace}
\newcommand{\XsfIccOdExmRdaRecall}{0.945\xspace}
\newcommand{\XsfIccOdExmRdaPrecision}{1.000\xspace}
\newcommand{\XsfIccOdExmRdaFone}{0.972\xspace}
\newcommand{\XsfIccOdExmRseRecall}{0.977\xspace}
\newcommand{\XsfIccOdExmRsePrecision}{1\xspace}
\newcommand{\XsfIccOdExmRseFone}{0.989\xspace}
\newcommand{\XsfIccOdLgtGT}{38359\xspace}
\newcommand{\XsfIccOdLgtLsRecall}{1\xspace}
\newcommand{\XsfIccOdLgtLsPrecision}{0.990\xspace}
\newcommand{\XsfIccOdLgtLsFone}{0.995\xspace}
\newcommand{\XsfIccOdLgtBapRecall}{0.875\xspace}
\newcommand{\XsfIccOdLgtBapPrecision}{0.998\xspace}
\newcommand{\XsfIccOdLgtBapFone}{0.932\xspace}
\newcommand{\XsfIccOdLgtGhiRecall}{0.999\xspace}
\newcommand{\XsfIccOdLgtGhiPrecision}{0.998\xspace}
\newcommand{\XsfIccOdLgtGhiFone}{0.999\xspace}
\newcommand{\XsfIccOdLgtRdaRecall}{0.937\xspace}
\newcommand{\XsfIccOdLgtRdaPrecision}{1.000\xspace}
\newcommand{\XsfIccOdLgtRdaFone}{0.968\xspace}
\newcommand{\XsfIccOdLgtRseRecall}{0.991\xspace}
\newcommand{\XsfIccOdLgtRsePrecision}{1\xspace}
\newcommand{\XsfIccOdLgtRseFone}{0.996\xspace}
\newcommand{\XsfIccOdBzpGT}{27088\xspace}
\newcommand{\XsfIccOdBzpLsRecall}{1\xspace}
\newcommand{\XsfIccOdBzpLsPrecision}{0.997\xspace}
\newcommand{\XsfIccOdBzpLsFone}{0.999\xspace}
\newcommand{\XsfIccOdBzpBapRecall}{0.830\xspace}
\newcommand{\XsfIccOdBzpBapPrecision}{1.000\xspace}
\newcommand{\XsfIccOdBzpBapFone}{0.907\xspace}
\newcommand{\XsfIccOdBzpGhiRecall}{1\xspace}
\newcommand{\XsfIccOdBzpGhiPrecision}{1.000\xspace}
\newcommand{\XsfIccOdBzpGhiFone}{1.000\xspace}
\newcommand{\XsfIccOdBzpRdaRecall}{0.978\xspace}
\newcommand{\XsfIccOdBzpRdaPrecision}{1\xspace}
\newcommand{\XsfIccOdBzpRdaFone}{0.989\xspace}
\newcommand{\XsfIccOdBzpRseRecall}{1\xspace}
\newcommand{\XsfIccOdBzpRsePrecision}{1\xspace}
\newcommand{\XsfIccOdBzpRseFone}{1\xspace}
\newcommand{\XsfIccOdGccGT}{3682510\xspace}
\newcommand{\XsfIccOdGccLsRecall}{1\xspace}
\newcommand{\XsfIccOdGccLsPrecision}{0.997\xspace}
\newcommand{\XsfIccOdGccLsFone}{0.999\xspace}
\newcommand{\XsfIccOdGccBapRecall}{0.885\xspace}
\newcommand{\XsfIccOdGccBapPrecision}{1.000\xspace}
\newcommand{\XsfIccOdGccBapFone}{0.939\xspace}
\newcommand{\XsfIccOdGccGhiRecall}{0.994\xspace}
\newcommand{\XsfIccOdGccGhiPrecision}{1.000\xspace}
\newcommand{\XsfIccOdGccGhiFone}{0.997\xspace}
\newcommand{\XsfIccOdGccRdaRecall}{0.938\xspace}
\newcommand{\XsfIccOdGccRdaPrecision}{1.000\xspace}
\newcommand{\XsfIccOdGccRdaFone}{0.968\xspace}
\newcommand{\XsfIccOdGccRseRecall}{0.995\xspace}
\newcommand{\XsfIccOdGccRsePrecision}{1.000\xspace}
\newcommand{\XsfIccOdGccRseFone}{0.998\xspace}
\newcommand{\XsfIccOdGzpGT}{28843\xspace}
\newcommand{\XsfIccOdGzpLsRecall}{1\xspace}
\newcommand{\XsfIccOdGzpLsPrecision}{0.997\xspace}
\newcommand{\XsfIccOdGzpLsFone}{0.998\xspace}
\newcommand{\XsfIccOdGzpBapRecall}{0.994\xspace}
\newcommand{\XsfIccOdGzpBapPrecision}{1.000\xspace}
\newcommand{\XsfIccOdGzpBapFone}{0.997\xspace}
\newcommand{\XsfIccOdGzpGhiRecall}{1\xspace}
\newcommand{\XsfIccOdGzpGhiPrecision}{1.000\xspace}
\newcommand{\XsfIccOdGzpGhiFone}{1.000\xspace}
\newcommand{\XsfIccOdGzpRdaRecall}{0.992\xspace}
\newcommand{\XsfIccOdGzpRdaPrecision}{1\xspace}
\newcommand{\XsfIccOdGzpRdaFone}{0.996\xspace}
\newcommand{\XsfIccOdGzpRseRecall}{1\xspace}
\newcommand{\XsfIccOdGzpRsePrecision}{1\xspace}
\newcommand{\XsfIccOdGzpRseFone}{1\xspace}
\newcommand{\XsfIccOdOggGT}{111485\xspace}
\newcommand{\XsfIccOdOggLsRecall}{1\xspace}
\newcommand{\XsfIccOdOggLsPrecision}{0.996\xspace}
\newcommand{\XsfIccOdOggLsFone}{0.998\xspace}
\newcommand{\XsfIccOdOggBapRecall}{0.980\xspace}
\newcommand{\XsfIccOdOggBapPrecision}{1.000\xspace}
\newcommand{\XsfIccOdOggBapFone}{0.990\xspace}
\newcommand{\XsfIccOdOggGhiRecall}{0.999\xspace}
\newcommand{\XsfIccOdOggGhiPrecision}{1.000\xspace}
\newcommand{\XsfIccOdOggGhiFone}{1.000\xspace}
\newcommand{\XsfIccOdOggRdaRecall}{0.992\xspace}
\newcommand{\XsfIccOdOggRdaPrecision}{1.000\xspace}
\newcommand{\XsfIccOdOggRdaFone}{0.996\xspace}
\newcommand{\XsfIccOdOggRseRecall}{0.992\xspace}
\newcommand{\XsfIccOdOggRsePrecision}{1\xspace}
\newcommand{\XsfIccOdOggRseFone}{0.996\xspace}
\newcommand{\XsfIccOdNgxGT}{154673\xspace}
\newcommand{\XsfIccOdNgxLsRecall}{1\xspace}
\newcommand{\XsfIccOdNgxLsPrecision}{0.992\xspace}
\newcommand{\XsfIccOdNgxLsFone}{0.996\xspace}
\newcommand{\XsfIccOdNgxBapRecall}{0.938\xspace}
\newcommand{\XsfIccOdNgxBapPrecision}{1.000\xspace}
\newcommand{\XsfIccOdNgxBapFone}{0.968\xspace}
\newcommand{\XsfIccOdNgxGhiRecall}{0.998\xspace}
\newcommand{\XsfIccOdNgxGhiPrecision}{1.000\xspace}
\newcommand{\XsfIccOdNgxGhiFone}{0.999\xspace}
\newcommand{\XsfIccOdNgxRdaRecall}{0.000\xspace}
\newcommand{\XsfIccOdNgxRdaPrecision}{NaN\xspace}
\newcommand{\XsfIccOdNgxRdaFone}{0.000\xspace}
\newcommand{\XsfIccOdNgxRseRecall}{0.966\xspace}
\newcommand{\XsfIccOdNgxRsePrecision}{1\xspace}
\newcommand{\XsfIccOdNgxRseFone}{0.983\xspace}
\newcommand{\XsfIccOdSshGT}{202902\xspace}
\newcommand{\XsfIccOdSshLsRecall}{1\xspace}
\newcommand{\XsfIccOdSshLsPrecision}{0.993\xspace}
\newcommand{\XsfIccOdSshLsFone}{0.996\xspace}
\newcommand{\XsfIccOdSshBapRecall}{0.927\xspace}
\newcommand{\XsfIccOdSshBapPrecision}{0.998\xspace}
\newcommand{\XsfIccOdSshBapFone}{0.961\xspace}
\newcommand{\XsfIccOdSshGhiRecall}{0.996\xspace}
\newcommand{\XsfIccOdSshGhiPrecision}{0.998\xspace}
\newcommand{\XsfIccOdSshGhiFone}{0.997\xspace}
\newcommand{\XsfIccOdSshRdaRecall}{0.885\xspace}
\newcommand{\XsfIccOdSshRdaPrecision}{0.998\xspace}
\newcommand{\XsfIccOdSshRdaFone}{0.938\xspace}
\newcommand{\XsfIccOdSshRseRecall}{0.978\xspace}
\newcommand{\XsfIccOdSshRsePrecision}{0.998\xspace}
\newcommand{\XsfIccOdSshRseFone}{0.988\xspace}
\newcommand{\XsfIccOdPcrGT}{5930\xspace}
\newcommand{\XsfIccOdPcrLsRecall}{1\xspace}
\newcommand{\XsfIccOdPcrLsPrecision}{0.996\xspace}
\newcommand{\XsfIccOdPcrLsFone}{0.998\xspace}
\newcommand{\XsfIccOdPcrBapRecall}{0.910\xspace}
\newcommand{\XsfIccOdPcrBapPrecision}{0.999\xspace}
\newcommand{\XsfIccOdPcrBapFone}{0.953\xspace}
\newcommand{\XsfIccOdPcrGhiRecall}{1\xspace}
\newcommand{\XsfIccOdPcrGhiPrecision}{0.999\xspace}
\newcommand{\XsfIccOdPcrGhiFone}{1.000\xspace}
\newcommand{\XsfIccOdPcrRdaRecall}{1\xspace}
\newcommand{\XsfIccOdPcrRdaPrecision}{1\xspace}
\newcommand{\XsfIccOdPcrRdaFone}{1\xspace}
\newcommand{\XsfIccOdPcrRseRecall}{1\xspace}
\newcommand{\XsfIccOdPcrRsePrecision}{1\xspace}
\newcommand{\XsfIccOdPcrRseFone}{1\xspace}
\newcommand{\XsfIccOdSqlGT}{410145\xspace}
\newcommand{\XsfIccOdSqlLsRecall}{1\xspace}
\newcommand{\XsfIccOdSqlLsPrecision}{0.995\xspace}
\newcommand{\XsfIccOdSqlLsFone}{0.998\xspace}
\newcommand{\XsfIccOdSqlBapRecall}{0.898\xspace}
\newcommand{\XsfIccOdSqlBapPrecision}{1.000\xspace}
\newcommand{\XsfIccOdSqlBapFone}{0.946\xspace}
\newcommand{\XsfIccOdSqlGhiRecall}{0.970\xspace}
\newcommand{\XsfIccOdSqlGhiPrecision}{1.000\xspace}
\newcommand{\XsfIccOdSqlGhiFone}{0.985\xspace}
\newcommand{\XsfIccOdSqlRdaRecall}{0.990\xspace}
\newcommand{\XsfIccOdSqlRdaPrecision}{1.000\xspace}
\newcommand{\XsfIccOdSqlRdaFone}{0.995\xspace}
\newcommand{\XsfIccOdSqlRseRecall}{0.992\xspace}
\newcommand{\XsfIccOdSqlRsePrecision}{1\xspace}
\newcommand{\XsfIccOdSqlRseFone}{0.996\xspace}
\newcommand{\XsfIccOdVimGT}{984307\xspace}
\newcommand{\XsfIccOdVimLsRecall}{1\xspace}
\newcommand{\XsfIccOdVimLsPrecision}{0.994\xspace}
\newcommand{\XsfIccOdVimLsFone}{0.997\xspace}
\newcommand{\XsfIccOdVimBapRecall}{0.915\xspace}
\newcommand{\XsfIccOdVimBapPrecision}{1.000\xspace}
\newcommand{\XsfIccOdVimBapFone}{0.956\xspace}
\newcommand{\XsfIccOdVimGhiRecall}{0.997\xspace}
\newcommand{\XsfIccOdVimGhiPrecision}{1.000\xspace}
\newcommand{\XsfIccOdVimGhiFone}{0.998\xspace}
\newcommand{\XsfIccOdVimRdaRecall}{0.975\xspace}
\newcommand{\XsfIccOdVimRdaPrecision}{1.000\xspace}
\newcommand{\XsfIccOdVimRdaFone}{0.988\xspace}
\newcommand{\XsfIccOdVimRseRecall}{0.998\xspace}
\newcommand{\XsfIccOdVimRsePrecision}{1\xspace}
\newcommand{\XsfIccOdVimRseFone}{0.999\xspace}
\newcommand{\XsfIccOdVsfGT}{25821\xspace}
\newcommand{\XsfIccOdVsfLsRecall}{1\xspace}
\newcommand{\XsfIccOdVsfLsPrecision}{0.972\xspace}
\newcommand{\XsfIccOdVsfLsFone}{0.986\xspace}
\newcommand{\XsfIccOdVsfBapRecall}{0.988\xspace}
\newcommand{\XsfIccOdVsfBapPrecision}{1.000\xspace}
\newcommand{\XsfIccOdVsfBapFone}{0.994\xspace}
\newcommand{\XsfIccOdVsfGhiRecall}{1\xspace}
\newcommand{\XsfIccOdVsfGhiPrecision}{1.000\xspace}
\newcommand{\XsfIccOdVsfGhiFone}{1.000\xspace}
\newcommand{\XsfIccOdVsfRdaRecall}{0.971\xspace}
\newcommand{\XsfIccOdVsfRdaPrecision}{0.998\xspace}
\newcommand{\XsfIccOdVsfRdaFone}{0.984\xspace}
\newcommand{\XsfIccOdVsfRseRecall}{1.000\xspace}
\newcommand{\XsfIccOdVsfRsePrecision}{1.000\xspace}
\newcommand{\XsfIccOdVsfRseFone}{1.000\xspace}
\newcommand{\XsfIccOsSzpGT}{11281\xspace}
\newcommand{\XsfIccOsSzpLsRecall}{1\xspace}
\newcommand{\XsfIccOsSzpLsPrecision}{1\xspace}
\newcommand{\XsfIccOsSzpLsFone}{1\xspace}
\newcommand{\XsfIccOsSzpBapRecall}{0.976\xspace}
\newcommand{\XsfIccOsSzpBapPrecision}{1\xspace}
\newcommand{\XsfIccOsSzpBapFone}{0.988\xspace}
\newcommand{\XsfIccOsSzpGhiRecall}{1\xspace}
\newcommand{\XsfIccOsSzpGhiPrecision}{1\xspace}
\newcommand{\XsfIccOsSzpGhiFone}{1\xspace}
\newcommand{\XsfIccOsSzpRdaRecall}{0.976\xspace}
\newcommand{\XsfIccOsSzpRdaPrecision}{1\xspace}
\newcommand{\XsfIccOsSzpRdaFone}{0.988\xspace}
\newcommand{\XsfIccOsSzpRseRecall}{1\xspace}
\newcommand{\XsfIccOsSzpRsePrecision}{1\xspace}
\newcommand{\XsfIccOsSzpRseFone}{1\xspace}
\newcommand{\XsfIccOsCapGT}{183544\xspace}
\newcommand{\XsfIccOsCapLsRecall}{1\xspace}
\newcommand{\XsfIccOsCapLsPrecision}{1\xspace}
\newcommand{\XsfIccOsCapLsFone}{1\xspace}
\newcommand{\XsfIccOsCapBapRecall}{0.274\xspace}
\newcommand{\XsfIccOsCapBapPrecision}{1\xspace}
\newcommand{\XsfIccOsCapBapFone}{0.430\xspace}
\newcommand{\XsfIccOsCapGhiRecall}{0.875\xspace}
\newcommand{\XsfIccOsCapGhiPrecision}{1\xspace}
\newcommand{\XsfIccOsCapGhiFone}{0.933\xspace}
\newcommand{\XsfIccOsCapRdaRecall}{0.000\xspace}
\newcommand{\XsfIccOsCapRdaPrecision}{NaN\xspace}
\newcommand{\XsfIccOsCapRdaFone}{0.000\xspace}
\newcommand{\XsfIccOsCapRseRecall}{0.668\xspace}
\newcommand{\XsfIccOsCapRsePrecision}{1\xspace}
\newcommand{\XsfIccOsCapRseFone}{0.801\xspace}
\newcommand{\XsfIccOsExmGT}{130443\xspace}
\newcommand{\XsfIccOsExmLsRecall}{1\xspace}
\newcommand{\XsfIccOsExmLsPrecision}{1\xspace}
\newcommand{\XsfIccOsExmLsFone}{1\xspace}
\newcommand{\XsfIccOsExmBapRecall}{0.829\xspace}
\newcommand{\XsfIccOsExmBapPrecision}{1\xspace}
\newcommand{\XsfIccOsExmBapFone}{0.907\xspace}
\newcommand{\XsfIccOsExmGhiRecall}{0.961\xspace}
\newcommand{\XsfIccOsExmGhiPrecision}{1\xspace}
\newcommand{\XsfIccOsExmGhiFone}{0.980\xspace}
\newcommand{\XsfIccOsExmRdaRecall}{0.801\xspace}
\newcommand{\XsfIccOsExmRdaPrecision}{1.000\xspace}
\newcommand{\XsfIccOsExmRdaFone}{0.890\xspace}
\newcommand{\XsfIccOsExmRseRecall}{0.972\xspace}
\newcommand{\XsfIccOsExmRsePrecision}{1\xspace}
\newcommand{\XsfIccOsExmRseFone}{0.986\xspace}
\newcommand{\XsfIccOsLgtGT}{24482\xspace}
\newcommand{\XsfIccOsLgtLsRecall}{1\xspace}
\newcommand{\XsfIccOsLgtLsPrecision}{1\xspace}
\newcommand{\XsfIccOsLgtLsFone}{1\xspace}
\newcommand{\XsfIccOsLgtBapRecall}{0.874\xspace}
\newcommand{\XsfIccOsLgtBapPrecision}{1\xspace}
\newcommand{\XsfIccOsLgtBapFone}{0.933\xspace}
\newcommand{\XsfIccOsLgtGhiRecall}{0.997\xspace}
\newcommand{\XsfIccOsLgtGhiPrecision}{1\xspace}
\newcommand{\XsfIccOsLgtGhiFone}{0.998\xspace}
\newcommand{\XsfIccOsLgtRdaRecall}{0.874\xspace}
\newcommand{\XsfIccOsLgtRdaPrecision}{1\xspace}
\newcommand{\XsfIccOsLgtRdaFone}{0.933\xspace}
\newcommand{\XsfIccOsLgtRseRecall}{0.994\xspace}
\newcommand{\XsfIccOsLgtRsePrecision}{1\xspace}
\newcommand{\XsfIccOsLgtRseFone}{0.997\xspace}
\newcommand{\XsfIccOsBzpGT}{11330\xspace}
\newcommand{\XsfIccOsBzpLsRecall}{1\xspace}
\newcommand{\XsfIccOsBzpLsPrecision}{1\xspace}
\newcommand{\XsfIccOsBzpLsFone}{1\xspace}
\newcommand{\XsfIccOsBzpBapRecall}{0.793\xspace}
\newcommand{\XsfIccOsBzpBapPrecision}{1\xspace}
\newcommand{\XsfIccOsBzpBapFone}{0.884\xspace}
\newcommand{\XsfIccOsBzpGhiRecall}{1\xspace}
\newcommand{\XsfIccOsBzpGhiPrecision}{1\xspace}
\newcommand{\XsfIccOsBzpGhiFone}{1\xspace}
\newcommand{\XsfIccOsBzpRdaRecall}{0.736\xspace}
\newcommand{\XsfIccOsBzpRdaPrecision}{1.000\xspace}
\newcommand{\XsfIccOsBzpRdaFone}{0.848\xspace}
\newcommand{\XsfIccOsBzpRseRecall}{0.917\xspace}
\newcommand{\XsfIccOsBzpRsePrecision}{1\xspace}
\newcommand{\XsfIccOsBzpRseFone}{0.957\xspace}
\newcommand{\XsfIccOsGccGT}{753051\xspace}
\newcommand{\XsfIccOsGccLsRecall}{1\xspace}
\newcommand{\XsfIccOsGccLsPrecision}{1\xspace}
\newcommand{\XsfIccOsGccLsFone}{1\xspace}
\newcommand{\XsfIccOsGccBapRecall}{0.754\xspace}
\newcommand{\XsfIccOsGccBapPrecision}{1.000\xspace}
\newcommand{\XsfIccOsGccBapFone}{0.860\xspace}
\newcommand{\XsfIccOsGccGhiRecall}{0.929\xspace}
\newcommand{\XsfIccOsGccGhiPrecision}{1\xspace}
\newcommand{\XsfIccOsGccGhiFone}{0.963\xspace}
\newcommand{\XsfIccOsGccRdaRecall}{0.734\xspace}
\newcommand{\XsfIccOsGccRdaPrecision}{1.000\xspace}
\newcommand{\XsfIccOsGccRdaFone}{0.846\xspace}
\newcommand{\XsfIccOsGccRseRecall}{0.960\xspace}
\newcommand{\XsfIccOsGccRsePrecision}{1\xspace}
\newcommand{\XsfIccOsGccRseFone}{0.980\xspace}
\newcommand{\XsfIccOsGzpGT}{9098\xspace}
\newcommand{\XsfIccOsGzpLsRecall}{1\xspace}
\newcommand{\XsfIccOsGzpLsPrecision}{1\xspace}
\newcommand{\XsfIccOsGzpLsFone}{1\xspace}
\newcommand{\XsfIccOsGzpBapRecall}{0.987\xspace}
\newcommand{\XsfIccOsGzpBapPrecision}{1\xspace}
\newcommand{\XsfIccOsGzpBapFone}{0.993\xspace}
\newcommand{\XsfIccOsGzpGhiRecall}{1\xspace}
\newcommand{\XsfIccOsGzpGhiPrecision}{1\xspace}
\newcommand{\XsfIccOsGzpGhiFone}{1\xspace}
\newcommand{\XsfIccOsGzpRdaRecall}{0.987\xspace}
\newcommand{\XsfIccOsGzpRdaPrecision}{1\xspace}
\newcommand{\XsfIccOsGzpRdaFone}{0.993\xspace}
\newcommand{\XsfIccOsGzpRseRecall}{0.998\xspace}
\newcommand{\XsfIccOsGzpRsePrecision}{1\xspace}
\newcommand{\XsfIccOsGzpRseFone}{0.999\xspace}
\newcommand{\XsfIccOsOggGT}{34251\xspace}
\newcommand{\XsfIccOsOggLsRecall}{1\xspace}
\newcommand{\XsfIccOsOggLsPrecision}{1\xspace}
\newcommand{\XsfIccOsOggLsFone}{1\xspace}
\newcommand{\XsfIccOsOggBapRecall}{0.972\xspace}
\newcommand{\XsfIccOsOggBapPrecision}{1\xspace}
\newcommand{\XsfIccOsOggBapFone}{0.986\xspace}
\newcommand{\XsfIccOsOggGhiRecall}{1.000\xspace}
\newcommand{\XsfIccOsOggGhiPrecision}{1\xspace}
\newcommand{\XsfIccOsOggGhiFone}{1.000\xspace}
\newcommand{\XsfIccOsOggRdaRecall}{0.972\xspace}
\newcommand{\XsfIccOsOggRdaPrecision}{1\xspace}
\newcommand{\XsfIccOsOggRdaFone}{0.986\xspace}
\newcommand{\XsfIccOsOggRseRecall}{0.982\xspace}
\newcommand{\XsfIccOsOggRsePrecision}{1\xspace}
\newcommand{\XsfIccOsOggRseFone}{0.991\xspace}
\newcommand{\XsfIccOsNgxGT}{96608\xspace}
\newcommand{\XsfIccOsNgxLsRecall}{1\xspace}
\newcommand{\XsfIccOsNgxLsPrecision}{1\xspace}
\newcommand{\XsfIccOsNgxLsFone}{1\xspace}
\newcommand{\XsfIccOsNgxBapRecall}{0.967\xspace}
\newcommand{\XsfIccOsNgxBapPrecision}{1.000\xspace}
\newcommand{\XsfIccOsNgxBapFone}{0.983\xspace}
\newcommand{\XsfIccOsNgxGhiRecall}{0.995\xspace}
\newcommand{\XsfIccOsNgxGhiPrecision}{1\xspace}
\newcommand{\XsfIccOsNgxGhiFone}{0.998\xspace}
\newcommand{\XsfIccOsNgxRdaRecall}{0.965\xspace}
\newcommand{\XsfIccOsNgxRdaPrecision}{1.000\xspace}
\newcommand{\XsfIccOsNgxRdaFone}{0.982\xspace}
\newcommand{\XsfIccOsNgxRseRecall}{0.993\xspace}
\newcommand{\XsfIccOsNgxRsePrecision}{1\xspace}
\newcommand{\XsfIccOsNgxRseFone}{0.997\xspace}
\newcommand{\XsfIccOsSshGT}{96073\xspace}
\newcommand{\XsfIccOsSshLsRecall}{1\xspace}
\newcommand{\XsfIccOsSshLsPrecision}{1\xspace}
\newcommand{\XsfIccOsSshLsFone}{1\xspace}
\newcommand{\XsfIccOsSshBapRecall}{0.938\xspace}
\newcommand{\XsfIccOsSshBapPrecision}{1\xspace}
\newcommand{\XsfIccOsSshBapFone}{0.968\xspace}
\newcommand{\XsfIccOsSshGhiRecall}{0.997\xspace}
\newcommand{\XsfIccOsSshGhiPrecision}{1\xspace}
\newcommand{\XsfIccOsSshGhiFone}{0.998\xspace}
\newcommand{\XsfIccOsSshRdaRecall}{0.927\xspace}
\newcommand{\XsfIccOsSshRdaPrecision}{0.999\xspace}
\newcommand{\XsfIccOsSshRdaFone}{0.962\xspace}
\newcommand{\XsfIccOsSshRseRecall}{0.971\xspace}
\newcommand{\XsfIccOsSshRsePrecision}{1\xspace}
\newcommand{\XsfIccOsSshRseFone}{0.985\xspace}
\newcommand{\XsfIccOsPcrGT}{4101\xspace}
\newcommand{\XsfIccOsPcrLsRecall}{1\xspace}
\newcommand{\XsfIccOsPcrLsPrecision}{1\xspace}
\newcommand{\XsfIccOsPcrLsFone}{1\xspace}
\newcommand{\XsfIccOsPcrBapRecall}{0.916\xspace}
\newcommand{\XsfIccOsPcrBapPrecision}{1\xspace}
\newcommand{\XsfIccOsPcrBapFone}{0.956\xspace}
\newcommand{\XsfIccOsPcrGhiRecall}{1\xspace}
\newcommand{\XsfIccOsPcrGhiPrecision}{1\xspace}
\newcommand{\XsfIccOsPcrGhiFone}{1\xspace}
\newcommand{\XsfIccOsPcrRdaRecall}{0.855\xspace}
\newcommand{\XsfIccOsPcrRdaPrecision}{1\xspace}
\newcommand{\XsfIccOsPcrRdaFone}{0.922\xspace}
\newcommand{\XsfIccOsPcrRseRecall}{0.995\xspace}
\newcommand{\XsfIccOsPcrRsePrecision}{1\xspace}
\newcommand{\XsfIccOsPcrRseFone}{0.998\xspace}
\newcommand{\XsfIccOsSqlGT}{145494\xspace}
\newcommand{\XsfIccOsSqlLsRecall}{1\xspace}
\newcommand{\XsfIccOsSqlLsPrecision}{1\xspace}
\newcommand{\XsfIccOsSqlLsFone}{1\xspace}
\newcommand{\XsfIccOsSqlBapRecall}{0.853\xspace}
\newcommand{\XsfIccOsSqlBapPrecision}{1\xspace}
\newcommand{\XsfIccOsSqlBapFone}{0.921\xspace}
\newcommand{\XsfIccOsSqlGhiRecall}{0.951\xspace}
\newcommand{\XsfIccOsSqlGhiPrecision}{1\xspace}
\newcommand{\XsfIccOsSqlGhiFone}{0.975\xspace}
\newcommand{\XsfIccOsSqlRdaRecall}{0.852\xspace}
\newcommand{\XsfIccOsSqlRdaPrecision}{1.000\xspace}
\newcommand{\XsfIccOsSqlRdaFone}{0.920\xspace}
\newcommand{\XsfIccOsSqlRseRecall}{0.929\xspace}
\newcommand{\XsfIccOsSqlRsePrecision}{1\xspace}
\newcommand{\XsfIccOsSqlRseFone}{0.963\xspace}
\newcommand{\XsfIccOsVimGT}{443486\xspace}
\newcommand{\XsfIccOsVimLsRecall}{1\xspace}
\newcommand{\XsfIccOsVimLsPrecision}{1\xspace}
\newcommand{\XsfIccOsVimLsFone}{1\xspace}
\newcommand{\XsfIccOsVimBapRecall}{0.912\xspace}
\newcommand{\XsfIccOsVimBapPrecision}{1.000\xspace}
\newcommand{\XsfIccOsVimBapFone}{0.954\xspace}
\newcommand{\XsfIccOsVimGhiRecall}{0.998\xspace}
\newcommand{\XsfIccOsVimGhiPrecision}{1\xspace}
\newcommand{\XsfIccOsVimGhiFone}{0.999\xspace}
\newcommand{\XsfIccOsVimRdaRecall}{0.901\xspace}
\newcommand{\XsfIccOsVimRdaPrecision}{1\xspace}
\newcommand{\XsfIccOsVimRdaFone}{0.948\xspace}
\newcommand{\XsfIccOsVimRseRecall}{0.982\xspace}
\newcommand{\XsfIccOsVimRsePrecision}{1.000\xspace}
\newcommand{\XsfIccOsVimRseFone}{0.991\xspace}
\newcommand{\XsfIccOsVsfGT}{17419\xspace}
\newcommand{\XsfIccOsVsfLsRecall}{1\xspace}
\newcommand{\XsfIccOsVsfLsPrecision}{1\xspace}
\newcommand{\XsfIccOsVsfLsFone}{1\xspace}
\newcommand{\XsfIccOsVsfBapRecall}{0.988\xspace}
\newcommand{\XsfIccOsVsfBapPrecision}{1\xspace}
\newcommand{\XsfIccOsVsfBapFone}{0.994\xspace}
\newcommand{\XsfIccOsVsfGhiRecall}{1\xspace}
\newcommand{\XsfIccOsVsfGhiPrecision}{1\xspace}
\newcommand{\XsfIccOsVsfGhiFone}{1\xspace}
\newcommand{\XsfIccOsVsfRdaRecall}{0.977\xspace}
\newcommand{\XsfIccOsVsfRdaPrecision}{0.998\xspace}
\newcommand{\XsfIccOsVsfRdaFone}{0.987\xspace}
\newcommand{\XsfIccOsVsfRseRecall}{1.000\xspace}
\newcommand{\XsfIccOsVsfRsePrecision}{1\xspace}
\newcommand{\XsfIccOsVsfRseFone}{1.000\xspace}
\newcommand{\XttGccFOoSzpGT}{21126\xspace}
\newcommand{\XttGccFOoSzpLsRecall}{1\xspace}
\newcommand{\XttGccFOoSzpLsPrecision}{1\xspace}
\newcommand{\XttGccFOoSzpLsFone}{1\xspace}
\newcommand{\XttGccFOoSzpBapRecall}{0.976\xspace}
\newcommand{\XttGccFOoSzpBapPrecision}{1\xspace}
\newcommand{\XttGccFOoSzpBapFone}{0.988\xspace}
\newcommand{\XttGccFOoSzpGhiRecall}{1\xspace}
\newcommand{\XttGccFOoSzpGhiPrecision}{1\xspace}
\newcommand{\XttGccFOoSzpGhiFone}{1\xspace}
\newcommand{\XttGccFOoSzpRdaRecall}{1\xspace}
\newcommand{\XttGccFOoSzpRdaPrecision}{1\xspace}
\newcommand{\XttGccFOoSzpRdaFone}{1\xspace}
\newcommand{\XttGccFOoSzpRseRecall}{1\xspace}
\newcommand{\XttGccFOoSzpRsePrecision}{1\xspace}
\newcommand{\XttGccFOoSzpRseFone}{1\xspace}
\newcommand{\XttGccFOoCapGT}{383603\xspace}
\newcommand{\XttGccFOoCapLsRecall}{1\xspace}
\newcommand{\XttGccFOoCapLsPrecision}{1\xspace}
\newcommand{\XttGccFOoCapLsFone}{1\xspace}
\newcommand{\XttGccFOoCapBapRecall}{0.446\xspace}
\newcommand{\XttGccFOoCapBapPrecision}{1\xspace}
\newcommand{\XttGccFOoCapBapFone}{0.617\xspace}
\newcommand{\XttGccFOoCapGhiRecall}{1.000\xspace}
\newcommand{\XttGccFOoCapGhiPrecision}{1\xspace}
\newcommand{\XttGccFOoCapGhiFone}{1.000\xspace}
\newcommand{\XttGccFOoCapRdaRecall}{0.480\xspace}
\newcommand{\XttGccFOoCapRdaPrecision}{1\xspace}
\newcommand{\XttGccFOoCapRdaFone}{0.649\xspace}
\newcommand{\XttGccFOoCapRseRecall}{0.612\xspace}
\newcommand{\XttGccFOoCapRsePrecision}{1.000\xspace}
\newcommand{\XttGccFOoCapRseFone}{0.760\xspace}
\newcommand{\XttGccFOoExmGT}{189184\xspace}
\newcommand{\XttGccFOoExmLsRecall}{1\xspace}
\newcommand{\XttGccFOoExmLsPrecision}{1\xspace}
\newcommand{\XttGccFOoExmLsFone}{1\xspace}
\newcommand{\XttGccFOoExmBapRecall}{0.853\xspace}
\newcommand{\XttGccFOoExmBapPrecision}{1\xspace}
\newcommand{\XttGccFOoExmBapFone}{0.921\xspace}
\newcommand{\XttGccFOoExmGhiRecall}{1\xspace}
\newcommand{\XttGccFOoExmGhiPrecision}{1\xspace}
\newcommand{\XttGccFOoExmGhiFone}{1\xspace}
\newcommand{\XttGccFOoExmRdaRecall}{0.888\xspace}
\newcommand{\XttGccFOoExmRdaPrecision}{1\xspace}
\newcommand{\XttGccFOoExmRdaFone}{0.941\xspace}
\newcommand{\XttGccFOoExmRseRecall}{0.959\xspace}
\newcommand{\XttGccFOoExmRsePrecision}{1.000\xspace}
\newcommand{\XttGccFOoExmRseFone}{0.979\xspace}
\newcommand{\XttGccFOoLgtGT}{41316\xspace}
\newcommand{\XttGccFOoLgtLsRecall}{1\xspace}
\newcommand{\XttGccFOoLgtLsPrecision}{1\xspace}
\newcommand{\XttGccFOoLgtLsFone}{1\xspace}
\newcommand{\XttGccFOoLgtBapRecall}{0.882\xspace}
\newcommand{\XttGccFOoLgtBapPrecision}{1\xspace}
\newcommand{\XttGccFOoLgtBapFone}{0.937\xspace}
\newcommand{\XttGccFOoLgtGhiRecall}{1\xspace}
\newcommand{\XttGccFOoLgtGhiPrecision}{1\xspace}
\newcommand{\XttGccFOoLgtGhiFone}{1\xspace}
\newcommand{\XttGccFOoLgtRdaRecall}{0.948\xspace}
\newcommand{\XttGccFOoLgtRdaPrecision}{1\xspace}
\newcommand{\XttGccFOoLgtRdaFone}{0.974\xspace}
\newcommand{\XttGccFOoLgtRseRecall}{0.974\xspace}
\newcommand{\XttGccFOoLgtRsePrecision}{1\xspace}
\newcommand{\XttGccFOoLgtRseFone}{0.987\xspace}
\newcommand{\XttGccFOoBzpGT}{23584\xspace}
\newcommand{\XttGccFOoBzpLsRecall}{1\xspace}
\newcommand{\XttGccFOoBzpLsPrecision}{1\xspace}
\newcommand{\XttGccFOoBzpLsFone}{1\xspace}
\newcommand{\XttGccFOoBzpBapRecall}{0.788\xspace}
\newcommand{\XttGccFOoBzpBapPrecision}{1\xspace}
\newcommand{\XttGccFOoBzpBapFone}{0.881\xspace}
\newcommand{\XttGccFOoBzpGhiRecall}{1\xspace}
\newcommand{\XttGccFOoBzpGhiPrecision}{1\xspace}
\newcommand{\XttGccFOoBzpGhiFone}{1\xspace}
\newcommand{\XttGccFOoBzpRdaRecall}{0.888\xspace}
\newcommand{\XttGccFOoBzpRdaPrecision}{1\xspace}
\newcommand{\XttGccFOoBzpRdaFone}{0.940\xspace}
\newcommand{\XttGccFOoBzpRseRecall}{1\xspace}
\newcommand{\XttGccFOoBzpRsePrecision}{1\xspace}
\newcommand{\XttGccFOoBzpRseFone}{1\xspace}
\newcommand{\XttGccFOoGccGT}{1341015\xspace}
\newcommand{\XttGccFOoGccLsRecall}{1\xspace}
\newcommand{\XttGccFOoGccLsPrecision}{1\xspace}
\newcommand{\XttGccFOoGccLsFone}{1\xspace}
\newcommand{\XttGccFOoGccBapRecall}{0.706\xspace}
\newcommand{\XttGccFOoGccBapPrecision}{1.000\xspace}
\newcommand{\XttGccFOoGccBapFone}{0.828\xspace}
\newcommand{\XttGccFOoGccGhiRecall}{0.998\xspace}
\newcommand{\XttGccFOoGccGhiPrecision}{1\xspace}
\newcommand{\XttGccFOoGccGhiFone}{0.999\xspace}
\newcommand{\XttGccFOoGccRdaRecall}{0.859\xspace}
\newcommand{\XttGccFOoGccRdaPrecision}{1\xspace}
\newcommand{\XttGccFOoGccRdaFone}{0.924\xspace}
\newcommand{\XttGccFOoGccRseRecall}{0.993\xspace}
\newcommand{\XttGccFOoGccRsePrecision}{1\xspace}
\newcommand{\XttGccFOoGccRseFone}{0.996\xspace}
\newcommand{\XttGccFOoGzpGT}{12768\xspace}
\newcommand{\XttGccFOoGzpLsRecall}{1\xspace}
\newcommand{\XttGccFOoGzpLsPrecision}{1\xspace}
\newcommand{\XttGccFOoGzpLsFone}{1\xspace}
\newcommand{\XttGccFOoGzpBapRecall}{0.990\xspace}
\newcommand{\XttGccFOoGzpBapPrecision}{1\xspace}
\newcommand{\XttGccFOoGzpBapFone}{0.995\xspace}
\newcommand{\XttGccFOoGzpGhiRecall}{1\xspace}
\newcommand{\XttGccFOoGzpGhiPrecision}{1\xspace}
\newcommand{\XttGccFOoGzpGhiFone}{1\xspace}
\newcommand{\XttGccFOoGzpRdaRecall}{1\xspace}
\newcommand{\XttGccFOoGzpRdaPrecision}{1\xspace}
\newcommand{\XttGccFOoGzpRdaFone}{1\xspace}
\newcommand{\XttGccFOoGzpRseRecall}{0.998\xspace}
\newcommand{\XttGccFOoGzpRsePrecision}{1\xspace}
\newcommand{\XttGccFOoGzpRseFone}{0.999\xspace}
\newcommand{\XttGccFOoOggGT}{58709\xspace}
\newcommand{\XttGccFOoOggLsRecall}{1\xspace}
\newcommand{\XttGccFOoOggLsPrecision}{1\xspace}
\newcommand{\XttGccFOoOggLsFone}{1\xspace}
\newcommand{\XttGccFOoOggBapRecall}{0.976\xspace}
\newcommand{\XttGccFOoOggBapPrecision}{1\xspace}
\newcommand{\XttGccFOoOggBapFone}{0.988\xspace}
\newcommand{\XttGccFOoOggGhiRecall}{1\xspace}
\newcommand{\XttGccFOoOggGhiPrecision}{1\xspace}
\newcommand{\XttGccFOoOggGhiFone}{1\xspace}
\newcommand{\XttGccFOoOggRdaRecall}{0.982\xspace}
\newcommand{\XttGccFOoOggRdaPrecision}{1\xspace}
\newcommand{\XttGccFOoOggRdaFone}{0.991\xspace}
\newcommand{\XttGccFOoOggRseRecall}{0.993\xspace}
\newcommand{\XttGccFOoOggRsePrecision}{1\xspace}
\newcommand{\XttGccFOoOggRseFone}{0.996\xspace}
\newcommand{\XttGccFOoNgxGT}{172051\xspace}
\newcommand{\XttGccFOoNgxLsRecall}{1\xspace}
\newcommand{\XttGccFOoNgxLsPrecision}{1\xspace}
\newcommand{\XttGccFOoNgxLsFone}{1\xspace}
\newcommand{\XttGccFOoNgxBapRecall}{0.972\xspace}
\newcommand{\XttGccFOoNgxBapPrecision}{1\xspace}
\newcommand{\XttGccFOoNgxBapFone}{0.986\xspace}
\newcommand{\XttGccFOoNgxGhiRecall}{1\xspace}
\newcommand{\XttGccFOoNgxGhiPrecision}{1\xspace}
\newcommand{\XttGccFOoNgxGhiFone}{1\xspace}
\newcommand{\XttGccFOoNgxRdaRecall}{0.976\xspace}
\newcommand{\XttGccFOoNgxRdaPrecision}{1\xspace}
\newcommand{\XttGccFOoNgxRdaFone}{0.988\xspace}
\newcommand{\XttGccFOoNgxRseRecall}{1.000\xspace}
\newcommand{\XttGccFOoNgxRsePrecision}{1\xspace}
\newcommand{\XttGccFOoNgxRseFone}{1.000\xspace}
\newcommand{\XttGccFOoSshGT}{177000\xspace}
\newcommand{\XttGccFOoSshLsRecall}{1\xspace}
\newcommand{\XttGccFOoSshLsPrecision}{1\xspace}
\newcommand{\XttGccFOoSshLsFone}{1\xspace}
\newcommand{\XttGccFOoSshBapRecall}{1\xspace}
\newcommand{\XttGccFOoSshBapPrecision}{1\xspace}
\newcommand{\XttGccFOoSshBapFone}{1\xspace}
\newcommand{\XttGccFOoSshGhiRecall}{0.274\xspace}
\newcommand{\XttGccFOoSshGhiPrecision}{0.985\xspace}
\newcommand{\XttGccFOoSshGhiFone}{0.429\xspace}
\newcommand{\XttGccFOoSshRdaRecall}{0.954\xspace}
\newcommand{\XttGccFOoSshRdaPrecision}{1.000\xspace}
\newcommand{\XttGccFOoSshRdaFone}{0.976\xspace}
\newcommand{\XttGccFOoSshRseRecall}{0.983\xspace}
\newcommand{\XttGccFOoSshRsePrecision}{1\xspace}
\newcommand{\XttGccFOoSshRseFone}{0.991\xspace}
\newcommand{\XttGccFOoPcrGT}{6340\xspace}
\newcommand{\XttGccFOoPcrLsRecall}{1\xspace}
\newcommand{\XttGccFOoPcrLsPrecision}{1\xspace}
\newcommand{\XttGccFOoPcrLsFone}{1\xspace}
\newcommand{\XttGccFOoPcrBapRecall}{0.915\xspace}
\newcommand{\XttGccFOoPcrBapPrecision}{1\xspace}
\newcommand{\XttGccFOoPcrBapFone}{0.956\xspace}
\newcommand{\XttGccFOoPcrGhiRecall}{1\xspace}
\newcommand{\XttGccFOoPcrGhiPrecision}{1\xspace}
\newcommand{\XttGccFOoPcrGhiFone}{1\xspace}
\newcommand{\XttGccFOoPcrRdaRecall}{0.988\xspace}
\newcommand{\XttGccFOoPcrRdaPrecision}{1\xspace}
\newcommand{\XttGccFOoPcrRdaFone}{0.994\xspace}
\newcommand{\XttGccFOoPcrRseRecall}{1\xspace}
\newcommand{\XttGccFOoPcrRsePrecision}{1\xspace}
\newcommand{\XttGccFOoPcrRseFone}{1\xspace}
\newcommand{\XttGccFOoSqlGT}{244320\xspace}
\newcommand{\XttGccFOoSqlLsRecall}{1\xspace}
\newcommand{\XttGccFOoSqlLsPrecision}{1\xspace}
\newcommand{\XttGccFOoSqlLsFone}{1\xspace}
\newcommand{\XttGccFOoSqlBapRecall}{0.882\xspace}
\newcommand{\XttGccFOoSqlBapPrecision}{1\xspace}
\newcommand{\XttGccFOoSqlBapFone}{0.937\xspace}
\newcommand{\XttGccFOoSqlGhiRecall}{1\xspace}
\newcommand{\XttGccFOoSqlGhiPrecision}{1\xspace}
\newcommand{\XttGccFOoSqlGhiFone}{1\xspace}
\newcommand{\XttGccFOoSqlRdaRecall}{0.964\xspace}
\newcommand{\XttGccFOoSqlRdaPrecision}{1\xspace}
\newcommand{\XttGccFOoSqlRdaFone}{0.982\xspace}
\newcommand{\XttGccFOoSqlRseRecall}{0.980\xspace}
\newcommand{\XttGccFOoSqlRsePrecision}{1\xspace}
\newcommand{\XttGccFOoSqlRseFone}{0.990\xspace}
\newcommand{\XttGccFOoVimGT}{674749\xspace}
\newcommand{\XttGccFOoVimLsRecall}{1\xspace}
\newcommand{\XttGccFOoVimLsPrecision}{1\xspace}
\newcommand{\XttGccFOoVimLsFone}{1\xspace}
\newcommand{\XttGccFOoVimBapRecall}{0.946\xspace}
\newcommand{\XttGccFOoVimBapPrecision}{1\xspace}
\newcommand{\XttGccFOoVimBapFone}{0.972\xspace}
\newcommand{\XttGccFOoVimGhiRecall}{1\xspace}
\newcommand{\XttGccFOoVimGhiPrecision}{1\xspace}
\newcommand{\XttGccFOoVimGhiFone}{1\xspace}
\newcommand{\XttGccFOoVimRdaRecall}{0.980\xspace}
\newcommand{\XttGccFOoVimRdaPrecision}{1\xspace}
\newcommand{\XttGccFOoVimRdaFone}{0.990\xspace}
\newcommand{\XttGccFOoVimRseRecall}{0.997\xspace}
\newcommand{\XttGccFOoVimRsePrecision}{1\xspace}
\newcommand{\XttGccFOoVimRseFone}{0.998\xspace}
\newcommand{\XttGccFOoVsfGT}{31376\xspace}
\newcommand{\XttGccFOoVsfLsRecall}{1\xspace}
\newcommand{\XttGccFOoVsfLsPrecision}{1\xspace}
\newcommand{\XttGccFOoVsfLsFone}{1\xspace}
\newcommand{\XttGccFOoVsfBapRecall}{1\xspace}
\newcommand{\XttGccFOoVsfBapPrecision}{1\xspace}
\newcommand{\XttGccFOoVsfBapFone}{1\xspace}
\newcommand{\XttGccFOoVsfGhiRecall}{0.150\xspace}
\newcommand{\XttGccFOoVsfGhiPrecision}{0.986\xspace}
\newcommand{\XttGccFOoVsfGhiFone}{0.261\xspace}
\newcommand{\XttGccFOoVsfRdaRecall}{0.988\xspace}
\newcommand{\XttGccFOoVsfRdaPrecision}{1\xspace}
\newcommand{\XttGccFOoVsfRdaFone}{0.994\xspace}
\newcommand{\XttGccFOoVsfRseRecall}{1.000\xspace}
\newcommand{\XttGccFOoVsfRsePrecision}{1\xspace}
\newcommand{\XttGccFOoVsfRseFone}{1.000\xspace}
\newcommand{\XttGccFOaSzpGT}{13850\xspace}
\newcommand{\XttGccFOaSzpLsRecall}{1\xspace}
\newcommand{\XttGccFOaSzpLsPrecision}{1\xspace}
\newcommand{\XttGccFOaSzpLsFone}{1\xspace}
\newcommand{\XttGccFOaSzpBapRecall}{0.975\xspace}
\newcommand{\XttGccFOaSzpBapPrecision}{1.000\xspace}
\newcommand{\XttGccFOaSzpBapFone}{0.987\xspace}
\newcommand{\XttGccFOaSzpGhiRecall}{1\xspace}
\newcommand{\XttGccFOaSzpGhiPrecision}{1\xspace}
\newcommand{\XttGccFOaSzpGhiFone}{1\xspace}
\newcommand{\XttGccFOaSzpRdaRecall}{0.988\xspace}
\newcommand{\XttGccFOaSzpRdaPrecision}{1\xspace}
\newcommand{\XttGccFOaSzpRdaFone}{0.994\xspace}
\newcommand{\XttGccFOaSzpRseRecall}{1\xspace}
\newcommand{\XttGccFOaSzpRsePrecision}{1\xspace}
\newcommand{\XttGccFOaSzpRseFone}{1\xspace}
\newcommand{\XttGccFOaCapGT}{271355\xspace}
\newcommand{\XttGccFOaCapLsRecall}{1\xspace}
\newcommand{\XttGccFOaCapLsPrecision}{1\xspace}
\newcommand{\XttGccFOaCapLsFone}{1\xspace}
\newcommand{\XttGccFOaCapBapRecall}{0.423\xspace}
\newcommand{\XttGccFOaCapBapPrecision}{1\xspace}
\newcommand{\XttGccFOaCapBapFone}{0.594\xspace}
\newcommand{\XttGccFOaCapGhiRecall}{0.868\xspace}
\newcommand{\XttGccFOaCapGhiPrecision}{1\xspace}
\newcommand{\XttGccFOaCapGhiFone}{0.929\xspace}
\newcommand{\XttGccFOaCapRdaRecall}{0.854\xspace}
\newcommand{\XttGccFOaCapRdaPrecision}{1\xspace}
\newcommand{\XttGccFOaCapRdaFone}{0.921\xspace}
\newcommand{\XttGccFOaCapRseRecall}{1.000\xspace}
\newcommand{\XttGccFOaCapRsePrecision}{1\xspace}
\newcommand{\XttGccFOaCapRseFone}{1.000\xspace}
\newcommand{\XttGccFOaExmGT}{150760\xspace}
\newcommand{\XttGccFOaExmLsRecall}{1\xspace}
\newcommand{\XttGccFOaExmLsPrecision}{1\xspace}
\newcommand{\XttGccFOaExmLsFone}{1\xspace}
\newcommand{\XttGccFOaExmBapRecall}{0.837\xspace}
\newcommand{\XttGccFOaExmBapPrecision}{1\xspace}
\newcommand{\XttGccFOaExmBapFone}{0.911\xspace}
\newcommand{\XttGccFOaExmGhiRecall}{1\xspace}
\newcommand{\XttGccFOaExmGhiPrecision}{1\xspace}
\newcommand{\XttGccFOaExmGhiFone}{1\xspace}
\newcommand{\XttGccFOaExmRdaRecall}{0.970\xspace}
\newcommand{\XttGccFOaExmRdaPrecision}{1\xspace}
\newcommand{\XttGccFOaExmRdaFone}{0.985\xspace}
\newcommand{\XttGccFOaExmRseRecall}{1.000\xspace}
\newcommand{\XttGccFOaExmRsePrecision}{1\xspace}
\newcommand{\XttGccFOaExmRseFone}{1.000\xspace}
\newcommand{\XttGccFOaLgtGT}{29171\xspace}
\newcommand{\XttGccFOaLgtLsRecall}{1\xspace}
\newcommand{\XttGccFOaLgtLsPrecision}{1\xspace}
\newcommand{\XttGccFOaLgtLsFone}{1\xspace}
\newcommand{\XttGccFOaLgtBapRecall}{0.873\xspace}
\newcommand{\XttGccFOaLgtBapPrecision}{1\xspace}
\newcommand{\XttGccFOaLgtBapFone}{0.932\xspace}
\newcommand{\XttGccFOaLgtGhiRecall}{1\xspace}
\newcommand{\XttGccFOaLgtGhiPrecision}{1\xspace}
\newcommand{\XttGccFOaLgtGhiFone}{1\xspace}
\newcommand{\XttGccFOaLgtRdaRecall}{0.985\xspace}
\newcommand{\XttGccFOaLgtRdaPrecision}{1\xspace}
\newcommand{\XttGccFOaLgtRdaFone}{0.992\xspace}
\newcommand{\XttGccFOaLgtRseRecall}{1\xspace}
\newcommand{\XttGccFOaLgtRsePrecision}{1\xspace}
\newcommand{\XttGccFOaLgtRseFone}{1\xspace}
\newcommand{\XttGccFOaBzpGT}{15208\xspace}
\newcommand{\XttGccFOaBzpLsRecall}{1\xspace}
\newcommand{\XttGccFOaBzpLsPrecision}{1\xspace}
\newcommand{\XttGccFOaBzpLsFone}{1\xspace}
\newcommand{\XttGccFOaBzpBapRecall}{0.836\xspace}
\newcommand{\XttGccFOaBzpBapPrecision}{1\xspace}
\newcommand{\XttGccFOaBzpBapFone}{0.911\xspace}
\newcommand{\XttGccFOaBzpGhiRecall}{1\xspace}
\newcommand{\XttGccFOaBzpGhiPrecision}{1\xspace}
\newcommand{\XttGccFOaBzpGhiFone}{1\xspace}
\newcommand{\XttGccFOaBzpRdaRecall}{0.857\xspace}
\newcommand{\XttGccFOaBzpRdaPrecision}{1\xspace}
\newcommand{\XttGccFOaBzpRdaFone}{0.923\xspace}
\newcommand{\XttGccFOaBzpRseRecall}{1\xspace}
\newcommand{\XttGccFOaBzpRsePrecision}{1\xspace}
\newcommand{\XttGccFOaBzpRseFone}{1\xspace}
\newcommand{\XttGccFOaGccGT}{909210\xspace}
\newcommand{\XttGccFOaGccLsRecall}{1\xspace}
\newcommand{\XttGccFOaGccLsPrecision}{1\xspace}
\newcommand{\XttGccFOaGccLsFone}{1\xspace}
\newcommand{\XttGccFOaGccBapRecall}{0.685\xspace}
\newcommand{\XttGccFOaGccBapPrecision}{1\xspace}
\newcommand{\XttGccFOaGccBapFone}{0.813\xspace}
\newcommand{\XttGccFOaGccGhiRecall}{0.987\xspace}
\newcommand{\XttGccFOaGccGhiPrecision}{1\xspace}
\newcommand{\XttGccFOaGccGhiFone}{0.993\xspace}
\newcommand{\XttGccFOaGccRdaRecall}{0.871\xspace}
\newcommand{\XttGccFOaGccRdaPrecision}{1\xspace}
\newcommand{\XttGccFOaGccRdaFone}{0.931\xspace}
\newcommand{\XttGccFOaGccRseRecall}{0.998\xspace}
\newcommand{\XttGccFOaGccRsePrecision}{1\xspace}
\newcommand{\XttGccFOaGccRseFone}{0.999\xspace}
\newcommand{\XttGccFOaGzpGT}{9925\xspace}
\newcommand{\XttGccFOaGzpLsRecall}{1\xspace}
\newcommand{\XttGccFOaGzpLsPrecision}{1\xspace}
\newcommand{\XttGccFOaGzpLsFone}{1\xspace}
\newcommand{\XttGccFOaGzpBapRecall}{0.985\xspace}
\newcommand{\XttGccFOaGzpBapPrecision}{1\xspace}
\newcommand{\XttGccFOaGzpBapFone}{0.992\xspace}
\newcommand{\XttGccFOaGzpGhiRecall}{1\xspace}
\newcommand{\XttGccFOaGzpGhiPrecision}{1\xspace}
\newcommand{\XttGccFOaGzpGhiFone}{1\xspace}
\newcommand{\XttGccFOaGzpRdaRecall}{1\xspace}
\newcommand{\XttGccFOaGzpRdaPrecision}{1\xspace}
\newcommand{\XttGccFOaGzpRdaFone}{1\xspace}
\newcommand{\XttGccFOaGzpRseRecall}{1\xspace}
\newcommand{\XttGccFOaGzpRsePrecision}{1\xspace}
\newcommand{\XttGccFOaGzpRseFone}{1\xspace}
\newcommand{\XttGccFOaOggGT}{41304\xspace}
\newcommand{\XttGccFOaOggLsRecall}{1\xspace}
\newcommand{\XttGccFOaOggLsPrecision}{1\xspace}
\newcommand{\XttGccFOaOggLsFone}{1\xspace}
\newcommand{\XttGccFOaOggBapRecall}{0.978\xspace}
\newcommand{\XttGccFOaOggBapPrecision}{1\xspace}
\newcommand{\XttGccFOaOggBapFone}{0.989\xspace}
\newcommand{\XttGccFOaOggGhiRecall}{1\xspace}
\newcommand{\XttGccFOaOggGhiPrecision}{1\xspace}
\newcommand{\XttGccFOaOggGhiFone}{1\xspace}
\newcommand{\XttGccFOaOggRdaRecall}{1\xspace}
\newcommand{\XttGccFOaOggRdaPrecision}{1\xspace}
\newcommand{\XttGccFOaOggRdaFone}{1\xspace}
\newcommand{\XttGccFOaOggRseRecall}{1\xspace}
\newcommand{\XttGccFOaOggRsePrecision}{1\xspace}
\newcommand{\XttGccFOaOggRseFone}{1\xspace}
\newcommand{\XttGccFOaNgxGT}{113361\xspace}
\newcommand{\XttGccFOaNgxLsRecall}{1\xspace}
\newcommand{\XttGccFOaNgxLsPrecision}{1\xspace}
\newcommand{\XttGccFOaNgxLsFone}{1\xspace}
\newcommand{\XttGccFOaNgxBapRecall}{0.965\xspace}
\newcommand{\XttGccFOaNgxBapPrecision}{1\xspace}
\newcommand{\XttGccFOaNgxBapFone}{0.982\xspace}
\newcommand{\XttGccFOaNgxGhiRecall}{1\xspace}
\newcommand{\XttGccFOaNgxGhiPrecision}{1\xspace}
\newcommand{\XttGccFOaNgxGhiFone}{1\xspace}
\newcommand{\XttGccFOaNgxRdaRecall}{1.000\xspace}
\newcommand{\XttGccFOaNgxRdaPrecision}{1\xspace}
\newcommand{\XttGccFOaNgxRdaFone}{1.000\xspace}
\newcommand{\XttGccFOaNgxRseRecall}{1\xspace}
\newcommand{\XttGccFOaNgxRsePrecision}{1\xspace}
\newcommand{\XttGccFOaNgxRseFone}{1\xspace}
\newcommand{\XttGccFOaSshGT}{132073\xspace}
\newcommand{\XttGccFOaSshLsRecall}{1\xspace}
\newcommand{\XttGccFOaSshLsPrecision}{1\xspace}
\newcommand{\XttGccFOaSshLsFone}{1\xspace}
\newcommand{\XttGccFOaSshBapRecall}{1\xspace}
\newcommand{\XttGccFOaSshBapPrecision}{1\xspace}
\newcommand{\XttGccFOaSshBapFone}{1\xspace}
\newcommand{\XttGccFOaSshGhiRecall}{0.327\xspace}
\newcommand{\XttGccFOaSshGhiPrecision}{0.986\xspace}
\newcommand{\XttGccFOaSshGhiFone}{0.491\xspace}
\newcommand{\XttGccFOaSshRdaRecall}{0.931\xspace}
\newcommand{\XttGccFOaSshRdaPrecision}{1\xspace}
\newcommand{\XttGccFOaSshRdaFone}{0.964\xspace}
\newcommand{\XttGccFOaSshRseRecall}{0.984\xspace}
\newcommand{\XttGccFOaSshRsePrecision}{1\xspace}
\newcommand{\XttGccFOaSshRseFone}{0.992\xspace}
\newcommand{\XttGccFOaPcrGT}{4852\xspace}
\newcommand{\XttGccFOaPcrLsRecall}{1\xspace}
\newcommand{\XttGccFOaPcrLsPrecision}{1\xspace}
\newcommand{\XttGccFOaPcrLsFone}{1\xspace}
\newcommand{\XttGccFOaPcrBapRecall}{0.889\xspace}
\newcommand{\XttGccFOaPcrBapPrecision}{1\xspace}
\newcommand{\XttGccFOaPcrBapFone}{0.941\xspace}
\newcommand{\XttGccFOaPcrGhiRecall}{1\xspace}
\newcommand{\XttGccFOaPcrGhiPrecision}{1\xspace}
\newcommand{\XttGccFOaPcrGhiFone}{1\xspace}
\newcommand{\XttGccFOaPcrRdaRecall}{1\xspace}
\newcommand{\XttGccFOaPcrRdaPrecision}{1\xspace}
\newcommand{\XttGccFOaPcrRdaFone}{1\xspace}
\newcommand{\XttGccFOaPcrRseRecall}{1\xspace}
\newcommand{\XttGccFOaPcrRsePrecision}{1\xspace}
\newcommand{\XttGccFOaPcrRseFone}{1\xspace}
\newcommand{\XttGccFOaSqlGT}{167679\xspace}
\newcommand{\XttGccFOaSqlLsRecall}{1\xspace}
\newcommand{\XttGccFOaSqlLsPrecision}{1\xspace}
\newcommand{\XttGccFOaSqlLsFone}{1\xspace}
\newcommand{\XttGccFOaSqlBapRecall}{0.869\xspace}
\newcommand{\XttGccFOaSqlBapPrecision}{1.000\xspace}
\newcommand{\XttGccFOaSqlBapFone}{0.930\xspace}
\newcommand{\XttGccFOaSqlGhiRecall}{1\xspace}
\newcommand{\XttGccFOaSqlGhiPrecision}{1\xspace}
\newcommand{\XttGccFOaSqlGhiFone}{1\xspace}
\newcommand{\XttGccFOaSqlRdaRecall}{0.998\xspace}
\newcommand{\XttGccFOaSqlRdaPrecision}{1\xspace}
\newcommand{\XttGccFOaSqlRdaFone}{0.999\xspace}
\newcommand{\XttGccFOaSqlRseRecall}{1\xspace}
\newcommand{\XttGccFOaSqlRsePrecision}{1\xspace}
\newcommand{\XttGccFOaSqlRseFone}{1\xspace}
\newcommand{\XttGccFOaVimGT}{517969\xspace}
\newcommand{\XttGccFOaVimLsRecall}{1\xspace}
\newcommand{\XttGccFOaVimLsPrecision}{1\xspace}
\newcommand{\XttGccFOaVimLsFone}{1\xspace}
\newcommand{\XttGccFOaVimBapRecall}{0.921\xspace}
\newcommand{\XttGccFOaVimBapPrecision}{1\xspace}
\newcommand{\XttGccFOaVimBapFone}{0.959\xspace}
\newcommand{\XttGccFOaVimGhiRecall}{1.000\xspace}
\newcommand{\XttGccFOaVimGhiPrecision}{1\xspace}
\newcommand{\XttGccFOaVimGhiFone}{1.000\xspace}
\newcommand{\XttGccFOaVimRdaRecall}{0.990\xspace}
\newcommand{\XttGccFOaVimRdaPrecision}{1\xspace}
\newcommand{\XttGccFOaVimRdaFone}{0.995\xspace}
\newcommand{\XttGccFOaVimRseRecall}{1.000\xspace}
\newcommand{\XttGccFOaVimRsePrecision}{1\xspace}
\newcommand{\XttGccFOaVimRseFone}{1.000\xspace}
\newcommand{\XttGccFOaVsfGT}{24186\xspace}
\newcommand{\XttGccFOaVsfLsRecall}{1\xspace}
\newcommand{\XttGccFOaVsfLsPrecision}{1\xspace}
\newcommand{\XttGccFOaVsfLsFone}{1\xspace}
\newcommand{\XttGccFOaVsfBapRecall}{1\xspace}
\newcommand{\XttGccFOaVsfBapPrecision}{1\xspace}
\newcommand{\XttGccFOaVsfBapFone}{1\xspace}
\newcommand{\XttGccFOaVsfGhiRecall}{0.152\xspace}
\newcommand{\XttGccFOaVsfGhiPrecision}{0.996\xspace}
\newcommand{\XttGccFOaVsfGhiFone}{0.264\xspace}
\newcommand{\XttGccFOaVsfRdaRecall}{0.987\xspace}
\newcommand{\XttGccFOaVsfRdaPrecision}{1\xspace}
\newcommand{\XttGccFOaVsfRdaFone}{0.993\xspace}
\newcommand{\XttGccFOaVsfRseRecall}{0.999\xspace}
\newcommand{\XttGccFOaVsfRsePrecision}{1\xspace}
\newcommand{\XttGccFOaVsfRseFone}{1.000\xspace}
\newcommand{\XttGccFObSzpGT}{13786\xspace}
\newcommand{\XttGccFObSzpLsRecall}{1\xspace}
\newcommand{\XttGccFObSzpLsPrecision}{1\xspace}
\newcommand{\XttGccFObSzpLsFone}{1\xspace}
\newcommand{\XttGccFObSzpBapRecall}{0.976\xspace}
\newcommand{\XttGccFObSzpBapPrecision}{1\xspace}
\newcommand{\XttGccFObSzpBapFone}{0.988\xspace}
\newcommand{\XttGccFObSzpGhiRecall}{1\xspace}
\newcommand{\XttGccFObSzpGhiPrecision}{1\xspace}
\newcommand{\XttGccFObSzpGhiFone}{1\xspace}
\newcommand{\XttGccFObSzpRdaRecall}{0.971\xspace}
\newcommand{\XttGccFObSzpRdaPrecision}{1\xspace}
\newcommand{\XttGccFObSzpRdaFone}{0.985\xspace}
\newcommand{\XttGccFObSzpRseRecall}{1\xspace}
\newcommand{\XttGccFObSzpRsePrecision}{1\xspace}
\newcommand{\XttGccFObSzpRseFone}{1\xspace}
\newcommand{\XttGccFObCapGT}{238243\xspace}
\newcommand{\XttGccFObCapLsRecall}{1\xspace}
\newcommand{\XttGccFObCapLsPrecision}{1\xspace}
\newcommand{\XttGccFObCapLsFone}{1\xspace}
\newcommand{\XttGccFObCapBapRecall}{0.487\xspace}
\newcommand{\XttGccFObCapBapPrecision}{1\xspace}
\newcommand{\XttGccFObCapBapFone}{0.655\xspace}
\newcommand{\XttGccFObCapGhiRecall}{0.854\xspace}
\newcommand{\XttGccFObCapGhiPrecision}{1\xspace}
\newcommand{\XttGccFObCapGhiFone}{0.921\xspace}
\newcommand{\XttGccFObCapRdaRecall}{0.822\xspace}
\newcommand{\XttGccFObCapRdaPrecision}{1\xspace}
\newcommand{\XttGccFObCapRdaFone}{0.902\xspace}
\newcommand{\XttGccFObCapRseRecall}{0.993\xspace}
\newcommand{\XttGccFObCapRsePrecision}{1\xspace}
\newcommand{\XttGccFObCapRseFone}{0.997\xspace}
\newcommand{\XttGccFObExmGT}{151690\xspace}
\newcommand{\XttGccFObExmLsRecall}{1\xspace}
\newcommand{\XttGccFObExmLsPrecision}{1\xspace}
\newcommand{\XttGccFObExmLsFone}{1\xspace}
\newcommand{\XttGccFObExmBapRecall}{0.863\xspace}
\newcommand{\XttGccFObExmBapPrecision}{1\xspace}
\newcommand{\XttGccFObExmBapFone}{0.926\xspace}
\newcommand{\XttGccFObExmGhiRecall}{0.988\xspace}
\newcommand{\XttGccFObExmGhiPrecision}{1\xspace}
\newcommand{\XttGccFObExmGhiFone}{0.994\xspace}
\newcommand{\XttGccFObExmRdaRecall}{0.947\xspace}
\newcommand{\XttGccFObExmRdaPrecision}{1\xspace}
\newcommand{\XttGccFObExmRdaFone}{0.973\xspace}
\newcommand{\XttGccFObExmRseRecall}{1.000\xspace}
\newcommand{\XttGccFObExmRsePrecision}{1\xspace}
\newcommand{\XttGccFObExmRseFone}{1.000\xspace}
\newcommand{\XttGccFObLgtGT}{29080\xspace}
\newcommand{\XttGccFObLgtLsRecall}{1\xspace}
\newcommand{\XttGccFObLgtLsPrecision}{1\xspace}
\newcommand{\XttGccFObLgtLsFone}{1\xspace}
\newcommand{\XttGccFObLgtBapRecall}{0.890\xspace}
\newcommand{\XttGccFObLgtBapPrecision}{1\xspace}
\newcommand{\XttGccFObLgtBapFone}{0.942\xspace}
\newcommand{\XttGccFObLgtGhiRecall}{0.970\xspace}
\newcommand{\XttGccFObLgtGhiPrecision}{1\xspace}
\newcommand{\XttGccFObLgtGhiFone}{0.985\xspace}
\newcommand{\XttGccFObLgtRdaRecall}{0.984\xspace}
\newcommand{\XttGccFObLgtRdaPrecision}{1\xspace}
\newcommand{\XttGccFObLgtRdaFone}{0.992\xspace}
\newcommand{\XttGccFObLgtRseRecall}{1\xspace}
\newcommand{\XttGccFObLgtRsePrecision}{1\xspace}
\newcommand{\XttGccFObLgtRseFone}{1\xspace}
\newcommand{\XttGccFObBzpGT}{15786\xspace}
\newcommand{\XttGccFObBzpLsRecall}{1\xspace}
\newcommand{\XttGccFObBzpLsPrecision}{1\xspace}
\newcommand{\XttGccFObBzpLsFone}{1\xspace}
\newcommand{\XttGccFObBzpBapRecall}{0.938\xspace}
\newcommand{\XttGccFObBzpBapPrecision}{1\xspace}
\newcommand{\XttGccFObBzpBapFone}{0.968\xspace}
\newcommand{\XttGccFObBzpGhiRecall}{1\xspace}
\newcommand{\XttGccFObBzpGhiPrecision}{1\xspace}
\newcommand{\XttGccFObBzpGhiFone}{1\xspace}
\newcommand{\XttGccFObBzpRdaRecall}{0.814\xspace}
\newcommand{\XttGccFObBzpRdaPrecision}{1\xspace}
\newcommand{\XttGccFObBzpRdaFone}{0.897\xspace}
\newcommand{\XttGccFObBzpRseRecall}{1\xspace}
\newcommand{\XttGccFObBzpRsePrecision}{1\xspace}
\newcommand{\XttGccFObBzpRseFone}{1\xspace}
\newcommand{\XttGccFObGccGT}{937181\xspace}
\newcommand{\XttGccFObGccLsRecall}{1\xspace}
\newcommand{\XttGccFObGccLsPrecision}{1\xspace}
\newcommand{\XttGccFObGccLsFone}{1\xspace}
\newcommand{\XttGccFObGccBapRecall}{0.713\xspace}
\newcommand{\XttGccFObGccBapPrecision}{1\xspace}
\newcommand{\XttGccFObGccBapFone}{0.832\xspace}
\newcommand{\XttGccFObGccGhiRecall}{0.985\xspace}
\newcommand{\XttGccFObGccGhiPrecision}{1\xspace}
\newcommand{\XttGccFObGccGhiFone}{0.992\xspace}
\newcommand{\XttGccFObGccRdaRecall}{0.845\xspace}
\newcommand{\XttGccFObGccRdaPrecision}{1\xspace}
\newcommand{\XttGccFObGccRdaFone}{0.916\xspace}
\newcommand{\XttGccFObGccRseRecall}{1.000\xspace}
\newcommand{\XttGccFObGccRsePrecision}{1\xspace}
\newcommand{\XttGccFObGccRseFone}{1.000\xspace}
\newcommand{\XttGccFObGzpGT}{10358\xspace}
\newcommand{\XttGccFObGzpLsRecall}{1\xspace}
\newcommand{\XttGccFObGzpLsPrecision}{1\xspace}
\newcommand{\XttGccFObGzpLsFone}{1\xspace}
\newcommand{\XttGccFObGzpBapRecall}{0.986\xspace}
\newcommand{\XttGccFObGzpBapPrecision}{1\xspace}
\newcommand{\XttGccFObGzpBapFone}{0.993\xspace}
\newcommand{\XttGccFObGzpGhiRecall}{1\xspace}
\newcommand{\XttGccFObGzpGhiPrecision}{1\xspace}
\newcommand{\XttGccFObGzpGhiFone}{1\xspace}
\newcommand{\XttGccFObGzpRdaRecall}{1\xspace}
\newcommand{\XttGccFObGzpRdaPrecision}{1\xspace}
\newcommand{\XttGccFObGzpRdaFone}{1\xspace}
\newcommand{\XttGccFObGzpRseRecall}{1\xspace}
\newcommand{\XttGccFObGzpRsePrecision}{1\xspace}
\newcommand{\XttGccFObGzpRseFone}{1\xspace}
\newcommand{\XttGccFObOggGT}{43662\xspace}
\newcommand{\XttGccFObOggLsRecall}{1\xspace}
\newcommand{\XttGccFObOggLsPrecision}{1\xspace}
\newcommand{\XttGccFObOggLsFone}{1\xspace}
\newcommand{\XttGccFObOggBapRecall}{0.980\xspace}
\newcommand{\XttGccFObOggBapPrecision}{1\xspace}
\newcommand{\XttGccFObOggBapFone}{0.990\xspace}
\newcommand{\XttGccFObOggGhiRecall}{1\xspace}
\newcommand{\XttGccFObOggGhiPrecision}{1\xspace}
\newcommand{\XttGccFObOggGhiFone}{1\xspace}
\newcommand{\XttGccFObOggRdaRecall}{1\xspace}
\newcommand{\XttGccFObOggRdaPrecision}{1\xspace}
\newcommand{\XttGccFObOggRdaFone}{1\xspace}
\newcommand{\XttGccFObOggRseRecall}{1\xspace}
\newcommand{\XttGccFObOggRsePrecision}{1\xspace}
\newcommand{\XttGccFObOggRseFone}{1\xspace}
\newcommand{\XttGccFObNgxGT}{113542\xspace}
\newcommand{\XttGccFObNgxLsRecall}{1\xspace}
\newcommand{\XttGccFObNgxLsPrecision}{1\xspace}
\newcommand{\XttGccFObNgxLsFone}{1\xspace}
\newcommand{\XttGccFObNgxBapRecall}{0.963\xspace}
\newcommand{\XttGccFObNgxBapPrecision}{1\xspace}
\newcommand{\XttGccFObNgxBapFone}{0.981\xspace}
\newcommand{\XttGccFObNgxGhiRecall}{1\xspace}
\newcommand{\XttGccFObNgxGhiPrecision}{1\xspace}
\newcommand{\XttGccFObNgxGhiFone}{1\xspace}
\newcommand{\XttGccFObNgxRdaRecall}{0.985\xspace}
\newcommand{\XttGccFObNgxRdaPrecision}{1\xspace}
\newcommand{\XttGccFObNgxRdaFone}{0.993\xspace}
\newcommand{\XttGccFObNgxRseRecall}{1\xspace}
\newcommand{\XttGccFObNgxRsePrecision}{1\xspace}
\newcommand{\XttGccFObNgxRseFone}{1\xspace}
\newcommand{\XttGccFObSshGT}{134096\xspace}
\newcommand{\XttGccFObSshLsRecall}{1\xspace}
\newcommand{\XttGccFObSshLsPrecision}{1\xspace}
\newcommand{\XttGccFObSshLsFone}{1\xspace}
\newcommand{\XttGccFObSshBapRecall}{1\xspace}
\newcommand{\XttGccFObSshBapPrecision}{1\xspace}
\newcommand{\XttGccFObSshBapFone}{1\xspace}
\newcommand{\XttGccFObSshGhiRecall}{0.409\xspace}
\newcommand{\XttGccFObSshGhiPrecision}{0.991\xspace}
\newcommand{\XttGccFObSshGhiFone}{0.579\xspace}
\newcommand{\XttGccFObSshRdaRecall}{0.946\xspace}
\newcommand{\XttGccFObSshRdaPrecision}{1.000\xspace}
\newcommand{\XttGccFObSshRdaFone}{0.972\xspace}
\newcommand{\XttGccFObSshRseRecall}{0.992\xspace}
\newcommand{\XttGccFObSshRsePrecision}{1\xspace}
\newcommand{\XttGccFObSshRseFone}{0.996\xspace}
\newcommand{\XttGccFObPcrGT}{4720\xspace}
\newcommand{\XttGccFObPcrLsRecall}{1\xspace}
\newcommand{\XttGccFObPcrLsPrecision}{1\xspace}
\newcommand{\XttGccFObPcrLsFone}{1\xspace}
\newcommand{\XttGccFObPcrBapRecall}{0.916\xspace}
\newcommand{\XttGccFObPcrBapPrecision}{1\xspace}
\newcommand{\XttGccFObPcrBapFone}{0.956\xspace}
\newcommand{\XttGccFObPcrGhiRecall}{1\xspace}
\newcommand{\XttGccFObPcrGhiPrecision}{1\xspace}
\newcommand{\XttGccFObPcrGhiFone}{1\xspace}
\newcommand{\XttGccFObPcrRdaRecall}{1\xspace}
\newcommand{\XttGccFObPcrRdaPrecision}{1\xspace}
\newcommand{\XttGccFObPcrRdaFone}{1\xspace}
\newcommand{\XttGccFObPcrRseRecall}{1\xspace}
\newcommand{\XttGccFObPcrRsePrecision}{1\xspace}
\newcommand{\XttGccFObPcrRseFone}{1\xspace}
\newcommand{\XttGccFObSqlGT}{188003\xspace}
\newcommand{\XttGccFObSqlLsRecall}{1\xspace}
\newcommand{\XttGccFObSqlLsPrecision}{1\xspace}
\newcommand{\XttGccFObSqlLsFone}{1\xspace}
\newcommand{\XttGccFObSqlBapRecall}{0.871\xspace}
\newcommand{\XttGccFObSqlBapPrecision}{1.000\xspace}
\newcommand{\XttGccFObSqlBapFone}{0.931\xspace}
\newcommand{\XttGccFObSqlGhiRecall}{0.951\xspace}
\newcommand{\XttGccFObSqlGhiPrecision}{1\xspace}
\newcommand{\XttGccFObSqlGhiFone}{0.975\xspace}
\newcommand{\XttGccFObSqlRdaRecall}{0.987\xspace}
\newcommand{\XttGccFObSqlRdaPrecision}{1\xspace}
\newcommand{\XttGccFObSqlRdaFone}{0.993\xspace}
\newcommand{\XttGccFObSqlRseRecall}{1\xspace}
\newcommand{\XttGccFObSqlRsePrecision}{1\xspace}
\newcommand{\XttGccFObSqlRseFone}{1\xspace}
\newcommand{\XttGccFObVimGT}{536743\xspace}
\newcommand{\XttGccFObVimLsRecall}{1\xspace}
\newcommand{\XttGccFObVimLsPrecision}{1\xspace}
\newcommand{\XttGccFObVimLsFone}{1\xspace}
\newcommand{\XttGccFObVimBapRecall}{0.928\xspace}
\newcommand{\XttGccFObVimBapPrecision}{1\xspace}
\newcommand{\XttGccFObVimBapFone}{0.962\xspace}
\newcommand{\XttGccFObVimGhiRecall}{1.000\xspace}
\newcommand{\XttGccFObVimGhiPrecision}{1\xspace}
\newcommand{\XttGccFObVimGhiFone}{1.000\xspace}
\newcommand{\XttGccFObVimRdaRecall}{0.962\xspace}
\newcommand{\XttGccFObVimRdaPrecision}{1.000\xspace}
\newcommand{\XttGccFObVimRdaFone}{0.981\xspace}
\newcommand{\XttGccFObVimRseRecall}{1.000\xspace}
\newcommand{\XttGccFObVimRsePrecision}{1\xspace}
\newcommand{\XttGccFObVimRseFone}{1.000\xspace}
\newcommand{\XttGccFObVsfGT}{25859\xspace}
\newcommand{\XttGccFObVsfLsRecall}{1\xspace}
\newcommand{\XttGccFObVsfLsPrecision}{1\xspace}
\newcommand{\XttGccFObVsfLsFone}{1\xspace}
\newcommand{\XttGccFObVsfBapRecall}{1\xspace}
\newcommand{\XttGccFObVsfBapPrecision}{1\xspace}
\newcommand{\XttGccFObVsfBapFone}{1\xspace}
\newcommand{\XttGccFObVsfGhiRecall}{0.138\xspace}
\newcommand{\XttGccFObVsfGhiPrecision}{0.993\xspace}
\newcommand{\XttGccFObVsfGhiFone}{0.242\xspace}
\newcommand{\XttGccFObVsfRdaRecall}{0.981\xspace}
\newcommand{\XttGccFObVsfRdaPrecision}{1\xspace}
\newcommand{\XttGccFObVsfRdaFone}{0.990\xspace}
\newcommand{\XttGccFObVsfRseRecall}{1\xspace}
\newcommand{\XttGccFObVsfRsePrecision}{1\xspace}
\newcommand{\XttGccFObVsfRseFone}{1\xspace}
\newcommand{\XttGccFOcSzpGT}{16290\xspace}
\newcommand{\XttGccFOcSzpLsRecall}{1\xspace}
\newcommand{\XttGccFOcSzpLsPrecision}{1\xspace}
\newcommand{\XttGccFOcSzpLsFone}{1\xspace}
\newcommand{\XttGccFOcSzpBapRecall}{0.962\xspace}
\newcommand{\XttGccFOcSzpBapPrecision}{1\xspace}
\newcommand{\XttGccFOcSzpBapFone}{0.981\xspace}
\newcommand{\XttGccFOcSzpGhiRecall}{1\xspace}
\newcommand{\XttGccFOcSzpGhiPrecision}{1\xspace}
\newcommand{\XttGccFOcSzpGhiFone}{1\xspace}
\newcommand{\XttGccFOcSzpRdaRecall}{0.966\xspace}
\newcommand{\XttGccFOcSzpRdaPrecision}{1\xspace}
\newcommand{\XttGccFOcSzpRdaFone}{0.983\xspace}
\newcommand{\XttGccFOcSzpRseRecall}{1\xspace}
\newcommand{\XttGccFOcSzpRsePrecision}{1\xspace}
\newcommand{\XttGccFOcSzpRseFone}{1\xspace}
\newcommand{\XttGccFOcCapGT}{260706\xspace}
\newcommand{\XttGccFOcCapLsRecall}{1\xspace}
\newcommand{\XttGccFOcCapLsPrecision}{1\xspace}
\newcommand{\XttGccFOcCapLsFone}{1\xspace}
\newcommand{\XttGccFOcCapBapRecall}{0.509\xspace}
\newcommand{\XttGccFOcCapBapPrecision}{1\xspace}
\newcommand{\XttGccFOcCapBapFone}{0.674\xspace}
\newcommand{\XttGccFOcCapGhiRecall}{0.866\xspace}
\newcommand{\XttGccFOcCapGhiPrecision}{1\xspace}
\newcommand{\XttGccFOcCapGhiFone}{0.928\xspace}
\newcommand{\XttGccFOcCapRdaRecall}{0.835\xspace}
\newcommand{\XttGccFOcCapRdaPrecision}{1\xspace}
\newcommand{\XttGccFOcCapRdaFone}{0.910\xspace}
\newcommand{\XttGccFOcCapRseRecall}{0.998\xspace}
\newcommand{\XttGccFOcCapRsePrecision}{1\xspace}
\newcommand{\XttGccFOcCapRseFone}{0.999\xspace}
\newcommand{\XttGccFOcExmGT}{182082\xspace}
\newcommand{\XttGccFOcExmLsRecall}{1\xspace}
\newcommand{\XttGccFOcExmLsPrecision}{1\xspace}
\newcommand{\XttGccFOcExmLsFone}{1\xspace}
\newcommand{\XttGccFOcExmBapRecall}{0.865\xspace}
\newcommand{\XttGccFOcExmBapPrecision}{1\xspace}
\newcommand{\XttGccFOcExmBapFone}{0.927\xspace}
\newcommand{\XttGccFOcExmGhiRecall}{0.990\xspace}
\newcommand{\XttGccFOcExmGhiPrecision}{1\xspace}
\newcommand{\XttGccFOcExmGhiFone}{0.995\xspace}
\newcommand{\XttGccFOcExmRdaRecall}{0.952\xspace}
\newcommand{\XttGccFOcExmRdaPrecision}{1\xspace}
\newcommand{\XttGccFOcExmRdaFone}{0.975\xspace}
\newcommand{\XttGccFOcExmRseRecall}{1.000\xspace}
\newcommand{\XttGccFOcExmRsePrecision}{1\xspace}
\newcommand{\XttGccFOcExmRseFone}{1.000\xspace}
\newcommand{\XttGccFOcLgtGT}{36210\xspace}
\newcommand{\XttGccFOcLgtLsRecall}{1\xspace}
\newcommand{\XttGccFOcLgtLsPrecision}{1\xspace}
\newcommand{\XttGccFOcLgtLsFone}{1\xspace}
\newcommand{\XttGccFOcLgtBapRecall}{0.899\xspace}
\newcommand{\XttGccFOcLgtBapPrecision}{1\xspace}
\newcommand{\XttGccFOcLgtBapFone}{0.947\xspace}
\newcommand{\XttGccFOcLgtGhiRecall}{0.976\xspace}
\newcommand{\XttGccFOcLgtGhiPrecision}{1\xspace}
\newcommand{\XttGccFOcLgtGhiFone}{0.988\xspace}
\newcommand{\XttGccFOcLgtRdaRecall}{0.973\xspace}
\newcommand{\XttGccFOcLgtRdaPrecision}{1\xspace}
\newcommand{\XttGccFOcLgtRdaFone}{0.986\xspace}
\newcommand{\XttGccFOcLgtRseRecall}{1\xspace}
\newcommand{\XttGccFOcLgtRsePrecision}{1\xspace}
\newcommand{\XttGccFOcLgtRseFone}{1\xspace}
\newcommand{\XttGccFOcBzpGT}{17897\xspace}
\newcommand{\XttGccFOcBzpLsRecall}{1\xspace}
\newcommand{\XttGccFOcBzpLsPrecision}{1\xspace}
\newcommand{\XttGccFOcBzpLsFone}{1\xspace}
\newcommand{\XttGccFOcBzpBapRecall}{0.945\xspace}
\newcommand{\XttGccFOcBzpBapPrecision}{1\xspace}
\newcommand{\XttGccFOcBzpBapFone}{0.972\xspace}
\newcommand{\XttGccFOcBzpGhiRecall}{1\xspace}
\newcommand{\XttGccFOcBzpGhiPrecision}{1\xspace}
\newcommand{\XttGccFOcBzpGhiFone}{1\xspace}
\newcommand{\XttGccFOcBzpRdaRecall}{0.818\xspace}
\newcommand{\XttGccFOcBzpRdaPrecision}{1\xspace}
\newcommand{\XttGccFOcBzpRdaFone}{0.900\xspace}
\newcommand{\XttGccFOcBzpRseRecall}{0.997\xspace}
\newcommand{\XttGccFOcBzpRsePrecision}{1\xspace}
\newcommand{\XttGccFOcBzpRseFone}{0.998\xspace}
\newcommand{\XttGccFOcGccGT}{1172779\xspace}
\newcommand{\XttGccFOcGccLsRecall}{1\xspace}
\newcommand{\XttGccFOcGccLsPrecision}{1\xspace}
\newcommand{\XttGccFOcGccLsFone}{1\xspace}
\newcommand{\XttGccFOcGccBapRecall}{0.760\xspace}
\newcommand{\XttGccFOcGccBapPrecision}{1\xspace}
\newcommand{\XttGccFOcGccBapFone}{0.863\xspace}
\newcommand{\XttGccFOcGccGhiRecall}{0.987\xspace}
\newcommand{\XttGccFOcGccGhiPrecision}{1\xspace}
\newcommand{\XttGccFOcGccGhiFone}{0.994\xspace}
\newcommand{\XttGccFOcGccRdaRecall}{0.863\xspace}
\newcommand{\XttGccFOcGccRdaPrecision}{1\xspace}
\newcommand{\XttGccFOcGccRdaFone}{0.926\xspace}
\newcommand{\XttGccFOcGccRseRecall}{1.000\xspace}
\newcommand{\XttGccFOcGccRsePrecision}{1\xspace}
\newcommand{\XttGccFOcGccRseFone}{1.000\xspace}
\newcommand{\XttGccFOcGzpGT}{17830\xspace}
\newcommand{\XttGccFOcGzpLsRecall}{1\xspace}
\newcommand{\XttGccFOcGzpLsPrecision}{1\xspace}
\newcommand{\XttGccFOcGzpLsFone}{1\xspace}
\newcommand{\XttGccFOcGzpBapRecall}{0.992\xspace}
\newcommand{\XttGccFOcGzpBapPrecision}{1\xspace}
\newcommand{\XttGccFOcGzpBapFone}{0.996\xspace}
\newcommand{\XttGccFOcGzpGhiRecall}{1\xspace}
\newcommand{\XttGccFOcGzpGhiPrecision}{1\xspace}
\newcommand{\XttGccFOcGzpGhiFone}{1\xspace}
\newcommand{\XttGccFOcGzpRdaRecall}{1\xspace}
\newcommand{\XttGccFOcGzpRdaPrecision}{1\xspace}
\newcommand{\XttGccFOcGzpRdaFone}{1\xspace}
\newcommand{\XttGccFOcGzpRseRecall}{1\xspace}
\newcommand{\XttGccFOcGzpRsePrecision}{1\xspace}
\newcommand{\XttGccFOcGzpRseFone}{1\xspace}
\newcommand{\XttGccFOcOggGT}{54879\xspace}
\newcommand{\XttGccFOcOggLsRecall}{1\xspace}
\newcommand{\XttGccFOcOggLsPrecision}{1\xspace}
\newcommand{\XttGccFOcOggLsFone}{1\xspace}
\newcommand{\XttGccFOcOggBapRecall}{0.983\xspace}
\newcommand{\XttGccFOcOggBapPrecision}{1\xspace}
\newcommand{\XttGccFOcOggBapFone}{0.992\xspace}
\newcommand{\XttGccFOcOggGhiRecall}{1\xspace}
\newcommand{\XttGccFOcOggGhiPrecision}{1\xspace}
\newcommand{\XttGccFOcOggGhiFone}{1\xspace}
\newcommand{\XttGccFOcOggRdaRecall}{1\xspace}
\newcommand{\XttGccFOcOggRdaPrecision}{1\xspace}
\newcommand{\XttGccFOcOggRdaFone}{1\xspace}
\newcommand{\XttGccFOcOggRseRecall}{1\xspace}
\newcommand{\XttGccFOcOggRsePrecision}{1\xspace}
\newcommand{\XttGccFOcOggRseFone}{1\xspace}
\newcommand{\XttGccFOcNgxGT}{126578\xspace}
\newcommand{\XttGccFOcNgxLsRecall}{1\xspace}
\newcommand{\XttGccFOcNgxLsPrecision}{1\xspace}
\newcommand{\XttGccFOcNgxLsFone}{1\xspace}
\newcommand{\XttGccFOcNgxBapRecall}{0.964\xspace}
\newcommand{\XttGccFOcNgxBapPrecision}{1\xspace}
\newcommand{\XttGccFOcNgxBapFone}{0.982\xspace}
\newcommand{\XttGccFOcNgxGhiRecall}{1\xspace}
\newcommand{\XttGccFOcNgxGhiPrecision}{1\xspace}
\newcommand{\XttGccFOcNgxGhiFone}{1\xspace}
\newcommand{\XttGccFOcNgxRdaRecall}{0.986\xspace}
\newcommand{\XttGccFOcNgxRdaPrecision}{1\xspace}
\newcommand{\XttGccFOcNgxRdaFone}{0.993\xspace}
\newcommand{\XttGccFOcNgxRseRecall}{1\xspace}
\newcommand{\XttGccFOcNgxRsePrecision}{1\xspace}
\newcommand{\XttGccFOcNgxRseFone}{1\xspace}
\newcommand{\XttGccFOcSshGT}{155421\xspace}
\newcommand{\XttGccFOcSshLsRecall}{1\xspace}
\newcommand{\XttGccFOcSshLsPrecision}{1\xspace}
\newcommand{\XttGccFOcSshLsFone}{1\xspace}
\newcommand{\XttGccFOcSshBapRecall}{1\xspace}
\newcommand{\XttGccFOcSshBapPrecision}{1\xspace}
\newcommand{\XttGccFOcSshBapFone}{1\xspace}
\newcommand{\XttGccFOcSshGhiRecall}{0.379\xspace}
\newcommand{\XttGccFOcSshGhiPrecision}{0.992\xspace}
\newcommand{\XttGccFOcSshGhiFone}{0.548\xspace}
\newcommand{\XttGccFOcSshRdaRecall}{0.938\xspace}
\newcommand{\XttGccFOcSshRdaPrecision}{1.000\xspace}
\newcommand{\XttGccFOcSshRdaFone}{0.968\xspace}
\newcommand{\XttGccFOcSshRseRecall}{0.993\xspace}
\newcommand{\XttGccFOcSshRsePrecision}{1\xspace}
\newcommand{\XttGccFOcSshRseFone}{0.997\xspace}
\newcommand{\XttGccFOcPcrGT}{5229\xspace}
\newcommand{\XttGccFOcPcrLsRecall}{1\xspace}
\newcommand{\XttGccFOcPcrLsPrecision}{1\xspace}
\newcommand{\XttGccFOcPcrLsFone}{1\xspace}
\newcommand{\XttGccFOcPcrBapRecall}{0.891\xspace}
\newcommand{\XttGccFOcPcrBapPrecision}{1\xspace}
\newcommand{\XttGccFOcPcrBapFone}{0.942\xspace}
\newcommand{\XttGccFOcPcrGhiRecall}{1\xspace}
\newcommand{\XttGccFOcPcrGhiPrecision}{1\xspace}
\newcommand{\XttGccFOcPcrGhiFone}{1\xspace}
\newcommand{\XttGccFOcPcrRdaRecall}{1\xspace}
\newcommand{\XttGccFOcPcrRdaPrecision}{1\xspace}
\newcommand{\XttGccFOcPcrRdaFone}{1\xspace}
\newcommand{\XttGccFOcPcrRseRecall}{1\xspace}
\newcommand{\XttGccFOcPcrRsePrecision}{1\xspace}
\newcommand{\XttGccFOcPcrRseFone}{1\xspace}
\newcommand{\XttGccFOcSqlGT}{230088\xspace}
\newcommand{\XttGccFOcSqlLsRecall}{1\xspace}
\newcommand{\XttGccFOcSqlLsPrecision}{1\xspace}
\newcommand{\XttGccFOcSqlLsFone}{1\xspace}
\newcommand{\XttGccFOcSqlBapRecall}{0.868\xspace}
\newcommand{\XttGccFOcSqlBapPrecision}{1\xspace}
\newcommand{\XttGccFOcSqlBapFone}{0.929\xspace}
\newcommand{\XttGccFOcSqlGhiRecall}{0.946\xspace}
\newcommand{\XttGccFOcSqlGhiPrecision}{1\xspace}
\newcommand{\XttGccFOcSqlGhiFone}{0.972\xspace}
\newcommand{\XttGccFOcSqlRdaRecall}{0.995\xspace}
\newcommand{\XttGccFOcSqlRdaPrecision}{1\xspace}
\newcommand{\XttGccFOcSqlRdaFone}{0.997\xspace}
\newcommand{\XttGccFOcSqlRseRecall}{1\xspace}
\newcommand{\XttGccFOcSqlRsePrecision}{1\xspace}
\newcommand{\XttGccFOcSqlRseFone}{1\xspace}
\newcommand{\XttGccFOcVimGT}{726641\xspace}
\newcommand{\XttGccFOcVimLsRecall}{1\xspace}
\newcommand{\XttGccFOcVimLsPrecision}{1\xspace}
\newcommand{\XttGccFOcVimLsFone}{1\xspace}
\newcommand{\XttGccFOcVimBapRecall}{0.938\xspace}
\newcommand{\XttGccFOcVimBapPrecision}{1\xspace}
\newcommand{\XttGccFOcVimBapFone}{0.968\xspace}
\newcommand{\XttGccFOcVimGhiRecall}{1\xspace}
\newcommand{\XttGccFOcVimGhiPrecision}{1\xspace}
\newcommand{\XttGccFOcVimGhiFone}{1\xspace}
\newcommand{\XttGccFOcVimRdaRecall}{0.962\xspace}
\newcommand{\XttGccFOcVimRdaPrecision}{1\xspace}
\newcommand{\XttGccFOcVimRdaFone}{0.981\xspace}
\newcommand{\XttGccFOcVimRseRecall}{1.000\xspace}
\newcommand{\XttGccFOcVimRsePrecision}{1\xspace}
\newcommand{\XttGccFOcVimRseFone}{1.000\xspace}
\newcommand{\XttGccFOcVsfGT}{33286\xspace}
\newcommand{\XttGccFOcVsfLsRecall}{1\xspace}
\newcommand{\XttGccFOcVsfLsPrecision}{1\xspace}
\newcommand{\XttGccFOcVsfLsFone}{1\xspace}
\newcommand{\XttGccFOcVsfBapRecall}{1\xspace}
\newcommand{\XttGccFOcVsfBapPrecision}{1\xspace}
\newcommand{\XttGccFOcVsfBapFone}{1\xspace}
\newcommand{\XttGccFOcVsfGhiRecall}{0.152\xspace}
\newcommand{\XttGccFOcVsfGhiPrecision}{0.984\xspace}
\newcommand{\XttGccFOcVsfGhiFone}{0.264\xspace}
\newcommand{\XttGccFOcVsfRdaRecall}{0.983\xspace}
\newcommand{\XttGccFOcVsfRdaPrecision}{1\xspace}
\newcommand{\XttGccFOcVsfRdaFone}{0.992\xspace}
\newcommand{\XttGccFOcVsfRseRecall}{1.000\xspace}
\newcommand{\XttGccFOcVsfRsePrecision}{1\xspace}
\newcommand{\XttGccFOcVsfRseFone}{1.000\xspace}
\newcommand{\XttGccFOdSzpGT}{16290\xspace}
\newcommand{\XttGccFOdSzpLsRecall}{1\xspace}
\newcommand{\XttGccFOdSzpLsPrecision}{1\xspace}
\newcommand{\XttGccFOdSzpLsFone}{1\xspace}
\newcommand{\XttGccFOdSzpBapRecall}{0.962\xspace}
\newcommand{\XttGccFOdSzpBapPrecision}{1\xspace}
\newcommand{\XttGccFOdSzpBapFone}{0.981\xspace}
\newcommand{\XttGccFOdSzpGhiRecall}{1\xspace}
\newcommand{\XttGccFOdSzpGhiPrecision}{1\xspace}
\newcommand{\XttGccFOdSzpGhiFone}{1\xspace}
\newcommand{\XttGccFOdSzpRdaRecall}{0.966\xspace}
\newcommand{\XttGccFOdSzpRdaPrecision}{1\xspace}
\newcommand{\XttGccFOdSzpRdaFone}{0.983\xspace}
\newcommand{\XttGccFOdSzpRseRecall}{1\xspace}
\newcommand{\XttGccFOdSzpRsePrecision}{1\xspace}
\newcommand{\XttGccFOdSzpRseFone}{1\xspace}
\newcommand{\XttGccFOdCapGT}{260706\xspace}
\newcommand{\XttGccFOdCapLsRecall}{1\xspace}
\newcommand{\XttGccFOdCapLsPrecision}{1\xspace}
\newcommand{\XttGccFOdCapLsFone}{1\xspace}
\newcommand{\XttGccFOdCapBapRecall}{0.509\xspace}
\newcommand{\XttGccFOdCapBapPrecision}{1\xspace}
\newcommand{\XttGccFOdCapBapFone}{0.674\xspace}
\newcommand{\XttGccFOdCapGhiRecall}{0.856\xspace}
\newcommand{\XttGccFOdCapGhiPrecision}{1\xspace}
\newcommand{\XttGccFOdCapGhiFone}{0.922\xspace}
\newcommand{\XttGccFOdCapRdaRecall}{0.835\xspace}
\newcommand{\XttGccFOdCapRdaPrecision}{1\xspace}
\newcommand{\XttGccFOdCapRdaFone}{0.910\xspace}
\newcommand{\XttGccFOdCapRseRecall}{0.998\xspace}
\newcommand{\XttGccFOdCapRsePrecision}{1\xspace}
\newcommand{\XttGccFOdCapRseFone}{0.999\xspace}
\newcommand{\XttGccFOdExmGT}{181979\xspace}
\newcommand{\XttGccFOdExmLsRecall}{1\xspace}
\newcommand{\XttGccFOdExmLsPrecision}{1\xspace}
\newcommand{\XttGccFOdExmLsFone}{1\xspace}
\newcommand{\XttGccFOdExmBapRecall}{0.865\xspace}
\newcommand{\XttGccFOdExmBapPrecision}{1\xspace}
\newcommand{\XttGccFOdExmBapFone}{0.927\xspace}
\newcommand{\XttGccFOdExmGhiRecall}{0.990\xspace}
\newcommand{\XttGccFOdExmGhiPrecision}{1\xspace}
\newcommand{\XttGccFOdExmGhiFone}{0.995\xspace}
\newcommand{\XttGccFOdExmRdaRecall}{0.952\xspace}
\newcommand{\XttGccFOdExmRdaPrecision}{1\xspace}
\newcommand{\XttGccFOdExmRdaFone}{0.975\xspace}
\newcommand{\XttGccFOdExmRseRecall}{1.000\xspace}
\newcommand{\XttGccFOdExmRsePrecision}{1\xspace}
\newcommand{\XttGccFOdExmRseFone}{1.000\xspace}
\newcommand{\XttGccFOdLgtGT}{36201\xspace}
\newcommand{\XttGccFOdLgtLsRecall}{1\xspace}
\newcommand{\XttGccFOdLgtLsPrecision}{1\xspace}
\newcommand{\XttGccFOdLgtLsFone}{1\xspace}
\newcommand{\XttGccFOdLgtBapRecall}{0.899\xspace}
\newcommand{\XttGccFOdLgtBapPrecision}{1\xspace}
\newcommand{\XttGccFOdLgtBapFone}{0.947\xspace}
\newcommand{\XttGccFOdLgtGhiRecall}{0.976\xspace}
\newcommand{\XttGccFOdLgtGhiPrecision}{1\xspace}
\newcommand{\XttGccFOdLgtGhiFone}{0.988\xspace}
\newcommand{\XttGccFOdLgtRdaRecall}{0.973\xspace}
\newcommand{\XttGccFOdLgtRdaPrecision}{1\xspace}
\newcommand{\XttGccFOdLgtRdaFone}{0.986\xspace}
\newcommand{\XttGccFOdLgtRseRecall}{1\xspace}
\newcommand{\XttGccFOdLgtRsePrecision}{1\xspace}
\newcommand{\XttGccFOdLgtRseFone}{1\xspace}
\newcommand{\XttGccFOdBzpGT}{17892\xspace}
\newcommand{\XttGccFOdBzpLsRecall}{1\xspace}
\newcommand{\XttGccFOdBzpLsPrecision}{1\xspace}
\newcommand{\XttGccFOdBzpLsFone}{1\xspace}
\newcommand{\XttGccFOdBzpBapRecall}{0.945\xspace}
\newcommand{\XttGccFOdBzpBapPrecision}{1\xspace}
\newcommand{\XttGccFOdBzpBapFone}{0.972\xspace}
\newcommand{\XttGccFOdBzpGhiRecall}{1\xspace}
\newcommand{\XttGccFOdBzpGhiPrecision}{1\xspace}
\newcommand{\XttGccFOdBzpGhiFone}{1\xspace}
\newcommand{\XttGccFOdBzpRdaRecall}{0.818\xspace}
\newcommand{\XttGccFOdBzpRdaPrecision}{1\xspace}
\newcommand{\XttGccFOdBzpRdaFone}{0.900\xspace}
\newcommand{\XttGccFOdBzpRseRecall}{0.997\xspace}
\newcommand{\XttGccFOdBzpRsePrecision}{1\xspace}
\newcommand{\XttGccFOdBzpRseFone}{0.998\xspace}
\newcommand{\XttGccFOdGccGT}{1171819\xspace}
\newcommand{\XttGccFOdGccLsRecall}{1\xspace}
\newcommand{\XttGccFOdGccLsPrecision}{1\xspace}
\newcommand{\XttGccFOdGccLsFone}{1\xspace}
\newcommand{\XttGccFOdGccBapRecall}{0.759\xspace}
\newcommand{\XttGccFOdGccBapPrecision}{1\xspace}
\newcommand{\XttGccFOdGccBapFone}{0.863\xspace}
\newcommand{\XttGccFOdGccGhiRecall}{0.987\xspace}
\newcommand{\XttGccFOdGccGhiPrecision}{1\xspace}
\newcommand{\XttGccFOdGccGhiFone}{0.994\xspace}
\newcommand{\XttGccFOdGccRdaRecall}{0.863\xspace}
\newcommand{\XttGccFOdGccRdaPrecision}{1\xspace}
\newcommand{\XttGccFOdGccRdaFone}{0.927\xspace}
\newcommand{\XttGccFOdGccRseRecall}{1.000\xspace}
\newcommand{\XttGccFOdGccRsePrecision}{1\xspace}
\newcommand{\XttGccFOdGccRseFone}{1.000\xspace}
\newcommand{\XttGccFOdGzpGT}{17741\xspace}
\newcommand{\XttGccFOdGzpLsRecall}{1\xspace}
\newcommand{\XttGccFOdGzpLsPrecision}{1\xspace}
\newcommand{\XttGccFOdGzpLsFone}{1\xspace}
\newcommand{\XttGccFOdGzpBapRecall}{0.992\xspace}
\newcommand{\XttGccFOdGzpBapPrecision}{1\xspace}
\newcommand{\XttGccFOdGzpBapFone}{0.996\xspace}
\newcommand{\XttGccFOdGzpGhiRecall}{1\xspace}
\newcommand{\XttGccFOdGzpGhiPrecision}{1\xspace}
\newcommand{\XttGccFOdGzpGhiFone}{1\xspace}
\newcommand{\XttGccFOdGzpRdaRecall}{1\xspace}
\newcommand{\XttGccFOdGzpRdaPrecision}{1\xspace}
\newcommand{\XttGccFOdGzpRdaFone}{1\xspace}
\newcommand{\XttGccFOdGzpRseRecall}{1\xspace}
\newcommand{\XttGccFOdGzpRsePrecision}{1\xspace}
\newcommand{\XttGccFOdGzpRseFone}{1\xspace}
\newcommand{\XttGccFOdOggGT}{53946\xspace}
\newcommand{\XttGccFOdOggLsRecall}{1\xspace}
\newcommand{\XttGccFOdOggLsPrecision}{1\xspace}
\newcommand{\XttGccFOdOggLsFone}{1\xspace}
\newcommand{\XttGccFOdOggBapRecall}{0.983\xspace}
\newcommand{\XttGccFOdOggBapPrecision}{1\xspace}
\newcommand{\XttGccFOdOggBapFone}{0.992\xspace}
\newcommand{\XttGccFOdOggGhiRecall}{1\xspace}
\newcommand{\XttGccFOdOggGhiPrecision}{1\xspace}
\newcommand{\XttGccFOdOggGhiFone}{1\xspace}
\newcommand{\XttGccFOdOggRdaRecall}{1\xspace}
\newcommand{\XttGccFOdOggRdaPrecision}{1\xspace}
\newcommand{\XttGccFOdOggRdaFone}{1\xspace}
\newcommand{\XttGccFOdOggRseRecall}{1\xspace}
\newcommand{\XttGccFOdOggRsePrecision}{1\xspace}
\newcommand{\XttGccFOdOggRseFone}{1\xspace}
\newcommand{\XttGccFOdNgxGT}{126485\xspace}
\newcommand{\XttGccFOdNgxLsRecall}{1\xspace}
\newcommand{\XttGccFOdNgxLsPrecision}{1\xspace}
\newcommand{\XttGccFOdNgxLsFone}{1\xspace}
\newcommand{\XttGccFOdNgxBapRecall}{0.965\xspace}
\newcommand{\XttGccFOdNgxBapPrecision}{1\xspace}
\newcommand{\XttGccFOdNgxBapFone}{0.982\xspace}
\newcommand{\XttGccFOdNgxGhiRecall}{1\xspace}
\newcommand{\XttGccFOdNgxGhiPrecision}{1\xspace}
\newcommand{\XttGccFOdNgxGhiFone}{1\xspace}
\newcommand{\XttGccFOdNgxRdaRecall}{0.986\xspace}
\newcommand{\XttGccFOdNgxRdaPrecision}{1\xspace}
\newcommand{\XttGccFOdNgxRdaFone}{0.993\xspace}
\newcommand{\XttGccFOdNgxRseRecall}{1\xspace}
\newcommand{\XttGccFOdNgxRsePrecision}{1\xspace}
\newcommand{\XttGccFOdNgxRseFone}{1\xspace}
\newcommand{\XttGccFOdSshGT}{155418\xspace}
\newcommand{\XttGccFOdSshLsRecall}{1\xspace}
\newcommand{\XttGccFOdSshLsPrecision}{1\xspace}
\newcommand{\XttGccFOdSshLsFone}{1\xspace}
\newcommand{\XttGccFOdSshBapRecall}{1\xspace}
\newcommand{\XttGccFOdSshBapPrecision}{1\xspace}
\newcommand{\XttGccFOdSshBapFone}{1\xspace}
\newcommand{\XttGccFOdSshGhiRecall}{0.376\xspace}
\newcommand{\XttGccFOdSshGhiPrecision}{0.992\xspace}
\newcommand{\XttGccFOdSshGhiFone}{0.545\xspace}
\newcommand{\XttGccFOdSshRdaRecall}{0.938\xspace}
\newcommand{\XttGccFOdSshRdaPrecision}{1.000\xspace}
\newcommand{\XttGccFOdSshRdaFone}{0.968\xspace}
\newcommand{\XttGccFOdSshRseRecall}{0.993\xspace}
\newcommand{\XttGccFOdSshRsePrecision}{1\xspace}
\newcommand{\XttGccFOdSshRseFone}{0.997\xspace}
\newcommand{\XttGccFOdPcrGT}{5229\xspace}
\newcommand{\XttGccFOdPcrLsRecall}{1\xspace}
\newcommand{\XttGccFOdPcrLsPrecision}{1\xspace}
\newcommand{\XttGccFOdPcrLsFone}{1\xspace}
\newcommand{\XttGccFOdPcrBapRecall}{0.891\xspace}
\newcommand{\XttGccFOdPcrBapPrecision}{1\xspace}
\newcommand{\XttGccFOdPcrBapFone}{0.942\xspace}
\newcommand{\XttGccFOdPcrGhiRecall}{1\xspace}
\newcommand{\XttGccFOdPcrGhiPrecision}{1\xspace}
\newcommand{\XttGccFOdPcrGhiFone}{1\xspace}
\newcommand{\XttGccFOdPcrRdaRecall}{1\xspace}
\newcommand{\XttGccFOdPcrRdaPrecision}{1\xspace}
\newcommand{\XttGccFOdPcrRdaFone}{1\xspace}
\newcommand{\XttGccFOdPcrRseRecall}{1\xspace}
\newcommand{\XttGccFOdPcrRsePrecision}{1\xspace}
\newcommand{\XttGccFOdPcrRseFone}{1\xspace}
\newcommand{\XttGccFOdSqlGT}{230018\xspace}
\newcommand{\XttGccFOdSqlLsRecall}{1\xspace}
\newcommand{\XttGccFOdSqlLsPrecision}{1\xspace}
\newcommand{\XttGccFOdSqlLsFone}{1\xspace}
\newcommand{\XttGccFOdSqlBapRecall}{0.869\xspace}
\newcommand{\XttGccFOdSqlBapPrecision}{1\xspace}
\newcommand{\XttGccFOdSqlBapFone}{0.930\xspace}
\newcommand{\XttGccFOdSqlGhiRecall}{0.945\xspace}
\newcommand{\XttGccFOdSqlGhiPrecision}{1\xspace}
\newcommand{\XttGccFOdSqlGhiFone}{0.972\xspace}
\newcommand{\XttGccFOdSqlRdaRecall}{0.995\xspace}
\newcommand{\XttGccFOdSqlRdaPrecision}{1\xspace}
\newcommand{\XttGccFOdSqlRdaFone}{0.997\xspace}
\newcommand{\XttGccFOdSqlRseRecall}{1\xspace}
\newcommand{\XttGccFOdSqlRsePrecision}{1\xspace}
\newcommand{\XttGccFOdSqlRseFone}{1\xspace}
\newcommand{\XttGccFOdVimGT}{726058\xspace}
\newcommand{\XttGccFOdVimLsRecall}{1\xspace}
\newcommand{\XttGccFOdVimLsPrecision}{1\xspace}
\newcommand{\XttGccFOdVimLsFone}{1\xspace}
\newcommand{\XttGccFOdVimBapRecall}{0.938\xspace}
\newcommand{\XttGccFOdVimBapPrecision}{1\xspace}
\newcommand{\XttGccFOdVimBapFone}{0.968\xspace}
\newcommand{\XttGccFOdVimGhiRecall}{1\xspace}
\newcommand{\XttGccFOdVimGhiPrecision}{1\xspace}
\newcommand{\XttGccFOdVimGhiFone}{1\xspace}
\newcommand{\XttGccFOdVimRdaRecall}{0.962\xspace}
\newcommand{\XttGccFOdVimRdaPrecision}{1.000\xspace}
\newcommand{\XttGccFOdVimRdaFone}{0.980\xspace}
\newcommand{\XttGccFOdVimRseRecall}{1.000\xspace}
\newcommand{\XttGccFOdVimRsePrecision}{1\xspace}
\newcommand{\XttGccFOdVimRseFone}{1.000\xspace}
\newcommand{\XttGccFOdVsfGT}{33288\xspace}
\newcommand{\XttGccFOdVsfLsRecall}{1\xspace}
\newcommand{\XttGccFOdVsfLsPrecision}{1\xspace}
\newcommand{\XttGccFOdVsfLsFone}{1\xspace}
\newcommand{\XttGccFOdVsfBapRecall}{1\xspace}
\newcommand{\XttGccFOdVsfBapPrecision}{1\xspace}
\newcommand{\XttGccFOdVsfBapFone}{1\xspace}
\newcommand{\XttGccFOdVsfGhiRecall}{0.157\xspace}
\newcommand{\XttGccFOdVsfGhiPrecision}{0.987\xspace}
\newcommand{\XttGccFOdVsfGhiFone}{0.270\xspace}
\newcommand{\XttGccFOdVsfRdaRecall}{0.983\xspace}
\newcommand{\XttGccFOdVsfRdaPrecision}{1\xspace}
\newcommand{\XttGccFOdVsfRdaFone}{0.992\xspace}
\newcommand{\XttGccFOdVsfRseRecall}{1.000\xspace}
\newcommand{\XttGccFOdVsfRsePrecision}{1\xspace}
\newcommand{\XttGccFOdVsfRseFone}{1.000\xspace}
\newcommand{\XttGccFOsSzpGT}{13048\xspace}
\newcommand{\XttGccFOsSzpLsRecall}{1\xspace}
\newcommand{\XttGccFOsSzpLsPrecision}{1\xspace}
\newcommand{\XttGccFOsSzpLsFone}{1\xspace}
\newcommand{\XttGccFOsSzpBapRecall}{0.978\xspace}
\newcommand{\XttGccFOsSzpBapPrecision}{1\xspace}
\newcommand{\XttGccFOsSzpBapFone}{0.989\xspace}
\newcommand{\XttGccFOsSzpGhiRecall}{1\xspace}
\newcommand{\XttGccFOsSzpGhiPrecision}{1\xspace}
\newcommand{\XttGccFOsSzpGhiFone}{1\xspace}
\newcommand{\XttGccFOsSzpRdaRecall}{0.958\xspace}
\newcommand{\XttGccFOsSzpRdaPrecision}{1\xspace}
\newcommand{\XttGccFOsSzpRdaFone}{0.979\xspace}
\newcommand{\XttGccFOsSzpRseRecall}{1\xspace}
\newcommand{\XttGccFOsSzpRsePrecision}{1\xspace}
\newcommand{\XttGccFOsSzpRseFone}{1\xspace}
\newcommand{\XttGccFOsCapGT}{232392\xspace}
\newcommand{\XttGccFOsCapLsRecall}{1\xspace}
\newcommand{\XttGccFOsCapLsPrecision}{1\xspace}
\newcommand{\XttGccFOsCapLsFone}{1\xspace}
\newcommand{\XttGccFOsCapBapRecall}{0.661\xspace}
\newcommand{\XttGccFOsCapBapPrecision}{1\xspace}
\newcommand{\XttGccFOsCapBapFone}{0.796\xspace}
\newcommand{\XttGccFOsCapGhiRecall}{0.997\xspace}
\newcommand{\XttGccFOsCapGhiPrecision}{1\xspace}
\newcommand{\XttGccFOsCapGhiFone}{0.998\xspace}
\newcommand{\XttGccFOsCapRdaRecall}{0.998\xspace}
\newcommand{\XttGccFOsCapRdaPrecision}{1\xspace}
\newcommand{\XttGccFOsCapRdaFone}{0.999\xspace}
\newcommand{\XttGccFOsCapRseRecall}{1\xspace}
\newcommand{\XttGccFOsCapRsePrecision}{1\xspace}
\newcommand{\XttGccFOsCapRseFone}{1\xspace}
\newcommand{\XttGccFOsExmGT}{139546\xspace}
\newcommand{\XttGccFOsExmLsRecall}{1\xspace}
\newcommand{\XttGccFOsExmLsPrecision}{1\xspace}
\newcommand{\XttGccFOsExmLsFone}{1\xspace}
\newcommand{\XttGccFOsExmBapRecall}{0.868\xspace}
\newcommand{\XttGccFOsExmBapPrecision}{1\xspace}
\newcommand{\XttGccFOsExmBapFone}{0.929\xspace}
\newcommand{\XttGccFOsExmGhiRecall}{0.997\xspace}
\newcommand{\XttGccFOsExmGhiPrecision}{1\xspace}
\newcommand{\XttGccFOsExmGhiFone}{0.999\xspace}
\newcommand{\XttGccFOsExmRdaRecall}{0.967\xspace}
\newcommand{\XttGccFOsExmRdaPrecision}{1\xspace}
\newcommand{\XttGccFOsExmRdaFone}{0.983\xspace}
\newcommand{\XttGccFOsExmRseRecall}{1.000\xspace}
\newcommand{\XttGccFOsExmRsePrecision}{1\xspace}
\newcommand{\XttGccFOsExmRseFone}{1.000\xspace}
\newcommand{\XttGccFOsLgtGT}{27708\xspace}
\newcommand{\XttGccFOsLgtLsRecall}{1\xspace}
\newcommand{\XttGccFOsLgtLsPrecision}{1\xspace}
\newcommand{\XttGccFOsLgtLsFone}{1\xspace}
\newcommand{\XttGccFOsLgtBapRecall}{0.913\xspace}
\newcommand{\XttGccFOsLgtBapPrecision}{1\xspace}
\newcommand{\XttGccFOsLgtBapFone}{0.955\xspace}
\newcommand{\XttGccFOsLgtGhiRecall}{1\xspace}
\newcommand{\XttGccFOsLgtGhiPrecision}{1\xspace}
\newcommand{\XttGccFOsLgtGhiFone}{1\xspace}
\newcommand{\XttGccFOsLgtRdaRecall}{0.995\xspace}
\newcommand{\XttGccFOsLgtRdaPrecision}{1\xspace}
\newcommand{\XttGccFOsLgtRdaFone}{0.998\xspace}
\newcommand{\XttGccFOsLgtRseRecall}{1\xspace}
\newcommand{\XttGccFOsLgtRsePrecision}{1\xspace}
\newcommand{\XttGccFOsLgtRseFone}{1\xspace}
\newcommand{\XttGccFOsBzpGT}{12157\xspace}
\newcommand{\XttGccFOsBzpLsRecall}{1\xspace}
\newcommand{\XttGccFOsBzpLsPrecision}{1\xspace}
\newcommand{\XttGccFOsBzpLsFone}{1\xspace}
\newcommand{\XttGccFOsBzpBapRecall}{0.810\xspace}
\newcommand{\XttGccFOsBzpBapPrecision}{1\xspace}
\newcommand{\XttGccFOsBzpBapFone}{0.895\xspace}
\newcommand{\XttGccFOsBzpGhiRecall}{1\xspace}
\newcommand{\XttGccFOsBzpGhiPrecision}{1\xspace}
\newcommand{\XttGccFOsBzpGhiFone}{1\xspace}
\newcommand{\XttGccFOsBzpRdaRecall}{0.993\xspace}
\newcommand{\XttGccFOsBzpRdaPrecision}{1\xspace}
\newcommand{\XttGccFOsBzpRdaFone}{0.997\xspace}
\newcommand{\XttGccFOsBzpRseRecall}{1\xspace}
\newcommand{\XttGccFOsBzpRsePrecision}{1\xspace}
\newcommand{\XttGccFOsBzpRseFone}{1\xspace}
\newcommand{\XttGccFOsGccGT}{852230\xspace}
\newcommand{\XttGccFOsGccLsRecall}{1\xspace}
\newcommand{\XttGccFOsGccLsPrecision}{1\xspace}
\newcommand{\XttGccFOsGccLsFone}{1\xspace}
\newcommand{\XttGccFOsGccBapRecall}{0.773\xspace}
\newcommand{\XttGccFOsGccBapPrecision}{1\xspace}
\newcommand{\XttGccFOsGccBapFone}{0.872\xspace}
\newcommand{\XttGccFOsGccGhiRecall}{0.989\xspace}
\newcommand{\XttGccFOsGccGhiPrecision}{1\xspace}
\newcommand{\XttGccFOsGccGhiFone}{0.995\xspace}
\newcommand{\XttGccFOsGccRdaRecall}{0.851\xspace}
\newcommand{\XttGccFOsGccRdaPrecision}{1\xspace}
\newcommand{\XttGccFOsGccRdaFone}{0.919\xspace}
\newcommand{\XttGccFOsGccRseRecall}{0.996\xspace}
\newcommand{\XttGccFOsGccRsePrecision}{1\xspace}
\newcommand{\XttGccFOsGccRseFone}{0.998\xspace}
\newcommand{\XttGccFOsGzpGT}{9178\xspace}
\newcommand{\XttGccFOsGzpLsRecall}{1\xspace}
\newcommand{\XttGccFOsGzpLsPrecision}{1\xspace}
\newcommand{\XttGccFOsGzpLsFone}{1\xspace}
\newcommand{\XttGccFOsGzpBapRecall}{0.985\xspace}
\newcommand{\XttGccFOsGzpBapPrecision}{1\xspace}
\newcommand{\XttGccFOsGzpBapFone}{0.993\xspace}
\newcommand{\XttGccFOsGzpGhiRecall}{1\xspace}
\newcommand{\XttGccFOsGzpGhiPrecision}{1\xspace}
\newcommand{\XttGccFOsGzpGhiFone}{1\xspace}
\newcommand{\XttGccFOsGzpRdaRecall}{1\xspace}
\newcommand{\XttGccFOsGzpRdaPrecision}{1\xspace}
\newcommand{\XttGccFOsGzpRdaFone}{1\xspace}
\newcommand{\XttGccFOsGzpRseRecall}{1\xspace}
\newcommand{\XttGccFOsGzpRsePrecision}{1\xspace}
\newcommand{\XttGccFOsGzpRseFone}{1\xspace}
\newcommand{\XttGccFOsOggGT}{37132\xspace}
\newcommand{\XttGccFOsOggLsRecall}{1\xspace}
\newcommand{\XttGccFOsOggLsPrecision}{1\xspace}
\newcommand{\XttGccFOsOggLsFone}{1\xspace}
\newcommand{\XttGccFOsOggBapRecall}{0.998\xspace}
\newcommand{\XttGccFOsOggBapPrecision}{1\xspace}
\newcommand{\XttGccFOsOggBapFone}{0.999\xspace}
\newcommand{\XttGccFOsOggGhiRecall}{1\xspace}
\newcommand{\XttGccFOsOggGhiPrecision}{1\xspace}
\newcommand{\XttGccFOsOggGhiFone}{1\xspace}
\newcommand{\XttGccFOsOggRdaRecall}{1\xspace}
\newcommand{\XttGccFOsOggRdaPrecision}{1\xspace}
\newcommand{\XttGccFOsOggRdaFone}{1\xspace}
\newcommand{\XttGccFOsOggRseRecall}{1\xspace}
\newcommand{\XttGccFOsOggRsePrecision}{1\xspace}
\newcommand{\XttGccFOsOggRseFone}{1\xspace}
\newcommand{\XttGccFOsNgxGT}{104325\xspace}
\newcommand{\XttGccFOsNgxLsRecall}{1\xspace}
\newcommand{\XttGccFOsNgxLsPrecision}{1\xspace}
\newcommand{\XttGccFOsNgxLsFone}{1\xspace}
\newcommand{\XttGccFOsNgxBapRecall}{0.978\xspace}
\newcommand{\XttGccFOsNgxBapPrecision}{1\xspace}
\newcommand{\XttGccFOsNgxBapFone}{0.989\xspace}
\newcommand{\XttGccFOsNgxGhiRecall}{1\xspace}
\newcommand{\XttGccFOsNgxGhiPrecision}{1\xspace}
\newcommand{\XttGccFOsNgxGhiFone}{1\xspace}
\newcommand{\XttGccFOsNgxRdaRecall}{0.992\xspace}
\newcommand{\XttGccFOsNgxRdaPrecision}{1\xspace}
\newcommand{\XttGccFOsNgxRdaFone}{0.996\xspace}
\newcommand{\XttGccFOsNgxRseRecall}{1\xspace}
\newcommand{\XttGccFOsNgxRsePrecision}{1\xspace}
\newcommand{\XttGccFOsNgxRseFone}{1\xspace}
\newcommand{\XttGccFOsSshGT}{126353\xspace}
\newcommand{\XttGccFOsSshLsRecall}{1\xspace}
\newcommand{\XttGccFOsSshLsPrecision}{1\xspace}
\newcommand{\XttGccFOsSshLsFone}{1\xspace}
\newcommand{\XttGccFOsSshBapRecall}{1\xspace}
\newcommand{\XttGccFOsSshBapPrecision}{1\xspace}
\newcommand{\XttGccFOsSshBapFone}{1\xspace}
\newcommand{\XttGccFOsSshGhiRecall}{0.330\xspace}
\newcommand{\XttGccFOsSshGhiPrecision}{0.990\xspace}
\newcommand{\XttGccFOsSshGhiFone}{0.495\xspace}
\newcommand{\XttGccFOsSshRdaRecall}{0.951\xspace}
\newcommand{\XttGccFOsSshRdaPrecision}{1\xspace}
\newcommand{\XttGccFOsSshRdaFone}{0.975\xspace}
\newcommand{\XttGccFOsSshRseRecall}{0.988\xspace}
\newcommand{\XttGccFOsSshRsePrecision}{1\xspace}
\newcommand{\XttGccFOsSshRseFone}{0.994\xspace}
\newcommand{\XttGccFOsPcrGT}{4364\xspace}
\newcommand{\XttGccFOsPcrLsRecall}{1\xspace}
\newcommand{\XttGccFOsPcrLsPrecision}{1\xspace}
\newcommand{\XttGccFOsPcrLsFone}{1\xspace}
\newcommand{\XttGccFOsPcrBapRecall}{0.947\xspace}
\newcommand{\XttGccFOsPcrBapPrecision}{1\xspace}
\newcommand{\XttGccFOsPcrBapFone}{0.973\xspace}
\newcommand{\XttGccFOsPcrGhiRecall}{1\xspace}
\newcommand{\XttGccFOsPcrGhiPrecision}{1\xspace}
\newcommand{\XttGccFOsPcrGhiFone}{1\xspace}
\newcommand{\XttGccFOsPcrRdaRecall}{1\xspace}
\newcommand{\XttGccFOsPcrRdaPrecision}{1\xspace}
\newcommand{\XttGccFOsPcrRdaFone}{1\xspace}
\newcommand{\XttGccFOsPcrRseRecall}{1\xspace}
\newcommand{\XttGccFOsPcrRsePrecision}{1\xspace}
\newcommand{\XttGccFOsPcrRseFone}{1\xspace}
\newcommand{\XttGccFOsSqlGT}{158314\xspace}
\newcommand{\XttGccFOsSqlLsRecall}{1\xspace}
\newcommand{\XttGccFOsSqlLsPrecision}{1\xspace}
\newcommand{\XttGccFOsSqlLsFone}{1\xspace}
\newcommand{\XttGccFOsSqlBapRecall}{0.876\xspace}
\newcommand{\XttGccFOsSqlBapPrecision}{1\xspace}
\newcommand{\XttGccFOsSqlBapFone}{0.934\xspace}
\newcommand{\XttGccFOsSqlGhiRecall}{1\xspace}
\newcommand{\XttGccFOsSqlGhiPrecision}{1\xspace}
\newcommand{\XttGccFOsSqlGhiFone}{1\xspace}
\newcommand{\XttGccFOsSqlRdaRecall}{0.997\xspace}
\newcommand{\XttGccFOsSqlRdaPrecision}{1\xspace}
\newcommand{\XttGccFOsSqlRdaFone}{0.998\xspace}
\newcommand{\XttGccFOsSqlRseRecall}{1\xspace}
\newcommand{\XttGccFOsSqlRsePrecision}{1\xspace}
\newcommand{\XttGccFOsSqlRseFone}{1\xspace}
\newcommand{\XttGccFOsVimGT}{488014\xspace}
\newcommand{\XttGccFOsVimLsRecall}{1\xspace}
\newcommand{\XttGccFOsVimLsPrecision}{1\xspace}
\newcommand{\XttGccFOsVimLsFone}{1\xspace}
\newcommand{\XttGccFOsVimBapRecall}{0.965\xspace}
\newcommand{\XttGccFOsVimBapPrecision}{1\xspace}
\newcommand{\XttGccFOsVimBapFone}{0.982\xspace}
\newcommand{\XttGccFOsVimGhiRecall}{1\xspace}
\newcommand{\XttGccFOsVimGhiPrecision}{1\xspace}
\newcommand{\XttGccFOsVimGhiFone}{1\xspace}
\newcommand{\XttGccFOsVimRdaRecall}{0.987\xspace}
\newcommand{\XttGccFOsVimRdaPrecision}{1\xspace}
\newcommand{\XttGccFOsVimRdaFone}{0.993\xspace}
\newcommand{\XttGccFOsVimRseRecall}{1.000\xspace}
\newcommand{\XttGccFOsVimRsePrecision}{1\xspace}
\newcommand{\XttGccFOsVimRseFone}{1.000\xspace}
\newcommand{\XttGccFOsVsfGT}{24872\xspace}
\newcommand{\XttGccFOsVsfLsRecall}{1\xspace}
\newcommand{\XttGccFOsVsfLsPrecision}{1\xspace}
\newcommand{\XttGccFOsVsfLsFone}{1\xspace}
\newcommand{\XttGccFOsVsfBapRecall}{1\xspace}
\newcommand{\XttGccFOsVsfBapPrecision}{1\xspace}
\newcommand{\XttGccFOsVsfBapFone}{1\xspace}
\newcommand{\XttGccFOsVsfGhiRecall}{0.192\xspace}
\newcommand{\XttGccFOsVsfGhiPrecision}{0.992\xspace}
\newcommand{\XttGccFOsVsfGhiFone}{0.322\xspace}
\newcommand{\XttGccFOsVsfRdaRecall}{0.988\xspace}
\newcommand{\XttGccFOsVsfRdaPrecision}{1\xspace}
\newcommand{\XttGccFOsVsfRdaFone}{0.994\xspace}
\newcommand{\XttGccFOsVsfRseRecall}{1.000\xspace}
\newcommand{\XttGccFOsVsfRsePrecision}{1\xspace}
\newcommand{\XttGccFOsVsfRseFone}{1.000\xspace}
\newcommand{\XttGccSOoSzpGT}{22107\xspace}
\newcommand{\XttGccSOoSzpLsRecall}{1\xspace}
\newcommand{\XttGccSOoSzpLsPrecision}{1\xspace}
\newcommand{\XttGccSOoSzpLsFone}{1\xspace}
\newcommand{\XttGccSOoSzpBapRecall}{1\xspace}
\newcommand{\XttGccSOoSzpBapPrecision}{1\xspace}
\newcommand{\XttGccSOoSzpBapFone}{1\xspace}
\newcommand{\XttGccSOoSzpGhiRecall}{0.409\xspace}
\newcommand{\XttGccSOoSzpGhiPrecision}{0.996\xspace}
\newcommand{\XttGccSOoSzpGhiFone}{0.580\xspace}
\newcommand{\XttGccSOoSzpRdaRecall}{0.977\xspace}
\newcommand{\XttGccSOoSzpRdaPrecision}{1\xspace}
\newcommand{\XttGccSOoSzpRdaFone}{0.989\xspace}
\newcommand{\XttGccSOoSzpRseRecall}{1\xspace}
\newcommand{\XttGccSOoSzpRsePrecision}{1\xspace}
\newcommand{\XttGccSOoSzpRseFone}{1\xspace}
\newcommand{\XttGccSOoCapGT}{404207\xspace}
\newcommand{\XttGccSOoCapLsRecall}{1\xspace}
\newcommand{\XttGccSOoCapLsPrecision}{1\xspace}
\newcommand{\XttGccSOoCapLsFone}{1\xspace}
\newcommand{\XttGccSOoCapBapRecall}{1\xspace}
\newcommand{\XttGccSOoCapBapPrecision}{1\xspace}
\newcommand{\XttGccSOoCapBapFone}{1\xspace}
\newcommand{\XttGccSOoCapGhiRecall}{0.248\xspace}
\newcommand{\XttGccSOoCapGhiPrecision}{0.988\xspace}
\newcommand{\XttGccSOoCapGhiFone}{0.397\xspace}
\newcommand{\XttGccSOoCapRdaRecall}{0.459\xspace}
\newcommand{\XttGccSOoCapRdaPrecision}{0.999\xspace}
\newcommand{\XttGccSOoCapRdaFone}{0.629\xspace}
\newcommand{\XttGccSOoCapRseRecall}{0.524\xspace}
\newcommand{\XttGccSOoCapRsePrecision}{1\xspace}
\newcommand{\XttGccSOoCapRseFone}{0.688\xspace}
\newcommand{\XttGccSOoExmGT}{205567\xspace}
\newcommand{\XttGccSOoExmLsRecall}{1\xspace}
\newcommand{\XttGccSOoExmLsPrecision}{1\xspace}
\newcommand{\XttGccSOoExmLsFone}{1\xspace}
\newcommand{\XttGccSOoExmBapRecall}{1\xspace}
\newcommand{\XttGccSOoExmBapPrecision}{1\xspace}
\newcommand{\XttGccSOoExmBapFone}{1\xspace}
\newcommand{\XttGccSOoExmGhiRecall}{0.472\xspace}
\newcommand{\XttGccSOoExmGhiPrecision}{0.996\xspace}
\newcommand{\XttGccSOoExmGhiFone}{0.641\xspace}
\newcommand{\XttGccSOoExmRdaRecall}{0.814\xspace}
\newcommand{\XttGccSOoExmRdaPrecision}{1.000\xspace}
\newcommand{\XttGccSOoExmRdaFone}{0.898\xspace}
\newcommand{\XttGccSOoExmRseRecall}{0.940\xspace}
\newcommand{\XttGccSOoExmRsePrecision}{1.000\xspace}
\newcommand{\XttGccSOoExmRseFone}{0.969\xspace}
\newcommand{\XttGccSOoLgtGT}{44784\xspace}
\newcommand{\XttGccSOoLgtLsRecall}{1\xspace}
\newcommand{\XttGccSOoLgtLsPrecision}{1\xspace}
\newcommand{\XttGccSOoLgtLsFone}{1\xspace}
\newcommand{\XttGccSOoLgtBapRecall}{1\xspace}
\newcommand{\XttGccSOoLgtBapPrecision}{1\xspace}
\newcommand{\XttGccSOoLgtBapFone}{1\xspace}
\newcommand{\XttGccSOoLgtGhiRecall}{0.355\xspace}
\newcommand{\XttGccSOoLgtGhiPrecision}{0.995\xspace}
\newcommand{\XttGccSOoLgtGhiFone}{0.524\xspace}
\newcommand{\XttGccSOoLgtRdaRecall}{0.884\xspace}
\newcommand{\XttGccSOoLgtRdaPrecision}{1\xspace}
\newcommand{\XttGccSOoLgtRdaFone}{0.938\xspace}
\newcommand{\XttGccSOoLgtRseRecall}{0.977\xspace}
\newcommand{\XttGccSOoLgtRsePrecision}{1\xspace}
\newcommand{\XttGccSOoLgtRseFone}{0.988\xspace}
\newcommand{\XttGccSOoBzpGT}{24242\xspace}
\newcommand{\XttGccSOoBzpLsRecall}{1\xspace}
\newcommand{\XttGccSOoBzpLsPrecision}{1\xspace}
\newcommand{\XttGccSOoBzpLsFone}{1\xspace}
\newcommand{\XttGccSOoBzpBapRecall}{1\xspace}
\newcommand{\XttGccSOoBzpBapPrecision}{1\xspace}
\newcommand{\XttGccSOoBzpBapFone}{1\xspace}
\newcommand{\XttGccSOoBzpGhiRecall}{0.895\xspace}
\newcommand{\XttGccSOoBzpGhiPrecision}{1\xspace}
\newcommand{\XttGccSOoBzpGhiFone}{0.945\xspace}
\newcommand{\XttGccSOoBzpRdaRecall}{0.715\xspace}
\newcommand{\XttGccSOoBzpRdaPrecision}{1\xspace}
\newcommand{\XttGccSOoBzpRdaFone}{0.834\xspace}
\newcommand{\XttGccSOoBzpRseRecall}{0.998\xspace}
\newcommand{\XttGccSOoBzpRsePrecision}{1\xspace}
\newcommand{\XttGccSOoBzpRseFone}{0.999\xspace}
\newcommand{\XttGccSOoGccGT}{1419054\xspace}
\newcommand{\XttGccSOoGccLsRecall}{1\xspace}
\newcommand{\XttGccSOoGccLsPrecision}{1\xspace}
\newcommand{\XttGccSOoGccLsFone}{1\xspace}
\newcommand{\XttGccSOoGccBapRecall}{1\xspace}
\newcommand{\XttGccSOoGccBapPrecision}{1\xspace}
\newcommand{\XttGccSOoGccBapFone}{1\xspace}
\newcommand{\XttGccSOoGccGhiRecall}{0.416\xspace}
\newcommand{\XttGccSOoGccGhiPrecision}{0.989\xspace}
\newcommand{\XttGccSOoGccGhiFone}{0.586\xspace}
\newcommand{\XttGccSOoGccRdaRecall}{0.647\xspace}
\newcommand{\XttGccSOoGccRdaPrecision}{1.000\xspace}
\newcommand{\XttGccSOoGccRdaFone}{0.785\xspace}
\newcommand{\XttGccSOoGccRseRecall}{0.845\xspace}
\newcommand{\XttGccSOoGccRsePrecision}{1.000\xspace}
\newcommand{\XttGccSOoGccRseFone}{0.916\xspace}
\newcommand{\XttGccSOoGzpGT}{14833\xspace}
\newcommand{\XttGccSOoGzpLsRecall}{1\xspace}
\newcommand{\XttGccSOoGzpLsPrecision}{1\xspace}
\newcommand{\XttGccSOoGzpLsFone}{1\xspace}
\newcommand{\XttGccSOoGzpBapRecall}{1\xspace}
\newcommand{\XttGccSOoGzpBapPrecision}{1\xspace}
\newcommand{\XttGccSOoGzpBapFone}{1\xspace}
\newcommand{\XttGccSOoGzpGhiRecall}{1\xspace}
\newcommand{\XttGccSOoGzpGhiPrecision}{1\xspace}
\newcommand{\XttGccSOoGzpGhiFone}{1\xspace}
\newcommand{\XttGccSOoGzpRdaRecall}{0.990\xspace}
\newcommand{\XttGccSOoGzpRdaPrecision}{1.000\xspace}
\newcommand{\XttGccSOoGzpRdaFone}{0.995\xspace}
\newcommand{\XttGccSOoGzpRseRecall}{0.995\xspace}
\newcommand{\XttGccSOoGzpRsePrecision}{1\xspace}
\newcommand{\XttGccSOoGzpRseFone}{0.998\xspace}
\newcommand{\XttGccSOoOggGT}{60598\xspace}
\newcommand{\XttGccSOoOggLsRecall}{1\xspace}
\newcommand{\XttGccSOoOggLsPrecision}{1\xspace}
\newcommand{\XttGccSOoOggLsFone}{1\xspace}
\newcommand{\XttGccSOoOggBapRecall}{1\xspace}
\newcommand{\XttGccSOoOggBapPrecision}{1\xspace}
\newcommand{\XttGccSOoOggBapFone}{1\xspace}
\newcommand{\XttGccSOoOggGhiRecall}{0.307\xspace}
\newcommand{\XttGccSOoOggGhiPrecision}{0.994\xspace}
\newcommand{\XttGccSOoOggGhiFone}{0.469\xspace}
\newcommand{\XttGccSOoOggRdaRecall}{0.973\xspace}
\newcommand{\XttGccSOoOggRdaPrecision}{1.000\xspace}
\newcommand{\XttGccSOoOggRdaFone}{0.986\xspace}
\newcommand{\XttGccSOoOggRseRecall}{0.994\xspace}
\newcommand{\XttGccSOoOggRsePrecision}{1\xspace}
\newcommand{\XttGccSOoOggRseFone}{0.997\xspace}
\newcommand{\XttGccSOoNgxGT}{180526\xspace}
\newcommand{\XttGccSOoNgxLsRecall}{1\xspace}
\newcommand{\XttGccSOoNgxLsPrecision}{1\xspace}
\newcommand{\XttGccSOoNgxLsFone}{1\xspace}
\newcommand{\XttGccSOoNgxBapRecall}{1\xspace}
\newcommand{\XttGccSOoNgxBapPrecision}{1\xspace}
\newcommand{\XttGccSOoNgxBapFone}{1\xspace}
\newcommand{\XttGccSOoNgxGhiRecall}{0.335\xspace}
\newcommand{\XttGccSOoNgxGhiPrecision}{0.986\xspace}
\newcommand{\XttGccSOoNgxGhiFone}{0.500\xspace}
\newcommand{\XttGccSOoNgxRdaRecall}{0.970\xspace}
\newcommand{\XttGccSOoNgxRdaPrecision}{1.000\xspace}
\newcommand{\XttGccSOoNgxRdaFone}{0.985\xspace}
\newcommand{\XttGccSOoNgxRseRecall}{0.993\xspace}
\newcommand{\XttGccSOoNgxRsePrecision}{1\xspace}
\newcommand{\XttGccSOoNgxRseFone}{0.997\xspace}
\newcommand{\XttGccSOoSshGT}{177312\xspace}
\newcommand{\XttGccSOoSshLsRecall}{1\xspace}
\newcommand{\XttGccSOoSshLsPrecision}{1\xspace}
\newcommand{\XttGccSOoSshLsFone}{1\xspace}
\newcommand{\XttGccSOoSshBapRecall}{1\xspace}
\newcommand{\XttGccSOoSshBapPrecision}{1\xspace}
\newcommand{\XttGccSOoSshBapFone}{1\xspace}
\newcommand{\XttGccSOoSshGhiRecall}{0.268\xspace}
\newcommand{\XttGccSOoSshGhiPrecision}{0.984\xspace}
\newcommand{\XttGccSOoSshGhiFone}{0.421\xspace}
\newcommand{\XttGccSOoSshRdaRecall}{0.953\xspace}
\newcommand{\XttGccSOoSshRdaPrecision}{1.000\xspace}
\newcommand{\XttGccSOoSshRdaFone}{0.976\xspace}
\newcommand{\XttGccSOoSshRseRecall}{0.982\xspace}
\newcommand{\XttGccSOoSshRsePrecision}{1\xspace}
\newcommand{\XttGccSOoSshRseFone}{0.991\xspace}
\newcommand{\XttGccSOoPcrGT}{6772\xspace}
\newcommand{\XttGccSOoPcrLsRecall}{1\xspace}
\newcommand{\XttGccSOoPcrLsPrecision}{1\xspace}
\newcommand{\XttGccSOoPcrLsFone}{1\xspace}
\newcommand{\XttGccSOoPcrBapRecall}{1\xspace}
\newcommand{\XttGccSOoPcrBapPrecision}{1\xspace}
\newcommand{\XttGccSOoPcrBapFone}{1\xspace}
\newcommand{\XttGccSOoPcrGhiRecall}{1\xspace}
\newcommand{\XttGccSOoPcrGhiPrecision}{1\xspace}
\newcommand{\XttGccSOoPcrGhiFone}{1\xspace}
\newcommand{\XttGccSOoPcrRdaRecall}{0.907\xspace}
\newcommand{\XttGccSOoPcrRdaPrecision}{1\xspace}
\newcommand{\XttGccSOoPcrRdaFone}{0.951\xspace}
\newcommand{\XttGccSOoPcrRseRecall}{0.988\xspace}
\newcommand{\XttGccSOoPcrRsePrecision}{1\xspace}
\newcommand{\XttGccSOoPcrRseFone}{0.994\xspace}
\newcommand{\XttGccSOoSqlGT}{255280\xspace}
\newcommand{\XttGccSOoSqlLsRecall}{1\xspace}
\newcommand{\XttGccSOoSqlLsPrecision}{1\xspace}
\newcommand{\XttGccSOoSqlLsFone}{1\xspace}
\newcommand{\XttGccSOoSqlBapRecall}{1\xspace}
\newcommand{\XttGccSOoSqlBapPrecision}{1\xspace}
\newcommand{\XttGccSOoSqlBapFone}{1\xspace}
\newcommand{\XttGccSOoSqlGhiRecall}{0.337\xspace}
\newcommand{\XttGccSOoSqlGhiPrecision}{0.984\xspace}
\newcommand{\XttGccSOoSqlGhiFone}{0.502\xspace}
\newcommand{\XttGccSOoSqlRdaRecall}{0.877\xspace}
\newcommand{\XttGccSOoSqlRdaPrecision}{1\xspace}
\newcommand{\XttGccSOoSqlRdaFone}{0.934\xspace}
\newcommand{\XttGccSOoSqlRseRecall}{0.916\xspace}
\newcommand{\XttGccSOoSqlRsePrecision}{1.000\xspace}
\newcommand{\XttGccSOoSqlRseFone}{0.956\xspace}
\newcommand{\XttGccSOoVimGT}{730975\xspace}
\newcommand{\XttGccSOoVimLsRecall}{1\xspace}
\newcommand{\XttGccSOoVimLsPrecision}{1\xspace}
\newcommand{\XttGccSOoVimLsFone}{1\xspace}
\newcommand{\XttGccSOoVimBapRecall}{1\xspace}
\newcommand{\XttGccSOoVimBapPrecision}{1\xspace}
\newcommand{\XttGccSOoVimBapFone}{1\xspace}
\newcommand{\XttGccSOoVimGhiRecall}{0.370\xspace}
\newcommand{\XttGccSOoVimGhiPrecision}{0.989\xspace}
\newcommand{\XttGccSOoVimGhiFone}{0.539\xspace}
\newcommand{\XttGccSOoVimRdaRecall}{0.938\xspace}
\newcommand{\XttGccSOoVimRdaPrecision}{1.000\xspace}
\newcommand{\XttGccSOoVimRdaFone}{0.968\xspace}
\newcommand{\XttGccSOoVimRseRecall}{0.974\xspace}
\newcommand{\XttGccSOoVimRsePrecision}{1.000\xspace}
\newcommand{\XttGccSOoVimRseFone}{0.987\xspace}
\newcommand{\XttGccSOoVsfGT}{31426\xspace}
\newcommand{\XttGccSOoVsfLsRecall}{1\xspace}
\newcommand{\XttGccSOoVsfLsPrecision}{1\xspace}
\newcommand{\XttGccSOoVsfLsFone}{1\xspace}
\newcommand{\XttGccSOoVsfBapRecall}{1\xspace}
\newcommand{\XttGccSOoVsfBapPrecision}{1\xspace}
\newcommand{\XttGccSOoVsfBapFone}{1\xspace}
\newcommand{\XttGccSOoVsfGhiRecall}{0.328\xspace}
\newcommand{\XttGccSOoVsfGhiPrecision}{0.995\xspace}
\newcommand{\XttGccSOoVsfGhiFone}{0.493\xspace}
\newcommand{\XttGccSOoVsfRdaRecall}{0.929\xspace}
\newcommand{\XttGccSOoVsfRdaPrecision}{1.000\xspace}
\newcommand{\XttGccSOoVsfRdaFone}{0.963\xspace}
\newcommand{\XttGccSOoVsfRseRecall}{1.000\xspace}
\newcommand{\XttGccSOoVsfRsePrecision}{1\xspace}
\newcommand{\XttGccSOoVsfRseFone}{1.000\xspace}
\newcommand{\XttGccSOaSzpGT}{13997\xspace}
\newcommand{\XttGccSOaSzpLsRecall}{1\xspace}
\newcommand{\XttGccSOaSzpLsPrecision}{1\xspace}
\newcommand{\XttGccSOaSzpLsFone}{1\xspace}
\newcommand{\XttGccSOaSzpBapRecall}{1\xspace}
\newcommand{\XttGccSOaSzpBapPrecision}{1\xspace}
\newcommand{\XttGccSOaSzpBapFone}{1\xspace}
\newcommand{\XttGccSOaSzpGhiRecall}{1\xspace}
\newcommand{\XttGccSOaSzpGhiPrecision}{1\xspace}
\newcommand{\XttGccSOaSzpGhiFone}{1\xspace}
\newcommand{\XttGccSOaSzpRdaRecall}{0.968\xspace}
\newcommand{\XttGccSOaSzpRdaPrecision}{1\xspace}
\newcommand{\XttGccSOaSzpRdaFone}{0.984\xspace}
\newcommand{\XttGccSOaSzpRseRecall}{0.993\xspace}
\newcommand{\XttGccSOaSzpRsePrecision}{1\xspace}
\newcommand{\XttGccSOaSzpRseFone}{0.997\xspace}
\newcommand{\XttGccSOaCapGT}{283991\xspace}
\newcommand{\XttGccSOaCapLsRecall}{1\xspace}
\newcommand{\XttGccSOaCapLsPrecision}{1\xspace}
\newcommand{\XttGccSOaCapLsFone}{1\xspace}
\newcommand{\XttGccSOaCapBapRecall}{1\xspace}
\newcommand{\XttGccSOaCapBapPrecision}{1\xspace}
\newcommand{\XttGccSOaCapBapFone}{1\xspace}
\newcommand{\XttGccSOaCapGhiRecall}{0.270\xspace}
\newcommand{\XttGccSOaCapGhiPrecision}{0.992\xspace}
\newcommand{\XttGccSOaCapGhiFone}{0.425\xspace}
\newcommand{\XttGccSOaCapRdaRecall}{0.399\xspace}
\newcommand{\XttGccSOaCapRdaPrecision}{0.999\xspace}
\newcommand{\XttGccSOaCapRdaFone}{0.570\xspace}
\newcommand{\XttGccSOaCapRseRecall}{0.480\xspace}
\newcommand{\XttGccSOaCapRsePrecision}{1.000\xspace}
\newcommand{\XttGccSOaCapRseFone}{0.649\xspace}
\newcommand{\XttGccSOaExmGT}{167488\xspace}
\newcommand{\XttGccSOaExmLsRecall}{1\xspace}
\newcommand{\XttGccSOaExmLsPrecision}{1\xspace}
\newcommand{\XttGccSOaExmLsFone}{1\xspace}
\newcommand{\XttGccSOaExmBapRecall}{1\xspace}
\newcommand{\XttGccSOaExmBapPrecision}{1\xspace}
\newcommand{\XttGccSOaExmBapFone}{1\xspace}
\newcommand{\XttGccSOaExmGhiRecall}{0.479\xspace}
\newcommand{\XttGccSOaExmGhiPrecision}{0.994\xspace}
\newcommand{\XttGccSOaExmGhiFone}{0.647\xspace}
\newcommand{\XttGccSOaExmRdaRecall}{0.782\xspace}
\newcommand{\XttGccSOaExmRdaPrecision}{1.000\xspace}
\newcommand{\XttGccSOaExmRdaFone}{0.878\xspace}
\newcommand{\XttGccSOaExmRseRecall}{0.983\xspace}
\newcommand{\XttGccSOaExmRsePrecision}{1\xspace}
\newcommand{\XttGccSOaExmRseFone}{0.992\xspace}
\newcommand{\XttGccSOaLgtGT}{32083\xspace}
\newcommand{\XttGccSOaLgtLsRecall}{1\xspace}
\newcommand{\XttGccSOaLgtLsPrecision}{1\xspace}
\newcommand{\XttGccSOaLgtLsFone}{1\xspace}
\newcommand{\XttGccSOaLgtBapRecall}{1\xspace}
\newcommand{\XttGccSOaLgtBapPrecision}{1\xspace}
\newcommand{\XttGccSOaLgtBapFone}{1\xspace}
\newcommand{\XttGccSOaLgtGhiRecall}{0.377\xspace}
\newcommand{\XttGccSOaLgtGhiPrecision}{0.999\xspace}
\newcommand{\XttGccSOaLgtGhiFone}{0.548\xspace}
\newcommand{\XttGccSOaLgtRdaRecall}{0.865\xspace}
\newcommand{\XttGccSOaLgtRdaPrecision}{1\xspace}
\newcommand{\XttGccSOaLgtRdaFone}{0.927\xspace}
\newcommand{\XttGccSOaLgtRseRecall}{0.975\xspace}
\newcommand{\XttGccSOaLgtRsePrecision}{1\xspace}
\newcommand{\XttGccSOaLgtRseFone}{0.987\xspace}
\newcommand{\XttGccSOaBzpGT}{16021\xspace}
\newcommand{\XttGccSOaBzpLsRecall}{1\xspace}
\newcommand{\XttGccSOaBzpLsPrecision}{1\xspace}
\newcommand{\XttGccSOaBzpLsFone}{1\xspace}
\newcommand{\XttGccSOaBzpBapRecall}{1\xspace}
\newcommand{\XttGccSOaBzpBapPrecision}{1\xspace}
\newcommand{\XttGccSOaBzpBapFone}{1\xspace}
\newcommand{\XttGccSOaBzpGhiRecall}{0.920\xspace}
\newcommand{\XttGccSOaBzpGhiPrecision}{1\xspace}
\newcommand{\XttGccSOaBzpGhiFone}{0.958\xspace}
\newcommand{\XttGccSOaBzpRdaRecall}{0.707\xspace}
\newcommand{\XttGccSOaBzpRdaPrecision}{1\xspace}
\newcommand{\XttGccSOaBzpRdaFone}{0.828\xspace}
\newcommand{\XttGccSOaBzpRseRecall}{0.998\xspace}
\newcommand{\XttGccSOaBzpRsePrecision}{1\xspace}
\newcommand{\XttGccSOaBzpRseFone}{0.999\xspace}
\newcommand{\XttGccSOaGccGT}{987798\xspace}
\newcommand{\XttGccSOaGccLsRecall}{1\xspace}
\newcommand{\XttGccSOaGccLsPrecision}{1\xspace}
\newcommand{\XttGccSOaGccLsFone}{1\xspace}
\newcommand{\XttGccSOaGccBapRecall}{1\xspace}
\newcommand{\XttGccSOaGccBapPrecision}{1\xspace}
\newcommand{\XttGccSOaGccBapFone}{1\xspace}
\newcommand{\XttGccSOaGccGhiRecall}{0.384\xspace}
\newcommand{\XttGccSOaGccGhiPrecision}{0.986\xspace}
\newcommand{\XttGccSOaGccGhiFone}{0.553\xspace}
\newcommand{\XttGccSOaGccRdaRecall}{0.603\xspace}
\newcommand{\XttGccSOaGccRdaPrecision}{1.000\xspace}
\newcommand{\XttGccSOaGccRdaFone}{0.752\xspace}
\newcommand{\XttGccSOaGccRseRecall}{0.853\xspace}
\newcommand{\XttGccSOaGccRsePrecision}{1.000\xspace}
\newcommand{\XttGccSOaGccRseFone}{0.920\xspace}
\newcommand{\XttGccSOaGzpGT}{11548\xspace}
\newcommand{\XttGccSOaGzpLsRecall}{1\xspace}
\newcommand{\XttGccSOaGzpLsPrecision}{1\xspace}
\newcommand{\XttGccSOaGzpLsFone}{1\xspace}
\newcommand{\XttGccSOaGzpBapRecall}{1\xspace}
\newcommand{\XttGccSOaGzpBapPrecision}{1\xspace}
\newcommand{\XttGccSOaGzpBapFone}{1\xspace}
\newcommand{\XttGccSOaGzpGhiRecall}{1\xspace}
\newcommand{\XttGccSOaGzpGhiPrecision}{1\xspace}
\newcommand{\XttGccSOaGzpGhiFone}{1\xspace}
\newcommand{\XttGccSOaGzpRdaRecall}{0.985\xspace}
\newcommand{\XttGccSOaGzpRdaPrecision}{1\xspace}
\newcommand{\XttGccSOaGzpRdaFone}{0.993\xspace}
\newcommand{\XttGccSOaGzpRseRecall}{0.993\xspace}
\newcommand{\XttGccSOaGzpRsePrecision}{1\xspace}
\newcommand{\XttGccSOaGzpRseFone}{0.997\xspace}
\newcommand{\XttGccSOaOggGT}{42747\xspace}
\newcommand{\XttGccSOaOggLsRecall}{1\xspace}
\newcommand{\XttGccSOaOggLsPrecision}{1\xspace}
\newcommand{\XttGccSOaOggLsFone}{1\xspace}
\newcommand{\XttGccSOaOggBapRecall}{1\xspace}
\newcommand{\XttGccSOaOggBapPrecision}{1\xspace}
\newcommand{\XttGccSOaOggBapFone}{1\xspace}
\newcommand{\XttGccSOaOggGhiRecall}{0.472\xspace}
\newcommand{\XttGccSOaOggGhiPrecision}{0.996\xspace}
\newcommand{\XttGccSOaOggGhiFone}{0.641\xspace}
\newcommand{\XttGccSOaOggRdaRecall}{0.976\xspace}
\newcommand{\XttGccSOaOggRdaPrecision}{1\xspace}
\newcommand{\XttGccSOaOggRdaFone}{0.988\xspace}
\newcommand{\XttGccSOaOggRseRecall}{0.993\xspace}
\newcommand{\XttGccSOaOggRsePrecision}{1\xspace}
\newcommand{\XttGccSOaOggRseFone}{0.997\xspace}
\newcommand{\XttGccSOaNgxGT}{120803\xspace}
\newcommand{\XttGccSOaNgxLsRecall}{1\xspace}
\newcommand{\XttGccSOaNgxLsPrecision}{1\xspace}
\newcommand{\XttGccSOaNgxLsFone}{1\xspace}
\newcommand{\XttGccSOaNgxBapRecall}{1\xspace}
\newcommand{\XttGccSOaNgxBapPrecision}{1\xspace}
\newcommand{\XttGccSOaNgxBapFone}{1\xspace}
\newcommand{\XttGccSOaNgxGhiRecall}{0.355\xspace}
\newcommand{\XttGccSOaNgxGhiPrecision}{0.986\xspace}
\newcommand{\XttGccSOaNgxGhiFone}{0.522\xspace}
\newcommand{\XttGccSOaNgxRdaRecall}{0.950\xspace}
\newcommand{\XttGccSOaNgxRdaPrecision}{1.000\xspace}
\newcommand{\XttGccSOaNgxRdaFone}{0.974\xspace}
\newcommand{\XttGccSOaNgxRseRecall}{0.993\xspace}
\newcommand{\XttGccSOaNgxRsePrecision}{1\xspace}
\newcommand{\XttGccSOaNgxRseFone}{0.997\xspace}
\newcommand{\XttGccSOaSshGT}{132033\xspace}
\newcommand{\XttGccSOaSshLsRecall}{1\xspace}
\newcommand{\XttGccSOaSshLsPrecision}{1\xspace}
\newcommand{\XttGccSOaSshLsFone}{1\xspace}
\newcommand{\XttGccSOaSshBapRecall}{1\xspace}
\newcommand{\XttGccSOaSshBapPrecision}{1\xspace}
\newcommand{\XttGccSOaSshBapFone}{1\xspace}
\newcommand{\XttGccSOaSshGhiRecall}{0.297\xspace}
\newcommand{\XttGccSOaSshGhiPrecision}{0.987\xspace}
\newcommand{\XttGccSOaSshGhiFone}{0.456\xspace}
\newcommand{\XttGccSOaSshRdaRecall}{0.925\xspace}
\newcommand{\XttGccSOaSshRdaPrecision}{1.000\xspace}
\newcommand{\XttGccSOaSshRdaFone}{0.961\xspace}
\newcommand{\XttGccSOaSshRseRecall}{0.992\xspace}
\newcommand{\XttGccSOaSshRsePrecision}{1\xspace}
\newcommand{\XttGccSOaSshRseFone}{0.996\xspace}
\newcommand{\XttGccSOaPcrGT}{5293\xspace}
\newcommand{\XttGccSOaPcrLsRecall}{1\xspace}
\newcommand{\XttGccSOaPcrLsPrecision}{1\xspace}
\newcommand{\XttGccSOaPcrLsFone}{1\xspace}
\newcommand{\XttGccSOaPcrBapRecall}{1\xspace}
\newcommand{\XttGccSOaPcrBapPrecision}{1\xspace}
\newcommand{\XttGccSOaPcrBapFone}{1\xspace}
\newcommand{\XttGccSOaPcrGhiRecall}{1\xspace}
\newcommand{\XttGccSOaPcrGhiPrecision}{1\xspace}
\newcommand{\XttGccSOaPcrGhiFone}{1\xspace}
\newcommand{\XttGccSOaPcrRdaRecall}{0.885\xspace}
\newcommand{\XttGccSOaPcrRdaPrecision}{1\xspace}
\newcommand{\XttGccSOaPcrRdaFone}{0.939\xspace}
\newcommand{\XttGccSOaPcrRseRecall}{0.987\xspace}
\newcommand{\XttGccSOaPcrRsePrecision}{1\xspace}
\newcommand{\XttGccSOaPcrRseFone}{0.993\xspace}
\newcommand{\XttGccSOaSqlGT}{173880\xspace}
\newcommand{\XttGccSOaSqlLsRecall}{1\xspace}
\newcommand{\XttGccSOaSqlLsPrecision}{1\xspace}
\newcommand{\XttGccSOaSqlLsFone}{1\xspace}
\newcommand{\XttGccSOaSqlBapRecall}{1\xspace}
\newcommand{\XttGccSOaSqlBapPrecision}{1\xspace}
\newcommand{\XttGccSOaSqlBapFone}{1\xspace}
\newcommand{\XttGccSOaSqlGhiRecall}{0.408\xspace}
\newcommand{\XttGccSOaSqlGhiPrecision}{0.991\xspace}
\newcommand{\XttGccSOaSqlGhiFone}{0.578\xspace}
\newcommand{\XttGccSOaSqlRdaRecall}{0.853\xspace}
\newcommand{\XttGccSOaSqlRdaPrecision}{1.000\xspace}
\newcommand{\XttGccSOaSqlRdaFone}{0.921\xspace}
\newcommand{\XttGccSOaSqlRseRecall}{0.931\xspace}
\newcommand{\XttGccSOaSqlRsePrecision}{1.000\xspace}
\newcommand{\XttGccSOaSqlRseFone}{0.964\xspace}
\newcommand{\XttGccSOaVimGT}{564162\xspace}
\newcommand{\XttGccSOaVimLsRecall}{1\xspace}
\newcommand{\XttGccSOaVimLsPrecision}{1\xspace}
\newcommand{\XttGccSOaVimLsFone}{1\xspace}
\newcommand{\XttGccSOaVimBapRecall}{1\xspace}
\newcommand{\XttGccSOaVimBapPrecision}{1\xspace}
\newcommand{\XttGccSOaVimBapFone}{1\xspace}
\newcommand{\XttGccSOaVimGhiRecall}{0.350\xspace}
\newcommand{\XttGccSOaVimGhiPrecision}{0.988\xspace}
\newcommand{\XttGccSOaVimGhiFone}{0.516\xspace}
\newcommand{\XttGccSOaVimRdaRecall}{0.903\xspace}
\newcommand{\XttGccSOaVimRdaPrecision}{1.000\xspace}
\newcommand{\XttGccSOaVimRdaFone}{0.949\xspace}
\newcommand{\XttGccSOaVimRseRecall}{0.985\xspace}
\newcommand{\XttGccSOaVimRsePrecision}{1.000\xspace}
\newcommand{\XttGccSOaVimRseFone}{0.993\xspace}
\newcommand{\XttGccSOaVsfGT}{24387\xspace}
\newcommand{\XttGccSOaVsfLsRecall}{1\xspace}
\newcommand{\XttGccSOaVsfLsPrecision}{1\xspace}
\newcommand{\XttGccSOaVsfLsFone}{1\xspace}
\newcommand{\XttGccSOaVsfBapRecall}{1\xspace}
\newcommand{\XttGccSOaVsfBapPrecision}{1\xspace}
\newcommand{\XttGccSOaVsfBapFone}{1\xspace}
\newcommand{\XttGccSOaVsfGhiRecall}{0.144\xspace}
\newcommand{\XttGccSOaVsfGhiPrecision}{0.994\xspace}
\newcommand{\XttGccSOaVsfGhiFone}{0.251\xspace}
\newcommand{\XttGccSOaVsfRdaRecall}{0.904\xspace}
\newcommand{\XttGccSOaVsfRdaPrecision}{1\xspace}
\newcommand{\XttGccSOaVsfRdaFone}{0.949\xspace}
\newcommand{\XttGccSOaVsfRseRecall}{1\xspace}
\newcommand{\XttGccSOaVsfRsePrecision}{1\xspace}
\newcommand{\XttGccSOaVsfRseFone}{1\xspace}
\newcommand{\XttGccSObSzpGT}{14303\xspace}
\newcommand{\XttGccSObSzpLsRecall}{1\xspace}
\newcommand{\XttGccSObSzpLsPrecision}{1\xspace}
\newcommand{\XttGccSObSzpLsFone}{1\xspace}
\newcommand{\XttGccSObSzpBapRecall}{1\xspace}
\newcommand{\XttGccSObSzpBapPrecision}{1\xspace}
\newcommand{\XttGccSObSzpBapFone}{1\xspace}
\newcommand{\XttGccSObSzpGhiRecall}{1\xspace}
\newcommand{\XttGccSObSzpGhiPrecision}{1\xspace}
\newcommand{\XttGccSObSzpGhiFone}{1\xspace}
\newcommand{\XttGccSObSzpRdaRecall}{0.976\xspace}
\newcommand{\XttGccSObSzpRdaPrecision}{1\xspace}
\newcommand{\XttGccSObSzpRdaFone}{0.988\xspace}
\newcommand{\XttGccSObSzpRseRecall}{1\xspace}
\newcommand{\XttGccSObSzpRsePrecision}{1\xspace}
\newcommand{\XttGccSObSzpRseFone}{1\xspace}
\newcommand{\XttGccSObCapGT}{237701\xspace}
\newcommand{\XttGccSObCapLsRecall}{1\xspace}
\newcommand{\XttGccSObCapLsPrecision}{1\xspace}
\newcommand{\XttGccSObCapLsFone}{1\xspace}
\newcommand{\XttGccSObCapBapRecall}{1\xspace}
\newcommand{\XttGccSObCapBapPrecision}{1\xspace}
\newcommand{\XttGccSObCapBapFone}{1\xspace}
\newcommand{\XttGccSObCapGhiRecall}{0.289\xspace}
\newcommand{\XttGccSObCapGhiPrecision}{0.993\xspace}
\newcommand{\XttGccSObCapGhiFone}{0.448\xspace}
\newcommand{\XttGccSObCapRdaRecall}{0.429\xspace}
\newcommand{\XttGccSObCapRdaPrecision}{0.999\xspace}
\newcommand{\XttGccSObCapRdaFone}{0.600\xspace}
\newcommand{\XttGccSObCapRseRecall}{0.650\xspace}
\newcommand{\XttGccSObCapRsePrecision}{1.000\xspace}
\newcommand{\XttGccSObCapRseFone}{0.788\xspace}
\newcommand{\XttGccSObExmGT}{170425\xspace}
\newcommand{\XttGccSObExmLsRecall}{1\xspace}
\newcommand{\XttGccSObExmLsPrecision}{1\xspace}
\newcommand{\XttGccSObExmLsFone}{1\xspace}
\newcommand{\XttGccSObExmBapRecall}{1\xspace}
\newcommand{\XttGccSObExmBapPrecision}{1\xspace}
\newcommand{\XttGccSObExmBapFone}{1\xspace}
\newcommand{\XttGccSObExmGhiRecall}{0.473\xspace}
\newcommand{\XttGccSObExmGhiPrecision}{0.996\xspace}
\newcommand{\XttGccSObExmGhiFone}{0.641\xspace}
\newcommand{\XttGccSObExmRdaRecall}{0.807\xspace}
\newcommand{\XttGccSObExmRdaPrecision}{1\xspace}
\newcommand{\XttGccSObExmRdaFone}{0.893\xspace}
\newcommand{\XttGccSObExmRseRecall}{0.996\xspace}
\newcommand{\XttGccSObExmRsePrecision}{1\xspace}
\newcommand{\XttGccSObExmRseFone}{0.998\xspace}
\newcommand{\XttGccSObLgtGT}{32272\xspace}
\newcommand{\XttGccSObLgtLsRecall}{1\xspace}
\newcommand{\XttGccSObLgtLsPrecision}{1\xspace}
\newcommand{\XttGccSObLgtLsFone}{1\xspace}
\newcommand{\XttGccSObLgtBapRecall}{1\xspace}
\newcommand{\XttGccSObLgtBapPrecision}{1\xspace}
\newcommand{\XttGccSObLgtBapFone}{1\xspace}
\newcommand{\XttGccSObLgtGhiRecall}{0.946\xspace}
\newcommand{\XttGccSObLgtGhiPrecision}{0.999\xspace}
\newcommand{\XttGccSObLgtGhiFone}{0.972\xspace}
\newcommand{\XttGccSObLgtRdaRecall}{0.872\xspace}
\newcommand{\XttGccSObLgtRdaPrecision}{1\xspace}
\newcommand{\XttGccSObLgtRdaFone}{0.931\xspace}
\newcommand{\XttGccSObLgtRseRecall}{0.995\xspace}
\newcommand{\XttGccSObLgtRsePrecision}{1\xspace}
\newcommand{\XttGccSObLgtRseFone}{0.997\xspace}
\newcommand{\XttGccSObBzpGT}{16242\xspace}
\newcommand{\XttGccSObBzpLsRecall}{1\xspace}
\newcommand{\XttGccSObBzpLsPrecision}{1\xspace}
\newcommand{\XttGccSObBzpLsFone}{1\xspace}
\newcommand{\XttGccSObBzpBapRecall}{1\xspace}
\newcommand{\XttGccSObBzpBapPrecision}{1\xspace}
\newcommand{\XttGccSObBzpBapFone}{1\xspace}
\newcommand{\XttGccSObBzpGhiRecall}{0.997\xspace}
\newcommand{\XttGccSObBzpGhiPrecision}{1\xspace}
\newcommand{\XttGccSObBzpGhiFone}{0.998\xspace}
\newcommand{\XttGccSObBzpRdaRecall}{0.616\xspace}
\newcommand{\XttGccSObBzpRdaPrecision}{1\xspace}
\newcommand{\XttGccSObBzpRdaFone}{0.763\xspace}
\newcommand{\XttGccSObBzpRseRecall}{1.000\xspace}
\newcommand{\XttGccSObBzpRsePrecision}{1\xspace}
\newcommand{\XttGccSObBzpRseFone}{1.000\xspace}
\newcommand{\XttGccSObGccGT}{1005179\xspace}
\newcommand{\XttGccSObGccLsRecall}{1\xspace}
\newcommand{\XttGccSObGccLsPrecision}{1\xspace}
\newcommand{\XttGccSObGccLsFone}{1\xspace}
\newcommand{\XttGccSObGccBapRecall}{1\xspace}
\newcommand{\XttGccSObGccBapPrecision}{1.000\xspace}
\newcommand{\XttGccSObGccBapFone}{1.000\xspace}
\newcommand{\XttGccSObGccGhiRecall}{0.382\xspace}
\newcommand{\XttGccSObGccGhiPrecision}{0.988\xspace}
\newcommand{\XttGccSObGccGhiFone}{0.551\xspace}
\newcommand{\XttGccSObGccRdaRecall}{0.626\xspace}
\newcommand{\XttGccSObGccRdaPrecision}{1.000\xspace}
\newcommand{\XttGccSObGccRdaFone}{0.770\xspace}
\newcommand{\XttGccSObGccRseRecall}{0.978\xspace}
\newcommand{\XttGccSObGccRsePrecision}{1.000\xspace}
\newcommand{\XttGccSObGccRseFone}{0.989\xspace}
\newcommand{\XttGccSObGzpGT}{11910\xspace}
\newcommand{\XttGccSObGzpLsRecall}{1\xspace}
\newcommand{\XttGccSObGzpLsPrecision}{1\xspace}
\newcommand{\XttGccSObGzpLsFone}{1\xspace}
\newcommand{\XttGccSObGzpBapRecall}{1\xspace}
\newcommand{\XttGccSObGzpBapPrecision}{1\xspace}
\newcommand{\XttGccSObGzpBapFone}{1\xspace}
\newcommand{\XttGccSObGzpGhiRecall}{1\xspace}
\newcommand{\XttGccSObGzpGhiPrecision}{1\xspace}
\newcommand{\XttGccSObGzpGhiFone}{1\xspace}
\newcommand{\XttGccSObGzpRdaRecall}{0.984\xspace}
\newcommand{\XttGccSObGzpRdaPrecision}{1\xspace}
\newcommand{\XttGccSObGzpRdaFone}{0.992\xspace}
\newcommand{\XttGccSObGzpRseRecall}{1\xspace}
\newcommand{\XttGccSObGzpRsePrecision}{1\xspace}
\newcommand{\XttGccSObGzpRseFone}{1\xspace}
\newcommand{\XttGccSObOggGT}{44745\xspace}
\newcommand{\XttGccSObOggLsRecall}{1\xspace}
\newcommand{\XttGccSObOggLsPrecision}{1\xspace}
\newcommand{\XttGccSObOggLsFone}{1\xspace}
\newcommand{\XttGccSObOggBapRecall}{1\xspace}
\newcommand{\XttGccSObOggBapPrecision}{1\xspace}
\newcommand{\XttGccSObOggBapFone}{1\xspace}
\newcommand{\XttGccSObOggGhiRecall}{0.637\xspace}
\newcommand{\XttGccSObOggGhiPrecision}{0.998\xspace}
\newcommand{\XttGccSObOggGhiFone}{0.777\xspace}
\newcommand{\XttGccSObOggRdaRecall}{0.978\xspace}
\newcommand{\XttGccSObOggRdaPrecision}{1\xspace}
\newcommand{\XttGccSObOggRdaFone}{0.989\xspace}
\newcommand{\XttGccSObOggRseRecall}{1\xspace}
\newcommand{\XttGccSObOggRsePrecision}{1\xspace}
\newcommand{\XttGccSObOggRseFone}{1\xspace}
\newcommand{\XttGccSObNgxGT}{121558\xspace}
\newcommand{\XttGccSObNgxLsRecall}{1\xspace}
\newcommand{\XttGccSObNgxLsPrecision}{1\xspace}
\newcommand{\XttGccSObNgxLsFone}{1\xspace}
\newcommand{\XttGccSObNgxBapRecall}{1\xspace}
\newcommand{\XttGccSObNgxBapPrecision}{1\xspace}
\newcommand{\XttGccSObNgxBapFone}{1\xspace}
\newcommand{\XttGccSObNgxGhiRecall}{0.358\xspace}
\newcommand{\XttGccSObNgxGhiPrecision}{0.986\xspace}
\newcommand{\XttGccSObNgxGhiFone}{0.525\xspace}
\newcommand{\XttGccSObNgxRdaRecall}{0.957\xspace}
\newcommand{\XttGccSObNgxRdaPrecision}{1\xspace}
\newcommand{\XttGccSObNgxRdaFone}{0.978\xspace}
\newcommand{\XttGccSObNgxRseRecall}{0.997\xspace}
\newcommand{\XttGccSObNgxRsePrecision}{1\xspace}
\newcommand{\XttGccSObNgxRseFone}{0.999\xspace}
\newcommand{\XttGccSObSshGT}{134393\xspace}
\newcommand{\XttGccSObSshLsRecall}{1\xspace}
\newcommand{\XttGccSObSshLsPrecision}{1\xspace}
\newcommand{\XttGccSObSshLsFone}{1\xspace}
\newcommand{\XttGccSObSshBapRecall}{1\xspace}
\newcommand{\XttGccSObSshBapPrecision}{1\xspace}
\newcommand{\XttGccSObSshBapFone}{1\xspace}
\newcommand{\XttGccSObSshGhiRecall}{0.391\xspace}
\newcommand{\XttGccSObSshGhiPrecision}{0.990\xspace}
\newcommand{\XttGccSObSshGhiFone}{0.561\xspace}
\newcommand{\XttGccSObSshRdaRecall}{0.858\xspace}
\newcommand{\XttGccSObSshRdaPrecision}{1\xspace}
\newcommand{\XttGccSObSshRdaFone}{0.924\xspace}
\newcommand{\XttGccSObSshRseRecall}{0.991\xspace}
\newcommand{\XttGccSObSshRsePrecision}{1\xspace}
\newcommand{\XttGccSObSshRseFone}{0.996\xspace}
\newcommand{\XttGccSObPcrGT}{5176\xspace}
\newcommand{\XttGccSObPcrLsRecall}{1\xspace}
\newcommand{\XttGccSObPcrLsPrecision}{1\xspace}
\newcommand{\XttGccSObPcrLsFone}{1\xspace}
\newcommand{\XttGccSObPcrBapRecall}{1\xspace}
\newcommand{\XttGccSObPcrBapPrecision}{1\xspace}
\newcommand{\XttGccSObPcrBapFone}{1\xspace}
\newcommand{\XttGccSObPcrGhiRecall}{1\xspace}
\newcommand{\XttGccSObPcrGhiPrecision}{1\xspace}
\newcommand{\XttGccSObPcrGhiFone}{1\xspace}
\newcommand{\XttGccSObPcrRdaRecall}{0.914\xspace}
\newcommand{\XttGccSObPcrRdaPrecision}{1\xspace}
\newcommand{\XttGccSObPcrRdaFone}{0.955\xspace}
\newcommand{\XttGccSObPcrRseRecall}{0.981\xspace}
\newcommand{\XttGccSObPcrRsePrecision}{1\xspace}
\newcommand{\XttGccSObPcrRseFone}{0.990\xspace}
\newcommand{\XttGccSObSqlGT}{195811\xspace}
\newcommand{\XttGccSObSqlLsRecall}{1\xspace}
\newcommand{\XttGccSObSqlLsPrecision}{1\xspace}
\newcommand{\XttGccSObSqlLsFone}{1\xspace}
\newcommand{\XttGccSObSqlBapRecall}{1\xspace}
\newcommand{\XttGccSObSqlBapPrecision}{1\xspace}
\newcommand{\XttGccSObSqlBapFone}{1\xspace}
\newcommand{\XttGccSObSqlGhiRecall}{0.457\xspace}
\newcommand{\XttGccSObSqlGhiPrecision}{0.993\xspace}
\newcommand{\XttGccSObSqlGhiFone}{0.626\xspace}
\newcommand{\XttGccSObSqlRdaRecall}{0.857\xspace}
\newcommand{\XttGccSObSqlRdaPrecision}{1.000\xspace}
\newcommand{\XttGccSObSqlRdaFone}{0.923\xspace}
\newcommand{\XttGccSObSqlRseRecall}{0.986\xspace}
\newcommand{\XttGccSObSqlRsePrecision}{1.000\xspace}
\newcommand{\XttGccSObSqlRseFone}{0.993\xspace}
\newcommand{\XttGccSObVimGT}{591753\xspace}
\newcommand{\XttGccSObVimLsRecall}{1\xspace}
\newcommand{\XttGccSObVimLsPrecision}{1\xspace}
\newcommand{\XttGccSObVimLsFone}{1\xspace}
\newcommand{\XttGccSObVimBapRecall}{1\xspace}
\newcommand{\XttGccSObVimBapPrecision}{1.000\xspace}
\newcommand{\XttGccSObVimBapFone}{1.000\xspace}
\newcommand{\XttGccSObVimGhiRecall}{0.344\xspace}
\newcommand{\XttGccSObVimGhiPrecision}{0.988\xspace}
\newcommand{\XttGccSObVimGhiFone}{0.510\xspace}
\newcommand{\XttGccSObVimRdaRecall}{0.914\xspace}
\newcommand{\XttGccSObVimRdaPrecision}{1.000\xspace}
\newcommand{\XttGccSObVimRdaFone}{0.955\xspace}
\newcommand{\XttGccSObVimRseRecall}{0.996\xspace}
\newcommand{\XttGccSObVimRsePrecision}{1.000\xspace}
\newcommand{\XttGccSObVimRseFone}{0.998\xspace}
\newcommand{\XttGccSObVsfGT}{25934\xspace}
\newcommand{\XttGccSObVsfLsRecall}{1\xspace}
\newcommand{\XttGccSObVsfLsPrecision}{1\xspace}
\newcommand{\XttGccSObVsfLsFone}{1\xspace}
\newcommand{\XttGccSObVsfBapRecall}{1\xspace}
\newcommand{\XttGccSObVsfBapPrecision}{1\xspace}
\newcommand{\XttGccSObVsfBapFone}{1\xspace}
\newcommand{\XttGccSObVsfGhiRecall}{0.153\xspace}
\newcommand{\XttGccSObVsfGhiPrecision}{0.991\xspace}
\newcommand{\XttGccSObVsfGhiFone}{0.265\xspace}
\newcommand{\XttGccSObVsfRdaRecall}{0.861\xspace}
\newcommand{\XttGccSObVsfRdaPrecision}{1\xspace}
\newcommand{\XttGccSObVsfRdaFone}{0.926\xspace}
\newcommand{\XttGccSObVsfRseRecall}{1\xspace}
\newcommand{\XttGccSObVsfRsePrecision}{1\xspace}
\newcommand{\XttGccSObVsfRseFone}{1\xspace}
\newcommand{\XttGccSOcSzpGT}{17392\xspace}
\newcommand{\XttGccSOcSzpLsRecall}{1\xspace}
\newcommand{\XttGccSOcSzpLsPrecision}{1\xspace}
\newcommand{\XttGccSOcSzpLsFone}{1\xspace}
\newcommand{\XttGccSOcSzpBapRecall}{1\xspace}
\newcommand{\XttGccSOcSzpBapPrecision}{1\xspace}
\newcommand{\XttGccSOcSzpBapFone}{1\xspace}
\newcommand{\XttGccSOcSzpGhiRecall}{0.986\xspace}
\newcommand{\XttGccSOcSzpGhiPrecision}{1\xspace}
\newcommand{\XttGccSOcSzpGhiFone}{0.993\xspace}
\newcommand{\XttGccSOcSzpRdaRecall}{0.957\xspace}
\newcommand{\XttGccSOcSzpRdaPrecision}{1\xspace}
\newcommand{\XttGccSOcSzpRdaFone}{0.978\xspace}
\newcommand{\XttGccSOcSzpRseRecall}{1\xspace}
\newcommand{\XttGccSOcSzpRsePrecision}{1\xspace}
\newcommand{\XttGccSOcSzpRseFone}{1\xspace}
\newcommand{\XttGccSOcCapGT}{271107\xspace}
\newcommand{\XttGccSOcCapLsRecall}{1\xspace}
\newcommand{\XttGccSOcCapLsPrecision}{1\xspace}
\newcommand{\XttGccSOcCapLsFone}{1\xspace}
\newcommand{\XttGccSOcCapBapRecall}{1\xspace}
\newcommand{\XttGccSOcCapBapPrecision}{1\xspace}
\newcommand{\XttGccSOcCapBapFone}{1\xspace}
\newcommand{\XttGccSOcCapGhiRecall}{0.240\xspace}
\newcommand{\XttGccSOcCapGhiPrecision}{0.991\xspace}
\newcommand{\XttGccSOcCapGhiFone}{0.386\xspace}
\newcommand{\XttGccSOcCapRdaRecall}{0.472\xspace}
\newcommand{\XttGccSOcCapRdaPrecision}{0.999\xspace}
\newcommand{\XttGccSOcCapRdaFone}{0.641\xspace}
\newcommand{\XttGccSOcCapRseRecall}{0.682\xspace}
\newcommand{\XttGccSOcCapRsePrecision}{1.000\xspace}
\newcommand{\XttGccSOcCapRseFone}{0.811\xspace}
\newcommand{\XttGccSOcExmGT}{206707\xspace}
\newcommand{\XttGccSOcExmLsRecall}{1\xspace}
\newcommand{\XttGccSOcExmLsPrecision}{1\xspace}
\newcommand{\XttGccSOcExmLsFone}{1\xspace}
\newcommand{\XttGccSOcExmBapRecall}{1\xspace}
\newcommand{\XttGccSOcExmBapPrecision}{1\xspace}
\newcommand{\XttGccSOcExmBapFone}{1\xspace}
\newcommand{\XttGccSOcExmGhiRecall}{0.483\xspace}
\newcommand{\XttGccSOcExmGhiPrecision}{0.997\xspace}
\newcommand{\XttGccSOcExmGhiFone}{0.650\xspace}
\newcommand{\XttGccSOcExmRdaRecall}{0.811\xspace}
\newcommand{\XttGccSOcExmRdaPrecision}{1\xspace}
\newcommand{\XttGccSOcExmRdaFone}{0.896\xspace}
\newcommand{\XttGccSOcExmRseRecall}{0.988\xspace}
\newcommand{\XttGccSOcExmRsePrecision}{1\xspace}
\newcommand{\XttGccSOcExmRseFone}{0.994\xspace}
\newcommand{\XttGccSOcLgtGT}{39269\xspace}
\newcommand{\XttGccSOcLgtLsRecall}{1\xspace}
\newcommand{\XttGccSOcLgtLsPrecision}{1\xspace}
\newcommand{\XttGccSOcLgtLsFone}{1\xspace}
\newcommand{\XttGccSOcLgtBapRecall}{1\xspace}
\newcommand{\XttGccSOcLgtBapPrecision}{1\xspace}
\newcommand{\XttGccSOcLgtBapFone}{1\xspace}
\newcommand{\XttGccSOcLgtGhiRecall}{0.505\xspace}
\newcommand{\XttGccSOcLgtGhiPrecision}{0.997\xspace}
\newcommand{\XttGccSOcLgtGhiFone}{0.671\xspace}
\newcommand{\XttGccSOcLgtRdaRecall}{0.886\xspace}
\newcommand{\XttGccSOcLgtRdaPrecision}{1\xspace}
\newcommand{\XttGccSOcLgtRdaFone}{0.940\xspace}
\newcommand{\XttGccSOcLgtRseRecall}{0.995\xspace}
\newcommand{\XttGccSOcLgtRsePrecision}{1\xspace}
\newcommand{\XttGccSOcLgtRseFone}{0.998\xspace}
\newcommand{\XttGccSOcBzpGT}{18507\xspace}
\newcommand{\XttGccSOcBzpLsRecall}{1\xspace}
\newcommand{\XttGccSOcBzpLsPrecision}{1\xspace}
\newcommand{\XttGccSOcBzpLsFone}{1\xspace}
\newcommand{\XttGccSOcBzpBapRecall}{1\xspace}
\newcommand{\XttGccSOcBzpBapPrecision}{1\xspace}
\newcommand{\XttGccSOcBzpBapFone}{1\xspace}
\newcommand{\XttGccSOcBzpGhiRecall}{0.953\xspace}
\newcommand{\XttGccSOcBzpGhiPrecision}{1\xspace}
\newcommand{\XttGccSOcBzpGhiFone}{0.976\xspace}
\newcommand{\XttGccSOcBzpRdaRecall}{0.696\xspace}
\newcommand{\XttGccSOcBzpRdaPrecision}{1\xspace}
\newcommand{\XttGccSOcBzpRdaFone}{0.821\xspace}
\newcommand{\XttGccSOcBzpRseRecall}{1.000\xspace}
\newcommand{\XttGccSOcBzpRsePrecision}{1\xspace}
\newcommand{\XttGccSOcBzpRseFone}{1.000\xspace}
\newcommand{\XttGccSOcGccGT}{1251290\xspace}
\newcommand{\XttGccSOcGccLsRecall}{1\xspace}
\newcommand{\XttGccSOcGccLsPrecision}{1\xspace}
\newcommand{\XttGccSOcGccLsFone}{1\xspace}
\newcommand{\XttGccSOcGccBapRecall}{1\xspace}
\newcommand{\XttGccSOcGccBapPrecision}{1\xspace}
\newcommand{\XttGccSOcGccBapFone}{1\xspace}
\newcommand{\XttGccSOcGccGhiRecall}{0.417\xspace}
\newcommand{\XttGccSOcGccGhiPrecision}{0.991\xspace}
\newcommand{\XttGccSOcGccGhiFone}{0.587\xspace}
\newcommand{\XttGccSOcGccRdaRecall}{0.664\xspace}
\newcommand{\XttGccSOcGccRdaPrecision}{1.000\xspace}
\newcommand{\XttGccSOcGccRdaFone}{0.798\xspace}
\newcommand{\XttGccSOcGccRseRecall}{0.980\xspace}
\newcommand{\XttGccSOcGccRsePrecision}{1.000\xspace}
\newcommand{\XttGccSOcGccRseFone}{0.990\xspace}
\newcommand{\XttGccSOcGzpGT}{19852\xspace}
\newcommand{\XttGccSOcGzpLsRecall}{1\xspace}
\newcommand{\XttGccSOcGzpLsPrecision}{1\xspace}
\newcommand{\XttGccSOcGzpLsFone}{1\xspace}
\newcommand{\XttGccSOcGzpBapRecall}{1\xspace}
\newcommand{\XttGccSOcGzpBapPrecision}{1\xspace}
\newcommand{\XttGccSOcGzpBapFone}{1\xspace}
\newcommand{\XttGccSOcGzpGhiRecall}{0.998\xspace}
\newcommand{\XttGccSOcGzpGhiPrecision}{1\xspace}
\newcommand{\XttGccSOcGzpGhiFone}{0.999\xspace}
\newcommand{\XttGccSOcGzpRdaRecall}{0.850\xspace}
\newcommand{\XttGccSOcGzpRdaPrecision}{1\xspace}
\newcommand{\XttGccSOcGzpRdaFone}{0.919\xspace}
\newcommand{\XttGccSOcGzpRseRecall}{1\xspace}
\newcommand{\XttGccSOcGzpRsePrecision}{1\xspace}
\newcommand{\XttGccSOcGzpRseFone}{1\xspace}
\newcommand{\XttGccSOcOggGT}{55883\xspace}
\newcommand{\XttGccSOcOggLsRecall}{1\xspace}
\newcommand{\XttGccSOcOggLsPrecision}{1\xspace}
\newcommand{\XttGccSOcOggLsFone}{1\xspace}
\newcommand{\XttGccSOcOggBapRecall}{1\xspace}
\newcommand{\XttGccSOcOggBapPrecision}{1\xspace}
\newcommand{\XttGccSOcOggBapFone}{1\xspace}
\newcommand{\XttGccSOcOggGhiRecall}{0.449\xspace}
\newcommand{\XttGccSOcOggGhiPrecision}{0.996\xspace}
\newcommand{\XttGccSOcOggGhiFone}{0.619\xspace}
\newcommand{\XttGccSOcOggRdaRecall}{0.982\xspace}
\newcommand{\XttGccSOcOggRdaPrecision}{1\xspace}
\newcommand{\XttGccSOcOggRdaFone}{0.991\xspace}
\newcommand{\XttGccSOcOggRseRecall}{1\xspace}
\newcommand{\XttGccSOcOggRsePrecision}{1\xspace}
\newcommand{\XttGccSOcOggRseFone}{1\xspace}
\newcommand{\XttGccSOcNgxGT}{141382\xspace}
\newcommand{\XttGccSOcNgxLsRecall}{1\xspace}
\newcommand{\XttGccSOcNgxLsPrecision}{1\xspace}
\newcommand{\XttGccSOcNgxLsFone}{1\xspace}
\newcommand{\XttGccSOcNgxBapRecall}{1\xspace}
\newcommand{\XttGccSOcNgxBapPrecision}{1\xspace}
\newcommand{\XttGccSOcNgxBapFone}{1\xspace}
\newcommand{\XttGccSOcNgxGhiRecall}{0.395\xspace}
\newcommand{\XttGccSOcNgxGhiPrecision}{0.991\xspace}
\newcommand{\XttGccSOcNgxGhiFone}{0.565\xspace}
\newcommand{\XttGccSOcNgxRdaRecall}{0.960\xspace}
\newcommand{\XttGccSOcNgxRdaPrecision}{1.000\xspace}
\newcommand{\XttGccSOcNgxRdaFone}{0.979\xspace}
\newcommand{\XttGccSOcNgxRseRecall}{0.996\xspace}
\newcommand{\XttGccSOcNgxRsePrecision}{1.000\xspace}
\newcommand{\XttGccSOcNgxRseFone}{0.998\xspace}
\newcommand{\XttGccSOcSshGT}{160129\xspace}
\newcommand{\XttGccSOcSshLsRecall}{1\xspace}
\newcommand{\XttGccSOcSshLsPrecision}{1\xspace}
\newcommand{\XttGccSOcSshLsFone}{1\xspace}
\newcommand{\XttGccSOcSshBapRecall}{1\xspace}
\newcommand{\XttGccSOcSshBapPrecision}{1\xspace}
\newcommand{\XttGccSOcSshBapFone}{1\xspace}
\newcommand{\XttGccSOcSshGhiRecall}{0.388\xspace}
\newcommand{\XttGccSOcSshGhiPrecision}{0.992\xspace}
\newcommand{\XttGccSOcSshGhiFone}{0.558\xspace}
\newcommand{\XttGccSOcSshRdaRecall}{0.827\xspace}
\newcommand{\XttGccSOcSshRdaPrecision}{1\xspace}
\newcommand{\XttGccSOcSshRdaFone}{0.906\xspace}
\newcommand{\XttGccSOcSshRseRecall}{0.993\xspace}
\newcommand{\XttGccSOcSshRsePrecision}{1\xspace}
\newcommand{\XttGccSOcSshRseFone}{0.996\xspace}
\newcommand{\XttGccSOcPcrGT}{6126\xspace}
\newcommand{\XttGccSOcPcrLsRecall}{1\xspace}
\newcommand{\XttGccSOcPcrLsPrecision}{1\xspace}
\newcommand{\XttGccSOcPcrLsFone}{1\xspace}
\newcommand{\XttGccSOcPcrBapRecall}{1\xspace}
\newcommand{\XttGccSOcPcrBapPrecision}{1\xspace}
\newcommand{\XttGccSOcPcrBapFone}{1\xspace}
\newcommand{\XttGccSOcPcrGhiRecall}{0.987\xspace}
\newcommand{\XttGccSOcPcrGhiPrecision}{1\xspace}
\newcommand{\XttGccSOcPcrGhiFone}{0.994\xspace}
\newcommand{\XttGccSOcPcrRdaRecall}{0.884\xspace}
\newcommand{\XttGccSOcPcrRdaPrecision}{1\xspace}
\newcommand{\XttGccSOcPcrRdaFone}{0.939\xspace}
\newcommand{\XttGccSOcPcrRseRecall}{0.992\xspace}
\newcommand{\XttGccSOcPcrRsePrecision}{1\xspace}
\newcommand{\XttGccSOcPcrRseFone}{0.996\xspace}
\newcommand{\XttGccSOcSqlGT}{236529\xspace}
\newcommand{\XttGccSOcSqlLsRecall}{1\xspace}
\newcommand{\XttGccSOcSqlLsPrecision}{1\xspace}
\newcommand{\XttGccSOcSqlLsFone}{1\xspace}
\newcommand{\XttGccSOcSqlBapRecall}{1\xspace}
\newcommand{\XttGccSOcSqlBapPrecision}{1\xspace}
\newcommand{\XttGccSOcSqlBapFone}{1\xspace}
\newcommand{\XttGccSOcSqlGhiRecall}{0.458\xspace}
\newcommand{\XttGccSOcSqlGhiPrecision}{0.994\xspace}
\newcommand{\XttGccSOcSqlGhiFone}{0.627\xspace}
\newcommand{\XttGccSOcSqlRdaRecall}{0.858\xspace}
\newcommand{\XttGccSOcSqlRdaPrecision}{1.000\xspace}
\newcommand{\XttGccSOcSqlRdaFone}{0.923\xspace}
\newcommand{\XttGccSOcSqlRseRecall}{0.985\xspace}
\newcommand{\XttGccSOcSqlRsePrecision}{1.000\xspace}
\newcommand{\XttGccSOcSqlRseFone}{0.993\xspace}
\newcommand{\XttGccSOcVimGT}{825701\xspace}
\newcommand{\XttGccSOcVimLsRecall}{1\xspace}
\newcommand{\XttGccSOcVimLsPrecision}{1\xspace}
\newcommand{\XttGccSOcVimLsFone}{1\xspace}
\newcommand{\XttGccSOcVimBapRecall}{1\xspace}
\newcommand{\XttGccSOcVimBapPrecision}{1.000\xspace}
\newcommand{\XttGccSOcVimBapFone}{1.000\xspace}
\newcommand{\XttGccSOcVimGhiRecall}{0.358\xspace}
\newcommand{\XttGccSOcVimGhiPrecision}{0.993\xspace}
\newcommand{\XttGccSOcVimGhiFone}{0.526\xspace}
\newcommand{\XttGccSOcVimRdaRecall}{0.914\xspace}
\newcommand{\XttGccSOcVimRdaPrecision}{1.000\xspace}
\newcommand{\XttGccSOcVimRdaFone}{0.955\xspace}
\newcommand{\XttGccSOcVimRseRecall}{0.994\xspace}
\newcommand{\XttGccSOcVimRsePrecision}{1\xspace}
\newcommand{\XttGccSOcVimRseFone}{0.997\xspace}
\newcommand{\XttGccSOcVsfGT}{33861\xspace}
\newcommand{\XttGccSOcVsfLsRecall}{1\xspace}
\newcommand{\XttGccSOcVsfLsPrecision}{1\xspace}
\newcommand{\XttGccSOcVsfLsFone}{1\xspace}
\newcommand{\XttGccSOcVsfBapRecall}{1\xspace}
\newcommand{\XttGccSOcVsfBapPrecision}{1\xspace}
\newcommand{\XttGccSOcVsfBapFone}{1\xspace}
\newcommand{\XttGccSOcVsfGhiRecall}{0.141\xspace}
\newcommand{\XttGccSOcVsfGhiPrecision}{0.985\xspace}
\newcommand{\XttGccSOcVsfGhiFone}{0.247\xspace}
\newcommand{\XttGccSOcVsfRdaRecall}{0.882\xspace}
\newcommand{\XttGccSOcVsfRdaPrecision}{1\xspace}
\newcommand{\XttGccSOcVsfRdaFone}{0.937\xspace}
\newcommand{\XttGccSOcVsfRseRecall}{1\xspace}
\newcommand{\XttGccSOcVsfRsePrecision}{1\xspace}
\newcommand{\XttGccSOcVsfRseFone}{1\xspace}
\newcommand{\XttGccSOdSzpGT}{17392\xspace}
\newcommand{\XttGccSOdSzpLsRecall}{1\xspace}
\newcommand{\XttGccSOdSzpLsPrecision}{1\xspace}
\newcommand{\XttGccSOdSzpLsFone}{1\xspace}
\newcommand{\XttGccSOdSzpBapRecall}{1\xspace}
\newcommand{\XttGccSOdSzpBapPrecision}{1\xspace}
\newcommand{\XttGccSOdSzpBapFone}{1\xspace}
\newcommand{\XttGccSOdSzpGhiRecall}{0.986\xspace}
\newcommand{\XttGccSOdSzpGhiPrecision}{1\xspace}
\newcommand{\XttGccSOdSzpGhiFone}{0.993\xspace}
\newcommand{\XttGccSOdSzpRdaRecall}{0.957\xspace}
\newcommand{\XttGccSOdSzpRdaPrecision}{1\xspace}
\newcommand{\XttGccSOdSzpRdaFone}{0.978\xspace}
\newcommand{\XttGccSOdSzpRseRecall}{1\xspace}
\newcommand{\XttGccSOdSzpRsePrecision}{1\xspace}
\newcommand{\XttGccSOdSzpRseFone}{1\xspace}
\newcommand{\XttGccSOdCapGT}{271107\xspace}
\newcommand{\XttGccSOdCapLsRecall}{1\xspace}
\newcommand{\XttGccSOdCapLsPrecision}{1\xspace}
\newcommand{\XttGccSOdCapLsFone}{1\xspace}
\newcommand{\XttGccSOdCapBapRecall}{1\xspace}
\newcommand{\XttGccSOdCapBapPrecision}{1\xspace}
\newcommand{\XttGccSOdCapBapFone}{1\xspace}
\newcommand{\XttGccSOdCapGhiRecall}{0.240\xspace}
\newcommand{\XttGccSOdCapGhiPrecision}{0.990\xspace}
\newcommand{\XttGccSOdCapGhiFone}{0.386\xspace}
\newcommand{\XttGccSOdCapRdaRecall}{0.472\xspace}
\newcommand{\XttGccSOdCapRdaPrecision}{0.999\xspace}
\newcommand{\XttGccSOdCapRdaFone}{0.642\xspace}
\newcommand{\XttGccSOdCapRseRecall}{0.682\xspace}
\newcommand{\XttGccSOdCapRsePrecision}{1.000\xspace}
\newcommand{\XttGccSOdCapRseFone}{0.811\xspace}
\newcommand{\XttGccSOdExmGT}{206609\xspace}
\newcommand{\XttGccSOdExmLsRecall}{1\xspace}
\newcommand{\XttGccSOdExmLsPrecision}{1\xspace}
\newcommand{\XttGccSOdExmLsFone}{1\xspace}
\newcommand{\XttGccSOdExmBapRecall}{1\xspace}
\newcommand{\XttGccSOdExmBapPrecision}{1\xspace}
\newcommand{\XttGccSOdExmBapFone}{1\xspace}
\newcommand{\XttGccSOdExmGhiRecall}{0.487\xspace}
\newcommand{\XttGccSOdExmGhiPrecision}{0.997\xspace}
\newcommand{\XttGccSOdExmGhiFone}{0.655\xspace}
\newcommand{\XttGccSOdExmRdaRecall}{0.812\xspace}
\newcommand{\XttGccSOdExmRdaPrecision}{1\xspace}
\newcommand{\XttGccSOdExmRdaFone}{0.896\xspace}
\newcommand{\XttGccSOdExmRseRecall}{0.988\xspace}
\newcommand{\XttGccSOdExmRsePrecision}{1\xspace}
\newcommand{\XttGccSOdExmRseFone}{0.994\xspace}
\newcommand{\XttGccSOdLgtGT}{39260\xspace}
\newcommand{\XttGccSOdLgtLsRecall}{1\xspace}
\newcommand{\XttGccSOdLgtLsPrecision}{1\xspace}
\newcommand{\XttGccSOdLgtLsFone}{1\xspace}
\newcommand{\XttGccSOdLgtBapRecall}{1\xspace}
\newcommand{\XttGccSOdLgtBapPrecision}{1\xspace}
\newcommand{\XttGccSOdLgtBapFone}{1\xspace}
\newcommand{\XttGccSOdLgtGhiRecall}{0.504\xspace}
\newcommand{\XttGccSOdLgtGhiPrecision}{0.997\xspace}
\newcommand{\XttGccSOdLgtGhiFone}{0.669\xspace}
\newcommand{\XttGccSOdLgtRdaRecall}{0.886\xspace}
\newcommand{\XttGccSOdLgtRdaPrecision}{1\xspace}
\newcommand{\XttGccSOdLgtRdaFone}{0.940\xspace}
\newcommand{\XttGccSOdLgtRseRecall}{0.995\xspace}
\newcommand{\XttGccSOdLgtRsePrecision}{1\xspace}
\newcommand{\XttGccSOdLgtRseFone}{0.998\xspace}
\newcommand{\XttGccSOdBzpGT}{18492\xspace}
\newcommand{\XttGccSOdBzpLsRecall}{1\xspace}
\newcommand{\XttGccSOdBzpLsPrecision}{1\xspace}
\newcommand{\XttGccSOdBzpLsFone}{1\xspace}
\newcommand{\XttGccSOdBzpBapRecall}{1\xspace}
\newcommand{\XttGccSOdBzpBapPrecision}{1\xspace}
\newcommand{\XttGccSOdBzpBapFone}{1\xspace}
\newcommand{\XttGccSOdBzpGhiRecall}{0.953\xspace}
\newcommand{\XttGccSOdBzpGhiPrecision}{1\xspace}
\newcommand{\XttGccSOdBzpGhiFone}{0.976\xspace}
\newcommand{\XttGccSOdBzpRdaRecall}{0.696\xspace}
\newcommand{\XttGccSOdBzpRdaPrecision}{1\xspace}
\newcommand{\XttGccSOdBzpRdaFone}{0.821\xspace}
\newcommand{\XttGccSOdBzpRseRecall}{1.000\xspace}
\newcommand{\XttGccSOdBzpRsePrecision}{1\xspace}
\newcommand{\XttGccSOdBzpRseFone}{1.000\xspace}
\newcommand{\XttGccSOdGccGT}{1250527\xspace}
\newcommand{\XttGccSOdGccLsRecall}{1\xspace}
\newcommand{\XttGccSOdGccLsPrecision}{1\xspace}
\newcommand{\XttGccSOdGccLsFone}{1\xspace}
\newcommand{\XttGccSOdGccBapRecall}{1\xspace}
\newcommand{\XttGccSOdGccBapPrecision}{1\xspace}
\newcommand{\XttGccSOdGccBapFone}{1\xspace}
\newcommand{\XttGccSOdGccGhiRecall}{0.415\xspace}
\newcommand{\XttGccSOdGccGhiPrecision}{0.991\xspace}
\newcommand{\XttGccSOdGccGhiFone}{0.585\xspace}
\newcommand{\XttGccSOdGccRdaRecall}{0.662\xspace}
\newcommand{\XttGccSOdGccRdaPrecision}{1.000\xspace}
\newcommand{\XttGccSOdGccRdaFone}{0.797\xspace}
\newcommand{\XttGccSOdGccRseRecall}{0.980\xspace}
\newcommand{\XttGccSOdGccRsePrecision}{1.000\xspace}
\newcommand{\XttGccSOdGccRseFone}{0.990\xspace}
\newcommand{\XttGccSOdGzpGT}{19757\xspace}
\newcommand{\XttGccSOdGzpLsRecall}{1\xspace}
\newcommand{\XttGccSOdGzpLsPrecision}{1\xspace}
\newcommand{\XttGccSOdGzpLsFone}{1\xspace}
\newcommand{\XttGccSOdGzpBapRecall}{1\xspace}
\newcommand{\XttGccSOdGzpBapPrecision}{1\xspace}
\newcommand{\XttGccSOdGzpBapFone}{1\xspace}
\newcommand{\XttGccSOdGzpGhiRecall}{0.991\xspace}
\newcommand{\XttGccSOdGzpGhiPrecision}{1\xspace}
\newcommand{\XttGccSOdGzpGhiFone}{0.995\xspace}
\newcommand{\XttGccSOdGzpRdaRecall}{0.849\xspace}
\newcommand{\XttGccSOdGzpRdaPrecision}{1\xspace}
\newcommand{\XttGccSOdGzpRdaFone}{0.919\xspace}
\newcommand{\XttGccSOdGzpRseRecall}{1\xspace}
\newcommand{\XttGccSOdGzpRsePrecision}{1\xspace}
\newcommand{\XttGccSOdGzpRseFone}{1\xspace}
\newcommand{\XttGccSOdOggGT}{55143\xspace}
\newcommand{\XttGccSOdOggLsRecall}{1\xspace}
\newcommand{\XttGccSOdOggLsPrecision}{1\xspace}
\newcommand{\XttGccSOdOggLsFone}{1\xspace}
\newcommand{\XttGccSOdOggBapRecall}{1\xspace}
\newcommand{\XttGccSOdOggBapPrecision}{1\xspace}
\newcommand{\XttGccSOdOggBapFone}{1\xspace}
\newcommand{\XttGccSOdOggGhiRecall}{0.432\xspace}
\newcommand{\XttGccSOdOggGhiPrecision}{0.995\xspace}
\newcommand{\XttGccSOdOggGhiFone}{0.602\xspace}
\newcommand{\XttGccSOdOggRdaRecall}{0.981\xspace}
\newcommand{\XttGccSOdOggRdaPrecision}{1\xspace}
\newcommand{\XttGccSOdOggRdaFone}{0.991\xspace}
\newcommand{\XttGccSOdOggRseRecall}{1\xspace}
\newcommand{\XttGccSOdOggRsePrecision}{1\xspace}
\newcommand{\XttGccSOdOggRseFone}{1\xspace}
\newcommand{\XttGccSOdNgxGT}{141292\xspace}
\newcommand{\XttGccSOdNgxLsRecall}{1\xspace}
\newcommand{\XttGccSOdNgxLsPrecision}{1\xspace}
\newcommand{\XttGccSOdNgxLsFone}{1\xspace}
\newcommand{\XttGccSOdNgxBapRecall}{1\xspace}
\newcommand{\XttGccSOdNgxBapPrecision}{1\xspace}
\newcommand{\XttGccSOdNgxBapFone}{1\xspace}
\newcommand{\XttGccSOdNgxGhiRecall}{0.382\xspace}
\newcommand{\XttGccSOdNgxGhiPrecision}{0.990\xspace}
\newcommand{\XttGccSOdNgxGhiFone}{0.552\xspace}
\newcommand{\XttGccSOdNgxRdaRecall}{0.962\xspace}
\newcommand{\XttGccSOdNgxRdaPrecision}{1.000\xspace}
\newcommand{\XttGccSOdNgxRdaFone}{0.980\xspace}
\newcommand{\XttGccSOdNgxRseRecall}{0.998\xspace}
\newcommand{\XttGccSOdNgxRsePrecision}{1\xspace}
\newcommand{\XttGccSOdNgxRseFone}{0.999\xspace}
\newcommand{\XttGccSOdSshGT}{160125\xspace}
\newcommand{\XttGccSOdSshLsRecall}{1\xspace}
\newcommand{\XttGccSOdSshLsPrecision}{1\xspace}
\newcommand{\XttGccSOdSshLsFone}{1\xspace}
\newcommand{\XttGccSOdSshBapRecall}{1\xspace}
\newcommand{\XttGccSOdSshBapPrecision}{1\xspace}
\newcommand{\XttGccSOdSshBapFone}{1\xspace}
\newcommand{\XttGccSOdSshGhiRecall}{0.387\xspace}
\newcommand{\XttGccSOdSshGhiPrecision}{0.992\xspace}
\newcommand{\XttGccSOdSshGhiFone}{0.557\xspace}
\newcommand{\XttGccSOdSshRdaRecall}{0.827\xspace}
\newcommand{\XttGccSOdSshRdaPrecision}{1\xspace}
\newcommand{\XttGccSOdSshRdaFone}{0.906\xspace}
\newcommand{\XttGccSOdSshRseRecall}{0.993\xspace}
\newcommand{\XttGccSOdSshRsePrecision}{1\xspace}
\newcommand{\XttGccSOdSshRseFone}{0.996\xspace}
\newcommand{\XttGccSOdPcrGT}{6126\xspace}
\newcommand{\XttGccSOdPcrLsRecall}{1\xspace}
\newcommand{\XttGccSOdPcrLsPrecision}{1\xspace}
\newcommand{\XttGccSOdPcrLsFone}{1\xspace}
\newcommand{\XttGccSOdPcrBapRecall}{1\xspace}
\newcommand{\XttGccSOdPcrBapPrecision}{1\xspace}
\newcommand{\XttGccSOdPcrBapFone}{1\xspace}
\newcommand{\XttGccSOdPcrGhiRecall}{0.987\xspace}
\newcommand{\XttGccSOdPcrGhiPrecision}{1\xspace}
\newcommand{\XttGccSOdPcrGhiFone}{0.994\xspace}
\newcommand{\XttGccSOdPcrRdaRecall}{0.884\xspace}
\newcommand{\XttGccSOdPcrRdaPrecision}{1\xspace}
\newcommand{\XttGccSOdPcrRdaFone}{0.939\xspace}
\newcommand{\XttGccSOdPcrRseRecall}{0.992\xspace}
\newcommand{\XttGccSOdPcrRsePrecision}{1\xspace}
\newcommand{\XttGccSOdPcrRseFone}{0.996\xspace}
\newcommand{\XttGccSOdSqlGT}{236418\xspace}
\newcommand{\XttGccSOdSqlLsRecall}{1\xspace}
\newcommand{\XttGccSOdSqlLsPrecision}{1\xspace}
\newcommand{\XttGccSOdSqlLsFone}{1\xspace}
\newcommand{\XttGccSOdSqlBapRecall}{1\xspace}
\newcommand{\XttGccSOdSqlBapPrecision}{1\xspace}
\newcommand{\XttGccSOdSqlBapFone}{1\xspace}
\newcommand{\XttGccSOdSqlGhiRecall}{0.507\xspace}
\newcommand{\XttGccSOdSqlGhiPrecision}{0.995\xspace}
\newcommand{\XttGccSOdSqlGhiFone}{0.671\xspace}
\newcommand{\XttGccSOdSqlRdaRecall}{0.859\xspace}
\newcommand{\XttGccSOdSqlRdaPrecision}{1.000\xspace}
\newcommand{\XttGccSOdSqlRdaFone}{0.924\xspace}
\newcommand{\XttGccSOdSqlRseRecall}{0.985\xspace}
\newcommand{\XttGccSOdSqlRsePrecision}{1.000\xspace}
\newcommand{\XttGccSOdSqlRseFone}{0.993\xspace}
\newcommand{\XttGccSOdVimGT}{825043\xspace}
\newcommand{\XttGccSOdVimLsRecall}{1\xspace}
\newcommand{\XttGccSOdVimLsPrecision}{1\xspace}
\newcommand{\XttGccSOdVimLsFone}{1\xspace}
\newcommand{\XttGccSOdVimBapRecall}{1\xspace}
\newcommand{\XttGccSOdVimBapPrecision}{1.000\xspace}
\newcommand{\XttGccSOdVimBapFone}{1.000\xspace}
\newcommand{\XttGccSOdVimGhiRecall}{0.361\xspace}
\newcommand{\XttGccSOdVimGhiPrecision}{0.993\xspace}
\newcommand{\XttGccSOdVimGhiFone}{0.530\xspace}
\newcommand{\XttGccSOdVimRdaRecall}{0.912\xspace}
\newcommand{\XttGccSOdVimRdaPrecision}{1.000\xspace}
\newcommand{\XttGccSOdVimRdaFone}{0.954\xspace}
\newcommand{\XttGccSOdVimRseRecall}{0.994\xspace}
\newcommand{\XttGccSOdVimRsePrecision}{1\xspace}
\newcommand{\XttGccSOdVimRseFone}{0.997\xspace}
\newcommand{\XttGccSOdVsfGT}{33866\xspace}
\newcommand{\XttGccSOdVsfLsRecall}{1\xspace}
\newcommand{\XttGccSOdVsfLsPrecision}{1\xspace}
\newcommand{\XttGccSOdVsfLsFone}{1\xspace}
\newcommand{\XttGccSOdVsfBapRecall}{1\xspace}
\newcommand{\XttGccSOdVsfBapPrecision}{1\xspace}
\newcommand{\XttGccSOdVsfBapFone}{1\xspace}
\newcommand{\XttGccSOdVsfGhiRecall}{0.253\xspace}
\newcommand{\XttGccSOdVsfGhiPrecision}{0.991\xspace}
\newcommand{\XttGccSOdVsfGhiFone}{0.403\xspace}
\newcommand{\XttGccSOdVsfRdaRecall}{0.882\xspace}
\newcommand{\XttGccSOdVsfRdaPrecision}{1\xspace}
\newcommand{\XttGccSOdVsfRdaFone}{0.937\xspace}
\newcommand{\XttGccSOdVsfRseRecall}{1\xspace}
\newcommand{\XttGccSOdVsfRsePrecision}{1\xspace}
\newcommand{\XttGccSOdVsfRseFone}{1\xspace}
\newcommand{\XttGccSOsSzpGT}{13493\xspace}
\newcommand{\XttGccSOsSzpLsRecall}{1\xspace}
\newcommand{\XttGccSOsSzpLsPrecision}{1\xspace}
\newcommand{\XttGccSOsSzpLsFone}{1\xspace}
\newcommand{\XttGccSOsSzpBapRecall}{1\xspace}
\newcommand{\XttGccSOsSzpBapPrecision}{1\xspace}
\newcommand{\XttGccSOsSzpBapFone}{1\xspace}
\newcommand{\XttGccSOsSzpGhiRecall}{1\xspace}
\newcommand{\XttGccSOsSzpGhiPrecision}{1\xspace}
\newcommand{\XttGccSOsSzpGhiFone}{1\xspace}
\newcommand{\XttGccSOsSzpRdaRecall}{0.953\xspace}
\newcommand{\XttGccSOsSzpRdaPrecision}{1\xspace}
\newcommand{\XttGccSOsSzpRdaFone}{0.976\xspace}
\newcommand{\XttGccSOsSzpRseRecall}{1\xspace}
\newcommand{\XttGccSOsSzpRsePrecision}{1\xspace}
\newcommand{\XttGccSOsSzpRseFone}{1\xspace}
\newcommand{\XttGccSOsCapGT}{234950\xspace}
\newcommand{\XttGccSOsCapLsRecall}{1\xspace}
\newcommand{\XttGccSOsCapLsPrecision}{1\xspace}
\newcommand{\XttGccSOsCapLsFone}{1\xspace}
\newcommand{\XttGccSOsCapBapRecall}{1\xspace}
\newcommand{\XttGccSOsCapBapPrecision}{1\xspace}
\newcommand{\XttGccSOsCapBapFone}{1\xspace}
\newcommand{\XttGccSOsCapGhiRecall}{0.305\xspace}
\newcommand{\XttGccSOsCapGhiPrecision}{0.994\xspace}
\newcommand{\XttGccSOsCapGhiFone}{0.467\xspace}
\newcommand{\XttGccSOsCapRdaRecall}{0.660\xspace}
\newcommand{\XttGccSOsCapRdaPrecision}{0.999\xspace}
\newcommand{\XttGccSOsCapRdaFone}{0.795\xspace}
\newcommand{\XttGccSOsCapRseRecall}{0.728\xspace}
\newcommand{\XttGccSOsCapRsePrecision}{1.000\xspace}
\newcommand{\XttGccSOsCapRseFone}{0.842\xspace}
\newcommand{\XttGccSOsExmGT}{155046\xspace}
\newcommand{\XttGccSOsExmLsRecall}{1\xspace}
\newcommand{\XttGccSOsExmLsPrecision}{1\xspace}
\newcommand{\XttGccSOsExmLsFone}{1\xspace}
\newcommand{\XttGccSOsExmBapRecall}{1\xspace}
\newcommand{\XttGccSOsExmBapPrecision}{1\xspace}
\newcommand{\XttGccSOsExmBapFone}{1\xspace}
\newcommand{\XttGccSOsExmGhiRecall}{0.591\xspace}
\newcommand{\XttGccSOsExmGhiPrecision}{0.998\xspace}
\newcommand{\XttGccSOsExmGhiFone}{0.743\xspace}
\newcommand{\XttGccSOsExmRdaRecall}{0.826\xspace}
\newcommand{\XttGccSOsExmRdaPrecision}{1.000\xspace}
\newcommand{\XttGccSOsExmRdaFone}{0.905\xspace}
\newcommand{\XttGccSOsExmRseRecall}{0.988\xspace}
\newcommand{\XttGccSOsExmRsePrecision}{1.000\xspace}
\newcommand{\XttGccSOsExmRseFone}{0.994\xspace}
\newcommand{\XttGccSOsLgtGT}{30599\xspace}
\newcommand{\XttGccSOsLgtLsRecall}{1\xspace}
\newcommand{\XttGccSOsLgtLsPrecision}{1\xspace}
\newcommand{\XttGccSOsLgtLsFone}{1\xspace}
\newcommand{\XttGccSOsLgtBapRecall}{1\xspace}
\newcommand{\XttGccSOsLgtBapPrecision}{1\xspace}
\newcommand{\XttGccSOsLgtBapFone}{1\xspace}
\newcommand{\XttGccSOsLgtGhiRecall}{0.998\xspace}
\newcommand{\XttGccSOsLgtGhiPrecision}{1\xspace}
\newcommand{\XttGccSOsLgtGhiFone}{0.999\xspace}
\newcommand{\XttGccSOsLgtRdaRecall}{0.902\xspace}
\newcommand{\XttGccSOsLgtRdaPrecision}{1\xspace}
\newcommand{\XttGccSOsLgtRdaFone}{0.948\xspace}
\newcommand{\XttGccSOsLgtRseRecall}{1.000\xspace}
\newcommand{\XttGccSOsLgtRsePrecision}{1\xspace}
\newcommand{\XttGccSOsLgtRseFone}{1.000\xspace}
\newcommand{\XttGccSOsBzpGT}{12882\xspace}
\newcommand{\XttGccSOsBzpLsRecall}{1\xspace}
\newcommand{\XttGccSOsBzpLsPrecision}{1\xspace}
\newcommand{\XttGccSOsBzpLsFone}{1\xspace}
\newcommand{\XttGccSOsBzpBapRecall}{1\xspace}
\newcommand{\XttGccSOsBzpBapPrecision}{1\xspace}
\newcommand{\XttGccSOsBzpBapFone}{1\xspace}
\newcommand{\XttGccSOsBzpGhiRecall}{0.980\xspace}
\newcommand{\XttGccSOsBzpGhiPrecision}{1\xspace}
\newcommand{\XttGccSOsBzpGhiFone}{0.990\xspace}
\newcommand{\XttGccSOsBzpRdaRecall}{0.813\xspace}
\newcommand{\XttGccSOsBzpRdaPrecision}{1\xspace}
\newcommand{\XttGccSOsBzpRdaFone}{0.897\xspace}
\newcommand{\XttGccSOsBzpRseRecall}{1\xspace}
\newcommand{\XttGccSOsBzpRsePrecision}{1\xspace}
\newcommand{\XttGccSOsBzpRseFone}{1\xspace}
\newcommand{\XttGccSOsGccGT}{911598\xspace}
\newcommand{\XttGccSOsGccLsRecall}{1\xspace}
\newcommand{\XttGccSOsGccLsPrecision}{1\xspace}
\newcommand{\XttGccSOsGccLsFone}{1\xspace}
\newcommand{\XttGccSOsGccBapRecall}{1\xspace}
\newcommand{\XttGccSOsGccBapPrecision}{1\xspace}
\newcommand{\XttGccSOsGccBapFone}{1\xspace}
\newcommand{\XttGccSOsGccGhiRecall}{0.395\xspace}
\newcommand{\XttGccSOsGccGhiPrecision}{0.988\xspace}
\newcommand{\XttGccSOsGccGhiFone}{0.565\xspace}
\newcommand{\XttGccSOsGccRdaRecall}{0.731\xspace}
\newcommand{\XttGccSOsGccRdaPrecision}{1.000\xspace}
\newcommand{\XttGccSOsGccRdaFone}{0.844\xspace}
\newcommand{\XttGccSOsGccRseRecall}{0.886\xspace}
\newcommand{\XttGccSOsGccRsePrecision}{1.000\xspace}
\newcommand{\XttGccSOsGccRseFone}{0.940\xspace}
\newcommand{\XttGccSOsGzpGT}{10694\xspace}
\newcommand{\XttGccSOsGzpLsRecall}{1\xspace}
\newcommand{\XttGccSOsGzpLsPrecision}{1\xspace}
\newcommand{\XttGccSOsGzpLsFone}{1\xspace}
\newcommand{\XttGccSOsGzpBapRecall}{1\xspace}
\newcommand{\XttGccSOsGzpBapPrecision}{1\xspace}
\newcommand{\XttGccSOsGzpBapFone}{1\xspace}
\newcommand{\XttGccSOsGzpGhiRecall}{1\xspace}
\newcommand{\XttGccSOsGzpGhiPrecision}{1\xspace}
\newcommand{\XttGccSOsGzpGhiFone}{1\xspace}
\newcommand{\XttGccSOsGzpRdaRecall}{0.986\xspace}
\newcommand{\XttGccSOsGzpRdaPrecision}{1\xspace}
\newcommand{\XttGccSOsGzpRdaFone}{0.993\xspace}
\newcommand{\XttGccSOsGzpRseRecall}{0.993\xspace}
\newcommand{\XttGccSOsGzpRsePrecision}{1\xspace}
\newcommand{\XttGccSOsGzpRseFone}{0.996\xspace}
\newcommand{\XttGccSOsOggGT}{38392\xspace}
\newcommand{\XttGccSOsOggLsRecall}{1\xspace}
\newcommand{\XttGccSOsOggLsPrecision}{1\xspace}
\newcommand{\XttGccSOsOggLsFone}{1\xspace}
\newcommand{\XttGccSOsOggBapRecall}{1\xspace}
\newcommand{\XttGccSOsOggBapPrecision}{1\xspace}
\newcommand{\XttGccSOsOggBapFone}{1\xspace}
\newcommand{\XttGccSOsOggGhiRecall}{0.357\xspace}
\newcommand{\XttGccSOsOggGhiPrecision}{0.997\xspace}
\newcommand{\XttGccSOsOggGhiFone}{0.526\xspace}
\newcommand{\XttGccSOsOggRdaRecall}{0.998\xspace}
\newcommand{\XttGccSOsOggRdaPrecision}{1\xspace}
\newcommand{\XttGccSOsOggRdaFone}{0.999\xspace}
\newcommand{\XttGccSOsOggRseRecall}{1\xspace}
\newcommand{\XttGccSOsOggRsePrecision}{1\xspace}
\newcommand{\XttGccSOsOggRseFone}{1\xspace}
\newcommand{\XttGccSOsNgxGT}{112866\xspace}
\newcommand{\XttGccSOsNgxLsRecall}{1\xspace}
\newcommand{\XttGccSOsNgxLsPrecision}{1\xspace}
\newcommand{\XttGccSOsNgxLsFone}{1\xspace}
\newcommand{\XttGccSOsNgxBapRecall}{1\xspace}
\newcommand{\XttGccSOsNgxBapPrecision}{1\xspace}
\newcommand{\XttGccSOsNgxBapFone}{1\xspace}
\newcommand{\XttGccSOsNgxGhiRecall}{0.410\xspace}
\newcommand{\XttGccSOsNgxGhiPrecision}{0.989\xspace}
\newcommand{\XttGccSOsNgxGhiFone}{0.579\xspace}
\newcommand{\XttGccSOsNgxRdaRecall}{0.978\xspace}
\newcommand{\XttGccSOsNgxRdaPrecision}{1\xspace}
\newcommand{\XttGccSOsNgxRdaFone}{0.989\xspace}
\newcommand{\XttGccSOsNgxRseRecall}{0.995\xspace}
\newcommand{\XttGccSOsNgxRsePrecision}{1.000\xspace}
\newcommand{\XttGccSOsNgxRseFone}{0.997\xspace}
\newcommand{\XttGccSOsSshGT}{125933\xspace}
\newcommand{\XttGccSOsSshLsRecall}{1\xspace}
\newcommand{\XttGccSOsSshLsPrecision}{1\xspace}
\newcommand{\XttGccSOsSshLsFone}{1\xspace}
\newcommand{\XttGccSOsSshBapRecall}{1\xspace}
\newcommand{\XttGccSOsSshBapPrecision}{1\xspace}
\newcommand{\XttGccSOsSshBapFone}{1\xspace}
\newcommand{\XttGccSOsSshGhiRecall}{0.420\xspace}
\newcommand{\XttGccSOsSshGhiPrecision}{0.992\xspace}
\newcommand{\XttGccSOsSshGhiFone}{0.590\xspace}
\newcommand{\XttGccSOsSshRdaRecall}{0.949\xspace}
\newcommand{\XttGccSOsSshRdaPrecision}{1\xspace}
\newcommand{\XttGccSOsSshRdaFone}{0.974\xspace}
\newcommand{\XttGccSOsSshRseRecall}{0.988\xspace}
\newcommand{\XttGccSOsSshRsePrecision}{1\xspace}
\newcommand{\XttGccSOsSshRseFone}{0.994\xspace}
\newcommand{\XttGccSOsPcrGT}{4732\xspace}
\newcommand{\XttGccSOsPcrLsRecall}{1\xspace}
\newcommand{\XttGccSOsPcrLsPrecision}{1\xspace}
\newcommand{\XttGccSOsPcrLsFone}{1\xspace}
\newcommand{\XttGccSOsPcrBapRecall}{1\xspace}
\newcommand{\XttGccSOsPcrBapPrecision}{1\xspace}
\newcommand{\XttGccSOsPcrBapFone}{1\xspace}
\newcommand{\XttGccSOsPcrGhiRecall}{1\xspace}
\newcommand{\XttGccSOsPcrGhiPrecision}{1\xspace}
\newcommand{\XttGccSOsPcrGhiFone}{1\xspace}
\newcommand{\XttGccSOsPcrRdaRecall}{0.947\xspace}
\newcommand{\XttGccSOsPcrRdaPrecision}{1\xspace}
\newcommand{\XttGccSOsPcrRdaFone}{0.973\xspace}
\newcommand{\XttGccSOsPcrRseRecall}{1\xspace}
\newcommand{\XttGccSOsPcrRsePrecision}{1\xspace}
\newcommand{\XttGccSOsPcrRseFone}{1\xspace}
\newcommand{\XttGccSOsSqlGT}{163827\xspace}
\newcommand{\XttGccSOsSqlLsRecall}{1\xspace}
\newcommand{\XttGccSOsSqlLsPrecision}{1\xspace}
\newcommand{\XttGccSOsSqlLsFone}{1\xspace}
\newcommand{\XttGccSOsSqlBapRecall}{1\xspace}
\newcommand{\XttGccSOsSqlBapPrecision}{1\xspace}
\newcommand{\XttGccSOsSqlBapFone}{1\xspace}
\newcommand{\XttGccSOsSqlGhiRecall}{0.554\xspace}
\newcommand{\XttGccSOsSqlGhiPrecision}{0.995\xspace}
\newcommand{\XttGccSOsSqlGhiFone}{0.712\xspace}
\newcommand{\XttGccSOsSqlRdaRecall}{0.876\xspace}
\newcommand{\XttGccSOsSqlRdaPrecision}{1.000\xspace}
\newcommand{\XttGccSOsSqlRdaFone}{0.934\xspace}
\newcommand{\XttGccSOsSqlRseRecall}{0.937\xspace}
\newcommand{\XttGccSOsSqlRsePrecision}{1.000\xspace}
\newcommand{\XttGccSOsSqlRseFone}{0.967\xspace}
\newcommand{\XttGccSOsVimGT}{534085\xspace}
\newcommand{\XttGccSOsVimLsRecall}{1\xspace}
\newcommand{\XttGccSOsVimLsPrecision}{1\xspace}
\newcommand{\XttGccSOsVimLsFone}{1\xspace}
\newcommand{\XttGccSOsVimBapRecall}{1\xspace}
\newcommand{\XttGccSOsVimBapPrecision}{1\xspace}
\newcommand{\XttGccSOsVimBapFone}{1\xspace}
\newcommand{\XttGccSOsVimGhiRecall}{0.376\xspace}
\newcommand{\XttGccSOsVimGhiPrecision}{0.989\xspace}
\newcommand{\XttGccSOsVimGhiFone}{0.545\xspace}
\newcommand{\XttGccSOsVimRdaRecall}{0.955\xspace}
\newcommand{\XttGccSOsVimRdaPrecision}{1.000\xspace}
\newcommand{\XttGccSOsVimRdaFone}{0.977\xspace}
\newcommand{\XttGccSOsVimRseRecall}{0.994\xspace}
\newcommand{\XttGccSOsVimRsePrecision}{1.000\xspace}
\newcommand{\XttGccSOsVimRseFone}{0.997\xspace}
\newcommand{\XttGccSOsVsfGT}{24805\xspace}
\newcommand{\XttGccSOsVsfLsRecall}{1\xspace}
\newcommand{\XttGccSOsVsfLsPrecision}{1\xspace}
\newcommand{\XttGccSOsVsfLsFone}{1\xspace}
\newcommand{\XttGccSOsVsfBapRecall}{1\xspace}
\newcommand{\XttGccSOsVsfBapPrecision}{1\xspace}
\newcommand{\XttGccSOsVsfBapFone}{1\xspace}
\newcommand{\XttGccSOsVsfGhiRecall}{0.181\xspace}
\newcommand{\XttGccSOsVsfGhiPrecision}{0.997\xspace}
\newcommand{\XttGccSOsVsfGhiFone}{0.307\xspace}
\newcommand{\XttGccSOsVsfRdaRecall}{0.915\xspace}
\newcommand{\XttGccSOsVsfRdaPrecision}{1\xspace}
\newcommand{\XttGccSOsVsfRdaFone}{0.956\xspace}
\newcommand{\XttGccSOsVsfRseRecall}{1.000\xspace}
\newcommand{\XttGccSOsVsfRsePrecision}{1\xspace}
\newcommand{\XttGccSOsVsfRseFone}{1.000\xspace}
\newcommand{\XttClangTOoSzpGT}{21921\xspace}
\newcommand{\XttClangTOoSzpLsRecall}{1\xspace}
\newcommand{\XttClangTOoSzpLsPrecision}{1\xspace}
\newcommand{\XttClangTOoSzpLsFone}{1\xspace}
\newcommand{\XttClangTOoSzpBapRecall}{0.980\xspace}
\newcommand{\XttClangTOoSzpBapPrecision}{1\xspace}
\newcommand{\XttClangTOoSzpBapFone}{0.990\xspace}
\newcommand{\XttClangTOoSzpGhiRecall}{1\xspace}
\newcommand{\XttClangTOoSzpGhiPrecision}{1\xspace}
\newcommand{\XttClangTOoSzpGhiFone}{1\xspace}
\newcommand{\XttClangTOoSzpRdaRecall}{0.980\xspace}
\newcommand{\XttClangTOoSzpRdaPrecision}{1\xspace}
\newcommand{\XttClangTOoSzpRdaFone}{0.990\xspace}
\newcommand{\XttClangTOoSzpRseRecall}{1\xspace}
\newcommand{\XttClangTOoSzpRsePrecision}{1\xspace}
\newcommand{\XttClangTOoSzpRseFone}{1\xspace}
\newcommand{\XttClangTOoCapGT}{416246\xspace}
\newcommand{\XttClangTOoCapLsRecall}{1\xspace}
\newcommand{\XttClangTOoCapLsPrecision}{1\xspace}
\newcommand{\XttClangTOoCapLsFone}{1\xspace}
\newcommand{\XttClangTOoCapBapRecall}{0.562\xspace}
\newcommand{\XttClangTOoCapBapPrecision}{1\xspace}
\newcommand{\XttClangTOoCapBapFone}{0.720\xspace}
\newcommand{\XttClangTOoCapGhiRecall}{1\xspace}
\newcommand{\XttClangTOoCapGhiPrecision}{1\xspace}
\newcommand{\XttClangTOoCapGhiFone}{1\xspace}
\newcommand{\XttClangTOoCapRdaRecall}{0.394\xspace}
\newcommand{\XttClangTOoCapRdaPrecision}{1\xspace}
\newcommand{\XttClangTOoCapRdaFone}{0.565\xspace}
\newcommand{\XttClangTOoCapRseRecall}{1\xspace}
\newcommand{\XttClangTOoCapRsePrecision}{1\xspace}
\newcommand{\XttClangTOoCapRseFone}{1\xspace}
\newcommand{\XttClangTOoExmGT}{204670\xspace}
\newcommand{\XttClangTOoExmLsRecall}{1\xspace}
\newcommand{\XttClangTOoExmLsPrecision}{1\xspace}
\newcommand{\XttClangTOoExmLsFone}{1\xspace}
\newcommand{\XttClangTOoExmBapRecall}{0.864\xspace}
\newcommand{\XttClangTOoExmBapPrecision}{1\xspace}
\newcommand{\XttClangTOoExmBapFone}{0.927\xspace}
\newcommand{\XttClangTOoExmGhiRecall}{0.986\xspace}
\newcommand{\XttClangTOoExmGhiPrecision}{1\xspace}
\newcommand{\XttClangTOoExmGhiFone}{0.993\xspace}
\newcommand{\XttClangTOoExmRdaRecall}{0.827\xspace}
\newcommand{\XttClangTOoExmRdaPrecision}{1\xspace}
\newcommand{\XttClangTOoExmRdaFone}{0.905\xspace}
\newcommand{\XttClangTOoExmRseRecall}{1.000\xspace}
\newcommand{\XttClangTOoExmRsePrecision}{1\xspace}
\newcommand{\XttClangTOoExmRseFone}{1.000\xspace}
\newcommand{\XttClangTOoLgtGT}{46912\xspace}
\newcommand{\XttClangTOoLgtLsRecall}{1\xspace}
\newcommand{\XttClangTOoLgtLsPrecision}{1\xspace}
\newcommand{\XttClangTOoLgtLsFone}{1\xspace}
\newcommand{\XttClangTOoLgtBapRecall}{0.914\xspace}
\newcommand{\XttClangTOoLgtBapPrecision}{1\xspace}
\newcommand{\XttClangTOoLgtBapFone}{0.955\xspace}
\newcommand{\XttClangTOoLgtGhiRecall}{1\xspace}
\newcommand{\XttClangTOoLgtGhiPrecision}{1\xspace}
\newcommand{\XttClangTOoLgtGhiFone}{1\xspace}
\newcommand{\XttClangTOoLgtRdaRecall}{0.914\xspace}
\newcommand{\XttClangTOoLgtRdaPrecision}{1\xspace}
\newcommand{\XttClangTOoLgtRdaFone}{0.955\xspace}
\newcommand{\XttClangTOoLgtRseRecall}{1\xspace}
\newcommand{\XttClangTOoLgtRsePrecision}{1\xspace}
\newcommand{\XttClangTOoLgtRseFone}{1\xspace}
\newcommand{\XttClangTOoBzpGT}{22167\xspace}
\newcommand{\XttClangTOoBzpLsRecall}{1\xspace}
\newcommand{\XttClangTOoBzpLsPrecision}{1\xspace}
\newcommand{\XttClangTOoBzpLsFone}{1\xspace}
\newcommand{\XttClangTOoBzpBapRecall}{0.795\xspace}
\newcommand{\XttClangTOoBzpBapPrecision}{1\xspace}
\newcommand{\XttClangTOoBzpBapFone}{0.886\xspace}
\newcommand{\XttClangTOoBzpGhiRecall}{0.980\xspace}
\newcommand{\XttClangTOoBzpGhiPrecision}{1\xspace}
\newcommand{\XttClangTOoBzpGhiFone}{0.990\xspace}
\newcommand{\XttClangTOoBzpRdaRecall}{0.788\xspace}
\newcommand{\XttClangTOoBzpRdaPrecision}{1\xspace}
\newcommand{\XttClangTOoBzpRdaFone}{0.881\xspace}
\newcommand{\XttClangTOoBzpRseRecall}{1\xspace}
\newcommand{\XttClangTOoBzpRsePrecision}{1\xspace}
\newcommand{\XttClangTOoBzpRseFone}{1\xspace}
\newcommand{\XttClangTOoGccGT}{1463639\xspace}
\newcommand{\XttClangTOoGccLsRecall}{1\xspace}
\newcommand{\XttClangTOoGccLsPrecision}{1\xspace}
\newcommand{\XttClangTOoGccLsFone}{1\xspace}
\newcommand{\XttClangTOoGccBapRecall}{0.799\xspace}
\newcommand{\XttClangTOoGccBapPrecision}{1\xspace}
\newcommand{\XttClangTOoGccBapFone}{0.888\xspace}
\newcommand{\XttClangTOoGccGhiRecall}{0.995\xspace}
\newcommand{\XttClangTOoGccGhiPrecision}{1\xspace}
\newcommand{\XttClangTOoGccGhiFone}{0.997\xspace}
\newcommand{\XttClangTOoGccRdaRecall}{0.743\xspace}
\newcommand{\XttClangTOoGccRdaPrecision}{1\xspace}
\newcommand{\XttClangTOoGccRdaFone}{0.853\xspace}
\newcommand{\XttClangTOoGccRseRecall}{1\xspace}
\newcommand{\XttClangTOoGccRsePrecision}{1\xspace}
\newcommand{\XttClangTOoGccRseFone}{1\xspace}
\newcommand{\XttClangTOoGzpGT}{13712\xspace}
\newcommand{\XttClangTOoGzpLsRecall}{1\xspace}
\newcommand{\XttClangTOoGzpLsPrecision}{1\xspace}
\newcommand{\XttClangTOoGzpLsFone}{1\xspace}
\newcommand{\XttClangTOoGzpBapRecall}{0.991\xspace}
\newcommand{\XttClangTOoGzpBapPrecision}{1\xspace}
\newcommand{\XttClangTOoGzpBapFone}{0.996\xspace}
\newcommand{\XttClangTOoGzpGhiRecall}{1\xspace}
\newcommand{\XttClangTOoGzpGhiPrecision}{1\xspace}
\newcommand{\XttClangTOoGzpGhiFone}{1\xspace}
\newcommand{\XttClangTOoGzpRdaRecall}{0.991\xspace}
\newcommand{\XttClangTOoGzpRdaPrecision}{1\xspace}
\newcommand{\XttClangTOoGzpRdaFone}{0.996\xspace}
\newcommand{\XttClangTOoGzpRseRecall}{0.999\xspace}
\newcommand{\XttClangTOoGzpRsePrecision}{1\xspace}
\newcommand{\XttClangTOoGzpRseFone}{0.999\xspace}
\newcommand{\XttClangTOoOggGT}{57131\xspace}
\newcommand{\XttClangTOoOggLsRecall}{1\xspace}
\newcommand{\XttClangTOoOggLsPrecision}{1\xspace}
\newcommand{\XttClangTOoOggLsFone}{1\xspace}
\newcommand{\XttClangTOoOggBapRecall}{0.996\xspace}
\newcommand{\XttClangTOoOggBapPrecision}{1\xspace}
\newcommand{\XttClangTOoOggBapFone}{0.998\xspace}
\newcommand{\XttClangTOoOggGhiRecall}{1\xspace}
\newcommand{\XttClangTOoOggGhiPrecision}{1\xspace}
\newcommand{\XttClangTOoOggGhiFone}{1\xspace}
\newcommand{\XttClangTOoOggRdaRecall}{0.996\xspace}
\newcommand{\XttClangTOoOggRdaPrecision}{1\xspace}
\newcommand{\XttClangTOoOggRdaFone}{0.998\xspace}
\newcommand{\XttClangTOoOggRseRecall}{1\xspace}
\newcommand{\XttClangTOoOggRsePrecision}{1\xspace}
\newcommand{\XttClangTOoOggRseFone}{1\xspace}
\newcommand{\XttClangTOoNgxGT}{179283\xspace}
\newcommand{\XttClangTOoNgxLsRecall}{1\xspace}
\newcommand{\XttClangTOoNgxLsPrecision}{1\xspace}
\newcommand{\XttClangTOoNgxLsFone}{1\xspace}
\newcommand{\XttClangTOoNgxBapRecall}{0.969\xspace}
\newcommand{\XttClangTOoNgxBapPrecision}{1\xspace}
\newcommand{\XttClangTOoNgxBapFone}{0.984\xspace}
\newcommand{\XttClangTOoNgxGhiRecall}{1.000\xspace}
\newcommand{\XttClangTOoNgxGhiPrecision}{1\xspace}
\newcommand{\XttClangTOoNgxGhiFone}{1.000\xspace}
\newcommand{\XttClangTOoNgxRdaRecall}{0.968\xspace}
\newcommand{\XttClangTOoNgxRdaPrecision}{1\xspace}
\newcommand{\XttClangTOoNgxRdaFone}{0.984\xspace}
\newcommand{\XttClangTOoNgxRseRecall}{1\xspace}
\newcommand{\XttClangTOoNgxRsePrecision}{1\xspace}
\newcommand{\XttClangTOoNgxRseFone}{1\xspace}
\newcommand{\XttClangTOoSshGT}{240212\xspace}
\newcommand{\XttClangTOoSshLsRecall}{1\xspace}
\newcommand{\XttClangTOoSshLsPrecision}{1\xspace}
\newcommand{\XttClangTOoSshLsFone}{1\xspace}
\newcommand{\XttClangTOoSshBapRecall}{0.967\xspace}
\newcommand{\XttClangTOoSshBapPrecision}{1\xspace}
\newcommand{\XttClangTOoSshBapFone}{0.983\xspace}
\newcommand{\XttClangTOoSshGhiRecall}{0.970\xspace}
\newcommand{\XttClangTOoSshGhiPrecision}{1\xspace}
\newcommand{\XttClangTOoSshGhiFone}{0.985\xspace}
\newcommand{\XttClangTOoSshRdaRecall}{0.804\xspace}
\newcommand{\XttClangTOoSshRdaPrecision}{1.000\xspace}
\newcommand{\XttClangTOoSshRdaFone}{0.892\xspace}
\newcommand{\XttClangTOoSshRseRecall}{0.989\xspace}
\newcommand{\XttClangTOoSshRsePrecision}{1\xspace}
\newcommand{\XttClangTOoSshRseFone}{0.995\xspace}
\newcommand{\XttClangTOoPcrGT}{6954\xspace}
\newcommand{\XttClangTOoPcrLsRecall}{1\xspace}
\newcommand{\XttClangTOoPcrLsPrecision}{1\xspace}
\newcommand{\XttClangTOoPcrLsFone}{1\xspace}
\newcommand{\XttClangTOoPcrBapRecall}{0.919\xspace}
\newcommand{\XttClangTOoPcrBapPrecision}{1\xspace}
\newcommand{\XttClangTOoPcrBapFone}{0.958\xspace}
\newcommand{\XttClangTOoPcrGhiRecall}{1\xspace}
\newcommand{\XttClangTOoPcrGhiPrecision}{1\xspace}
\newcommand{\XttClangTOoPcrGhiFone}{1\xspace}
\newcommand{\XttClangTOoPcrRdaRecall}{0.919\xspace}
\newcommand{\XttClangTOoPcrRdaPrecision}{1\xspace}
\newcommand{\XttClangTOoPcrRdaFone}{0.958\xspace}
\newcommand{\XttClangTOoPcrRseRecall}{1\xspace}
\newcommand{\XttClangTOoPcrRsePrecision}{1\xspace}
\newcommand{\XttClangTOoPcrRseFone}{1\xspace}
\newcommand{\XttClangTOoSqlGT}{259096\xspace}
\newcommand{\XttClangTOoSqlLsRecall}{1\xspace}
\newcommand{\XttClangTOoSqlLsPrecision}{1\xspace}
\newcommand{\XttClangTOoSqlLsFone}{1\xspace}
\newcommand{\XttClangTOoSqlBapRecall}{0.900\xspace}
\newcommand{\XttClangTOoSqlBapPrecision}{1\xspace}
\newcommand{\XttClangTOoSqlBapFone}{0.947\xspace}
\newcommand{\XttClangTOoSqlGhiRecall}{1.000\xspace}
\newcommand{\XttClangTOoSqlGhiPrecision}{1\xspace}
\newcommand{\XttClangTOoSqlGhiFone}{1.000\xspace}
\newcommand{\XttClangTOoSqlRdaRecall}{0.892\xspace}
\newcommand{\XttClangTOoSqlRdaPrecision}{1\xspace}
\newcommand{\XttClangTOoSqlRdaFone}{0.943\xspace}
\newcommand{\XttClangTOoSqlRseRecall}{1\xspace}
\newcommand{\XttClangTOoSqlRsePrecision}{1\xspace}
\newcommand{\XttClangTOoSqlRseFone}{1\xspace}
\newcommand{\XttClangTOoVimGT}{704361\xspace}
\newcommand{\XttClangTOoVimLsRecall}{1\xspace}
\newcommand{\XttClangTOoVimLsPrecision}{1\xspace}
\newcommand{\XttClangTOoVimLsFone}{1\xspace}
\newcommand{\XttClangTOoVimBapRecall}{0.965\xspace}
\newcommand{\XttClangTOoVimBapPrecision}{1.000\xspace}
\newcommand{\XttClangTOoVimBapFone}{0.982\xspace}
\newcommand{\XttClangTOoVimGhiRecall}{1.000\xspace}
\newcommand{\XttClangTOoVimGhiPrecision}{1\xspace}
\newcommand{\XttClangTOoVimGhiFone}{1.000\xspace}
\newcommand{\XttClangTOoVimRdaRecall}{0.927\xspace}
\newcommand{\XttClangTOoVimRdaPrecision}{1\xspace}
\newcommand{\XttClangTOoVimRdaFone}{0.962\xspace}
\newcommand{\XttClangTOoVimRseRecall}{1.000\xspace}
\newcommand{\XttClangTOoVimRsePrecision}{1\xspace}
\newcommand{\XttClangTOoVimRseFone}{1.000\xspace}
\newcommand{\XttClangTOoVsfGT}{38215\xspace}
\newcommand{\XttClangTOoVsfLsRecall}{1\xspace}
\newcommand{\XttClangTOoVsfLsPrecision}{1\xspace}
\newcommand{\XttClangTOoVsfLsFone}{1\xspace}
\newcommand{\XttClangTOoVsfBapRecall}{0.993\xspace}
\newcommand{\XttClangTOoVsfBapPrecision}{1\xspace}
\newcommand{\XttClangTOoVsfBapFone}{0.997\xspace}
\newcommand{\XttClangTOoVsfGhiRecall}{0.994\xspace}
\newcommand{\XttClangTOoVsfGhiPrecision}{1\xspace}
\newcommand{\XttClangTOoVsfGhiFone}{0.997\xspace}
\newcommand{\XttClangTOoVsfRdaRecall}{0.986\xspace}
\newcommand{\XttClangTOoVsfRdaPrecision}{1\xspace}
\newcommand{\XttClangTOoVsfRdaFone}{0.993\xspace}
\newcommand{\XttClangTOoVsfRseRecall}{1.000\xspace}
\newcommand{\XttClangTOoVsfRsePrecision}{1\xspace}
\newcommand{\XttClangTOoVsfRseFone}{1.000\xspace}
\newcommand{\XttClangTOaSzpGT}{13516\xspace}
\newcommand{\XttClangTOaSzpLsRecall}{1\xspace}
\newcommand{\XttClangTOaSzpLsPrecision}{1\xspace}
\newcommand{\XttClangTOaSzpLsFone}{1\xspace}
\newcommand{\XttClangTOaSzpBapRecall}{0.979\xspace}
\newcommand{\XttClangTOaSzpBapPrecision}{1\xspace}
\newcommand{\XttClangTOaSzpBapFone}{0.990\xspace}
\newcommand{\XttClangTOaSzpGhiRecall}{1\xspace}
\newcommand{\XttClangTOaSzpGhiPrecision}{1\xspace}
\newcommand{\XttClangTOaSzpGhiFone}{1\xspace}
\newcommand{\XttClangTOaSzpRdaRecall}{1\xspace}
\newcommand{\XttClangTOaSzpRdaPrecision}{1\xspace}
\newcommand{\XttClangTOaSzpRdaFone}{1\xspace}
\newcommand{\XttClangTOaSzpRseRecall}{1\xspace}
\newcommand{\XttClangTOaSzpRsePrecision}{1\xspace}
\newcommand{\XttClangTOaSzpRseFone}{1\xspace}
\newcommand{\XttClangTOaCapGT}{193218\xspace}
\newcommand{\XttClangTOaCapLsRecall}{1\xspace}
\newcommand{\XttClangTOaCapLsPrecision}{1\xspace}
\newcommand{\XttClangTOaCapLsFone}{1\xspace}
\newcommand{\XttClangTOaCapBapRecall}{0.385\xspace}
\newcommand{\XttClangTOaCapBapPrecision}{1\xspace}
\newcommand{\XttClangTOaCapBapFone}{0.556\xspace}
\newcommand{\XttClangTOaCapGhiRecall}{0.996\xspace}
\newcommand{\XttClangTOaCapGhiPrecision}{1\xspace}
\newcommand{\XttClangTOaCapGhiFone}{0.998\xspace}
\newcommand{\XttClangTOaCapRdaRecall}{1.000\xspace}
\newcommand{\XttClangTOaCapRdaPrecision}{1\xspace}
\newcommand{\XttClangTOaCapRdaFone}{1.000\xspace}
\newcommand{\XttClangTOaCapRseRecall}{1.000\xspace}
\newcommand{\XttClangTOaCapRsePrecision}{1\xspace}
\newcommand{\XttClangTOaCapRseFone}{1.000\xspace}
\newcommand{\XttClangTOaExmGT}{131530\xspace}
\newcommand{\XttClangTOaExmLsRecall}{1\xspace}
\newcommand{\XttClangTOaExmLsPrecision}{1\xspace}
\newcommand{\XttClangTOaExmLsFone}{1\xspace}
\newcommand{\XttClangTOaExmBapRecall}{0.857\xspace}
\newcommand{\XttClangTOaExmBapPrecision}{1\xspace}
\newcommand{\XttClangTOaExmBapFone}{0.923\xspace}
\newcommand{\XttClangTOaExmGhiRecall}{1\xspace}
\newcommand{\XttClangTOaExmGhiPrecision}{1\xspace}
\newcommand{\XttClangTOaExmGhiFone}{1\xspace}
\newcommand{\XttClangTOaExmRdaRecall}{0.977\xspace}
\newcommand{\XttClangTOaExmRdaPrecision}{1\xspace}
\newcommand{\XttClangTOaExmRdaFone}{0.988\xspace}
\newcommand{\XttClangTOaExmRseRecall}{1.000\xspace}
\newcommand{\XttClangTOaExmRsePrecision}{1\xspace}
\newcommand{\XttClangTOaExmRseFone}{1.000\xspace}
\newcommand{\XttClangTOaLgtGT}{25622\xspace}
\newcommand{\XttClangTOaLgtLsRecall}{1\xspace}
\newcommand{\XttClangTOaLgtLsPrecision}{1\xspace}
\newcommand{\XttClangTOaLgtLsFone}{1\xspace}
\newcommand{\XttClangTOaLgtBapRecall}{0.909\xspace}
\newcommand{\XttClangTOaLgtBapPrecision}{1\xspace}
\newcommand{\XttClangTOaLgtBapFone}{0.952\xspace}
\newcommand{\XttClangTOaLgtGhiRecall}{1\xspace}
\newcommand{\XttClangTOaLgtGhiPrecision}{1\xspace}
\newcommand{\XttClangTOaLgtGhiFone}{1\xspace}
\newcommand{\XttClangTOaLgtRdaRecall}{0.990\xspace}
\newcommand{\XttClangTOaLgtRdaPrecision}{1\xspace}
\newcommand{\XttClangTOaLgtRdaFone}{0.995\xspace}
\newcommand{\XttClangTOaLgtRseRecall}{1\xspace}
\newcommand{\XttClangTOaLgtRsePrecision}{1\xspace}
\newcommand{\XttClangTOaLgtRseFone}{1\xspace}
\newcommand{\XttClangTOaBzpGT}{12400\xspace}
\newcommand{\XttClangTOaBzpLsRecall}{1\xspace}
\newcommand{\XttClangTOaBzpLsPrecision}{1\xspace}
\newcommand{\XttClangTOaBzpLsFone}{1\xspace}
\newcommand{\XttClangTOaBzpBapRecall}{0.770\xspace}
\newcommand{\XttClangTOaBzpBapPrecision}{1\xspace}
\newcommand{\XttClangTOaBzpBapFone}{0.870\xspace}
\newcommand{\XttClangTOaBzpGhiRecall}{1\xspace}
\newcommand{\XttClangTOaBzpGhiPrecision}{1\xspace}
\newcommand{\XttClangTOaBzpGhiFone}{1\xspace}
\newcommand{\XttClangTOaBzpRdaRecall}{0.975\xspace}
\newcommand{\XttClangTOaBzpRdaPrecision}{1\xspace}
\newcommand{\XttClangTOaBzpRdaFone}{0.987\xspace}
\newcommand{\XttClangTOaBzpRseRecall}{1.000\xspace}
\newcommand{\XttClangTOaBzpRsePrecision}{1\xspace}
\newcommand{\XttClangTOaBzpRseFone}{1.000\xspace}
\newcommand{\XttClangTOaGccGT}{807753\xspace}
\newcommand{\XttClangTOaGccLsRecall}{1\xspace}
\newcommand{\XttClangTOaGccLsPrecision}{1\xspace}
\newcommand{\XttClangTOaGccLsFone}{1\xspace}
\newcommand{\XttClangTOaGccBapRecall}{0.760\xspace}
\newcommand{\XttClangTOaGccBapPrecision}{1\xspace}
\newcommand{\XttClangTOaGccBapFone}{0.863\xspace}
\newcommand{\XttClangTOaGccGhiRecall}{0.993\xspace}
\newcommand{\XttClangTOaGccGhiPrecision}{1\xspace}
\newcommand{\XttClangTOaGccGhiFone}{0.996\xspace}
\newcommand{\XttClangTOaGccRdaRecall}{0.866\xspace}
\newcommand{\XttClangTOaGccRdaPrecision}{1\xspace}
\newcommand{\XttClangTOaGccRdaFone}{0.928\xspace}
\newcommand{\XttClangTOaGccRseRecall}{1.000\xspace}
\newcommand{\XttClangTOaGccRsePrecision}{1\xspace}
\newcommand{\XttClangTOaGccRseFone}{1.000\xspace}
\newcommand{\XttClangTOaGzpGT}{9131\xspace}
\newcommand{\XttClangTOaGzpLsRecall}{1\xspace}
\newcommand{\XttClangTOaGzpLsPrecision}{1\xspace}
\newcommand{\XttClangTOaGzpLsFone}{1\xspace}
\newcommand{\XttClangTOaGzpBapRecall}{0.991\xspace}
\newcommand{\XttClangTOaGzpBapPrecision}{1\xspace}
\newcommand{\XttClangTOaGzpBapFone}{0.995\xspace}
\newcommand{\XttClangTOaGzpGhiRecall}{1\xspace}
\newcommand{\XttClangTOaGzpGhiPrecision}{1\xspace}
\newcommand{\XttClangTOaGzpGhiFone}{1\xspace}
\newcommand{\XttClangTOaGzpRdaRecall}{1\xspace}
\newcommand{\XttClangTOaGzpRdaPrecision}{1\xspace}
\newcommand{\XttClangTOaGzpRdaFone}{1\xspace}
\newcommand{\XttClangTOaGzpRseRecall}{0.998\xspace}
\newcommand{\XttClangTOaGzpRsePrecision}{1\xspace}
\newcommand{\XttClangTOaGzpRseFone}{0.999\xspace}
\newcommand{\XttClangTOaOggGT}{36252\xspace}
\newcommand{\XttClangTOaOggLsRecall}{1\xspace}
\newcommand{\XttClangTOaOggLsPrecision}{1\xspace}
\newcommand{\XttClangTOaOggLsFone}{1\xspace}
\newcommand{\XttClangTOaOggBapRecall}{0.982\xspace}
\newcommand{\XttClangTOaOggBapPrecision}{1\xspace}
\newcommand{\XttClangTOaOggBapFone}{0.991\xspace}
\newcommand{\XttClangTOaOggGhiRecall}{1\xspace}
\newcommand{\XttClangTOaOggGhiPrecision}{1\xspace}
\newcommand{\XttClangTOaOggGhiFone}{1\xspace}
\newcommand{\XttClangTOaOggRdaRecall}{1\xspace}
\newcommand{\XttClangTOaOggRdaPrecision}{1\xspace}
\newcommand{\XttClangTOaOggRdaFone}{1\xspace}
\newcommand{\XttClangTOaOggRseRecall}{1\xspace}
\newcommand{\XttClangTOaOggRsePrecision}{1\xspace}
\newcommand{\XttClangTOaOggRseFone}{1\xspace}
\newcommand{\XttClangTOaNgxGT}{101768\xspace}
\newcommand{\XttClangTOaNgxLsRecall}{1\xspace}
\newcommand{\XttClangTOaNgxLsPrecision}{1\xspace}
\newcommand{\XttClangTOaNgxLsFone}{1\xspace}
\newcommand{\XttClangTOaNgxBapRecall}{0.971\xspace}
\newcommand{\XttClangTOaNgxBapPrecision}{1\xspace}
\newcommand{\XttClangTOaNgxBapFone}{0.985\xspace}
\newcommand{\XttClangTOaNgxGhiRecall}{1\xspace}
\newcommand{\XttClangTOaNgxGhiPrecision}{1\xspace}
\newcommand{\XttClangTOaNgxGhiFone}{1\xspace}
\newcommand{\XttClangTOaNgxRdaRecall}{0.997\xspace}
\newcommand{\XttClangTOaNgxRdaPrecision}{1\xspace}
\newcommand{\XttClangTOaNgxRdaFone}{0.999\xspace}
\newcommand{\XttClangTOaNgxRseRecall}{1\xspace}
\newcommand{\XttClangTOaNgxRsePrecision}{1\xspace}
\newcommand{\XttClangTOaNgxRseFone}{1\xspace}
\newcommand{\XttClangTOaSshGT}{116742\xspace}
\newcommand{\XttClangTOaSshLsRecall}{1\xspace}
\newcommand{\XttClangTOaSshLsPrecision}{1\xspace}
\newcommand{\XttClangTOaSshLsFone}{1\xspace}
\newcommand{\XttClangTOaSshBapRecall}{0.960\xspace}
\newcommand{\XttClangTOaSshBapPrecision}{1\xspace}
\newcommand{\XttClangTOaSshBapFone}{0.980\xspace}
\newcommand{\XttClangTOaSshGhiRecall}{1.000\xspace}
\newcommand{\XttClangTOaSshGhiPrecision}{1\xspace}
\newcommand{\XttClangTOaSshGhiFone}{1.000\xspace}
\newcommand{\XttClangTOaSshRdaRecall}{0.957\xspace}
\newcommand{\XttClangTOaSshRdaPrecision}{1\xspace}
\newcommand{\XttClangTOaSshRdaFone}{0.978\xspace}
\newcommand{\XttClangTOaSshRseRecall}{0.991\xspace}
\newcommand{\XttClangTOaSshRsePrecision}{1\xspace}
\newcommand{\XttClangTOaSshRseFone}{0.995\xspace}
\newcommand{\XttClangTOaPcrGT}{4420\xspace}
\newcommand{\XttClangTOaPcrLsRecall}{1\xspace}
\newcommand{\XttClangTOaPcrLsPrecision}{1\xspace}
\newcommand{\XttClangTOaPcrLsFone}{1\xspace}
\newcommand{\XttClangTOaPcrBapRecall}{0.891\xspace}
\newcommand{\XttClangTOaPcrBapPrecision}{1\xspace}
\newcommand{\XttClangTOaPcrBapFone}{0.942\xspace}
\newcommand{\XttClangTOaPcrGhiRecall}{1\xspace}
\newcommand{\XttClangTOaPcrGhiPrecision}{1\xspace}
\newcommand{\XttClangTOaPcrGhiFone}{1\xspace}
\newcommand{\XttClangTOaPcrRdaRecall}{1\xspace}
\newcommand{\XttClangTOaPcrRdaPrecision}{1\xspace}
\newcommand{\XttClangTOaPcrRdaFone}{1\xspace}
\newcommand{\XttClangTOaPcrRseRecall}{1\xspace}
\newcommand{\XttClangTOaPcrRsePrecision}{1\xspace}
\newcommand{\XttClangTOaPcrRseFone}{1\xspace}
\newcommand{\XttClangTOaSqlGT}{162011\xspace}
\newcommand{\XttClangTOaSqlLsRecall}{1\xspace}
\newcommand{\XttClangTOaSqlLsPrecision}{1\xspace}
\newcommand{\XttClangTOaSqlLsFone}{1\xspace}
\newcommand{\XttClangTOaSqlBapRecall}{0.884\xspace}
\newcommand{\XttClangTOaSqlBapPrecision}{1\xspace}
\newcommand{\XttClangTOaSqlBapFone}{0.938\xspace}
\newcommand{\XttClangTOaSqlGhiRecall}{1\xspace}
\newcommand{\XttClangTOaSqlGhiPrecision}{1\xspace}
\newcommand{\XttClangTOaSqlGhiFone}{1\xspace}
\newcommand{\XttClangTOaSqlRdaRecall}{1\xspace}
\newcommand{\XttClangTOaSqlRdaPrecision}{1\xspace}
\newcommand{\XttClangTOaSqlRdaFone}{1\xspace}
\newcommand{\XttClangTOaSqlRseRecall}{0.999\xspace}
\newcommand{\XttClangTOaSqlRsePrecision}{1\xspace}
\newcommand{\XttClangTOaSqlRseFone}{1.000\xspace}
\newcommand{\XttClangTOaVimGT}{461319\xspace}
\newcommand{\XttClangTOaVimLsRecall}{1\xspace}
\newcommand{\XttClangTOaVimLsPrecision}{1\xspace}
\newcommand{\XttClangTOaVimLsFone}{1\xspace}
\newcommand{\XttClangTOaVimBapRecall}{0.928\xspace}
\newcommand{\XttClangTOaVimBapPrecision}{1\xspace}
\newcommand{\XttClangTOaVimBapFone}{0.963\xspace}
\newcommand{\XttClangTOaVimGhiRecall}{1\xspace}
\newcommand{\XttClangTOaVimGhiPrecision}{1\xspace}
\newcommand{\XttClangTOaVimGhiFone}{1\xspace}
\newcommand{\XttClangTOaVimRdaRecall}{0.990\xspace}
\newcommand{\XttClangTOaVimRdaPrecision}{1\xspace}
\newcommand{\XttClangTOaVimRdaFone}{0.995\xspace}
\newcommand{\XttClangTOaVimRseRecall}{1.000\xspace}
\newcommand{\XttClangTOaVimRsePrecision}{1\xspace}
\newcommand{\XttClangTOaVimRseFone}{1.000\xspace}
\newcommand{\XttClangTOaVsfGT}{22751\xspace}
\newcommand{\XttClangTOaVsfLsRecall}{1\xspace}
\newcommand{\XttClangTOaVsfLsPrecision}{1\xspace}
\newcommand{\XttClangTOaVsfLsFone}{1\xspace}
\newcommand{\XttClangTOaVsfBapRecall}{0.991\xspace}
\newcommand{\XttClangTOaVsfBapPrecision}{1\xspace}
\newcommand{\XttClangTOaVsfBapFone}{0.995\xspace}
\newcommand{\XttClangTOaVsfGhiRecall}{1\xspace}
\newcommand{\XttClangTOaVsfGhiPrecision}{1\xspace}
\newcommand{\XttClangTOaVsfGhiFone}{1\xspace}
\newcommand{\XttClangTOaVsfRdaRecall}{0.984\xspace}
\newcommand{\XttClangTOaVsfRdaPrecision}{1\xspace}
\newcommand{\XttClangTOaVsfRdaFone}{0.992\xspace}
\newcommand{\XttClangTOaVsfRseRecall}{0.997\xspace}
\newcommand{\XttClangTOaVsfRsePrecision}{1\xspace}
\newcommand{\XttClangTOaVsfRseFone}{0.999\xspace}
\newcommand{\XttClangTObSzpGT}{15452\xspace}
\newcommand{\XttClangTObSzpLsRecall}{1\xspace}
\newcommand{\XttClangTObSzpLsPrecision}{1\xspace}
\newcommand{\XttClangTObSzpLsFone}{1\xspace}
\newcommand{\XttClangTObSzpBapRecall}{0.966\xspace}
\newcommand{\XttClangTObSzpBapPrecision}{1\xspace}
\newcommand{\XttClangTObSzpBapFone}{0.983\xspace}
\newcommand{\XttClangTObSzpGhiRecall}{0.976\xspace}
\newcommand{\XttClangTObSzpGhiPrecision}{1\xspace}
\newcommand{\XttClangTObSzpGhiFone}{0.988\xspace}
\newcommand{\XttClangTObSzpRdaRecall}{1.000\xspace}
\newcommand{\XttClangTObSzpRdaPrecision}{1\xspace}
\newcommand{\XttClangTObSzpRdaFone}{1.000\xspace}
\newcommand{\XttClangTObSzpRseRecall}{1\xspace}
\newcommand{\XttClangTObSzpRsePrecision}{1\xspace}
\newcommand{\XttClangTObSzpRseFone}{1\xspace}
\newcommand{\XttClangTObCapGT}{183059\xspace}
\newcommand{\XttClangTObCapLsRecall}{1\xspace}
\newcommand{\XttClangTObCapLsPrecision}{1\xspace}
\newcommand{\XttClangTObCapLsFone}{1\xspace}
\newcommand{\XttClangTObCapBapRecall}{0.363\xspace}
\newcommand{\XttClangTObCapBapPrecision}{1\xspace}
\newcommand{\XttClangTObCapBapFone}{0.533\xspace}
\newcommand{\XttClangTObCapGhiRecall}{0.962\xspace}
\newcommand{\XttClangTObCapGhiPrecision}{1\xspace}
\newcommand{\XttClangTObCapGhiFone}{0.981\xspace}
\newcommand{\XttClangTObCapRdaRecall}{0.976\xspace}
\newcommand{\XttClangTObCapRdaPrecision}{1\xspace}
\newcommand{\XttClangTObCapRdaFone}{0.988\xspace}
\newcommand{\XttClangTObCapRseRecall}{1\xspace}
\newcommand{\XttClangTObCapRsePrecision}{1\xspace}
\newcommand{\XttClangTObCapRseFone}{1\xspace}
\newcommand{\XttClangTObExmGT}{151461\xspace}
\newcommand{\XttClangTObExmLsRecall}{1\xspace}
\newcommand{\XttClangTObExmLsPrecision}{1\xspace}
\newcommand{\XttClangTObExmLsFone}{1\xspace}
\newcommand{\XttClangTObExmBapRecall}{0.843\xspace}
\newcommand{\XttClangTObExmBapPrecision}{1\xspace}
\newcommand{\XttClangTObExmBapFone}{0.915\xspace}
\newcommand{\XttClangTObExmGhiRecall}{0.969\xspace}
\newcommand{\XttClangTObExmGhiPrecision}{1\xspace}
\newcommand{\XttClangTObExmGhiFone}{0.984\xspace}
\newcommand{\XttClangTObExmRdaRecall}{0.965\xspace}
\newcommand{\XttClangTObExmRdaPrecision}{1\xspace}
\newcommand{\XttClangTObExmRdaFone}{0.982\xspace}
\newcommand{\XttClangTObExmRseRecall}{1.000\xspace}
\newcommand{\XttClangTObExmRsePrecision}{1\xspace}
\newcommand{\XttClangTObExmRseFone}{1.000\xspace}
\newcommand{\XttClangTObLgtGT}{27731\xspace}
\newcommand{\XttClangTObLgtLsRecall}{1\xspace}
\newcommand{\XttClangTObLgtLsPrecision}{1\xspace}
\newcommand{\XttClangTObLgtLsFone}{1\xspace}
\newcommand{\XttClangTObLgtBapRecall}{0.899\xspace}
\newcommand{\XttClangTObLgtBapPrecision}{1\xspace}
\newcommand{\XttClangTObLgtBapFone}{0.947\xspace}
\newcommand{\XttClangTObLgtGhiRecall}{1\xspace}
\newcommand{\XttClangTObLgtGhiPrecision}{1\xspace}
\newcommand{\XttClangTObLgtGhiFone}{1\xspace}
\newcommand{\XttClangTObLgtRdaRecall}{1\xspace}
\newcommand{\XttClangTObLgtRdaPrecision}{1\xspace}
\newcommand{\XttClangTObLgtRdaFone}{1\xspace}
\newcommand{\XttClangTObLgtRseRecall}{1\xspace}
\newcommand{\XttClangTObLgtRsePrecision}{1\xspace}
\newcommand{\XttClangTObLgtRseFone}{1\xspace}
\newcommand{\XttClangTObBzpGT}{15871\xspace}
\newcommand{\XttClangTObBzpLsRecall}{1\xspace}
\newcommand{\XttClangTObBzpLsPrecision}{1\xspace}
\newcommand{\XttClangTObBzpLsFone}{1\xspace}
\newcommand{\XttClangTObBzpBapRecall}{0.976\xspace}
\newcommand{\XttClangTObBzpBapPrecision}{1\xspace}
\newcommand{\XttClangTObBzpBapFone}{0.988\xspace}
\newcommand{\XttClangTObBzpGhiRecall}{1\xspace}
\newcommand{\XttClangTObBzpGhiPrecision}{1\xspace}
\newcommand{\XttClangTObBzpGhiFone}{1\xspace}
\newcommand{\XttClangTObBzpRdaRecall}{0.912\xspace}
\newcommand{\XttClangTObBzpRdaPrecision}{1\xspace}
\newcommand{\XttClangTObBzpRdaFone}{0.954\xspace}
\newcommand{\XttClangTObBzpRseRecall}{1\xspace}
\newcommand{\XttClangTObBzpRsePrecision}{1\xspace}
\newcommand{\XttClangTObBzpRseFone}{1\xspace}
\newcommand{\XttClangTObGccGT}{1131858\xspace}
\newcommand{\XttClangTObGccLsRecall}{1\xspace}
\newcommand{\XttClangTObGccLsPrecision}{1\xspace}
\newcommand{\XttClangTObGccLsFone}{1\xspace}
\newcommand{\XttClangTObGccBapRecall}{0.729\xspace}
\newcommand{\XttClangTObGccBapPrecision}{1\xspace}
\newcommand{\XttClangTObGccBapFone}{0.843\xspace}
\newcommand{\XttClangTObGccGhiRecall}{0.965\xspace}
\newcommand{\XttClangTObGccGhiPrecision}{1\xspace}
\newcommand{\XttClangTObGccGhiFone}{0.982\xspace}
\newcommand{\XttClangTObGccRdaRecall}{0.878\xspace}
\newcommand{\XttClangTObGccRdaPrecision}{1\xspace}
\newcommand{\XttClangTObGccRdaFone}{0.935\xspace}
\newcommand{\XttClangTObGccRseRecall}{1.000\xspace}
\newcommand{\XttClangTObGccRsePrecision}{1\xspace}
\newcommand{\XttClangTObGccRseFone}{1.000\xspace}
\newcommand{\XttClangTObGzpGT}{12034\xspace}
\newcommand{\XttClangTObGzpLsRecall}{1\xspace}
\newcommand{\XttClangTObGzpLsPrecision}{1\xspace}
\newcommand{\XttClangTObGzpLsFone}{1\xspace}
\newcommand{\XttClangTObGzpBapRecall}{0.985\xspace}
\newcommand{\XttClangTObGzpBapPrecision}{1\xspace}
\newcommand{\XttClangTObGzpBapFone}{0.992\xspace}
\newcommand{\XttClangTObGzpGhiRecall}{1\xspace}
\newcommand{\XttClangTObGzpGhiPrecision}{1\xspace}
\newcommand{\XttClangTObGzpGhiFone}{1\xspace}
\newcommand{\XttClangTObGzpRdaRecall}{1\xspace}
\newcommand{\XttClangTObGzpRdaPrecision}{1\xspace}
\newcommand{\XttClangTObGzpRdaFone}{1\xspace}
\newcommand{\XttClangTObGzpRseRecall}{0.999\xspace}
\newcommand{\XttClangTObGzpRsePrecision}{1\xspace}
\newcommand{\XttClangTObGzpRseFone}{0.999\xspace}
\newcommand{\XttClangTObOggGT}{46919\xspace}
\newcommand{\XttClangTObOggLsRecall}{1\xspace}
\newcommand{\XttClangTObOggLsPrecision}{1\xspace}
\newcommand{\XttClangTObOggLsFone}{1\xspace}
\newcommand{\XttClangTObOggBapRecall}{0.984\xspace}
\newcommand{\XttClangTObOggBapPrecision}{1\xspace}
\newcommand{\XttClangTObOggBapFone}{0.992\xspace}
\newcommand{\XttClangTObOggGhiRecall}{1\xspace}
\newcommand{\XttClangTObOggGhiPrecision}{1\xspace}
\newcommand{\XttClangTObOggGhiFone}{1\xspace}
\newcommand{\XttClangTObOggRdaRecall}{1\xspace}
\newcommand{\XttClangTObOggRdaPrecision}{1\xspace}
\newcommand{\XttClangTObOggRdaFone}{1\xspace}
\newcommand{\XttClangTObOggRseRecall}{1\xspace}
\newcommand{\XttClangTObOggRsePrecision}{1\xspace}
\newcommand{\XttClangTObOggRseFone}{1\xspace}
\newcommand{\XttClangTObNgxGT}{105180\xspace}
\newcommand{\XttClangTObNgxLsRecall}{1\xspace}
\newcommand{\XttClangTObNgxLsPrecision}{1\xspace}
\newcommand{\XttClangTObNgxLsFone}{1\xspace}
\newcommand{\XttClangTObNgxBapRecall}{0.968\xspace}
\newcommand{\XttClangTObNgxBapPrecision}{1\xspace}
\newcommand{\XttClangTObNgxBapFone}{0.984\xspace}
\newcommand{\XttClangTObNgxGhiRecall}{1\xspace}
\newcommand{\XttClangTObNgxGhiPrecision}{1\xspace}
\newcommand{\XttClangTObNgxGhiFone}{1\xspace}
\newcommand{\XttClangTObNgxRdaRecall}{0.998\xspace}
\newcommand{\XttClangTObNgxRdaPrecision}{1\xspace}
\newcommand{\XttClangTObNgxRdaFone}{0.999\xspace}
\newcommand{\XttClangTObNgxRseRecall}{1\xspace}
\newcommand{\XttClangTObNgxRsePrecision}{1\xspace}
\newcommand{\XttClangTObNgxRseFone}{1\xspace}
\newcommand{\XttClangTObSshGT}{126250\xspace}
\newcommand{\XttClangTObSshLsRecall}{1\xspace}
\newcommand{\XttClangTObSshLsPrecision}{1\xspace}
\newcommand{\XttClangTObSshLsFone}{1\xspace}
\newcommand{\XttClangTObSshBapRecall}{0.948\xspace}
\newcommand{\XttClangTObSshBapPrecision}{1\xspace}
\newcommand{\XttClangTObSshBapFone}{0.973\xspace}
\newcommand{\XttClangTObSshGhiRecall}{1.000\xspace}
\newcommand{\XttClangTObSshGhiPrecision}{1\xspace}
\newcommand{\XttClangTObSshGhiFone}{1.000\xspace}
\newcommand{\XttClangTObSshRdaRecall}{0.943\xspace}
\newcommand{\XttClangTObSshRdaPrecision}{1.000\xspace}
\newcommand{\XttClangTObSshRdaFone}{0.971\xspace}
\newcommand{\XttClangTObSshRseRecall}{0.990\xspace}
\newcommand{\XttClangTObSshRsePrecision}{1\xspace}
\newcommand{\XttClangTObSshRseFone}{0.995\xspace}
\newcommand{\XttClangTObPcrGT}{4882\xspace}
\newcommand{\XttClangTObPcrLsRecall}{1\xspace}
\newcommand{\XttClangTObPcrLsPrecision}{1\xspace}
\newcommand{\XttClangTObPcrLsFone}{1\xspace}
\newcommand{\XttClangTObPcrBapRecall}{0.845\xspace}
\newcommand{\XttClangTObPcrBapPrecision}{1\xspace}
\newcommand{\XttClangTObPcrBapFone}{0.916\xspace}
\newcommand{\XttClangTObPcrGhiRecall}{1\xspace}
\newcommand{\XttClangTObPcrGhiPrecision}{1\xspace}
\newcommand{\XttClangTObPcrGhiFone}{1\xspace}
\newcommand{\XttClangTObPcrRdaRecall}{1\xspace}
\newcommand{\XttClangTObPcrRdaPrecision}{1\xspace}
\newcommand{\XttClangTObPcrRdaFone}{1\xspace}
\newcommand{\XttClangTObPcrRseRecall}{1\xspace}
\newcommand{\XttClangTObPcrRsePrecision}{1\xspace}
\newcommand{\XttClangTObPcrRseFone}{1\xspace}
\newcommand{\XttClangTObSqlGT}{266078\xspace}
\newcommand{\XttClangTObSqlLsRecall}{1\xspace}
\newcommand{\XttClangTObSqlLsPrecision}{1\xspace}
\newcommand{\XttClangTObSqlLsFone}{1\xspace}
\newcommand{\XttClangTObSqlBapRecall}{0.813\xspace}
\newcommand{\XttClangTObSqlBapPrecision}{1\xspace}
\newcommand{\XttClangTObSqlBapFone}{0.897\xspace}
\newcommand{\XttClangTObSqlGhiRecall}{0.939\xspace}
\newcommand{\XttClangTObSqlGhiPrecision}{1\xspace}
\newcommand{\XttClangTObSqlGhiFone}{0.969\xspace}
\newcommand{\XttClangTObSqlRdaRecall}{0.975\xspace}
\newcommand{\XttClangTObSqlRdaPrecision}{1\xspace}
\newcommand{\XttClangTObSqlRdaFone}{0.987\xspace}
\newcommand{\XttClangTObSqlRseRecall}{0.996\xspace}
\newcommand{\XttClangTObSqlRsePrecision}{1.000\xspace}
\newcommand{\XttClangTObSqlRseFone}{0.998\xspace}
\newcommand{\XttClangTObVimGT}{559661\xspace}
\newcommand{\XttClangTObVimLsRecall}{1\xspace}
\newcommand{\XttClangTObVimLsPrecision}{1\xspace}
\newcommand{\XttClangTObVimLsFone}{1\xspace}
\newcommand{\XttClangTObVimBapRecall}{0.870\xspace}
\newcommand{\XttClangTObVimBapPrecision}{1\xspace}
\newcommand{\XttClangTObVimBapFone}{0.930\xspace}
\newcommand{\XttClangTObVimGhiRecall}{1.000\xspace}
\newcommand{\XttClangTObVimGhiPrecision}{1\xspace}
\newcommand{\XttClangTObVimGhiFone}{1.000\xspace}
\newcommand{\XttClangTObVimRdaRecall}{0.974\xspace}
\newcommand{\XttClangTObVimRdaPrecision}{1\xspace}
\newcommand{\XttClangTObVimRdaFone}{0.987\xspace}
\newcommand{\XttClangTObVimRseRecall}{1.000\xspace}
\newcommand{\XttClangTObVimRsePrecision}{1\xspace}
\newcommand{\XttClangTObVimRseFone}{1.000\xspace}
\newcommand{\XttClangTObVsfGT}{23537\xspace}
\newcommand{\XttClangTObVsfLsRecall}{1\xspace}
\newcommand{\XttClangTObVsfLsPrecision}{1\xspace}
\newcommand{\XttClangTObVsfLsFone}{1\xspace}
\newcommand{\XttClangTObVsfBapRecall}{0.990\xspace}
\newcommand{\XttClangTObVsfBapPrecision}{1\xspace}
\newcommand{\XttClangTObVsfBapFone}{0.995\xspace}
\newcommand{\XttClangTObVsfGhiRecall}{1\xspace}
\newcommand{\XttClangTObVsfGhiPrecision}{1\xspace}
\newcommand{\XttClangTObVsfGhiFone}{1\xspace}
\newcommand{\XttClangTObVsfRdaRecall}{0.984\xspace}
\newcommand{\XttClangTObVsfRdaPrecision}{1\xspace}
\newcommand{\XttClangTObVsfRdaFone}{0.992\xspace}
\newcommand{\XttClangTObVsfRseRecall}{1.000\xspace}
\newcommand{\XttClangTObVsfRsePrecision}{1\xspace}
\newcommand{\XttClangTObVsfRseFone}{1.000\xspace}
\newcommand{\XttClangTOcSzpGT}{15753\xspace}
\newcommand{\XttClangTOcSzpLsRecall}{1\xspace}
\newcommand{\XttClangTOcSzpLsPrecision}{1\xspace}
\newcommand{\XttClangTOcSzpLsFone}{1\xspace}
\newcommand{\XttClangTOcSzpBapRecall}{0.967\xspace}
\newcommand{\XttClangTOcSzpBapPrecision}{1\xspace}
\newcommand{\XttClangTOcSzpBapFone}{0.983\xspace}
\newcommand{\XttClangTOcSzpGhiRecall}{0.976\xspace}
\newcommand{\XttClangTOcSzpGhiPrecision}{1\xspace}
\newcommand{\XttClangTOcSzpGhiFone}{0.988\xspace}
\newcommand{\XttClangTOcSzpRdaRecall}{0.998\xspace}
\newcommand{\XttClangTOcSzpRdaPrecision}{1\xspace}
\newcommand{\XttClangTOcSzpRdaFone}{0.999\xspace}
\newcommand{\XttClangTOcSzpRseRecall}{1\xspace}
\newcommand{\XttClangTOcSzpRsePrecision}{1\xspace}
\newcommand{\XttClangTOcSzpRseFone}{1\xspace}
\newcommand{\XttClangTOcCapGT}{186119\xspace}
\newcommand{\XttClangTOcCapLsRecall}{1\xspace}
\newcommand{\XttClangTOcCapLsPrecision}{1\xspace}
\newcommand{\XttClangTOcCapLsFone}{1\xspace}
\newcommand{\XttClangTOcCapBapRecall}{0.359\xspace}
\newcommand{\XttClangTOcCapBapPrecision}{1\xspace}
\newcommand{\XttClangTOcCapBapFone}{0.529\xspace}
\newcommand{\XttClangTOcCapGhiRecall}{0.963\xspace}
\newcommand{\XttClangTOcCapGhiPrecision}{1\xspace}
\newcommand{\XttClangTOcCapGhiFone}{0.981\xspace}
\newcommand{\XttClangTOcCapRdaRecall}{0.977\xspace}
\newcommand{\XttClangTOcCapRdaPrecision}{1\xspace}
\newcommand{\XttClangTOcCapRdaFone}{0.988\xspace}
\newcommand{\XttClangTOcCapRseRecall}{1\xspace}
\newcommand{\XttClangTOcCapRsePrecision}{1\xspace}
\newcommand{\XttClangTOcCapRseFone}{1\xspace}
\newcommand{\XttClangTOcExmGT}{160122\xspace}
\newcommand{\XttClangTOcExmLsRecall}{1\xspace}
\newcommand{\XttClangTOcExmLsPrecision}{1\xspace}
\newcommand{\XttClangTOcExmLsFone}{1\xspace}
\newcommand{\XttClangTOcExmBapRecall}{0.837\xspace}
\newcommand{\XttClangTOcExmBapPrecision}{1\xspace}
\newcommand{\XttClangTOcExmBapFone}{0.911\xspace}
\newcommand{\XttClangTOcExmGhiRecall}{0.967\xspace}
\newcommand{\XttClangTOcExmGhiPrecision}{1\xspace}
\newcommand{\XttClangTOcExmGhiFone}{0.983\xspace}
\newcommand{\XttClangTOcExmRdaRecall}{0.964\xspace}
\newcommand{\XttClangTOcExmRdaPrecision}{1\xspace}
\newcommand{\XttClangTOcExmRdaFone}{0.982\xspace}
\newcommand{\XttClangTOcExmRseRecall}{1.000\xspace}
\newcommand{\XttClangTOcExmRsePrecision}{1\xspace}
\newcommand{\XttClangTOcExmRseFone}{1.000\xspace}
\newcommand{\XttClangTOcLgtGT}{28402\xspace}
\newcommand{\XttClangTOcLgtLsRecall}{1\xspace}
\newcommand{\XttClangTOcLgtLsPrecision}{1\xspace}
\newcommand{\XttClangTOcLgtLsFone}{1\xspace}
\newcommand{\XttClangTOcLgtBapRecall}{0.896\xspace}
\newcommand{\XttClangTOcLgtBapPrecision}{1\xspace}
\newcommand{\XttClangTOcLgtBapFone}{0.945\xspace}
\newcommand{\XttClangTOcLgtGhiRecall}{1\xspace}
\newcommand{\XttClangTOcLgtGhiPrecision}{1\xspace}
\newcommand{\XttClangTOcLgtGhiFone}{1\xspace}
\newcommand{\XttClangTOcLgtRdaRecall}{1\xspace}
\newcommand{\XttClangTOcLgtRdaPrecision}{1\xspace}
\newcommand{\XttClangTOcLgtRdaFone}{1\xspace}
\newcommand{\XttClangTOcLgtRseRecall}{1\xspace}
\newcommand{\XttClangTOcLgtRsePrecision}{1\xspace}
\newcommand{\XttClangTOcLgtRseFone}{1\xspace}
\newcommand{\XttClangTOcBzpGT}{16408\xspace}
\newcommand{\XttClangTOcBzpLsRecall}{1\xspace}
\newcommand{\XttClangTOcBzpLsPrecision}{1\xspace}
\newcommand{\XttClangTOcBzpLsFone}{1\xspace}
\newcommand{\XttClangTOcBzpBapRecall}{0.977\xspace}
\newcommand{\XttClangTOcBzpBapPrecision}{1\xspace}
\newcommand{\XttClangTOcBzpBapFone}{0.988\xspace}
\newcommand{\XttClangTOcBzpGhiRecall}{1\xspace}
\newcommand{\XttClangTOcBzpGhiPrecision}{1\xspace}
\newcommand{\XttClangTOcBzpGhiFone}{1\xspace}
\newcommand{\XttClangTOcBzpRdaRecall}{0.918\xspace}
\newcommand{\XttClangTOcBzpRdaPrecision}{1\xspace}
\newcommand{\XttClangTOcBzpRdaFone}{0.957\xspace}
\newcommand{\XttClangTOcBzpRseRecall}{1\xspace}
\newcommand{\XttClangTOcBzpRsePrecision}{1\xspace}
\newcommand{\XttClangTOcBzpRseFone}{1\xspace}
\newcommand{\XttClangTOcGccGT}{1229470\xspace}
\newcommand{\XttClangTOcGccLsRecall}{1\xspace}
\newcommand{\XttClangTOcGccLsPrecision}{1\xspace}
\newcommand{\XttClangTOcGccLsFone}{1\xspace}
\newcommand{\XttClangTOcGccBapRecall}{0.725\xspace}
\newcommand{\XttClangTOcGccBapPrecision}{1\xspace}
\newcommand{\XttClangTOcGccBapFone}{0.841\xspace}
\newcommand{\XttClangTOcGccGhiRecall}{0.969\xspace}
\newcommand{\XttClangTOcGccGhiPrecision}{1\xspace}
\newcommand{\XttClangTOcGccGhiFone}{0.984\xspace}
\newcommand{\XttClangTOcGccRdaRecall}{0.873\xspace}
\newcommand{\XttClangTOcGccRdaPrecision}{1.000\xspace}
\newcommand{\XttClangTOcGccRdaFone}{0.932\xspace}
\newcommand{\XttClangTOcGccRseRecall}{1.000\xspace}
\newcommand{\XttClangTOcGccRsePrecision}{1\xspace}
\newcommand{\XttClangTOcGccRseFone}{1.000\xspace}
\newcommand{\XttClangTOcGzpGT}{14431\xspace}
\newcommand{\XttClangTOcGzpLsRecall}{1\xspace}
\newcommand{\XttClangTOcGzpLsPrecision}{1\xspace}
\newcommand{\XttClangTOcGzpLsFone}{1\xspace}
\newcommand{\XttClangTOcGzpBapRecall}{0.987\xspace}
\newcommand{\XttClangTOcGzpBapPrecision}{1\xspace}
\newcommand{\XttClangTOcGzpBapFone}{0.994\xspace}
\newcommand{\XttClangTOcGzpGhiRecall}{1\xspace}
\newcommand{\XttClangTOcGzpGhiPrecision}{1\xspace}
\newcommand{\XttClangTOcGzpGhiFone}{1\xspace}
\newcommand{\XttClangTOcGzpRdaRecall}{1\xspace}
\newcommand{\XttClangTOcGzpRdaPrecision}{1\xspace}
\newcommand{\XttClangTOcGzpRdaFone}{1\xspace}
\newcommand{\XttClangTOcGzpRseRecall}{0.999\xspace}
\newcommand{\XttClangTOcGzpRsePrecision}{1\xspace}
\newcommand{\XttClangTOcGzpRseFone}{0.999\xspace}
\newcommand{\XttClangTOcOggGT}{55754\xspace}
\newcommand{\XttClangTOcOggLsRecall}{1\xspace}
\newcommand{\XttClangTOcOggLsPrecision}{1\xspace}
\newcommand{\XttClangTOcOggLsFone}{1\xspace}
\newcommand{\XttClangTOcOggBapRecall}{0.987\xspace}
\newcommand{\XttClangTOcOggBapPrecision}{1\xspace}
\newcommand{\XttClangTOcOggBapFone}{0.993\xspace}
\newcommand{\XttClangTOcOggGhiRecall}{1\xspace}
\newcommand{\XttClangTOcOggGhiPrecision}{1\xspace}
\newcommand{\XttClangTOcOggGhiFone}{1\xspace}
\newcommand{\XttClangTOcOggRdaRecall}{1\xspace}
\newcommand{\XttClangTOcOggRdaPrecision}{1\xspace}
\newcommand{\XttClangTOcOggRdaFone}{1\xspace}
\newcommand{\XttClangTOcOggRseRecall}{1\xspace}
\newcommand{\XttClangTOcOggRsePrecision}{1\xspace}
\newcommand{\XttClangTOcOggRseFone}{1\xspace}
\newcommand{\XttClangTOcNgxGT}{109110\xspace}
\newcommand{\XttClangTOcNgxLsRecall}{1\xspace}
\newcommand{\XttClangTOcNgxLsPrecision}{1\xspace}
\newcommand{\XttClangTOcNgxLsFone}{1\xspace}
\newcommand{\XttClangTOcNgxBapRecall}{0.967\xspace}
\newcommand{\XttClangTOcNgxBapPrecision}{1\xspace}
\newcommand{\XttClangTOcNgxBapFone}{0.983\xspace}
\newcommand{\XttClangTOcNgxGhiRecall}{1\xspace}
\newcommand{\XttClangTOcNgxGhiPrecision}{1\xspace}
\newcommand{\XttClangTOcNgxGhiFone}{1\xspace}
\newcommand{\XttClangTOcNgxRdaRecall}{0.998\xspace}
\newcommand{\XttClangTOcNgxRdaPrecision}{1\xspace}
\newcommand{\XttClangTOcNgxRdaFone}{0.999\xspace}
\newcommand{\XttClangTOcNgxRseRecall}{1\xspace}
\newcommand{\XttClangTOcNgxRsePrecision}{1\xspace}
\newcommand{\XttClangTOcNgxRseFone}{1\xspace}
\newcommand{\XttClangTOcSshGT}{135672\xspace}
\newcommand{\XttClangTOcSshLsRecall}{1\xspace}
\newcommand{\XttClangTOcSshLsPrecision}{1\xspace}
\newcommand{\XttClangTOcSshLsFone}{1\xspace}
\newcommand{\XttClangTOcSshBapRecall}{0.950\xspace}
\newcommand{\XttClangTOcSshBapPrecision}{1\xspace}
\newcommand{\XttClangTOcSshBapFone}{0.975\xspace}
\newcommand{\XttClangTOcSshGhiRecall}{1.000\xspace}
\newcommand{\XttClangTOcSshGhiPrecision}{1\xspace}
\newcommand{\XttClangTOcSshGhiFone}{1.000\xspace}
\newcommand{\XttClangTOcSshRdaRecall}{0.904\xspace}
\newcommand{\XttClangTOcSshRdaPrecision}{1\xspace}
\newcommand{\XttClangTOcSshRdaFone}{0.950\xspace}
\newcommand{\XttClangTOcSshRseRecall}{0.991\xspace}
\newcommand{\XttClangTOcSshRsePrecision}{1\xspace}
\newcommand{\XttClangTOcSshRseFone}{0.995\xspace}
\newcommand{\XttClangTOcPcrGT}{5611\xspace}
\newcommand{\XttClangTOcPcrLsRecall}{1\xspace}
\newcommand{\XttClangTOcPcrLsPrecision}{1\xspace}
\newcommand{\XttClangTOcPcrLsFone}{1\xspace}
\newcommand{\XttClangTOcPcrBapRecall}{0.833\xspace}
\newcommand{\XttClangTOcPcrBapPrecision}{1\xspace}
\newcommand{\XttClangTOcPcrBapFone}{0.909\xspace}
\newcommand{\XttClangTOcPcrGhiRecall}{1\xspace}
\newcommand{\XttClangTOcPcrGhiPrecision}{1\xspace}
\newcommand{\XttClangTOcPcrGhiFone}{1\xspace}
\newcommand{\XttClangTOcPcrRdaRecall}{1\xspace}
\newcommand{\XttClangTOcPcrRdaPrecision}{1\xspace}
\newcommand{\XttClangTOcPcrRdaFone}{1\xspace}
\newcommand{\XttClangTOcPcrRseRecall}{1\xspace}
\newcommand{\XttClangTOcPcrRsePrecision}{1\xspace}
\newcommand{\XttClangTOcPcrRseFone}{1\xspace}
\newcommand{\XttClangTOcSqlGT}{310170\xspace}
\newcommand{\XttClangTOcSqlLsRecall}{1\xspace}
\newcommand{\XttClangTOcSqlLsPrecision}{1\xspace}
\newcommand{\XttClangTOcSqlLsFone}{1\xspace}
\newcommand{\XttClangTOcSqlBapRecall}{0.810\xspace}
\newcommand{\XttClangTOcSqlBapPrecision}{1\xspace}
\newcommand{\XttClangTOcSqlBapFone}{0.895\xspace}
\newcommand{\XttClangTOcSqlGhiRecall}{0.937\xspace}
\newcommand{\XttClangTOcSqlGhiPrecision}{1\xspace}
\newcommand{\XttClangTOcSqlGhiFone}{0.967\xspace}
\newcommand{\XttClangTOcSqlRdaRecall}{0.967\xspace}
\newcommand{\XttClangTOcSqlRdaPrecision}{1\xspace}
\newcommand{\XttClangTOcSqlRdaFone}{0.983\xspace}
\newcommand{\XttClangTOcSqlRseRecall}{0.997\xspace}
\newcommand{\XttClangTOcSqlRsePrecision}{1\xspace}
\newcommand{\XttClangTOcSqlRseFone}{0.999\xspace}
\newcommand{\XttClangTOcVimGT}{598038\xspace}
\newcommand{\XttClangTOcVimLsRecall}{1\xspace}
\newcommand{\XttClangTOcVimLsPrecision}{1\xspace}
\newcommand{\XttClangTOcVimLsFone}{1\xspace}
\newcommand{\XttClangTOcVimBapRecall}{0.874\xspace}
\newcommand{\XttClangTOcVimBapPrecision}{1\xspace}
\newcommand{\XttClangTOcVimBapFone}{0.933\xspace}
\newcommand{\XttClangTOcVimGhiRecall}{0.999\xspace}
\newcommand{\XttClangTOcVimGhiPrecision}{1\xspace}
\newcommand{\XttClangTOcVimGhiFone}{1.000\xspace}
\newcommand{\XttClangTOcVimRdaRecall}{0.973\xspace}
\newcommand{\XttClangTOcVimRdaPrecision}{1\xspace}
\newcommand{\XttClangTOcVimRdaFone}{0.986\xspace}
\newcommand{\XttClangTOcVimRseRecall}{1.000\xspace}
\newcommand{\XttClangTOcVimRsePrecision}{1\xspace}
\newcommand{\XttClangTOcVimRseFone}{1.000\xspace}
\newcommand{\XttClangTOcVsfGT}{24160\xspace}
\newcommand{\XttClangTOcVsfLsRecall}{1\xspace}
\newcommand{\XttClangTOcVsfLsPrecision}{1\xspace}
\newcommand{\XttClangTOcVsfLsFone}{1\xspace}
\newcommand{\XttClangTOcVsfBapRecall}{0.990\xspace}
\newcommand{\XttClangTOcVsfBapPrecision}{1\xspace}
\newcommand{\XttClangTOcVsfBapFone}{0.995\xspace}
\newcommand{\XttClangTOcVsfGhiRecall}{0.995\xspace}
\newcommand{\XttClangTOcVsfGhiPrecision}{1.000\xspace}
\newcommand{\XttClangTOcVsfGhiFone}{0.997\xspace}
\newcommand{\XttClangTOcVsfRdaRecall}{0.984\xspace}
\newcommand{\XttClangTOcVsfRdaPrecision}{1\xspace}
\newcommand{\XttClangTOcVsfRdaFone}{0.992\xspace}
\newcommand{\XttClangTOcVsfRseRecall}{1.000\xspace}
\newcommand{\XttClangTOcVsfRsePrecision}{1\xspace}
\newcommand{\XttClangTOcVsfRseFone}{1.000\xspace}
\newcommand{\XttClangTOdSzpGT}{15753\xspace}
\newcommand{\XttClangTOdSzpLsRecall}{1\xspace}
\newcommand{\XttClangTOdSzpLsPrecision}{1\xspace}
\newcommand{\XttClangTOdSzpLsFone}{1\xspace}
\newcommand{\XttClangTOdSzpBapRecall}{0.967\xspace}
\newcommand{\XttClangTOdSzpBapPrecision}{1\xspace}
\newcommand{\XttClangTOdSzpBapFone}{0.983\xspace}
\newcommand{\XttClangTOdSzpGhiRecall}{0.976\xspace}
\newcommand{\XttClangTOdSzpGhiPrecision}{1\xspace}
\newcommand{\XttClangTOdSzpGhiFone}{0.988\xspace}
\newcommand{\XttClangTOdSzpRdaRecall}{0.998\xspace}
\newcommand{\XttClangTOdSzpRdaPrecision}{1\xspace}
\newcommand{\XttClangTOdSzpRdaFone}{0.999\xspace}
\newcommand{\XttClangTOdSzpRseRecall}{1\xspace}
\newcommand{\XttClangTOdSzpRsePrecision}{1\xspace}
\newcommand{\XttClangTOdSzpRseFone}{1\xspace}
\newcommand{\XttClangTOdCapGT}{186119\xspace}
\newcommand{\XttClangTOdCapLsRecall}{1\xspace}
\newcommand{\XttClangTOdCapLsPrecision}{1\xspace}
\newcommand{\XttClangTOdCapLsFone}{1\xspace}
\newcommand{\XttClangTOdCapBapRecall}{0.359\xspace}
\newcommand{\XttClangTOdCapBapPrecision}{1\xspace}
\newcommand{\XttClangTOdCapBapFone}{0.529\xspace}
\newcommand{\XttClangTOdCapGhiRecall}{0.963\xspace}
\newcommand{\XttClangTOdCapGhiPrecision}{1\xspace}
\newcommand{\XttClangTOdCapGhiFone}{0.981\xspace}
\newcommand{\XttClangTOdCapRdaRecall}{0.977\xspace}
\newcommand{\XttClangTOdCapRdaPrecision}{1\xspace}
\newcommand{\XttClangTOdCapRdaFone}{0.988\xspace}
\newcommand{\XttClangTOdCapRseRecall}{1\xspace}
\newcommand{\XttClangTOdCapRsePrecision}{1\xspace}
\newcommand{\XttClangTOdCapRseFone}{1\xspace}
\newcommand{\XttClangTOdExmGT}{160116\xspace}
\newcommand{\XttClangTOdExmLsRecall}{1\xspace}
\newcommand{\XttClangTOdExmLsPrecision}{1\xspace}
\newcommand{\XttClangTOdExmLsFone}{1\xspace}
\newcommand{\XttClangTOdExmBapRecall}{0.837\xspace}
\newcommand{\XttClangTOdExmBapPrecision}{1\xspace}
\newcommand{\XttClangTOdExmBapFone}{0.912\xspace}
\newcommand{\XttClangTOdExmGhiRecall}{0.931\xspace}
\newcommand{\XttClangTOdExmGhiPrecision}{1\xspace}
\newcommand{\XttClangTOdExmGhiFone}{0.964\xspace}
\newcommand{\XttClangTOdExmRdaRecall}{0.964\xspace}
\newcommand{\XttClangTOdExmRdaPrecision}{1\xspace}
\newcommand{\XttClangTOdExmRdaFone}{0.982\xspace}
\newcommand{\XttClangTOdExmRseRecall}{1.000\xspace}
\newcommand{\XttClangTOdExmRsePrecision}{1\xspace}
\newcommand{\XttClangTOdExmRseFone}{1.000\xspace}
\newcommand{\XttClangTOdLgtGT}{28402\xspace}
\newcommand{\XttClangTOdLgtLsRecall}{1\xspace}
\newcommand{\XttClangTOdLgtLsPrecision}{1\xspace}
\newcommand{\XttClangTOdLgtLsFone}{1\xspace}
\newcommand{\XttClangTOdLgtBapRecall}{0.896\xspace}
\newcommand{\XttClangTOdLgtBapPrecision}{1\xspace}
\newcommand{\XttClangTOdLgtBapFone}{0.945\xspace}
\newcommand{\XttClangTOdLgtGhiRecall}{1\xspace}
\newcommand{\XttClangTOdLgtGhiPrecision}{1\xspace}
\newcommand{\XttClangTOdLgtGhiFone}{1\xspace}
\newcommand{\XttClangTOdLgtRdaRecall}{1\xspace}
\newcommand{\XttClangTOdLgtRdaPrecision}{1\xspace}
\newcommand{\XttClangTOdLgtRdaFone}{1\xspace}
\newcommand{\XttClangTOdLgtRseRecall}{1\xspace}
\newcommand{\XttClangTOdLgtRsePrecision}{1\xspace}
\newcommand{\XttClangTOdLgtRseFone}{1\xspace}
\newcommand{\XttClangTOdBzpGT}{16408\xspace}
\newcommand{\XttClangTOdBzpLsRecall}{1\xspace}
\newcommand{\XttClangTOdBzpLsPrecision}{1\xspace}
\newcommand{\XttClangTOdBzpLsFone}{1\xspace}
\newcommand{\XttClangTOdBzpBapRecall}{0.977\xspace}
\newcommand{\XttClangTOdBzpBapPrecision}{1\xspace}
\newcommand{\XttClangTOdBzpBapFone}{0.988\xspace}
\newcommand{\XttClangTOdBzpGhiRecall}{1\xspace}
\newcommand{\XttClangTOdBzpGhiPrecision}{1\xspace}
\newcommand{\XttClangTOdBzpGhiFone}{1\xspace}
\newcommand{\XttClangTOdBzpRdaRecall}{0.918\xspace}
\newcommand{\XttClangTOdBzpRdaPrecision}{1\xspace}
\newcommand{\XttClangTOdBzpRdaFone}{0.957\xspace}
\newcommand{\XttClangTOdBzpRseRecall}{1\xspace}
\newcommand{\XttClangTOdBzpRsePrecision}{1\xspace}
\newcommand{\XttClangTOdBzpRseFone}{1\xspace}
\newcommand{\XttClangTOdGccGT}{1229469\xspace}
\newcommand{\XttClangTOdGccLsRecall}{1\xspace}
\newcommand{\XttClangTOdGccLsPrecision}{1\xspace}
\newcommand{\XttClangTOdGccLsFone}{1\xspace}
\newcommand{\XttClangTOdGccBapRecall}{0.725\xspace}
\newcommand{\XttClangTOdGccBapPrecision}{1\xspace}
\newcommand{\XttClangTOdGccBapFone}{0.841\xspace}
\newcommand{\XttClangTOdGccGhiRecall}{0.969\xspace}
\newcommand{\XttClangTOdGccGhiPrecision}{1\xspace}
\newcommand{\XttClangTOdGccGhiFone}{0.984\xspace}
\newcommand{\XttClangTOdGccRdaRecall}{0.873\xspace}
\newcommand{\XttClangTOdGccRdaPrecision}{1\xspace}
\newcommand{\XttClangTOdGccRdaFone}{0.932\xspace}
\newcommand{\XttClangTOdGccRseRecall}{1.000\xspace}
\newcommand{\XttClangTOdGccRsePrecision}{1\xspace}
\newcommand{\XttClangTOdGccRseFone}{1.000\xspace}
\newcommand{\XttClangTOdGzpGT}{14431\xspace}
\newcommand{\XttClangTOdGzpLsRecall}{1\xspace}
\newcommand{\XttClangTOdGzpLsPrecision}{1\xspace}
\newcommand{\XttClangTOdGzpLsFone}{1\xspace}
\newcommand{\XttClangTOdGzpBapRecall}{0.987\xspace}
\newcommand{\XttClangTOdGzpBapPrecision}{1\xspace}
\newcommand{\XttClangTOdGzpBapFone}{0.994\xspace}
\newcommand{\XttClangTOdGzpGhiRecall}{1\xspace}
\newcommand{\XttClangTOdGzpGhiPrecision}{1\xspace}
\newcommand{\XttClangTOdGzpGhiFone}{1\xspace}
\newcommand{\XttClangTOdGzpRdaRecall}{1\xspace}
\newcommand{\XttClangTOdGzpRdaPrecision}{1\xspace}
\newcommand{\XttClangTOdGzpRdaFone}{1\xspace}
\newcommand{\XttClangTOdGzpRseRecall}{0.999\xspace}
\newcommand{\XttClangTOdGzpRsePrecision}{1\xspace}
\newcommand{\XttClangTOdGzpRseFone}{0.999\xspace}
\newcommand{\XttClangTOdOggGT}{56073\xspace}
\newcommand{\XttClangTOdOggLsRecall}{1\xspace}
\newcommand{\XttClangTOdOggLsPrecision}{1\xspace}
\newcommand{\XttClangTOdOggLsFone}{1\xspace}
\newcommand{\XttClangTOdOggBapRecall}{0.987\xspace}
\newcommand{\XttClangTOdOggBapPrecision}{1\xspace}
\newcommand{\XttClangTOdOggBapFone}{0.993\xspace}
\newcommand{\XttClangTOdOggGhiRecall}{1\xspace}
\newcommand{\XttClangTOdOggGhiPrecision}{1\xspace}
\newcommand{\XttClangTOdOggGhiFone}{1\xspace}
\newcommand{\XttClangTOdOggRdaRecall}{1\xspace}
\newcommand{\XttClangTOdOggRdaPrecision}{1\xspace}
\newcommand{\XttClangTOdOggRdaFone}{1\xspace}
\newcommand{\XttClangTOdOggRseRecall}{1\xspace}
\newcommand{\XttClangTOdOggRsePrecision}{1\xspace}
\newcommand{\XttClangTOdOggRseFone}{1\xspace}
\newcommand{\XttClangTOdNgxGT}{109111\xspace}
\newcommand{\XttClangTOdNgxLsRecall}{1\xspace}
\newcommand{\XttClangTOdNgxLsPrecision}{1\xspace}
\newcommand{\XttClangTOdNgxLsFone}{1\xspace}
\newcommand{\XttClangTOdNgxBapRecall}{0.967\xspace}
\newcommand{\XttClangTOdNgxBapPrecision}{1\xspace}
\newcommand{\XttClangTOdNgxBapFone}{0.983\xspace}
\newcommand{\XttClangTOdNgxGhiRecall}{1\xspace}
\newcommand{\XttClangTOdNgxGhiPrecision}{1\xspace}
\newcommand{\XttClangTOdNgxGhiFone}{1\xspace}
\newcommand{\XttClangTOdNgxRdaRecall}{0.998\xspace}
\newcommand{\XttClangTOdNgxRdaPrecision}{1\xspace}
\newcommand{\XttClangTOdNgxRdaFone}{0.999\xspace}
\newcommand{\XttClangTOdNgxRseRecall}{1\xspace}
\newcommand{\XttClangTOdNgxRsePrecision}{1\xspace}
\newcommand{\XttClangTOdNgxRseFone}{1\xspace}
\newcommand{\XttClangTOdSshGT}{135672\xspace}
\newcommand{\XttClangTOdSshLsRecall}{1\xspace}
\newcommand{\XttClangTOdSshLsPrecision}{1\xspace}
\newcommand{\XttClangTOdSshLsFone}{1\xspace}
\newcommand{\XttClangTOdSshBapRecall}{0.950\xspace}
\newcommand{\XttClangTOdSshBapPrecision}{1\xspace}
\newcommand{\XttClangTOdSshBapFone}{0.975\xspace}
\newcommand{\XttClangTOdSshGhiRecall}{1.000\xspace}
\newcommand{\XttClangTOdSshGhiPrecision}{1\xspace}
\newcommand{\XttClangTOdSshGhiFone}{1.000\xspace}
\newcommand{\XttClangTOdSshRdaRecall}{0.904\xspace}
\newcommand{\XttClangTOdSshRdaPrecision}{1\xspace}
\newcommand{\XttClangTOdSshRdaFone}{0.950\xspace}
\newcommand{\XttClangTOdSshRseRecall}{0.991\xspace}
\newcommand{\XttClangTOdSshRsePrecision}{1\xspace}
\newcommand{\XttClangTOdSshRseFone}{0.995\xspace}
\newcommand{\XttClangTOdPcrGT}{5611\xspace}
\newcommand{\XttClangTOdPcrLsRecall}{1\xspace}
\newcommand{\XttClangTOdPcrLsPrecision}{1\xspace}
\newcommand{\XttClangTOdPcrLsFone}{1\xspace}
\newcommand{\XttClangTOdPcrBapRecall}{0.833\xspace}
\newcommand{\XttClangTOdPcrBapPrecision}{1\xspace}
\newcommand{\XttClangTOdPcrBapFone}{0.909\xspace}
\newcommand{\XttClangTOdPcrGhiRecall}{1\xspace}
\newcommand{\XttClangTOdPcrGhiPrecision}{1\xspace}
\newcommand{\XttClangTOdPcrGhiFone}{1\xspace}
\newcommand{\XttClangTOdPcrRdaRecall}{1\xspace}
\newcommand{\XttClangTOdPcrRdaPrecision}{1\xspace}
\newcommand{\XttClangTOdPcrRdaFone}{1\xspace}
\newcommand{\XttClangTOdPcrRseRecall}{1\xspace}
\newcommand{\XttClangTOdPcrRsePrecision}{1\xspace}
\newcommand{\XttClangTOdPcrRseFone}{1\xspace}
\newcommand{\XttClangTOdSqlGT}{310297\xspace}
\newcommand{\XttClangTOdSqlLsRecall}{1\xspace}
\newcommand{\XttClangTOdSqlLsPrecision}{1\xspace}
\newcommand{\XttClangTOdSqlLsFone}{1\xspace}
\newcommand{\XttClangTOdSqlBapRecall}{0.809\xspace}
\newcommand{\XttClangTOdSqlBapPrecision}{1\xspace}
\newcommand{\XttClangTOdSqlBapFone}{0.895\xspace}
\newcommand{\XttClangTOdSqlGhiRecall}{0.937\xspace}
\newcommand{\XttClangTOdSqlGhiPrecision}{1\xspace}
\newcommand{\XttClangTOdSqlGhiFone}{0.967\xspace}
\newcommand{\XttClangTOdSqlRdaRecall}{0.967\xspace}
\newcommand{\XttClangTOdSqlRdaPrecision}{1\xspace}
\newcommand{\XttClangTOdSqlRdaFone}{0.983\xspace}
\newcommand{\XttClangTOdSqlRseRecall}{0.997\xspace}
\newcommand{\XttClangTOdSqlRsePrecision}{1\xspace}
\newcommand{\XttClangTOdSqlRseFone}{0.999\xspace}
\newcommand{\XttClangTOdVimGT}{598043\xspace}
\newcommand{\XttClangTOdVimLsRecall}{1\xspace}
\newcommand{\XttClangTOdVimLsPrecision}{1\xspace}
\newcommand{\XttClangTOdVimLsFone}{1\xspace}
\newcommand{\XttClangTOdVimBapRecall}{0.874\xspace}
\newcommand{\XttClangTOdVimBapPrecision}{1\xspace}
\newcommand{\XttClangTOdVimBapFone}{0.933\xspace}
\newcommand{\XttClangTOdVimGhiRecall}{0.999\xspace}
\newcommand{\XttClangTOdVimGhiPrecision}{1\xspace}
\newcommand{\XttClangTOdVimGhiFone}{1.000\xspace}
\newcommand{\XttClangTOdVimRdaRecall}{0.973\xspace}
\newcommand{\XttClangTOdVimRdaPrecision}{1\xspace}
\newcommand{\XttClangTOdVimRdaFone}{0.986\xspace}
\newcommand{\XttClangTOdVimRseRecall}{1.000\xspace}
\newcommand{\XttClangTOdVimRsePrecision}{1\xspace}
\newcommand{\XttClangTOdVimRseFone}{1.000\xspace}
\newcommand{\XttClangTOdVsfGT}{24152\xspace}
\newcommand{\XttClangTOdVsfLsRecall}{1\xspace}
\newcommand{\XttClangTOdVsfLsPrecision}{1\xspace}
\newcommand{\XttClangTOdVsfLsFone}{1\xspace}
\newcommand{\XttClangTOdVsfBapRecall}{0.990\xspace}
\newcommand{\XttClangTOdVsfBapPrecision}{1\xspace}
\newcommand{\XttClangTOdVsfBapFone}{0.995\xspace}
\newcommand{\XttClangTOdVsfGhiRecall}{0.999\xspace}
\newcommand{\XttClangTOdVsfGhiPrecision}{1.000\xspace}
\newcommand{\XttClangTOdVsfGhiFone}{1.000\xspace}
\newcommand{\XttClangTOdVsfRdaRecall}{0.984\xspace}
\newcommand{\XttClangTOdVsfRdaPrecision}{1\xspace}
\newcommand{\XttClangTOdVsfRdaFone}{0.992\xspace}
\newcommand{\XttClangTOdVsfRseRecall}{1.000\xspace}
\newcommand{\XttClangTOdVsfRsePrecision}{1\xspace}
\newcommand{\XttClangTOdVsfRseFone}{1.000\xspace}
\newcommand{\XttClangTOsSzpGT}{13742\xspace}
\newcommand{\XttClangTOsSzpLsRecall}{1\xspace}
\newcommand{\XttClangTOsSzpLsPrecision}{1\xspace}
\newcommand{\XttClangTOsSzpLsFone}{1\xspace}
\newcommand{\XttClangTOsSzpBapRecall}{0.976\xspace}
\newcommand{\XttClangTOsSzpBapPrecision}{1\xspace}
\newcommand{\XttClangTOsSzpBapFone}{0.988\xspace}
\newcommand{\XttClangTOsSzpGhiRecall}{1\xspace}
\newcommand{\XttClangTOsSzpGhiPrecision}{1\xspace}
\newcommand{\XttClangTOsSzpGhiFone}{1\xspace}
\newcommand{\XttClangTOsSzpRdaRecall}{1\xspace}
\newcommand{\XttClangTOsSzpRdaPrecision}{1\xspace}
\newcommand{\XttClangTOsSzpRdaFone}{1\xspace}
\newcommand{\XttClangTOsSzpRseRecall}{1\xspace}
\newcommand{\XttClangTOsSzpRsePrecision}{1\xspace}
\newcommand{\XttClangTOsSzpRseFone}{1\xspace}
\newcommand{\XttClangTOsCapGT}{189502\xspace}
\newcommand{\XttClangTOsCapLsRecall}{1\xspace}
\newcommand{\XttClangTOsCapLsPrecision}{1\xspace}
\newcommand{\XttClangTOsCapLsFone}{1\xspace}
\newcommand{\XttClangTOsCapBapRecall}{0.349\xspace}
\newcommand{\XttClangTOsCapBapPrecision}{1\xspace}
\newcommand{\XttClangTOsCapBapFone}{0.517\xspace}
\newcommand{\XttClangTOsCapGhiRecall}{0.855\xspace}
\newcommand{\XttClangTOsCapGhiPrecision}{1\xspace}
\newcommand{\XttClangTOsCapGhiFone}{0.922\xspace}
\newcommand{\XttClangTOsCapRdaRecall}{0.977\xspace}
\newcommand{\XttClangTOsCapRdaPrecision}{1\xspace}
\newcommand{\XttClangTOsCapRdaFone}{0.988\xspace}
\newcommand{\XttClangTOsCapRseRecall}{1\xspace}
\newcommand{\XttClangTOsCapRsePrecision}{1\xspace}
\newcommand{\XttClangTOsCapRseFone}{1\xspace}
\newcommand{\XttClangTOsExmGT}{148782\xspace}
\newcommand{\XttClangTOsExmLsRecall}{1\xspace}
\newcommand{\XttClangTOsExmLsPrecision}{1\xspace}
\newcommand{\XttClangTOsExmLsFone}{1\xspace}
\newcommand{\XttClangTOsExmBapRecall}{0.826\xspace}
\newcommand{\XttClangTOsExmBapPrecision}{1\xspace}
\newcommand{\XttClangTOsExmBapFone}{0.905\xspace}
\newcommand{\XttClangTOsExmGhiRecall}{0.984\xspace}
\newcommand{\XttClangTOsExmGhiPrecision}{1\xspace}
\newcommand{\XttClangTOsExmGhiFone}{0.992\xspace}
\newcommand{\XttClangTOsExmRdaRecall}{0.966\xspace}
\newcommand{\XttClangTOsExmRdaPrecision}{1\xspace}
\newcommand{\XttClangTOsExmRdaFone}{0.983\xspace}
\newcommand{\XttClangTOsExmRseRecall}{0.999\xspace}
\newcommand{\XttClangTOsExmRsePrecision}{1\xspace}
\newcommand{\XttClangTOsExmRseFone}{1.000\xspace}
\newcommand{\XttClangTOsLgtGT}{28756\xspace}
\newcommand{\XttClangTOsLgtLsRecall}{1\xspace}
\newcommand{\XttClangTOsLgtLsPrecision}{1\xspace}
\newcommand{\XttClangTOsLgtLsFone}{1\xspace}
\newcommand{\XttClangTOsLgtBapRecall}{0.895\xspace}
\newcommand{\XttClangTOsLgtBapPrecision}{1\xspace}
\newcommand{\XttClangTOsLgtBapFone}{0.945\xspace}
\newcommand{\XttClangTOsLgtGhiRecall}{0.995\xspace}
\newcommand{\XttClangTOsLgtGhiPrecision}{1\xspace}
\newcommand{\XttClangTOsLgtGhiFone}{0.997\xspace}
\newcommand{\XttClangTOsLgtRdaRecall}{1.000\xspace}
\newcommand{\XttClangTOsLgtRdaPrecision}{1\xspace}
\newcommand{\XttClangTOsLgtRdaFone}{1.000\xspace}
\newcommand{\XttClangTOsLgtRseRecall}{0.997\xspace}
\newcommand{\XttClangTOsLgtRsePrecision}{1\xspace}
\newcommand{\XttClangTOsLgtRseFone}{0.999\xspace}
\newcommand{\XttClangTOsBzpGT}{15215\xspace}
\newcommand{\XttClangTOsBzpLsRecall}{1\xspace}
\newcommand{\XttClangTOsBzpLsPrecision}{1\xspace}
\newcommand{\XttClangTOsBzpLsFone}{1\xspace}
\newcommand{\XttClangTOsBzpBapRecall}{0.979\xspace}
\newcommand{\XttClangTOsBzpBapPrecision}{1\xspace}
\newcommand{\XttClangTOsBzpBapFone}{0.990\xspace}
\newcommand{\XttClangTOsBzpGhiRecall}{0.998\xspace}
\newcommand{\XttClangTOsBzpGhiPrecision}{1\xspace}
\newcommand{\XttClangTOsBzpGhiFone}{0.999\xspace}
\newcommand{\XttClangTOsBzpRdaRecall}{0.826\xspace}
\newcommand{\XttClangTOsBzpRdaPrecision}{1\xspace}
\newcommand{\XttClangTOsBzpRdaFone}{0.904\xspace}
\newcommand{\XttClangTOsBzpRseRecall}{0.999\xspace}
\newcommand{\XttClangTOsBzpRsePrecision}{1\xspace}
\newcommand{\XttClangTOsBzpRseFone}{1.000\xspace}
\newcommand{\XttClangTOsGccGT}{917942\xspace}
\newcommand{\XttClangTOsGccLsRecall}{1\xspace}
\newcommand{\XttClangTOsGccLsPrecision}{1\xspace}
\newcommand{\XttClangTOsGccLsFone}{1\xspace}
\newcommand{\XttClangTOsGccBapRecall}{0.722\xspace}
\newcommand{\XttClangTOsGccBapPrecision}{1\xspace}
\newcommand{\XttClangTOsGccBapFone}{0.839\xspace}
\newcommand{\XttClangTOsGccGhiRecall}{0.969\xspace}
\newcommand{\XttClangTOsGccGhiPrecision}{1\xspace}
\newcommand{\XttClangTOsGccGhiFone}{0.984\xspace}
\newcommand{\XttClangTOsGccRdaRecall}{0.880\xspace}
\newcommand{\XttClangTOsGccRdaPrecision}{1.000\xspace}
\newcommand{\XttClangTOsGccRdaFone}{0.936\xspace}
\newcommand{\XttClangTOsGccRseRecall}{1.000\xspace}
\newcommand{\XttClangTOsGccRsePrecision}{1\xspace}
\newcommand{\XttClangTOsGccRseFone}{1.000\xspace}
\newcommand{\XttClangTOsGzpGT}{9811\xspace}
\newcommand{\XttClangTOsGzpLsRecall}{1\xspace}
\newcommand{\XttClangTOsGzpLsPrecision}{1\xspace}
\newcommand{\XttClangTOsGzpLsFone}{1\xspace}
\newcommand{\XttClangTOsGzpBapRecall}{0.984\xspace}
\newcommand{\XttClangTOsGzpBapPrecision}{1\xspace}
\newcommand{\XttClangTOsGzpBapFone}{0.992\xspace}
\newcommand{\XttClangTOsGzpGhiRecall}{0.999\xspace}
\newcommand{\XttClangTOsGzpGhiPrecision}{1\xspace}
\newcommand{\XttClangTOsGzpGhiFone}{1.000\xspace}
\newcommand{\XttClangTOsGzpRdaRecall}{1\xspace}
\newcommand{\XttClangTOsGzpRdaPrecision}{1\xspace}
\newcommand{\XttClangTOsGzpRdaFone}{1\xspace}
\newcommand{\XttClangTOsGzpRseRecall}{0.995\xspace}
\newcommand{\XttClangTOsGzpRsePrecision}{1\xspace}
\newcommand{\XttClangTOsGzpRseFone}{0.998\xspace}
\newcommand{\XttClangTOsOggGT}{39283\xspace}
\newcommand{\XttClangTOsOggLsRecall}{1\xspace}
\newcommand{\XttClangTOsOggLsPrecision}{1\xspace}
\newcommand{\XttClangTOsOggLsFone}{1\xspace}
\newcommand{\XttClangTOsOggBapRecall}{0.982\xspace}
\newcommand{\XttClangTOsOggBapPrecision}{1\xspace}
\newcommand{\XttClangTOsOggBapFone}{0.991\xspace}
\newcommand{\XttClangTOsOggGhiRecall}{0.999\xspace}
\newcommand{\XttClangTOsOggGhiPrecision}{1\xspace}
\newcommand{\XttClangTOsOggGhiFone}{0.999\xspace}
\newcommand{\XttClangTOsOggRdaRecall}{0.999\xspace}
\newcommand{\XttClangTOsOggRdaPrecision}{1\xspace}
\newcommand{\XttClangTOsOggRdaFone}{0.999\xspace}
\newcommand{\XttClangTOsOggRseRecall}{1.000\xspace}
\newcommand{\XttClangTOsOggRsePrecision}{1\xspace}
\newcommand{\XttClangTOsOggRseFone}{1.000\xspace}
\newcommand{\XttClangTOsNgxGT}{107868\xspace}
\newcommand{\XttClangTOsNgxLsRecall}{1\xspace}
\newcommand{\XttClangTOsNgxLsPrecision}{1\xspace}
\newcommand{\XttClangTOsNgxLsFone}{1\xspace}
\newcommand{\XttClangTOsNgxBapRecall}{0.972\xspace}
\newcommand{\XttClangTOsNgxBapPrecision}{1\xspace}
\newcommand{\XttClangTOsNgxBapFone}{0.986\xspace}
\newcommand{\XttClangTOsNgxGhiRecall}{0.999\xspace}
\newcommand{\XttClangTOsNgxGhiPrecision}{1\xspace}
\newcommand{\XttClangTOsNgxGhiFone}{0.999\xspace}
\newcommand{\XttClangTOsNgxRdaRecall}{0.997\xspace}
\newcommand{\XttClangTOsNgxRdaPrecision}{1\xspace}
\newcommand{\XttClangTOsNgxRdaFone}{0.999\xspace}
\newcommand{\XttClangTOsNgxRseRecall}{1.000\xspace}
\newcommand{\XttClangTOsNgxRsePrecision}{1\xspace}
\newcommand{\XttClangTOsNgxRseFone}{1.000\xspace}
\newcommand{\XttClangTOsSshGT}{124133\xspace}
\newcommand{\XttClangTOsSshLsRecall}{1\xspace}
\newcommand{\XttClangTOsSshLsPrecision}{1\xspace}
\newcommand{\XttClangTOsSshLsFone}{1\xspace}
\newcommand{\XttClangTOsSshBapRecall}{0.950\xspace}
\newcommand{\XttClangTOsSshBapPrecision}{1\xspace}
\newcommand{\XttClangTOsSshBapFone}{0.974\xspace}
\newcommand{\XttClangTOsSshGhiRecall}{0.997\xspace}
\newcommand{\XttClangTOsSshGhiPrecision}{1\xspace}
\newcommand{\XttClangTOsSshGhiFone}{0.998\xspace}
\newcommand{\XttClangTOsSshRdaRecall}{0.943\xspace}
\newcommand{\XttClangTOsSshRdaPrecision}{1\xspace}
\newcommand{\XttClangTOsSshRdaFone}{0.971\xspace}
\newcommand{\XttClangTOsSshRseRecall}{0.977\xspace}
\newcommand{\XttClangTOsSshRsePrecision}{1\xspace}
\newcommand{\XttClangTOsSshRseFone}{0.988\xspace}
\newcommand{\XttClangTOsPcrGT}{4677\xspace}
\newcommand{\XttClangTOsPcrLsRecall}{1\xspace}
\newcommand{\XttClangTOsPcrLsPrecision}{1\xspace}
\newcommand{\XttClangTOsPcrLsFone}{1\xspace}
\newcommand{\XttClangTOsPcrBapRecall}{0.877\xspace}
\newcommand{\XttClangTOsPcrBapPrecision}{1\xspace}
\newcommand{\XttClangTOsPcrBapFone}{0.935\xspace}
\newcommand{\XttClangTOsPcrGhiRecall}{1.000\xspace}
\newcommand{\XttClangTOsPcrGhiPrecision}{1\xspace}
\newcommand{\XttClangTOsPcrGhiFone}{1.000\xspace}
\newcommand{\XttClangTOsPcrRdaRecall}{0.967\xspace}
\newcommand{\XttClangTOsPcrRdaPrecision}{1\xspace}
\newcommand{\XttClangTOsPcrRdaFone}{0.983\xspace}
\newcommand{\XttClangTOsPcrRseRecall}{1.000\xspace}
\newcommand{\XttClangTOsPcrRsePrecision}{1\xspace}
\newcommand{\XttClangTOsPcrRseFone}{1.000\xspace}
\newcommand{\XttClangTOsSqlGT}{186599\xspace}
\newcommand{\XttClangTOsSqlLsRecall}{1\xspace}
\newcommand{\XttClangTOsSqlLsPrecision}{1\xspace}
\newcommand{\XttClangTOsSqlLsFone}{1\xspace}
\newcommand{\XttClangTOsSqlBapRecall}{0.841\xspace}
\newcommand{\XttClangTOsSqlBapPrecision}{1\xspace}
\newcommand{\XttClangTOsSqlBapFone}{0.914\xspace}
\newcommand{\XttClangTOsSqlGhiRecall}{0.945\xspace}
\newcommand{\XttClangTOsSqlGhiPrecision}{1\xspace}
\newcommand{\XttClangTOsSqlGhiFone}{0.971\xspace}
\newcommand{\XttClangTOsSqlRdaRecall}{0.996\xspace}
\newcommand{\XttClangTOsSqlRdaPrecision}{1\xspace}
\newcommand{\XttClangTOsSqlRdaFone}{0.998\xspace}
\newcommand{\XttClangTOsSqlRseRecall}{0.968\xspace}
\newcommand{\XttClangTOsSqlRsePrecision}{1\xspace}
\newcommand{\XttClangTOsSqlRseFone}{0.984\xspace}
\newcommand{\XttClangTOsVimGT}{496912\xspace}
\newcommand{\XttClangTOsVimLsRecall}{1\xspace}
\newcommand{\XttClangTOsVimLsPrecision}{1\xspace}
\newcommand{\XttClangTOsVimLsFone}{1\xspace}
\newcommand{\XttClangTOsVimBapRecall}{0.909\xspace}
\newcommand{\XttClangTOsVimBapPrecision}{1\xspace}
\newcommand{\XttClangTOsVimBapFone}{0.952\xspace}
\newcommand{\XttClangTOsVimGhiRecall}{1.000\xspace}
\newcommand{\XttClangTOsVimGhiPrecision}{1\xspace}
\newcommand{\XttClangTOsVimGhiFone}{1.000\xspace}
\newcommand{\XttClangTOsVimRdaRecall}{0.985\xspace}
\newcommand{\XttClangTOsVimRdaPrecision}{1\xspace}
\newcommand{\XttClangTOsVimRdaFone}{0.993\xspace}
\newcommand{\XttClangTOsVimRseRecall}{0.999\xspace}
\newcommand{\XttClangTOsVimRsePrecision}{1.000\xspace}
\newcommand{\XttClangTOsVimRseFone}{1.000\xspace}
\newcommand{\XttClangTOsVsfGT}{24046\xspace}
\newcommand{\XttClangTOsVsfLsRecall}{1\xspace}
\newcommand{\XttClangTOsVsfLsPrecision}{1\xspace}
\newcommand{\XttClangTOsVsfLsFone}{1\xspace}
\newcommand{\XttClangTOsVsfBapRecall}{0.990\xspace}
\newcommand{\XttClangTOsVsfBapPrecision}{1\xspace}
\newcommand{\XttClangTOsVsfBapFone}{0.995\xspace}
\newcommand{\XttClangTOsVsfGhiRecall}{1.000\xspace}
\newcommand{\XttClangTOsVsfGhiPrecision}{1\xspace}
\newcommand{\XttClangTOsVsfGhiFone}{1.000\xspace}
\newcommand{\XttClangTOsVsfRdaRecall}{0.983\xspace}
\newcommand{\XttClangTOsVsfRdaPrecision}{1\xspace}
\newcommand{\XttClangTOsVsfRdaFone}{0.992\xspace}
\newcommand{\XttClangTOsVsfRseRecall}{1.000\xspace}
\newcommand{\XttClangTOsVsfRsePrecision}{1\xspace}
\newcommand{\XttClangTOsVsfRseFone}{1.000\xspace}
\newcommand{\XttClangSOoSzpGT}{22207\xspace}
\newcommand{\XttClangSOoSzpLsRecall}{1\xspace}
\newcommand{\XttClangSOoSzpLsPrecision}{1\xspace}
\newcommand{\XttClangSOoSzpLsFone}{1\xspace}
\newcommand{\XttClangSOoSzpBapRecall}{0.980\xspace}
\newcommand{\XttClangSOoSzpBapPrecision}{1\xspace}
\newcommand{\XttClangSOoSzpBapFone}{0.990\xspace}
\newcommand{\XttClangSOoSzpGhiRecall}{1\xspace}
\newcommand{\XttClangSOoSzpGhiPrecision}{1\xspace}
\newcommand{\XttClangSOoSzpGhiFone}{1\xspace}
\newcommand{\XttClangSOoSzpRdaRecall}{0.980\xspace}
\newcommand{\XttClangSOoSzpRdaPrecision}{1\xspace}
\newcommand{\XttClangSOoSzpRdaFone}{0.990\xspace}
\newcommand{\XttClangSOoSzpRseRecall}{1\xspace}
\newcommand{\XttClangSOoSzpRsePrecision}{1\xspace}
\newcommand{\XttClangSOoSzpRseFone}{1\xspace}
\newcommand{\XttClangSOoCapGT}{417367\xspace}
\newcommand{\XttClangSOoCapLsRecall}{1\xspace}
\newcommand{\XttClangSOoCapLsPrecision}{1\xspace}
\newcommand{\XttClangSOoCapLsFone}{1\xspace}
\newcommand{\XttClangSOoCapBapRecall}{0.446\xspace}
\newcommand{\XttClangSOoCapBapPrecision}{1\xspace}
\newcommand{\XttClangSOoCapBapFone}{0.617\xspace}
\newcommand{\XttClangSOoCapGhiRecall}{1\xspace}
\newcommand{\XttClangSOoCapGhiPrecision}{1\xspace}
\newcommand{\XttClangSOoCapGhiFone}{1\xspace}
\newcommand{\XttClangSOoCapRdaRecall}{0.377\xspace}
\newcommand{\XttClangSOoCapRdaPrecision}{1\xspace}
\newcommand{\XttClangSOoCapRdaFone}{0.547\xspace}
\newcommand{\XttClangSOoCapRseRecall}{1\xspace}
\newcommand{\XttClangSOoCapRsePrecision}{1\xspace}
\newcommand{\XttClangSOoCapRseFone}{1\xspace}
\newcommand{\XttClangSOoExmGT}{204189\xspace}
\newcommand{\XttClangSOoExmLsRecall}{1\xspace}
\newcommand{\XttClangSOoExmLsPrecision}{1\xspace}
\newcommand{\XttClangSOoExmLsFone}{1\xspace}
\newcommand{\XttClangSOoExmBapRecall}{0.852\xspace}
\newcommand{\XttClangSOoExmBapPrecision}{1\xspace}
\newcommand{\XttClangSOoExmBapFone}{0.920\xspace}
\newcommand{\XttClangSOoExmGhiRecall}{0.985\xspace}
\newcommand{\XttClangSOoExmGhiPrecision}{1\xspace}
\newcommand{\XttClangSOoExmGhiFone}{0.992\xspace}
\newcommand{\XttClangSOoExmRdaRecall}{0.815\xspace}
\newcommand{\XttClangSOoExmRdaPrecision}{1\xspace}
\newcommand{\XttClangSOoExmRdaFone}{0.898\xspace}
\newcommand{\XttClangSOoExmRseRecall}{1.000\xspace}
\newcommand{\XttClangSOoExmRsePrecision}{1\xspace}
\newcommand{\XttClangSOoExmRseFone}{1.000\xspace}
\newcommand{\XttClangSOoLgtGT}{46904\xspace}
\newcommand{\XttClangSOoLgtLsRecall}{1\xspace}
\newcommand{\XttClangSOoLgtLsPrecision}{1\xspace}
\newcommand{\XttClangSOoLgtLsFone}{1\xspace}
\newcommand{\XttClangSOoLgtBapRecall}{0.876\xspace}
\newcommand{\XttClangSOoLgtBapPrecision}{1\xspace}
\newcommand{\XttClangSOoLgtBapFone}{0.934\xspace}
\newcommand{\XttClangSOoLgtGhiRecall}{1\xspace}
\newcommand{\XttClangSOoLgtGhiPrecision}{1\xspace}
\newcommand{\XttClangSOoLgtGhiFone}{1\xspace}
\newcommand{\XttClangSOoLgtRdaRecall}{0.876\xspace}
\newcommand{\XttClangSOoLgtRdaPrecision}{1\xspace}
\newcommand{\XttClangSOoLgtRdaFone}{0.934\xspace}
\newcommand{\XttClangSOoLgtRseRecall}{1\xspace}
\newcommand{\XttClangSOoLgtRsePrecision}{1\xspace}
\newcommand{\XttClangSOoLgtRseFone}{1\xspace}
\newcommand{\XttClangSOoBzpGT}{22074\xspace}
\newcommand{\XttClangSOoBzpLsRecall}{1\xspace}
\newcommand{\XttClangSOoBzpLsPrecision}{1\xspace}
\newcommand{\XttClangSOoBzpLsFone}{1\xspace}
\newcommand{\XttClangSOoBzpBapRecall}{0.793\xspace}
\newcommand{\XttClangSOoBzpBapPrecision}{1\xspace}
\newcommand{\XttClangSOoBzpBapFone}{0.884\xspace}
\newcommand{\XttClangSOoBzpGhiRecall}{0.980\xspace}
\newcommand{\XttClangSOoBzpGhiPrecision}{1\xspace}
\newcommand{\XttClangSOoBzpGhiFone}{0.990\xspace}
\newcommand{\XttClangSOoBzpRdaRecall}{0.785\xspace}
\newcommand{\XttClangSOoBzpRdaPrecision}{1\xspace}
\newcommand{\XttClangSOoBzpRdaFone}{0.880\xspace}
\newcommand{\XttClangSOoBzpRseRecall}{1\xspace}
\newcommand{\XttClangSOoBzpRsePrecision}{1\xspace}
\newcommand{\XttClangSOoBzpRseFone}{1\xspace}
\newcommand{\XttClangSOoGccGT}{1450359\xspace}
\newcommand{\XttClangSOoGccLsRecall}{1\xspace}
\newcommand{\XttClangSOoGccLsPrecision}{1\xspace}
\newcommand{\XttClangSOoGccLsFone}{1\xspace}
\newcommand{\XttClangSOoGccBapRecall}{0.721\xspace}
\newcommand{\XttClangSOoGccBapPrecision}{1\xspace}
\newcommand{\XttClangSOoGccBapFone}{0.838\xspace}
\newcommand{\XttClangSOoGccGhiRecall}{0.994\xspace}
\newcommand{\XttClangSOoGccGhiPrecision}{1\xspace}
\newcommand{\XttClangSOoGccGhiFone}{0.997\xspace}
\newcommand{\XttClangSOoGccRdaRecall}{0.649\xspace}
\newcommand{\XttClangSOoGccRdaPrecision}{1\xspace}
\newcommand{\XttClangSOoGccRdaFone}{0.787\xspace}
\newcommand{\XttClangSOoGccRseRecall}{1\xspace}
\newcommand{\XttClangSOoGccRsePrecision}{1\xspace}
\newcommand{\XttClangSOoGccRseFone}{1\xspace}
\newcommand{\XttClangSOoGzpGT}{13869\xspace}
\newcommand{\XttClangSOoGzpLsRecall}{1\xspace}
\newcommand{\XttClangSOoGzpLsPrecision}{1\xspace}
\newcommand{\XttClangSOoGzpLsFone}{1\xspace}
\newcommand{\XttClangSOoGzpBapRecall}{0.991\xspace}
\newcommand{\XttClangSOoGzpBapPrecision}{1\xspace}
\newcommand{\XttClangSOoGzpBapFone}{0.996\xspace}
\newcommand{\XttClangSOoGzpGhiRecall}{1\xspace}
\newcommand{\XttClangSOoGzpGhiPrecision}{1\xspace}
\newcommand{\XttClangSOoGzpGhiFone}{1\xspace}
\newcommand{\XttClangSOoGzpRdaRecall}{0.991\xspace}
\newcommand{\XttClangSOoGzpRdaPrecision}{1\xspace}
\newcommand{\XttClangSOoGzpRdaFone}{0.996\xspace}
\newcommand{\XttClangSOoGzpRseRecall}{0.998\xspace}
\newcommand{\XttClangSOoGzpRsePrecision}{1\xspace}
\newcommand{\XttClangSOoGzpRseFone}{0.999\xspace}
\newcommand{\XttClangSOoOggGT}{57937\xspace}
\newcommand{\XttClangSOoOggLsRecall}{1\xspace}
\newcommand{\XttClangSOoOggLsPrecision}{1\xspace}
\newcommand{\XttClangSOoOggLsFone}{1\xspace}
\newcommand{\XttClangSOoOggBapRecall}{0.978\xspace}
\newcommand{\XttClangSOoOggBapPrecision}{1\xspace}
\newcommand{\XttClangSOoOggBapFone}{0.989\xspace}
\newcommand{\XttClangSOoOggGhiRecall}{1\xspace}
\newcommand{\XttClangSOoOggGhiPrecision}{1\xspace}
\newcommand{\XttClangSOoOggGhiFone}{1\xspace}
\newcommand{\XttClangSOoOggRdaRecall}{0.978\xspace}
\newcommand{\XttClangSOoOggRdaPrecision}{1\xspace}
\newcommand{\XttClangSOoOggRdaFone}{0.989\xspace}
\newcommand{\XttClangSOoOggRseRecall}{1\xspace}
\newcommand{\XttClangSOoOggRsePrecision}{1\xspace}
\newcommand{\XttClangSOoOggRseFone}{1\xspace}
\newcommand{\XttClangSOoNgxGT}{179476\xspace}
\newcommand{\XttClangSOoNgxLsRecall}{1\xspace}
\newcommand{\XttClangSOoNgxLsPrecision}{1\xspace}
\newcommand{\XttClangSOoNgxLsFone}{1\xspace}
\newcommand{\XttClangSOoNgxBapRecall}{0.967\xspace}
\newcommand{\XttClangSOoNgxBapPrecision}{1\xspace}
\newcommand{\XttClangSOoNgxBapFone}{0.983\xspace}
\newcommand{\XttClangSOoNgxGhiRecall}{1.000\xspace}
\newcommand{\XttClangSOoNgxGhiPrecision}{1\xspace}
\newcommand{\XttClangSOoNgxGhiFone}{1.000\xspace}
\newcommand{\XttClangSOoNgxRdaRecall}{0.966\xspace}
\newcommand{\XttClangSOoNgxRdaPrecision}{1\xspace}
\newcommand{\XttClangSOoNgxRdaFone}{0.982\xspace}
\newcommand{\XttClangSOoNgxRseRecall}{1\xspace}
\newcommand{\XttClangSOoNgxRsePrecision}{1\xspace}
\newcommand{\XttClangSOoNgxRseFone}{1\xspace}
\newcommand{\XttClangSOoSshGT}{197106\xspace}
\newcommand{\XttClangSOoSshLsRecall}{1\xspace}
\newcommand{\XttClangSOoSshLsPrecision}{1\xspace}
\newcommand{\XttClangSOoSshLsFone}{1\xspace}
\newcommand{\XttClangSOoSshBapRecall}{0.952\xspace}
\newcommand{\XttClangSOoSshBapPrecision}{1\xspace}
\newcommand{\XttClangSOoSshBapFone}{0.975\xspace}
\newcommand{\XttClangSOoSshGhiRecall}{0.955\xspace}
\newcommand{\XttClangSOoSshGhiPrecision}{1\xspace}
\newcommand{\XttClangSOoSshGhiFone}{0.977\xspace}
\newcommand{\XttClangSOoSshRdaRecall}{0.937\xspace}
\newcommand{\XttClangSOoSshRdaPrecision}{1.000\xspace}
\newcommand{\XttClangSOoSshRdaFone}{0.968\xspace}
\newcommand{\XttClangSOoSshRseRecall}{0.982\xspace}
\newcommand{\XttClangSOoSshRsePrecision}{1\xspace}
\newcommand{\XttClangSOoSshRseFone}{0.991\xspace}
\newcommand{\XttClangSOoPcrGT}{6827\xspace}
\newcommand{\XttClangSOoPcrLsRecall}{1\xspace}
\newcommand{\XttClangSOoPcrLsPrecision}{1\xspace}
\newcommand{\XttClangSOoPcrLsFone}{1\xspace}
\newcommand{\XttClangSOoPcrBapRecall}{0.902\xspace}
\newcommand{\XttClangSOoPcrBapPrecision}{1\xspace}
\newcommand{\XttClangSOoPcrBapFone}{0.949\xspace}
\newcommand{\XttClangSOoPcrGhiRecall}{1\xspace}
\newcommand{\XttClangSOoPcrGhiPrecision}{1\xspace}
\newcommand{\XttClangSOoPcrGhiFone}{1\xspace}
\newcommand{\XttClangSOoPcrRdaRecall}{0.902\xspace}
\newcommand{\XttClangSOoPcrRdaPrecision}{1\xspace}
\newcommand{\XttClangSOoPcrRdaFone}{0.949\xspace}
\newcommand{\XttClangSOoPcrRseRecall}{1\xspace}
\newcommand{\XttClangSOoPcrRsePrecision}{1\xspace}
\newcommand{\XttClangSOoPcrRseFone}{1\xspace}
\newcommand{\XttClangSOoSqlGT}{261123\xspace}
\newcommand{\XttClangSOoSqlLsRecall}{1\xspace}
\newcommand{\XttClangSOoSqlLsPrecision}{1\xspace}
\newcommand{\XttClangSOoSqlLsFone}{1\xspace}
\newcommand{\XttClangSOoSqlBapRecall}{0.883\xspace}
\newcommand{\XttClangSOoSqlBapPrecision}{1\xspace}
\newcommand{\XttClangSOoSqlBapFone}{0.938\xspace}
\newcommand{\XttClangSOoSqlGhiRecall}{0.999\xspace}
\newcommand{\XttClangSOoSqlGhiPrecision}{1\xspace}
\newcommand{\XttClangSOoSqlGhiFone}{1.000\xspace}
\newcommand{\XttClangSOoSqlRdaRecall}{0.877\xspace}
\newcommand{\XttClangSOoSqlRdaPrecision}{1\xspace}
\newcommand{\XttClangSOoSqlRdaFone}{0.935\xspace}
\newcommand{\XttClangSOoSqlRseRecall}{1\xspace}
\newcommand{\XttClangSOoSqlRsePrecision}{1\xspace}
\newcommand{\XttClangSOoSqlRseFone}{1\xspace}
\newcommand{\XttClangSOoVimGT}{704704\xspace}
\newcommand{\XttClangSOoVimLsRecall}{1\xspace}
\newcommand{\XttClangSOoVimLsPrecision}{1\xspace}
\newcommand{\XttClangSOoVimLsFone}{1\xspace}
\newcommand{\XttClangSOoVimBapRecall}{0.948\xspace}
\newcommand{\XttClangSOoVimBapPrecision}{1\xspace}
\newcommand{\XttClangSOoVimBapFone}{0.973\xspace}
\newcommand{\XttClangSOoVimGhiRecall}{1.000\xspace}
\newcommand{\XttClangSOoVimGhiPrecision}{1\xspace}
\newcommand{\XttClangSOoVimGhiFone}{1.000\xspace}
\newcommand{\XttClangSOoVimRdaRecall}{0.915\xspace}
\newcommand{\XttClangSOoVimRdaPrecision}{1\xspace}
\newcommand{\XttClangSOoVimRdaFone}{0.956\xspace}
\newcommand{\XttClangSOoVimRseRecall}{1.000\xspace}
\newcommand{\XttClangSOoVimRsePrecision}{1\xspace}
\newcommand{\XttClangSOoVimRseFone}{1.000\xspace}
\newcommand{\XttClangSOoVsfGT}{38298\xspace}
\newcommand{\XttClangSOoVsfLsRecall}{1\xspace}
\newcommand{\XttClangSOoVsfLsPrecision}{1\xspace}
\newcommand{\XttClangSOoVsfLsFone}{1\xspace}
\newcommand{\XttClangSOoVsfBapRecall}{0.993\xspace}
\newcommand{\XttClangSOoVsfBapPrecision}{1\xspace}
\newcommand{\XttClangSOoVsfBapFone}{0.997\xspace}
\newcommand{\XttClangSOoVsfGhiRecall}{0.994\xspace}
\newcommand{\XttClangSOoVsfGhiPrecision}{1\xspace}
\newcommand{\XttClangSOoVsfGhiFone}{0.997\xspace}
\newcommand{\XttClangSOoVsfRdaRecall}{0.943\xspace}
\newcommand{\XttClangSOoVsfRdaPrecision}{1\xspace}
\newcommand{\XttClangSOoVsfRdaFone}{0.971\xspace}
\newcommand{\XttClangSOoVsfRseRecall}{1.000\xspace}
\newcommand{\XttClangSOoVsfRsePrecision}{1\xspace}
\newcommand{\XttClangSOoVsfRseFone}{1.000\xspace}
\newcommand{\XttClangSOaSzpGT}{13584\xspace}
\newcommand{\XttClangSOaSzpLsRecall}{1\xspace}
\newcommand{\XttClangSOaSzpLsPrecision}{1\xspace}
\newcommand{\XttClangSOaSzpLsFone}{1\xspace}
\newcommand{\XttClangSOaSzpBapRecall}{0.980\xspace}
\newcommand{\XttClangSOaSzpBapPrecision}{1\xspace}
\newcommand{\XttClangSOaSzpBapFone}{0.990\xspace}
\newcommand{\XttClangSOaSzpGhiRecall}{1\xspace}
\newcommand{\XttClangSOaSzpGhiPrecision}{1\xspace}
\newcommand{\XttClangSOaSzpGhiFone}{1\xspace}
\newcommand{\XttClangSOaSzpRdaRecall}{1\xspace}
\newcommand{\XttClangSOaSzpRdaPrecision}{1\xspace}
\newcommand{\XttClangSOaSzpRdaFone}{1\xspace}
\newcommand{\XttClangSOaSzpRseRecall}{1\xspace}
\newcommand{\XttClangSOaSzpRsePrecision}{1\xspace}
\newcommand{\XttClangSOaSzpRseFone}{1\xspace}
\newcommand{\XttClangSOaCapGT}{193981\xspace}
\newcommand{\XttClangSOaCapLsRecall}{1\xspace}
\newcommand{\XttClangSOaCapLsPrecision}{1\xspace}
\newcommand{\XttClangSOaCapLsFone}{1\xspace}
\newcommand{\XttClangSOaCapBapRecall}{0.326\xspace}
\newcommand{\XttClangSOaCapBapPrecision}{1\xspace}
\newcommand{\XttClangSOaCapBapFone}{0.492\xspace}
\newcommand{\XttClangSOaCapGhiRecall}{1.000\xspace}
\newcommand{\XttClangSOaCapGhiPrecision}{1\xspace}
\newcommand{\XttClangSOaCapGhiFone}{1.000\xspace}
\newcommand{\XttClangSOaCapRdaRecall}{0.857\xspace}
\newcommand{\XttClangSOaCapRdaPrecision}{1\xspace}
\newcommand{\XttClangSOaCapRdaFone}{0.923\xspace}
\newcommand{\XttClangSOaCapRseRecall}{1.000\xspace}
\newcommand{\XttClangSOaCapRsePrecision}{1\xspace}
\newcommand{\XttClangSOaCapRseFone}{1.000\xspace}
\newcommand{\XttClangSOaExmGT}{144944\xspace}
\newcommand{\XttClangSOaExmLsRecall}{1\xspace}
\newcommand{\XttClangSOaExmLsPrecision}{1\xspace}
\newcommand{\XttClangSOaExmLsFone}{1\xspace}
\newcommand{\XttClangSOaExmBapRecall}{0.850\xspace}
\newcommand{\XttClangSOaExmBapPrecision}{1\xspace}
\newcommand{\XttClangSOaExmBapFone}{0.919\xspace}
\newcommand{\XttClangSOaExmGhiRecall}{0.984\xspace}
\newcommand{\XttClangSOaExmGhiPrecision}{1\xspace}
\newcommand{\XttClangSOaExmGhiFone}{0.992\xspace}
\newcommand{\XttClangSOaExmRdaRecall}{0.965\xspace}
\newcommand{\XttClangSOaExmRdaPrecision}{1\xspace}
\newcommand{\XttClangSOaExmRdaFone}{0.982\xspace}
\newcommand{\XttClangSOaExmRseRecall}{0.999\xspace}
\newcommand{\XttClangSOaExmRsePrecision}{1\xspace}
\newcommand{\XttClangSOaExmRseFone}{1.000\xspace}
\newcommand{\XttClangSOaLgtGT}{28237\xspace}
\newcommand{\XttClangSOaLgtLsRecall}{1\xspace}
\newcommand{\XttClangSOaLgtLsPrecision}{1\xspace}
\newcommand{\XttClangSOaLgtLsFone}{1\xspace}
\newcommand{\XttClangSOaLgtBapRecall}{0.881\xspace}
\newcommand{\XttClangSOaLgtBapPrecision}{1\xspace}
\newcommand{\XttClangSOaLgtBapFone}{0.937\xspace}
\newcommand{\XttClangSOaLgtGhiRecall}{0.995\xspace}
\newcommand{\XttClangSOaLgtGhiPrecision}{1\xspace}
\newcommand{\XttClangSOaLgtGhiFone}{0.997\xspace}
\newcommand{\XttClangSOaLgtRdaRecall}{1.000\xspace}
\newcommand{\XttClangSOaLgtRdaPrecision}{1\xspace}
\newcommand{\XttClangSOaLgtRdaFone}{1.000\xspace}
\newcommand{\XttClangSOaLgtRseRecall}{0.997\xspace}
\newcommand{\XttClangSOaLgtRsePrecision}{1\xspace}
\newcommand{\XttClangSOaLgtRseFone}{0.998\xspace}
\newcommand{\XttClangSOaBzpGT}{12469\xspace}
\newcommand{\XttClangSOaBzpLsRecall}{1\xspace}
\newcommand{\XttClangSOaBzpLsPrecision}{1\xspace}
\newcommand{\XttClangSOaBzpLsFone}{1\xspace}
\newcommand{\XttClangSOaBzpBapRecall}{0.917\xspace}
\newcommand{\XttClangSOaBzpBapPrecision}{1\xspace}
\newcommand{\XttClangSOaBzpBapFone}{0.956\xspace}
\newcommand{\XttClangSOaBzpGhiRecall}{0.998\xspace}
\newcommand{\XttClangSOaBzpGhiPrecision}{1\xspace}
\newcommand{\XttClangSOaBzpGhiFone}{0.999\xspace}
\newcommand{\XttClangSOaBzpRdaRecall}{0.968\xspace}
\newcommand{\XttClangSOaBzpRdaPrecision}{1\xspace}
\newcommand{\XttClangSOaBzpRdaFone}{0.984\xspace}
\newcommand{\XttClangSOaBzpRseRecall}{0.999\xspace}
\newcommand{\XttClangSOaBzpRsePrecision}{1\xspace}
\newcommand{\XttClangSOaBzpRseFone}{1.000\xspace}
\newcommand{\XttClangSOaGccGT}{844427\xspace}
\newcommand{\XttClangSOaGccLsRecall}{1\xspace}
\newcommand{\XttClangSOaGccLsPrecision}{1\xspace}
\newcommand{\XttClangSOaGccLsFone}{1\xspace}
\newcommand{\XttClangSOaGccBapRecall}{0.726\xspace}
\newcommand{\XttClangSOaGccBapPrecision}{1.000\xspace}
\newcommand{\XttClangSOaGccBapFone}{0.841\xspace}
\newcommand{\XttClangSOaGccGhiRecall}{0.997\xspace}
\newcommand{\XttClangSOaGccGhiPrecision}{1\xspace}
\newcommand{\XttClangSOaGccGhiFone}{0.999\xspace}
\newcommand{\XttClangSOaGccRdaRecall}{0.852\xspace}
\newcommand{\XttClangSOaGccRdaPrecision}{1\xspace}
\newcommand{\XttClangSOaGccRdaFone}{0.920\xspace}
\newcommand{\XttClangSOaGccRseRecall}{0.999\xspace}
\newcommand{\XttClangSOaGccRsePrecision}{1\xspace}
\newcommand{\XttClangSOaGccRseFone}{1.000\xspace}
\newcommand{\XttClangSOaGzpGT}{9426\xspace}
\newcommand{\XttClangSOaGzpLsRecall}{1\xspace}
\newcommand{\XttClangSOaGzpLsPrecision}{1\xspace}
\newcommand{\XttClangSOaGzpLsFone}{1\xspace}
\newcommand{\XttClangSOaGzpBapRecall}{0.983\xspace}
\newcommand{\XttClangSOaGzpBapPrecision}{1\xspace}
\newcommand{\XttClangSOaGzpBapFone}{0.991\xspace}
\newcommand{\XttClangSOaGzpGhiRecall}{0.998\xspace}
\newcommand{\XttClangSOaGzpGhiPrecision}{1\xspace}
\newcommand{\XttClangSOaGzpGhiFone}{0.999\xspace}
\newcommand{\XttClangSOaGzpRdaRecall}{1.000\xspace}
\newcommand{\XttClangSOaGzpRdaPrecision}{1\xspace}
\newcommand{\XttClangSOaGzpRdaFone}{1.000\xspace}
\newcommand{\XttClangSOaGzpRseRecall}{0.997\xspace}
\newcommand{\XttClangSOaGzpRsePrecision}{1\xspace}
\newcommand{\XttClangSOaGzpRseFone}{0.998\xspace}
\newcommand{\XttClangSOaOggGT}{36395\xspace}
\newcommand{\XttClangSOaOggLsRecall}{1\xspace}
\newcommand{\XttClangSOaOggLsPrecision}{1\xspace}
\newcommand{\XttClangSOaOggLsFone}{1\xspace}
\newcommand{\XttClangSOaOggBapRecall}{0.974\xspace}
\newcommand{\XttClangSOaOggBapPrecision}{1\xspace}
\newcommand{\XttClangSOaOggBapFone}{0.987\xspace}
\newcommand{\XttClangSOaOggGhiRecall}{1.000\xspace}
\newcommand{\XttClangSOaOggGhiPrecision}{1\xspace}
\newcommand{\XttClangSOaOggGhiFone}{1.000\xspace}
\newcommand{\XttClangSOaOggRdaRecall}{1.000\xspace}
\newcommand{\XttClangSOaOggRdaPrecision}{1\xspace}
\newcommand{\XttClangSOaOggRdaFone}{1.000\xspace}
\newcommand{\XttClangSOaOggRseRecall}{1.000\xspace}
\newcommand{\XttClangSOaOggRsePrecision}{1\xspace}
\newcommand{\XttClangSOaOggRseFone}{1.000\xspace}
\newcommand{\XttClangSOaNgxGT}{105977\xspace}
\newcommand{\XttClangSOaNgxLsRecall}{1\xspace}
\newcommand{\XttClangSOaNgxLsPrecision}{1\xspace}
\newcommand{\XttClangSOaNgxLsFone}{1\xspace}
\newcommand{\XttClangSOaNgxBapRecall}{0.970\xspace}
\newcommand{\XttClangSOaNgxBapPrecision}{1\xspace}
\newcommand{\XttClangSOaNgxBapFone}{0.985\xspace}
\newcommand{\XttClangSOaNgxGhiRecall}{1.000\xspace}
\newcommand{\XttClangSOaNgxGhiPrecision}{1\xspace}
\newcommand{\XttClangSOaNgxGhiFone}{1.000\xspace}
\newcommand{\XttClangSOaNgxRdaRecall}{0.996\xspace}
\newcommand{\XttClangSOaNgxRdaPrecision}{1\xspace}
\newcommand{\XttClangSOaNgxRdaFone}{0.998\xspace}
\newcommand{\XttClangSOaNgxRseRecall}{1.000\xspace}
\newcommand{\XttClangSOaNgxRsePrecision}{1\xspace}
\newcommand{\XttClangSOaNgxRseFone}{1.000\xspace}
\newcommand{\XttClangSOaSshGT}{123295\xspace}
\newcommand{\XttClangSOaSshLsRecall}{1\xspace}
\newcommand{\XttClangSOaSshLsPrecision}{1\xspace}
\newcommand{\XttClangSOaSshLsFone}{1\xspace}
\newcommand{\XttClangSOaSshBapRecall}{0.953\xspace}
\newcommand{\XttClangSOaSshBapPrecision}{1.000\xspace}
\newcommand{\XttClangSOaSshBapFone}{0.976\xspace}
\newcommand{\XttClangSOaSshGhiRecall}{0.996\xspace}
\newcommand{\XttClangSOaSshGhiPrecision}{1\xspace}
\newcommand{\XttClangSOaSshGhiFone}{0.998\xspace}
\newcommand{\XttClangSOaSshRdaRecall}{0.932\xspace}
\newcommand{\XttClangSOaSshRdaPrecision}{1\xspace}
\newcommand{\XttClangSOaSshRdaFone}{0.965\xspace}
\newcommand{\XttClangSOaSshRseRecall}{0.985\xspace}
\newcommand{\XttClangSOaSshRsePrecision}{1\xspace}
\newcommand{\XttClangSOaSshRseFone}{0.993\xspace}
\newcommand{\XttClangSOaPcrGT}{4639\xspace}
\newcommand{\XttClangSOaPcrLsRecall}{1\xspace}
\newcommand{\XttClangSOaPcrLsPrecision}{1\xspace}
\newcommand{\XttClangSOaPcrLsFone}{1\xspace}
\newcommand{\XttClangSOaPcrBapRecall}{0.893\xspace}
\newcommand{\XttClangSOaPcrBapPrecision}{1\xspace}
\newcommand{\XttClangSOaPcrBapFone}{0.944\xspace}
\newcommand{\XttClangSOaPcrGhiRecall}{1.000\xspace}
\newcommand{\XttClangSOaPcrGhiPrecision}{1\xspace}
\newcommand{\XttClangSOaPcrGhiFone}{1.000\xspace}
\newcommand{\XttClangSOaPcrRdaRecall}{0.970\xspace}
\newcommand{\XttClangSOaPcrRdaPrecision}{1\xspace}
\newcommand{\XttClangSOaPcrRdaFone}{0.985\xspace}
\newcommand{\XttClangSOaPcrRseRecall}{1.000\xspace}
\newcommand{\XttClangSOaPcrRsePrecision}{1\xspace}
\newcommand{\XttClangSOaPcrRseFone}{1.000\xspace}
\newcommand{\XttClangSOaSqlGT}{166639\xspace}
\newcommand{\XttClangSOaSqlLsRecall}{1\xspace}
\newcommand{\XttClangSOaSqlLsPrecision}{1\xspace}
\newcommand{\XttClangSOaSqlLsFone}{1\xspace}
\newcommand{\XttClangSOaSqlBapRecall}{0.875\xspace}
\newcommand{\XttClangSOaSqlBapPrecision}{1\xspace}
\newcommand{\XttClangSOaSqlBapFone}{0.933\xspace}
\newcommand{\XttClangSOaSqlGhiRecall}{0.962\xspace}
\newcommand{\XttClangSOaSqlGhiPrecision}{1\xspace}
\newcommand{\XttClangSOaSqlGhiFone}{0.981\xspace}
\newcommand{\XttClangSOaSqlRdaRecall}{0.962\xspace}
\newcommand{\XttClangSOaSqlRdaPrecision}{1\xspace}
\newcommand{\XttClangSOaSqlRdaFone}{0.980\xspace}
\newcommand{\XttClangSOaSqlRseRecall}{0.999\xspace}
\newcommand{\XttClangSOaSqlRsePrecision}{1\xspace}
\newcommand{\XttClangSOaSqlRseFone}{1.000\xspace}
\newcommand{\XttClangSOaVimGT}{472940\xspace}
\newcommand{\XttClangSOaVimLsRecall}{1\xspace}
\newcommand{\XttClangSOaVimLsPrecision}{1\xspace}
\newcommand{\XttClangSOaVimLsFone}{1\xspace}
\newcommand{\XttClangSOaVimBapRecall}{0.919\xspace}
\newcommand{\XttClangSOaVimBapPrecision}{1\xspace}
\newcommand{\XttClangSOaVimBapFone}{0.958\xspace}
\newcommand{\XttClangSOaVimGhiRecall}{0.997\xspace}
\newcommand{\XttClangSOaVimGhiPrecision}{1\xspace}
\newcommand{\XttClangSOaVimGhiFone}{0.998\xspace}
\newcommand{\XttClangSOaVimRdaRecall}{0.987\xspace}
\newcommand{\XttClangSOaVimRdaPrecision}{1\xspace}
\newcommand{\XttClangSOaVimRdaFone}{0.993\xspace}
\newcommand{\XttClangSOaVimRseRecall}{0.998\xspace}
\newcommand{\XttClangSOaVimRsePrecision}{1\xspace}
\newcommand{\XttClangSOaVimRseFone}{0.999\xspace}
\newcommand{\XttClangSOaVsfGT}{24301\xspace}
\newcommand{\XttClangSOaVsfLsRecall}{1\xspace}
\newcommand{\XttClangSOaVsfLsPrecision}{1\xspace}
\newcommand{\XttClangSOaVsfLsFone}{1\xspace}
\newcommand{\XttClangSOaVsfBapRecall}{0.991\xspace}
\newcommand{\XttClangSOaVsfBapPrecision}{1\xspace}
\newcommand{\XttClangSOaVsfBapFone}{0.995\xspace}
\newcommand{\XttClangSOaVsfGhiRecall}{1.000\xspace}
\newcommand{\XttClangSOaVsfGhiPrecision}{1\xspace}
\newcommand{\XttClangSOaVsfGhiFone}{1.000\xspace}
\newcommand{\XttClangSOaVsfRdaRecall}{0.969\xspace}
\newcommand{\XttClangSOaVsfRdaPrecision}{1\xspace}
\newcommand{\XttClangSOaVsfRdaFone}{0.984\xspace}
\newcommand{\XttClangSOaVsfRseRecall}{0.999\xspace}
\newcommand{\XttClangSOaVsfRsePrecision}{1\xspace}
\newcommand{\XttClangSOaVsfRseFone}{0.999\xspace}
\newcommand{\XttClangSObSzpGT}{15661\xspace}
\newcommand{\XttClangSObSzpLsRecall}{1\xspace}
\newcommand{\XttClangSObSzpLsPrecision}{1\xspace}
\newcommand{\XttClangSObSzpLsFone}{1\xspace}
\newcommand{\XttClangSObSzpBapRecall}{0.966\xspace}
\newcommand{\XttClangSObSzpBapPrecision}{1\xspace}
\newcommand{\XttClangSObSzpBapFone}{0.983\xspace}
\newcommand{\XttClangSObSzpGhiRecall}{1\xspace}
\newcommand{\XttClangSObSzpGhiPrecision}{1\xspace}
\newcommand{\XttClangSObSzpGhiFone}{1\xspace}
\newcommand{\XttClangSObSzpRdaRecall}{0.993\xspace}
\newcommand{\XttClangSObSzpRdaPrecision}{1\xspace}
\newcommand{\XttClangSObSzpRdaFone}{0.997\xspace}
\newcommand{\XttClangSObSzpRseRecall}{1\xspace}
\newcommand{\XttClangSObSzpRsePrecision}{1\xspace}
\newcommand{\XttClangSObSzpRseFone}{1\xspace}
\newcommand{\XttClangSObCapGT}{206294\xspace}
\newcommand{\XttClangSObCapLsRecall}{1\xspace}
\newcommand{\XttClangSObCapLsPrecision}{1\xspace}
\newcommand{\XttClangSObCapLsFone}{1\xspace}
\newcommand{\XttClangSObCapBapRecall}{0.276\xspace}
\newcommand{\XttClangSObCapBapPrecision}{1\xspace}
\newcommand{\XttClangSObCapBapFone}{0.433\xspace}
\newcommand{\XttClangSObCapGhiRecall}{0.864\xspace}
\newcommand{\XttClangSObCapGhiPrecision}{1\xspace}
\newcommand{\XttClangSObCapGhiFone}{0.927\xspace}
\newcommand{\XttClangSObCapRdaRecall}{0.861\xspace}
\newcommand{\XttClangSObCapRdaPrecision}{1\xspace}
\newcommand{\XttClangSObCapRdaFone}{0.925\xspace}
\newcommand{\XttClangSObCapRseRecall}{1\xspace}
\newcommand{\XttClangSObCapRsePrecision}{1\xspace}
\newcommand{\XttClangSObCapRseFone}{1\xspace}
\newcommand{\XttClangSObExmGT}{166527\xspace}
\newcommand{\XttClangSObExmLsRecall}{1\xspace}
\newcommand{\XttClangSObExmLsPrecision}{1\xspace}
\newcommand{\XttClangSObExmLsFone}{1\xspace}
\newcommand{\XttClangSObExmBapRecall}{0.821\xspace}
\newcommand{\XttClangSObExmBapPrecision}{1\xspace}
\newcommand{\XttClangSObExmBapFone}{0.902\xspace}
\newcommand{\XttClangSObExmGhiRecall}{0.962\xspace}
\newcommand{\XttClangSObExmGhiPrecision}{1\xspace}
\newcommand{\XttClangSObExmGhiFone}{0.981\xspace}
\newcommand{\XttClangSObExmRdaRecall}{0.957\xspace}
\newcommand{\XttClangSObExmRdaPrecision}{1\xspace}
\newcommand{\XttClangSObExmRdaFone}{0.978\xspace}
\newcommand{\XttClangSObExmRseRecall}{0.998\xspace}
\newcommand{\XttClangSObExmRsePrecision}{1\xspace}
\newcommand{\XttClangSObExmRseFone}{0.999\xspace}
\newcommand{\XttClangSObLgtGT}{30586\xspace}
\newcommand{\XttClangSObLgtLsRecall}{1\xspace}
\newcommand{\XttClangSObLgtLsPrecision}{1\xspace}
\newcommand{\XttClangSObLgtLsFone}{1\xspace}
\newcommand{\XttClangSObLgtBapRecall}{0.872\xspace}
\newcommand{\XttClangSObLgtBapPrecision}{1\xspace}
\newcommand{\XttClangSObLgtBapFone}{0.932\xspace}
\newcommand{\XttClangSObLgtGhiRecall}{0.993\xspace}
\newcommand{\XttClangSObLgtGhiPrecision}{1\xspace}
\newcommand{\XttClangSObLgtGhiFone}{0.997\xspace}
\newcommand{\XttClangSObLgtRdaRecall}{0.989\xspace}
\newcommand{\XttClangSObLgtRdaPrecision}{1\xspace}
\newcommand{\XttClangSObLgtRdaFone}{0.995\xspace}
\newcommand{\XttClangSObLgtRseRecall}{0.997\xspace}
\newcommand{\XttClangSObLgtRsePrecision}{1\xspace}
\newcommand{\XttClangSObLgtRseFone}{0.999\xspace}
\newcommand{\XttClangSObBzpGT}{18111\xspace}
\newcommand{\XttClangSObBzpLsRecall}{1\xspace}
\newcommand{\XttClangSObBzpLsPrecision}{1\xspace}
\newcommand{\XttClangSObBzpLsFone}{1\xspace}
\newcommand{\XttClangSObBzpBapRecall}{0.988\xspace}
\newcommand{\XttClangSObBzpBapPrecision}{1\xspace}
\newcommand{\XttClangSObBzpBapFone}{0.994\xspace}
\newcommand{\XttClangSObBzpGhiRecall}{0.998\xspace}
\newcommand{\XttClangSObBzpGhiPrecision}{1\xspace}
\newcommand{\XttClangSObBzpGhiFone}{0.999\xspace}
\newcommand{\XttClangSObBzpRdaRecall}{0.861\xspace}
\newcommand{\XttClangSObBzpRdaPrecision}{1\xspace}
\newcommand{\XttClangSObBzpRdaFone}{0.925\xspace}
\newcommand{\XttClangSObBzpRseRecall}{1.000\xspace}
\newcommand{\XttClangSObBzpRsePrecision}{1\xspace}
\newcommand{\XttClangSObBzpRseFone}{1.000\xspace}
\newcommand{\XttClangSObGccGT}{1211100\xspace}
\newcommand{\XttClangSObGccLsRecall}{1\xspace}
\newcommand{\XttClangSObGccLsPrecision}{1\xspace}
\newcommand{\XttClangSObGccLsFone}{1\xspace}
\newcommand{\XttClangSObGccBapRecall}{0.703\xspace}
\newcommand{\XttClangSObGccBapPrecision}{1\xspace}
\newcommand{\XttClangSObGccBapFone}{0.825\xspace}
\newcommand{\XttClangSObGccGhiRecall}{0.968\xspace}
\newcommand{\XttClangSObGccGhiPrecision}{1\xspace}
\newcommand{\XttClangSObGccGhiFone}{0.984\xspace}
\newcommand{\XttClangSObGccRdaRecall}{0.817\xspace}
\newcommand{\XttClangSObGccRdaPrecision}{1\xspace}
\newcommand{\XttClangSObGccRdaFone}{0.899\xspace}
\newcommand{\XttClangSObGccRseRecall}{0.999\xspace}
\newcommand{\XttClangSObGccRsePrecision}{1\xspace}
\newcommand{\XttClangSObGccRseFone}{0.999\xspace}
\newcommand{\XttClangSObGzpGT}{12988\xspace}
\newcommand{\XttClangSObGzpLsRecall}{1\xspace}
\newcommand{\XttClangSObGzpLsPrecision}{1\xspace}
\newcommand{\XttClangSObGzpLsFone}{1\xspace}
\newcommand{\XttClangSObGzpBapRecall}{0.987\xspace}
\newcommand{\XttClangSObGzpBapPrecision}{1\xspace}
\newcommand{\XttClangSObGzpBapFone}{0.994\xspace}
\newcommand{\XttClangSObGzpGhiRecall}{0.998\xspace}
\newcommand{\XttClangSObGzpGhiPrecision}{1\xspace}
\newcommand{\XttClangSObGzpGhiFone}{0.999\xspace}
\newcommand{\XttClangSObGzpRdaRecall}{1.000\xspace}
\newcommand{\XttClangSObGzpRdaPrecision}{1\xspace}
\newcommand{\XttClangSObGzpRdaFone}{1.000\xspace}
\newcommand{\XttClangSObGzpRseRecall}{0.998\xspace}
\newcommand{\XttClangSObGzpRsePrecision}{1\xspace}
\newcommand{\XttClangSObGzpRseFone}{0.999\xspace}
\newcommand{\XttClangSObOggGT}{56849\xspace}
\newcommand{\XttClangSObOggLsRecall}{1\xspace}
\newcommand{\XttClangSObOggLsPrecision}{1\xspace}
\newcommand{\XttClangSObOggLsFone}{1\xspace}
\newcommand{\XttClangSObOggBapRecall}{0.966\xspace}
\newcommand{\XttClangSObOggBapPrecision}{1\xspace}
\newcommand{\XttClangSObOggBapFone}{0.983\xspace}
\newcommand{\XttClangSObOggGhiRecall}{1.000\xspace}
\newcommand{\XttClangSObOggGhiPrecision}{1\xspace}
\newcommand{\XttClangSObOggGhiFone}{1.000\xspace}
\newcommand{\XttClangSObOggRdaRecall}{1.000\xspace}
\newcommand{\XttClangSObOggRdaPrecision}{1\xspace}
\newcommand{\XttClangSObOggRdaFone}{1.000\xspace}
\newcommand{\XttClangSObOggRseRecall}{1.000\xspace}
\newcommand{\XttClangSObOggRsePrecision}{1\xspace}
\newcommand{\XttClangSObOggRseFone}{1.000\xspace}
\newcommand{\XttClangSObNgxGT}{113469\xspace}
\newcommand{\XttClangSObNgxLsRecall}{1\xspace}
\newcommand{\XttClangSObNgxLsPrecision}{1\xspace}
\newcommand{\XttClangSObNgxLsFone}{1\xspace}
\newcommand{\XttClangSObNgxBapRecall}{0.970\xspace}
\newcommand{\XttClangSObNgxBapPrecision}{1\xspace}
\newcommand{\XttClangSObNgxBapFone}{0.985\xspace}
\newcommand{\XttClangSObNgxGhiRecall}{1.000\xspace}
\newcommand{\XttClangSObNgxGhiPrecision}{1\xspace}
\newcommand{\XttClangSObNgxGhiFone}{1.000\xspace}
\newcommand{\XttClangSObNgxRdaRecall}{0.996\xspace}
\newcommand{\XttClangSObNgxRdaPrecision}{1\xspace}
\newcommand{\XttClangSObNgxRdaFone}{0.998\xspace}
\newcommand{\XttClangSObNgxRseRecall}{1.000\xspace}
\newcommand{\XttClangSObNgxRsePrecision}{1\xspace}
\newcommand{\XttClangSObNgxRseFone}{1.000\xspace}
\newcommand{\XttClangSObSshGT}{141690\xspace}
\newcommand{\XttClangSObSshLsRecall}{1\xspace}
\newcommand{\XttClangSObSshLsPrecision}{1\xspace}
\newcommand{\XttClangSObSshLsFone}{1\xspace}
\newcommand{\XttClangSObSshBapRecall}{0.936\xspace}
\newcommand{\XttClangSObSshBapPrecision}{1\xspace}
\newcommand{\XttClangSObSshBapFone}{0.967\xspace}
\newcommand{\XttClangSObSshGhiRecall}{0.996\xspace}
\newcommand{\XttClangSObSshGhiPrecision}{1\xspace}
\newcommand{\XttClangSObSshGhiFone}{0.998\xspace}
\newcommand{\XttClangSObSshRdaRecall}{0.870\xspace}
\newcommand{\XttClangSObSshRdaPrecision}{1\xspace}
\newcommand{\XttClangSObSshRdaFone}{0.931\xspace}
\newcommand{\XttClangSObSshRseRecall}{0.983\xspace}
\newcommand{\XttClangSObSshRsePrecision}{1\xspace}
\newcommand{\XttClangSObSshRseFone}{0.991\xspace}
\newcommand{\XttClangSObPcrGT}{5419\xspace}
\newcommand{\XttClangSObPcrLsRecall}{1\xspace}
\newcommand{\XttClangSObPcrLsPrecision}{1\xspace}
\newcommand{\XttClangSObPcrLsFone}{1\xspace}
\newcommand{\XttClangSObPcrBapRecall}{0.861\xspace}
\newcommand{\XttClangSObPcrBapPrecision}{1\xspace}
\newcommand{\XttClangSObPcrBapFone}{0.925\xspace}
\newcommand{\XttClangSObPcrGhiRecall}{1.000\xspace}
\newcommand{\XttClangSObPcrGhiPrecision}{1\xspace}
\newcommand{\XttClangSObPcrGhiFone}{1.000\xspace}
\newcommand{\XttClangSObPcrRdaRecall}{1.000\xspace}
\newcommand{\XttClangSObPcrRdaPrecision}{1\xspace}
\newcommand{\XttClangSObPcrRdaFone}{1.000\xspace}
\newcommand{\XttClangSObPcrRseRecall}{1.000\xspace}
\newcommand{\XttClangSObPcrRsePrecision}{1\xspace}
\newcommand{\XttClangSObPcrRseFone}{1.000\xspace}
\newcommand{\XttClangSObSqlGT}{289121\xspace}
\newcommand{\XttClangSObSqlLsRecall}{1\xspace}
\newcommand{\XttClangSObSqlLsPrecision}{1\xspace}
\newcommand{\XttClangSObSqlLsFone}{1\xspace}
\newcommand{\XttClangSObSqlBapRecall}{0.810\xspace}
\newcommand{\XttClangSObSqlBapPrecision}{1\xspace}
\newcommand{\XttClangSObSqlBapFone}{0.895\xspace}
\newcommand{\XttClangSObSqlGhiRecall}{0.929\xspace}
\newcommand{\XttClangSObSqlGhiPrecision}{1\xspace}
\newcommand{\XttClangSObSqlGhiFone}{0.963\xspace}
\newcommand{\XttClangSObSqlRdaRecall}{0.907\xspace}
\newcommand{\XttClangSObSqlRdaPrecision}{1\xspace}
\newcommand{\XttClangSObSqlRdaFone}{0.951\xspace}
\newcommand{\XttClangSObSqlRseRecall}{0.965\xspace}
\newcommand{\XttClangSObSqlRsePrecision}{1.000\xspace}
\newcommand{\XttClangSObSqlRseFone}{0.982\xspace}
\newcommand{\XttClangSObVimGT}{580720\xspace}
\newcommand{\XttClangSObVimLsRecall}{1\xspace}
\newcommand{\XttClangSObVimLsPrecision}{1\xspace}
\newcommand{\XttClangSObVimLsFone}{1\xspace}
\newcommand{\XttClangSObVimBapRecall}{0.853\xspace}
\newcommand{\XttClangSObVimBapPrecision}{1\xspace}
\newcommand{\XttClangSObVimBapFone}{0.921\xspace}
\newcommand{\XttClangSObVimGhiRecall}{0.995\xspace}
\newcommand{\XttClangSObVimGhiPrecision}{1\xspace}
\newcommand{\XttClangSObVimGhiFone}{0.997\xspace}
\newcommand{\XttClangSObVimRdaRecall}{0.971\xspace}
\newcommand{\XttClangSObVimRdaPrecision}{1.000\xspace}
\newcommand{\XttClangSObVimRdaFone}{0.985\xspace}
\newcommand{\XttClangSObVimRseRecall}{0.998\xspace}
\newcommand{\XttClangSObVimRsePrecision}{1\xspace}
\newcommand{\XttClangSObVimRseFone}{0.999\xspace}
\newcommand{\XttClangSObVsfGT}{24987\xspace}
\newcommand{\XttClangSObVsfLsRecall}{1\xspace}
\newcommand{\XttClangSObVsfLsPrecision}{1\xspace}
\newcommand{\XttClangSObVsfLsFone}{1\xspace}
\newcommand{\XttClangSObVsfBapRecall}{0.990\xspace}
\newcommand{\XttClangSObVsfBapPrecision}{1\xspace}
\newcommand{\XttClangSObVsfBapFone}{0.995\xspace}
\newcommand{\XttClangSObVsfGhiRecall}{1.000\xspace}
\newcommand{\XttClangSObVsfGhiPrecision}{1\xspace}
\newcommand{\XttClangSObVsfGhiFone}{1.000\xspace}
\newcommand{\XttClangSObVsfRdaRecall}{0.966\xspace}
\newcommand{\XttClangSObVsfRdaPrecision}{1\xspace}
\newcommand{\XttClangSObVsfRdaFone}{0.983\xspace}
\newcommand{\XttClangSObVsfRseRecall}{1.000\xspace}
\newcommand{\XttClangSObVsfRsePrecision}{1\xspace}
\newcommand{\XttClangSObVsfRseFone}{1.000\xspace}
\newcommand{\XttClangSOcSzpGT}{16419\xspace}
\newcommand{\XttClangSOcSzpLsRecall}{1\xspace}
\newcommand{\XttClangSOcSzpLsPrecision}{1\xspace}
\newcommand{\XttClangSOcSzpLsFone}{1\xspace}
\newcommand{\XttClangSOcSzpBapRecall}{0.967\xspace}
\newcommand{\XttClangSOcSzpBapPrecision}{1\xspace}
\newcommand{\XttClangSOcSzpBapFone}{0.983\xspace}
\newcommand{\XttClangSOcSzpGhiRecall}{1\xspace}
\newcommand{\XttClangSOcSzpGhiPrecision}{1\xspace}
\newcommand{\XttClangSOcSzpGhiFone}{1\xspace}
\newcommand{\XttClangSOcSzpRdaRecall}{0.993\xspace}
\newcommand{\XttClangSOcSzpRdaPrecision}{1\xspace}
\newcommand{\XttClangSOcSzpRdaFone}{0.997\xspace}
\newcommand{\XttClangSOcSzpRseRecall}{1\xspace}
\newcommand{\XttClangSOcSzpRsePrecision}{1\xspace}
\newcommand{\XttClangSOcSzpRseFone}{1\xspace}
\newcommand{\XttClangSOcCapGT}{210578\xspace}
\newcommand{\XttClangSOcCapLsRecall}{1\xspace}
\newcommand{\XttClangSOcCapLsPrecision}{1\xspace}
\newcommand{\XttClangSOcCapLsFone}{1\xspace}
\newcommand{\XttClangSOcCapBapRecall}{0.273\xspace}
\newcommand{\XttClangSOcCapBapPrecision}{1\xspace}
\newcommand{\XttClangSOcCapBapFone}{0.428\xspace}
\newcommand{\XttClangSOcCapGhiRecall}{0.866\xspace}
\newcommand{\XttClangSOcCapGhiPrecision}{1\xspace}
\newcommand{\XttClangSOcCapGhiFone}{0.928\xspace}
\newcommand{\XttClangSOcCapRdaRecall}{0.863\xspace}
\newcommand{\XttClangSOcCapRdaPrecision}{1\xspace}
\newcommand{\XttClangSOcCapRdaFone}{0.927\xspace}
\newcommand{\XttClangSOcCapRseRecall}{1\xspace}
\newcommand{\XttClangSOcCapRsePrecision}{1\xspace}
\newcommand{\XttClangSOcCapRseFone}{1\xspace}
\newcommand{\XttClangSOcExmGT}{178828\xspace}
\newcommand{\XttClangSOcExmLsRecall}{1\xspace}
\newcommand{\XttClangSOcExmLsPrecision}{1\xspace}
\newcommand{\XttClangSOcExmLsFone}{1\xspace}
\newcommand{\XttClangSOcExmBapRecall}{0.819\xspace}
\newcommand{\XttClangSOcExmBapPrecision}{1\xspace}
\newcommand{\XttClangSOcExmBapFone}{0.900\xspace}
\newcommand{\XttClangSOcExmGhiRecall}{0.919\xspace}
\newcommand{\XttClangSOcExmGhiPrecision}{1\xspace}
\newcommand{\XttClangSOcExmGhiFone}{0.958\xspace}
\newcommand{\XttClangSOcExmRdaRecall}{0.957\xspace}
\newcommand{\XttClangSOcExmRdaPrecision}{1\xspace}
\newcommand{\XttClangSOcExmRdaFone}{0.978\xspace}
\newcommand{\XttClangSOcExmRseRecall}{1.000\xspace}
\newcommand{\XttClangSOcExmRsePrecision}{1\xspace}
\newcommand{\XttClangSOcExmRseFone}{1.000\xspace}
\newcommand{\XttClangSOcLgtGT}{32726\xspace}
\newcommand{\XttClangSOcLgtLsRecall}{1\xspace}
\newcommand{\XttClangSOcLgtLsPrecision}{1\xspace}
\newcommand{\XttClangSOcLgtLsFone}{1\xspace}
\newcommand{\XttClangSOcLgtBapRecall}{0.860\xspace}
\newcommand{\XttClangSOcLgtBapPrecision}{1\xspace}
\newcommand{\XttClangSOcLgtBapFone}{0.925\xspace}
\newcommand{\XttClangSOcLgtGhiRecall}{0.993\xspace}
\newcommand{\XttClangSOcLgtGhiPrecision}{1\xspace}
\newcommand{\XttClangSOcLgtGhiFone}{0.997\xspace}
\newcommand{\XttClangSOcLgtRdaRecall}{0.990\xspace}
\newcommand{\XttClangSOcLgtRdaPrecision}{1\xspace}
\newcommand{\XttClangSOcLgtRdaFone}{0.995\xspace}
\newcommand{\XttClangSOcLgtRseRecall}{0.997\xspace}
\newcommand{\XttClangSOcLgtRsePrecision}{1\xspace}
\newcommand{\XttClangSOcLgtRseFone}{0.999\xspace}
\newcommand{\XttClangSOcBzpGT}{19345\xspace}
\newcommand{\XttClangSOcBzpLsRecall}{1\xspace}
\newcommand{\XttClangSOcBzpLsPrecision}{1\xspace}
\newcommand{\XttClangSOcBzpLsFone}{1\xspace}
\newcommand{\XttClangSOcBzpBapRecall}{0.988\xspace}
\newcommand{\XttClangSOcBzpBapPrecision}{1\xspace}
\newcommand{\XttClangSOcBzpBapFone}{0.994\xspace}
\newcommand{\XttClangSOcBzpGhiRecall}{0.998\xspace}
\newcommand{\XttClangSOcBzpGhiPrecision}{1\xspace}
\newcommand{\XttClangSOcBzpGhiFone}{0.999\xspace}
\newcommand{\XttClangSOcBzpRdaRecall}{0.909\xspace}
\newcommand{\XttClangSOcBzpRdaPrecision}{1\xspace}
\newcommand{\XttClangSOcBzpRdaFone}{0.952\xspace}
\newcommand{\XttClangSOcBzpRseRecall}{1.000\xspace}
\newcommand{\XttClangSOcBzpRsePrecision}{1\xspace}
\newcommand{\XttClangSOcBzpRseFone}{1.000\xspace}
\newcommand{\XttClangSOcGccGT}{1302031\xspace}
\newcommand{\XttClangSOcGccLsRecall}{1\xspace}
\newcommand{\XttClangSOcGccLsPrecision}{1\xspace}
\newcommand{\XttClangSOcGccLsFone}{1\xspace}
\newcommand{\XttClangSOcGccBapRecall}{0.711\xspace}
\newcommand{\XttClangSOcGccBapPrecision}{1\xspace}
\newcommand{\XttClangSOcGccBapFone}{0.831\xspace}
\newcommand{\XttClangSOcGccGhiRecall}{0.962\xspace}
\newcommand{\XttClangSOcGccGhiPrecision}{1\xspace}
\newcommand{\XttClangSOcGccGhiFone}{0.981\xspace}
\newcommand{\XttClangSOcGccRdaRecall}{0.817\xspace}
\newcommand{\XttClangSOcGccRdaPrecision}{1\xspace}
\newcommand{\XttClangSOcGccRdaFone}{0.899\xspace}
\newcommand{\XttClangSOcGccRseRecall}{0.999\xspace}
\newcommand{\XttClangSOcGccRsePrecision}{1.000\xspace}
\newcommand{\XttClangSOcGccRseFone}{0.999\xspace}
\newcommand{\XttClangSOcGzpGT}{15610\xspace}
\newcommand{\XttClangSOcGzpLsRecall}{1\xspace}
\newcommand{\XttClangSOcGzpLsPrecision}{1\xspace}
\newcommand{\XttClangSOcGzpLsFone}{1\xspace}
\newcommand{\XttClangSOcGzpBapRecall}{0.989\xspace}
\newcommand{\XttClangSOcGzpBapPrecision}{1\xspace}
\newcommand{\XttClangSOcGzpBapFone}{0.994\xspace}
\newcommand{\XttClangSOcGzpGhiRecall}{0.999\xspace}
\newcommand{\XttClangSOcGzpGhiPrecision}{1\xspace}
\newcommand{\XttClangSOcGzpGhiFone}{0.999\xspace}
\newcommand{\XttClangSOcGzpRdaRecall}{1.000\xspace}
\newcommand{\XttClangSOcGzpRdaPrecision}{1\xspace}
\newcommand{\XttClangSOcGzpRdaFone}{1.000\xspace}
\newcommand{\XttClangSOcGzpRseRecall}{0.998\xspace}
\newcommand{\XttClangSOcGzpRsePrecision}{1\xspace}
\newcommand{\XttClangSOcGzpRseFone}{0.999\xspace}
\newcommand{\XttClangSOcOggGT}{60080\xspace}
\newcommand{\XttClangSOcOggLsRecall}{1\xspace}
\newcommand{\XttClangSOcOggLsPrecision}{1\xspace}
\newcommand{\XttClangSOcOggLsFone}{1\xspace}
\newcommand{\XttClangSOcOggBapRecall}{0.968\xspace}
\newcommand{\XttClangSOcOggBapPrecision}{1\xspace}
\newcommand{\XttClangSOcOggBapFone}{0.984\xspace}
\newcommand{\XttClangSOcOggGhiRecall}{1.000\xspace}
\newcommand{\XttClangSOcOggGhiPrecision}{1\xspace}
\newcommand{\XttClangSOcOggGhiFone}{1.000\xspace}
\newcommand{\XttClangSOcOggRdaRecall}{0.998\xspace}
\newcommand{\XttClangSOcOggRdaPrecision}{1\xspace}
\newcommand{\XttClangSOcOggRdaFone}{0.999\xspace}
\newcommand{\XttClangSOcOggRseRecall}{1.000\xspace}
\newcommand{\XttClangSOcOggRsePrecision}{1\xspace}
\newcommand{\XttClangSOcOggRseFone}{1.000\xspace}
\newcommand{\XttClangSOcNgxGT}{119681\xspace}
\newcommand{\XttClangSOcNgxLsRecall}{1\xspace}
\newcommand{\XttClangSOcNgxLsPrecision}{1\xspace}
\newcommand{\XttClangSOcNgxLsFone}{1\xspace}
\newcommand{\XttClangSOcNgxBapRecall}{0.964\xspace}
\newcommand{\XttClangSOcNgxBapPrecision}{1\xspace}
\newcommand{\XttClangSOcNgxBapFone}{0.982\xspace}
\newcommand{\XttClangSOcNgxGhiRecall}{0.999\xspace}
\newcommand{\XttClangSOcNgxGhiPrecision}{1\xspace}
\newcommand{\XttClangSOcNgxGhiFone}{0.999\xspace}
\newcommand{\XttClangSOcNgxRdaRecall}{0.997\xspace}
\newcommand{\XttClangSOcNgxRdaPrecision}{1\xspace}
\newcommand{\XttClangSOcNgxRdaFone}{0.998\xspace}
\newcommand{\XttClangSOcNgxRseRecall}{1.000\xspace}
\newcommand{\XttClangSOcNgxRsePrecision}{1\xspace}
\newcommand{\XttClangSOcNgxRseFone}{1.000\xspace}
\newcommand{\XttClangSOcSshGT}{150494\xspace}
\newcommand{\XttClangSOcSshLsRecall}{1\xspace}
\newcommand{\XttClangSOcSshLsPrecision}{1\xspace}
\newcommand{\XttClangSOcSshLsFone}{1\xspace}
\newcommand{\XttClangSOcSshBapRecall}{0.923\xspace}
\newcommand{\XttClangSOcSshBapPrecision}{1\xspace}
\newcommand{\XttClangSOcSshBapFone}{0.960\xspace}
\newcommand{\XttClangSOcSshGhiRecall}{0.989\xspace}
\newcommand{\XttClangSOcSshGhiPrecision}{1\xspace}
\newcommand{\XttClangSOcSshGhiFone}{0.994\xspace}
\newcommand{\XttClangSOcSshRdaRecall}{0.875\xspace}
\newcommand{\XttClangSOcSshRdaPrecision}{1\xspace}
\newcommand{\XttClangSOcSshRdaFone}{0.934\xspace}
\newcommand{\XttClangSOcSshRseRecall}{0.966\xspace}
\newcommand{\XttClangSOcSshRsePrecision}{1\xspace}
\newcommand{\XttClangSOcSshRseFone}{0.983\xspace}
\newcommand{\XttClangSOcPcrGT}{6135\xspace}
\newcommand{\XttClangSOcPcrLsRecall}{1\xspace}
\newcommand{\XttClangSOcPcrLsPrecision}{1\xspace}
\newcommand{\XttClangSOcPcrLsFone}{1\xspace}
\newcommand{\XttClangSOcPcrBapRecall}{0.851\xspace}
\newcommand{\XttClangSOcPcrBapPrecision}{1\xspace}
\newcommand{\XttClangSOcPcrBapFone}{0.919\xspace}
\newcommand{\XttClangSOcPcrGhiRecall}{0.970\xspace}
\newcommand{\XttClangSOcPcrGhiPrecision}{1\xspace}
\newcommand{\XttClangSOcPcrGhiFone}{0.985\xspace}
\newcommand{\XttClangSOcPcrRdaRecall}{1.000\xspace}
\newcommand{\XttClangSOcPcrRdaPrecision}{1\xspace}
\newcommand{\XttClangSOcPcrRdaFone}{1.000\xspace}
\newcommand{\XttClangSOcPcrRseRecall}{1.000\xspace}
\newcommand{\XttClangSOcPcrRsePrecision}{1\xspace}
\newcommand{\XttClangSOcPcrRseFone}{1.000\xspace}
\newcommand{\XttClangSOcSqlGT}{334790\xspace}
\newcommand{\XttClangSOcSqlLsRecall}{1\xspace}
\newcommand{\XttClangSOcSqlLsPrecision}{1\xspace}
\newcommand{\XttClangSOcSqlLsFone}{1\xspace}
\newcommand{\XttClangSOcSqlBapRecall}{0.792\xspace}
\newcommand{\XttClangSOcSqlBapPrecision}{1.000\xspace}
\newcommand{\XttClangSOcSqlBapFone}{0.884\xspace}
\newcommand{\XttClangSOcSqlGhiRecall}{0.943\xspace}
\newcommand{\XttClangSOcSqlGhiPrecision}{1\xspace}
\newcommand{\XttClangSOcSqlGhiFone}{0.971\xspace}
\newcommand{\XttClangSOcSqlRdaRecall}{0.910\xspace}
\newcommand{\XttClangSOcSqlRdaPrecision}{1\xspace}
\newcommand{\XttClangSOcSqlRdaFone}{0.953\xspace}
\newcommand{\XttClangSOcSqlRseRecall}{0.967\xspace}
\newcommand{\XttClangSOcSqlRsePrecision}{1.000\xspace}
\newcommand{\XttClangSOcSqlRseFone}{0.983\xspace}
\newcommand{\XttClangSOcVimGT}{641448\xspace}
\newcommand{\XttClangSOcVimLsRecall}{1\xspace}
\newcommand{\XttClangSOcVimLsPrecision}{1\xspace}
\newcommand{\XttClangSOcVimLsFone}{1\xspace}
\newcommand{\XttClangSOcVimBapRecall}{0.856\xspace}
\newcommand{\XttClangSOcVimBapPrecision}{1\xspace}
\newcommand{\XttClangSOcVimBapFone}{0.923\xspace}
\newcommand{\XttClangSOcVimGhiRecall}{0.995\xspace}
\newcommand{\XttClangSOcVimGhiPrecision}{1\xspace}
\newcommand{\XttClangSOcVimGhiFone}{0.997\xspace}
\newcommand{\XttClangSOcVimRdaRecall}{0.966\xspace}
\newcommand{\XttClangSOcVimRdaPrecision}{1\xspace}
\newcommand{\XttClangSOcVimRdaFone}{0.983\xspace}
\newcommand{\XttClangSOcVimRseRecall}{0.999\xspace}
\newcommand{\XttClangSOcVimRsePrecision}{1\xspace}
\newcommand{\XttClangSOcVimRseFone}{1.000\xspace}
\newcommand{\XttClangSOcVsfGT}{26776\xspace}
\newcommand{\XttClangSOcVsfLsRecall}{1\xspace}
\newcommand{\XttClangSOcVsfLsPrecision}{1\xspace}
\newcommand{\XttClangSOcVsfLsFone}{1\xspace}
\newcommand{\XttClangSOcVsfBapRecall}{0.990\xspace}
\newcommand{\XttClangSOcVsfBapPrecision}{1\xspace}
\newcommand{\XttClangSOcVsfBapFone}{0.995\xspace}
\newcommand{\XttClangSOcVsfGhiRecall}{1.000\xspace}
\newcommand{\XttClangSOcVsfGhiPrecision}{1\xspace}
\newcommand{\XttClangSOcVsfGhiFone}{1.000\xspace}
\newcommand{\XttClangSOcVsfRdaRecall}{0.958\xspace}
\newcommand{\XttClangSOcVsfRdaPrecision}{1\xspace}
\newcommand{\XttClangSOcVsfRdaFone}{0.979\xspace}
\newcommand{\XttClangSOcVsfRseRecall}{1.000\xspace}
\newcommand{\XttClangSOcVsfRsePrecision}{1\xspace}
\newcommand{\XttClangSOcVsfRseFone}{1.000\xspace}
\newcommand{\XttClangSOdSzpGT}{16419\xspace}
\newcommand{\XttClangSOdSzpLsRecall}{1\xspace}
\newcommand{\XttClangSOdSzpLsPrecision}{1\xspace}
\newcommand{\XttClangSOdSzpLsFone}{1\xspace}
\newcommand{\XttClangSOdSzpBapRecall}{0.967\xspace}
\newcommand{\XttClangSOdSzpBapPrecision}{1\xspace}
\newcommand{\XttClangSOdSzpBapFone}{0.983\xspace}
\newcommand{\XttClangSOdSzpGhiRecall}{1\xspace}
\newcommand{\XttClangSOdSzpGhiPrecision}{1\xspace}
\newcommand{\XttClangSOdSzpGhiFone}{1\xspace}
\newcommand{\XttClangSOdSzpRdaRecall}{0.993\xspace}
\newcommand{\XttClangSOdSzpRdaPrecision}{1\xspace}
\newcommand{\XttClangSOdSzpRdaFone}{0.997\xspace}
\newcommand{\XttClangSOdSzpRseRecall}{1\xspace}
\newcommand{\XttClangSOdSzpRsePrecision}{1\xspace}
\newcommand{\XttClangSOdSzpRseFone}{1\xspace}
\newcommand{\XttClangSOdCapGT}{210578\xspace}
\newcommand{\XttClangSOdCapLsRecall}{1\xspace}
\newcommand{\XttClangSOdCapLsPrecision}{1\xspace}
\newcommand{\XttClangSOdCapLsFone}{1\xspace}
\newcommand{\XttClangSOdCapBapRecall}{0.273\xspace}
\newcommand{\XttClangSOdCapBapPrecision}{1\xspace}
\newcommand{\XttClangSOdCapBapFone}{0.428\xspace}
\newcommand{\XttClangSOdCapGhiRecall}{0.866\xspace}
\newcommand{\XttClangSOdCapGhiPrecision}{1\xspace}
\newcommand{\XttClangSOdCapGhiFone}{0.928\xspace}
\newcommand{\XttClangSOdCapRdaRecall}{0.863\xspace}
\newcommand{\XttClangSOdCapRdaPrecision}{1\xspace}
\newcommand{\XttClangSOdCapRdaFone}{0.927\xspace}
\newcommand{\XttClangSOdCapRseRecall}{1\xspace}
\newcommand{\XttClangSOdCapRsePrecision}{1\xspace}
\newcommand{\XttClangSOdCapRseFone}{1\xspace}
\newcommand{\XttClangSOdExmGT}{178823\xspace}
\newcommand{\XttClangSOdExmLsRecall}{1\xspace}
\newcommand{\XttClangSOdExmLsPrecision}{1\xspace}
\newcommand{\XttClangSOdExmLsFone}{1\xspace}
\newcommand{\XttClangSOdExmBapRecall}{0.819\xspace}
\newcommand{\XttClangSOdExmBapPrecision}{1\xspace}
\newcommand{\XttClangSOdExmBapFone}{0.900\xspace}
\newcommand{\XttClangSOdExmGhiRecall}{0.896\xspace}
\newcommand{\XttClangSOdExmGhiPrecision}{1\xspace}
\newcommand{\XttClangSOdExmGhiFone}{0.945\xspace}
\newcommand{\XttClangSOdExmRdaRecall}{0.957\xspace}
\newcommand{\XttClangSOdExmRdaPrecision}{1\xspace}
\newcommand{\XttClangSOdExmRdaFone}{0.978\xspace}
\newcommand{\XttClangSOdExmRseRecall}{1.000\xspace}
\newcommand{\XttClangSOdExmRsePrecision}{1\xspace}
\newcommand{\XttClangSOdExmRseFone}{1.000\xspace}
\newcommand{\XttClangSOdLgtGT}{32726\xspace}
\newcommand{\XttClangSOdLgtLsRecall}{1\xspace}
\newcommand{\XttClangSOdLgtLsPrecision}{1\xspace}
\newcommand{\XttClangSOdLgtLsFone}{1\xspace}
\newcommand{\XttClangSOdLgtBapRecall}{0.860\xspace}
\newcommand{\XttClangSOdLgtBapPrecision}{1\xspace}
\newcommand{\XttClangSOdLgtBapFone}{0.925\xspace}
\newcommand{\XttClangSOdLgtGhiRecall}{0.993\xspace}
\newcommand{\XttClangSOdLgtGhiPrecision}{1\xspace}
\newcommand{\XttClangSOdLgtGhiFone}{0.997\xspace}
\newcommand{\XttClangSOdLgtRdaRecall}{0.990\xspace}
\newcommand{\XttClangSOdLgtRdaPrecision}{1\xspace}
\newcommand{\XttClangSOdLgtRdaFone}{0.995\xspace}
\newcommand{\XttClangSOdLgtRseRecall}{0.997\xspace}
\newcommand{\XttClangSOdLgtRsePrecision}{1\xspace}
\newcommand{\XttClangSOdLgtRseFone}{0.999\xspace}
\newcommand{\XttClangSOdBzpGT}{19342\xspace}
\newcommand{\XttClangSOdBzpLsRecall}{1\xspace}
\newcommand{\XttClangSOdBzpLsPrecision}{1\xspace}
\newcommand{\XttClangSOdBzpLsFone}{1\xspace}
\newcommand{\XttClangSOdBzpBapRecall}{0.988\xspace}
\newcommand{\XttClangSOdBzpBapPrecision}{1\xspace}
\newcommand{\XttClangSOdBzpBapFone}{0.994\xspace}
\newcommand{\XttClangSOdBzpGhiRecall}{0.998\xspace}
\newcommand{\XttClangSOdBzpGhiPrecision}{1\xspace}
\newcommand{\XttClangSOdBzpGhiFone}{0.999\xspace}
\newcommand{\XttClangSOdBzpRdaRecall}{0.909\xspace}
\newcommand{\XttClangSOdBzpRdaPrecision}{1\xspace}
\newcommand{\XttClangSOdBzpRdaFone}{0.952\xspace}
\newcommand{\XttClangSOdBzpRseRecall}{1.000\xspace}
\newcommand{\XttClangSOdBzpRsePrecision}{1\xspace}
\newcommand{\XttClangSOdBzpRseFone}{1.000\xspace}
\newcommand{\XttClangSOdGccGT}{1302091\xspace}
\newcommand{\XttClangSOdGccLsRecall}{1\xspace}
\newcommand{\XttClangSOdGccLsPrecision}{1\xspace}
\newcommand{\XttClangSOdGccLsFone}{1\xspace}
\newcommand{\XttClangSOdGccBapRecall}{0.711\xspace}
\newcommand{\XttClangSOdGccBapPrecision}{1\xspace}
\newcommand{\XttClangSOdGccBapFone}{0.831\xspace}
\newcommand{\XttClangSOdGccGhiRecall}{0.959\xspace}
\newcommand{\XttClangSOdGccGhiPrecision}{1\xspace}
\newcommand{\XttClangSOdGccGhiFone}{0.979\xspace}
\newcommand{\XttClangSOdGccRdaRecall}{0.817\xspace}
\newcommand{\XttClangSOdGccRdaPrecision}{1\xspace}
\newcommand{\XttClangSOdGccRdaFone}{0.899\xspace}
\newcommand{\XttClangSOdGccRseRecall}{0.999\xspace}
\newcommand{\XttClangSOdGccRsePrecision}{1.000\xspace}
\newcommand{\XttClangSOdGccRseFone}{0.999\xspace}
\newcommand{\XttClangSOdGzpGT}{15610\xspace}
\newcommand{\XttClangSOdGzpLsRecall}{1\xspace}
\newcommand{\XttClangSOdGzpLsPrecision}{1\xspace}
\newcommand{\XttClangSOdGzpLsFone}{1\xspace}
\newcommand{\XttClangSOdGzpBapRecall}{0.989\xspace}
\newcommand{\XttClangSOdGzpBapPrecision}{1\xspace}
\newcommand{\XttClangSOdGzpBapFone}{0.994\xspace}
\newcommand{\XttClangSOdGzpGhiRecall}{0.999\xspace}
\newcommand{\XttClangSOdGzpGhiPrecision}{1\xspace}
\newcommand{\XttClangSOdGzpGhiFone}{0.999\xspace}
\newcommand{\XttClangSOdGzpRdaRecall}{1.000\xspace}
\newcommand{\XttClangSOdGzpRdaPrecision}{1\xspace}
\newcommand{\XttClangSOdGzpRdaFone}{1.000\xspace}
\newcommand{\XttClangSOdGzpRseRecall}{0.998\xspace}
\newcommand{\XttClangSOdGzpRsePrecision}{1\xspace}
\newcommand{\XttClangSOdGzpRseFone}{0.999\xspace}
\newcommand{\XttClangSOdOggGT}{60571\xspace}
\newcommand{\XttClangSOdOggLsRecall}{1\xspace}
\newcommand{\XttClangSOdOggLsPrecision}{1\xspace}
\newcommand{\XttClangSOdOggLsFone}{1\xspace}
\newcommand{\XttClangSOdOggBapRecall}{0.968\xspace}
\newcommand{\XttClangSOdOggBapPrecision}{1\xspace}
\newcommand{\XttClangSOdOggBapFone}{0.984\xspace}
\newcommand{\XttClangSOdOggGhiRecall}{1.000\xspace}
\newcommand{\XttClangSOdOggGhiPrecision}{1\xspace}
\newcommand{\XttClangSOdOggGhiFone}{1.000\xspace}
\newcommand{\XttClangSOdOggRdaRecall}{0.996\xspace}
\newcommand{\XttClangSOdOggRdaPrecision}{1\xspace}
\newcommand{\XttClangSOdOggRdaFone}{0.998\xspace}
\newcommand{\XttClangSOdOggRseRecall}{1.000\xspace}
\newcommand{\XttClangSOdOggRsePrecision}{1\xspace}
\newcommand{\XttClangSOdOggRseFone}{1.000\xspace}
\newcommand{\XttClangSOdNgxGT}{119682\xspace}
\newcommand{\XttClangSOdNgxLsRecall}{1\xspace}
\newcommand{\XttClangSOdNgxLsPrecision}{1\xspace}
\newcommand{\XttClangSOdNgxLsFone}{1\xspace}
\newcommand{\XttClangSOdNgxBapRecall}{0.964\xspace}
\newcommand{\XttClangSOdNgxBapPrecision}{1\xspace}
\newcommand{\XttClangSOdNgxBapFone}{0.982\xspace}
\newcommand{\XttClangSOdNgxGhiRecall}{0.999\xspace}
\newcommand{\XttClangSOdNgxGhiPrecision}{1\xspace}
\newcommand{\XttClangSOdNgxGhiFone}{0.999\xspace}
\newcommand{\XttClangSOdNgxRdaRecall}{0.997\xspace}
\newcommand{\XttClangSOdNgxRdaPrecision}{1\xspace}
\newcommand{\XttClangSOdNgxRdaFone}{0.998\xspace}
\newcommand{\XttClangSOdNgxRseRecall}{1.000\xspace}
\newcommand{\XttClangSOdNgxRsePrecision}{1\xspace}
\newcommand{\XttClangSOdNgxRseFone}{1.000\xspace}
\newcommand{\XttClangSOdSshGT}{150494\xspace}
\newcommand{\XttClangSOdSshLsRecall}{1\xspace}
\newcommand{\XttClangSOdSshLsPrecision}{1\xspace}
\newcommand{\XttClangSOdSshLsFone}{1\xspace}
\newcommand{\XttClangSOdSshBapRecall}{0.923\xspace}
\newcommand{\XttClangSOdSshBapPrecision}{1\xspace}
\newcommand{\XttClangSOdSshBapFone}{0.960\xspace}
\newcommand{\XttClangSOdSshGhiRecall}{0.989\xspace}
\newcommand{\XttClangSOdSshGhiPrecision}{1\xspace}
\newcommand{\XttClangSOdSshGhiFone}{0.994\xspace}
\newcommand{\XttClangSOdSshRdaRecall}{0.875\xspace}
\newcommand{\XttClangSOdSshRdaPrecision}{1\xspace}
\newcommand{\XttClangSOdSshRdaFone}{0.934\xspace}
\newcommand{\XttClangSOdSshRseRecall}{0.966\xspace}
\newcommand{\XttClangSOdSshRsePrecision}{1\xspace}
\newcommand{\XttClangSOdSshRseFone}{0.983\xspace}
\newcommand{\XttClangSOdPcrGT}{6135\xspace}
\newcommand{\XttClangSOdPcrLsRecall}{1\xspace}
\newcommand{\XttClangSOdPcrLsPrecision}{1\xspace}
\newcommand{\XttClangSOdPcrLsFone}{1\xspace}
\newcommand{\XttClangSOdPcrBapRecall}{0.851\xspace}
\newcommand{\XttClangSOdPcrBapPrecision}{1\xspace}
\newcommand{\XttClangSOdPcrBapFone}{0.919\xspace}
\newcommand{\XttClangSOdPcrGhiRecall}{0.970\xspace}
\newcommand{\XttClangSOdPcrGhiPrecision}{1\xspace}
\newcommand{\XttClangSOdPcrGhiFone}{0.985\xspace}
\newcommand{\XttClangSOdPcrRdaRecall}{1.000\xspace}
\newcommand{\XttClangSOdPcrRdaPrecision}{1\xspace}
\newcommand{\XttClangSOdPcrRdaFone}{1.000\xspace}
\newcommand{\XttClangSOdPcrRseRecall}{1.000\xspace}
\newcommand{\XttClangSOdPcrRsePrecision}{1\xspace}
\newcommand{\XttClangSOdPcrRseFone}{1.000\xspace}
\newcommand{\XttClangSOdSqlGT}{334807\xspace}
\newcommand{\XttClangSOdSqlLsRecall}{1\xspace}
\newcommand{\XttClangSOdSqlLsPrecision}{1\xspace}
\newcommand{\XttClangSOdSqlLsFone}{1\xspace}
\newcommand{\XttClangSOdSqlBapRecall}{0.792\xspace}
\newcommand{\XttClangSOdSqlBapPrecision}{1.000\xspace}
\newcommand{\XttClangSOdSqlBapFone}{0.884\xspace}
\newcommand{\XttClangSOdSqlGhiRecall}{0.943\xspace}
\newcommand{\XttClangSOdSqlGhiPrecision}{1\xspace}
\newcommand{\XttClangSOdSqlGhiFone}{0.971\xspace}
\newcommand{\XttClangSOdSqlRdaRecall}{0.910\xspace}
\newcommand{\XttClangSOdSqlRdaPrecision}{1\xspace}
\newcommand{\XttClangSOdSqlRdaFone}{0.953\xspace}
\newcommand{\XttClangSOdSqlRseRecall}{0.967\xspace}
\newcommand{\XttClangSOdSqlRsePrecision}{1.000\xspace}
\newcommand{\XttClangSOdSqlRseFone}{0.983\xspace}
\newcommand{\XttClangSOdVimGT}{641475\xspace}
\newcommand{\XttClangSOdVimLsRecall}{1\xspace}
\newcommand{\XttClangSOdVimLsPrecision}{1\xspace}
\newcommand{\XttClangSOdVimLsFone}{1\xspace}
\newcommand{\XttClangSOdVimBapRecall}{0.856\xspace}
\newcommand{\XttClangSOdVimBapPrecision}{1\xspace}
\newcommand{\XttClangSOdVimBapFone}{0.923\xspace}
\newcommand{\XttClangSOdVimGhiRecall}{0.995\xspace}
\newcommand{\XttClangSOdVimGhiPrecision}{1\xspace}
\newcommand{\XttClangSOdVimGhiFone}{0.997\xspace}
\newcommand{\XttClangSOdVimRdaRecall}{0.966\xspace}
\newcommand{\XttClangSOdVimRdaPrecision}{1\xspace}
\newcommand{\XttClangSOdVimRdaFone}{0.983\xspace}
\newcommand{\XttClangSOdVimRseRecall}{0.999\xspace}
\newcommand{\XttClangSOdVimRsePrecision}{1\xspace}
\newcommand{\XttClangSOdVimRseFone}{1.000\xspace}
\newcommand{\XttClangSOdVsfGT}{26770\xspace}
\newcommand{\XttClangSOdVsfLsRecall}{1\xspace}
\newcommand{\XttClangSOdVsfLsPrecision}{1\xspace}
\newcommand{\XttClangSOdVsfLsFone}{1\xspace}
\newcommand{\XttClangSOdVsfBapRecall}{0.990\xspace}
\newcommand{\XttClangSOdVsfBapPrecision}{1\xspace}
\newcommand{\XttClangSOdVsfBapFone}{0.995\xspace}
\newcommand{\XttClangSOdVsfGhiRecall}{1.000\xspace}
\newcommand{\XttClangSOdVsfGhiPrecision}{1\xspace}
\newcommand{\XttClangSOdVsfGhiFone}{1.000\xspace}
\newcommand{\XttClangSOdVsfRdaRecall}{0.958\xspace}
\newcommand{\XttClangSOdVsfRdaPrecision}{1\xspace}
\newcommand{\XttClangSOdVsfRdaFone}{0.979\xspace}
\newcommand{\XttClangSOdVsfRseRecall}{1.000\xspace}
\newcommand{\XttClangSOdVsfRsePrecision}{1\xspace}
\newcommand{\XttClangSOdVsfRseFone}{1.000\xspace}
\newcommand{\XttClangSOsSzpGT}{13673\xspace}
\newcommand{\XttClangSOsSzpLsRecall}{1\xspace}
\newcommand{\XttClangSOsSzpLsPrecision}{1\xspace}
\newcommand{\XttClangSOsSzpLsFone}{1\xspace}
\newcommand{\XttClangSOsSzpBapRecall}{0.975\xspace}
\newcommand{\XttClangSOsSzpBapPrecision}{1\xspace}
\newcommand{\XttClangSOsSzpBapFone}{0.987\xspace}
\newcommand{\XttClangSOsSzpGhiRecall}{1\xspace}
\newcommand{\XttClangSOsSzpGhiPrecision}{1\xspace}
\newcommand{\XttClangSOsSzpGhiFone}{1\xspace}
\newcommand{\XttClangSOsSzpRdaRecall}{1\xspace}
\newcommand{\XttClangSOsSzpRdaPrecision}{1\xspace}
\newcommand{\XttClangSOsSzpRdaFone}{1\xspace}
\newcommand{\XttClangSOsSzpRseRecall}{1\xspace}
\newcommand{\XttClangSOsSzpRsePrecision}{1\xspace}
\newcommand{\XttClangSOsSzpRseFone}{1\xspace}
\newcommand{\XttClangSOsCapGT}{183596\xspace}
\newcommand{\XttClangSOsCapLsRecall}{1\xspace}
\newcommand{\XttClangSOsCapLsPrecision}{1\xspace}
\newcommand{\XttClangSOsCapLsFone}{1\xspace}
\newcommand{\XttClangSOsCapBapRecall}{0.354\xspace}
\newcommand{\XttClangSOsCapBapPrecision}{1\xspace}
\newcommand{\XttClangSOsCapBapFone}{0.523\xspace}
\newcommand{\XttClangSOsCapGhiRecall}{0.805\xspace}
\newcommand{\XttClangSOsCapGhiPrecision}{1\xspace}
\newcommand{\XttClangSOsCapGhiFone}{0.892\xspace}
\newcommand{\XttClangSOsCapRdaRecall}{0.998\xspace}
\newcommand{\XttClangSOsCapRdaPrecision}{1\xspace}
\newcommand{\XttClangSOsCapRdaFone}{0.999\xspace}
\newcommand{\XttClangSOsCapRseRecall}{1\xspace}
\newcommand{\XttClangSOsCapRsePrecision}{1\xspace}
\newcommand{\XttClangSOsCapRseFone}{1\xspace}
\newcommand{\XttClangSOsExmGT}{147463\xspace}
\newcommand{\XttClangSOsExmLsRecall}{1\xspace}
\newcommand{\XttClangSOsExmLsPrecision}{1\xspace}
\newcommand{\XttClangSOsExmLsFone}{1\xspace}
\newcommand{\XttClangSOsExmBapRecall}{0.831\xspace}
\newcommand{\XttClangSOsExmBapPrecision}{1\xspace}
\newcommand{\XttClangSOsExmBapFone}{0.908\xspace}
\newcommand{\XttClangSOsExmGhiRecall}{0.984\xspace}
\newcommand{\XttClangSOsExmGhiPrecision}{1\xspace}
\newcommand{\XttClangSOsExmGhiFone}{0.992\xspace}
\newcommand{\XttClangSOsExmRdaRecall}{0.961\xspace}
\newcommand{\XttClangSOsExmRdaPrecision}{1\xspace}
\newcommand{\XttClangSOsExmRdaFone}{0.980\xspace}
\newcommand{\XttClangSOsExmRseRecall}{1.000\xspace}
\newcommand{\XttClangSOsExmRsePrecision}{1\xspace}
\newcommand{\XttClangSOsExmRseFone}{1.000\xspace}
\newcommand{\XttClangSOsLgtGT}{28392\xspace}
\newcommand{\XttClangSOsLgtLsRecall}{1\xspace}
\newcommand{\XttClangSOsLgtLsPrecision}{1\xspace}
\newcommand{\XttClangSOsLgtLsFone}{1\xspace}
\newcommand{\XttClangSOsLgtBapRecall}{0.897\xspace}
\newcommand{\XttClangSOsLgtBapPrecision}{1\xspace}
\newcommand{\XttClangSOsLgtBapFone}{0.946\xspace}
\newcommand{\XttClangSOsLgtGhiRecall}{0.995\xspace}
\newcommand{\XttClangSOsLgtGhiPrecision}{1\xspace}
\newcommand{\XttClangSOsLgtGhiFone}{0.997\xspace}
\newcommand{\XttClangSOsLgtRdaRecall}{0.971\xspace}
\newcommand{\XttClangSOsLgtRdaPrecision}{1\xspace}
\newcommand{\XttClangSOsLgtRdaFone}{0.986\xspace}
\newcommand{\XttClangSOsLgtRseRecall}{0.997\xspace}
\newcommand{\XttClangSOsLgtRsePrecision}{1\xspace}
\newcommand{\XttClangSOsLgtRseFone}{0.999\xspace}
\newcommand{\XttClangSOsBzpGT}{15670\xspace}
\newcommand{\XttClangSOsBzpLsRecall}{1\xspace}
\newcommand{\XttClangSOsBzpLsPrecision}{1\xspace}
\newcommand{\XttClangSOsBzpLsFone}{1\xspace}
\newcommand{\XttClangSOsBzpBapRecall}{0.985\xspace}
\newcommand{\XttClangSOsBzpBapPrecision}{1\xspace}
\newcommand{\XttClangSOsBzpBapFone}{0.993\xspace}
\newcommand{\XttClangSOsBzpGhiRecall}{0.998\xspace}
\newcommand{\XttClangSOsBzpGhiPrecision}{1\xspace}
\newcommand{\XttClangSOsBzpGhiFone}{0.999\xspace}
\newcommand{\XttClangSOsBzpRdaRecall}{0.806\xspace}
\newcommand{\XttClangSOsBzpRdaPrecision}{1\xspace}
\newcommand{\XttClangSOsBzpRdaFone}{0.893\xspace}
\newcommand{\XttClangSOsBzpRseRecall}{0.999\xspace}
\newcommand{\XttClangSOsBzpRsePrecision}{1\xspace}
\newcommand{\XttClangSOsBzpRseFone}{1.000\xspace}
\newcommand{\XttClangSOsGccGT}{901039\xspace}
\newcommand{\XttClangSOsGccLsRecall}{1\xspace}
\newcommand{\XttClangSOsGccLsPrecision}{1\xspace}
\newcommand{\XttClangSOsGccLsFone}{1\xspace}
\newcommand{\XttClangSOsGccBapRecall}{0.728\xspace}
\newcommand{\XttClangSOsGccBapPrecision}{1.000\xspace}
\newcommand{\XttClangSOsGccBapFone}{0.843\xspace}
\newcommand{\XttClangSOsGccGhiRecall}{0.967\xspace}
\newcommand{\XttClangSOsGccGhiPrecision}{1\xspace}
\newcommand{\XttClangSOsGccGhiFone}{0.983\xspace}
\newcommand{\XttClangSOsGccRdaRecall}{0.875\xspace}
\newcommand{\XttClangSOsGccRdaPrecision}{1\xspace}
\newcommand{\XttClangSOsGccRdaFone}{0.933\xspace}
\newcommand{\XttClangSOsGccRseRecall}{0.999\xspace}
\newcommand{\XttClangSOsGccRsePrecision}{1\xspace}
\newcommand{\XttClangSOsGccRseFone}{1.000\xspace}
\newcommand{\XttClangSOsGzpGT}{9541\xspace}
\newcommand{\XttClangSOsGzpLsRecall}{1\xspace}
\newcommand{\XttClangSOsGzpLsPrecision}{1\xspace}
\newcommand{\XttClangSOsGzpLsFone}{1\xspace}
\newcommand{\XttClangSOsGzpBapRecall}{0.990\xspace}
\newcommand{\XttClangSOsGzpBapPrecision}{1\xspace}
\newcommand{\XttClangSOsGzpBapFone}{0.995\xspace}
\newcommand{\XttClangSOsGzpGhiRecall}{0.998\xspace}
\newcommand{\XttClangSOsGzpGhiPrecision}{1\xspace}
\newcommand{\XttClangSOsGzpGhiFone}{0.999\xspace}
\newcommand{\XttClangSOsGzpRdaRecall}{1.000\xspace}
\newcommand{\XttClangSOsGzpRdaPrecision}{1\xspace}
\newcommand{\XttClangSOsGzpRdaFone}{1.000\xspace}
\newcommand{\XttClangSOsGzpRseRecall}{0.999\xspace}
\newcommand{\XttClangSOsGzpRsePrecision}{1\xspace}
\newcommand{\XttClangSOsGzpRseFone}{0.999\xspace}
\newcommand{\XttClangSOsOggGT}{38132\xspace}
\newcommand{\XttClangSOsOggLsRecall}{1\xspace}
\newcommand{\XttClangSOsOggLsPrecision}{1\xspace}
\newcommand{\XttClangSOsOggLsFone}{1\xspace}
\newcommand{\XttClangSOsOggBapRecall}{0.984\xspace}
\newcommand{\XttClangSOsOggBapPrecision}{1\xspace}
\newcommand{\XttClangSOsOggBapFone}{0.992\xspace}
\newcommand{\XttClangSOsOggGhiRecall}{1.000\xspace}
\newcommand{\XttClangSOsOggGhiPrecision}{1\xspace}
\newcommand{\XttClangSOsOggGhiFone}{1.000\xspace}
\newcommand{\XttClangSOsOggRdaRecall}{1.000\xspace}
\newcommand{\XttClangSOsOggRdaPrecision}{1\xspace}
\newcommand{\XttClangSOsOggRdaFone}{1.000\xspace}
\newcommand{\XttClangSOsOggRseRecall}{0.999\xspace}
\newcommand{\XttClangSOsOggRsePrecision}{1\xspace}
\newcommand{\XttClangSOsOggRseFone}{0.999\xspace}
\newcommand{\XttClangSOsNgxGT}{106786\xspace}
\newcommand{\XttClangSOsNgxLsRecall}{1\xspace}
\newcommand{\XttClangSOsNgxLsPrecision}{1\xspace}
\newcommand{\XttClangSOsNgxLsFone}{1\xspace}
\newcommand{\XttClangSOsNgxBapRecall}{0.973\xspace}
\newcommand{\XttClangSOsNgxBapPrecision}{1\xspace}
\newcommand{\XttClangSOsNgxBapFone}{0.986\xspace}
\newcommand{\XttClangSOsNgxGhiRecall}{1.000\xspace}
\newcommand{\XttClangSOsNgxGhiPrecision}{1\xspace}
\newcommand{\XttClangSOsNgxGhiFone}{1.000\xspace}
\newcommand{\XttClangSOsNgxRdaRecall}{0.997\xspace}
\newcommand{\XttClangSOsNgxRdaPrecision}{1\xspace}
\newcommand{\XttClangSOsNgxRdaFone}{0.998\xspace}
\newcommand{\XttClangSOsNgxRseRecall}{1.000\xspace}
\newcommand{\XttClangSOsNgxRsePrecision}{1\xspace}
\newcommand{\XttClangSOsNgxRseFone}{1.000\xspace}
\newcommand{\XttClangSOsSshGT}{122455\xspace}
\newcommand{\XttClangSOsSshLsRecall}{1\xspace}
\newcommand{\XttClangSOsSshLsPrecision}{1\xspace}
\newcommand{\XttClangSOsSshLsFone}{1\xspace}
\newcommand{\XttClangSOsSshBapRecall}{0.947\xspace}
\newcommand{\XttClangSOsSshBapPrecision}{1\xspace}
\newcommand{\XttClangSOsSshBapFone}{0.973\xspace}
\newcommand{\XttClangSOsSshGhiRecall}{0.995\xspace}
\newcommand{\XttClangSOsSshGhiPrecision}{1\xspace}
\newcommand{\XttClangSOsSshGhiFone}{0.997\xspace}
\newcommand{\XttClangSOsSshRdaRecall}{0.918\xspace}
\newcommand{\XttClangSOsSshRdaPrecision}{1\xspace}
\newcommand{\XttClangSOsSshRdaFone}{0.957\xspace}
\newcommand{\XttClangSOsSshRseRecall}{0.988\xspace}
\newcommand{\XttClangSOsSshRsePrecision}{1\xspace}
\newcommand{\XttClangSOsSshRseFone}{0.994\xspace}
\newcommand{\XttClangSOsPcrGT}{4638\xspace}
\newcommand{\XttClangSOsPcrLsRecall}{1\xspace}
\newcommand{\XttClangSOsPcrLsPrecision}{1\xspace}
\newcommand{\XttClangSOsPcrLsFone}{1\xspace}
\newcommand{\XttClangSOsPcrBapRecall}{0.889\xspace}
\newcommand{\XttClangSOsPcrBapPrecision}{1\xspace}
\newcommand{\XttClangSOsPcrBapFone}{0.941\xspace}
\newcommand{\XttClangSOsPcrGhiRecall}{1.000\xspace}
\newcommand{\XttClangSOsPcrGhiPrecision}{1\xspace}
\newcommand{\XttClangSOsPcrGhiFone}{1.000\xspace}
\newcommand{\XttClangSOsPcrRdaRecall}{0.970\xspace}
\newcommand{\XttClangSOsPcrRdaPrecision}{1\xspace}
\newcommand{\XttClangSOsPcrRdaFone}{0.985\xspace}
\newcommand{\XttClangSOsPcrRseRecall}{1.000\xspace}
\newcommand{\XttClangSOsPcrRsePrecision}{1\xspace}
\newcommand{\XttClangSOsPcrRseFone}{1.000\xspace}
\newcommand{\XttClangSOsSqlGT}{183494\xspace}
\newcommand{\XttClangSOsSqlLsRecall}{1\xspace}
\newcommand{\XttClangSOsSqlLsPrecision}{1\xspace}
\newcommand{\XttClangSOsSqlLsFone}{1\xspace}
\newcommand{\XttClangSOsSqlBapRecall}{0.848\xspace}
\newcommand{\XttClangSOsSqlBapPrecision}{1\xspace}
\newcommand{\XttClangSOsSqlBapFone}{0.918\xspace}
\newcommand{\XttClangSOsSqlGhiRecall}{0.947\xspace}
\newcommand{\XttClangSOsSqlGhiPrecision}{1\xspace}
\newcommand{\XttClangSOsSqlGhiFone}{0.973\xspace}
\newcommand{\XttClangSOsSqlRdaRecall}{0.944\xspace}
\newcommand{\XttClangSOsSqlRdaPrecision}{1\xspace}
\newcommand{\XttClangSOsSqlRdaFone}{0.971\xspace}
\newcommand{\XttClangSOsSqlRseRecall}{1.000\xspace}
\newcommand{\XttClangSOsSqlRsePrecision}{1\xspace}
\newcommand{\XttClangSOsSqlRseFone}{1.000\xspace}
\newcommand{\XttClangSOsVimGT}{489499\xspace}
\newcommand{\XttClangSOsVimLsRecall}{1\xspace}
\newcommand{\XttClangSOsVimLsPrecision}{1\xspace}
\newcommand{\XttClangSOsVimLsFone}{1\xspace}
\newcommand{\XttClangSOsVimBapRecall}{0.913\xspace}
\newcommand{\XttClangSOsVimBapPrecision}{1.000\xspace}
\newcommand{\XttClangSOsVimBapFone}{0.954\xspace}
\newcommand{\XttClangSOsVimGhiRecall}{1.000\xspace}
\newcommand{\XttClangSOsVimGhiPrecision}{1\xspace}
\newcommand{\XttClangSOsVimGhiFone}{1.000\xspace}
\newcommand{\XttClangSOsVimRdaRecall}{0.981\xspace}
\newcommand{\XttClangSOsVimRdaPrecision}{1\xspace}
\newcommand{\XttClangSOsVimRdaFone}{0.991\xspace}
\newcommand{\XttClangSOsVimRseRecall}{1.000\xspace}
\newcommand{\XttClangSOsVimRsePrecision}{1\xspace}
\newcommand{\XttClangSOsVimRseFone}{1.000\xspace}
\newcommand{\XttClangSOsVsfGT}{23899\xspace}
\newcommand{\XttClangSOsVsfLsRecall}{1\xspace}
\newcommand{\XttClangSOsVsfLsPrecision}{1\xspace}
\newcommand{\XttClangSOsVsfLsFone}{1\xspace}
\newcommand{\XttClangSOsVsfBapRecall}{0.990\xspace}
\newcommand{\XttClangSOsVsfBapPrecision}{1\xspace}
\newcommand{\XttClangSOsVsfBapFone}{0.995\xspace}
\newcommand{\XttClangSOsVsfGhiRecall}{1.000\xspace}
\newcommand{\XttClangSOsVsfGhiPrecision}{1\xspace}
\newcommand{\XttClangSOsVsfGhiFone}{1.000\xspace}
\newcommand{\XttClangSOsVsfRdaRecall}{0.967\xspace}
\newcommand{\XttClangSOsVsfRdaPrecision}{1\xspace}
\newcommand{\XttClangSOsVsfRdaFone}{0.983\xspace}
\newcommand{\XttClangSOsVsfRseRecall}{1.000\xspace}
\newcommand{\XttClangSOsVsfRsePrecision}{1\xspace}
\newcommand{\XttClangSOsVsfRseFone}{1.000\xspace}
\newcommand{\XttIccOoSzpGT}{24319\xspace}
\newcommand{\XttIccOoSzpLsRecall}{1\xspace}
\newcommand{\XttIccOoSzpLsPrecision}{1\xspace}
\newcommand{\XttIccOoSzpLsFone}{1\xspace}
\newcommand{\XttIccOoSzpBapRecall}{0.975\xspace}
\newcommand{\XttIccOoSzpBapPrecision}{1\xspace}
\newcommand{\XttIccOoSzpBapFone}{0.987\xspace}
\newcommand{\XttIccOoSzpGhiRecall}{1\xspace}
\newcommand{\XttIccOoSzpGhiPrecision}{1\xspace}
\newcommand{\XttIccOoSzpGhiFone}{1\xspace}
\newcommand{\XttIccOoSzpRdaRecall}{1\xspace}
\newcommand{\XttIccOoSzpRdaPrecision}{1\xspace}
\newcommand{\XttIccOoSzpRdaFone}{1\xspace}
\newcommand{\XttIccOoSzpRseRecall}{1\xspace}
\newcommand{\XttIccOoSzpRsePrecision}{1\xspace}
\newcommand{\XttIccOoSzpRseFone}{1\xspace}
\newcommand{\XttIccOoCapGT}{522487\xspace}
\newcommand{\XttIccOoCapLsRecall}{1\xspace}
\newcommand{\XttIccOoCapLsPrecision}{1\xspace}
\newcommand{\XttIccOoCapLsFone}{1\xspace}
\newcommand{\XttIccOoCapBapRecall}{0.232\xspace}
\newcommand{\XttIccOoCapBapPrecision}{1\xspace}
\newcommand{\XttIccOoCapBapFone}{0.377\xspace}
\newcommand{\XttIccOoCapGhiRecall}{0.987\xspace}
\newcommand{\XttIccOoCapGhiPrecision}{1\xspace}
\newcommand{\XttIccOoCapGhiFone}{0.994\xspace}
\newcommand{\XttIccOoCapRdaRecall}{0.854\xspace}
\newcommand{\XttIccOoCapRdaPrecision}{1\xspace}
\newcommand{\XttIccOoCapRdaFone}{0.921\xspace}
\newcommand{\XttIccOoCapRseRecall}{1\xspace}
\newcommand{\XttIccOoCapRsePrecision}{1\xspace}
\newcommand{\XttIccOoCapRseFone}{1\xspace}
\newcommand{\XttIccOoExmGT}{229423\xspace}
\newcommand{\XttIccOoExmLsRecall}{1\xspace}
\newcommand{\XttIccOoExmLsPrecision}{1\xspace}
\newcommand{\XttIccOoExmLsFone}{1\xspace}
\newcommand{\XttIccOoExmBapRecall}{0.851\xspace}
\newcommand{\XttIccOoExmBapPrecision}{1\xspace}
\newcommand{\XttIccOoExmBapFone}{0.920\xspace}
\newcommand{\XttIccOoExmGhiRecall}{0.948\xspace}
\newcommand{\XttIccOoExmGhiPrecision}{1\xspace}
\newcommand{\XttIccOoExmGhiFone}{0.973\xspace}
\newcommand{\XttIccOoExmRdaRecall}{0.960\xspace}
\newcommand{\XttIccOoExmRdaPrecision}{1\xspace}
\newcommand{\XttIccOoExmRdaFone}{0.980\xspace}
\newcommand{\XttIccOoExmRseRecall}{1.000\xspace}
\newcommand{\XttIccOoExmRsePrecision}{1\xspace}
\newcommand{\XttIccOoExmRseFone}{1.000\xspace}
\newcommand{\XttIccOoLgtGT}{49048\xspace}
\newcommand{\XttIccOoLgtLsRecall}{1\xspace}
\newcommand{\XttIccOoLgtLsPrecision}{1\xspace}
\newcommand{\XttIccOoLgtLsFone}{1\xspace}
\newcommand{\XttIccOoLgtBapRecall}{0.889\xspace}
\newcommand{\XttIccOoLgtBapPrecision}{1\xspace}
\newcommand{\XttIccOoLgtBapFone}{0.941\xspace}
\newcommand{\XttIccOoLgtGhiRecall}{1\xspace}
\newcommand{\XttIccOoLgtGhiPrecision}{1\xspace}
\newcommand{\XttIccOoLgtGhiFone}{1\xspace}
\newcommand{\XttIccOoLgtRdaRecall}{1\xspace}
\newcommand{\XttIccOoLgtRdaPrecision}{1\xspace}
\newcommand{\XttIccOoLgtRdaFone}{1\xspace}
\newcommand{\XttIccOoLgtRseRecall}{1.000\xspace}
\newcommand{\XttIccOoLgtRsePrecision}{1\xspace}
\newcommand{\XttIccOoLgtRseFone}{1.000\xspace}
\newcommand{\XttIccOoBzpGT}{26730\xspace}
\newcommand{\XttIccOoBzpLsRecall}{1\xspace}
\newcommand{\XttIccOoBzpLsPrecision}{1\xspace}
\newcommand{\XttIccOoBzpLsFone}{1\xspace}
\newcommand{\XttIccOoBzpBapRecall}{0.797\xspace}
\newcommand{\XttIccOoBzpBapPrecision}{1\xspace}
\newcommand{\XttIccOoBzpBapFone}{0.887\xspace}
\newcommand{\XttIccOoBzpGhiRecall}{1\xspace}
\newcommand{\XttIccOoBzpGhiPrecision}{1\xspace}
\newcommand{\XttIccOoBzpGhiFone}{1\xspace}
\newcommand{\XttIccOoBzpRdaRecall}{0.958\xspace}
\newcommand{\XttIccOoBzpRdaPrecision}{1\xspace}
\newcommand{\XttIccOoBzpRdaFone}{0.978\xspace}
\newcommand{\XttIccOoBzpRseRecall}{1\xspace}
\newcommand{\XttIccOoBzpRsePrecision}{1\xspace}
\newcommand{\XttIccOoBzpRseFone}{1\xspace}
\newcommand{\XttIccOoGccGT}{1720499\xspace}
\newcommand{\XttIccOoGccLsRecall}{1\xspace}
\newcommand{\XttIccOoGccLsPrecision}{1\xspace}
\newcommand{\XttIccOoGccLsFone}{1\xspace}
\newcommand{\XttIccOoGccBapRecall}{0.711\xspace}
\newcommand{\XttIccOoGccBapPrecision}{1\xspace}
\newcommand{\XttIccOoGccBapFone}{0.831\xspace}
\newcommand{\XttIccOoGccGhiRecall}{0.982\xspace}
\newcommand{\XttIccOoGccGhiPrecision}{1\xspace}
\newcommand{\XttIccOoGccGhiFone}{0.991\xspace}
\newcommand{\XttIccOoGccRdaRecall}{0.870\xspace}
\newcommand{\XttIccOoGccRdaPrecision}{1\xspace}
\newcommand{\XttIccOoGccRdaFone}{0.931\xspace}
\newcommand{\XttIccOoGccRseRecall}{0.998\xspace}
\newcommand{\XttIccOoGccRsePrecision}{1.000\xspace}
\newcommand{\XttIccOoGccRseFone}{0.999\xspace}
\newcommand{\XttIccOoGzpGT}{16523\xspace}
\newcommand{\XttIccOoGzpLsRecall}{1\xspace}
\newcommand{\XttIccOoGzpLsPrecision}{1\xspace}
\newcommand{\XttIccOoGzpLsFone}{1\xspace}
\newcommand{\XttIccOoGzpBapRecall}{0.991\xspace}
\newcommand{\XttIccOoGzpBapPrecision}{1\xspace}
\newcommand{\XttIccOoGzpBapFone}{0.995\xspace}
\newcommand{\XttIccOoGzpGhiRecall}{1\xspace}
\newcommand{\XttIccOoGzpGhiPrecision}{1\xspace}
\newcommand{\XttIccOoGzpGhiFone}{1\xspace}
\newcommand{\XttIccOoGzpRdaRecall}{1\xspace}
\newcommand{\XttIccOoGzpRdaPrecision}{1\xspace}
\newcommand{\XttIccOoGzpRdaFone}{1\xspace}
\newcommand{\XttIccOoGzpRseRecall}{0.999\xspace}
\newcommand{\XttIccOoGzpRsePrecision}{1\xspace}
\newcommand{\XttIccOoGzpRseFone}{1.000\xspace}
\newcommand{\XttIccOoOggGT}{64368\xspace}
\newcommand{\XttIccOoOggLsRecall}{1\xspace}
\newcommand{\XttIccOoOggLsPrecision}{1\xspace}
\newcommand{\XttIccOoOggLsFone}{1\xspace}
\newcommand{\XttIccOoOggBapRecall}{0.982\xspace}
\newcommand{\XttIccOoOggBapPrecision}{1\xspace}
\newcommand{\XttIccOoOggBapFone}{0.991\xspace}
\newcommand{\XttIccOoOggGhiRecall}{1\xspace}
\newcommand{\XttIccOoOggGhiPrecision}{1\xspace}
\newcommand{\XttIccOoOggGhiFone}{1\xspace}
\newcommand{\XttIccOoOggRdaRecall}{1\xspace}
\newcommand{\XttIccOoOggRdaPrecision}{1\xspace}
\newcommand{\XttIccOoOggRdaFone}{1\xspace}
\newcommand{\XttIccOoOggRseRecall}{1\xspace}
\newcommand{\XttIccOoOggRsePrecision}{1\xspace}
\newcommand{\XttIccOoOggRseFone}{1\xspace}
\newcommand{\XttIccOoNgxGT}{196559\xspace}
\newcommand{\XttIccOoNgxLsRecall}{1\xspace}
\newcommand{\XttIccOoNgxLsPrecision}{1\xspace}
\newcommand{\XttIccOoNgxLsFone}{1\xspace}
\newcommand{\XttIccOoNgxBapRecall}{0.969\xspace}
\newcommand{\XttIccOoNgxBapPrecision}{1\xspace}
\newcommand{\XttIccOoNgxBapFone}{0.984\xspace}
\newcommand{\XttIccOoNgxGhiRecall}{1\xspace}
\newcommand{\XttIccOoNgxGhiPrecision}{1\xspace}
\newcommand{\XttIccOoNgxGhiFone}{1\xspace}
\newcommand{\XttIccOoNgxRdaRecall}{0.998\xspace}
\newcommand{\XttIccOoNgxRdaPrecision}{1\xspace}
\newcommand{\XttIccOoNgxRdaFone}{0.999\xspace}
\newcommand{\XttIccOoNgxRseRecall}{1.000\xspace}
\newcommand{\XttIccOoNgxRsePrecision}{1\xspace}
\newcommand{\XttIccOoNgxRseFone}{1.000\xspace}
\newcommand{\XttIccOoSshGT}{236299\xspace}
\newcommand{\XttIccOoSshLsRecall}{1\xspace}
\newcommand{\XttIccOoSshLsPrecision}{1\xspace}
\newcommand{\XttIccOoSshLsFone}{1\xspace}
\newcommand{\XttIccOoSshBapRecall}{0.955\xspace}
\newcommand{\XttIccOoSshBapPrecision}{1\xspace}
\newcommand{\XttIccOoSshBapFone}{0.977\xspace}
\newcommand{\XttIccOoSshGhiRecall}{0.962\xspace}
\newcommand{\XttIccOoSshGhiPrecision}{1.000\xspace}
\newcommand{\XttIccOoSshGhiFone}{0.981\xspace}
\newcommand{\XttIccOoSshRdaRecall}{0.949\xspace}
\newcommand{\XttIccOoSshRdaPrecision}{0.999\xspace}
\newcommand{\XttIccOoSshRdaFone}{0.973\xspace}
\newcommand{\XttIccOoSshRseRecall}{0.983\xspace}
\newcommand{\XttIccOoSshRsePrecision}{1\xspace}
\newcommand{\XttIccOoSshRseFone}{0.991\xspace}
\newcommand{\XttIccOoPcrGT}{7487\xspace}
\newcommand{\XttIccOoPcrLsRecall}{1\xspace}
\newcommand{\XttIccOoPcrLsPrecision}{1\xspace}
\newcommand{\XttIccOoPcrLsFone}{1\xspace}
\newcommand{\XttIccOoPcrBapRecall}{0.942\xspace}
\newcommand{\XttIccOoPcrBapPrecision}{1\xspace}
\newcommand{\XttIccOoPcrBapFone}{0.970\xspace}
\newcommand{\XttIccOoPcrGhiRecall}{1\xspace}
\newcommand{\XttIccOoPcrGhiPrecision}{1\xspace}
\newcommand{\XttIccOoPcrGhiFone}{1\xspace}
\newcommand{\XttIccOoPcrRdaRecall}{0.997\xspace}
\newcommand{\XttIccOoPcrRdaPrecision}{1\xspace}
\newcommand{\XttIccOoPcrRdaFone}{0.999\xspace}
\newcommand{\XttIccOoPcrRseRecall}{1\xspace}
\newcommand{\XttIccOoPcrRsePrecision}{1\xspace}
\newcommand{\XttIccOoPcrRseFone}{1\xspace}
\newcommand{\XttIccOoSqlGT}{284617\xspace}
\newcommand{\XttIccOoSqlLsRecall}{1\xspace}
\newcommand{\XttIccOoSqlLsPrecision}{1\xspace}
\newcommand{\XttIccOoSqlLsFone}{1\xspace}
\newcommand{\XttIccOoSqlBapRecall}{0.881\xspace}
\newcommand{\XttIccOoSqlBapPrecision}{1\xspace}
\newcommand{\XttIccOoSqlBapFone}{0.937\xspace}
\newcommand{\XttIccOoSqlGhiRecall}{0.963\xspace}
\newcommand{\XttIccOoSqlGhiPrecision}{1\xspace}
\newcommand{\XttIccOoSqlGhiFone}{0.981\xspace}
\newcommand{\XttIccOoSqlRdaRecall}{0.996\xspace}
\newcommand{\XttIccOoSqlRdaPrecision}{1\xspace}
\newcommand{\XttIccOoSqlRdaFone}{0.998\xspace}
\newcommand{\XttIccOoSqlRseRecall}{0.999\xspace}
\newcommand{\XttIccOoSqlRsePrecision}{1.000\xspace}
\newcommand{\XttIccOoSqlRseFone}{0.999\xspace}
\newcommand{\XttIccOoVimGT}{815603\xspace}
\newcommand{\XttIccOoVimLsRecall}{1\xspace}
\newcommand{\XttIccOoVimLsPrecision}{1\xspace}
\newcommand{\XttIccOoVimLsFone}{1\xspace}
\newcommand{\XttIccOoVimBapRecall}{0.915\xspace}
\newcommand{\XttIccOoVimBapPrecision}{1\xspace}
\newcommand{\XttIccOoVimBapFone}{0.955\xspace}
\newcommand{\XttIccOoVimGhiRecall}{0.995\xspace}
\newcommand{\XttIccOoVimGhiPrecision}{1\xspace}
\newcommand{\XttIccOoVimGhiFone}{0.997\xspace}
\newcommand{\XttIccOoVimRdaRecall}{0.985\xspace}
\newcommand{\XttIccOoVimRdaPrecision}{1\xspace}
\newcommand{\XttIccOoVimRdaFone}{0.992\xspace}
\newcommand{\XttIccOoVimRseRecall}{1\xspace}
\newcommand{\XttIccOoVimRsePrecision}{1\xspace}
\newcommand{\XttIccOoVimRseFone}{1\xspace}
\newcommand{\XttIccOoVsfGT}{44735\xspace}
\newcommand{\XttIccOoVsfLsRecall}{1\xspace}
\newcommand{\XttIccOoVsfLsPrecision}{1\xspace}
\newcommand{\XttIccOoVsfLsFone}{1\xspace}
\newcommand{\XttIccOoVsfBapRecall}{0.997\xspace}
\newcommand{\XttIccOoVsfBapPrecision}{1\xspace}
\newcommand{\XttIccOoVsfBapFone}{0.998\xspace}
\newcommand{\XttIccOoVsfGhiRecall}{0.997\xspace}
\newcommand{\XttIccOoVsfGhiPrecision}{1.000\xspace}
\newcommand{\XttIccOoVsfGhiFone}{0.999\xspace}
\newcommand{\XttIccOoVsfRdaRecall}{0.971\xspace}
\newcommand{\XttIccOoVsfRdaPrecision}{0.999\xspace}
\newcommand{\XttIccOoVsfRdaFone}{0.985\xspace}
\newcommand{\XttIccOoVsfRseRecall}{1.000\xspace}
\newcommand{\XttIccOoVsfRsePrecision}{1\xspace}
\newcommand{\XttIccOoVsfRseFone}{1.000\xspace}
\newcommand{\XttIccOaSzpGT}{12917\xspace}
\newcommand{\XttIccOaSzpLsRecall}{1\xspace}
\newcommand{\XttIccOaSzpLsPrecision}{1\xspace}
\newcommand{\XttIccOaSzpLsFone}{1\xspace}
\newcommand{\XttIccOaSzpBapRecall}{0.977\xspace}
\newcommand{\XttIccOaSzpBapPrecision}{1\xspace}
\newcommand{\XttIccOaSzpBapFone}{0.989\xspace}
\newcommand{\XttIccOaSzpGhiRecall}{1\xspace}
\newcommand{\XttIccOaSzpGhiPrecision}{1\xspace}
\newcommand{\XttIccOaSzpGhiFone}{1\xspace}
\newcommand{\XttIccOaSzpRdaRecall}{0.977\xspace}
\newcommand{\XttIccOaSzpRdaPrecision}{1\xspace}
\newcommand{\XttIccOaSzpRdaFone}{0.989\xspace}
\newcommand{\XttIccOaSzpRseRecall}{1\xspace}
\newcommand{\XttIccOaSzpRsePrecision}{1\xspace}
\newcommand{\XttIccOaSzpRseFone}{1\xspace}
\newcommand{\XttIccOaCapGT}{224453\xspace}
\newcommand{\XttIccOaCapLsRecall}{1\xspace}
\newcommand{\XttIccOaCapLsPrecision}{1\xspace}
\newcommand{\XttIccOaCapLsFone}{1\xspace}
\newcommand{\XttIccOaCapBapRecall}{0.259\xspace}
\newcommand{\XttIccOaCapBapPrecision}{1\xspace}
\newcommand{\XttIccOaCapBapFone}{0.411\xspace}
\newcommand{\XttIccOaCapGhiRecall}{0.761\xspace}
\newcommand{\XttIccOaCapGhiPrecision}{1.000\xspace}
\newcommand{\XttIccOaCapGhiFone}{0.864\xspace}
\newcommand{\XttIccOaCapRdaRecall}{0.348\xspace}
\newcommand{\XttIccOaCapRdaPrecision}{1\xspace}
\newcommand{\XttIccOaCapRdaFone}{0.516\xspace}
\newcommand{\XttIccOaCapRseRecall}{0.792\xspace}
\newcommand{\XttIccOaCapRsePrecision}{1.000\xspace}
\newcommand{\XttIccOaCapRseFone}{0.884\xspace}
\newcommand{\XttIccOaExmGT}{139234\xspace}
\newcommand{\XttIccOaExmLsRecall}{1\xspace}
\newcommand{\XttIccOaExmLsPrecision}{1\xspace}
\newcommand{\XttIccOaExmLsFone}{1\xspace}
\newcommand{\XttIccOaExmBapRecall}{0.835\xspace}
\newcommand{\XttIccOaExmBapPrecision}{1\xspace}
\newcommand{\XttIccOaExmBapFone}{0.910\xspace}
\newcommand{\XttIccOaExmGhiRecall}{0.948\xspace}
\newcommand{\XttIccOaExmGhiPrecision}{1\xspace}
\newcommand{\XttIccOaExmGhiFone}{0.973\xspace}
\newcommand{\XttIccOaExmRdaRecall}{0.798\xspace}
\newcommand{\XttIccOaExmRdaPrecision}{1\xspace}
\newcommand{\XttIccOaExmRdaFone}{0.888\xspace}
\newcommand{\XttIccOaExmRseRecall}{0.986\xspace}
\newcommand{\XttIccOaExmRsePrecision}{1.000\xspace}
\newcommand{\XttIccOaExmRseFone}{0.993\xspace}
\newcommand{\XttIccOaLgtGT}{27333\xspace}
\newcommand{\XttIccOaLgtLsRecall}{1\xspace}
\newcommand{\XttIccOaLgtLsPrecision}{1\xspace}
\newcommand{\XttIccOaLgtLsFone}{1\xspace}
\newcommand{\XttIccOaLgtBapRecall}{0.881\xspace}
\newcommand{\XttIccOaLgtBapPrecision}{1\xspace}
\newcommand{\XttIccOaLgtBapFone}{0.937\xspace}
\newcommand{\XttIccOaLgtGhiRecall}{0.998\xspace}
\newcommand{\XttIccOaLgtGhiPrecision}{1\xspace}
\newcommand{\XttIccOaLgtGhiFone}{0.999\xspace}
\newcommand{\XttIccOaLgtRdaRecall}{0.876\xspace}
\newcommand{\XttIccOaLgtRdaPrecision}{1\xspace}
\newcommand{\XttIccOaLgtRdaFone}{0.934\xspace}
\newcommand{\XttIccOaLgtRseRecall}{0.997\xspace}
\newcommand{\XttIccOaLgtRsePrecision}{1\xspace}
\newcommand{\XttIccOaLgtRseFone}{0.999\xspace}
\newcommand{\XttIccOaBzpGT}{11857\xspace}
\newcommand{\XttIccOaBzpLsRecall}{1\xspace}
\newcommand{\XttIccOaBzpLsPrecision}{1\xspace}
\newcommand{\XttIccOaBzpLsFone}{1\xspace}
\newcommand{\XttIccOaBzpBapRecall}{0.798\xspace}
\newcommand{\XttIccOaBzpBapPrecision}{1\xspace}
\newcommand{\XttIccOaBzpBapFone}{0.887\xspace}
\newcommand{\XttIccOaBzpGhiRecall}{1\xspace}
\newcommand{\XttIccOaBzpGhiPrecision}{1\xspace}
\newcommand{\XttIccOaBzpGhiFone}{1\xspace}
\newcommand{\XttIccOaBzpRdaRecall}{0.788\xspace}
\newcommand{\XttIccOaBzpRdaPrecision}{1\xspace}
\newcommand{\XttIccOaBzpRdaFone}{0.882\xspace}
\newcommand{\XttIccOaBzpRseRecall}{0.924\xspace}
\newcommand{\XttIccOaBzpRsePrecision}{1\xspace}
\newcommand{\XttIccOaBzpRseFone}{0.960\xspace}
\newcommand{\XttIccOaGccGT}{786775\xspace}
\newcommand{\XttIccOaGccLsRecall}{1\xspace}
\newcommand{\XttIccOaGccLsPrecision}{1\xspace}
\newcommand{\XttIccOaGccLsFone}{1\xspace}
\newcommand{\XttIccOaGccBapRecall}{0.761\xspace}
\newcommand{\XttIccOaGccBapPrecision}{1\xspace}
\newcommand{\XttIccOaGccBapFone}{0.865\xspace}
\newcommand{\XttIccOaGccGhiRecall}{0.988\xspace}
\newcommand{\XttIccOaGccGhiPrecision}{1\xspace}
\newcommand{\XttIccOaGccGhiFone}{0.994\xspace}
\newcommand{\XttIccOaGccRdaRecall}{0.713\xspace}
\newcommand{\XttIccOaGccRdaPrecision}{1.000\xspace}
\newcommand{\XttIccOaGccRdaFone}{0.833\xspace}
\newcommand{\XttIccOaGccRseRecall}{0.969\xspace}
\newcommand{\XttIccOaGccRsePrecision}{1.000\xspace}
\newcommand{\XttIccOaGccRseFone}{0.984\xspace}
\newcommand{\XttIccOaGzpGT}{9190\xspace}
\newcommand{\XttIccOaGzpLsRecall}{1\xspace}
\newcommand{\XttIccOaGzpLsPrecision}{1\xspace}
\newcommand{\XttIccOaGzpLsFone}{1\xspace}
\newcommand{\XttIccOaGzpBapRecall}{0.987\xspace}
\newcommand{\XttIccOaGzpBapPrecision}{1\xspace}
\newcommand{\XttIccOaGzpBapFone}{0.993\xspace}
\newcommand{\XttIccOaGzpGhiRecall}{1\xspace}
\newcommand{\XttIccOaGzpGhiPrecision}{1\xspace}
\newcommand{\XttIccOaGzpGhiFone}{1\xspace}
\newcommand{\XttIccOaGzpRdaRecall}{0.987\xspace}
\newcommand{\XttIccOaGzpRdaPrecision}{1\xspace}
\newcommand{\XttIccOaGzpRdaFone}{0.993\xspace}
\newcommand{\XttIccOaGzpRseRecall}{0.997\xspace}
\newcommand{\XttIccOaGzpRsePrecision}{1\xspace}
\newcommand{\XttIccOaGzpRseFone}{0.998\xspace}
\newcommand{\XttIccOaOggGT}{35971\xspace}
\newcommand{\XttIccOaOggLsRecall}{1\xspace}
\newcommand{\XttIccOaOggLsPrecision}{1\xspace}
\newcommand{\XttIccOaOggLsFone}{1\xspace}
\newcommand{\XttIccOaOggBapRecall}{0.983\xspace}
\newcommand{\XttIccOaOggBapPrecision}{1\xspace}
\newcommand{\XttIccOaOggBapFone}{0.992\xspace}
\newcommand{\XttIccOaOggGhiRecall}{1\xspace}
\newcommand{\XttIccOaOggGhiPrecision}{1\xspace}
\newcommand{\XttIccOaOggGhiFone}{1\xspace}
\newcommand{\XttIccOaOggRdaRecall}{0.983\xspace}
\newcommand{\XttIccOaOggRdaPrecision}{1\xspace}
\newcommand{\XttIccOaOggRdaFone}{0.992\xspace}
\newcommand{\XttIccOaOggRseRecall}{0.994\xspace}
\newcommand{\XttIccOaOggRsePrecision}{1\xspace}
\newcommand{\XttIccOaOggRseFone}{0.997\xspace}
\newcommand{\XttIccOaNgxGT}{106043\xspace}
\newcommand{\XttIccOaNgxLsRecall}{1\xspace}
\newcommand{\XttIccOaNgxLsPrecision}{1\xspace}
\newcommand{\XttIccOaNgxLsFone}{1\xspace}
\newcommand{\XttIccOaNgxBapRecall}{0.972\xspace}
\newcommand{\XttIccOaNgxBapPrecision}{1\xspace}
\newcommand{\XttIccOaNgxBapFone}{0.986\xspace}
\newcommand{\XttIccOaNgxGhiRecall}{0.999\xspace}
\newcommand{\XttIccOaNgxGhiPrecision}{1\xspace}
\newcommand{\XttIccOaNgxGhiFone}{0.999\xspace}
\newcommand{\XttIccOaNgxRdaRecall}{0.971\xspace}
\newcommand{\XttIccOaNgxRdaPrecision}{1\xspace}
\newcommand{\XttIccOaNgxRdaFone}{0.985\xspace}
\newcommand{\XttIccOaNgxRseRecall}{0.998\xspace}
\newcommand{\XttIccOaNgxRsePrecision}{1\xspace}
\newcommand{\XttIccOaNgxRseFone}{0.999\xspace}
\newcommand{\XttIccOaSshGT}{127720\xspace}
\newcommand{\XttIccOaSshLsRecall}{1\xspace}
\newcommand{\XttIccOaSshLsPrecision}{1\xspace}
\newcommand{\XttIccOaSshLsFone}{1\xspace}
\newcommand{\XttIccOaSshBapRecall}{0.954\xspace}
\newcommand{\XttIccOaSshBapPrecision}{1\xspace}
\newcommand{\XttIccOaSshBapFone}{0.977\xspace}
\newcommand{\XttIccOaSshGhiRecall}{0.982\xspace}
\newcommand{\XttIccOaSshGhiPrecision}{1\xspace}
\newcommand{\XttIccOaSshGhiFone}{0.991\xspace}
\newcommand{\XttIccOaSshRdaRecall}{0.945\xspace}
\newcommand{\XttIccOaSshRdaPrecision}{0.999\xspace}
\newcommand{\XttIccOaSshRdaFone}{0.971\xspace}
\newcommand{\XttIccOaSshRseRecall}{0.984\xspace}
\newcommand{\XttIccOaSshRsePrecision}{1\xspace}
\newcommand{\XttIccOaSshRseFone}{0.992\xspace}
\newcommand{\XttIccOaPcrGT}{4375\xspace}
\newcommand{\XttIccOaPcrLsRecall}{1\xspace}
\newcommand{\XttIccOaPcrLsPrecision}{1\xspace}
\newcommand{\XttIccOaPcrLsFone}{1\xspace}
\newcommand{\XttIccOaPcrBapRecall}{0.913\xspace}
\newcommand{\XttIccOaPcrBapPrecision}{1\xspace}
\newcommand{\XttIccOaPcrBapFone}{0.955\xspace}
\newcommand{\XttIccOaPcrGhiRecall}{1\xspace}
\newcommand{\XttIccOaPcrGhiPrecision}{1\xspace}
\newcommand{\XttIccOaPcrGhiFone}{1\xspace}
\newcommand{\XttIccOaPcrRdaRecall}{0.850\xspace}
\newcommand{\XttIccOaPcrRdaPrecision}{1\xspace}
\newcommand{\XttIccOaPcrRdaFone}{0.919\xspace}
\newcommand{\XttIccOaPcrRseRecall}{0.995\xspace}
\newcommand{\XttIccOaPcrRsePrecision}{1\xspace}
\newcommand{\XttIccOaPcrRseFone}{0.997\xspace}
\newcommand{\XttIccOaSqlGT}{159351\xspace}
\newcommand{\XttIccOaSqlLsRecall}{1\xspace}
\newcommand{\XttIccOaSqlLsPrecision}{1\xspace}
\newcommand{\XttIccOaSqlLsFone}{1\xspace}
\newcommand{\XttIccOaSqlBapRecall}{0.857\xspace}
\newcommand{\XttIccOaSqlBapPrecision}{1\xspace}
\newcommand{\XttIccOaSqlBapFone}{0.923\xspace}
\newcommand{\XttIccOaSqlGhiRecall}{0.952\xspace}
\newcommand{\XttIccOaSqlGhiPrecision}{1\xspace}
\newcommand{\XttIccOaSqlGhiFone}{0.975\xspace}
\newcommand{\XttIccOaSqlRdaRecall}{0.857\xspace}
\newcommand{\XttIccOaSqlRdaPrecision}{1\xspace}
\newcommand{\XttIccOaSqlRdaFone}{0.923\xspace}
\newcommand{\XttIccOaSqlRseRecall}{0.940\xspace}
\newcommand{\XttIccOaSqlRsePrecision}{1.000\xspace}
\newcommand{\XttIccOaSqlRseFone}{0.969\xspace}
\newcommand{\XttIccOaVimGT}{490419\xspace}
\newcommand{\XttIccOaVimLsRecall}{1\xspace}
\newcommand{\XttIccOaVimLsPrecision}{1\xspace}
\newcommand{\XttIccOaVimLsFone}{1\xspace}
\newcommand{\XttIccOaVimBapRecall}{0.914\xspace}
\newcommand{\XttIccOaVimBapPrecision}{1\xspace}
\newcommand{\XttIccOaVimBapFone}{0.955\xspace}
\newcommand{\XttIccOaVimGhiRecall}{0.997\xspace}
\newcommand{\XttIccOaVimGhiPrecision}{1\xspace}
\newcommand{\XttIccOaVimGhiFone}{0.999\xspace}
\newcommand{\XttIccOaVimRdaRecall}{0.902\xspace}
\newcommand{\XttIccOaVimRdaPrecision}{1\xspace}
\newcommand{\XttIccOaVimRdaFone}{0.949\xspace}
\newcommand{\XttIccOaVimRseRecall}{0.982\xspace}
\newcommand{\XttIccOaVimRsePrecision}{1.000\xspace}
\newcommand{\XttIccOaVimRseFone}{0.991\xspace}
\newcommand{\XttIccOaVsfGT}{24509\xspace}
\newcommand{\XttIccOaVsfLsRecall}{1\xspace}
\newcommand{\XttIccOaVsfLsPrecision}{1\xspace}
\newcommand{\XttIccOaVsfLsFone}{1\xspace}
\newcommand{\XttIccOaVsfBapRecall}{0.987\xspace}
\newcommand{\XttIccOaVsfBapPrecision}{1\xspace}
\newcommand{\XttIccOaVsfBapFone}{0.993\xspace}
\newcommand{\XttIccOaVsfGhiRecall}{1\xspace}
\newcommand{\XttIccOaVsfGhiPrecision}{1\xspace}
\newcommand{\XttIccOaVsfGhiFone}{1\xspace}
\newcommand{\XttIccOaVsfRdaRecall}{0.971\xspace}
\newcommand{\XttIccOaVsfRdaPrecision}{0.992\xspace}
\newcommand{\XttIccOaVsfRdaFone}{0.981\xspace}
\newcommand{\XttIccOaVsfRseRecall}{1.000\xspace}
\newcommand{\XttIccOaVsfRsePrecision}{1\xspace}
\newcommand{\XttIccOaVsfRseFone}{1.000\xspace}
\newcommand{\XttIccObSzpGT}{23426\xspace}
\newcommand{\XttIccObSzpLsRecall}{1\xspace}
\newcommand{\XttIccObSzpLsPrecision}{0.993\xspace}
\newcommand{\XttIccObSzpLsFone}{0.997\xspace}
\newcommand{\XttIccObSzpBapRecall}{0.971\xspace}
\newcommand{\XttIccObSzpBapPrecision}{1\xspace}
\newcommand{\XttIccObSzpBapFone}{0.986\xspace}
\newcommand{\XttIccObSzpGhiRecall}{1\xspace}
\newcommand{\XttIccObSzpGhiPrecision}{1\xspace}
\newcommand{\XttIccObSzpGhiFone}{1\xspace}
\newcommand{\XttIccObSzpRdaRecall}{0.969\xspace}
\newcommand{\XttIccObSzpRdaPrecision}{1\xspace}
\newcommand{\XttIccObSzpRdaFone}{0.984\xspace}
\newcommand{\XttIccObSzpRseRecall}{1\xspace}
\newcommand{\XttIccObSzpRsePrecision}{1\xspace}
\newcommand{\XttIccObSzpRseFone}{1\xspace}
\newcommand{\XttIccObCapGT}{294067\xspace}
\newcommand{\XttIccObCapLsRecall}{1\xspace}
\newcommand{\XttIccObCapLsPrecision}{0.995\xspace}
\newcommand{\XttIccObCapLsFone}{0.998\xspace}
\newcommand{\XttIccObCapBapRecall}{0.266\xspace}
\newcommand{\XttIccObCapBapPrecision}{1\xspace}
\newcommand{\XttIccObCapBapFone}{0.421\xspace}
\newcommand{\XttIccObCapGhiRecall}{0.909\xspace}
\newcommand{\XttIccObCapGhiPrecision}{1.000\xspace}
\newcommand{\XttIccObCapGhiFone}{0.952\xspace}
\newcommand{\XttIccObCapRdaRecall}{0.502\xspace}
\newcommand{\XttIccObCapRdaPrecision}{1\xspace}
\newcommand{\XttIccObCapRdaFone}{0.668\xspace}
\newcommand{\XttIccObCapRseRecall}{0.718\xspace}
\newcommand{\XttIccObCapRsePrecision}{1.000\xspace}
\newcommand{\XttIccObCapRseFone}{0.836\xspace}
\newcommand{\XttIccObExmGT}{212288\xspace}
\newcommand{\XttIccObExmLsRecall}{1\xspace}
\newcommand{\XttIccObExmLsPrecision}{0.995\xspace}
\newcommand{\XttIccObExmLsFone}{0.998\xspace}
\newcommand{\XttIccObExmBapRecall}{0.850\xspace}
\newcommand{\XttIccObExmBapPrecision}{1.000\xspace}
\newcommand{\XttIccObExmBapFone}{0.919\xspace}
\newcommand{\XttIccObExmGhiRecall}{0.939\xspace}
\newcommand{\XttIccObExmGhiPrecision}{1.000\xspace}
\newcommand{\XttIccObExmGhiFone}{0.969\xspace}
\newcommand{\XttIccObExmRdaRecall}{0.817\xspace}
\newcommand{\XttIccObExmRdaPrecision}{1\xspace}
\newcommand{\XttIccObExmRdaFone}{0.899\xspace}
\newcommand{\XttIccObExmRseRecall}{0.981\xspace}
\newcommand{\XttIccObExmRsePrecision}{1\xspace}
\newcommand{\XttIccObExmRseFone}{0.991\xspace}
\newcommand{\XttIccObLgtGT}{37864\xspace}
\newcommand{\XttIccObLgtLsRecall}{1\xspace}
\newcommand{\XttIccObLgtLsPrecision}{0.989\xspace}
\newcommand{\XttIccObLgtLsFone}{0.995\xspace}
\newcommand{\XttIccObLgtBapRecall}{0.876\xspace}
\newcommand{\XttIccObLgtBapPrecision}{0.998\xspace}
\newcommand{\XttIccObLgtBapFone}{0.933\xspace}
\newcommand{\XttIccObLgtGhiRecall}{0.998\xspace}
\newcommand{\XttIccObLgtGhiPrecision}{0.999\xspace}
\newcommand{\XttIccObLgtGhiFone}{0.998\xspace}
\newcommand{\XttIccObLgtRdaRecall}{0.807\xspace}
\newcommand{\XttIccObLgtRdaPrecision}{1\xspace}
\newcommand{\XttIccObLgtRdaFone}{0.893\xspace}
\newcommand{\XttIccObLgtRseRecall}{0.990\xspace}
\newcommand{\XttIccObLgtRsePrecision}{1\xspace}
\newcommand{\XttIccObLgtRseFone}{0.995\xspace}
\newcommand{\XttIccObBzpGT}{24443\xspace}
\newcommand{\XttIccObBzpLsRecall}{1\xspace}
\newcommand{\XttIccObBzpLsPrecision}{0.997\xspace}
\newcommand{\XttIccObBzpLsFone}{0.998\xspace}
\newcommand{\XttIccObBzpBapRecall}{0.836\xspace}
\newcommand{\XttIccObBzpBapPrecision}{1.000\xspace}
\newcommand{\XttIccObBzpBapFone}{0.911\xspace}
\newcommand{\XttIccObBzpGhiRecall}{1\xspace}
\newcommand{\XttIccObBzpGhiPrecision}{1.000\xspace}
\newcommand{\XttIccObBzpGhiFone}{1.000\xspace}
\newcommand{\XttIccObBzpRdaRecall}{0.829\xspace}
\newcommand{\XttIccObBzpRdaPrecision}{1\xspace}
\newcommand{\XttIccObBzpRdaFone}{0.906\xspace}
\newcommand{\XttIccObBzpRseRecall}{0.908\xspace}
\newcommand{\XttIccObBzpRsePrecision}{1\xspace}
\newcommand{\XttIccObBzpRseFone}{0.952\xspace}
\newcommand{\XttIccObGzpGT}{21185\xspace}
\newcommand{\XttIccObGzpLsRecall}{1\xspace}
\newcommand{\XttIccObGzpLsPrecision}{0.996\xspace}
\newcommand{\XttIccObGzpLsFone}{0.998\xspace}
\newcommand{\XttIccObGzpBapRecall}{0.992\xspace}
\newcommand{\XttIccObGzpBapPrecision}{1.000\xspace}
\newcommand{\XttIccObGzpBapFone}{0.996\xspace}
\newcommand{\XttIccObGzpGhiRecall}{1\xspace}
\newcommand{\XttIccObGzpGhiPrecision}{1.000\xspace}
\newcommand{\XttIccObGzpGhiFone}{1.000\xspace}
\newcommand{\XttIccObGzpRdaRecall}{0.980\xspace}
\newcommand{\XttIccObGzpRdaPrecision}{1\xspace}
\newcommand{\XttIccObGzpRdaFone}{0.990\xspace}
\newcommand{\XttIccObGzpRseRecall}{0.998\xspace}
\newcommand{\XttIccObGzpRsePrecision}{1\xspace}
\newcommand{\XttIccObGzpRseFone}{0.999\xspace}
\newcommand{\XttIccObOggGT}{98145\xspace}
\newcommand{\XttIccObOggLsRecall}{1\xspace}
\newcommand{\XttIccObOggLsPrecision}{0.996\xspace}
\newcommand{\XttIccObOggLsFone}{0.998\xspace}
\newcommand{\XttIccObOggBapRecall}{0.986\xspace}
\newcommand{\XttIccObOggBapPrecision}{1.000\xspace}
\newcommand{\XttIccObOggBapFone}{0.993\xspace}
\newcommand{\XttIccObOggGhiRecall}{1\xspace}
\newcommand{\XttIccObOggGhiPrecision}{1.000\xspace}
\newcommand{\XttIccObOggGhiFone}{1.000\xspace}
\newcommand{\XttIccObOggRdaRecall}{0.985\xspace}
\newcommand{\XttIccObOggRdaPrecision}{1\xspace}
\newcommand{\XttIccObOggRdaFone}{0.993\xspace}
\newcommand{\XttIccObOggRseRecall}{0.998\xspace}
\newcommand{\XttIccObOggRsePrecision}{1\xspace}
\newcommand{\XttIccObOggRseFone}{0.999\xspace}
\newcommand{\XttIccObNgxGT}{147072\xspace}
\newcommand{\XttIccObNgxLsRecall}{1\xspace}
\newcommand{\XttIccObNgxLsPrecision}{0.991\xspace}
\newcommand{\XttIccObNgxLsFone}{0.996\xspace}
\newcommand{\XttIccObNgxBapRecall}{0.974\xspace}
\newcommand{\XttIccObNgxBapPrecision}{1.000\xspace}
\newcommand{\XttIccObNgxBapFone}{0.987\xspace}
\newcommand{\XttIccObNgxGhiRecall}{0.999\xspace}
\newcommand{\XttIccObNgxGhiPrecision}{1.000\xspace}
\newcommand{\XttIccObNgxGhiFone}{0.999\xspace}
\newcommand{\XttIccObNgxRdaRecall}{0.965\xspace}
\newcommand{\XttIccObNgxRdaPrecision}{1\xspace}
\newcommand{\XttIccObNgxRdaFone}{0.982\xspace}
\newcommand{\XttIccObNgxRseRecall}{0.996\xspace}
\newcommand{\XttIccObNgxRsePrecision}{1\xspace}
\newcommand{\XttIccObNgxRseFone}{0.998\xspace}
\newcommand{\XttIccObSshGT}{203999\xspace}
\newcommand{\XttIccObSshLsRecall}{1\xspace}
\newcommand{\XttIccObSshLsPrecision}{0.993\xspace}
\newcommand{\XttIccObSshLsFone}{0.996\xspace}
\newcommand{\XttIccObSshBapRecall}{0.949\xspace}
\newcommand{\XttIccObSshBapPrecision}{0.998\xspace}
\newcommand{\XttIccObSshBapFone}{0.973\xspace}
\newcommand{\XttIccObSshGhiRecall}{0.987\xspace}
\newcommand{\XttIccObSshGhiPrecision}{0.998\xspace}
\newcommand{\XttIccObSshGhiFone}{0.993\xspace}
\newcommand{\XttIccObSshRdaRecall}{0.935\xspace}
\newcommand{\XttIccObSshRdaPrecision}{0.998\xspace}
\newcommand{\XttIccObSshRdaFone}{0.965\xspace}
\newcommand{\XttIccObSshRseRecall}{0.991\xspace}
\newcommand{\XttIccObSshRsePrecision}{0.998\xspace}
\newcommand{\XttIccObSshRseFone}{0.995\xspace}
\newcommand{\XttIccObPcrGT}{5947\xspace}
\newcommand{\XttIccObPcrLsRecall}{1\xspace}
\newcommand{\XttIccObPcrLsPrecision}{0.996\xspace}
\newcommand{\XttIccObPcrLsFone}{0.998\xspace}
\newcommand{\XttIccObPcrBapRecall}{0.902\xspace}
\newcommand{\XttIccObPcrBapPrecision}{0.999\xspace}
\newcommand{\XttIccObPcrBapFone}{0.948\xspace}
\newcommand{\XttIccObPcrGhiRecall}{1\xspace}
\newcommand{\XttIccObPcrGhiPrecision}{1.000\xspace}
\newcommand{\XttIccObPcrGhiFone}{1.000\xspace}
\newcommand{\XttIccObPcrRdaRecall}{0.900\xspace}
\newcommand{\XttIccObPcrRdaPrecision}{1\xspace}
\newcommand{\XttIccObPcrRdaFone}{0.947\xspace}
\newcommand{\XttIccObPcrRseRecall}{0.992\xspace}
\newcommand{\XttIccObPcrRsePrecision}{1\xspace}
\newcommand{\XttIccObPcrRseFone}{0.996\xspace}
\newcommand{\XttIccObSqlGT}{387906\xspace}
\newcommand{\XttIccObSqlLsRecall}{1\xspace}
\newcommand{\XttIccObSqlLsPrecision}{0.995\xspace}
\newcommand{\XttIccObSqlLsFone}{0.998\xspace}
\newcommand{\XttIccObSqlBapRecall}{0.907\xspace}
\newcommand{\XttIccObSqlBapPrecision}{1.000\xspace}
\newcommand{\XttIccObSqlBapFone}{0.951\xspace}
\newcommand{\XttIccObSqlGhiRecall}{0.971\xspace}
\newcommand{\XttIccObSqlGhiPrecision}{1.000\xspace}
\newcommand{\XttIccObSqlGhiFone}{0.985\xspace}
\newcommand{\XttIccObSqlRdaRecall}{0.904\xspace}
\newcommand{\XttIccObSqlRdaPrecision}{1\xspace}
\newcommand{\XttIccObSqlRdaFone}{0.950\xspace}
\newcommand{\XttIccObSqlRseRecall}{0.963\xspace}
\newcommand{\XttIccObSqlRsePrecision}{1\xspace}
\newcommand{\XttIccObSqlRseFone}{0.981\xspace}
\newcommand{\XttIccObVimGT}{854267\xspace}
\newcommand{\XttIccObVimLsRecall}{1\xspace}
\newcommand{\XttIccObVimLsPrecision}{0.993\xspace}
\newcommand{\XttIccObVimLsFone}{0.997\xspace}
\newcommand{\XttIccObVimBapRecall}{0.909\xspace}
\newcommand{\XttIccObVimBapPrecision}{1.000\xspace}
\newcommand{\XttIccObVimBapFone}{0.952\xspace}
\newcommand{\XttIccObVimGhiRecall}{0.991\xspace}
\newcommand{\XttIccObVimGhiPrecision}{1.000\xspace}
\newcommand{\XttIccObVimGhiFone}{0.995\xspace}
\newcommand{\XttIccObVimRdaRecall}{0.888\xspace}
\newcommand{\XttIccObVimRdaPrecision}{1.000\xspace}
\newcommand{\XttIccObVimRdaFone}{0.941\xspace}
\newcommand{\XttIccObVimRseRecall}{0.985\xspace}
\newcommand{\XttIccObVimRsePrecision}{1.000\xspace}
\newcommand{\XttIccObVimRseFone}{0.992\xspace}
\newcommand{\XttIccObVsfGT}{33257\xspace}
\newcommand{\XttIccObVsfLsRecall}{1\xspace}
\newcommand{\XttIccObVsfLsPrecision}{0.980\xspace}
\newcommand{\XttIccObVsfLsFone}{0.990\xspace}
\newcommand{\XttIccObVsfBapRecall}{0.987\xspace}
\newcommand{\XttIccObVsfBapPrecision}{1.000\xspace}
\newcommand{\XttIccObVsfBapFone}{0.993\xspace}
\newcommand{\XttIccObVsfGhiRecall}{1\xspace}
\newcommand{\XttIccObVsfGhiPrecision}{1.000\xspace}
\newcommand{\XttIccObVsfGhiFone}{1.000\xspace}
\newcommand{\XttIccObVsfRdaRecall}{0.916\xspace}
\newcommand{\XttIccObVsfRdaPrecision}{0.998\xspace}
\newcommand{\XttIccObVsfRdaFone}{0.956\xspace}
\newcommand{\XttIccObVsfRseRecall}{1.000\xspace}
\newcommand{\XttIccObVsfRsePrecision}{1.000\xspace}
\newcommand{\XttIccObVsfRseFone}{1.000\xspace}
\newcommand{\XttIccOcSzpGT}{24238\xspace}
\newcommand{\XttIccOcSzpLsRecall}{1\xspace}
\newcommand{\XttIccOcSzpLsPrecision}{0.993\xspace}
\newcommand{\XttIccOcSzpLsFone}{0.997\xspace}
\newcommand{\XttIccOcSzpBapRecall}{0.970\xspace}
\newcommand{\XttIccOcSzpBapPrecision}{1\xspace}
\newcommand{\XttIccOcSzpBapFone}{0.985\xspace}
\newcommand{\XttIccOcSzpGhiRecall}{1\xspace}
\newcommand{\XttIccOcSzpGhiPrecision}{1\xspace}
\newcommand{\XttIccOcSzpGhiFone}{1\xspace}
\newcommand{\XttIccOcSzpRdaRecall}{0.968\xspace}
\newcommand{\XttIccOcSzpRdaPrecision}{1\xspace}
\newcommand{\XttIccOcSzpRdaFone}{0.984\xspace}
\newcommand{\XttIccOcSzpRseRecall}{1\xspace}
\newcommand{\XttIccOcSzpRsePrecision}{1\xspace}
\newcommand{\XttIccOcSzpRseFone}{1\xspace}
\newcommand{\XttIccOcCapGT}{296396\xspace}
\newcommand{\XttIccOcCapLsRecall}{1\xspace}
\newcommand{\XttIccOcCapLsPrecision}{0.995\xspace}
\newcommand{\XttIccOcCapLsFone}{0.998\xspace}
\newcommand{\XttIccOcCapBapRecall}{0.265\xspace}
\newcommand{\XttIccOcCapBapPrecision}{1\xspace}
\newcommand{\XttIccOcCapBapFone}{0.418\xspace}
\newcommand{\XttIccOcCapGhiRecall}{0.830\xspace}
\newcommand{\XttIccOcCapGhiPrecision}{1.000\xspace}
\newcommand{\XttIccOcCapGhiFone}{0.907\xspace}
\newcommand{\XttIccOcCapRdaRecall}{0.498\xspace}
\newcommand{\XttIccOcCapRdaPrecision}{1\xspace}
\newcommand{\XttIccOcCapRdaFone}{0.665\xspace}
\newcommand{\XttIccOcCapRseRecall}{0.713\xspace}
\newcommand{\XttIccOcCapRsePrecision}{1.000\xspace}
\newcommand{\XttIccOcCapRseFone}{0.832\xspace}
\newcommand{\XttIccOcExmGT}{216944\xspace}
\newcommand{\XttIccOcExmLsRecall}{1\xspace}
\newcommand{\XttIccOcExmLsPrecision}{0.996\xspace}
\newcommand{\XttIccOcExmLsFone}{0.998\xspace}
\newcommand{\XttIccOcExmBapRecall}{0.852\xspace}
\newcommand{\XttIccOcExmBapPrecision}{1.000\xspace}
\newcommand{\XttIccOcExmBapFone}{0.920\xspace}
\newcommand{\XttIccOcExmGhiRecall}{0.940\xspace}
\newcommand{\XttIccOcExmGhiPrecision}{1.000\xspace}
\newcommand{\XttIccOcExmGhiFone}{0.969\xspace}
\newcommand{\XttIccOcExmRdaRecall}{0.820\xspace}
\newcommand{\XttIccOcExmRdaPrecision}{1\xspace}
\newcommand{\XttIccOcExmRdaFone}{0.901\xspace}
\newcommand{\XttIccOcExmRseRecall}{0.981\xspace}
\newcommand{\XttIccOcExmRsePrecision}{1\xspace}
\newcommand{\XttIccOcExmRseFone}{0.990\xspace}
\newcommand{\XttIccOcLgtGT}{38485\xspace}
\newcommand{\XttIccOcLgtLsRecall}{1\xspace}
\newcommand{\XttIccOcLgtLsPrecision}{0.990\xspace}
\newcommand{\XttIccOcLgtLsFone}{0.995\xspace}
\newcommand{\XttIccOcLgtBapRecall}{0.878\xspace}
\newcommand{\XttIccOcLgtBapPrecision}{0.998\xspace}
\newcommand{\XttIccOcLgtBapFone}{0.934\xspace}
\newcommand{\XttIccOcLgtGhiRecall}{0.998\xspace}
\newcommand{\XttIccOcLgtGhiPrecision}{0.998\xspace}
\newcommand{\XttIccOcLgtGhiFone}{0.998\xspace}
\newcommand{\XttIccOcLgtRdaRecall}{0.810\xspace}
\newcommand{\XttIccOcLgtRdaPrecision}{1\xspace}
\newcommand{\XttIccOcLgtRdaFone}{0.895\xspace}
\newcommand{\XttIccOcLgtRseRecall}{0.991\xspace}
\newcommand{\XttIccOcLgtRsePrecision}{1\xspace}
\newcommand{\XttIccOcLgtRseFone}{0.995\xspace}
\newcommand{\XttIccOcBzpGT}{26479\xspace}
\newcommand{\XttIccOcBzpLsRecall}{1\xspace}
\newcommand{\XttIccOcBzpLsPrecision}{0.997\xspace}
\newcommand{\XttIccOcBzpLsFone}{0.998\xspace}
\newcommand{\XttIccOcBzpBapRecall}{0.833\xspace}
\newcommand{\XttIccOcBzpBapPrecision}{1.000\xspace}
\newcommand{\XttIccOcBzpBapFone}{0.909\xspace}
\newcommand{\XttIccOcBzpGhiRecall}{1\xspace}
\newcommand{\XttIccOcBzpGhiPrecision}{1.000\xspace}
\newcommand{\XttIccOcBzpGhiFone}{1.000\xspace}
\newcommand{\XttIccOcBzpRdaRecall}{0.826\xspace}
\newcommand{\XttIccOcBzpRdaPrecision}{1\xspace}
\newcommand{\XttIccOcBzpRdaFone}{0.905\xspace}
\newcommand{\XttIccOcBzpRseRecall}{0.910\xspace}
\newcommand{\XttIccOcBzpRsePrecision}{1\xspace}
\newcommand{\XttIccOcBzpRseFone}{0.953\xspace}
\newcommand{\XttIccOcGzpGT}{23409\xspace}
\newcommand{\XttIccOcGzpLsRecall}{1\xspace}
\newcommand{\XttIccOcGzpLsPrecision}{0.996\xspace}
\newcommand{\XttIccOcGzpLsFone}{0.998\xspace}
\newcommand{\XttIccOcGzpBapRecall}{0.993\xspace}
\newcommand{\XttIccOcGzpBapPrecision}{1.000\xspace}
\newcommand{\XttIccOcGzpBapFone}{0.996\xspace}
\newcommand{\XttIccOcGzpGhiRecall}{1\xspace}
\newcommand{\XttIccOcGzpGhiPrecision}{1.000\xspace}
\newcommand{\XttIccOcGzpGhiFone}{1.000\xspace}
\newcommand{\XttIccOcGzpRdaRecall}{0.982\xspace}
\newcommand{\XttIccOcGzpRdaPrecision}{1\xspace}
\newcommand{\XttIccOcGzpRdaFone}{0.991\xspace}
\newcommand{\XttIccOcGzpRseRecall}{0.998\xspace}
\newcommand{\XttIccOcGzpRsePrecision}{1\xspace}
\newcommand{\XttIccOcGzpRseFone}{0.999\xspace}
\newcommand{\XttIccOcOggGT}{102238\xspace}
\newcommand{\XttIccOcOggLsRecall}{1\xspace}
\newcommand{\XttIccOcOggLsPrecision}{0.996\xspace}
\newcommand{\XttIccOcOggLsFone}{0.998\xspace}
\newcommand{\XttIccOcOggBapRecall}{0.986\xspace}
\newcommand{\XttIccOcOggBapPrecision}{1.000\xspace}
\newcommand{\XttIccOcOggBapFone}{0.993\xspace}
\newcommand{\XttIccOcOggGhiRecall}{1\xspace}
\newcommand{\XttIccOcOggGhiPrecision}{1.000\xspace}
\newcommand{\XttIccOcOggGhiFone}{1.000\xspace}
\newcommand{\XttIccOcOggRdaRecall}{0.986\xspace}
\newcommand{\XttIccOcOggRdaPrecision}{1\xspace}
\newcommand{\XttIccOcOggRdaFone}{0.993\xspace}
\newcommand{\XttIccOcOggRseRecall}{0.998\xspace}
\newcommand{\XttIccOcOggRsePrecision}{1\xspace}
\newcommand{\XttIccOcOggRseFone}{0.999\xspace}
\newcommand{\XttIccOcNgxGT}{150688\xspace}
\newcommand{\XttIccOcNgxLsRecall}{1\xspace}
\newcommand{\XttIccOcNgxLsPrecision}{0.991\xspace}
\newcommand{\XttIccOcNgxLsFone}{0.996\xspace}
\newcommand{\XttIccOcNgxBapRecall}{0.973\xspace}
\newcommand{\XttIccOcNgxBapPrecision}{1.000\xspace}
\newcommand{\XttIccOcNgxBapFone}{0.986\xspace}
\newcommand{\XttIccOcNgxGhiRecall}{0.999\xspace}
\newcommand{\XttIccOcNgxGhiPrecision}{1.000\xspace}
\newcommand{\XttIccOcNgxGhiFone}{0.999\xspace}
\newcommand{\XttIccOcNgxRdaRecall}{0.965\xspace}
\newcommand{\XttIccOcNgxRdaPrecision}{1\xspace}
\newcommand{\XttIccOcNgxRdaFone}{0.982\xspace}
\newcommand{\XttIccOcNgxRseRecall}{0.996\xspace}
\newcommand{\XttIccOcNgxRsePrecision}{1\xspace}
\newcommand{\XttIccOcNgxRseFone}{0.998\xspace}
\newcommand{\XttIccOcSshGT}{205965\xspace}
\newcommand{\XttIccOcSshLsRecall}{1\xspace}
\newcommand{\XttIccOcSshLsPrecision}{0.993\xspace}
\newcommand{\XttIccOcSshLsFone}{0.996\xspace}
\newcommand{\XttIccOcSshBapRecall}{0.949\xspace}
\newcommand{\XttIccOcSshBapPrecision}{0.998\xspace}
\newcommand{\XttIccOcSshBapFone}{0.973\xspace}
\newcommand{\XttIccOcSshGhiRecall}{0.990\xspace}
\newcommand{\XttIccOcSshGhiPrecision}{0.998\xspace}
\newcommand{\XttIccOcSshGhiFone}{0.994\xspace}
\newcommand{\XttIccOcSshRdaRecall}{0.936\xspace}
\newcommand{\XttIccOcSshRdaPrecision}{0.998\xspace}
\newcommand{\XttIccOcSshRdaFone}{0.966\xspace}
\newcommand{\XttIccOcSshRseRecall}{0.991\xspace}
\newcommand{\XttIccOcSshRsePrecision}{0.998\xspace}
\newcommand{\XttIccOcSshRseFone}{0.995\xspace}
\newcommand{\XttIccOcPcrGT}{6171\xspace}
\newcommand{\XttIccOcPcrLsRecall}{1\xspace}
\newcommand{\XttIccOcPcrLsPrecision}{0.996\xspace}
\newcommand{\XttIccOcPcrLsFone}{0.998\xspace}
\newcommand{\XttIccOcPcrBapRecall}{0.906\xspace}
\newcommand{\XttIccOcPcrBapPrecision}{0.999\xspace}
\newcommand{\XttIccOcPcrBapFone}{0.950\xspace}
\newcommand{\XttIccOcPcrGhiRecall}{1\xspace}
\newcommand{\XttIccOcPcrGhiPrecision}{1.000\xspace}
\newcommand{\XttIccOcPcrGhiFone}{1.000\xspace}
\newcommand{\XttIccOcPcrRdaRecall}{0.904\xspace}
\newcommand{\XttIccOcPcrRdaPrecision}{1\xspace}
\newcommand{\XttIccOcPcrRdaFone}{0.949\xspace}
\newcommand{\XttIccOcPcrRseRecall}{0.992\xspace}
\newcommand{\XttIccOcPcrRsePrecision}{1\xspace}
\newcommand{\XttIccOcPcrRseFone}{0.996\xspace}
\newcommand{\XttIccOcSqlGT}{399162\xspace}
\newcommand{\XttIccOcSqlLsRecall}{1\xspace}
\newcommand{\XttIccOcSqlLsPrecision}{0.995\xspace}
\newcommand{\XttIccOcSqlLsFone}{0.998\xspace}
\newcommand{\XttIccOcSqlBapRecall}{0.909\xspace}
\newcommand{\XttIccOcSqlBapPrecision}{1.000\xspace}
\newcommand{\XttIccOcSqlBapFone}{0.952\xspace}
\newcommand{\XttIccOcSqlGhiRecall}{0.971\xspace}
\newcommand{\XttIccOcSqlGhiPrecision}{1.000\xspace}
\newcommand{\XttIccOcSqlGhiFone}{0.985\xspace}
\newcommand{\XttIccOcSqlRdaRecall}{0.904\xspace}
\newcommand{\XttIccOcSqlRdaPrecision}{1\xspace}
\newcommand{\XttIccOcSqlRdaFone}{0.949\xspace}
\newcommand{\XttIccOcSqlRseRecall}{0.962\xspace}
\newcommand{\XttIccOcSqlRsePrecision}{1\xspace}
\newcommand{\XttIccOcSqlRseFone}{0.981\xspace}
\newcommand{\XttIccOcVimGT}{870861\xspace}
\newcommand{\XttIccOcVimLsRecall}{1\xspace}
\newcommand{\XttIccOcVimLsPrecision}{0.993\xspace}
\newcommand{\XttIccOcVimLsFone}{0.997\xspace}
\newcommand{\XttIccOcVimBapRecall}{0.911\xspace}
\newcommand{\XttIccOcVimBapPrecision}{1.000\xspace}
\newcommand{\XttIccOcVimBapFone}{0.953\xspace}
\newcommand{\XttIccOcVimGhiRecall}{0.993\xspace}
\newcommand{\XttIccOcVimGhiPrecision}{1.000\xspace}
\newcommand{\XttIccOcVimGhiFone}{0.997\xspace}
\newcommand{\XttIccOcVimRdaRecall}{0.889\xspace}
\newcommand{\XttIccOcVimRdaPrecision}{1.000\xspace}
\newcommand{\XttIccOcVimRdaFone}{0.941\xspace}
\newcommand{\XttIccOcVimRseRecall}{0.985\xspace}
\newcommand{\XttIccOcVimRsePrecision}{1.000\xspace}
\newcommand{\XttIccOcVimRseFone}{0.993\xspace}
\newcommand{\XttIccOcVsfGT}{33364\xspace}
\newcommand{\XttIccOcVsfLsRecall}{1\xspace}
\newcommand{\XttIccOcVsfLsPrecision}{0.980\xspace}
\newcommand{\XttIccOcVsfLsFone}{0.990\xspace}
\newcommand{\XttIccOcVsfBapRecall}{0.987\xspace}
\newcommand{\XttIccOcVsfBapPrecision}{1.000\xspace}
\newcommand{\XttIccOcVsfBapFone}{0.993\xspace}
\newcommand{\XttIccOcVsfGhiRecall}{1\xspace}
\newcommand{\XttIccOcVsfGhiPrecision}{1.000\xspace}
\newcommand{\XttIccOcVsfGhiFone}{1.000\xspace}
\newcommand{\XttIccOcVsfRdaRecall}{0.918\xspace}
\newcommand{\XttIccOcVsfRdaPrecision}{0.998\xspace}
\newcommand{\XttIccOcVsfRdaFone}{0.956\xspace}
\newcommand{\XttIccOcVsfRseRecall}{1.000\xspace}
\newcommand{\XttIccOcVsfRsePrecision}{1.000\xspace}
\newcommand{\XttIccOcVsfRseFone}{1.000\xspace}
\newcommand{\XttIccOdSzpGT}{24238\xspace}
\newcommand{\XttIccOdSzpLsRecall}{1\xspace}
\newcommand{\XttIccOdSzpLsPrecision}{0.993\xspace}
\newcommand{\XttIccOdSzpLsFone}{0.997\xspace}
\newcommand{\XttIccOdSzpBapRecall}{0.970\xspace}
\newcommand{\XttIccOdSzpBapPrecision}{1\xspace}
\newcommand{\XttIccOdSzpBapFone}{0.985\xspace}
\newcommand{\XttIccOdSzpGhiRecall}{1\xspace}
\newcommand{\XttIccOdSzpGhiPrecision}{1\xspace}
\newcommand{\XttIccOdSzpGhiFone}{1\xspace}
\newcommand{\XttIccOdSzpRdaRecall}{0.968\xspace}
\newcommand{\XttIccOdSzpRdaPrecision}{1\xspace}
\newcommand{\XttIccOdSzpRdaFone}{0.984\xspace}
\newcommand{\XttIccOdSzpRseRecall}{1\xspace}
\newcommand{\XttIccOdSzpRsePrecision}{1\xspace}
\newcommand{\XttIccOdSzpRseFone}{1\xspace}
\newcommand{\XttIccOdCapGT}{296396\xspace}
\newcommand{\XttIccOdCapLsRecall}{1\xspace}
\newcommand{\XttIccOdCapLsPrecision}{0.995\xspace}
\newcommand{\XttIccOdCapLsFone}{0.998\xspace}
\newcommand{\XttIccOdCapBapRecall}{0.265\xspace}
\newcommand{\XttIccOdCapBapPrecision}{1\xspace}
\newcommand{\XttIccOdCapBapFone}{0.418\xspace}
\newcommand{\XttIccOdCapGhiRecall}{0.886\xspace}
\newcommand{\XttIccOdCapGhiPrecision}{1.000\xspace}
\newcommand{\XttIccOdCapGhiFone}{0.940\xspace}
\newcommand{\XttIccOdCapRdaRecall}{0.498\xspace}
\newcommand{\XttIccOdCapRdaPrecision}{1\xspace}
\newcommand{\XttIccOdCapRdaFone}{0.665\xspace}
\newcommand{\XttIccOdCapRseRecall}{0.713\xspace}
\newcommand{\XttIccOdCapRsePrecision}{1.000\xspace}
\newcommand{\XttIccOdCapRseFone}{0.832\xspace}
\newcommand{\XttIccOdExmGT}{216948\xspace}
\newcommand{\XttIccOdExmLsRecall}{1\xspace}
\newcommand{\XttIccOdExmLsPrecision}{0.996\xspace}
\newcommand{\XttIccOdExmLsFone}{0.998\xspace}
\newcommand{\XttIccOdExmBapRecall}{0.852\xspace}
\newcommand{\XttIccOdExmBapPrecision}{1.000\xspace}
\newcommand{\XttIccOdExmBapFone}{0.920\xspace}
\newcommand{\XttIccOdExmGhiRecall}{0.941\xspace}
\newcommand{\XttIccOdExmGhiPrecision}{1.000\xspace}
\newcommand{\XttIccOdExmGhiFone}{0.969\xspace}
\newcommand{\XttIccOdExmRdaRecall}{0.820\xspace}
\newcommand{\XttIccOdExmRdaPrecision}{1\xspace}
\newcommand{\XttIccOdExmRdaFone}{0.901\xspace}
\newcommand{\XttIccOdExmRseRecall}{0.981\xspace}
\newcommand{\XttIccOdExmRsePrecision}{1\xspace}
\newcommand{\XttIccOdExmRseFone}{0.990\xspace}
\newcommand{\XttIccOdLgtGT}{38485\xspace}
\newcommand{\XttIccOdLgtLsRecall}{1\xspace}
\newcommand{\XttIccOdLgtLsPrecision}{0.990\xspace}
\newcommand{\XttIccOdLgtLsFone}{0.995\xspace}
\newcommand{\XttIccOdLgtBapRecall}{0.878\xspace}
\newcommand{\XttIccOdLgtBapPrecision}{0.998\xspace}
\newcommand{\XttIccOdLgtBapFone}{0.934\xspace}
\newcommand{\XttIccOdLgtGhiRecall}{0.998\xspace}
\newcommand{\XttIccOdLgtGhiPrecision}{0.998\xspace}
\newcommand{\XttIccOdLgtGhiFone}{0.998\xspace}
\newcommand{\XttIccOdLgtRdaRecall}{0.810\xspace}
\newcommand{\XttIccOdLgtRdaPrecision}{1\xspace}
\newcommand{\XttIccOdLgtRdaFone}{0.895\xspace}
\newcommand{\XttIccOdLgtRseRecall}{0.991\xspace}
\newcommand{\XttIccOdLgtRsePrecision}{1\xspace}
\newcommand{\XttIccOdLgtRseFone}{0.995\xspace}
\newcommand{\XttIccOdBzpGT}{26481\xspace}
\newcommand{\XttIccOdBzpLsRecall}{1\xspace}
\newcommand{\XttIccOdBzpLsPrecision}{0.997\xspace}
\newcommand{\XttIccOdBzpLsFone}{0.998\xspace}
\newcommand{\XttIccOdBzpBapRecall}{0.833\xspace}
\newcommand{\XttIccOdBzpBapPrecision}{1.000\xspace}
\newcommand{\XttIccOdBzpBapFone}{0.909\xspace}
\newcommand{\XttIccOdBzpGhiRecall}{1\xspace}
\newcommand{\XttIccOdBzpGhiPrecision}{1.000\xspace}
\newcommand{\XttIccOdBzpGhiFone}{1.000\xspace}
\newcommand{\XttIccOdBzpRdaRecall}{0.826\xspace}
\newcommand{\XttIccOdBzpRdaPrecision}{1\xspace}
\newcommand{\XttIccOdBzpRdaFone}{0.905\xspace}
\newcommand{\XttIccOdBzpRseRecall}{0.910\xspace}
\newcommand{\XttIccOdBzpRsePrecision}{1\xspace}
\newcommand{\XttIccOdBzpRseFone}{0.953\xspace}
\newcommand{\XttIccOdGzpGT}{23410\xspace}
\newcommand{\XttIccOdGzpLsRecall}{1\xspace}
\newcommand{\XttIccOdGzpLsPrecision}{0.996\xspace}
\newcommand{\XttIccOdGzpLsFone}{0.998\xspace}
\newcommand{\XttIccOdGzpBapRecall}{0.993\xspace}
\newcommand{\XttIccOdGzpBapPrecision}{1.000\xspace}
\newcommand{\XttIccOdGzpBapFone}{0.996\xspace}
\newcommand{\XttIccOdGzpGhiRecall}{1\xspace}
\newcommand{\XttIccOdGzpGhiPrecision}{1.000\xspace}
\newcommand{\XttIccOdGzpGhiFone}{1.000\xspace}
\newcommand{\XttIccOdGzpRdaRecall}{0.982\xspace}
\newcommand{\XttIccOdGzpRdaPrecision}{1\xspace}
\newcommand{\XttIccOdGzpRdaFone}{0.991\xspace}
\newcommand{\XttIccOdGzpRseRecall}{0.998\xspace}
\newcommand{\XttIccOdGzpRsePrecision}{1\xspace}
\newcommand{\XttIccOdGzpRseFone}{0.999\xspace}
\newcommand{\XttIccOdOggGT}{102491\xspace}
\newcommand{\XttIccOdOggLsRecall}{1\xspace}
\newcommand{\XttIccOdOggLsPrecision}{0.996\xspace}
\newcommand{\XttIccOdOggLsFone}{0.998\xspace}
\newcommand{\XttIccOdOggBapRecall}{0.986\xspace}
\newcommand{\XttIccOdOggBapPrecision}{1.000\xspace}
\newcommand{\XttIccOdOggBapFone}{0.993\xspace}
\newcommand{\XttIccOdOggGhiRecall}{1\xspace}
\newcommand{\XttIccOdOggGhiPrecision}{1.000\xspace}
\newcommand{\XttIccOdOggGhiFone}{1.000\xspace}
\newcommand{\XttIccOdOggRdaRecall}{0.986\xspace}
\newcommand{\XttIccOdOggRdaPrecision}{1\xspace}
\newcommand{\XttIccOdOggRdaFone}{0.993\xspace}
\newcommand{\XttIccOdOggRseRecall}{0.998\xspace}
\newcommand{\XttIccOdOggRsePrecision}{1\xspace}
\newcommand{\XttIccOdOggRseFone}{0.999\xspace}
\newcommand{\XttIccOdNgxGT}{150688\xspace}
\newcommand{\XttIccOdNgxLsRecall}{1\xspace}
\newcommand{\XttIccOdNgxLsPrecision}{0.991\xspace}
\newcommand{\XttIccOdNgxLsFone}{0.996\xspace}
\newcommand{\XttIccOdNgxBapRecall}{0.973\xspace}
\newcommand{\XttIccOdNgxBapPrecision}{1.000\xspace}
\newcommand{\XttIccOdNgxBapFone}{0.986\xspace}
\newcommand{\XttIccOdNgxGhiRecall}{0.999\xspace}
\newcommand{\XttIccOdNgxGhiPrecision}{1.000\xspace}
\newcommand{\XttIccOdNgxGhiFone}{0.999\xspace}
\newcommand{\XttIccOdNgxRdaRecall}{0.965\xspace}
\newcommand{\XttIccOdNgxRdaPrecision}{1\xspace}
\newcommand{\XttIccOdNgxRdaFone}{0.982\xspace}
\newcommand{\XttIccOdNgxRseRecall}{0.996\xspace}
\newcommand{\XttIccOdNgxRsePrecision}{1\xspace}
\newcommand{\XttIccOdNgxRseFone}{0.998\xspace}
\newcommand{\XttIccOdSshGT}{205965\xspace}
\newcommand{\XttIccOdSshLsRecall}{1\xspace}
\newcommand{\XttIccOdSshLsPrecision}{0.993\xspace}
\newcommand{\XttIccOdSshLsFone}{0.996\xspace}
\newcommand{\XttIccOdSshBapRecall}{0.949\xspace}
\newcommand{\XttIccOdSshBapPrecision}{0.998\xspace}
\newcommand{\XttIccOdSshBapFone}{0.973\xspace}
\newcommand{\XttIccOdSshGhiRecall}{0.990\xspace}
\newcommand{\XttIccOdSshGhiPrecision}{0.998\xspace}
\newcommand{\XttIccOdSshGhiFone}{0.994\xspace}
\newcommand{\XttIccOdSshRdaRecall}{0.936\xspace}
\newcommand{\XttIccOdSshRdaPrecision}{0.998\xspace}
\newcommand{\XttIccOdSshRdaFone}{0.966\xspace}
\newcommand{\XttIccOdSshRseRecall}{0.991\xspace}
\newcommand{\XttIccOdSshRsePrecision}{0.998\xspace}
\newcommand{\XttIccOdSshRseFone}{0.995\xspace}
\newcommand{\XttIccOdPcrGT}{6171\xspace}
\newcommand{\XttIccOdPcrLsRecall}{1\xspace}
\newcommand{\XttIccOdPcrLsPrecision}{0.996\xspace}
\newcommand{\XttIccOdPcrLsFone}{0.998\xspace}
\newcommand{\XttIccOdPcrBapRecall}{0.906\xspace}
\newcommand{\XttIccOdPcrBapPrecision}{0.999\xspace}
\newcommand{\XttIccOdPcrBapFone}{0.950\xspace}
\newcommand{\XttIccOdPcrGhiRecall}{1\xspace}
\newcommand{\XttIccOdPcrGhiPrecision}{1.000\xspace}
\newcommand{\XttIccOdPcrGhiFone}{1.000\xspace}
\newcommand{\XttIccOdPcrRdaRecall}{0.904\xspace}
\newcommand{\XttIccOdPcrRdaPrecision}{1\xspace}
\newcommand{\XttIccOdPcrRdaFone}{0.949\xspace}
\newcommand{\XttIccOdPcrRseRecall}{0.992\xspace}
\newcommand{\XttIccOdPcrRsePrecision}{1\xspace}
\newcommand{\XttIccOdPcrRseFone}{0.996\xspace}
\newcommand{\XttIccOdSqlGT}{399092\xspace}
\newcommand{\XttIccOdSqlLsRecall}{1\xspace}
\newcommand{\XttIccOdSqlLsPrecision}{0.995\xspace}
\newcommand{\XttIccOdSqlLsFone}{0.998\xspace}
\newcommand{\XttIccOdSqlBapRecall}{0.909\xspace}
\newcommand{\XttIccOdSqlBapPrecision}{1.000\xspace}
\newcommand{\XttIccOdSqlBapFone}{0.952\xspace}
\newcommand{\XttIccOdSqlGhiRecall}{0.971\xspace}
\newcommand{\XttIccOdSqlGhiPrecision}{1.000\xspace}
\newcommand{\XttIccOdSqlGhiFone}{0.985\xspace}
\newcommand{\XttIccOdSqlRdaRecall}{0.904\xspace}
\newcommand{\XttIccOdSqlRdaPrecision}{1\xspace}
\newcommand{\XttIccOdSqlRdaFone}{0.949\xspace}
\newcommand{\XttIccOdSqlRseRecall}{0.962\xspace}
\newcommand{\XttIccOdSqlRsePrecision}{1\xspace}
\newcommand{\XttIccOdSqlRseFone}{0.981\xspace}
\newcommand{\XttIccOdVimGT}{870867\xspace}
\newcommand{\XttIccOdVimLsRecall}{1\xspace}
\newcommand{\XttIccOdVimLsPrecision}{0.993\xspace}
\newcommand{\XttIccOdVimLsFone}{0.997\xspace}
\newcommand{\XttIccOdVimBapRecall}{0.911\xspace}
\newcommand{\XttIccOdVimBapPrecision}{1.000\xspace}
\newcommand{\XttIccOdVimBapFone}{0.953\xspace}
\newcommand{\XttIccOdVimGhiRecall}{0.993\xspace}
\newcommand{\XttIccOdVimGhiPrecision}{1.000\xspace}
\newcommand{\XttIccOdVimGhiFone}{0.996\xspace}
\newcommand{\XttIccOdVimRdaRecall}{0.889\xspace}
\newcommand{\XttIccOdVimRdaPrecision}{1.000\xspace}
\newcommand{\XttIccOdVimRdaFone}{0.941\xspace}
\newcommand{\XttIccOdVimRseRecall}{0.985\xspace}
\newcommand{\XttIccOdVimRsePrecision}{1.000\xspace}
\newcommand{\XttIccOdVimRseFone}{0.993\xspace}
\newcommand{\XttIccOdVsfGT}{33365\xspace}
\newcommand{\XttIccOdVsfLsRecall}{1\xspace}
\newcommand{\XttIccOdVsfLsPrecision}{0.980\xspace}
\newcommand{\XttIccOdVsfLsFone}{0.990\xspace}
\newcommand{\XttIccOdVsfBapRecall}{0.987\xspace}
\newcommand{\XttIccOdVsfBapPrecision}{1.000\xspace}
\newcommand{\XttIccOdVsfBapFone}{0.993\xspace}
\newcommand{\XttIccOdVsfGhiRecall}{1\xspace}
\newcommand{\XttIccOdVsfGhiPrecision}{1.000\xspace}
\newcommand{\XttIccOdVsfGhiFone}{1.000\xspace}
\newcommand{\XttIccOdVsfRdaRecall}{0.917\xspace}
\newcommand{\XttIccOdVsfRdaPrecision}{0.998\xspace}
\newcommand{\XttIccOdVsfRdaFone}{0.956\xspace}
\newcommand{\XttIccOdVsfRseRecall}{1.000\xspace}
\newcommand{\XttIccOdVsfRsePrecision}{1.000\xspace}
\newcommand{\XttIccOdVsfRseFone}{1.000\xspace}
\newcommand{\XttIccOsSzpGT}{13104\xspace}
\newcommand{\XttIccOsSzpLsRecall}{1\xspace}
\newcommand{\XttIccOsSzpLsPrecision}{1\xspace}
\newcommand{\XttIccOsSzpLsFone}{1\xspace}
\newcommand{\XttIccOsSzpBapRecall}{0.978\xspace}
\newcommand{\XttIccOsSzpBapPrecision}{1\xspace}
\newcommand{\XttIccOsSzpBapFone}{0.989\xspace}
\newcommand{\XttIccOsSzpGhiRecall}{1\xspace}
\newcommand{\XttIccOsSzpGhiPrecision}{1\xspace}
\newcommand{\XttIccOsSzpGhiFone}{1\xspace}
\newcommand{\XttIccOsSzpRdaRecall}{0.978\xspace}
\newcommand{\XttIccOsSzpRdaPrecision}{1\xspace}
\newcommand{\XttIccOsSzpRdaFone}{0.989\xspace}
\newcommand{\XttIccOsSzpRseRecall}{1\xspace}
\newcommand{\XttIccOsSzpRsePrecision}{1\xspace}
\newcommand{\XttIccOsSzpRseFone}{1\xspace}
\newcommand{\XttIccOsCapGT}{224426\xspace}
\newcommand{\XttIccOsCapLsRecall}{1\xspace}
\newcommand{\XttIccOsCapLsPrecision}{1\xspace}
\newcommand{\XttIccOsCapLsFone}{1\xspace}
\newcommand{\XttIccOsCapBapRecall}{0.259\xspace}
\newcommand{\XttIccOsCapBapPrecision}{1\xspace}
\newcommand{\XttIccOsCapBapFone}{0.411\xspace}
\newcommand{\XttIccOsCapGhiRecall}{0.851\xspace}
\newcommand{\XttIccOsCapGhiPrecision}{1.000\xspace}
\newcommand{\XttIccOsCapGhiFone}{0.919\xspace}
\newcommand{\XttIccOsCapRdaRecall}{0.347\xspace}
\newcommand{\XttIccOsCapRdaPrecision}{1\xspace}
\newcommand{\XttIccOsCapRdaFone}{0.515\xspace}
\newcommand{\XttIccOsCapRseRecall}{0.792\xspace}
\newcommand{\XttIccOsCapRsePrecision}{1.000\xspace}
\newcommand{\XttIccOsCapRseFone}{0.884\xspace}
\newcommand{\XttIccOsExmGT}{139700\xspace}
\newcommand{\XttIccOsExmLsRecall}{1\xspace}
\newcommand{\XttIccOsExmLsPrecision}{1\xspace}
\newcommand{\XttIccOsExmLsFone}{1\xspace}
\newcommand{\XttIccOsExmBapRecall}{0.836\xspace}
\newcommand{\XttIccOsExmBapPrecision}{1\xspace}
\newcommand{\XttIccOsExmBapFone}{0.911\xspace}
\newcommand{\XttIccOsExmGhiRecall}{0.947\xspace}
\newcommand{\XttIccOsExmGhiPrecision}{1\xspace}
\newcommand{\XttIccOsExmGhiFone}{0.973\xspace}
\newcommand{\XttIccOsExmRdaRecall}{0.802\xspace}
\newcommand{\XttIccOsExmRdaPrecision}{1\xspace}
\newcommand{\XttIccOsExmRdaFone}{0.890\xspace}
\newcommand{\XttIccOsExmRseRecall}{0.984\xspace}
\newcommand{\XttIccOsExmRsePrecision}{1\xspace}
\newcommand{\XttIccOsExmRseFone}{0.992\xspace}
\newcommand{\XttIccOsLgtGT}{27134\xspace}
\newcommand{\XttIccOsLgtLsRecall}{1\xspace}
\newcommand{\XttIccOsLgtLsPrecision}{1\xspace}
\newcommand{\XttIccOsLgtLsFone}{1\xspace}
\newcommand{\XttIccOsLgtBapRecall}{0.881\xspace}
\newcommand{\XttIccOsLgtBapPrecision}{1\xspace}
\newcommand{\XttIccOsLgtBapFone}{0.937\xspace}
\newcommand{\XttIccOsLgtGhiRecall}{0.998\xspace}
\newcommand{\XttIccOsLgtGhiPrecision}{1\xspace}
\newcommand{\XttIccOsLgtGhiFone}{0.999\xspace}
\newcommand{\XttIccOsLgtRdaRecall}{0.875\xspace}
\newcommand{\XttIccOsLgtRdaPrecision}{1\xspace}
\newcommand{\XttIccOsLgtRdaFone}{0.933\xspace}
\newcommand{\XttIccOsLgtRseRecall}{0.997\xspace}
\newcommand{\XttIccOsLgtRsePrecision}{1\xspace}
\newcommand{\XttIccOsLgtRseFone}{0.999\xspace}
\newcommand{\XttIccOsBzpGT}{12085\xspace}
\newcommand{\XttIccOsBzpLsRecall}{1\xspace}
\newcommand{\XttIccOsBzpLsPrecision}{1\xspace}
\newcommand{\XttIccOsBzpLsFone}{1\xspace}
\newcommand{\XttIccOsBzpBapRecall}{0.802\xspace}
\newcommand{\XttIccOsBzpBapPrecision}{1\xspace}
\newcommand{\XttIccOsBzpBapFone}{0.890\xspace}
\newcommand{\XttIccOsBzpGhiRecall}{1\xspace}
\newcommand{\XttIccOsBzpGhiPrecision}{1\xspace}
\newcommand{\XttIccOsBzpGhiFone}{1\xspace}
\newcommand{\XttIccOsBzpRdaRecall}{0.792\xspace}
\newcommand{\XttIccOsBzpRdaPrecision}{1\xspace}
\newcommand{\XttIccOsBzpRdaFone}{0.884\xspace}
\newcommand{\XttIccOsBzpRseRecall}{0.924\xspace}
\newcommand{\XttIccOsBzpRsePrecision}{1\xspace}
\newcommand{\XttIccOsBzpRseFone}{0.961\xspace}
\newcommand{\XttIccOsGccGT}{788948\xspace}
\newcommand{\XttIccOsGccLsRecall}{1\xspace}
\newcommand{\XttIccOsGccLsPrecision}{1\xspace}
\newcommand{\XttIccOsGccLsFone}{1\xspace}
\newcommand{\XttIccOsGccBapRecall}{0.762\xspace}
\newcommand{\XttIccOsGccBapPrecision}{1\xspace}
\newcommand{\XttIccOsGccBapFone}{0.865\xspace}
\newcommand{\XttIccOsGccGhiRecall}{0.988\xspace}
\newcommand{\XttIccOsGccGhiPrecision}{1\xspace}
\newcommand{\XttIccOsGccGhiFone}{0.994\xspace}
\newcommand{\XttIccOsGccRdaRecall}{0.714\xspace}
\newcommand{\XttIccOsGccRdaPrecision}{1.000\xspace}
\newcommand{\XttIccOsGccRdaFone}{0.833\xspace}
\newcommand{\XttIccOsGccRseRecall}{0.963\xspace}
\newcommand{\XttIccOsGccRsePrecision}{1\xspace}
\newcommand{\XttIccOsGccRseFone}{0.981\xspace}
\newcommand{\XttIccOsGzpGT}{9399\xspace}
\newcommand{\XttIccOsGzpLsRecall}{1\xspace}
\newcommand{\XttIccOsGzpLsPrecision}{1\xspace}
\newcommand{\XttIccOsGzpLsFone}{1\xspace}
\newcommand{\XttIccOsGzpBapRecall}{0.987\xspace}
\newcommand{\XttIccOsGzpBapPrecision}{1\xspace}
\newcommand{\XttIccOsGzpBapFone}{0.994\xspace}
\newcommand{\XttIccOsGzpGhiRecall}{1\xspace}
\newcommand{\XttIccOsGzpGhiPrecision}{1\xspace}
\newcommand{\XttIccOsGzpGhiFone}{1\xspace}
\newcommand{\XttIccOsGzpRdaRecall}{0.987\xspace}
\newcommand{\XttIccOsGzpRdaPrecision}{1\xspace}
\newcommand{\XttIccOsGzpRdaFone}{0.994\xspace}
\newcommand{\XttIccOsGzpRseRecall}{0.998\xspace}
\newcommand{\XttIccOsGzpRsePrecision}{1\xspace}
\newcommand{\XttIccOsGzpRseFone}{0.999\xspace}
\newcommand{\XttIccOsOggGT}{38078\xspace}
\newcommand{\XttIccOsOggLsRecall}{1\xspace}
\newcommand{\XttIccOsOggLsPrecision}{1\xspace}
\newcommand{\XttIccOsOggLsFone}{1\xspace}
\newcommand{\XttIccOsOggBapRecall}{0.983\xspace}
\newcommand{\XttIccOsOggBapPrecision}{1\xspace}
\newcommand{\XttIccOsOggBapFone}{0.992\xspace}
\newcommand{\XttIccOsOggGhiRecall}{1\xspace}
\newcommand{\XttIccOsOggGhiPrecision}{1\xspace}
\newcommand{\XttIccOsOggGhiFone}{1\xspace}
\newcommand{\XttIccOsOggRdaRecall}{0.983\xspace}
\newcommand{\XttIccOsOggRdaPrecision}{1\xspace}
\newcommand{\XttIccOsOggRdaFone}{0.992\xspace}
\newcommand{\XttIccOsOggRseRecall}{0.995\xspace}
\newcommand{\XttIccOsOggRsePrecision}{1\xspace}
\newcommand{\XttIccOsOggRseFone}{0.998\xspace}
\newcommand{\XttIccOsNgxGT}{106323\xspace}
\newcommand{\XttIccOsNgxLsRecall}{1\xspace}
\newcommand{\XttIccOsNgxLsPrecision}{1\xspace}
\newcommand{\XttIccOsNgxLsFone}{1\xspace}
\newcommand{\XttIccOsNgxBapRecall}{0.972\xspace}
\newcommand{\XttIccOsNgxBapPrecision}{1\xspace}
\newcommand{\XttIccOsNgxBapFone}{0.986\xspace}
\newcommand{\XttIccOsNgxGhiRecall}{0.999\xspace}
\newcommand{\XttIccOsNgxGhiPrecision}{1\xspace}
\newcommand{\XttIccOsNgxGhiFone}{0.999\xspace}
\newcommand{\XttIccOsNgxRdaRecall}{0.971\xspace}
\newcommand{\XttIccOsNgxRdaPrecision}{1\xspace}
\newcommand{\XttIccOsNgxRdaFone}{0.985\xspace}
\newcommand{\XttIccOsNgxRseRecall}{0.998\xspace}
\newcommand{\XttIccOsNgxRsePrecision}{1\xspace}
\newcommand{\XttIccOsNgxRseFone}{0.999\xspace}
\newcommand{\XttIccOsSshGT}{127715\xspace}
\newcommand{\XttIccOsSshLsRecall}{1\xspace}
\newcommand{\XttIccOsSshLsPrecision}{1\xspace}
\newcommand{\XttIccOsSshLsFone}{1\xspace}
\newcommand{\XttIccOsSshBapRecall}{0.954\xspace}
\newcommand{\XttIccOsSshBapPrecision}{1\xspace}
\newcommand{\XttIccOsSshBapFone}{0.976\xspace}
\newcommand{\XttIccOsSshGhiRecall}{0.996\xspace}
\newcommand{\XttIccOsSshGhiPrecision}{1\xspace}
\newcommand{\XttIccOsSshGhiFone}{0.998\xspace}
\newcommand{\XttIccOsSshRdaRecall}{0.944\xspace}
\newcommand{\XttIccOsSshRdaPrecision}{0.999\xspace}
\newcommand{\XttIccOsSshRdaFone}{0.971\xspace}
\newcommand{\XttIccOsSshRseRecall}{0.985\xspace}
\newcommand{\XttIccOsSshRsePrecision}{1\xspace}
\newcommand{\XttIccOsSshRseFone}{0.992\xspace}
\newcommand{\XttIccOsPcrGT}{4386\xspace}
\newcommand{\XttIccOsPcrLsRecall}{1\xspace}
\newcommand{\XttIccOsPcrLsPrecision}{1\xspace}
\newcommand{\XttIccOsPcrLsFone}{1\xspace}
\newcommand{\XttIccOsPcrBapRecall}{0.913\xspace}
\newcommand{\XttIccOsPcrBapPrecision}{1\xspace}
\newcommand{\XttIccOsPcrBapFone}{0.955\xspace}
\newcommand{\XttIccOsPcrGhiRecall}{1\xspace}
\newcommand{\XttIccOsPcrGhiPrecision}{1\xspace}
\newcommand{\XttIccOsPcrGhiFone}{1\xspace}
\newcommand{\XttIccOsPcrRdaRecall}{0.849\xspace}
\newcommand{\XttIccOsPcrRdaPrecision}{1\xspace}
\newcommand{\XttIccOsPcrRdaFone}{0.918\xspace}
\newcommand{\XttIccOsPcrRseRecall}{0.995\xspace}
\newcommand{\XttIccOsPcrRsePrecision}{1\xspace}
\newcommand{\XttIccOsPcrRseFone}{0.997\xspace}
\newcommand{\XttIccOsSqlGT}{159621\xspace}
\newcommand{\XttIccOsSqlLsRecall}{1\xspace}
\newcommand{\XttIccOsSqlLsPrecision}{1\xspace}
\newcommand{\XttIccOsSqlLsFone}{1\xspace}
\newcommand{\XttIccOsSqlBapRecall}{0.857\xspace}
\newcommand{\XttIccOsSqlBapPrecision}{1\xspace}
\newcommand{\XttIccOsSqlBapFone}{0.923\xspace}
\newcommand{\XttIccOsSqlGhiRecall}{0.951\xspace}
\newcommand{\XttIccOsSqlGhiPrecision}{1\xspace}
\newcommand{\XttIccOsSqlGhiFone}{0.975\xspace}
\newcommand{\XttIccOsSqlRdaRecall}{0.856\xspace}
\newcommand{\XttIccOsSqlRdaPrecision}{1\xspace}
\newcommand{\XttIccOsSqlRdaFone}{0.923\xspace}
\newcommand{\XttIccOsSqlRseRecall}{0.939\xspace}
\newcommand{\XttIccOsSqlRsePrecision}{1.000\xspace}
\newcommand{\XttIccOsSqlRseFone}{0.969\xspace}
\newcommand{\XttIccOsVimGT}{492274\xspace}
\newcommand{\XttIccOsVimLsRecall}{1\xspace}
\newcommand{\XttIccOsVimLsPrecision}{1\xspace}
\newcommand{\XttIccOsVimLsFone}{1\xspace}
\newcommand{\XttIccOsVimBapRecall}{0.914\xspace}
\newcommand{\XttIccOsVimBapPrecision}{1\xspace}
\newcommand{\XttIccOsVimBapFone}{0.955\xspace}
\newcommand{\XttIccOsVimGhiRecall}{0.999\xspace}
\newcommand{\XttIccOsVimGhiPrecision}{1\xspace}
\newcommand{\XttIccOsVimGhiFone}{0.999\xspace}
\newcommand{\XttIccOsVimRdaRecall}{0.902\xspace}
\newcommand{\XttIccOsVimRdaPrecision}{1\xspace}
\newcommand{\XttIccOsVimRdaFone}{0.948\xspace}
\newcommand{\XttIccOsVimRseRecall}{0.984\xspace}
\newcommand{\XttIccOsVimRsePrecision}{1\xspace}
\newcommand{\XttIccOsVimRseFone}{0.992\xspace}
\newcommand{\XttIccOsVsfGT}{24534\xspace}
\newcommand{\XttIccOsVsfLsRecall}{1\xspace}
\newcommand{\XttIccOsVsfLsPrecision}{1\xspace}
\newcommand{\XttIccOsVsfLsFone}{1\xspace}
\newcommand{\XttIccOsVsfBapRecall}{0.987\xspace}
\newcommand{\XttIccOsVsfBapPrecision}{1\xspace}
\newcommand{\XttIccOsVsfBapFone}{0.993\xspace}
\newcommand{\XttIccOsVsfGhiRecall}{1\xspace}
\newcommand{\XttIccOsVsfGhiPrecision}{1\xspace}
\newcommand{\XttIccOsVsfGhiFone}{1\xspace}
\newcommand{\XttIccOsVsfRdaRecall}{0.976\xspace}
\newcommand{\XttIccOsVsfRdaPrecision}{0.997\xspace}
\newcommand{\XttIccOsVsfRdaFone}{0.986\xspace}
\newcommand{\XttIccOsVsfRseRecall}{1.000\xspace}
\newcommand{\XttIccOsVsfRsePrecision}{1\xspace}
\newcommand{\XttIccOsVsfRseFone}{1.000\xspace}
\newcommand{\XsfClOdSzpGT}{18244\xspace}
\newcommand{\XsfClOdSzpLsRecall}{1\xspace}
\newcommand{\XsfClOdSzpLsPrecision}{0.999\xspace}
\newcommand{\XsfClOdSzpLsFone}{1.000\xspace}
\newcommand{\XsfClOdSzpBapRecall}{0.952\xspace}
\newcommand{\XsfClOdSzpBapPrecision}{1\xspace}
\newcommand{\XsfClOdSzpBapFone}{0.976\xspace}
\newcommand{\XsfClOdSzpGhiRecall}{1.000\xspace}
\newcommand{\XsfClOdSzpGhiPrecision}{1\xspace}
\newcommand{\XsfClOdSzpGhiFone}{1.000\xspace}
\newcommand{\XsfClOdSzpRdaRecall}{0.966\xspace}
\newcommand{\XsfClOdSzpRdaPrecision}{1\xspace}
\newcommand{\XsfClOdSzpRdaFone}{0.982\xspace}
\newcommand{\XsfClOdSzpRseRecall}{0.966\xspace}
\newcommand{\XsfClOdSzpRsePrecision}{1\xspace}
\newcommand{\XsfClOdSzpRseFone}{0.983\xspace}
\newcommand{\XsfClOdCapGT}{282530\xspace}
\newcommand{\XsfClOdCapLsRecall}{1\xspace}
\newcommand{\XsfClOdCapLsPrecision}{0.955\xspace}
\newcommand{\XsfClOdCapLsFone}{0.977\xspace}
\newcommand{\XsfClOdCapBapRecall}{0.244\xspace}
\newcommand{\XsfClOdCapBapPrecision}{1\xspace}
\newcommand{\XsfClOdCapBapFone}{0.392\xspace}
\newcommand{\XsfClOdCapGhiRecall}{0.733\xspace}
\newcommand{\XsfClOdCapGhiPrecision}{1\xspace}
\newcommand{\XsfClOdCapGhiFone}{0.846\xspace}
\newcommand{\XsfClOdCapRdaRecall}{0.734\xspace}
\newcommand{\XsfClOdCapRdaPrecision}{1.000\xspace}
\newcommand{\XsfClOdCapRdaFone}{0.846\xspace}
\newcommand{\XsfClOdCapRseRecall}{0.400\xspace}
\newcommand{\XsfClOdCapRsePrecision}{1\xspace}
\newcommand{\XsfClOdCapRseFone}{0.572\xspace}
\newcommand{\XsfClOdBzpGT}{22765\xspace}
\newcommand{\XsfClOdBzpLsRecall}{1\xspace}
\newcommand{\XsfClOdBzpLsPrecision}{0.993\xspace}
\newcommand{\XsfClOdBzpLsFone}{0.997\xspace}
\newcommand{\XsfClOdBzpBapRecall}{0.756\xspace}
\newcommand{\XsfClOdBzpBapPrecision}{1\xspace}
\newcommand{\XsfClOdBzpBapFone}{0.861\xspace}
\newcommand{\XsfClOdBzpGhiRecall}{0.988\xspace}
\newcommand{\XsfClOdBzpGhiPrecision}{1\xspace}
\newcommand{\XsfClOdBzpGhiFone}{0.994\xspace}
\newcommand{\XsfClOdBzpRdaRecall}{0.758\xspace}
\newcommand{\XsfClOdBzpRdaPrecision}{1\xspace}
\newcommand{\XsfClOdBzpRdaFone}{0.862\xspace}
\newcommand{\XsfClOdBzpRseRecall}{0.984\xspace}
\newcommand{\XsfClOdBzpRsePrecision}{1\xspace}
\newcommand{\XsfClOdBzpRseFone}{0.992\xspace}
\newcommand{\XsfClOdPcrGT}{69111\xspace}
\newcommand{\XsfClOdPcrLsRecall}{1\xspace}
\newcommand{\XsfClOdPcrLsPrecision}{0.972\xspace}
\newcommand{\XsfClOdPcrLsFone}{0.986\xspace}
\newcommand{\XsfClOdPcrBapRecall}{0.328\xspace}
\newcommand{\XsfClOdPcrBapPrecision}{1\xspace}
\newcommand{\XsfClOdPcrBapFone}{0.493\xspace}
\newcommand{\XsfClOdPcrGhiRecall}{0.234\xspace}
\newcommand{\XsfClOdPcrGhiPrecision}{1\xspace}
\newcommand{\XsfClOdPcrGhiFone}{0.379\xspace}
\newcommand{\XsfClOdPcrRdaRecall}{0.568\xspace}
\newcommand{\XsfClOdPcrRdaPrecision}{1\xspace}
\newcommand{\XsfClOdPcrRdaFone}{0.724\xspace}
\newcommand{\XsfClOdPcrRseRecall}{0.575\xspace}
\newcommand{\XsfClOdPcrRsePrecision}{1.000\xspace}
\newcommand{\XsfClOdPcrRseFone}{0.730\xspace}
\newcommand{\XsfClOdPutGT}{199714\xspace}
\newcommand{\XsfClOdPutLsRecall}{1.000\xspace}
\newcommand{\XsfClOdPutLsPrecision}{0.988\xspace}
\newcommand{\XsfClOdPutLsFone}{0.994\xspace}
\newcommand{\XsfClOdPutBapRecall}{0.722\xspace}
\newcommand{\XsfClOdPutBapPrecision}{1\xspace}
\newcommand{\XsfClOdPutBapFone}{0.839\xspace}
\newcommand{\XsfClOdPutGhiRecall}{0.967\xspace}
\newcommand{\XsfClOdPutGhiPrecision}{1\xspace}
\newcommand{\XsfClOdPutGhiFone}{0.983\xspace}
\newcommand{\XsfClOdPutRdaRecall}{0.720\xspace}
\newcommand{\XsfClOdPutRdaPrecision}{1\xspace}
\newcommand{\XsfClOdPutRdaFone}{0.837\xspace}
\newcommand{\XsfClOdPutRseRecall}{0.732\xspace}
\newcommand{\XsfClOdPutRsePrecision}{1.000\xspace}
\newcommand{\XsfClOdPutRseFone}{0.845\xspace}
\newcommand{\XsfClOdSqlGT}{201527\xspace}
\newcommand{\XsfClOdSqlLsRecall}{1\xspace}
\newcommand{\XsfClOdSqlLsPrecision}{0.992\xspace}
\newcommand{\XsfClOdSqlLsFone}{0.996\xspace}
\newcommand{\XsfClOdSqlBapRecall}{0.785\xspace}
\newcommand{\XsfClOdSqlBapPrecision}{1\xspace}
\newcommand{\XsfClOdSqlBapFone}{0.879\xspace}
\newcommand{\XsfClOdSqlGhiRecall}{0.906\xspace}
\newcommand{\XsfClOdSqlGhiPrecision}{1\xspace}
\newcommand{\XsfClOdSqlGhiFone}{0.951\xspace}
\newcommand{\XsfClOdSqlRdaRecall}{0.824\xspace}
\newcommand{\XsfClOdSqlRdaPrecision}{1\xspace}
\newcommand{\XsfClOdSqlRdaFone}{0.903\xspace}
\newcommand{\XsfClOdSqlRseRecall}{0.814\xspace}
\newcommand{\XsfClOdSqlRsePrecision}{1.000\xspace}
\newcommand{\XsfClOdSqlRseFone}{0.897\xspace}
\newcommand{\XsfClOdVimGT}{592322\xspace}
\newcommand{\XsfClOdVimLsRecall}{1\xspace}
\newcommand{\XsfClOdVimLsPrecision}{0.986\xspace}
\newcommand{\XsfClOdVimLsFone}{0.993\xspace}
\newcommand{\XsfClOdVimBapRecall}{0.812\xspace}
\newcommand{\XsfClOdVimBapPrecision}{1.000\xspace}
\newcommand{\XsfClOdVimBapFone}{0.896\xspace}
\newcommand{\XsfClOdVimGhiRecall}{0.988\xspace}
\newcommand{\XsfClOdVimGhiPrecision}{1\xspace}
\newcommand{\XsfClOdVimGhiFone}{0.994\xspace}
\newcommand{\XsfClOdVimRdaRecall}{0.871\xspace}
\newcommand{\XsfClOdVimRdaPrecision}{1.000\xspace}
\newcommand{\XsfClOdVimRdaFone}{0.931\xspace}
\newcommand{\XsfClOdVimRseRecall}{0.864\xspace}
\newcommand{\XsfClOdVimRsePrecision}{1.000\xspace}
\newcommand{\XsfClOdVimRseFone}{0.927\xspace}
\newcommand{\XsfClOaSzpGT}{10500\xspace}
\newcommand{\XsfClOaSzpLsRecall}{1\xspace}
\newcommand{\XsfClOaSzpLsPrecision}{1\xspace}
\newcommand{\XsfClOaSzpLsFone}{1\xspace}
\newcommand{\XsfClOaSzpBapRecall}{0.964\xspace}
\newcommand{\XsfClOaSzpBapPrecision}{1\xspace}
\newcommand{\XsfClOaSzpBapFone}{0.981\xspace}
\newcommand{\XsfClOaSzpGhiRecall}{1\xspace}
\newcommand{\XsfClOaSzpGhiPrecision}{1\xspace}
\newcommand{\XsfClOaSzpGhiFone}{1\xspace}
\newcommand{\XsfClOaSzpRdaRecall}{0.964\xspace}
\newcommand{\XsfClOaSzpRdaPrecision}{1\xspace}
\newcommand{\XsfClOaSzpRdaFone}{0.981\xspace}
\newcommand{\XsfClOaSzpRseRecall}{0.968\xspace}
\newcommand{\XsfClOaSzpRsePrecision}{1\xspace}
\newcommand{\XsfClOaSzpRseFone}{0.984\xspace}
\newcommand{\XsfClOaCapGT}{185701\xspace}
\newcommand{\XsfClOaCapLsRecall}{1\xspace}
\newcommand{\XsfClOaCapLsPrecision}{0.981\xspace}
\newcommand{\XsfClOaCapLsFone}{0.991\xspace}
\newcommand{\XsfClOaCapBapRecall}{0.784\xspace}
\newcommand{\XsfClOaCapBapPrecision}{1.000\xspace}
\newcommand{\XsfClOaCapBapFone}{0.879\xspace}
\newcommand{\XsfClOaCapGhiRecall}{0.982\xspace}
\newcommand{\XsfClOaCapGhiPrecision}{1\xspace}
\newcommand{\XsfClOaCapGhiFone}{0.991\xspace}
\newcommand{\XsfClOaCapRdaRecall}{0.878\xspace}
\newcommand{\XsfClOaCapRdaPrecision}{1.000\xspace}
\newcommand{\XsfClOaCapRdaFone}{0.935\xspace}
\newcommand{\XsfClOaCapRseRecall}{0.776\xspace}
\newcommand{\XsfClOaCapRsePrecision}{1.000\xspace}
\newcommand{\XsfClOaCapRseFone}{0.874\xspace}
\newcommand{\XsfClOaBzpGT}{12628\xspace}
\newcommand{\XsfClOaBzpLsRecall}{1\xspace}
\newcommand{\XsfClOaBzpLsPrecision}{1\xspace}
\newcommand{\XsfClOaBzpLsFone}{1\xspace}
\newcommand{\XsfClOaBzpBapRecall}{0.952\xspace}
\newcommand{\XsfClOaBzpBapPrecision}{1\xspace}
\newcommand{\XsfClOaBzpBapFone}{0.975\xspace}
\newcommand{\XsfClOaBzpGhiRecall}{0.968\xspace}
\newcommand{\XsfClOaBzpGhiPrecision}{1\xspace}
\newcommand{\XsfClOaBzpGhiFone}{0.984\xspace}
\newcommand{\XsfClOaBzpRdaRecall}{0.937\xspace}
\newcommand{\XsfClOaBzpRdaPrecision}{1\xspace}
\newcommand{\XsfClOaBzpRdaFone}{0.967\xspace}
\newcommand{\XsfClOaBzpRseRecall}{0.952\xspace}
\newcommand{\XsfClOaBzpRsePrecision}{1\xspace}
\newcommand{\XsfClOaBzpRseFone}{0.975\xspace}
\newcommand{\XsfClOaPcrGT}{37856\xspace}
\newcommand{\XsfClOaPcrLsRecall}{1\xspace}
\newcommand{\XsfClOaPcrLsPrecision}{0.990\xspace}
\newcommand{\XsfClOaPcrLsFone}{0.995\xspace}
\newcommand{\XsfClOaPcrBapRecall}{0.570\xspace}
\newcommand{\XsfClOaPcrBapPrecision}{1\xspace}
\newcommand{\XsfClOaPcrBapFone}{0.726\xspace}
\newcommand{\XsfClOaPcrGhiRecall}{0.611\xspace}
\newcommand{\XsfClOaPcrGhiPrecision}{1\xspace}
\newcommand{\XsfClOaPcrGhiFone}{0.758\xspace}
\newcommand{\XsfClOaPcrRdaRecall}{0.962\xspace}
\newcommand{\XsfClOaPcrRdaPrecision}{1\xspace}
\newcommand{\XsfClOaPcrRdaFone}{0.981\xspace}
\newcommand{\XsfClOaPcrRseRecall}{0.972\xspace}
\newcommand{\XsfClOaPcrRsePrecision}{1.000\xspace}
\newcommand{\XsfClOaPcrRseFone}{0.986\xspace}
\newcommand{\XsfClOaPutGT}{136907\xspace}
\newcommand{\XsfClOaPutLsRecall}{1.000\xspace}
\newcommand{\XsfClOaPutLsPrecision}{0.999\xspace}
\newcommand{\XsfClOaPutLsFone}{1.000\xspace}
\newcommand{\XsfClOaPutBapRecall}{0.794\xspace}
\newcommand{\XsfClOaPutBapPrecision}{1\xspace}
\newcommand{\XsfClOaPutBapFone}{0.885\xspace}
\newcommand{\XsfClOaPutGhiRecall}{0.939\xspace}
\newcommand{\XsfClOaPutGhiPrecision}{1\xspace}
\newcommand{\XsfClOaPutGhiFone}{0.968\xspace}
\newcommand{\XsfClOaPutRdaRecall}{0.669\xspace}
\newcommand{\XsfClOaPutRdaPrecision}{1\xspace}
\newcommand{\XsfClOaPutRdaFone}{0.802\xspace}
\newcommand{\XsfClOaPutRseRecall}{0.692\xspace}
\newcommand{\XsfClOaPutRsePrecision}{1.000\xspace}
\newcommand{\XsfClOaPutRseFone}{0.818\xspace}
\newcommand{\XsfClOaSqlGT}{165250\xspace}
\newcommand{\XsfClOaSqlLsRecall}{1\xspace}
\newcommand{\XsfClOaSqlLsPrecision}{0.995\xspace}
\newcommand{\XsfClOaSqlLsFone}{0.997\xspace}
\newcommand{\XsfClOaSqlBapRecall}{0.821\xspace}
\newcommand{\XsfClOaSqlBapPrecision}{1\xspace}
\newcommand{\XsfClOaSqlBapFone}{0.901\xspace}
\newcommand{\XsfClOaSqlGhiRecall}{0.971\xspace}
\newcommand{\XsfClOaSqlGhiPrecision}{1\xspace}
\newcommand{\XsfClOaSqlGhiFone}{0.985\xspace}
\newcommand{\XsfClOaSqlRdaRecall}{0.775\xspace}
\newcommand{\XsfClOaSqlRdaPrecision}{1\xspace}
\newcommand{\XsfClOaSqlRdaFone}{0.873\xspace}
\newcommand{\XsfClOaSqlRseRecall}{0.808\xspace}
\newcommand{\XsfClOaSqlRsePrecision}{1.000\xspace}
\newcommand{\XsfClOaSqlRseFone}{0.894\xspace}
\newcommand{\XsfClOaVimGT}{433615\xspace}
\newcommand{\XsfClOaVimLsRecall}{1.000\xspace}
\newcommand{\XsfClOaVimLsPrecision}{0.997\xspace}
\newcommand{\XsfClOaVimLsFone}{0.998\xspace}
\newcommand{\XsfClOaVimBapRecall}{0.915\xspace}
\newcommand{\XsfClOaVimBapPrecision}{1.000\xspace}
\newcommand{\XsfClOaVimBapFone}{0.956\xspace}
\newcommand{\XsfClOaVimGhiRecall}{0.993\xspace}
\newcommand{\XsfClOaVimGhiPrecision}{1\xspace}
\newcommand{\XsfClOaVimGhiFone}{0.996\xspace}
\newcommand{\XsfClOaVimRdaRecall}{0.871\xspace}
\newcommand{\XsfClOaVimRdaPrecision}{1\xspace}
\newcommand{\XsfClOaVimRdaFone}{0.931\xspace}
\newcommand{\XsfClOaVimRseRecall}{0.881\xspace}
\newcommand{\XsfClOaVimRsePrecision}{1\xspace}
\newcommand{\XsfClOaVimRseFone}{0.937\xspace}
\newcommand{\XsfClObSzpGT}{11026\xspace}
\newcommand{\XsfClObSzpLsRecall}{1\xspace}
\newcommand{\XsfClObSzpLsPrecision}{0.997\xspace}
\newcommand{\XsfClObSzpLsFone}{0.999\xspace}
\newcommand{\XsfClObSzpBapRecall}{0.934\xspace}
\newcommand{\XsfClObSzpBapPrecision}{1\xspace}
\newcommand{\XsfClObSzpBapFone}{0.966\xspace}
\newcommand{\XsfClObSzpGhiRecall}{1\xspace}
\newcommand{\XsfClObSzpGhiPrecision}{1\xspace}
\newcommand{\XsfClObSzpGhiFone}{1\xspace}
\newcommand{\XsfClObSzpRdaRecall}{0.960\xspace}
\newcommand{\XsfClObSzpRdaPrecision}{1\xspace}
\newcommand{\XsfClObSzpRdaFone}{0.980\xspace}
\newcommand{\XsfClObSzpRseRecall}{0.956\xspace}
\newcommand{\XsfClObSzpRsePrecision}{1\xspace}
\newcommand{\XsfClObSzpRseFone}{0.977\xspace}
\newcommand{\XsfClObCapGT}{200703\xspace}
\newcommand{\XsfClObCapLsRecall}{1\xspace}
\newcommand{\XsfClObCapLsPrecision}{0.934\xspace}
\newcommand{\XsfClObCapLsFone}{0.966\xspace}
\newcommand{\XsfClObCapBapRecall}{0.220\xspace}
\newcommand{\XsfClObCapBapPrecision}{1\xspace}
\newcommand{\XsfClObCapBapFone}{0.361\xspace}
\newcommand{\XsfClObCapGhiRecall}{0.847\xspace}
\newcommand{\XsfClObCapGhiPrecision}{1\xspace}
\newcommand{\XsfClObCapGhiFone}{0.917\xspace}
\newcommand{\XsfClObCapRdaRecall}{0.522\xspace}
\newcommand{\XsfClObCapRdaPrecision}{1.000\xspace}
\newcommand{\XsfClObCapRdaFone}{0.686\xspace}
\newcommand{\XsfClObCapRseRecall}{0.441\xspace}
\newcommand{\XsfClObCapRsePrecision}{0.993\xspace}
\newcommand{\XsfClObCapRseFone}{0.611\xspace}
\newcommand{\XsfClObBzpGT}{17103\xspace}
\newcommand{\XsfClObBzpLsRecall}{1\xspace}
\newcommand{\XsfClObBzpLsPrecision}{0.994\xspace}
\newcommand{\XsfClObBzpLsFone}{0.997\xspace}
\newcommand{\XsfClObBzpBapRecall}{0.731\xspace}
\newcommand{\XsfClObBzpBapPrecision}{1\xspace}
\newcommand{\XsfClObBzpBapFone}{0.845\xspace}
\newcommand{\XsfClObBzpGhiRecall}{0.957\xspace}
\newcommand{\XsfClObBzpGhiPrecision}{1\xspace}
\newcommand{\XsfClObBzpGhiFone}{0.978\xspace}
\newcommand{\XsfClObBzpRdaRecall}{0.645\xspace}
\newcommand{\XsfClObBzpRdaPrecision}{1\xspace}
\newcommand{\XsfClObBzpRdaFone}{0.784\xspace}
\newcommand{\XsfClObBzpRseRecall}{0.862\xspace}
\newcommand{\XsfClObBzpRsePrecision}{1\xspace}
\newcommand{\XsfClObBzpRseFone}{0.926\xspace}
\newcommand{\XsfClObPcrGT}{39830\xspace}
\newcommand{\XsfClObPcrLsRecall}{1\xspace}
\newcommand{\XsfClObPcrLsPrecision}{0.947\xspace}
\newcommand{\XsfClObPcrLsFone}{0.973\xspace}
\newcommand{\XsfClObPcrBapRecall}{0.396\xspace}
\newcommand{\XsfClObPcrBapPrecision}{1\xspace}
\newcommand{\XsfClObPcrBapFone}{0.567\xspace}
\newcommand{\XsfClObPcrGhiRecall}{0.629\xspace}
\newcommand{\XsfClObPcrGhiPrecision}{1\xspace}
\newcommand{\XsfClObPcrGhiFone}{0.772\xspace}
\newcommand{\XsfClObPcrRdaRecall}{0.441\xspace}
\newcommand{\XsfClObPcrRdaPrecision}{1\xspace}
\newcommand{\XsfClObPcrRdaFone}{0.612\xspace}
\newcommand{\XsfClObPcrRseRecall}{0.709\xspace}
\newcommand{\XsfClObPcrRsePrecision}{0.997\xspace}
\newcommand{\XsfClObPcrRseFone}{0.829\xspace}
\newcommand{\XsfClObPutGT}{157942\xspace}
\newcommand{\XsfClObPutLsRecall}{1.000\xspace}
\newcommand{\XsfClObPutLsPrecision}{0.986\xspace}
\newcommand{\XsfClObPutLsFone}{0.993\xspace}
\newcommand{\XsfClObPutBapRecall}{0.645\xspace}
\newcommand{\XsfClObPutBapPrecision}{1.000\xspace}
\newcommand{\XsfClObPutBapFone}{0.784\xspace}
\newcommand{\XsfClObPutGhiRecall}{0.953\xspace}
\newcommand{\XsfClObPutGhiPrecision}{1\xspace}
\newcommand{\XsfClObPutGhiFone}{0.976\xspace}
\newcommand{\XsfClObPutRdaRecall}{0.603\xspace}
\newcommand{\XsfClObPutRdaPrecision}{1\xspace}
\newcommand{\XsfClObPutRdaFone}{0.753\xspace}
\newcommand{\XsfClObPutRseRecall}{0.650\xspace}
\newcommand{\XsfClObPutRsePrecision}{1.000\xspace}
\newcommand{\XsfClObPutRseFone}{0.788\xspace}
\newcommand{\XsfClObSqlGT}{235347\xspace}
\newcommand{\XsfClObSqlLsRecall}{1\xspace}
\newcommand{\XsfClObSqlLsPrecision}{0.992\xspace}
\newcommand{\XsfClObSqlLsFone}{0.996\xspace}
\newcommand{\XsfClObSqlBapRecall}{0.756\xspace}
\newcommand{\XsfClObSqlBapPrecision}{1.000\xspace}
\newcommand{\XsfClObSqlBapFone}{0.861\xspace}
\newcommand{\XsfClObSqlGhiRecall}{0.958\xspace}
\newcommand{\XsfClObSqlGhiPrecision}{1\xspace}
\newcommand{\XsfClObSqlGhiFone}{0.978\xspace}
\newcommand{\XsfClObSqlRdaRecall}{0.714\xspace}
\newcommand{\XsfClObSqlRdaPrecision}{1\xspace}
\newcommand{\XsfClObSqlRdaFone}{0.833\xspace}
\newcommand{\XsfClObSqlRseRecall}{0.755\xspace}
\newcommand{\XsfClObSqlRsePrecision}{1.000\xspace}
\newcommand{\XsfClObSqlRseFone}{0.860\xspace}
\newcommand{\XsfClObVimGT}{511380\xspace}
\newcommand{\XsfClObVimLsRecall}{1.000\xspace}
\newcommand{\XsfClObVimLsPrecision}{0.984\xspace}
\newcommand{\XsfClObVimLsFone}{0.992\xspace}
\newcommand{\XsfClObVimBapRecall}{0.813\xspace}
\newcommand{\XsfClObVimBapPrecision}{1.000\xspace}
\newcommand{\XsfClObVimBapFone}{0.897\xspace}
\newcommand{\XsfClObVimGhiRecall}{0.996\xspace}
\newcommand{\XsfClObVimGhiPrecision}{1\xspace}
\newcommand{\XsfClObVimGhiFone}{0.998\xspace}
\newcommand{\XsfClObVimRdaRecall}{0.832\xspace}
\newcommand{\XsfClObVimRdaPrecision}{1\xspace}
\newcommand{\XsfClObVimRdaFone}{0.908\xspace}
\newcommand{\XsfClObVimRseRecall}{0.841\xspace}
\newcommand{\XsfClObVimRsePrecision}{1.000\xspace}
\newcommand{\XsfClObVimRseFone}{0.914\xspace}
\newcommand{\XttClOdSzpGT}{17774\xspace}
\newcommand{\XttClOdSzpLsRecall}{1\xspace}
\newcommand{\XttClOdSzpLsPrecision}{0.999\xspace}
\newcommand{\XttClOdSzpLsFone}{1.000\xspace}
\newcommand{\XttClOdSzpBapRecall}{0.942\xspace}
\newcommand{\XttClOdSzpBapPrecision}{1\xspace}
\newcommand{\XttClOdSzpBapFone}{0.970\xspace}
\newcommand{\XttClOdSzpGhiRecall}{1.000\xspace}
\newcommand{\XttClOdSzpGhiPrecision}{1\xspace}
\newcommand{\XttClOdSzpGhiFone}{1.000\xspace}
\newcommand{\XttClOdSzpRdaRecall}{0.965\xspace}
\newcommand{\XttClOdSzpRdaPrecision}{1\xspace}
\newcommand{\XttClOdSzpRdaFone}{0.982\xspace}
\newcommand{\XttClOdSzpRseRecall}{1\xspace}
\newcommand{\XttClOdSzpRsePrecision}{1\xspace}
\newcommand{\XttClOdSzpRseFone}{1\xspace}
\newcommand{\XttClOdCapGT}{388495\xspace}
\newcommand{\XttClOdCapLsRecall}{1.000\xspace}
\newcommand{\XttClOdCapLsPrecision}{0.957\xspace}
\newcommand{\XttClOdCapLsFone}{0.978\xspace}
\newcommand{\XttClOdCapBapRecall}{0.207\xspace}
\newcommand{\XttClOdCapBapPrecision}{1\xspace}
\newcommand{\XttClOdCapBapFone}{0.343\xspace}
\newcommand{\XttClOdCapGhiRecall}{0.998\xspace}
\newcommand{\XttClOdCapGhiPrecision}{1\xspace}
\newcommand{\XttClOdCapGhiFone}{0.999\xspace}
\newcommand{\XttClOdCapRdaRecall}{0.949\xspace}
\newcommand{\XttClOdCapRdaPrecision}{1.000\xspace}
\newcommand{\XttClOdCapRdaFone}{0.974\xspace}
\newcommand{\XttClOdCapRseRecall}{1.000\xspace}
\newcommand{\XttClOdCapRsePrecision}{0.998\xspace}
\newcommand{\XttClOdCapRseFone}{0.999\xspace}
\newcommand{\XttClOdBzpGT}{22637\xspace}
\newcommand{\XttClOdBzpLsRecall}{1.000\xspace}
\newcommand{\XttClOdBzpLsPrecision}{0.993\xspace}
\newcommand{\XttClOdBzpLsFone}{0.996\xspace}
\newcommand{\XttClOdBzpBapRecall}{0.758\xspace}
\newcommand{\XttClOdBzpBapPrecision}{1\xspace}
\newcommand{\XttClOdBzpBapFone}{0.863\xspace}
\newcommand{\XttClOdBzpGhiRecall}{0.999\xspace}
\newcommand{\XttClOdBzpGhiPrecision}{1\xspace}
\newcommand{\XttClOdBzpGhiFone}{1.000\xspace}
\newcommand{\XttClOdBzpRdaRecall}{0.973\xspace}
\newcommand{\XttClOdBzpRdaPrecision}{1\xspace}
\newcommand{\XttClOdBzpRdaFone}{0.986\xspace}
\newcommand{\XttClOdBzpRseRecall}{1\xspace}
\newcommand{\XttClOdBzpRsePrecision}{1\xspace}
\newcommand{\XttClOdBzpRseFone}{1\xspace}
\newcommand{\XttClOdPcrGT}{67973\xspace}
\newcommand{\XttClOdPcrLsRecall}{1.000\xspace}
\newcommand{\XttClOdPcrLsPrecision}{0.958\xspace}
\newcommand{\XttClOdPcrLsFone}{0.978\xspace}
\newcommand{\XttClOdPcrBapRecall}{0.320\xspace}
\newcommand{\XttClOdPcrBapPrecision}{1\xspace}
\newcommand{\XttClOdPcrBapFone}{0.485\xspace}
\newcommand{\XttClOdPcrGhiRecall}{0.997\xspace}
\newcommand{\XttClOdPcrGhiPrecision}{1\xspace}
\newcommand{\XttClOdPcrGhiFone}{0.998\xspace}
\newcommand{\XttClOdPcrRdaRecall}{0.985\xspace}
\newcommand{\XttClOdPcrRdaPrecision}{1\xspace}
\newcommand{\XttClOdPcrRdaFone}{0.993\xspace}
\newcommand{\XttClOdPcrRseRecall}{1\xspace}
\newcommand{\XttClOdPcrRsePrecision}{1.000\xspace}
\newcommand{\XttClOdPcrRseFone}{1.000\xspace}
\newcommand{\XttClOdPutGT}{224349\xspace}
\newcommand{\XttClOdPutLsRecall}{1.000\xspace}
\newcommand{\XttClOdPutLsPrecision}{0.983\xspace}
\newcommand{\XttClOdPutLsFone}{0.991\xspace}
\newcommand{\XttClOdPutBapRecall}{0.716\xspace}
\newcommand{\XttClOdPutBapPrecision}{1\xspace}
\newcommand{\XttClOdPutBapFone}{0.834\xspace}
\newcommand{\XttClOdPutGhiRecall}{0.999\xspace}
\newcommand{\XttClOdPutGhiPrecision}{1\xspace}
\newcommand{\XttClOdPutGhiFone}{0.999\xspace}
\newcommand{\XttClOdPutRdaRecall}{0.729\xspace}
\newcommand{\XttClOdPutRdaPrecision}{1\xspace}
\newcommand{\XttClOdPutRdaFone}{0.843\xspace}
\newcommand{\XttClOdPutRseRecall}{0.992\xspace}
\newcommand{\XttClOdPutRsePrecision}{1.000\xspace}
\newcommand{\XttClOdPutRseFone}{0.996\xspace}
\newcommand{\XttClOdSqlGT}{237869\xspace}
\newcommand{\XttClOdSqlLsRecall}{1.000\xspace}
\newcommand{\XttClOdSqlLsPrecision}{0.989\xspace}
\newcommand{\XttClOdSqlLsFone}{0.994\xspace}
\newcommand{\XttClOdSqlBapRecall}{0.789\xspace}
\newcommand{\XttClOdSqlBapPrecision}{1\xspace}
\newcommand{\XttClOdSqlBapFone}{0.882\xspace}
\newcommand{\XttClOdSqlGhiRecall}{0.999\xspace}
\newcommand{\XttClOdSqlGhiPrecision}{1\xspace}
\newcommand{\XttClOdSqlGhiFone}{1.000\xspace}
\newcommand{\XttClOdSqlRdaRecall}{0.891\xspace}
\newcommand{\XttClOdSqlRdaPrecision}{1\xspace}
\newcommand{\XttClOdSqlRdaFone}{0.942\xspace}
\newcommand{\XttClOdSqlRseRecall}{1\xspace}
\newcommand{\XttClOdSqlRsePrecision}{0.999\xspace}
\newcommand{\XttClOdSqlRseFone}{1.000\xspace}
\newcommand{\XttClOdVimGT}{651318\xspace}
\newcommand{\XttClOdVimLsRecall}{1.000\xspace}
\newcommand{\XttClOdVimLsPrecision}{0.981\xspace}
\newcommand{\XttClOdVimLsFone}{0.990\xspace}
\newcommand{\XttClOdVimBapRecall}{0.823\xspace}
\newcommand{\XttClOdVimBapPrecision}{1\xspace}
\newcommand{\XttClOdVimBapFone}{0.903\xspace}
\newcommand{\XttClOdVimGhiRecall}{0.999\xspace}
\newcommand{\XttClOdVimGhiPrecision}{1\xspace}
\newcommand{\XttClOdVimGhiFone}{1.000\xspace}
\newcommand{\XttClOdVimRdaRecall}{0.878\xspace}
\newcommand{\XttClOdVimRdaPrecision}{1.000\xspace}
\newcommand{\XttClOdVimRdaFone}{0.935\xspace}
\newcommand{\XttClOdVimRseRecall}{1.000\xspace}
\newcommand{\XttClOdVimRsePrecision}{1.000\xspace}
\newcommand{\XttClOdVimRseFone}{1.000\xspace}
\newcommand{\XttClOaSzpGT}{11430\xspace}
\newcommand{\XttClOaSzpLsRecall}{1\xspace}
\newcommand{\XttClOaSzpLsPrecision}{1\xspace}
\newcommand{\XttClOaSzpLsFone}{1\xspace}
\newcommand{\XttClOaSzpBapRecall}{0.964\xspace}
\newcommand{\XttClOaSzpBapPrecision}{1\xspace}
\newcommand{\XttClOaSzpBapFone}{0.982\xspace}
\newcommand{\XttClOaSzpGhiRecall}{1\xspace}
\newcommand{\XttClOaSzpGhiPrecision}{1\xspace}
\newcommand{\XttClOaSzpGhiFone}{1\xspace}
\newcommand{\XttClOaSzpRdaRecall}{0.964\xspace}
\newcommand{\XttClOaSzpRdaPrecision}{1\xspace}
\newcommand{\XttClOaSzpRdaFone}{0.982\xspace}
\newcommand{\XttClOaSzpRseRecall}{0.998\xspace}
\newcommand{\XttClOaSzpRsePrecision}{1\xspace}
\newcommand{\XttClOaSzpRseFone}{0.999\xspace}
\newcommand{\XttClOaCapGT}{218002\xspace}
\newcommand{\XttClOaCapLsRecall}{1\xspace}
\newcommand{\XttClOaCapLsPrecision}{0.953\xspace}
\newcommand{\XttClOaCapLsFone}{0.976\xspace}
\newcommand{\XttClOaCapBapRecall}{0.505\xspace}
\newcommand{\XttClOaCapBapPrecision}{1\xspace}
\newcommand{\XttClOaCapBapFone}{0.671\xspace}
\newcommand{\XttClOaCapGhiRecall}{0.841\xspace}
\newcommand{\XttClOaCapGhiPrecision}{1\xspace}
\newcommand{\XttClOaCapGhiFone}{0.913\xspace}
\newcommand{\XttClOaCapRdaRecall}{0.928\xspace}
\newcommand{\XttClOaCapRdaPrecision}{1.000\xspace}
\newcommand{\XttClOaCapRdaFone}{0.963\xspace}
\newcommand{\XttClOaCapRseRecall}{0.944\xspace}
\newcommand{\XttClOaCapRsePrecision}{1.000\xspace}
\newcommand{\XttClOaCapRseFone}{0.971\xspace}
\newcommand{\XttClOaBzpGT}{13153\xspace}
\newcommand{\XttClOaBzpLsRecall}{1\xspace}
\newcommand{\XttClOaBzpLsPrecision}{0.992\xspace}
\newcommand{\XttClOaBzpLsFone}{0.996\xspace}
\newcommand{\XttClOaBzpBapRecall}{0.780\xspace}
\newcommand{\XttClOaBzpBapPrecision}{0.990\xspace}
\newcommand{\XttClOaBzpBapFone}{0.873\xspace}
\newcommand{\XttClOaBzpGhiRecall}{0.969\xspace}
\newcommand{\XttClOaBzpGhiPrecision}{0.992\xspace}
\newcommand{\XttClOaBzpGhiFone}{0.981\xspace}
\newcommand{\XttClOaBzpRdaRecall}{0.958\xspace}
\newcommand{\XttClOaBzpRdaPrecision}{0.999\xspace}
\newcommand{\XttClOaBzpRdaFone}{0.978\xspace}
\newcommand{\XttClOaBzpRseRecall}{0.975\xspace}
\newcommand{\XttClOaBzpRsePrecision}{0.992\xspace}
\newcommand{\XttClOaBzpRseFone}{0.983\xspace}
\newcommand{\XttClOaPcrGT}{38804\xspace}
\newcommand{\XttClOaPcrLsRecall}{1\xspace}
\newcommand{\XttClOaPcrLsPrecision}{0.973\xspace}
\newcommand{\XttClOaPcrLsFone}{0.986\xspace}
\newcommand{\XttClOaPcrBapRecall}{0.556\xspace}
\newcommand{\XttClOaPcrBapPrecision}{0.999\xspace}
\newcommand{\XttClOaPcrBapFone}{0.714\xspace}
\newcommand{\XttClOaPcrGhiRecall}{0.989\xspace}
\newcommand{\XttClOaPcrGhiPrecision}{0.999\xspace}
\newcommand{\XttClOaPcrGhiFone}{0.994\xspace}
\newcommand{\XttClOaPcrRdaRecall}{0.909\xspace}
\newcommand{\XttClOaPcrRdaPrecision}{1.000\xspace}
\newcommand{\XttClOaPcrRdaFone}{0.952\xspace}
\newcommand{\XttClOaPcrRseRecall}{0.983\xspace}
\newcommand{\XttClOaPcrRsePrecision}{0.998\xspace}
\newcommand{\XttClOaPcrRseFone}{0.991\xspace}
\newcommand{\XttClOaPutGT}{145490\xspace}
\newcommand{\XttClOaPutLsRecall}{1.000\xspace}
\newcommand{\XttClOaPutLsPrecision}{0.990\xspace}
\newcommand{\XttClOaPutLsFone}{0.995\xspace}
\newcommand{\XttClOaPutBapRecall}{0.691\xspace}
\newcommand{\XttClOaPutBapPrecision}{1.000\xspace}
\newcommand{\XttClOaPutBapFone}{0.817\xspace}
\newcommand{\XttClOaPutGhiRecall}{0.912\xspace}
\newcommand{\XttClOaPutGhiPrecision}{1.000\xspace}
\newcommand{\XttClOaPutGhiFone}{0.954\xspace}
\newcommand{\XttClOaPutRdaRecall}{0.696\xspace}
\newcommand{\XttClOaPutRdaPrecision}{1.000\xspace}
\newcommand{\XttClOaPutRdaFone}{0.821\xspace}
\newcommand{\XttClOaPutRseRecall}{0.791\xspace}
\newcommand{\XttClOaPutRsePrecision}{1.000\xspace}
\newcommand{\XttClOaPutRseFone}{0.883\xspace}
\newcommand{\XttClOaSqlGT}{172160\xspace}
\newcommand{\XttClOaSqlLsRecall}{1\xspace}
\newcommand{\XttClOaSqlLsPrecision}{0.992\xspace}
\newcommand{\XttClOaSqlLsFone}{0.996\xspace}
\newcommand{\XttClOaSqlBapRecall}{0.778\xspace}
\newcommand{\XttClOaSqlBapPrecision}{1.000\xspace}
\newcommand{\XttClOaSqlBapFone}{0.875\xspace}
\newcommand{\XttClOaSqlGhiRecall}{0.963\xspace}
\newcommand{\XttClOaSqlGhiPrecision}{1.000\xspace}
\newcommand{\XttClOaSqlGhiFone}{0.981\xspace}
\newcommand{\XttClOaSqlRdaRecall}{0.864\xspace}
\newcommand{\XttClOaSqlRdaPrecision}{1.000\xspace}
\newcommand{\XttClOaSqlRdaFone}{0.927\xspace}
\newcommand{\XttClOaSqlRseRecall}{0.886\xspace}
\newcommand{\XttClOaSqlRsePrecision}{1.000\xspace}
\newcommand{\XttClOaSqlRseFone}{0.940\xspace}
\newcommand{\XttClOaVimGT}{475102\xspace}
\newcommand{\XttClOaVimLsRecall}{1.000\xspace}
\newcommand{\XttClOaVimLsPrecision}{0.990\xspace}
\newcommand{\XttClOaVimLsFone}{0.995\xspace}
\newcommand{\XttClOaVimBapRecall}{0.849\xspace}
\newcommand{\XttClOaVimBapPrecision}{1\xspace}
\newcommand{\XttClOaVimBapFone}{0.919\xspace}
\newcommand{\XttClOaVimGhiRecall}{0.932\xspace}
\newcommand{\XttClOaVimGhiPrecision}{1.000\xspace}
\newcommand{\XttClOaVimGhiFone}{0.965\xspace}
\newcommand{\XttClOaVimRdaRecall}{0.883\xspace}
\newcommand{\XttClOaVimRdaPrecision}{1.000\xspace}
\newcommand{\XttClOaVimRdaFone}{0.938\xspace}
\newcommand{\XttClOaVimRseRecall}{0.921\xspace}
\newcommand{\XttClOaVimRsePrecision}{0.999\xspace}
\newcommand{\XttClOaVimRseFone}{0.959\xspace}
\newcommand{\XttClObSzpGT}{12438\xspace}
\newcommand{\XttClObSzpLsRecall}{1\xspace}
\newcommand{\XttClObSzpLsPrecision}{0.999\xspace}
\newcommand{\XttClObSzpLsFone}{1.000\xspace}
\newcommand{\XttClObSzpBapRecall}{0.937\xspace}
\newcommand{\XttClObSzpBapPrecision}{1\xspace}
\newcommand{\XttClObSzpBapFone}{0.968\xspace}
\newcommand{\XttClObSzpGhiRecall}{1\xspace}
\newcommand{\XttClObSzpGhiPrecision}{1\xspace}
\newcommand{\XttClObSzpGhiFone}{1\xspace}
\newcommand{\XttClObSzpRdaRecall}{0.963\xspace}
\newcommand{\XttClObSzpRdaPrecision}{1\xspace}
\newcommand{\XttClObSzpRdaFone}{0.981\xspace}
\newcommand{\XttClObSzpRseRecall}{0.963\xspace}
\newcommand{\XttClObSzpRsePrecision}{1\xspace}
\newcommand{\XttClObSzpRseFone}{0.981\xspace}
\newcommand{\XttClObCapGT}{237348\xspace}
\newcommand{\XttClObCapLsRecall}{1\xspace}
\newcommand{\XttClObCapLsPrecision}{0.925\xspace}
\newcommand{\XttClObCapLsFone}{0.961\xspace}
\newcommand{\XttClObCapBapRecall}{0.183\xspace}
\newcommand{\XttClObCapBapPrecision}{1\xspace}
\newcommand{\XttClObCapBapFone}{0.309\xspace}
\newcommand{\XttClObCapGhiRecall}{0.909\xspace}
\newcommand{\XttClObCapGhiPrecision}{1\xspace}
\newcommand{\XttClObCapGhiFone}{0.952\xspace}
\newcommand{\XttClObCapRdaRecall}{0.643\xspace}
\newcommand{\XttClObCapRdaPrecision}{1.000\xspace}
\newcommand{\XttClObCapRdaFone}{0.783\xspace}
\newcommand{\XttClObCapRseRecall}{0.875\xspace}
\newcommand{\XttClObCapRsePrecision}{1.000\xspace}
\newcommand{\XttClObCapRseFone}{0.933\xspace}
\newcommand{\XttClObBzpGT}{15854\xspace}
\newcommand{\XttClObBzpLsRecall}{1\xspace}
\newcommand{\XttClObBzpLsPrecision}{0.989\xspace}
\newcommand{\XttClObBzpLsFone}{0.994\xspace}
\newcommand{\XttClObBzpBapRecall}{0.740\xspace}
\newcommand{\XttClObBzpBapPrecision}{0.987\xspace}
\newcommand{\XttClObBzpBapFone}{0.846\xspace}
\newcommand{\XttClObBzpGhiRecall}{0.910\xspace}
\newcommand{\XttClObBzpGhiPrecision}{0.990\xspace}
\newcommand{\XttClObBzpGhiFone}{0.948\xspace}
\newcommand{\XttClObBzpRdaRecall}{0.886\xspace}
\newcommand{\XttClObBzpRdaPrecision}{0.996\xspace}
\newcommand{\XttClObBzpRdaFone}{0.938\xspace}
\newcommand{\XttClObBzpRseRecall}{0.911\xspace}
\newcommand{\XttClObBzpRsePrecision}{0.990\xspace}
\newcommand{\XttClObBzpRseFone}{0.949\xspace}
\newcommand{\XttClObPcrGT}{41597\xspace}
\newcommand{\XttClObPcrLsRecall}{1\xspace}
\newcommand{\XttClObPcrLsPrecision}{0.930\xspace}
\newcommand{\XttClObPcrLsFone}{0.964\xspace}
\newcommand{\XttClObPcrBapRecall}{0.385\xspace}
\newcommand{\XttClObPcrBapPrecision}{1\xspace}
\newcommand{\XttClObPcrBapFone}{0.556\xspace}
\newcommand{\XttClObPcrGhiRecall}{0.971\xspace}
\newcommand{\XttClObPcrGhiPrecision}{0.997\xspace}
\newcommand{\XttClObPcrGhiFone}{0.984\xspace}
\newcommand{\XttClObPcrRdaRecall}{0.880\xspace}
\newcommand{\XttClObPcrRdaPrecision}{0.997\xspace}
\newcommand{\XttClObPcrRdaFone}{0.935\xspace}
\newcommand{\XttClObPcrRseRecall}{0.962\xspace}
\newcommand{\XttClObPcrRsePrecision}{0.994\xspace}
\newcommand{\XttClObPcrRseFone}{0.977\xspace}
\newcommand{\XttClObPutGT}{171862\xspace}
\newcommand{\XttClObPutLsRecall}{1.000\xspace}
\newcommand{\XttClObPutLsPrecision}{0.977\xspace}
\newcommand{\XttClObPutLsFone}{0.988\xspace}
\newcommand{\XttClObPutBapRecall}{0.606\xspace}
\newcommand{\XttClObPutBapPrecision}{0.999\xspace}
\newcommand{\XttClObPutBapFone}{0.754\xspace}
\newcommand{\XttClObPutGhiRecall}{0.906\xspace}
\newcommand{\XttClObPutGhiPrecision}{0.999\xspace}
\newcommand{\XttClObPutGhiFone}{0.950\xspace}
\newcommand{\XttClObPutRdaRecall}{0.633\xspace}
\newcommand{\XttClObPutRdaPrecision}{1.000\xspace}
\newcommand{\XttClObPutRdaFone}{0.775\xspace}
\newcommand{\XttClObPutRseRecall}{0.692\xspace}
\newcommand{\XttClObPutRsePrecision}{0.999\xspace}
\newcommand{\XttClObPutRseFone}{0.818\xspace}
\newcommand{\XttClObSqlGT}{244247\xspace}
\newcommand{\XttClObSqlLsRecall}{1\xspace}
\newcommand{\XttClObSqlLsPrecision}{0.989\xspace}
\newcommand{\XttClObSqlLsFone}{0.995\xspace}
\newcommand{\XttClObSqlBapRecall}{0.707\xspace}
\newcommand{\XttClObSqlBapPrecision}{1.000\xspace}
\newcommand{\XttClObSqlBapFone}{0.828\xspace}
\newcommand{\XttClObSqlGhiRecall}{0.922\xspace}
\newcommand{\XttClObSqlGhiPrecision}{1.000\xspace}
\newcommand{\XttClObSqlGhiFone}{0.960\xspace}
\newcommand{\XttClObSqlRdaRecall}{0.801\xspace}
\newcommand{\XttClObSqlRdaPrecision}{1.000\xspace}
\newcommand{\XttClObSqlRdaFone}{0.889\xspace}
\newcommand{\XttClObSqlRseRecall}{0.820\xspace}
\newcommand{\XttClObSqlRsePrecision}{0.999\xspace}
\newcommand{\XttClObSqlRseFone}{0.901\xspace}
\newcommand{\XttClObVimGT}{540064\xspace}
\newcommand{\XttClObVimLsRecall}{1.000\xspace}
\newcommand{\XttClObVimLsPrecision}{0.976\xspace}
\newcommand{\XttClObVimLsFone}{0.988\xspace}
\newcommand{\XttClObVimBapRecall}{0.793\xspace}
\newcommand{\XttClObVimBapPrecision}{0.999\xspace}
\newcommand{\XttClObVimBapFone}{0.884\xspace}
\newcommand{\XttClObVimGhiRecall}{0.961\xspace}
\newcommand{\XttClObVimGhiPrecision}{1.000\xspace}
\newcommand{\XttClObVimGhiFone}{0.980\xspace}
\newcommand{\XttClObVimRdaRecall}{0.858\xspace}
\newcommand{\XttClObVimRdaPrecision}{0.999\xspace}
\newcommand{\XttClObVimRdaFone}{0.923\xspace}
\newcommand{\XttClObVimRseRecall}{0.877\xspace}
\newcommand{\XttClObVimRsePrecision}{0.998\xspace}
\newcommand{\XttClObVimRseFone}{0.933\xspace}

%% file: abstract.tex
\begin{abstract}

  When a software transformation or software security task needs to analyze a
  given program binary, the first step is often disassembly.
  Since many modern disassemblers have become highly accurate on many binaries,
  we believe reliable disassembler benchmarking requires standardizing the set
  of binaries used and the disassembly ground truth about these binaries.
  This paper presents (i) a first version of our work-in-progress disassembly
  benchmark suite, which comprises $879$ binaries from diverse projects compiled
  with multiple compilers and optimization settings, and
  (ii) a novel disassembly ground truth generator leveraging the notion of
  ``listing files'', which has broad support by Clang, GCC, ICC, and MSVC.
  In additional, it presents our evaluation of four prominent open-source
  disassemblers using this benchmark suite and a custom evaluation system.
  Our entire system and all generated data are maintained openly on GitHub to
  encourage community adoption.

\end{abstract}

%% file: introduction.tex
\section{Introduction}
\label{sec:intro}

Many scenarios in software transformation and software security require us to
analyze or operate on a given program binary.
The most common example is when we do not have access to the source code of the
binary.
But even when we do, we may not have access to the toolchain or the environment
needed to compile the transformed source, or we may be interested in analyses
that depend on the actual machine code in the binary such as the amount of
padding around functions or buffers.
In these applications, the first step of the analysis is usually to
\emph{disassemble} the binary, which refers to the process of translating its
machine code into assembly code.

The study of \emph{disassembly} and the corresponding tool \emph{disassemblers}
dates back to at least 1980~\cite{Horspool+Marovac:1980:AnApproach}.
In recent years, the accuracy of many disassemblers is approaching or even
exceeding $99\percent$ on many real-world binaries---see,
e.g.,~\cite{Andriesse+etal:2016:in-depth}
and~\cite{Flores-Montoya+Schulte:2020:ddisasm}.
Unfortunately, due to well-known problems in binary analysis such as
indirect jump resolution and %
function start identification, %
the proverbial ``last $1$\percent'' remains a challenge in disassembly.
In response,
our community has continued to invent innovative disassembly algorithms, with
new work and improvements appearing frequently---see,
e.g.,~\cite{Flores-Montoya+Schulte:2020:ddisasm},
\cite{Miller+etal:2019:probabilistic}, and new releases of various existing
disassemblers.

Comparing highly-accurate disassemblers is difficult for two reasons.
First, since different disassemblers may have different weaknesses, a small
change to the set of binaries used to evaluate disassemblers may change their
accuracy ranking significantly.
Unfortunately, so far the publications in the disassembly literature tend to use
different target sets; see, e.g., ~\cite{Andriesse+etal:2016:in-depth} vs.
\cite{Flores-Montoya+Schulte:2020:ddisasm}.
Second, since the accuracy of current disassemblers are exceeding $99\percent$
on many binaries, we need essentially $100\percent$-accurate ground truth to
rank the disassemblers reliably.
However, as this paper will present, while previous research tended not to
discuss their ground truth generation methods at any length, our own experience
suggests that accurate ground truth generation has much complexity and is thus
not easy to reproduce without a detailed description.

Due to the above reasons, we therefore believe it is high time for our community
to start standardizing on a set of community-accepted binaries for benchmarking
disassemblers on their accuracy and, for the sake of practicality, their running
time and memory usage.
To help jumpstart this process, this paper presents a first version of our
work-in-progress disassembly benchmark suite.
We will discuss the design of our suite from two aspects:
(a) the set of binaries included, and
(b) our method to generate accurate disassembly ground truth on the instructions
in these binaries.

For aspect (a), we believe a good disassembly benchmark suite should have these
properties:
\begin{enumerate*}[(i)]

  \item The included binaries should be \emph{diverse} in
  size,
  type (e.g., editor vs. web server),
  compilers used, %
  and optimization settings used. %
  Achieving these would allow the binaries to better capture the complexities in
  real-world binaries.

  \item The number of included binaries should be \emph{moderate}, i.e., large
  enough but not too large.
  This would ensure the practicality of evaluating disassemblers over every
  included binary, whose number grows multiplicatively in the number of included
  compilers and optimization settings in each included ISA-OS pair.

\end{enumerate*}
At present, our benchmark suite is organized by a notion of ``project names'',
which comprises an open-source project and a specific version (e.g.,
\verb!openssh-7.1p2!).
For each project name, we specify a specific target binary inside (e.g.,
\verb!sshd!) and the set of ISA-OS pairs (e.g., \{x86-Linux, x64-Linux\}) that
we include when building this project.
In addition, for each ISA-OS pair, we specify the list of compiler-version pairs
and the list of optimization settings we include during the build, which will be
detailed in~\S\ref{sec:current-benchmark}.
To encourage adoption, we maintain our benchmark suite openly on GitHub
(\url{https://github.com/pangine/disasm-benchmark}), with the hope that it will
evolve over time in ways similar to the SPEC CPU benchmark~\cite{misc:speccpu}
due to community inputs and future investigations.

For aspect (b), while one may believe that the generation of disassembly ground
truth is simply a matter of having the compiler save the assembly code generated
(e.g., \verb!gcc -S!), we find this to be an over-simplification in practice.
First, even assuming we have all the generated assembly files, in truth the
assembly statements (instructions/directives) in these files do \emph{not} form
a complete description of the machine code in the binary because some assembly
statements admit multiple machine code encodings.
Second, during the development of our current benchmark suite, we had discovered
many corner cases that needed to be handled when
collecting the generated assembly files, %
extracting information from them, %
and representing such information. %
In~\S\ref{sec:gt}, we will discuss these challenges and our solutions to them.

Finally, using our work-in-progress benchmark suite and our (already mature)
ground truth generator and disassembler evaluator, we will present
in~\S\ref{sec:eval} our findings on four prominent open-source disassemblers:
BAP~\cite{misc:bap}, Ghidra~\cite{misc:ghidra}, Radare2~\cite{misc:radaretwo},
and ROSE~\cite{misc:rose}.
Our findings largely agree with~\cite{Andriesse+etal:2016:in-depth}:
(i) function start identification remains a major issue that inhibits accurate
disassembly,
and (ii) linear-sweep disassembly~\cite{Schwarz+etal:2002:Disassembly} can be
\emph{incidentally} highly accurate, even though it also has no hope of
guaranteeing perfect accuracy and thus cannot be relied upon as a long-term
solution.

In summary, our contributions in this work are:
\begin{enumerate*}

  \item We propose to start standardizing on a set of binaries from diverse
  projects compiled with multiple compilers and optimization settings for
  disassembler evaluation and present our work-in-progress suite as a starting
  point for community discussion.

  \item We developed a new system based on a broadly-supported compiler
  toolchain feature known as ``listing files'' to generate accurate disassembly
  ground truth.
  We also developed support programs to evaluate four prominent open-source
  disassemblers against our ground truth and we present our findings.

  \item We open-source all our code and data to enable future studies and to
  encourage community adoption of and contribution to our system and our
  benchmark suite.

\end{enumerate*}

%% file: related.tex
\section{Background and Related Work}
\label{sec:bg+related}

\paragraph{Disassembly}

The problem of disassembly has a long history that dates back to at least
1980~\cite{Horspool+Marovac:1980:AnApproach}.
Nowadays, the most common disassembly algorithms are usually based on either
linear sweep or recursive traversal as presented
in~\cite{Schwarz+etal:2002:Disassembly}, often with substantial enhancements.
For more information on disassembly in general, we refer the reader to more
recent works~\cite{Khadra+Stoffel+Kunz:2016:speculative,
  Bauman+Lin+Hamlen:2018:superset, Miller+etal:2019:probabilistic,
  Flores-Montoya+Schulte:2020:ddisasm} and the references therein.
In this paper, we will be evaluating four prominent open-source disassemblers:
BAP, Ghidra, Radare2, and ROSE.
(Note that ROSE comes with multiple disassemblers and we used its recursive
traversal implementation.)
To the best of our knowledge based on reading their source code and their
documentation, we believe the evaluated disassemblers are all based on recursive
traversal.

\paragraph{Prior Work on Disassembly Ground Truth.}

The topic of disassembly ground truth generation has received surprisingly
little space in the literature.
With the exception of the work by Andriesse
\etal~\cite{Andriesse+etal:2016:in-depth}, most publications are relatively
brief in their description in this aspect.
Here we present a few examples to show three common approaches.
\begin{enumerate*}

  \item Debug info: Miller \etal~\cite{Miller+etal:2019:probabilistic} derived
  their ground truth from
  symbol information for ELF (\cite[\S5]{Miller+etal:2019:probabilistic}) and
  PDB for COFF (\cite[\S5.3]{Miller+etal:2019:probabilistic}).

  \item IDA Pro: Wartell \etal~\cite[\S3]{Wartell+etal:2011:differentiating}
  obtained their ground truth using IDA Pro, but they also specifically
  mentioned that they needed manual effort to compare the disassembly results
  because they noted inaccuracies in the IDA Pro output.

  \item Objdump: Khadra \etal~\cite[\S5.2]{Khadra+Stoffel+Kunz:2016:speculative}
  developed a custom disassembler for ground truth generation and they mentioned
  validating its result with objdump. %

\end{enumerate*}

In contrast to the brief descriptions inside the above examples, Andriesse
\etal~\cite[\S2.5]{Andriesse+etal:2016:in-depth} dedicated almost an entire
column to disassembly ground truth generation.
In their paper, they studied $981$ real-world x86 and x64 binaries from C/C++
projects compiled using GCC v5.1, Clang v3.7, and Visual C++ (MSVC) 2015 with
various optimization settings.
For each Linux binary, they used a custom LLVM pass to collect source-level
information such as source lines belonging to functions and switch statements.
Then they used DWARF to link this information to binary offsets and to extract
function starts and signatures.
Finally, they used a conservative linear-sweep to obtain the ground truth on
over $98\percent$ of the code bytes.
As for the Windows binaries, they briefly mentioned that their ground truth
extraction relied on PDB.

\newcommand{\BosGTTotal}{665\xspace}
\newcommand{\BosGTXesClangOa}{12\xspace}
\newcommand{\BosGTGccTotal}{321\xspace}
\newcommand{\BosGTClangTotal}{200\xspace}
\newcommand{\BosGTVscTotal}{144\xspace}

Our investigation of their released data and documentation reveals a few
limitations.
\begin{enumerate*}

  \item{Not extendable by others:} Although Andriesse \etal have generously
  released their build information and ground truth data in full, they did not
  release their tools and thus their benchmark suite is currently not extendable
  by others.
  Our work aims to overcome this.

  \item{Not fully-automated:} Since their ground truth generator did not cover
  the last $2\percent$ of the code bytes in their Linux binaries, Andriesse
  \etal relied on manual analysis to obtain ground truth on those remaining
  bytes.
  Since we anticipate our benchmark suite will change over time, we seek a fully
  automatic approach to increase efficiency and reliability.

  \item{Dependent on \LLVM:} Since their ground truth generator used an \LLVM
  pass to read source-level informations, this restricts the benchmark suite to
  binaries written in languages that have compiler frontends capable of using
  LLVM as a backend.
  Although our current suite contains C projects only, a dependency on LLVM
  would prohibit future extension to include binaries that do not fit the above
  criteria (current examples include Go and OCaml).

\end{enumerate*}

%% file: groundtruth.tex
\section{Ground Truth Generation}
\label{sec:gt}

In this section, we present our disassembly ground truth generation method,
which is based on the parsing and manipulation of compiler-generated assembly
files, object files, listing files, debug information, and the actual binaries.
Our description here is Linux-centric even though the targeted compilers in our
work include MSVC on Windows.
This is made possible because, as part of this project, we have matured
the technology of using Wine to run \verb!cl.exe! and related Windows tools
inside Linux to a degree that is sufficient for our Windows tasks.

\paragraph{Scope.}
The instructions in a binary can be classified into four categories:
\begin{enumerate*}[(i)]

  \item\label{enm:gt:gttypes:source} instructions emitted due to the source code
  of the binary,

  \item\label{enm:gt:gttypes:static} instructions from statically-linked
  libraries used by the binary,

  \item\label{enm:gt:gttypes:padding} \nop instructions inserted for
  alignment,

  \item\label{enm:gt:gttypes:runtime} other instructions inserted by the
  compiler toolchain, e.g., those inside \verb!_start!.

\end{enumerate*}
Our system (and thus our ground truth data) currently targets instructions of
types~\ref*{enm:gt:gttypes:source} and~\ref*{enm:gt:gttypes:padding} only.
In addition, while our system supports inline assembly, our current
investigation does not consider malware.

\begin{figure}[t]
  \centering
  \includegraphics[width=\columnwidth]{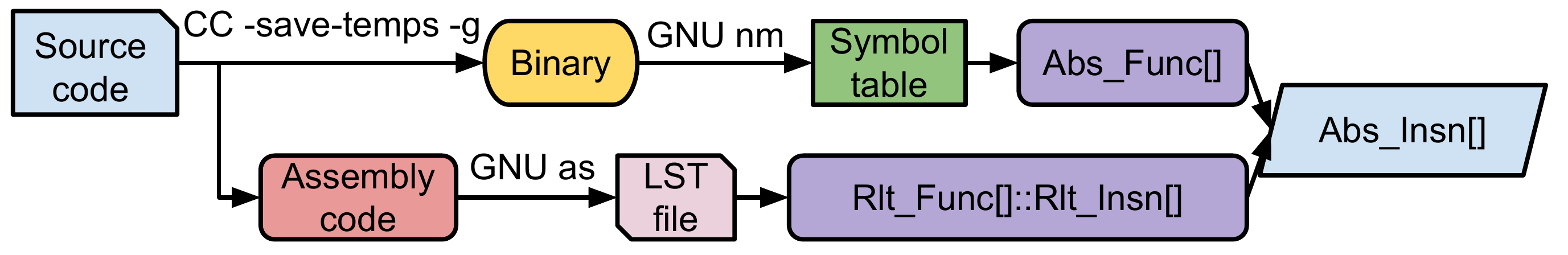}
  \caption{%
    Our disassembly ground truth generation pipeline depicted for Linux
    Clang/GCC/ICC.
    For Windows MSVC, \texttt{cl.exe} directly generates listing files and
    \texttt{dumpbin.exe} replaces \texttt{nm}.
  }
  \label{fig:gt:approach}
\end{figure}

\subsection{Our Approach}
\label{sec:gt:approach}

Our system assumes the targeted compiler toolchains are benign (not malicious)
but can have bugs.
The latter and the possibility of bugs in our own code motivate the need for the
automatic checks in~\S\ref{sec:gt:check}.
Figure~\ref{fig:gt:approach} shows the entire pipeline and the tools used for
x86/x64 Linux C programs with Clang/GCC/ICC.
The process for x86/x64 Windows with MSVC is similar, except for the specific
file formats and tools involved.

To generate a target binary and its associated ground truth data using one of
the targeted compilers and optimization settings, our system starts by using the
targeted compiler to generate the target binary along with the corresponding
symbol table and assembly files.
In Linux, these assembly files are then converted into listing files by the GNU
Assembler (\verb!as!);
in Windows, the listing files are generated directly by \verb!cl.exe! during
compilation.
The two listing file formats differ, but they both specify the size and the
offset of every instruction in every function.
We have developed a extractor for each format, and from this point on we will
call both types of listing files ``LSTs''.

From the symbol table, our system retrieves the absolute offset of each function
in the binary.
By combining this information with the offsets from LSTs, our system generates
the absolute offset of every instruction in the binary by adding the absolute
offset of its containing function and the relative offset of the instruction
within the function, i.e., $Abs\_Insn = Abs\_Func + (Rlt\_Insn - Rlt\_Func)$.

Our reliance on LSTs have both pros and cons.
Since LST generation happens to be an existing feature in all our included
compiler toolchains, our approach has the benefit of
(i) a symmetry on Linux and Windows,
(ii) the potential to support other languages with compiler toolchains that can
generate assembly and listing files (of which there are plenty in Linux, though
admittedly not in Windows), and
(iii) the avoidance to depend on compiler instrumentation, which may be
impossible for closed-source toolchains. %
In regards to (iii), we remark that the ability to add instrumentation does not
mean ground truth generation becomes trivial.
Specifically, in~\cite{Andriesse+etal:2016:in-depth}, Andriesse \etal marked all
ground truth for their Windows MSVC binaries as fully certain, but they did not
do so for their Linux binaries.
On the other hand, our LST-based approach has a shortcoming when compared to
instrumentation.
Specifically, since LSTs are generated before linking, our approach does not
support binaries compiled with link-time optimizations (LTO).
Although in principle we can instrument the linker, this would not be possible
for closed-source toolchains.
We remark that~\cite{Andriesse+etal:2016:in-depth} studied LTO binaries from x64
GCC only and they did not explain how/if their method differs on these LTO
binaries.

\subsection{Implementation Details}
\label{sec:gt:implementation}

\input{impldetail}

\subsection{Correctness Check on Our Ground Truth}
\label{sec:gt:check}

To detect potential bugs in both our ground truth generator and the targeted
compiler toolchains, we have also built an automatic correctness checker.
Recall from~\S\ref{sec:gt:me}, our system iteratively modifies an assembly file
(and thus its LST) to determine the byte-length of instructions that have
multiple encodings.
Our checker verifies that, at the end of ground truth generation, there is a
$1$-$1$ correspondence between every instruction in the final set of LSTs and in
the ground truth data.
Not counting bugs in our own system, this process has helped us uncover two bugs
in the GNU Assembler.
We have reported these bugs and one of the bugs has already been fixed
upstream.\footnote{%
  \url{https://sourceware.org/bugzilla/show_bug.cgi?id=X}, X $\in$ \{25125,
  25621\}
}

%% file: impldetail.tex
In this section, we present several challenges we had met and solved when
implementing our ground truth generator.
The content of this section has been extended significantly in comparison to our
FEAST 2020 publication.

\subsubsection{Multiple Encoding Problem}
\label{sec:gt:me}

\input{impldetail.table.mc.tex}

Some x86/x64 assembly instructions, e.g, \verb!jmp!, can be encoded into
different machine code, usually with different lengths.
In our current benchmark suite, we have observed that using different compilers
and/or different optimization settings can indeed lead to such instructions
getting different encodings.
Table~\ref{tbl:impl:mc:insn} shows all the x86/x64 multiple-encoding
instructions we have encountered so far.
We have identified two major reasons for these occurrences.
First, although both Clang and ICC can emit assembly files, they use their own
internal assemblers to produce the target binaries.
Since we pass the assembly files from Clang/ICC to the GNU Assembler to generate
LSTs, we often run into situations where the binaries from Clang/ICC disagree
with the LSTs generated by the GNU Assembler because the former and the latter
choose different encodings for some instructions.
This situation is especially common for binaries compiled by Clang with
\verb!-O0!.
This is because the internal assembler of Clang, at this optimization level,
would use a longer encoding even when a shorter encoding is available whereas
the GNU Assembler would use the shorter encoding.
Second, due to our current system design, we have
(i) one Docker image per OS-compiler pair (e.g., Ubuntu 18.04 + GCC v7.5.0) to
generate the target binaries and their associated intermediate files and
(ii) a unified Docker image to generate our ground truth data using these files.
Thus, in our design it is possible for a Linux binary in our suite to be
compiled with one version of GCC (with its corresponding version of the GNU
Assembler) but have its LSTs generated with a different version of the GNU
Assembler.

Although this multiple encoding problem could potentially be solved by manually
maintaining a multiple-encoding equivalence table per ISA and using these tables
to canonicalize the ground truth representation %
and the comparison functions during accuracy evaluation, we have opted for an
automatic and more generic solution instead.
Specifically, our solution is to \emph{iteratively} modify each generated
assembly file by replacing the first mismatched assembly instruction in each
function with the encoding obtained from the actual binary and represent a
mismatched instruction using \verb!.byte! directives explicitly.
Then we regenerate the LST with the modified assembly file and re-extract
offsets from the new LST, which will either result in full agreement in the
function or let us identify the next multiply-encoded instruction in it.

\subsubsection{Alignment Directive Translation Problem}
\label{sec:gt:align}

Due to performance reasons, machine code is often aligned at various powers of
$2$. %
With the GNU Assembler, an assembly file can use alignment directives such as
\texttt{.align} and \texttt{.p2align} to inform the assembler about the desired
alignment.
Since each alignment directive can be implemented using any valid combination of
various \nop instructions, the actual alignment instruction(s) inserted into
LSTs can be different from those in the binary.
As a result, our ground truth generator may often detect mismatches between the
LSTs and the binary at alignment locations.

Our solution to this problem is to introduce the concept of ``\nop regions'' in
our ground truth representation when dealing with alignment directives.
Specifically, a \nop region in our ground truth records only the offset and the
size of an alignment.
If the corresponding offsets in the binary are all \nop instructions known to us
and the total size of these instructions equals to the region size, then our
system considers this a match instead.

\subsubsection{As-Data Instructions Problem}
\label{sec:gt:dri}

In assembly languages, instructions and data have different syntaxes.
However, compilers can choose to represent an instruction using the data syntax.
For example, we have observed ICC spelling out the instruction \texttt{nopl
  (\%rax)} using the \texttt{.byte} directive.
Since these ``as-data instructions'' are not declared to be instructions in the
assembly files and thus also not in the LSTs, without special treatment our
ground truth generator would miss them in the generated ground truth data,
leading to ``false false-positives'' during accuracy evaluations.

Our current solution to this problem is to introduce the concept of ``optional
instructions'' in our ground truth representation and use the following
post-processing algorithm to detect and save these additional instructions into
the ground truth after we finish the first pass of ground truth generation,
i.e., after the set of instructions expressed in the instruction syntax in the
LSTs has been recorded.
We define the set of conservative successors of an instruction $i$ to be the set
of instructions that our system \emph{knows} where the control flow \emph{may}
continue at.
Our algorithm keeps track of the set of conservative successors for every
instruction in the LSTs.
For every instruction $i$, this set starts empty and it is computed with the
following steps.

First, if $i$ is a non-control-flow instruction or a direct conditional jump,
our algorithm adds the instruction at the offset following $i$ to the set.
Second, if $i$ is a direct jump (both conditional and unconditional), our
algorithm adds the instruction at the target address to the set.
Third, if $i$ is an indirect jump (both conditional and unconditional; note the
latter of which includes the return instruction), our algorithm records the
empty set because our system currently does not perform indirect jump resolution
during ground truth generation.
Finally, if $i$ is any other control-flow instruction that our current system
supports (e.g., \texttt{call}), our algorithm records the empty set.
We remark that if $i$ is a control-flow instruction not supported by our current
system (e.g., \texttt{syscall}), our algorithm would have erroneously treated
$i$ as a non-control-flow instruction back in the first step above.
Incidentally, when this happens in practice, the set of correct conservative
successors computed is often actually correct.

After computing the set of conservative successors for every instruction
discovered from the LSTs, our algorithm processes these sets as follows in an
attempt to discover as many as-data instructions as it can.
For every instruction $j$ in a conservative successor set, if $j$ is already
recorded in the ground truth data, then our algorithm will skip over it.
Otherwise, our algorithm starts a conservative recursive traversal (RT) from
$j$, which is a RT that uses the conservative successor relation described
above.
During this conservative RT, our algorithm will record every
previously-unrecorded instruction in the RT output as an optional instruction in
the ground truth.
In addition, we also modified our accuracy evaluator to cope with these optional
instructions by not counting any true positive or false negative that involves
an optional instruction.

\begin{figure}[tbp]
  \centering
  \includegraphics[width=0.5\columnwidth]{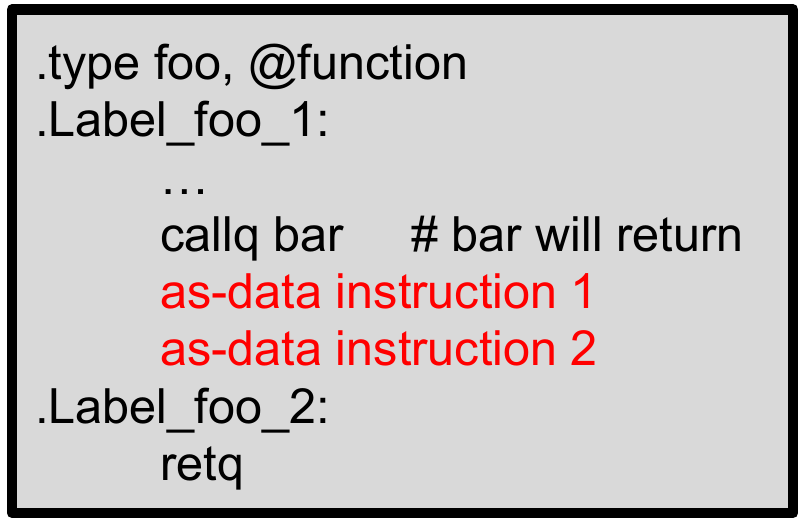}
  \caption{%
    An example of a failure to discover as-data instructions with our current
    solution.
  }
  \label{fig:gt:asdata}
\end{figure}

Our current solution is an effective albeit technically-flawed heuristic and we
are aware of a specific class of failure scenarios that we have identified from
our benchmark suite.
In particular, it may fail to discover any sequence of as-data instruction(s)
that starts at the target address of an indirect jump.
The reason is because our conservative RT does not continue past an indirect
jump (see the third step above), and therefore an as-data instruction $j$ that
is the successor of an indirect jump $i$ will not be discovered unless $j$ is
discovered as a conservative successor of an instruction other than $i$.
As a result of not discovering $j$, all as-data instructions that immediately
follow $j$ will also not get discovered.
A simplified example of this is given in Figure~\ref{fig:gt:asdata}.
In the example, as-data instructions 1 and 2 will not be discovered by our
algorithm because the only predecessor of these two instructions are
respectively the return instruction in the function ``bar'' and instruction 1
and thus our conservative RT cannot reach these two instructions.
Additional examples can also be given using other types of indirect jumps such
as implementations of switch tables.

At present, we believe the effect of this weakness in our current solution is
small on our evaluation because
\begin{enumerate*}[(i)]

  \item all optional instructions discovered in our current benchmark suite are
  in fact \nop instructions used for alignments, and

  \item we have observed as-data instructions only with ICC at high optimization
  levels (\verb!-O2!, \verb!-O3!, and \verb!-Ofast!).

\end{enumerate*}
When presenting the evaluation in~\S\ref{sec:gt-gen-method}, we will give an
over-approximation on the number of possible ``false false-positives'' by
counting the number of known \nop instructions that appear as false positives in
the binaries produced by ICC at high optimization levels.

We leave it as future work to improve our current solution.
For example, we observe that some compilers output a list of target addresses
for some indirect jumps in the assembly comment.
A future system may parse these comments to extend the conservative RT, thereby
increasing the set of instructions that can be discovered.

\subsubsection{Instruction Prefix Problem}
\label{sec:gt:prefix}

Both x86 and x64 ISAs feature the concept of an instruction prefix, which refers
to a piece of machine code that augments the effect of the next instruction in
the instruction stream.
For our purpose, complexity arises because a single instruction can have
multiple prefixes and these prefixes can come in any order.
For example, a repeat store instruction with a $16$-bit operand size \texttt{rep
  stosw} could be encoded as either ``0x66 0xf3 0xab'' or ``0xf3 0x66 0xab'',
where ``0x66'' specifies the operand size and ``0xf3''' specifies the repeat
operation in \texttt{rep stos[m16|m32]}.
Different assemblers and disassemblers may use different policies in regards to
whether a prefix is an instruction by itself or is part of the next instruction.
Since there is no standard in this, if the compiler used to generate a binary
adopts one prefix policy (which affects how prefixes are represented in our
ground truth data) and a disassembler adopts a different policy, the
disassembler accuracy evaluation could result in many unjustified mismatches.

To solve this problem, we decided to adopt a uniform prefix policy and
canonicalize all outputs during both ground truth generation and evaluation.
The policy we chose is ``most coarse-grained'', meaning our ground truth records
all prefixes to an instruction $i$ as an unordered set attached to $i$ and our
accuracy evaluator uses set comparison to check the equality of the prefixes of
each instruction.
Our reason to prefer this policy is because compiler toolchains and
disassemblers can emit multiple prefixes in any order and therefore we need to
implement the set comparison semantics even if each prefix is to be treated as a
separate instruction.

\subsubsection{Duplicated Function Name Problem}
\label{sec:gt:dupfunc}

\begin{figure}[tbp]
  \centering
  \includegraphics[width=\columnwidth]{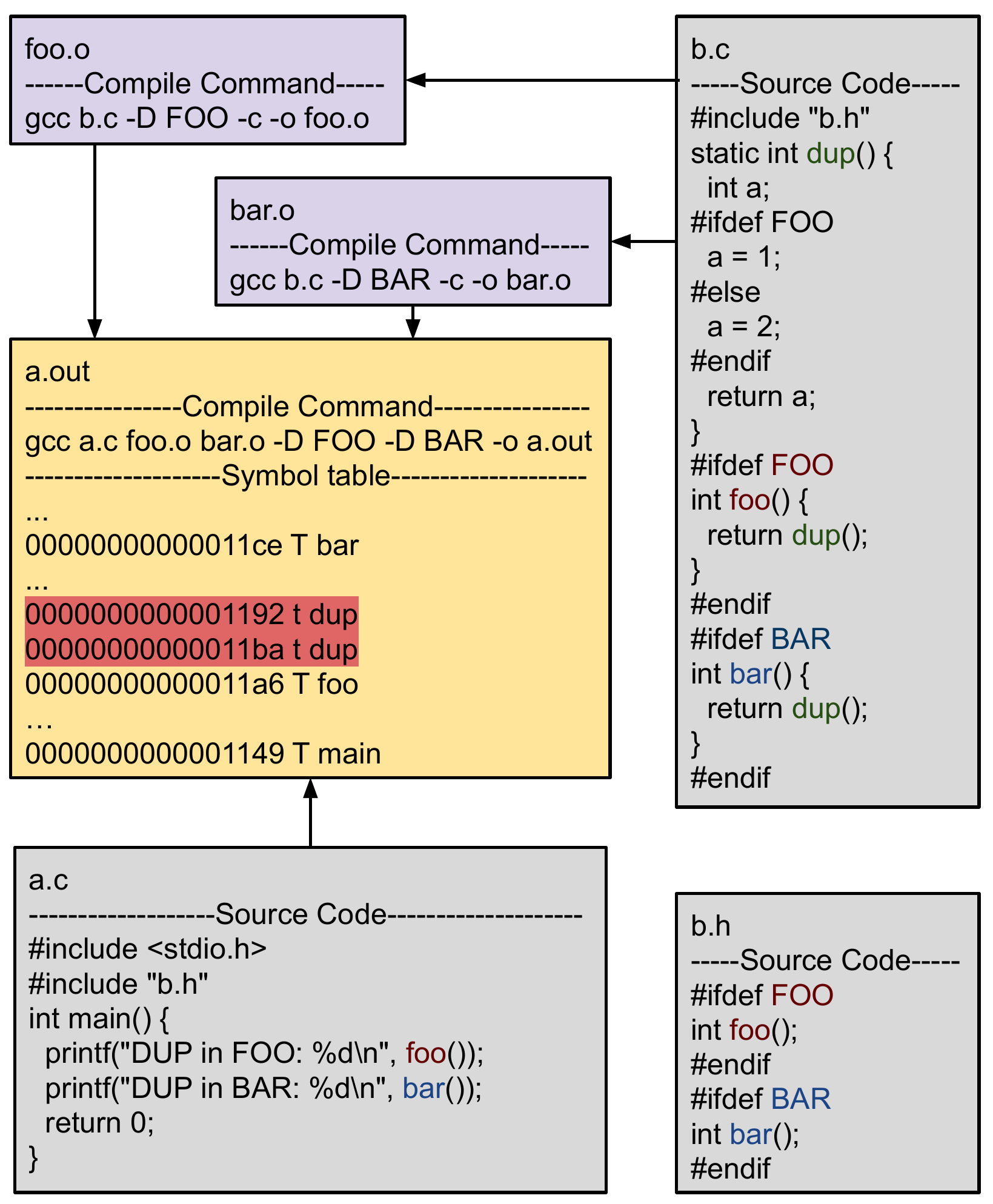}
  \caption{%
    An example where a function name ``dup'' appears twice in the symbol table
    of a binary.
  }
  \label{fig:gt:dup}
\end{figure}

In C, a function declared in a header file and included in multiple source files
can be compiled into functions with potentially different content but with the
same name in the symbol table.
We have observed this in our benchmark suite and it is due to unexported
functions that contain \texttt{\#ifdef}s getting compiled under different
\texttt{\#define}s in different source files.
Therefore, our ground truth generator at times needs to figure out a matching
between identically-named functions in a binary and their counterparts in the
LSTs.

Figure~\ref{fig:gt:dup} is a simplified example that shows how to make a
function name appear twice in the symbol table of a binary.
Three functions ``dup'', ``foo'', and ``bar'' are defined in b.c and the
functions ``foo'' and ``bar'' are both called from the ``main'' function in a.c.
First, we compile b.c into two object files foo.o and bar.o using different
\texttt{\#define}s.
Notice that foo.o does not contain ``bar'', bar.o does not contain ``foo'', and
both object files contain an unexported (static) function ``dup'' even though
their ``dup'' functions have different bodies due to the \texttt{\#ifdef} in its
source.
Then, we compile a.c and link it with foo.o and bar.o to produce a.out.
As shown in the figure, the symbol ``dup'' appears twice as functions with
different offsets in the symbol table of a.out.

Our solution to solve the duplicated function name problem is to match functions
greedily as follows.
Our ground truth generator first groups all functions in the LSTs into
equivalence classes by their names, with the understanding that every member of
each class needs to be matched with an actual function of the same name in the
binary.
Then, for every function in the binary, our system greedily identifies the first
byte-by-byte identical function from the equivalence class of that name and
declares it to be the match.
This takes advantage of the fact that if two functions in a binary have
identical names and identical contents, then it does not matter how they are
matched to their LST counterparts.

\subsubsection{Overwritten Assembly File Problem}
\label{sec:gt:asmoverwritten}

Some build processes that generate multiple build targets may compile the same
source file multiple times, often with different headers and/or macros.
This was a setback for our approach initially.
This is because some compilers by default name the generated assembly file after
the first source file in the command line, and so a build process with the above
characteristic could result in the situation where some assembly files have
already been overwritten if we wait till the end of the build process to capture
the assembly files for LST generation.

Our solution to this problem is to build a system that aims to record every
assembly file that was generated during a build.
In our current implementation, we use compiler wrapper scripts that call Git
after every compiler invocation and let Git track the content of the entire
build directory over time.
After the build finishes, the recorded Git history is processed by one of our
programs to extract every generated assembly file in chronological order.

%% file: impldetail.table.mc.tex
\setlength{\tabcolsep}{2pt}
\begin{table*}[tb]
  \caption{Supported x86/x64 Instructions That Admit Multiple Encodings}
  \footnotesize
  \begin{center}
    \begin{tabular}{|l l|l l|l|}
      \hline
      \multicolumn{2}{|l|}{\bf Encoding 1} & \multicolumn{2}{l|}{\bf Encoding 2} & \bf Explanation                                                                                                                 \\ \hline
      jmp                                  & rel8                                & jmp                & rel32                   & Jump to a relative location that is within {[}-128, 127{]}.                      \\ \hline
      jcc                                  & rel8                                & jcc                & rel32                   & Conditionally jump to a relative location that is within {[}-128, 127{]}.        \\ \hhline{|==|==|=|}
      BOP$^{\mathrm{a}}$                   & AL, imm8                            & BOP$^{\mathrm{a}}$ & r8, imm8                & Binary operation on an imm8 and the register AL.                                 \\ \hline
      BOP$^{\mathrm{a}}$                   & AX, imm16                           & BOP$^{\mathrm{a}}$ & r16, imm16              & Binary operation on an imm16 and the register AX.                                \\ \hline
      BOP$^{\mathrm{a}}$                   & EAX, imm32                          & BOP$^{\mathrm{a}}$ & r32, imm32              & Binary operation on an imm32 and the register EAX.                               \\ \hline
      BOP$^{\mathrm{a}}$                   & RAX, imm32                          & BOP$^{\mathrm{a}}$ & r64, imm32              & Binary operation on an imm32 and the register RAX.                               \\ \hhline{|==|==|=|}
      BOP$^{\mathrm{a}}$                   & r16, imm16                          & BOP$^{\mathrm{a}}$ & r16, imm8               & Binary operation on a 16-bit register and an imm within {[}-128, 127{]}.         \\ \hline
      BOP$^{\mathrm{a}}$                   & r32, imm32                          & BOP$^{\mathrm{a}}$ & r32, imm8               & Binary operation on a 32-bit register and an imm within {[}-128, 127{]}.         \\ \hline
      BOP$^{\mathrm{a}}$                   & r64, imm32                          & BOP$^{\mathrm{a}}$ & r64, imm8               & Binary operation on a 64-bit register and an imm within {[}-128, 127{]}.         \\ \hhline{|==|==|=|}
      imul                                 & r16, r/m16, imm16                   & imul               & r16, r/m16, imm8        & Signed-multiply of a 16-bit register/address with an imm within {[}-128, 127{]}. \\ \hline
      imul                                 & r32, r/m32, imm32                   & imul               & r32, r/m32, imm8        & Signed-multiply of a 32-bit register/address with an imm within {[}-128, 127{]}. \\ \hline
      imul                                 & r64, r/m64, imm32                   & imul               & r64, r/m64, imm8        & Signed-multiply of a 32-bit register/address with an imm within {[}-128, 127{]}. \\ \hhline{|==|==|=|}
      SHF$^{\mathrm{b}}$                   & r8, 1                               & SHF$^{\mathrm{b}}$ & r8, imm8                & Shift an 8-bit register by 1 bit.                                                \\ \hline
      SHF$^{\mathrm{b}}$                   & r16, 1                              & SHF$^{\mathrm{b}}$ & r16, imm8               & Shift a 16-bit register by 1 bit.                                                \\ \hline
      SHF$^{\mathrm{b}}$                   & r32, 1                              & SHF$^{\mathrm{b}}$ & r32, imm8               & Shift a 32-bit register by 1 bit.                                                \\ \hline
      SHF$^{\mathrm{b}}$                   & r64, 1                              & SHF$^{\mathrm{b}}$ & r64, imm8               & Shift a 64-bit register by 1 bit.                                                \\ \hhline{|==|==|=|}
      mov                                  & AL, moffs8$^{\mathrm{c}}$           & mov                & r8, m8$^{\mathrm{c}}$   & Move 8-bit data between the register AL and a memory address.                    \\ \hline
      mov                                  & AX, moffs16$^{\mathrm{c}}$          & mov                & r16, m16$^{\mathrm{c}}$ & Move 16-bit data between the register AX and a memory address.                   \\ \hline
      mov                                  & EAX, moffs32$^{\mathrm{c}}$         & mov                & r32, m32$^{\mathrm{c}}$ & Move 32-bit data between the register EAX and a memory address.                  \\ \hhline{|==|==|=|}
      \multicolumn{5}{|r|}{\begin{tabular}[c]{@{}r@{}}$^{\mathrm{a}}$\,BOP represents the general binary operations as represented by Intel~\cite[V1 \S5.1.2 and V1 \S5.1.4]{misc:intel}, \\
          including \textrm{add}, \textrm{adc}, \textrm{sub}, \textrm{sbb}, \textrm{cmp}, \textrm{and}, \textrm{or}, and \textrm{xor}.\end{tabular}}
      \\\hline
      \multicolumn{5}{|r|}{\begin{tabular}[c]{@{}r@{}}$^{\mathrm{b}}$\,SHF represents the shifting operations as represented by Intel~\cite[V1 \S5.1.5]{misc:intel}, \\
          including \textrm{sar}, \textrm{shr}, and \textrm{sal/shl}\end{tabular}}
      \\\hline
      \multicolumn{5}{|r|}{\begin{tabular}[c]{@{}r@{}}$^{\mathrm{c}}$\,The register and the memory address are exchangeable, meaning that both the instructions that \\
          move data from a register to the memory and from the memory to a register support multiple encodings.\end{tabular}}
      \\\hline
    \end{tabular}
  \end{center}
  \label{tbl:impl:mc:insn}
\end{table*}

%% file: evaluation.tex
\section{Evaluation}
\label{sec:eval}

In this section, we will first present our benchmark suite
(\S\ref{sec:current-benchmark}) and then our evaluation to the following
questions:
\begin{enumerate}[leftmargin=*,label=RQ\arabic*:]

  \item\label{en:eva:qs:gt} Can our ground truth generator work with various
  compilers and optimization settings and generate ground truth data on our
  current benchmark suite? Are there surprises? (\S\ref{sec:gt-gen-method})

  \item\label{en:eva:qs:char} What are the characteristics of the ground truth
  data generated using our current benchmark suite? (\S\ref{sec:gt-gen-char})

  \item\label{en:eva:qs:cmp} What is the accuracy of the four included
  open-source disassemblers according to our ground truth? (\S\ref{sec:eva:acc})

\end{enumerate}

\subsection{Our Benchmark Suite}
\label{sec:current-benchmark}

\begin{table}[t]
  \centering
  \caption{Projects and Binaries in Our Benchmark Suite}
  \label{tbl:eva:gt:pkg}
  \footnotesize
  \begin{tabular}{|l|r|c|c|c|}
    \hline
    \multicolumn{1}{|c|}{Project} & \multicolumn{1}{c|}{Version} & Binary    & Linux & Windows \\ \hline
    7zip~\cite{misc:7zip}         & 19.00                        & 7zDec     & Y     & Y       \\ \hline
    capstone~\cite{misc:capstone} & 4.0.2                        & cstool    & Y     & Y       \\ \hline
    exim~\cite{misc:exim}         & 4.86                         & exim      & Y     &         \\ \hline
    lighttpd~\cite{misc:lighttpd} & 1.4.39                       & lighttpd  & Y     &         \\ \hline
    mit-bzip2~\cite{misc:mit4}    & (2006-01-11)                 & bzip2     & Y     & Y       \\ \hline
    mit-gcc~\cite{misc:mit4}      & (2006-01-11)                 & gcc       & Y     &         \\ \hline
    mit-gzip~\cite{misc:mit4}     & (2006-01-11)                 & gzip      & Y     &         \\ \hline
    mit-oggenc~\cite{misc:mit4}   & (2006-01-11)                 & oggenc    & Y     &         \\ \hline
    nginx~\cite{misc:nginx}       & 1.8.0                        & nginx     & Y     &         \\ \hline
    openssh~\cite{misc:openssh}   & 7.1p2                        & sshd      & Y     &         \\ \hline
    pcre2~\cite{misc:pcre}        & 10.35                        & pcre2grep & Y     & Y       \\ \hline
    putty~\cite{misc:putty}       & 0.73                         & putty     &       & Y       \\ \hline
    sqlite~\cite{misc:sqlite}     & 3.30.1                       & sqlite3   & Y     & Y       \\ \hline
    vim~\cite{misc:vim}           & 8.2.0821                     & vim       & Y     & Y       \\ \hline
    vsftpd~\cite{misc:vsftpd}     & 3.0.3                        & vsftpd    & Y     &         \\ \hline
  \end{tabular}
\end{table}

\paragraph{Projects.}
Our current benchmark suite comprises $15$ open-source projects and so far we
have considered the x86 and x64 ISAs only.
Table~\ref{tbl:eva:gt:pkg} shows detailed information of the included projects,
the target binary in each project, and whether that project is included in our
Linux / Windows sub-suite.
Admittedly our current suite is biased towards Linux due to its history.
Its initial composition includes the five Linux programs used
in~\cite{Andriesse+etal:2016:in-depth} and then we added four MIT-produced
amalgamations of common Unix programs, one of which happens to be compilable on
Windows.
On top of these, we added five commonly-recognizable projects that support both
Linux and Windows (7zip, capstone, pcre2, sqlite, vim) and one that supports
Windows-only (putty).
We must caution that we anticipate the membership of our suite will change over
time due to community inputs or future investigations.
In particular, we believe it would be a very interesting scientific study on how
to put together a ``best'' benchmark suite in view of the competing goals to
control the number of included binaries and to increase the complexity exhibited
by these binaries.

\paragraph{Toolchains \& Settings.}
On Linux, we support GCC v5.4.0 and v7.5.0, Clang v3.8.0 and v6.0.0, and ICC
v19.1.1.219.
For these compilers, we support six settings: \verb!-O0!, \verb!-O1!,
\verb!-O2!, \verb!-O3!, \verb!-Ofast!, and \verb!-Os!.
On Windows, we support MSVC v19.26.28806 with three settings: \verb!/Od!,
\verb!/O1!, and \verb!/O2!.
The versions of GCC and Clang used are the ones distributed in Ubuntu 16.04 LTS
and 18.04 LTS and the versions of ICC and MSVC are both the latest as of
2020-07-01.
The GCC and Clang versions we used are slightly newer than the ones used
in~\cite{Andriesse+etal:2016:in-depth} because currently we can afford to
support only LTS.
We leave it as future work to use our ground truth generator on the older
compiler versions used in~\cite{Andriesse+etal:2016:in-depth} and to measure the
accuracy of their released ground truth.

\subsection{Evaluating Our Ground Truth Generator}
\label{sec:gt-gen-method}

We have run our ground truth generator over our benchmark suite to ensure
compatibility and robustness.
Our generator is bundled as a set of programs and Docker images and is written
with scripting in mind.
Specifically, for each project, our system expects a project directory with
specific subdirectory names and specific build scripts at hard-coded locations.
Given such a directory, one of our programs will produce a Docker image (for
full reproducibility) and run the image to obtain an archive containing the
build artifacts in our special Git-based format.
Finally, we launch another Docker image to process the collected data and
produce the ground truth files.

With our current suite, all compilations succeeded with $3$ exceptions: ICC
failed to compile mit-gcc with \verb!-O2!, \verb!-O3!, and \verb!-Ofast! into
x86 ELF. %
The error message is ``\textrm{internal error: 04010022\_1238}'' and we have
already reported this bug to
Intel.\footnote{\url{https://community.intel.com/t5/Intel-C-Compiler/Failed-to-compile-the-MIT-amalgamated-gcc-c-into-a-x86-ELF/m-p/1196443}}
In total, we obtained
\begin{enumerate*}[(i)]
  \item $420$ x64 Linux ELF,
  \item $417$ x86 Linux ELF,
  \item $21$ x64 Windows COFF, and
  \item $21$ x86 Windows COFF.
\end{enumerate*}
All generated ground truth passed the check described in~\S\ref{sec:gt:check}.

Unfortunately, our current algorithm to gather ``optional instructions''
(\S\ref{sec:gt:dri}) can miss any instruction targeted by an indirect jump if
the target instruction is encoded as data by the compiler.
When we performed the disassembler accuracy evaluation that will be presented
in~\S\ref{sec:eva:acc}, we observed this happening with ICC at \verb!-O2!,
\verb!-O3!, and \verb!-Ofast!, which accounted for under $1500$ and $3000$
``false false-positives'' in our x86 and x64 sub-suites respectively.
So far, this is the only source of error that we are aware of in our ground
truth data and we are investigating how to fix this.

\subsection{Characteristics of Our Benchmark Suite}
\label{sec:gt-gen-char}

\input{evaluation.stat.tex}

Table~\ref{tbl:eva:char} presents the characteristics of the binaries and their
associated ground truth data files in each of the ISA-OS pairs we currently
support.
The first two rows show the total download sizes for users who trust us and do
not want to regenerate the benchmark binaries and the ground truth data from
scratch.

\paragraph{Statistics.}
The binaries in our benchmark suite has a wide range in size---from
\XsfELFMinBin~KB to \XsfELFMaxBin~KB.
To give a sense of scale/complexity from the perspective of a disassembler, we
also counted the number of functions/instructions/indirect jumps and, using
LLVM-MC v8.0.0, the number distinct mnemonics in these binaries.
We stress that the number of indirect jumps is only a proxy to estimate the
actual complexity faced by a disassembler because the resolution complexity of
each indirect jump can vary greatly.

\paragraph{Code-Data Interleave.}
The last row of Table~\ref{tbl:eva:char} was generated by checking whether a
linear-sweep disassembler achieves $100\percent$ recall and precision against
our ground truth.
We performed this experiment to repeat part of the study by Andriesse
\etal~\cite{Andriesse+etal:2016:in-depth} and we arrived at the same conclusion:
modern Linux compilers rarely create interleaving code and data.
In the machine code due to functions present in the source code of our Linux
sub-suite, we did not discover any code-data interleave and therefore a
linear-sweep disassembler would achieve perfect disassembly when compared with
our ground truth.
However, we note that code-data interleaves actually exist in some
highly-optimized (\verb!-O2!, \verb!-O3!, \verb!-Ofast!) ICC-generated x64
binaries in our suite.
The reason they do not result in disassembly errors here is because they are due
to functions from statically-linked libraries (e.g.,
\verb!__intel_mic_avx512f_memcpy!) and our current ground truth data and our
evaluator both do not account for instructions in this category.
On the other hand, we observed that MSVC in both x86 and x64 can generate
code-date interleaves where data is present at the end of some functions,
including those appearing in the source code.
Since linear sweep does not know where a function ends, it outputted many false
positives (disassembling data as code) and false negatives (at mismatched
instruction boundaries).
Finally, we remark that code-data interleaves happen even in modern Linux
binaries; see, e.g.,~\cite[Figure~1]{Miller+etal:2019:probabilistic}.
Also, we are aware that code-data interleaves can be common in other ISAs such
as ARM. %
Therefore, we believe continued research in advanced disassembly algorithms (as
opposed to settling on linear sweep) is warranted.

\subsection{Evaluating The Included Disassemblers}
\label{sec:eva:acc}

We have used our benchmark suite and our own custom scripts to evaluate four
prominent open-source disassemblers:
\begin{enumerate*}[(i)]
  \item BAP v2.1.0 (2020-05-29),
  \item Ghidra v9.1.2 (2020-02-12),
  \item Radare2 v4.4.0 (2020-04-13), and
  \item ROSE v0.10.4.3 (2020-05-05).
\end{enumerate*}
While we specifically excluded commercial disassemblers such as IDA Pro and
Binary Ninja due to licensing and their limit to API access in their free
versions, we believe these vendors can publish their own numbers using our
benchmark suite to enable comparisons.

\paragraph{Unstripped Binaries.}
As strange as this may sound, for this paper we tested the included
disassemblers with the \emph{unstripped} version of the binaries from our suite
to simulate an experiment where we use stripped binaries but provide each
disassembler with perfect function starts.
Even though disassembly is arguably much more often conducted on stripped
binaries and our simulation is not without caveats, there are two reasons behind
this decision:
\begin{enumerate*}[(i)]

  \item After an initial testing with various stripped binaries, we discovered
  that we did not want to test function start identification (FSI), which as
  identified in~\cite{Andriesse+etal:2016:in-depth} is a much less well-solved
  challenge in disassembly.
  For example, the instruction recall of Ghidra on the unstriped x86 vim ELF
  binary compiled with ICC \verb!-O2! is $99.059\percent$, and stripping drops
  this to $81.429\percent$.

  \item Our current workstation has $128$GB of memory and it proves to be
  insufficient for the stripped experiment without heavily relying on swapping.
  For example, when running ROSE on the x86 mit-gcc binary compiled by Clang
  v3.8.0 with \verb!-Ofast!, ROSE consumed over $125$GB of memory and was soon
  killed by the OOM-killer when the stack-delta analysis stage was at
  $88\percent$.

\end{enumerate*}
We believe an interesting future work would be to provide FSI hints to each
disassembler and then also test them with stripped binaries.

\paragraph{Invocations.}
For fairness, we followed the documentation of each disassembler on how to run
it.
Even though we recognize that expert users may run a disassembler with various
non-default flags and/or third-party plugins, we feel our method better mimics
the experience of a typical user.
With \verb!${BIN}! denoting an input binary and \verb!${RST}! denoting the
output file, our method to run the disassemblers were:
\begin{enumerate*}[(i)]

  \item BAP: \verb!bap ${BIN} -d asm >${RST}!;

  \item Ghidra: we run \texttt{analyzeHeadless} with a Java program we bundled
  to output all instruction offsets using \verb!currentProgram.getListing().!
  \verb!getInstructions(true)! without changing any other configuration;

  \item Radare2: \verb!r2 -Aqc 'pdr @@f >${RST}' ${BIN}!;

  \item ROSE: \verb!rose-recursive-disassemble ${BIN} >${RST}!.

\end{enumerate*}

\paragraph{Crashes.}
We observed a number of crashes in our experiment.
In particular, Radare2 seg-faulted on $11$ binaries, which we noticed were all
large-size binaries including cstool, mit-gcc, nginx, and vim compiled by
various compilers.
In these cases, we decided to consider the Radare2 output as empty in our
accuracy evaluation and we have already filed a bug with the Radare2
developers.\footnote{\url{https://github.com/radareorg/radare2/issues/17388}}

\paragraph{Accuracy.}
Using our bundled wrappers for the included disassemblers, we collected their
outputs on each binary and computed their true positive, false positive, and
false negative counts against our ground truth.
These counts are summarized into the common Receiver Operating Characteristic
(ROC) metrics of precision, recall, and F1 scores.
In this paper, we divide the binaries into groups according to the compiler and
optimization setting used and by the ISA-OS pair.
Since we have $5$ compilers and $6$ optimization settings in Linux, there are
$30$ groups in each ISA-Linux pair and there are $14$ binaries inside each of
these groups (see Table~\ref{tbl:eva:gt:pkg}).
Similarly, the corresponding number for each ISA-Windows pair is $1 \times 3 =
  3$ and each of these groups contains $7$ binaries.

To summarize the accuracy of a disassembler in a group, let $N$ be the number of
binaries in the group and define the following weights for each binary in the
group to adjust for the different number of instructions in each binary:
\begin{gather*}
  W^{Recall}_{project_i} = \frac{
    TP_i + FN_i
  }{
    \sum_{j=1}^N TP_j + FN_j
  }
  \quad\quad
  W^{Precision}_{project_i} = \frac{
    TP_i + FP_i
  }{
    \sum_{j=1}^N TP_j + FP_j
  }
\end{gather*}

With these weights, we computed the weighted precision, recall, and F1 score for
each disassembler over each group in each ISA-OS pair.
This in turn allowed us to count the number of times when a disassembler has the
highest precision/recall/F1 score in each ISA-OS pair, i.e., the number of
``wins'' in each ISA-OS pair for each ROC metric.
These numbers are presented in the left sub-column under each disassembler in
Table~\ref{tbl:eva:acc:roc}.
In addition, to provide a summary of the per-group weighted precision/recall/F1
score, we also computed the harmonic means of the per-group metrics for each
disassembler for each ISA-OS pair.
These numbers are presented in the corresponding right sub-column in the table.

\input{evaluation.tables.tex}

Adopting F1 scores as our metric for accuracy, ROSE and Ghidra are the most
accurate disassemblers for respectively the Linux and Windows binaries in our
suite.
Overall, every included disassembler achieved a high precision that exceeds
$97\percent$ in our suite, but the recall is comparably lower.
This reflects the common design choice in disassemblers where soundness (every
outputted instruction is true) is valued over completeness (every instruction is
outputted).
A more detailed version of the above analysis using our Linux binaries will be
presented in Appendix~\ref{sec:in-depth-Linux}.

\input{evaluation.perf.tex}

\paragraph{Resource Consumption.}
To compare the performance characteristics of the included disassemblers, each
disassembler invocation in our experiment was run with \verb!/usr/bin/time -v!.
Ideally, we would present the measurements with a breakdown by each binary,
similar to how SPEC CPU results are typically presented.
However, due to space, in this paper we have selected to present with sqlite
only.
Our suite supports sqlite on all four OS-ISA pairs we support and there are $66$
binaries in total ($60$ Linux + $6$ Windows).
We ran each disassembler sequentially on each binary on an otherwise-idle Intel
i9-9900K machine with $128$GB memory.
Table~\ref{tbl:eva:acc:time} shows the time and memory consumption over the
entire sequence.
For sqlite, BAP occupied the most amount of memory and ROSE used the most CPU
(User+Sys) time.
However, Ghidra and ROSE both use multi-core and BAP was in fact slower in
wall-clock (Real) time.

%% file: evaluation.stat.tex
\begin{table}[t]
  \centering
  \caption{Characteristics of Benchmark Binaries and Ground Truth Data per
    ISA-OS pair}
  \label{tbl:eva:char}
  \footnotesize
  \begin{tabular}{|c|r|r|r|r|}
    \hline
    ISA \& OS              & \multicolumn{1}{c|}{x64 Linux} & \multicolumn{1}{c|}{x86 Linux} & \multicolumn{1}{c|}{x64 Windows} & \multicolumn{1}{c|}{x86 Windows} \\ \hline
    GT Total Size (MB)     & \XsfELFGTSize                  & \XttELFGTSize                  & \XsfCOFFGTSize                   & \XttCOFFGTSize                   \\ \hline
    Total Binary Size (MB) & \XsfELFBinSize                 & \XttELFBinSize                 & \XsfCOFFBinSize                  & \XttCOFFBinSize                  \\ \hline
    Max Binary Size (KB)   & \XsfELFMaxBin                  & \XttELFMaxBin                  & \XsfCOFFMaxBin                   & \XttCOFFMaxBin                   \\ \hline
    Min Binary Size (KB)   & \XsfELFMinBin                  & \XttELFMinBin                  & \XsfCOFFMinBin                   & \XttCOFFMinBin                   \\ \hline
    \# Functions           & \XsfELFFunc                    & \XttELFFunc                    & \XsfCOFFFunc                     & \XttCOFFFunc                     \\ \hline
    \# Instructions        & \XsfELFInsn                    & \XttELFInsn                    & \XsfCOFFInsn                     & \XttCOFFInsn                     \\ \hline
    \# Indirect Jumps      & \XsfELFIndJ                    & \XttELFIndJ                    & \XsfCOFFIndJ                     & \XttCOFFIndJ                     \\ \hline
    \# Distinct Mnemonics  & \XsfELFMnem                    & \XttELFMnem                    & \XsfCOFFMnem                     & \XttCOFFMnem                     \\ \hline
    Code-Data Interleave   & \multicolumn{1}{c|}{N}         & \multicolumn{1}{c|}{N}         & \multicolumn{1}{c|}{Y}           & \multicolumn{1}{c|}{Y}           \\ \hline
  \end{tabular}
\end{table}

%% file: evaluation.tables.tex
\begin{table}
  \centering
  \caption{\#Wins \& Harmonic Means of Each ROC Metric of Each Disassembler
    (Highest F1 in Each ISA-OS Shaded)}
  \label{tbl:eva:acc:roc}
  \footnotesize
  \begin{tabular}{|c|c|r|r|r|r|r|r|r|r|}
    \hline
    \multicolumn{2}{|c|}{Disassemblers} & \multicolumn{2}{c|}{\bap} & \multicolumn{2}{c|}{\ghidra} & \multicolumn{2}{c|}{\radare} & \multicolumn{2}{c|}{\rose}                                                                                                                                                            \\ \hline
    \multirow{3}{*}{\rotatebox{90}{\parbox{2.5em}{x64 Linux}}}
                                        & Recall                    & \LnxXsfWinBapRecall          & \LnxXsfAvgBapRecall          & \LnxXsfWinGhiRecall        & \LnxXsfAvgGhiRecall                   & \LnxXsfWinRdaRecall    & \LnxXsfAvgRdaRecall    & \LnxXsfWinRseRecall    & \LnxXsfAvgRseRecall                   \\ \cline{2-10}
                                        & Prec                      & \LnxXsfWinBapPrecision       & \LnxXsfAvgBapPrecision       & \LnxXsfWinGhiPrecision     & \LnxXsfAvgGhiPrecision                & \LnxXsfWinRdaPrecision & \LnxXsfAvgRdaPrecision & \LnxXsfWinRsePrecision & \LnxXsfAvgRsePrecision                \\ \cline{2-10}
                                        & F1                        & \LnxXsfWinBapFone            & \LnxXsfAvgBapFone            & \LnxXsfWinGhiFone          & \LnxXsfAvgGhiFone                     & \LnxXsfWinRdaFone      & \LnxXsfAvgRdaFone      & \LnxXsfWinRseFone      & \cellcolor{green!25}\LnxXsfAvgRseFone \\ \hhline{|=|=|=|=|=|=|=|=|=|=|}
    \multirow{3}{*}{\rotatebox{90}{\parbox{2.5em}{x86 Linux}}}
                                        & Recall                    & \LnxXttWinBapRecall          & \LnxXttAvgBapRecall          & \LnxXttWinGhiRecall        & \LnxXttAvgGhiRecall                   & \LnxXttWinRdaRecall    & \LnxXttAvgRdaRecall    & \LnxXttWinRseRecall    & \LnxXttAvgRseRecall                   \\  \cline{2-10}
                                        & Prec                      & \LnxXttWinBapPrecision       & \LnxXttAvgBapPrecision       & \LnxXttWinGhiPrecision     & \LnxXttAvgGhiPrecision                & \LnxXttWinRdaPrecision & \LnxXttAvgRdaPrecision & \LnxXttWinRsePrecision & \LnxXttAvgRsePrecision                \\  \cline{2-10}
                                        & F1                        & \LnxXttWinBapFone            & \LnxXttAvgBapFone            & \LnxXttWinGhiFone          & \LnxXttAvgGhiFone                     & \LnxXttWinRdaFone      & \LnxXttAvgRdaFone      & \LnxXttWinRseFone      & \cellcolor{green!25}\LnxXttAvgRseFone \\  \hhline{|=|=|=|=|=|=|=|=|=|=|}
    \multirow{3}{*}{\rotatebox{90}{\parbox{2.5em}{x64 Windows}}}
                                        & Recall                    & \WinXsfWinBapRecall          & \WinXsfAvgBapRecall          & \WinXsfWinGhiRecall        & \WinXsfAvgGhiRecall                   & \WinXsfWinRdaRecall    & \WinXsfAvgRdaRecall    & \WinXsfWinRseRecall    & \WinXsfAvgRseRecall                   \\  \cline{2-10}
                                        & Prec                      & \WinXsfWinBapPrecision       & \WinXsfAvgBapPrecision       & \WinXsfWinGhiPrecision     & \WinXsfAvgGhiPrecision                & \WinXsfWinRdaPrecision & \WinXsfAvgRdaPrecision & \WinXsfWinRsePrecision & \WinXsfAvgRsePrecision                \\  \cline{2-10}
                                        & F1                        & \WinXsfWinBapFone            & \WinXsfAvgBapFone            & \WinXsfWinGhiFone          & \cellcolor{green!25}\WinXsfAvgGhiFone & \WinXsfWinRdaFone      & \WinXsfAvgRdaFone      & \WinXsfWinRseFone      & \WinXsfAvgRseFone                     \\  \hhline{|=|=|=|=|=|=|=|=|=|=|}
    \multirow{3}{*}{\rotatebox{90}{\parbox{2.5em}{x86 Windows}}}
                                        & Recall                    & \WinXttWinBapRecall          & \WinXttAvgBapRecall          & \WinXttWinGhiRecall        & \WinXttAvgGhiRecall                   & \WinXttWinRdaRecall    & \WinXttAvgRdaRecall    & \WinXttWinRseRecall    & \WinXttAvgRseRecall                   \\  \cline{2-10}
                                        & Prec                      & \WinXttWinBapPrecision       & \WinXttAvgBapPrecision       & \WinXttWinGhiPrecision     & \WinXttAvgGhiPrecision                & \WinXttWinRdaPrecision & \WinXttAvgRdaPrecision & \WinXttWinRsePrecision & \WinXttAvgRsePrecision                \\  \cline{2-10}
                                        & F1                        & \WinXttWinBapFone            & \WinXttAvgBapFone            & \WinXttWinGhiFone          & \cellcolor{green!25}\WinXttAvgGhiFone & \WinXttWinRdaFone      & \WinXttAvgRdaFone      & \WinXttWinRseFone      & \WinXttAvgRseFone                     \\  \hline
  \end{tabular}
\end{table}

%% file: evaluation.perf.tex
\begin{table}[t]
  \centering
  \caption{Time \& Memory Consumption of Each Disassembler Over All $66$ sqlite3
    Binaries (\{x86, x64\} $\times$ \{ELF, COFF\})}
  \label{tbl:eva:acc:time}
  \footnotesize
  \begin{tabular}{|c|r|r|r|r|}
    \hline
    Disassemblers         & \multicolumn{1}{c|}{\bap} & \multicolumn{1}{c|}{\ghidra} & \multicolumn{1}{c|}{\radare} & \multicolumn{1}{c|}{\rose} \\ \hline
    Total User Time (min) & \SqlUsrTimeBap            & \SqlUsrTimeGhi               & \SqlUsrTimeRda               & \SqlUsrTimeRse             \\
    \hline
    Total Sys Time (min)  & \SqlSysTimeBap            & \SqlSysTimeGhi               & \SqlSysTimeRda               & \SqlSysTimeRse             \\
    \hline
    Total Real Time (min) & \SqlClkTimeBap            & \SqlClkTimeGhi               & \SqlClkTimeRda               & \SqlClkTimeRse             \\
    \hline
    Max Resident Mem (KB) & \SqlSpaceBap              & \SqlSpaceGhi                 & \SqlSpaceRda                 & \SqlSpaceRse               \\
    \hline
    Use Multi-core        & \multicolumn{1}{c|}{N}    & \multicolumn{1}{c|}{Y}
                          & \multicolumn{1}{c|}{N}    & \multicolumn{1}{c|}{Y}                                                                   \\
    \hline
  \end{tabular}
\end{table}

%% file: conclusion.tex
\section{Concluding Remarks}
\label{sec:concluding}

In this paper, we proposed to start standardizing the set of binaries used for
future disassembler evaluations and presented our work-in-progress benchmark
suite.
We presented our ground truth generator and evaluated four prominent open-source
disassemblers using our ground truth data.
Our project is in active development in multiple directions.
Aside from fixing the algorithm used in~\S\ref{sec:gt:dri}, we believe the
most important future work is to investigate what binaries should be included in
the benchmark suite.
Even limiting to C programs only, we would like to more formally define and
increase the complexity captured by the benchmark binaries.
For example, while our ground truth representation allows the presence of
overlapping instructions, none of the programs in our current suite contains
this feature.
Other future work includes:
\begin{enumerate*}[(i)]

  \item adding the ability to provide function start hints to each included
  disassembler,
  \item adding instructions from statically-linked libraries and data
  declarations to the ground truth,
  \item adding ICC support on Windows,
  \item adding projects from other languages (e.g., C++), and
  \item adding other ISAs (e.g., ARM).

\end{enumerate*}
We sincerely hope the community will provide us with inputs or even pull
requests to evolve our benchmark suite into a community standard for future
disassembler evaluations.

\paragraph{Acknowledgement.}
This work was supported by ONR Award No. N00014-17-1-2892.
Any opinions, findings, and conclusions or recommendations expressed in this
material are those of the authors and do not necessarily reflect the views of
the Office of Naval Research.
In addition, we thank the anonymous reviewers for their many suggestions and
feedback.

%% file: appendix.tex
\section{In-depth Evaluation Using Our Linux Benchmark Suite}
\label{sec:in-depth-Linux}

In~\S\ref{sec:eva:acc}, we presented Table~\ref{tbl:eva:acc:roc} to summarize
the precision, recall, and F1 scores of each included disassembler on our entire
benchmark suite by projecting these metrics into the sub-space defined by ISA-OS
pairs.
In this section, we provide additional analysis on the accuracy of each included
disassembler by projecting their F1 scores into several other sub-spaces that
are of potential interest.
Note that, as indicated in the section heading, we have opted to limit the
analysis in this section to only the Linux binaries in our benchmark suite.
Our justification is as follows:
\begin{enumerate}[(i)]

  \item As indicated in Table~\ref{tbl:eva:gt:pkg}, our current benchmark suite
  have a large discrepancy in the set of projects across the two included OSes.
  Therefore, any comparison of disassembler accuracy that mixes our benchmark
  binaries for different OSes incurs a substantial risk of bias due to the
  differences rooted in the project discrepancy.

  \item We have considered limiting ourselves to the $6$ projects shared across
  the two OSes in our current suite.
  However, we feel this number is too small.

\end{enumerate}

\begin{figure}[t]
  \centering
  \includegraphics[width=\columnwidth]{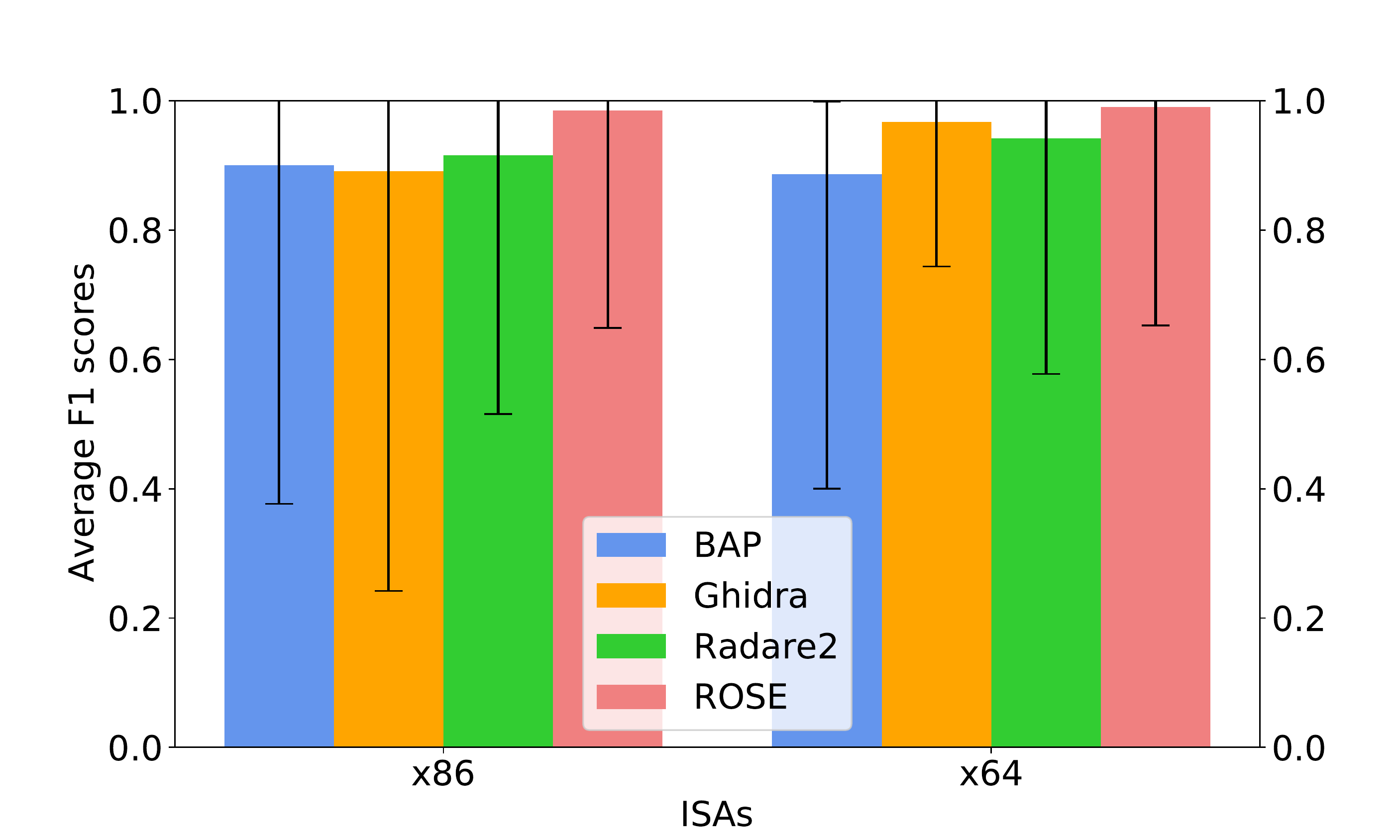}
  \caption{Average F1 Scores of Included Disassemblers on Our Linux Benchmark
    Binaries, Grouped by ISAs.}
  \label{fig:app:byisa}
\end{figure}

\begin{figure}[t]
  \centering
  \includegraphics[width=\columnwidth]{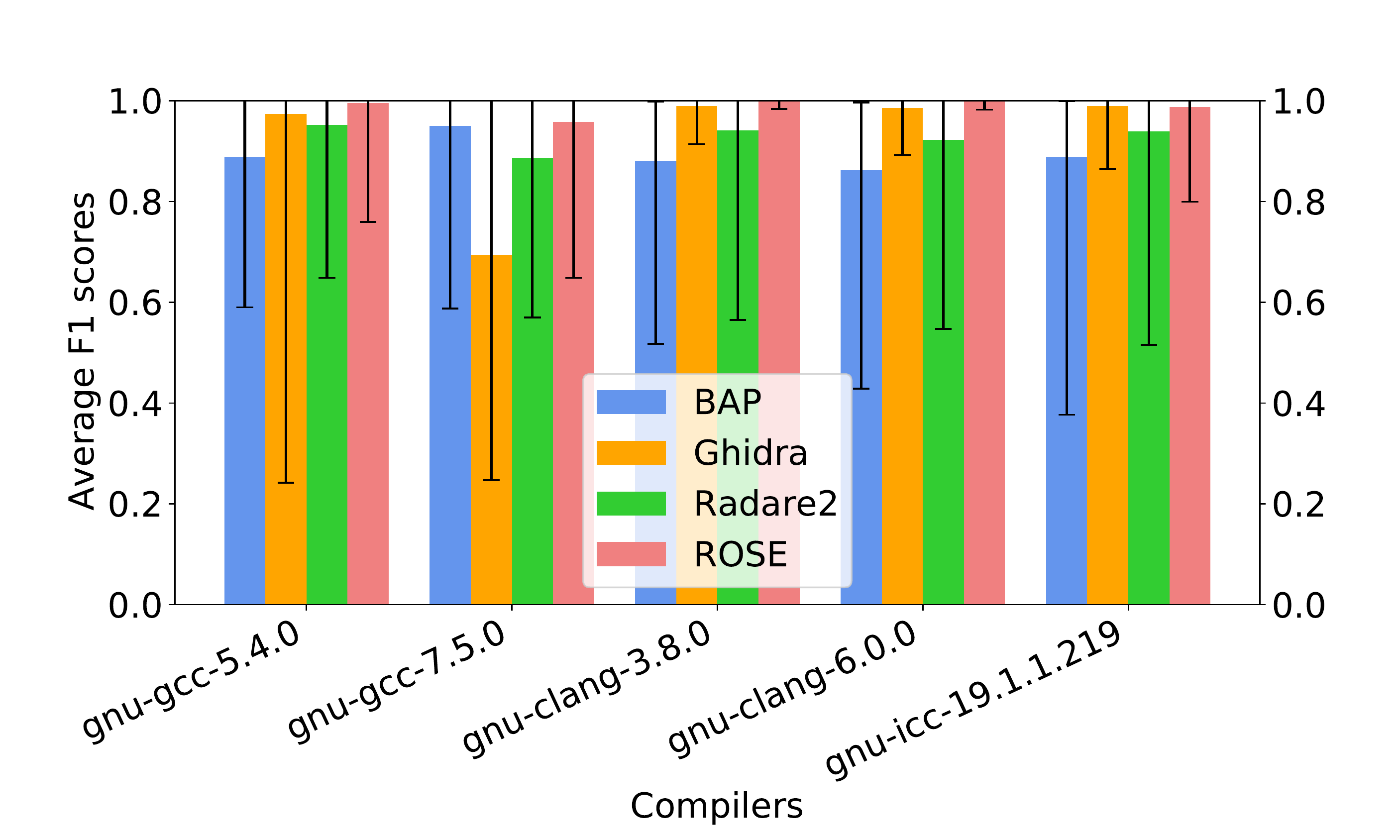}
  \caption{Average F1 Scores of Included Disassemblers On Our Linux Benchmark
    Binaries, Grouped by Compilers.}
  \label{fig:app:bycpl}
\end{figure}

\begin{figure*}[t]
  \centering
  \includegraphics[width=\textwidth]{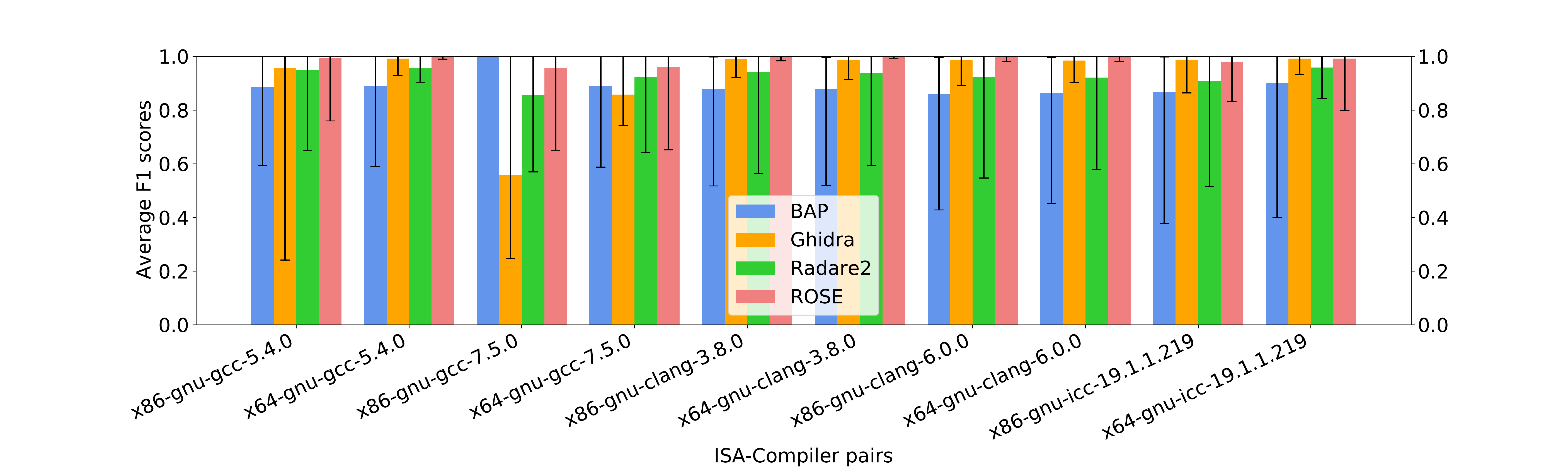}
  \caption{Average F1 Scores of Included Disassemblers on Our Linux Benchmark
    Binaries, Grouped by ISA-Compiler Pairs.}
  \label{fig:app:bycplisa}
\end{figure*}

\subsection{Groupings}
\label{sec:in-depth-Linux-groupings}

To start, we consider four different groupings of the binaries: by ISAs, by
compilers, by optimization levels, and by projects.
Within a grouping (e.g., by ISAs), we summarize the F1 scores of each
disassembler on the binaries in each group (e.g., x86 and x64) using the
weighted arithmetic mean, with the weight being the number of ground truth
instructions in the binaries in the group.
Figures~\ref{fig:app:byisa},~\ref{fig:app:bycpl},~\ref{fig:app:byopt},
and~\ref{fig:app:bypkg} are the corresponding bar charts that show these
summarized scores along with error bars to indicate the minimum
non-zero\footnote{%
  As we explained in~\S\ref{sec:eva:acc}, the version of Radare2 we tested
  crashes on $11$ of our benchmark binaries and we consider the output of
  Radare2 on these binaries to be empty.
  However, to make the error bars in the figures in this section more useful, we
  have excluded these zero values before we plot.
} scores and the maximum scores attained by each disassembler in each specific
group.
In addition, we also consider the grouping by ISA-compiler pairs, which yields
Figure~\ref{fig:app:bycplisa}, because it illustrates one of the major findings
in our analysis below.

\subsection{Major Findings}
\label{sec:in-depth-Linux-findings}

Below we will present and discuss each of our major findings in its own section.

\subsubsection{Ghidra Performs Poorly Against x86 Linux Binaries Compiled With
  GCC v7.5.0}
\label{sec:app:acc:ghi}

As seen in Figure~\ref{fig:app:byisa}, Ghidra is highly accurate on our x64
Linux binaries, with an average F1 score close to that of the forerunner ROSE.
In addition, it a highest minimum F1 score among the four included
disassemblers.
However, if we move to consider our x86 Linux binaries, the average F1 score of
Ghidra drops to become the lowest and, unlike the other three disassemblers,
Ghidra has a very noticeable drop in its minimum F1 score.
By inspecting Figures~\ref{fig:app:bycpl} and~\ref{fig:app:bycplisa}, we
discovered that the inaccuracy in the Ghidra output can be attributed to
binaries from the two included versions of GCC.
For our GCC v5.4.0 binaries, we were able to verify that Ghidra only outputted
inaccurate results in just a few binaries; therefore, while the minimum F1 score
of Ghidra in this group is severely affected, the corresponding average F1 score
of Ghidra remains high.
But for our GCC v7.5.0 binaries, we observed that Ghidra missed a large number
of instructions and that was why the average F1 score of Ghidra drops below
$0.7$ in this group.

According to our investigation, the low accuracy of Ghidra on our GCC v7.5.0
binaries can be attributed mostly to its failure in resolving switch cases in
these binaries.
In particular, we observed that the implementation of switch in GCC v7.5.0
binaries often involves separating the relevant instructions in a switch into
different basic blocks, with unrelated instructions appearing in-between.
Among the included compilers, this pattern is unique to GCC v7.5.0 and it
appears to stumble the version of Ghidra tested.
Since Ghidra has a low success rate in retrieving all cases in this switch
implementation, it misses a large amount of instructions in our x86 Linux
binaries that were compiled with GCC v7.5.0.
Furthermore, since the reason behind this phenomenon is not specific to the
binaries in our benchmark suite, we believe Ghidra would likely suffer a lower
accuracy against any real-world x86 Linux binaries compiled with GCC v7.5.0.

\subsubsection{All Four Included Disassemblers Achieve Similar Accuracy Across
  Every Included Compiler Except GCC v7.5.0}
\label{sec:app:acc:cpl}

One interesting phenomenon revealed by our analysis is that the accuracy of
every included disassembler on our benchmark binaries does not vary greatly
across the included compilers except for GCC v7.5.0.
This can be seen in Figure~\ref{fig:app:bycpl}.
We observe that, if we exclude GCC v7.5.0, varying the compiler among the other
four compilers affects the average F1 score of an included disassembler by less
than $\pm 0.05$.
As for GCC v7.5.0, we see the average F1 scores of Ghidra, Radare2, and ROSE all
drop with larger amounts while that of BAP actually rises.

Based on our current investigation, we hypothesize that the aforementioned F1
score variations are likely best explained by the difference in the switch case
handling capability of the included disassemblers due to the following reasons.
First, we have not identified any significant difference in any other aspect in
the code generated by the included compilers.
Second, we observed that the two included versions of Clang use almost the same
switch table implementations as GCC v5.4.0, where the switch cases are short and
contiguous.
Third, even though the included version of ICC uses a different implementation
than the above three compilers, the switch cases generated by this version of
ICC remain contiguous.
Fourth, as has been discussed in~\S\ref{sec:app:acc:ghi}, GCC v7.5.0 is an
exception here because its switch implementation can be discontiguous.
Combining these reasons, it is conceivable that disassemblers that assume switch
cases to be contiguous would achieve a lower accuracy on binaries compiled with
GCC v7.5.0 than on binaries compiled with the other four included compilers.
As for the relative increase in accuracy of BAP on our GCC v7.5.0 binaries, we
hypothesize it is likely because BAP has been tuned with binaries generated by
GCC v7.5.0 or other close-by versions that share the same switch implementation
strategy.
Therefore, its accuracy would drop on binaries compiled with the other four
included compilers due to the substantially different strategy.
An interesting future work would be to study the validity of the two hypotheses
above.

\subsubsection{All Four Included Disassemblers Achieve Similar Accuracy Across
  Every Included Optimization Level}
\label{sec:app:acc:opt}

\begin{figure}[t]
  \centering
  \includegraphics[width=\columnwidth]{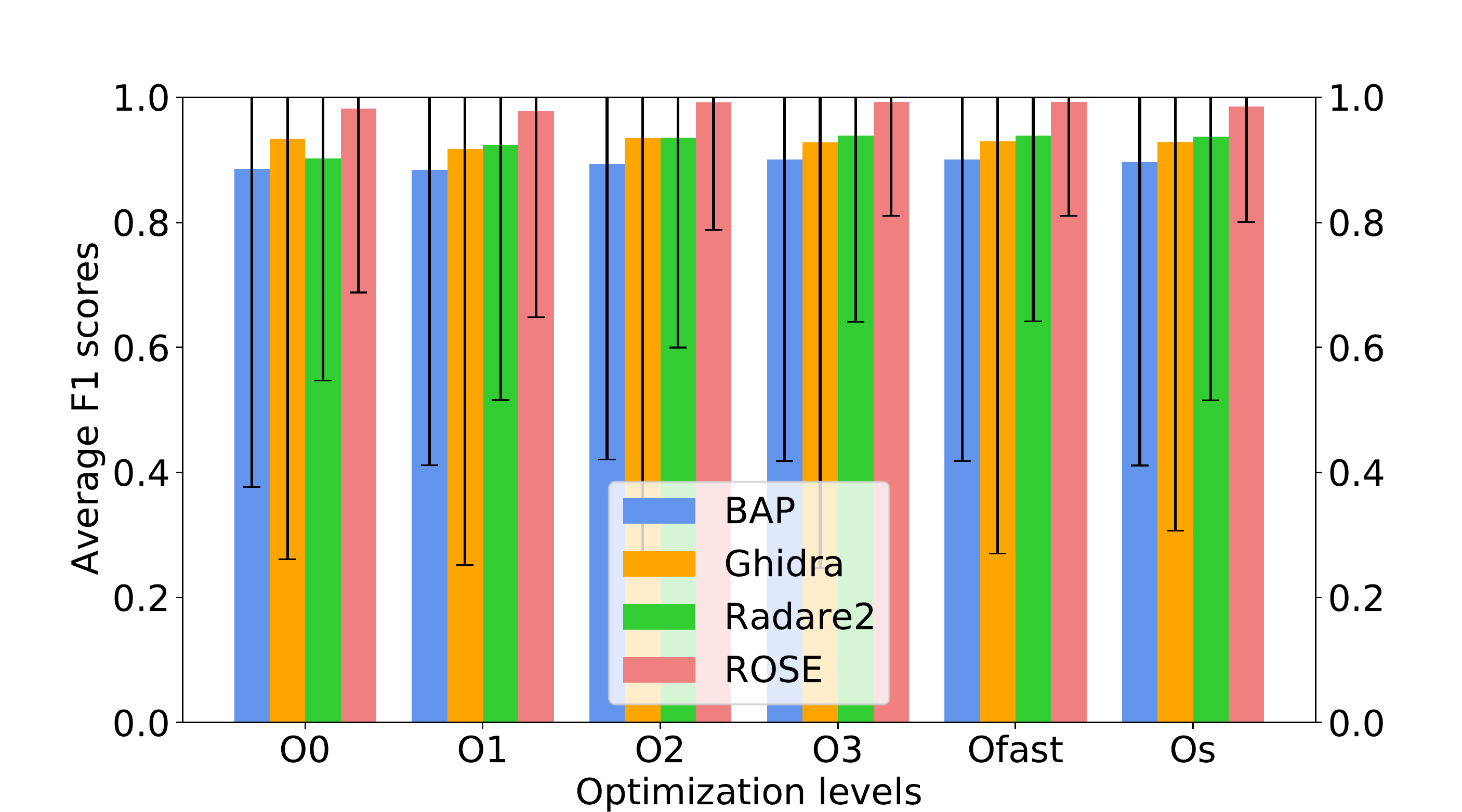}
  \caption{Average F1 Scores of Included Disassemblers on Our Linux Benchmark
    Binaries, Grouped by Optimization Levels.}
  \label{fig:app:byopt}
\end{figure}

Aside from varying the compiler, we have also investigated how the optimization
level used during compilation affects disassembly accuracy.
A common wisdom in binary analysis is that optimized binaries can be more
difficult to analyze than unoptimized binaries because optimized binaries may
contain more complex code constructs.
However, while these constructs have the potential to severely reduce
disassembly accuracy, they also appear to be rare in real-world binaries
according to the study by Andriesse \etal~\cite{Andriesse+etal:2016:in-depth}.
With our current benchmark binaries, we observed that the average F1 score of
each of the four included disassemblers fluctuates within $\pm 0.05$ when
varying across the six included optimization levels.
This can be seen in Figure~\ref{fig:app:byopt}.
In addition, we also observed that the minimum F1 scores of BAP, Radare2, and
ROSE on our \verb!-O0! binaries are actually lower than the corresponding scores
on our \verb!-O2! binaries.
This is interesting because one may expect the opposite trend to occur and we
believe a more thorough investigation to explain the above fluctuations should
be conducted.

\subsubsection{Binaries from the Capstone Project Stumble All Four Included
  Disassemblers}
\label{sec:app:acc:pkg}

\begin{figure*}[tb]
  \centering
  \includegraphics[width=\textwidth]{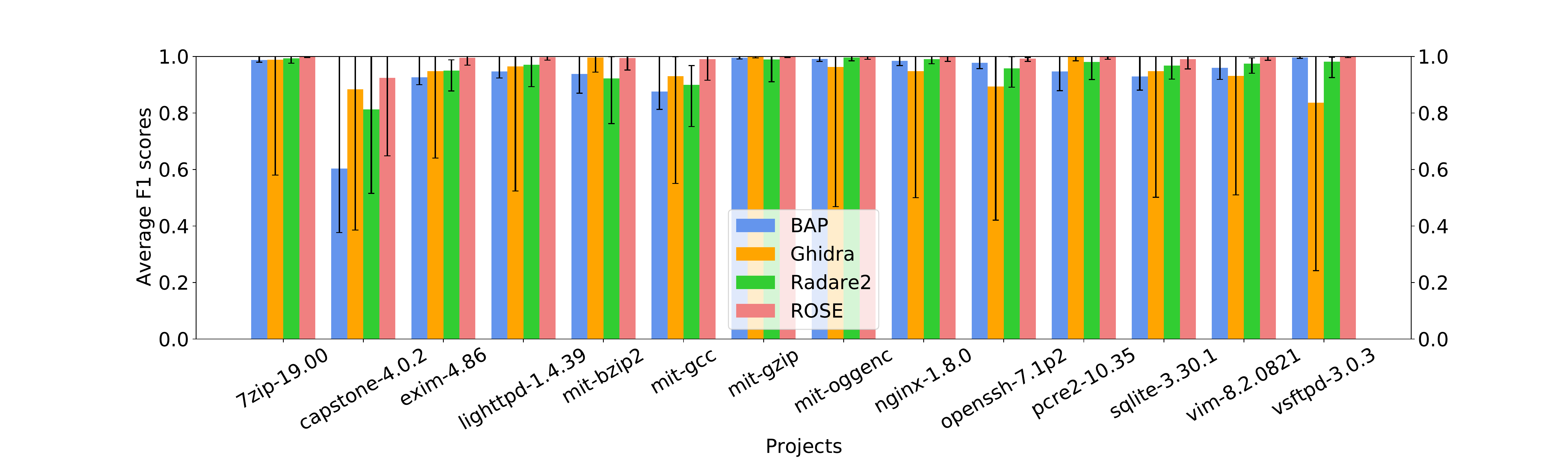}
  \caption{Average F1 Scores of Included Disassemblers on Our Linux Benchmark
    Binaries, Grouped by Projects.}
  \label{fig:app:bypkg}
\end{figure*}

Finally, we have also investigated disassembly accuracy from the perspective of
the projects in our current benchmark.
Figure~\ref{fig:app:bypkg} shows the average F1 scores of the included
disassemblers.
From the figure, we made the following observations:

(1) The average F1 scores of every included disassembler are distinctively low
on capstone.
In fact, disassembling the capstone binaries leads to the lowest average F1
scores for BAP, Radare2, and ROSE and the second lowest for Ghidra.
Since we already know indirect jump resolution can substantially affect
disassembly accuracy, our first attempt to investigate our observation is to
look into the number of indirect jumps in the capstone binaries.
\input{appendix.table.pkgindj.tex}
Table~\ref{tbl:app:indj} shows the total number of indirect jumps in each
project in our Linux benchmark binaries.
As it turns out, the capstone binaries actually do not contain significantly
more indirect jumps than binaries from the other included projects.
Unfortunately, since at present we lack the tools to automatically study the
indirect jumps appearing in the benchmark binaries more deeply, we do not know
whether some of these indirect jumps are the primary source that causes the low
disassembly accuracy in the capstone binaries.

(2) We noticed that Ghidra achieves the lowest minimum F1 score on many of the
included projects even though its corresponding average F1 score is competitive
against the other disassemblers.
This is consistent with the observation in~\S\ref{sec:app:acc:ghi}, where we
showed that Ghidra stumbles on our x86 binaries compiled with GCC v7.5.0 more
often than the other included disassemblers.
In contrast, Figure~\ref{fig:app:bypkg} lets us observe that BAP stumbles only
on the capstone binaries even though BAP also achieves some of the lowest
minimum F1 scores in Figures~\ref{fig:app:byisa}, \ref{fig:app:bycpl},
and~\ref{fig:app:bycplisa}.

%% file: appendix.table.pkgindj.tex
\begin{table}[t]
  \centering
  \caption{Total Number of Indirect Jumps in Each Project in Our Linux Benchmark
  Binaries.}
  \label{tbl:app:indj}
  \footnotesize
  \begin{tabular}{|c|r|r|}
    \hline
    Packages        & \multicolumn{1}{c|}{x86-Linux} & \multicolumn{1}{c|}{x64-Linux} \\ \hline
    7zip-19.00      & 3,709                          & 3,704                          \\ \hline
    capstone-4.0.2  & 11,128                         & 11,137                         \\ \hline
    exim-4.86       & 5,250                          & 8,865                          \\ \hline
    lighttpd-1.4.39 & 2,517                          & 2,646                          \\ \hline
    mit-bzip2       & 1,273                          & 1,401                          \\ \hline
    mit-gcc         & 69,992                         & 140,748                        \\ \hline
    mit-gzip        & 176                            & 448                            \\ \hline
    mit-oggenc      & 1,779                          & 2,383                          \\ \hline
    nginx-1.8.0     & 10,751                         & 11,235                         \\ \hline
    openssh-7.1p2   & 5,884                          & 7,254                          \\ \hline
    pcre2-10.35     & 126                            & 187                            \\ \hline
    sqlite-3.30.1   & 68,051                         & 72,963                         \\ \hline
    vim-8.2.0821    & 43,851                         & 56,011                         \\ \hline
    vsftpd-3.0.3    & 786                            & 675                            \\ \hline
  \end{tabular}
\end{table}